\documentclass[a4paper,10pt,
colorlinks,urlcolor=black,linkcolor=black,citecolor=black,
]{article}
\usepackage{amsmath,amssymb}
\usepackage[dvips]{color}
\usepackage{amsmath,amssymb}
\usepackage[dvips]{color}
\usepackage{eepic}
\usepackage{epic}
\usepackage{amsmath}
\usepackage{amssymb}
\usepackage{amscd}
\usepackage{amsthm}
\usepackage{ascmac}
\usepackage{fancybox}
\usepackage[dvipdfmx,%
 bookmarks=true,%
 bookmarksnumbered=true,%
 colorlinks=true,%
 setpagesize=false,%
 pdfkeywords={TeX; dvipdfmx; hyperref; color;}]{hyperref}
\def\roman{\rm}
\def\bigstimes{\boxtimes}
\def\Cal{\cal}
\def\SD{\;{}^\circ \text{C}}
\def\cases{\left\{\begin{array}{ll}}
\def\endcases{\end{array}\right.}
\def\bigtimes{\mathop{\mbox{\Large $\times$}}}
\def\FND{\scriptscriptstyle{\roman{EXA}}}

\def\cases{\left\{\begin{array}{ll}}
\def\endcases{\end{array}\right.}
\def\text{\mbox}
\def\roman{\rm}
\def\bigtimes{\mathop{\mbox{\Large $\times$}}}
\def\Cal{\cal}
\def\LL{{{\lq\lq}}}
\def\RR{{{\rq\rq}}}
\def\bold{\bf}
\def\text{\mbox}
\def\BBBC{{\Bbb C}}
\def\BA{\begin{align*}}
\def\EA{\end{align*}}

\def\bigtimes{\mathop{\mbox{\Large $\times$}}}

\def\bigstimes{\text{\large $\: \boxtimes \,$}}
\def\FIN{{\roman{(exa)}}}
\def\EXI{{\roman{(exi)}}}
\newcommand{\qp}{\overset{{\rm qp}}{\pmb{\pmb\times}}}

\begin{document}
%\ssetcounter{page}{35}
%\twocolumn[
\vskip0.5cm
\centerline{\LARGE \bf 
Measurement Theory in the Philosophy of Science
%The Linguistic Interpretation of Quantum Mechanics}
%\\
%\centerline{\LARGE \bf in Quantum Philosopy
%%nterpretation of quantum mechanics
%
}
\vskip0.5cm
\begin{center}
{\rm
\large
Shiro Ishikawa
}
\\
\vskip0.2cm
\rm
%\email{ishikawa@math.keio.ac.jp }
\it
Department of Mathematics, Faculty of Science and Technology,
Keio University, 
\\
3-14-1 Hiyoshi, Kohoku-ku, Yokohama, 223-8522 Japan ({E-Mail:
ishikawa@math.keio.ac.jp})
\\
\end{center}
\par
\rm
\vskip0.1cm
\par
\noindent
%{\bf Abstract}
{}
\normalsize
\par
%Abstract:
The philosophy of science is a discipline concerning
the metaphysical aspect of science.
Recently, I proposed measurement theory,
which is characterized as the metaphysical and linguistic interpretation
of quantum mechanics.
I assert
that
this theory is 
one of the most fundamental languages in science,
%scientific language as well as the world-view,
and thus,
it
is located at the central position in science. 
%should play a central role
%in
%science
%as well as the philosophy of science.
This assertion will be examined throughout this preprint,
which is written as the draft of my future book
(concerning the philosophy of science).
Hence,
I hope to hear various opinions about this draft.
%This is the draft of our future book.
\par
\vskip0.3cm
\par
\noindent
%{\bf Keywords}: the Copenhagen interpretation, quantum and classical measurement theory, 
%the law of large numbers,
%maximum likelihood 
%estimation,
%Kolmogorov extension theorem,
%wavefunction collapse,
%Bell's inequality
\par
\vskip1.0cm
\par
%]
\def\LL{\lq\lq}
\def\RR{\rq\rq {$\;$}}

\par
\noindent
{\bf
\large
Guide to read this preprint}:
%%\section{
\par \noindent
Contents are as follows.
\begin{align*}
\left.\begin{array}{ll}
{}
\text{Chap. 1: Measurement theory born from quantum mechanics}
& \text{\pageref{Chap1}}
%\pageref{}
\\
\text{Chap. 2 :{Axiom${}_{\text{\scriptsize c}}^{\text{\scriptsize p}}$ 1
---
{{measurement}}}}
& \pageref{Chap2}
\\
\text{Chap. 3 :{From Quantum Mechanics to Measurement Theory}}
& \pageref{Chap3}
\\
\text{Chap. 4 :{Fisher {statistics}\ I}}
& \pageref{Chap4}
\\
\text{Chap. 5 :{{{Practical Logic}}}}
& \pageref{Chap5}
\\
\text{Chap. 6 :Axiom${}_{\text{\scriptsize c}}^{\text{\scriptsize pm}}$ 2 - causality}  
& \pageref{Chap6}
\\
\text{Chap. 7 :Fisher {statistics}\ II}
& \pageref{Chap7}
\\
\text{Chap. 8 :Reconsideration of traditional philosophies
in measurement theory}
& \pageref{Chap8}
\\
\text{Chap. 9 :Equilibrium statistical mechanics}
& \pageref{Chap9}
\\
\text{Chap. 10:Axiom${}_{\text{\scriptsize b}}^{\text{\scriptsize p}}$ 1
---
{{measurement  (bounded type)}}}
& \pageref{Chap10}
\\
\text{Chap. 11:Axiom${}_{\text{\scriptsize b}}^{\text{\scriptsize pm}}$ 2 - causality  (bounded type)}
& \pageref{Chap11}
\\
\text{Chap. 12:Realistic world-view and Linguistic world-view}
& \pageref{Chap12}
\\
\text{Chap. 13:Conclusions}
& \pageref{Chap13}
\end{array}\right.
\end{align*}

\par
\noindent
Measurement theory is composed of two  axioms as follows.
%formulated as follows:
\begin{align*}
\underset{\text{\scriptsize (scientific language)}}{\text{{} $\fbox{{{measurement theory}}}$}}
:=
{
%\overset{\text{\scriptsize [Axiom 1\textcolor{black}{(Sec. \REF{2secAxiom 1})}]}}
\overset{\text{\scriptsize [Axiom
%${}_{\text{\scriptsize c}}^{\text{\scriptsize p}}$
 1]}}
{
\underset{\text{\scriptsize
[probabilistic interpretation]}}{\text{{} $\fbox{{{measurement}}}$}}}
}
+
{
%\overset{\text{\scriptsize [Axiom 2\textcolor{black}{(Sec.\REF{6secAxiom 2})}]}}
\overset{\text{\scriptsize [Axiom
%${}_{\text{\scriptsize c}}^{\text{\scriptsize p}}$
 2]}}
{
\underset{\text{\scriptsize [{{the Heisenberg picture}}]}}
{\text{{}$\fbox{ causality }$}}
}
}
\end{align*}
And it has the following classification:
\par
\noindent
$\;\;$
$\underset{\text{\scriptsize Chap. 1}}{\text{(Y)}}$
$
%\underset{(quantum mechanics)}
\underset{\text{\scriptsize (scientific language)}}{\text{
%\footnotesize
{{{measurement theory}}}}}
%{\footnotesize {{{measurement}}}}
\cases
\underset{}{\text{
%\footnotesize
quantum {{{measurement}}}}}{\textcircled{\scriptsize 6}}:
\text{\footnotesize \textcolor{black}{{Chap. 3}}}
\\
\underset{}{\text{
%\footnotesize
classical {{{measurement}}}}}
%\footnotemark
\cases
\!\!\!
\text{\small continuous}
\cases
\!\!
\text{\footnotesize pure type}{\textcircled{\scriptsize 1}}
%\longleftarrow
:
\text{\textcolor{black}{\footnotesize (Chaps. 2$\text{--}$9)}}\\
\!\!
\text{\footnotesize mixed type}
{\textcircled{\scriptsize 2}}:
\text{\footnotesize
\textcolor{black}{{{{}}}{Sec.4.4}}
}
\\
\endcases
\\
\\
\!\!\!
\text{\small bounded}
\cases
\!\!
\text{\footnotesize pure type}{\textcircled{\scriptsize 3}}
:
\text{\textcolor{black}{\footnotesize Chaps. 10, 11}}\\
\!\!
\text{\footnotesize mixed type}{\textcircled{\scriptsize 4}}
:
\text{\footnotesize
\textcolor{black}{{Notes }10.4, 11.3}
}
\\
\endcases
\endcases
\endcases
$
%\END{itemize}

%AAAAAAAAAAAAAAAAAAAAAAAAAAAAAAAAAAA
\newpage

For the section in which each axiom is explained, see the following table:
\begin{table}[h]%%b h(here) t p
\begin{center}
\begin{tabular}{l|l|l}
               &  Axiom 1 &$\quad$ Axiom 2\\
\hline
%%%%%
{\textcircled{\scriptsize 1}}:continuous${{\cdot}}$pure &  \ {Sec.$\;$}2.2 (\textcolor{black}{Axiom${}_{\text{\scriptsize c}}^{\text{\scriptsize p}}$} 1)     &$\quad$  \ {Sec.$\;$}6.4	(\textcolor{black}{Axiom${}_{\text{\scriptsize c}}^{\text{\scriptsize pm}}$} 2)   	\\
%%%%%
{\textcircled{\scriptsize 2}}:continuous${{\cdot}}$mixed &  \ {Sec.$\;$}4.4  (\textcolor{black}{Axiom${}_{\text{\scriptsize c}}^{\text{\scriptsize m}}$} 1)      &$\quad$ 	\ ditto			\\
%%%%%
{\textcircled{\scriptsize 3}}:bounded${{\cdot}}$pure &  {Sec.$\;$}10.1  (\textcolor{black}{Axiom${}_{\text{\scriptsize b}}^{\text{\scriptsize m}}$} 1)       &$\quad$  {Sec.$\;$}11.3    (\textcolor{black}{Axiom${}_{\text{\scriptsize b}}^{\text{\scriptsize pm}}$} 2)     \\
%%%%%
{\textcircled{\scriptsize 4}}:bounded${{\cdot}}$mixed &  {Sec.$\;$}10.4   (\textcolor{black}{Axiom${}_{\text{\scriptsize b}}^{\text{\scriptsize p}}$} 1)      &$\quad$ 	ditto			\\
%%%%%
{\textcircled{\scriptsize 5}}:trial         &  \ {Sec.$\;$}4.1  (\textcolor{black}{Axiom(T)}1)    &$\quad$ 	many opinions			\\
%%%%%%
{\textcircled{\scriptsize 6}}:quantum      &  \ {Sec.$\;$}3.1  (\textcolor{black}{Axiom(Q)}1)      &$\quad$ 	(cf. \cite {Neum})			\\
\end{tabular}
%%\caption{Axiom}
%\BEGIN{center}
%	0.1: Axiom
%\END{center}
\end{center}
\end{table}
In this preprint,
we mainly devote ourselves to classical pure measurement theory
({\textcircled{\scriptsize 1}} and {\textcircled{\scriptsize 3}}).
\par\noindent

%111111111111111111111111111111111111111111

%\newpage

\vskip0.5cm
\noindent
\par
\noindent
%\ssection{quantum mechanics and %\textcolor{black}{NZ}{{{measurement theory}}} 
%111111111111111111111111111111111111111111
\section{{{Measurement Theory born from Quantum Mechanics}}  \label{Chap1}}  %
%%\vspace{-0.8cm}
\noindent
\begin{itemize}
\item[{}]
{
\small
\par%[Abstract].
\rm
{{Measurement theory}}
is characterized as a language
created in order to
describe ordinary phenomena numerically,
though it is oriented in quantum mechanics.
In this chapter,
the outline
of
two axioms
(i.e.,
Axiom 1 (in Chap. 2)
and
Axiom 2 (in Chap. 5)
)
will be introduced.
}
\end{itemize}

%\ssubsection{{world-description}%\textcolor{black}{NZ}
%
%quantum mechanics%\textcolor{black}{NZ}?
%}

%\cite{IIJ}
%\BEGIN{document}
\baselineskip=18pt
\subsection{Why do we start from quantum mechanics?}%{Sec.1.1}
\par
We think that readers, who read the abstract, ask the following:
\begin{itemize}
\item[({}A$_1$)]
Why is
quantum mechanics
(i.e.,
the physics concerning
the microscopic world)
applicable to
the description of
ordinary phenomena
(i.e.,
economics,
psychology,
electric engineering,
etc.
)
?
$\;\;$
Or,
why
is the quantum mechanical approach
(i.e.,
measurement theory)
indispensable?
$\;\;$
That is,
why
does
{{measurement theory}}
---
quantum mechanical language
---
hold?
$\;\;$
Further,
is there another approach?
\end{itemize}
Although the problem will be answered here and there
in this print,
our answers are not sufficient.
However,
we are convinced that
the readers,
who read the whole of this print,
agree to
the following
opinion:
\begin{itemize}
\item[({}A$_2$)]
Measurement theory
---
quantum mechanical approach
---
is most natural.
And,
there is no powerful rival
against measurement theory.
\end{itemize}
The worst understanding of
measurement theory
is
to consider
measurement theory
as
"an ass in a lion's skin".
In this preprint,
we consistently
assert
(cf. \textcolor{black}{\cite{INewi}})
that
{{measurement theory}}
is more fundamental than quantum mechanics
(cf. \textcolor{black}{{Sec.9.3}}).
%\END{document}
%EEEEEEEEEEE:::::

\subsubsection{The classical mechanical world-view}
Before
we explain the quantum mechanical world-view
(i.e.,
{{measurement theory}}
),
in this section
we discuss
the classical mechanical approach
and its problems.

The law is most essential in physics,
and moreover,
there is no physics
without the law.
However,
the language should be prepared before
the is declared.
In this sense,
physics has the following structure{:}
\begin{itemize}
\item[(B)]
$\quad
%\qquad
$
$
\text{[{physics}]}:=\text{[language]}\;+\;\text{[{law}]}
\;\;
\Big(
\;
=
%\overset{\text{\scriptsize }}{\underset{({physics})}{\text{
%\fbox{Newtonian mechanics}}}}
\overset{\text{\scriptsize (law)}}{\underset{}{\text{
\fbox{language}}}}
\;
\Big)
$
\end{itemize}
For example,
consider Newtonian mechanics
If we
do not know
how to use
the term
"causality"
(i.e.,
"time",
"space",
"velocity",
"acceleration"
and so on),
then
we can not declare
"the law of Newton kinetic equation".
That is,
Newtonian mechanics has the following structure{:}
\begin{itemize}
\item[(C)]
$\quad
%\qquad
$
$
\overset{\text{\scriptsize }}{\underset{}{\text{
\fbox{{{Newtonian}} mechanics}}}}
:=
\overset{\text{\scriptsize ({{Newton}}kinetic equation)}}{\underset{}{\text{
\fbox{how to use the term "causality"}}}}
\;
$
\end{itemize}
This means that
the linguistic aspect is
valid
without the test of the law.
Consider
{{Newton}} mechanics\textcolor{black}{({}C)}.
the
time variable differential equation
is valid
independently whether the law is true or not.
That is,
\begin{itemize}
\item[({}D)]
%$%\quad
%\qquad
%$
$
\overset{\text{\scriptsize }}{\underset{}{\text{
\fbox{Newton mechanics}}}}
\xrightarrow[\text{linguistic aspect}]{\text{without verification experiment}}
\overset{\text{\scriptsize }}{\underset{[\text{differential equation}]}{\text{\fbox{
how to use "causality"}}}}
$
\end{itemize}
And,
applying
{differential equation}
to lots of phenomena,
you get the belief
such that
\begin{itemize}
\item[]
{differential equation} is quite applicable to
sciences and engineering.
(In this preprint,
we consider that
"sciences = engineering"(cf. {Note }1.2, {Note }1.11, {Sec.8.1}(m)). ).
\end{itemize}
If you get this belief,
it may be called
{\bf {the causal world-view}}.
%index{causal world-view@{the causal world-view}}
The general form of {differential equation}
is represented by
the system of {differential equation}
(which is called the
{\bf {state equation}}):
\begin{align*}
\text{state equation}
:=
\cases
\frac{d\omega_1}{dt}{} (t)=v_1(\omega_1(t),\omega_2(t),\ldots,\omega_n(t), t)
\\
\frac{d\omega_2}{dt}{} (t)=v_2(\omega_1(t),\omega_2(t),\ldots,\omega_n(t), t)
\\
\cdots \cdots
\\
\frac{d\omega_n}{dt}{} (t)=v_n (\omega_1(t),\omega_2(t),\ldots,\omega_n(t), t)
\endcases
\tag*{{\color{black}$\displaystyle{\mathop{(1.1)}}$}}
\end{align*}
%\BEGIN{document}
\baselineskip=18pt
Therefore,
\begin{itemize}
\item[(E$_1$)]
%index{state equation@{state equation}}
%index{state equation method@{state equation method}}
{the causal world-view}
({\bf {state equation method}})
is the spirit that
every motion phenomena
should be represented by
{state equation}\textcolor{black}{(1.1)},
or
the spirit
to believe that,
if
a certain phenomenon
is represented by
{state equation}\textcolor{black}{(1.1)},
it has the causality
(cf. {Sec.6.1.2}(c))
\end{itemize}

And further,
the state equation method
acquires the concept of
"probability",
we get
%%\textcolor{black}{NZ}
%{the causal world-view}%\textcolor{black}{NZ}
%, %\textcolor{black}{NZ},
%{\bf probability }
%
%{state equation} and probability 
%
%,
%index{@{dynamical system theory}}
%index{ and @{statistics}}
\begin{itemize}
\item[(E$_2$)]
$\quad$
\bf
{dynamical system theory}(={statistics})
\end{itemize}
\rm
(Here,
we consider that
"dynamical system theory = statistics"(cf. Sec.7.1). ).
\rm
This is the
{\bf classical mechanical world-view method},
whose starting point is Newtonian mechanics.

This classical mechanical {world-view} method,
in which Newtonian mechanics is regarded as the starting point,
we have acquired a great success.
Since there is no science without
{statistics}(\:that is, {differential equation}, probability ).
the classical mechanical {world-view}
(={statistics}${{\cdot}}${dynamical system theory})
is a base supporting present age science

\par
However,
there are points of uncertainty in
the classical mechanical method.
For example,
the following problems
\textcolor{black}{(F$_1$)--(F$_5$)}
are considered:.
\begin{itemize}
\item[(F$_1$)]
What kind of theory
is statistics?
That is,
is it
"mathematics",
"mathematical method",
"applied mathematics",
"world-view",
"language"
(and so on)
?
\item[(F$_2$)]
Why
is
the concept of
"probability"
added to
the state equation method?
That is,
why
does classical world-view
have
two
birthplaces
(i.e.,
Newtonian mechanics
and
gamble)?
%
% and , %\textcolor{black}{NZ}2%\textcolor{black}{NZ}%\textcolor{black}{NZ}%\textcolor{black}{NZ},
%%\textcolor{black}{NZ}
\item[(F$_3$)]
The concept of
"probability"
could be added
to the state equation method.
If it be so,
some may want to add another concept
(e.g.,
fuzzy, chaos, etc.).
Is it possible?
%probability  and %\textcolor{black}{NZ},
%\renewcommand{\footnoterule}{%
%  \vspace{1mm}                      % %\textcolor{black}{NZ}
%  \noindent\rule{\textwidth}{0.4pt}   % %\textcolor{black}{NZ}, 
%  \vspace{-0mm}
%}
%%\textcolor{black}{NZ}%\textcolor{black}{NZ}%\textcolor{black}{NZ}?\footnote{
%, %\textcolor{black}{NZ} and ,
%%\BEGIN{align*}
%,
%%\textcolor{black}{{\cite{{Zade}}}},
%${{\cdot}}$%\textcolor{black}{NZ}
%%%\END{align*}
%.
%%\textcolor{black}{NZ}, %\textcolor{black}{NZ}{{{{h}}}} and , %\textcolor{black}{NZ},
% and  and .
%, 
% and .
%,
%%\textcolor{black}{NZ}%\textcolor{black}{NZ}
%%\textcolor{black}{NZ}{}
%%\textcolor{black}{NZ}, {statistics}${{\cdot}}${dynamical system theory}%\textcolor{black}{NZ}
%
%,
%, , 
%.
% and  and 
%% and 
%{world-description}%\textcolor{black}{NZ}%\textcolor{black}{NZ}.
%}
\end{itemize}
In other words,
\begin{itemize}
\item[(F$_4$)]
Is the development
(from the state equation method to statistics)
%{statistics}
inevitable?
\end{itemize}
which is outstanding point.
And further,
we have the following most fundamental question:
\begin{itemize}
\item[(F$_5$)]
Why is
mathematical theories
differential equation
and probability
applicable to
the description of ordinary phenomena?
\end{itemize}
%\newpage
%\par
%\noindent
%\par
%\noindent
%BBBBBBBBBBBBBBBBBB%SBSBSBS
\par
\noindent
{\small%%{\footnotesize
%\vspace{0.2cm}
\begin{itemize}
\item[$\spadesuit$] \bf {{}}{Note }1.1{{}} \rm
Since language has not fully been prepared yet, this question (F$_5$)
may be a vague expression.
But,
there is a reason to consider that
\begin{itemize}
\item[($\sharp_1$)]
the useful mathematical theory that has the reason to be useful.
That is,
the powerful mechanical world view
is hidden
behind
a useful mathematical theory.
\end{itemize}
Because
mathematic itself is independent of
world.
Thus,
{world-description method}
is indispensable.
For example,

%\BEGIN{table}[h]%%b h(here) t p
%\BEGIN{center}
%\BEGIN{tabular}{l|ll}
%%               &  mathematics & $\quad$ world-view\\
%%\hline
%%%%%%%
%%continuous${{\cdot}}$pure &  \ {Sec.$\;$}2.2 (\textcolor{black}{Axiom${}_{\text{\scriptsize c}}^{\text{\scriptsize p}}$} 1)     
%%&$\quad$  
%\ {Sec.$\;$}6.4	(\textcolor{black}{Axiom${}_{\text{\scriptsize c}}^{\text{\scriptsize pm}}$} 2)   	\\
%%%%%%
%continuous${{\cdot}}$mixed &  \ {Sec.$\;$}4.4  (\textcolor{black}{Axiom${}_{\text{\scriptsize c}}^{\text{\scriptsize m}}$} 1)      
%&$\quad$ 	\ ditto			\\
%%%%%%
%bounded${{\cdot}}$pure &  {Sec.$\;$}10.1  (\textcolor{black}{Axiom${}_{\text{\scriptsize b}}^{\text{\scriptsize m}}$} 1)       
%&$\quad$  {Sec.$\;$}11.3    (\textcolor{black}{Axiom${}_{\text{\scriptsize b}}^{\text{\scriptsize pm}}$} 2)     \\
%%%%%%
%bounded${{\cdot}}$mixed &  {Sec.$\;$}10.4   (\textcolor{black}{Axiom${}_{\text{\scriptsize b}}^{\text{\scriptsize p}}$} 1)      
%&$\quad$ 	ditto			\\
%%%%%%
%trial         &  \ {Sec.$\;$}4.1  (\textcolor{black}{Axiom(T)}1)    &$\quad$ 	many opinions			\\
%%%%%%%
%quantum      &  \ {Sec.$\;$}3.1  (\textcolor{black}{Axiom(Q)}1)      &$\quad$ 	omit			\\
%\END{tabular}
%%%\caption{Axiom}
%%\BEGIN{center}
%%	0.1: Axiom
%%\END{center}
%\END{center}
%\END{table}
%
%

\begin{table}[h]%%b h(here) t p
\begin{center}
\begin{tabular}{l|ll}
       mathematics        &  & $\quad$  world-description method \\
\hline
%%%%
\text{differential geometry}
&
&$\quad$ 
\text{the theory of relativity}
\\
\text{differential equation}
&
&$\quad$ 
\text{{{Newton}} mechanics,
electromagnetism 
}
\\
\text{Hilbert space}
&
&$\quad$
\text{quantum mechanics}
	\\
\end{tabular}
%%\caption{Axiom}
%\BEGIN{center}
%	0.1: Axiom
%\END{center}
\end{center}
\end{table}
%\END{itemize}
Problem (F$_5$)
is equivalent to the following:
\begin{itemize}
\item[($\sharp_2$)]
Behind hat kind of world-description
is probability theory?
hidden
\end{itemize}
\end{itemize}
}

%%BBBBBBBBBBBBBBBBBB%SBSBSBSS

\subsubsection{Start from quantum mechanics and not Newtonian mechanics
}%{Sec. 1.1.2}

When we start from Newtonian mechanics,
we have nuisances such as
\textcolor{black}{(F$_1$)--(F$_5$)}:
Thus,
we start from quantum mechanics.
That is,
\begin{itemize}
\item[({}G)]
{{measurement theory}}
is a scientific language
modeled on quantum mechanics.
%(=)
%
%quantum mechanics%\textcolor{black}{NZ}
%
%
\end{itemize}
Now, let us explain it.

As mentioned later
(in \textcolor{black}{{Chap. 3}}),
quantum mechanics, which was discovered
by
Heisenberg,
Schr\"odinger,
Born
in 1925--1927,
is physics for
%the physics concerning
the microscopic world.
%
%
%%\textcolor{black}{NZ}{physics},
%Einstein%\textcolor{black}{NZ}{the theory of relativity} and 20%\textcolor{black}{NZ}2
quantum mechanics
is composed of two laws
(i.e.,
"Born's probabilistic interpretation of quantum mechanics"
and
"quantum kinetic equation
(due to Heisenberg and
Schr\"odinger)"
).
%index{@{quantum kinetic equation}}
That is,
\begin{itemize}
\item[(H)]
$ \qquad
\underset{\text{\scriptsize ({physics})}}{\fbox{quantum mechanics}}
:= \underset{\text{\scriptsize (Born's probabilistic interpretation)}}{\fbox{{{measurement}}}}
+ \underset{\text{\scriptsize (
{quantum kinetic equation})}}{\fbox{causality }}
$
\end{itemize}
%
%\BEGIN{itemize}
%\item[({}H)]
%$\;\;
%$
%$
%{
%\overset{\text{\scriptsize ({physics})}}
%{
%}
%\underset{\text{\scriptsize [{physics}]}}{\text{{} $\fbox{quantum mechanics}$}}
%:=
%{
%\overset{\text{\scriptsize (Born%\textcolor{black}{NZ}probability ;\textcolor{black}{{Sec. 3.1.1}})}}
%{
%\underset{\text{\scriptsize }}{\text{{} $\fbox{{{measurement}}{}}$}}}
%}
%+
%{
%\overset{\text{\scriptsize ({quantum kinetic equation};\textcolor{black}{{Note }10.2})}}
%{
%\underset{\text{\scriptsize }}
%{\text{{}$\fbox{causality }$}}
%}
%}
%$
%\

As the linguistic turn of
quantum mechanics,
we get measurement theory
---
{quantum mechanical {world-view}}\textcolor{black}{({}G)}
---
as follows:
%((D) and )%\textcolor{black}{NZ}{linguistic aspect}, That is,
\begin{itemize}
\item[(I)]
$ \qquad
\underset{\text{\scriptsize ({physics})}}{\fbox{quantum mechanics}}
\xrightarrow[\text{\scriptsize linguistic turn}]{\text{
\scriptsize verbalizing}}
\underset{\text{\scriptsize (scientific language)}}{\fbox{{{measurement theory}}}}
$
\end{itemize}
%\BEGIN{itemize}
%\item[(I)]
%$
%\qquad
%\quad
%$
%$
%\overset{\text{\scriptsize ({physics})}}{\underset{}{\text{
%\fbox{quantum mechanics}}}}
%\xrightarrow[{linguistic aspect}]{{physical law}%\textcolor{black}{NZ}
%}
%\overset{\text{\scriptsize }}{\underset{({{measurement}} and causality  and %\textcolor{black}{NZ})}{\text{\fbox{{{measurement theory}}}}}}
%$
%\END{itemize}
Quantum mechanics is physics for microscopic
phenomena.
However,
its linguistic turn
(=measurement theory)
has a power to describe
phenomena
in our usual world.

Measurement theory is quite simple language,
which has two
key-words
(i.e.,
"measurement"
and
"causality").
% and ,
%{ordinary language}%\textcolor{black}{NZ}%\textcolor{black}{NZ}.
%, {},
%{{measurement theory}}%\textcolor{black}{NZ}
%(,  and ,
%quantum mechanics%\textcolor{black}{NZ}){\bf {{measurement}}} and {\bf causality }%\textcolor{black}{NZ}2
%,
%"measurement"
%{{measurement theory}}%\textcolor{black}{NZ},
%%:
\begin{itemize}
\item[({{}}J)]
\begin{itemize}
\item[({{}}J$_1$)]
{\bf {{measurement}} (}
use its items;
{\bf
{{observer}}, {measuring object}, {{state}},
observable ($\approx${measuring instrument}), measured value , probability )
}
following Axiom 1\textcolor{black}{(Sec. 2.2)}
\item[({{}}J$_2$)]
{\bf causality (}
use the words:
{\bf
{{semi-ordered tree}}, {{causal operator}}
)
}
following
Axiom 2\textcolor{black}{(Sec. 6.4)}
\end{itemize}
\end{itemize}
Writing
diagrammatically,
\begin{itemize}
\item[(K)]
\qquad
$
\underset{\text{\scriptsize (scientific language)}}{\fbox{{{measurement theory}}}}
:=
\overset{\text{\scriptsize [Axiom 1]}}{\fbox{{{measurement}} (J$_1$)}}
+
\overset{\text{\scriptsize [Axiom 2]}}{\fbox{ causality  (J$_2$)}}
$
\end{itemize}
%
%\BEGIN{itemize}
%\item[({}K)]
%$
%\quad
%$
%$
%{
%\overset{\text{\scriptsize (scientific language)}}{$\fbox{{{measurement theory}}}$}
%{
%\underset{\text{\scriptsize (quantum mechanical {world-view}%\textcolor{black}{NZ})}}{\text{{} $\fbox{{{measurement theory}}{}}$}}}
%$\fbox{{{measurement theory}}}$
%
%%\dashbox{5}
%\overset{\text{\scriptsize (scientific language)}}{\text{{} $\fbox{{{{measurement theory}}}}$}}
%:=
%{
%\overset{\text{\scriptsize [Axiom 1\textcolor{black}{(SCe. \REF{2secAxiom 1})}]}}
%{
%\underset{\text{\scriptsize
%}}{\text{{} $\fbox{{{measurement}}\text{({{}}J$_1$)}}$}}}
%}
%+
%{
%\overset{\text{\scriptsize [Axiom 2\textcolor{black}{(Sec. \REF{6secAxiom 2})}]}}
%{
%\underset{\text{}}
%{\text{{}$\fbox{ causality \text{({{}}J$_2$)}}$}}
%}
%}
%$
%\END{itemize}
where
{\bf scientific language}
must be distinguished A from B
{ordinary language} and mathematics
(=mathematical language).
%index{@scientific language}

Here,
the
quantum mechanical {world-view} method
(=measurement theory)
clarifies
classical mechanical problems
\textcolor{black}{(F$_1$)--(F$_5$)}
in what follows(
\textcolor{black}{(F$'_1$)--(F$'_5$)}):
\begin{itemize}
\item[(F$'_1$)]
{{measurement theory}}is
a {world-description} language
(i.e.,
scientific language
)
\item[(F$'_2$)]
The source of measurement theory is
quantum mechanics.
\item[(F$'_3$)]
%{{measurement theory}}
If some try to add a basic concept to measurement theory,
they must start to add the  basic concept to
quantum mechanics.
Therefore,
we can assure that
the trial is impossible.
\item[(F$'_4$)]
As mentioned in
{\textcolor{black}{Fig.}$\;$}8.2
in \textcolor{black}{{Chap.{\;}}8},
we assert:
\begin{itemize}
\item[]
the development from
{the causal world-view}(E$_1$)
to classical mechanical {world-view}(E$_2$)
is not inevitable
\end{itemize}
Also,
we can do well without "gamble".
% and .
\end{itemize}

The reason that the problems
\textcolor{black}{(F$_1$)--(F$_4$)}
are solved
is due to the fact that
quantum mechanics itself
possesses
the concept "probability".

If it be so,
there may be a reason to choose
{{measurement theory}}
---
quantum mechanical {world-view} method
---,
however,
it is a matter of course that
\begin{itemize}
\item[({}L$_1$)]
in every theory,
the most important thing is to
determine the starting point.
\end{itemize}
Thus,
an immediate conclusion.
should be avoided.
However,
from the above
\textcolor{black}{(F$'_1$)--(F$'_4$)}
we assert that
%%\textcolor{black}{NZ}
\begin{itemize}
\item[({}L$_2$)]
%{{measurement theory}},
%{statistics}${{\cdot}}${dynamical system theory}%\textcolor{black}{NZ}
%${{\cdot}}${}
%,
{statistics}${{\cdot}}${dynamical system theory}
(classical mechanical {world-view})
is the abbreviation of
{{measurement theory}}
(quantum mechanical {world-view} method)
%(, \textcolor{black}{{FIG.$\;$}8.2})
\end{itemize}
And thus,
The question
$(\sharp_2)$
in
\textcolor{black}{(F$_5$)(={Note }1.1})
---
Why are the mathematical theories
(i.e.,
{differential equation}
and
probability theory)
are useful?
---
is answered as follows
\begin{itemize}
\item[(F$'_5$)]
measurement theory
is hidden behind
these mathematical theories
\end{itemize}

\par
\noindent
\par
\noindent
%BBBBBBBBBBBBBBBBBB%SBSBSBSPOIUYTREWQ
\par
\noindent
\renewcommand{\footnoterule}{
  \vspace{2mm}                      % %\textcolor{black}{NZ}
  \noindent\rule{\textwidth}{0.4pt}  
  \vspace{-5mm}
}
{\small%%{\footnotesize
\vspace{0.2cm}
\begin{itemize}
\item[$\spadesuit$] \bf {{}}{Note }1.2{{}} \rm
In this print, we assert that
%$$
%[]=[ and {FIG.$\;$}]=[]
%$$
% and ,
%%,
\begin{itemize}
\item[]
$
\qquad
\quad
\text{\bf {{measurement theory}}= the language of engineering (or, sciences)}
%\;\;\;\;
%\tex
$
\end{itemize}
%({Chap.{\;}}8{}(l))
That is, (L$_2$)
---
{statistics}${{\cdot}}${dynamical system theory}
is immature,
and
%,
{{measurement theory}} is mature
---
says that
%\BEGIN{itemize},
\begin{itemize}
\item[]
{{measurement theory}}
makes
engineering (or, sciences)
mature
\end{itemize}
(\textcolor{black}{{Note }1.12}).
\end{itemize}
}
%%BBBBBBBBBBBBBBBBBB%SBSBSBSS

\subsubsection{Why does measurement theory hold?}%{Sec. 1.1.3}
%index{@{realistic method}(=realistic{world-description method})}
%index{@{linguistic method}(={linguistic world-description method})}
\par
If we believe in
the argument
mentioned in the previous section
(i.e.,
the
quantum mechanical {world-view}(K)
is superior
to
classical mechanical {world-view}(E$_2$)
),
increasingly
we want
to answer
(A$_1$)
:
\begin{itemize}
\item[({}A$_1$)]
Why is
quantum mechanics
(i.e.,
the physics concerning
the microscopic world)
applicable to
the description of
ordinary phenomena
(i.e.,
economics,
psychology,
electric engineering,
etc.
)
?
$\;\;$
Or,
why
is the quantum mechanical approach
(i.e.,
measurement theory)
indispensable?
$\;\;$
That is,
why
does
{{measurement theory}}
---
quantum mechanical language
---
hold?
$\;\;$
Further,
is there another approach?
\end{itemize}

Of course,
we do not have
the absolute answer.
Thus
we want to add the following discussion:
For example,
consider
\begin{align*}
\text{
the statement such as
{\bf "even monkeys fall from trees"}
}
\end{align*}
This is the famous proverb in Japan.
This is the same as
the proverb
"Even Homer sometimes nods"
or
"A good swimmer is not safe against drowning".
%%index{monkey@ Even monkeys fall from trees}
The statement
"Even monkeys fall from trees",
which must have described the actual phenomenon,
is isolated from reality.
And the statement becomes
the proverb
"Even monkeys fall from trees".
Writing
diagrammatically,
%%index{monkey@ Even monkeys fall from trees}
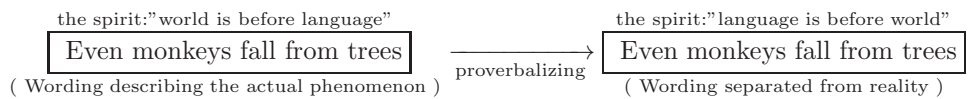
\begin{align*}
\overset{\text{\scriptsize the spirit:{"world is before language"}}}{
\underset{\text{\scriptsize (
Wording describing the actual phenomenon
)}}{\text{
\fbox{ Even monkeys fall from trees}}}}
\xrightarrow[\text{\scriptsize proverbalizing}]{}
\overset{\text{\scriptsize the spirit:"language is before world"}}{\underset{
\text{\scriptsize
(
Wording separated from reality
)}}{\text{\fbox{
 Even monkeys fall from trees}}}}
\end{align*}
%\END{itemize}
As in the above,
"proverbalizing"
means
that
the order of the "world" and "language" is reversed.
That is,
the proverb
"Even monkeys fall from trees
"
can be applicable to
different world
(which is not related to
"monkey"
nor
"tree").
This is

\begin{itemize}
\item[({}M$_1$)]
$\quad
%\qquad
\qquad
$
$
\text{
\bf
Wonder of man's linguistic competence
}
$
\end{itemize}
This cannot but accept as a fact.

Thus,
{{measurement theory}}
is characterize as
the linguistic turn.
That is, rewriting
(I):
\begin{itemize}
\item[$\underset{\text{\scriptsize (={(I)})}}{\text{({}M$_2$)}}$]
$\quad
%\qquad
$
$
\overset{\text{\scriptsize (the terms in (J)
connect reality
)}}{\underset{\text{\scriptsize({physics})}}{\text{
\fbox{quantum mechanics(H)}}}}
\xrightarrow[\text{\scriptsize the linguistic turn}]{\text{\scriptsize proverbalizing}}
\overset{\text{\scriptsize (the terms in (J) have no reality)}}{\underset{\text{\scriptsize(scientific language)}}{\text{\fbox{
{{measurement theory}}(K)}}}}
$
\end{itemize}
Quantum mechanics is physics for microscopic world,
however,
its linguistic turn
(i.e.,
measurement theory)
can describe
ordinary world.

\subsubsection{Two {world-description method}
---
realistic vs. linguistic}%1.1.4
%\ssubsubsection{realistic{world-description method} and
%{linguistic world-description method}(idealism)}
\par
It is a matter of course that
\begin{itemize}
\item[({}M$_3$)]
%$\quad
%%\qquad
%$
%$
%\text
{\bf
The essence of the proverb
is not
{\LL}experimental verification{\RR}
but
{\LL}usability{\RR}
}
\end{itemize}
%,
%{{measurement theory}}
%is not based on
%
%
%(That is, {world is before language})
%%\textcolor{black}{NZ}
%,
And further,
\begin{itemize}
\item[({{}}N)]
{{measurement theory}}
is based on the spirit of "language is before world"
%{metaphysics}
\end{itemize}

\par
\noindent
\par
\noindent
%BBBBBBBBBBBBBBBBBB%SBSBSBS
\par
\noindent
{\small%%{\footnotesize
\vspace{0.2cm}
\begin{itemize}
\item[$\spadesuit$] \bf {{}}{Note }1.3{{}} \rm
%index{@{metaphysics}}
{\bf {metaphysics}}
is an academic discipline
concerning the propositions in which empirical validation is impossible.
Lord Kelvin(1824--1907{)} said that
\begin{align*}
\text{
Mathematics is
the only good metaphysics.
}
\end{align*}
This is very persuasive saying.
However,
Our purpose is
\begin{itemize}
\item[$(\sharp)$]
{\bf
to establish metaphysics(called measurement theory)
as a discipline which forms the base of science
}
\end{itemize}
% and . , %\textcolor{black}{NZ}{metaphysics},
%{metaphysics}
%(={metaphysics}),
%${{\cdot}}$${{\cdot}}$.
%%index{@}
\end{itemize}
}
%%BBBBBBBBBBBBBBBBBB%SBSBSBSS

If it be so,
the world-description
is classified
as follows.
% and 
%(That is,
%{metaphysics}%\textcolor{black}{NZ} and %\textcolor{black}{NZ}),
%{world-description}2%\textcolor{black}{NZ}
%
%{realistic method}(${{\cdot}}$) and {linguistic method}(${{\cdot}}$)
%
% and .
%That is,
\begin{itemize}
\item[({{}}O)]
{world-description}
$\cases
&
\!\!\!\!\!\!
{\textcircled{\scriptsize 1}}
%\underset{(${{\cdot}}$)\qquad \qquad \quad}
{\text{\bf  {realistic method}(In the beginning was the "world")}}
%[realistic{world-description method}};{realistic world-view};
%realisticscientific language]
\\
&
\text{physical phenomena are directly described in mathematics.}
\\
&
\text{{physics}
is created by this method.}
\\
&
\text{The law is main}
\\
& %\qquad
\text{"world is before language". }
\\
\\
&
\!\!\!\!\!\!
{\textcircled{\scriptsize 2}}
%\u
{\text{\bf  {linguistic method}({In the beginning was the "word"})}}
%[{linguistic world-description method}};linguistic world-view;
%{linguistic scientific language}]
\\
&
\text{Phenomena are described in a scientific language.}
\\
&
\text{Variuos sciences are created by this method}
\\
&
\text{The scientific law and the scientific laguage are main}
\\
&
\text{As a scientific language,
measurent thorey is adoptted}
\\
&
\text{"language is before world". }
\endcases
$
\end{itemize}

The {linguistic method}
is, for the first time,
established my
{{measurement theory}},
and thus it is a new
{world-description method}.
Thus,
for completeness,
we add the following two notes.

%%\textcolor{black}{NZ}\textcolor{black}{({}O)}%\textcolor{black}{NZ}, %\textcolor{black}{NZ},
%%\textcolor{black}{NZ}2%\textcolor{black}{NZ}{Note }.
%\textcolor{black}{({}O)}%\textcolor{black}{NZ} and ,  and %\textcolor{black}{NZ},
%.
\par
\noindent
\par
\noindent
%BBBBBBBBBBBBBBBBBB%SBSBSBS
\par
\noindent
{\small%%{\footnotesize
\vspace{0.1cm}
\begin{itemize}
\item[$\spadesuit$] \bf {{}}{Note }1.4{{}} \rm
For example,
assume that you want to understand some economical phenomenon $P$.
For this, consider the following four methods
(a)--(d):
\begin{itemize}
\item[(a)]
you exactly measure
the economical phenomenon $P$,
and
represent it mathematically.
Then, you can create a certain economical theory $T_a$.
%%\textcolor{black}{NZ}{realistic method} and ,
%%\textcolor{black}{NZ}(e), (That is, {linguistic method}%\textcolor{black}{NZ})
\item[(b)]
First
we decide use the mathematical theory
(differential equation,
probability theory).
you exactly measure
the economical phenomenon $P$,
and
represent it by the above mathematics.
Then, you can create a certain economical theory $T_b$.
\item[(c)]
First
we decide use statistics
(=dynamical system theory).
you exactly measure
the economical language before world $P$,
and
represent it by the above mathematics.
Then, you can create a certain economical theory $T_c$.
\item[(d)]
First
we decide use measurement theory.
you exactly measure
the economical phenomenon $P$,
and
describe it by measurement theory.
Then, you can create a certain economical theory $T_d$.
This is the linguistic method.
\end{itemize}
Note that the economical phenomenon $P$
is common.
Thus,
if
each (a)--(d) is the fully considered theory,
we can expect that
$T_a=T_b=T_c=T_d$.
%
%
%
%
%
%, $P$%\textcolor{black}{NZ}, ,
%,
%$T_a=T_b=T_c=T_d$ and {}
\begin{itemize}
\item[(e)]
if $T_a=T_b=T_c=T_d$,
then
we can consider that
it is created by (d).
\end{itemize}
However, the theory of relativity can not be understood
in measurement theory
i.e.,
in the linguistic method).
Also,
as seen in \textcolor{black}{{Chap.{\;}}9{}},
it is interesting to see that
$T'_c \not= T'_d$
in equilibrium statistical statistics.
In this case,
we assert that
$T'_d$ should be adopted.
%
%,
%, {{measurement theory}}{linguistic method}({}d),
%({{{{h}}}}%\textcolor{black}{NZ})equilibrium statistical mechanics
%$T'_d$
%,
%{{w}} and ,
%$T'_c \not= T'_d$ and .
%%\textcolor{black}{NZ}, ,
%$T'_c$(=(c)
%equilibrium statistical mechanics
%)
% and ,
%$ T'_d$
%.
\end{itemize}
}
%%BBBBBBBBBBBBBBBBBB%SBSBSBSS

\par
\noindent

%BBBBBBBBBBBBBBBBBB%SBSBSBS
\par
\noindent
{\small%%{\footnotesize
\vspace{0.1cm}
\begin{itemize}
\item[$\spadesuit$] \bf {{}}{Note }1.5{{}} \rm
%language is before world and {metaphysics} and ,
%, \textcolor{black}{{Note }1.3}%\textcolor{black}{NZ}$(\sharp)$%\textcolor{black}{NZ},
%That is,
%\BEGIN{itemize}
%\item[$(\sharp_1)$]
%(%\textcolor{black}{NZ}{metaphysics}){{measurement theory}}
%%\textcolor{black}{NZ} and  and
%\END{itemize}
%, {}
%
If some may regard
"{realistic method} vs. {linguistic method}"
(the world-description classification \textcolor{black}{({}O)})
as
"materialism vs. idealism" in philosophy,
they never accept
"linguistic method".
However,
we think that
"linguistic method"
is acceptable for everyone.
As mentioned in
\textcolor{black}{{Chap.{\;}}8{}},
we think that
% and , ,
\begin{itemize}
\item[$(\sharp)$]
"idealism = linguistic method"
\end{itemize}
In this sense,
the idealism can not be understood without
measurement theory.
\end{itemize}
}
%%BBBBBBBBBBBBBBBBBB%SBSBSBSS

\subsection{Monism and dualism}%{Sec.1.2}
\par
%index{@monism}
%index{@dualism}

\subsubsection{Monism[="matter"]  and dualism[="mind and matter"]}%1.2.1
\par
In the previous section,
we discuss the world-description classification
\textcolor{black}{(O)}.
In this section,
we introduce anther
world-description classification,
i,e.,
monism and dualism.
In monism,
we consider that
"world"
=
"matter",
and
in dualism,
"world"
=
"I(=mind)"+"matter".
That is,
That is,
\begin{itemize}
\item[({{}}P$_1$)]
{world-description}
$\cases
\text{
monism=[
"matter"])
}
\\
\\
\text{
dualism=["I(=mind)"+"matter"])
}
\endcases
$
\end{itemize}

\par
Newtonian mechanics
and
{the theory of relativity},
which are formulated in monism,
acquire a great success.

If we are concerned with the troublesome thing
such as "I(=mind)",
objectivity is spoiled,
and thus,
we are without science.
However,
quantum mechanics,
which are formulated in dualism,
acquires a great success.

Therefore,
the dualism in this not is the dualism inspired from quantum mechanics.
\par
\noindent
\par
\noindent
%BBBBBBBBBBBBBBBBBB%SBSBSBS
\par\noindent
{\small%%{\footnotesize
%\vspace{0.2cm}
\begin{itemize}
\item[$\spadesuit$] \bf {{}}{Note }1.6{{}} \rm
The world-description
mentioned in this print
is always quantitative.
And thus,
mathematics is always fundamental and essential in our
world-description.
However, it should be noted that
\begin{itemize}
\item[$(\sharp)$]
mathematics itself is independent of
world.
That is,
mathematics
exists without world.
\end{itemize}
That is,
mathematics is
the origin learning (before science,
that is,
before the world-description classifications
\textcolor{black}{(O) and (P$_1$)}).
%
%
%(That is,
%)%\textcolor{black}{NZ}{}
%%\textcolor{black}{NZ} and ,
%Newtonian mechanics(monism)
%
%quantum mechanics(dualism), {}
\end{itemize}
}
%%BBBBBBBBBBBBBBBBBB%SBSBSBSS

%, {linguistic world-description method}%\textcolor{black}{NZ},
Considering (O) and (P$_1$),
we get the following classification:
\begin{itemize}
\item[({{}}P$_2$)]
{world-description}
$\cases
{\textcircled{\scriptsize 1}}\
\text{ realistic method}
\cases
\text{
monism {$\cdots$ \;}classical mechanics,...
}
\\
\text{
dualism{$\cdots$ \;}quantum mechanics}
\endcases
\\
\\
{\textcircled{\scriptsize 2}}\ \text{linguistic method}
\cases
\text{monism{$\cdots$ \;}(cf. {Sec.1.2.2})}
\\
\text{
dualism{$\cdots$ \;}(measurement theory, {Sec.1.2.3})
}
\endcases
\endcases
$
\end{itemize}
%n%\textcolor{black}{NZ},
%%\textcolor{black}{NZ}.

\subsubsection{{Linguistic world-description in monism}({state equation method})}%{Sec.1.2.2}
\par
The realistic world-description in monism
is
well known as physics
(i.e.,
Newtonian mechanics, etc.).
Thus,
in this section,
we devote ourselves to
the linguistic world-description in monism.
%method, %\textcolor{black}{NZ}(Newtonian mechanics{the theory of relativity})%\textcolor{black}{NZ}
% and ,
%.
Although
this may not be authorized yet,
there may be a reason to
consider that
it is the same as
the {state equation method},
which is characterized as the linguistic turn
of Newtonian mechanics
(see the (D)).
%monism{linguistic world-description method},
% and (\textcolor{black}{{Note }11.6}), ,
%(D), Newtonian mechanics%\textcolor{black}{NZ}{linguistic aspect}
%{state equation method}
%
% and .
%%\textcolor{black}{NZ}dualism{linguistic world-description method} and %\textcolor{black}{NZ}%\textcolor{black}{NZ},
%%\textcolor{black}{NZ} and .
The essence of the state equation method
is only
"causality".
Thus,
the state equation method
without causality
is quite simple.
That is,
%
%
%without causality
%%\textcolor{black}{NZ}, \textcolor{black}{{Chap.{\;}}2{}}%\textcolor{black}{NZ}\textcolor{black}{Axiom 1}({{measurement}}{)}
%%\textcolor{black}{NZ}
%%\textcolor{black}{NZ},
%causality 
%RT.
%{state equation method}
%%\textcolor{black}{NZ}, causality 
%%\textcolor{black}{NZ}, ,
%monism{linguistic world-description method}%\textcolor{black}{NZ}%\textcolor{black}{NZ}.
\par
\noindent
\par
The key-words are as follows:
%,
\begin{itemize}
\item[({{}}Q)]
$\qquad
\qquad
\text{\bf object{(}=matter{)}}
\text{ and }
\text{\bf {{state}}{(}=property{)}}
$
\end{itemize}
\rm
And thus,
the statements
(in {linguistic world-description method})
are as follows:

%\BEGIN{itembox}[c]
\par
\noindent
\begin{center}
{\bf {Linguistic world-description method in monism}}
\end{center}
\par
\noindent
%\vskip0.1cm
\par
\noindent
\fbox{\parbox{150mm}{
$\quad$
\\
$\qquad$
 An {\bf object} has a {\bf {{state}}} $\omega$
\\
}
}
%\END{picture}
%%%%%%%%%%%%%%%%%center
%\BEGIN{FIGUre*}[htbp]
%%%%%%%%%%%%%%%%%%
%\vskip0.3cm
%\caption{
%Which is the hidden urn,
%$U_1$ or $U_2$?
%%(= FIG. 4.2)
%}
%\END{figure*}
%%%%%%%%%%%%%%%%%%

\par
\vskip0.5cm
\par
This can be easily understood in the following simple examples:
\begin{itemize}
\item[({{}}R$_1$)]
$\;\;$The temperature of water in this cup is 5$\SD$
\\
$\Rightarrow$
[The object (i.e., the water in this cup )] has a state such as [5$\SD$]
% and {{state}}
\item[({{}}R$_2$)]
$\;\;$John has 500 dollars in his purchase
%%\textcolor{black}{NZ}500
\\
$\Rightarrow$
[The object (i.e., John's purchase)] has a state such as [500 dollars]
%[A%\textcolor{black}{NZ} and ], [500] and {{state}}.
%,
%[B%\textcolor{black}{NZ}%\textcolor{black}{NZ}] and , [300] and {{state}}.
\item[({{}}R$_3$)]
$\;\;$This flower is red
\\
$\Rightarrow$
[The object (i.e., this flower)] has a state such as [red]
%[%\textcolor{black}{NZ} and ], [] and {{state}}
\item[({{}}R$_4$)]
$\;\;$The water in this cup is cold
%$\;\;$%\textcolor{black}{NZ}{cup}%\textcolor{black}{NZ}%\textcolor{black}{NZ}{{{{c}}}}
\\
$\Rightarrow$
[The object (i.e., the water in this cup )] has a state such as [cold]
%[%\textcolor{black}{NZ}{cup}%\textcolor{black}{NZ}%\textcolor{black}{NZ} and ], [{{{{c}}}}] and {{state}}
\end{itemize}
Here,
\textcolor{black}{({{}}R$_1$)}
and
\textcolor{black}{({{}}R$_2$)}
are important.
That is,
\textcolor{black}{({{}}R$_3$)}
and
\textcolor{black}{({{}}R$_4$)}
are
regarded
as the preparation of
Sec. 1.2.3
(The linguistic world-description in dualism).

\par
There is a possibility that
a {{state}} is $\omega$.
or
$\omega'$,...
Put
$\Omega=\{\omega, \omega', .... \}$,
which is a state space.
For example,
\begin{itemize}
\item[({{}}R$'_1$)]
In the case of
\textcolor{black}{({{}}R$_1$)},
we see that
the state space
$\Omega$
$
=
\{ \omega \;|\;
\omega
\SD
$
is the temperature of water
$\}=[0,100]$.
\item[({{}}R$'_2$)]
In the \textcolor{black}{({{}}R$_2$)},
we consider the possibility such that
$1$ cent, $2$ cents, ....
Therefore,
The state space $\Omega$
$=$
$\{0,1,2,\ldots\}$
\end{itemize}
% and .

\vskip0.3cm

%%BBBBBBBBBBBBBBBBBB%SBSBSBS{\mathsf O}
\par
\noindent
{\small%%{\footnotesize
\begin{itemize}
\item[$\spadesuit$] \bf {{}}{Note }1.7{{}} \rm
Most scientists may not be familiar  with the following questions:,
\begin{itemize}
\item[$(\sharp)$]
"{realistic method}"
or
"{linguistic method}"?
$\qquad$
"monism"
or
dualism"?
\end{itemize}
As mentioned in
(F$_1$),
the reason is due to
the fact
that
these questions
have been been kept ambiguous
in
{statistics}
{(}={dynamical system theory}).
That is,
as mentioned in (L$_2$),
{statistics}
(={dynamical system theory})
does not have a power
to clarify
the questions
$(\sharp)$.
Therefore,
the argument in this section
is
not authorized,
and thus it should be regarded as
the preparation
for
the following
% and (That is,
%,
%{statistics}${{\cdot}}${dynamical system theory},
%%\textcolor{black}{NZ}$(\sharp)$
% and ),
%
%monismlinguistic
% and .
%, %\textcolor{black}{NZ}
%{state equation method} and ,
% and (\textcolor{black}{{Note }11.6}).
%%,
%%\textcolor{black}{NZ},
\textcolor{black}{({Sec.1.2.3})}
(Linguistic method in dualism).
%%\textcolor{black}{NZ}dualism{linguistic method}(={{measurement theory}})%\textcolor{black}{NZ}
%%\textcolor{black}{NZ}%\textcolor{black}{NZ}%\textcolor{black}{NZ} and .
%%(\textcolor{black}{{Chap.$\;$1}%\textcolor{black}{NZ}(F$'_3$)}).
\end{itemize}
}
%%BBBBBBBBBBBBBBBBBB%SBSBSBSS
%
\subsubsection{{Linguistic world-description method in dualism}(= the Copenhagen interpretation of {{measurement theory}})}%{Sec.1.2.3}
\par
Quantum mechanics
is dualistic {physics}
(i.e.,
dualistic and realistic {world-description),
and,
{{measurement theory}}
is
the linguistic turn
(
verbalizing,
proverbalizing)
of
quantum mechanics.
And therefore,
{{measurement theory}}
is
dualistic and {linguistic world-description}.
The monastic and {linguistic world-description}
mentioned in the previous section
is too simple,
but
the
dualism and {linguistic world-description}
{(}i.e.,
{{{measurement theory}}}{)}
is rather
troublesome.

As mentioned
in
(K)
and
(J$_1$),
{{measurement theory}}
has the following structure:
\begin{itemize}
\item[(K)]
\qquad
$
\underset{\text{\scriptsize (scientific language)}}{\fbox{{{measurement theory}}}}
:=
\overset{\text{\scriptsize [Axiom 1]}}{\fbox{{{measurement}} (J$_1$)}}
+
\overset{\text{\scriptsize [Axiom 2]}}{\fbox{ causality  (J$_2$)}}
$
\end{itemize}
%
%
%
%\BEGIN{itemize}
%\item[(K)]
%$
%\qquad
%\qquad
%$
%$
%\overset{\text{\scriptsize [scientific language]}}{\text{{} $\fbox{{{{measurement theory}}}}$}}
%:=
%{
%\overset{\text{\scriptsize [Axiom 1\textcolor{black}{(\REF{2secAxiom 1})}]}}
%{
%\overset{\text{\scriptsize
%}}{\text{{} $\fbox{{{measurement}}}$}}}
%}
%+
%{
%\overset{\text{\scriptsize [Axiom 2\textcolor{black}{(\REF{6secAxiom 2})}]}}
%{
%\overset{\text{}}
%{\text{{}$\fbox{ causality }$}}
%}
%}
%$
%\END{itemize}
in which
we see two key-words
(i.e.,
{{"measurement"}}
and
"causality".
Thus,
even if we omit
"causality",
we have "measurement"
which is composed of
the following terms:
\begin{itemize}
\item[(J$_1$)]
{\bf {{measurement}} (}
the items:
{\bf
{{observer}}, {measuring object}, {{state}},
observable ($\approx${measuring instrument}), measured value , probability ).
}
\end{itemize}
Thus,
compared with monism,
dualism has many key-words.
\par
The concept
of
"{{measurement}}"
can be,
for the first time,
understood
in dualism.
Let us explain it.
The image
of
"{{measurement}}"
is as shown in
\textcolor{black}{Fig. 1.1}.

\par
%\noindent
%\BEGIN{FIGUre}[htbp]
%
%\noindent
%\unitlength=0.7mm
%\BEGIN{picture}(200,62)(10,0)
%{{{
%\allinethickness{0.5mm}
%\drawline[-40](70,0)(70,52)(30,52)(30,0)%%\textcolor{black}{NZ}
%\drawline[-40](115,0)(115,52)(155,52)(155,0)%%\textcolor{black}{NZ}
%\path(20,0)(160,0)%(100,0)(100,70)(20,70)(20,0)
%\put(20,70){\bold{}}
%\put(14,-5){
%\put(32,50){$\bullet$}
%}%
%\put(100,0){
%\put(14,-5){
%}
%}
%\put(45,25){\ellipse{17}{25}}%
%\put(45,44){\ellipse{10}{13}}%
%\put(2,34){\put(43,25){\bf \footnotesize{}}
%\put(39,20){\scriptsize{(observer)}}
%}
%\put(7,7){\path(41,27)(50,20)(53,20)}%
%\path(43,13)(42,0)(44,0)(45,13)%
%\path(46,13)(47,0)(49,0)(48,13)%
%\put(0,26){
%\put(131,33){\bf \footnotesize }
%\put(117,28){\scriptsize ({measuring object}, )}
%}
%\path(135,0)(135,20)(148,20)(133,50)(118,20)(131,20)(131,0)%
%\put(10,0){
%}
%\allinethickness{0.2mm}
%\put(0,-5){
%\put(112,39){\vector(-1,0){43}}%
%\put(70,43){\vector(1,0){43}}%
%\put(75,51){\bf \scriptsize [observable ${{\cdot}}${measuring instrument}]}
%\put(52,50){\bf \scriptsize [measured value ]}
%\put(51,46){\scriptsize }
%\put(75,46){\scriptsize \textcircled{\scriptsize a} }
%\put(75,33){\scriptsize \textcircled{\scriptsize b} }
%\put(117,48){\bf \scriptsize [{{state}}]}
%}
%\put(-30,0){
%}
%}}}
%\END{picture}
%\caption{{{measurement}}%\textcolor{black}{NZ}{FIG.$\;$}}
%\END{figure}
%
%

%
\par
\noindent
%\vskip-0.4cm%%%
%\begin{figure}[htbp]
\noindent
\unitlength=0.7mm
\begin{picture}(200,60)(10,0)
%\put(0,-40)
{{{
\allinethickness{0.5mm}
\drawline[-40](80,0)(80,52)(30,52)(30,0)
\drawline[-40](130,0)(130,52)(170,52)(170,0)
%\drawline[-20](10,-13)(240,-13)(240,80)(10,80)(10,-13)
\path(20,0)(200,0)%y{}$B<J{}(B(100,0)(100,70)(20,70)(20,0)
\put(20,70){\bold{}}
\put(14,-5){
\put(37,50){$\bullet$}
}
\put(100,0){
\put(14,-5){
}
}
\put(50,25){\ellipse{17}{25}}%zV
\put(50,44){\ellipse{10}{13}}%zI
\put(0,34){\put(43,25){\bf \footnotesize{observer}}
\put(42,20){\scriptsize{(I(=mind))}}
}
%\put(48,25){\bf \footnotesize{u{}$B%!!+{}(B}
%\put(42,20){\scriptsize{(measurement{}$B<T{}(B)}}
\put(7,7){\path(46,27)(55,20)(58,20)}
\path(48,13)(47,0)(49,0)(50,13)%yJ
\path(51,13)(52,0)(54,0)(53,13)%yJ
%\put(0.20){
%\put(147,33){\bf \footnotesize {G}
%\put(121,28){\s
%%\put(140,23){\scriptsize {(}measurementymxz{)}}
%}
\put(0,26){
\put(147,33){\bf \footnotesize system}
\put(148,28){\scriptsize (matter)}
%\put(140,23){\scriptsize {(}measurementymxz{)}}
}
\path(152,0)(152,20)(165,20)(150,50)(135,20)(148,20)(148,0)%{{}$B<X{}(B
%\put(130,55){\bf \footnotesize [$(\sharp_1)$\textcolor{black}{v{}$B<>{}(Bv{}$B~U{}(B*qzr*}2(t{}$B.X{}(Biu{}$B<V{}(Bv{}$B<W{}(B)]}
\put(10,0){
}
\allinethickness{0.2mm}
\put(0,-5){
\put(130,39){\vector(-1,0){60}}
\put(70,43){\vector(1,0){60}}
\put(92,51){\bf \scriptsize [observable]}
\put(58,50){\bf \scriptsize }
\put(57,48){\bf \scriptsize [measured value]}
\put(91,44){\scriptsize \textcircled{\scriptsize a}interfere}
\put(91,33){\scriptsize \textcircled{\scriptsize b}perceive a reaction}
\put(130,48){\bf \scriptsize [state]}
}
\put(-30,0){
%\put(75,60){\bf \footnotesize [\textcolor{black}{v{}$B<>{}(Bv{}$B~U{}(B*qzr*}1{(}measurement)]}
%
}
}}}
\end{picture}
%\hfill{\rm [FN]}\footnotemark
\begin{center}{Figure 1.1: The image of {\lq\lq}measurement(=\textcircled{a}+\textcircled{b})"
in dualism
}
\end{center}
\par
\noindent

In the above,
\begin{itemize}
\item[({}S$_1$)]
\textcircled{\scriptsize a}:
it suffices to understand that
"interfere"
is,
for example,
"apply light".
\\
\textcircled{\scriptsize b}:
perceive the reaction.
\end{itemize}
That is,
"measurement"
is characterized as
the interaction between
"{{observer}}"
and
"{measuring object}".
However,
\begin{itemize}
\item[({}S$_2$)]
In {{{measurement theory}}},
"interaction"
must not be emphasized.
\end{itemize}
Therefore,
in order to avoid confusion,
it might better to omit
the interaction
"\textcircled{\scriptsize a}
and
\textcircled{\scriptsize b}"
in
\textcolor{black}{Fig.} \textcolor{black}{1.1}.

Before we mention several
rules
[\textcolor{black}{({{}}U$_1$)--({{}}U$_7$)}]
(which is called "the Copenhagen interpretation"),
we must say
so called
"the Copenhagen interpretation"
in that follows.
\begin{itemize}
\item[(T)]
The Copenhagen interpretation is one of the earliest and most commonly taught interpretations of quantum mechanics.
The essential concepts of the Copenhagen interpretation
were devised by Niels Bohr, Werner Heisenberg and others.
And finally
it was accomplished by
von Neumann
\textcolor{black}{\cite{Neum}}(1932).
\end{itemize}
This Copenhagen interpretation in (T) is
within physics.
On the other,
our
"Copenhagen interpretation
[\textcolor{black}{({{}}U$_1$)--({{}}U$_7$)}]"
is not physics.
Thus the two are completely different.
In this sense,
there may be an opinion that
"the Copenhagen interpretation",
Hence
in this book
it
should have been called
"the linguistic interpretation"
or
"the linguistic Copenhagen interpretation".

\par
\vskip0.3cm
\par
\noindent {\bf The Copenhagen inteppretation}

%index{@}
%index{@the Copenhagen interpretation}
%index{@von Neumann}
\par
\noindent
%\vskip0.3cm
%BFBF

\begin{itemize}
\item[({{}}U$_1$)]
Consider the dualism composed of {\lq\lq}observer{\rq\rq} and {\lq\lq}system( =measuring object){\rq\rq}.
And therefore,
{\lq\lq}observer{\rq\rq} and {\lq\lq}system{\rq\rq}
must be absolutely separated.
(\textcolor{black}{(this section, {Sec.5.3.2})}.
If it says for a metaphor,
we say
"Audience should not be up to the stage"
%\END{itemize}
\item[({{}}U$_2$)]
Of course,
"matter(={measuring object})"
has
the space-time.
%%\textcolor{black}{NZ}, ${{\cdot}}$,
On the other hand, the observer
does not have
the space-time.
Thus,
the question:
{\lq\lq}When and where is a measured value obtained?{\rq\rq}
is out of measurement theory,
Thus, there is no tense in measurement theory.
This implies that
there is no tense in science.
\textcolor{black}{({Sec. 2.3.3}, {Sec.6.4.2})}.
\item[$\underset{\text{\scriptsize (=(\text{S$_2$}))}}{\text{({{}}U$_3$)}}$]
In {{{measurement theory}}},
"interaction"
must not be emphasized\textcolor{black}{(this section, {Sec.3.4})}.
\item[({{}}U$_4$)]
Only one measurement is permitted\textcolor{black}{({Sec.2.5})}.
Thus,
the state after measurement is meaningless.
\item[({{}}U$_5$)]
There is no probability without
measurement\textcolor{black}{({Sec.2.2}, {Note }4.7)}.
\item[({{}}U$_6$)]
observable is before {{state}},
Or,
observable
is superior to state\textcolor{black}{({Sec.2.6.2})}.
\item[({{}}U$_7$)]
State never moves(\textcolor{black}{{}{{{}}}{Sec.6.4.1}})
\end{itemize}
and so on.
Also, see
{Sec.3.3.1}(e), {Sec.$\;$}6.4.3(f$_3$).

%, ${{\cdot}}$ and Axiom%\textcolor{black}{NZ}%\textcolor{black}{NZ}
%{}

\par
%{
\vskip0.3cm
%BFBF
\par
\noindent

%dualism{(=}{{{measurement theory}}}{)}{{measurement}}Axiom,
%%[\textcolor{black}{({{}}S$_2$)}]
%the Copenhagen interpretation%\textcolor{black}{NZ},
\rm

Now we can state
the Prototype
of
Axiom 1({{measurement}})
as follows
(for the precise version, see
\textcolor{black}{Axiom${}_{\text{\scriptsize c}}^{\text{\scriptsize p}}$ 1(page \pageref{axiomcp1})})".

%\BEGIN{itembox}[c]
\par
\noindent
\begin{center}
{\bf {
Axiom 1({{measurement}}) Prototype
%Linguistic world-description method in monism
}}
\label{proto}
\end{center}
\par
\noindent
%\vskip0.1cm
\par
\noindent
\fbox{\parbox{155mm}{
\begin{itemize}
%\item[(i)]
%{{measurement}}
\item[]
{
When
an {\bf {{observer}}}
takes a measurement
of an {\bf observable ${\mathsf O}$}
(or,
by a {\bf {measuring instrument}${\mathsf O}$}
)
for a {\bf {measuring object}}
with a
{\bf {{state}}}
$\omega$,
the probability
that
a {\bf measured value } $x$
is obtained
is given by
$P $.}
\end{itemize}
}
}
\par
\vskip0.5cm
\par
\noindent

%\END{picture}

%\BEGIN{itembox}[c]{
%\textcolor{black}{
%{\bf
%Axiom 1({{measurement}}) Prototype
%}
%}
%}
%\label{rule102}
%\BEGIN{itemize}
%%\item[(i)]
%%{{measurement}}
%\item[]
%{
%When
%an {\bf {{observer}}}
%takes a measurement
%of an {\bf observable ${\mathsf O}$}
%(or,
%by a {\bf {measuring instrument}${\mathsf O}$}
%)
%for a {\bf {measuring object}}
%with a
%{\bf {{state}}}
%$\omega$,
%the probability
%that
%a {\bf measured value } $x$
%is obtained
%is given by
%$P $.}
%\END{itemize}
%\END{itembox}
%%,
%%{{observer}}==
%%{(}That is,
%%{{observer}}%\textcolor{black}{NZ},  and {)}{Remark }.
%%%, (i),
%%%\textcolor{black}{NZ}\textcolor{black}{({{}}U$_4$)} and , %\textcolor{black}{NZ}, .

Here,
measurement theory assert that
\begin{itemize}
\item[]
Describe every phenomenon
modeled on
Axiom 1
under the direction of
the Copenhagen interpretation
[\textcolor{black}{({{}}U$_1$)--({{}}U$_7$)}]
%[\textcolor{black}{({{}}U$_1$)--({{}}U$_7$)}]}
\end{itemize}
% and {{measurement theory}}.

\par
\noindent
\par
\noindent
%BBBBBBBBBBBBBBBBBB%SBSBSBS
\par
\noindent
{\small%%{\footnotesize
\vspace{0.1cm}
\begin{itemize}
\item[$\spadesuit$] \bf {{}}{Note }1.8{{}} \rm
Summing up
\textcolor{black}{({}M$_2$)},
we see
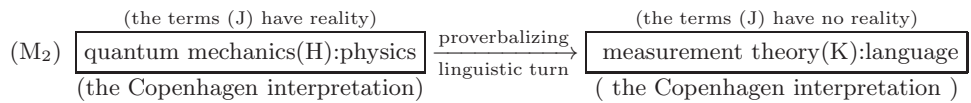
\begin{itemize}
\item[({}M$_2$)]
$
\overset{\text{\scriptsize (the terms (J) have reality)}}{\underset{
\text{(the Copenhagen interpretation)}}
%the Copenhagen interpretation\text{\scriptsize \textcolor{black}{[({{}}U$_1$)--({{}}U$_7$)]}})}
{\text{\fbox{quantum mechanics(H):{physics}}}}}
\xrightarrow[\text{\scriptsize linguistic turn}]{\text{\scriptsize proverbalizing}}
\overset{\text{\scriptsize (the terms (J) have no reality)}}{\underset{\text{(
the Copenhagen interpretation
%\text{\scriptsize \textcolor{black}{[({{}}U$_1$)--({{}}U$_7$)]}}
)}}{\text{\fbox{
{{measurement theory}}(K):language}}}}
$
\end{itemize}
Therefore,
\begin{itemize}
\item[$(\sharp)$]
the Copenhagen interpretation[\textcolor{black}{({{}}U$_1$)--({{}}U$_7$)}]
is common in
both
quantum mechanics(H)
and
{{measurement theory}}(K).
\end{itemize}

\end{itemize}
}
%%BBBBBBBBBBBBBBBBBB%SBSBSBSS

\par
For example,
the following
is one of the simplest dualistic statements.
%, dualism{linguistic world-description method}%\textcolor{black}{NZ} and ,
\begin{itemize}
\item[({{}}V)]
When an {\bf observer}
measures by the exact ruler
(i.e.,
{\bf {measuring instrument}}
)
for the
pencil
(i.e.,
{\bf {measuring object}}
)
with
the {\bf state}
(
the length 15.3 cm
),
the
{\bf probability}
that
{\bf measured value }[15.3 cm]
is obtained
is
$1$.
\end{itemize}

Here,
the region in which
there is a possibility that
a state $\omega$ [resp.
a measured value $x$]
exists
is called a state space
[resp.
measured value space],
and denoted by
$\Omega$
[resp.
$X$
].

In the above
(V),
we may consider that
$\Omega =[5,30]$.
Also,
if the rule's length
is 25cm,
we may put
$X=[ 0, 25]$.

%%\textcolor{black}{NZ},
%5 cm
%
%30 cm
% and ,
%state space $\Omega=$
%$ [5,30]$
%(=
%$5{{\; \leqq \;}}\omega {{\; \leqq \;}}30$)
% and .
%, 0 cm
%
%50 cm
%{{measurement}},
%measured value 
%$X= [0,50]$
% and .
\par

\par
\vskip0.5cm

\par
%%\textcolor{black}{NZ}(V){{measurement}}%\textcolor{black}{NZ},
%{{measurement theory}}, {{measurement}} and
%.
%%\textcolor{black}{NZ}.
%%\textcolor{black}{NZ}%\textcolor{black}{NZ},
% and .

The following example is typical:
\par
\noindent
\par
\noindent
{\bf Example 1.1}
{\bf [The water is cold or hot?]}$\;\;$%POPOPO
\rm
Let testees drink water with various temperature
$\omega \SD$
$(0 {{\; \leqq \;}}\omega $
${{\; \leqq \;}}100)$.
And
you ask them {\lq\lq}cold"
or {\lq\lq}hot"
alternatively.
Gather the data,
(
for example,
$g_{c}(\omega)$ persons say {\lq\lq}cold",
$g_{h}(\omega)$ persons say {\lq\lq}hot")
and
normalize them,
that is,
get
the polygonal lines
such that
\begin{align*}
&
f_{c}(\omega)= \frac{g_{c}(\omega)}{\text{the numbers of testees}}
\\
&f_{h}(\omega)=\frac{g_{h}(\omega)}{\text{the numbers of testees}}
\end{align*}
And
\begin{itemize}
\item[]
$
\qquad
\qquad
f_{c} (\omega)
=
\cases
1 & \quad (0 {{\; \leqq \;}}\omega {{\; \leqq \;}}10 ) \\
\frac{70- \omega}{60}  & \quad (10 {{\; \leqq \;}}\omega {{\; \leqq \;}}70 ) \\
0 & \quad (70 {{\; \leqq \;}}\omega {{\; \leqq \;}}100 )
\endcases,
$
$
\qquad
f_{h} (\omega) = 1- f_{c} (\omega)
%%P%%P%\TAG{2.4}
$
\end{itemize}
\par
%\vskip5.9cm
\par
\noindent
\noindent
%%\vskip-0.4cm%%%
%\begin{figure}[htbp]
%%%%eepic
\unitlength=0.28mm
\begin{picture}(500,100)
\put(-80,0){{{
\thicklines
\path(100,2)(597,2)
\path(100,2)(100,100)
\path(97,70)(104,70)%
\put(87,70){1}
\put(60,0){
\path(40,70)(85,70)(382,3)(537,4)
\path(40,4)(85,4)(382,70)(537,70)
}
\put(170,90){$f_{c}$}
\put(445,90){$f_{h}$}
\multiput(98,0)(50,0){11}{
\dottedline{2}(0,0)(0,10)
}
\multiputlist(98,-10)(50,0)
{
%{\footnotesize BC700},
{\footnotesize 0},{\footnotesize 10},
{\footnotesize 20},{\footnotesize 30},
{\footnotesize 40},{\footnotesize 50},
{\footnotesize 60},{\footnotesize 70},
{\footnotesize 80},{\footnotesize 90},
{\footnotesize 100}
%,{\footnotesize 70},
}
%%%%%%%%%%%%%%%
}}}
\end{picture}
\vskip0.3cm
\begin{center}{Figure 1.2:
Cold or hot?}
\end{center}
\par
\noindent
Therefore, for example,
\begin{itemize}
\item[({{}}W$_1$)]
You choose one person from the testees,
and
you ask him/her {\lq\lq}cold"
or {\lq\lq}hot"
alternatively.
Then the probability that
he/she
says
$
\left[\begin{array}{ll}
{}
\text{"cold"}
\\
%[{\mathsf O}]
{}
\text{"hot"}
\end{array}\right]
$
is
given by
\rm
$
\left[\begin{array}{ll}
{}f_{\text \rm c}(55)=0.25
\\
{}
f_{\text \rm h}(55)=0.75
\end{array}\right]
$
\end{itemize}

This is described in terms of
Axiom 1
in what follows.
%\BEGIN{itemize}

Define the state space
$\Omega$
such that
$\Omega$
$=$ interval $[0, 100](\subset {\mathbb R})$
and measured value space $X=\{c, h\}$.
Here, consider
the
"[C-H]-thermometer"
such that
\begin{itemize}
\item[($W_2$)]
for water with $\omega \SD$,
[C-H]-thermometer
presents
$
\left[\begin{array}{ll}
{}
\text{{{{{c}}}}}
%[{\mathsf O}]
%
%\text{
\\
{}
\text{{{{{h}}}}}
%[{\mathsf O}]
%
%\text{
%{}[{\mathsf O}] \text{ and {measuring instrument}}
\end{array}\right]
$
with probability
\rm
$
{\roman
\left[\begin{array}{ll}
{}
f_{\text \rm {{{{c}}}}}(\omega)
%[{\mathsf O}]
%
%\text{
\\
{}
f_{\text \rm {{{{h}}}}}(\omega)
%[{\mathsf O}]
%
%\text{
%{}[{\mathsf O}] \text{ and {measuring instrument}}
\end{array}\right]
}
$.
This [C-H]-thermometer
is denoted by
${\mathsf O} =$
$(f_{{c}},f_{{h}})$
%\\
%\\
\end{itemize}
Note that
this [C-H]-thermometer
can be easily realized
by
"random number generator".

Here, we have the following identification:
\begin{itemize}
\item[({{}}W$_3$)]
$\qquad$
({{}}W$_1$)
$\Longleftrightarrow$
({{}}W$_2$)
%
%%\textcolor{black}{NZ}%\textcolor{black}{NZ}, $55$%\textcolor{black}{NZ}{{{{cup}}}}%\textcolor{black}{NZ}
%{{{h}}}?{{{{c}}}}?
% and {{measurement}} and ,
%{{{{c}}}}{{{{h}}}}-{measuring instrument}${\mathsf O}${{measurement}}
% and {}
%That is,
%%\textcolor{black}{NZ}%\textcolor{black}{NZ}%\textcolor{black}{NZ}{$\cdots \cdots$}
%%, $55$%\textcolor{black}{NZ}{{{{cup}}}}%\textcolor{black}{NZ}
%%{{{h}}}?{{{{c}}}}?
% and %\textcolor{black}{NZ},
%,
%{{{{c}}}}{{{{h}}}}-{measuring instrument}${\mathsf O}$
%{{measurement}} and %\textcolor{black}{NZ} and  and
%.
\end{itemize}
Therefore,
the statement
\textcolor{black}{({{}}W$_1$)}
in ordinary language
can be represented
in terms of measurement theory
as follows.
\begin{itemize}
\item[({{}}W$_4$)]
When
an {\bf {{observer}}}
takes a measurement
by
$
{\underset{{ \scriptsize
\text{{\bf {measuring instrument}}}
{{\mathsf O} =(f_{\text {{{{c}}}}},f_{\text {{{{h}}}}})}
}}
%{AAA}
{\text{[[C-H]-{instrument}]}}
}
$
%by a {\bf {measuring instrument}${\mathsf O}$}
%)
\\
\\
for
$
\underset{\text{ \scriptsize  ({\bf System {(}{measuring object})})}}{\text{[{{{{water}}}}]}}
$
with
$
{\underset{\text{ \scriptsize ({\bf {{state}}}$(=\omega \in \Omega)$
)}}{\text{[55$\SD$]}}}
$,
the probability
that
{\bf measured}
{\bf value }
\\
\\
\\
$
\left[\begin{array}{ll}
{}
\text{{{{{c}}}}}
%[{\mathsf O}]
%
%\text{
\\
{}
\text{{{{{h}}}}}
\end{array}\right]
$
is obtained
is given
by
\rm
$
\left[\begin{array}{ll}
{}
f_{\text \rm {{{{c}}}}}(55)=0.25
\\
{}
f_{\text \rm {{{{h}}}}}(55)=0.75
\end{array}\right]
$
\end{itemize}
In the above,
note that,
\begin{itemize}
\item[({{}}W$_5^{\roman a}$)]
The terms in
(J$_1$)
are used
%Axiom 1%\textcolor{black}{NZ} and
\item[({{}}W$_5^{\roman b}$)]
It is due to
the Copenhagen interpretation
[\textcolor{black}{({{}}U$_1$)--({{}}U$_7$)}].
(although
((U$_2$), (U$_6$), (U$_7$)
are not used,
these will be used
in the relation with
Axiom 2({Chap.{\;}}6{}).
% and %\textcolor{black}{NZ}))
\end{itemize}

\par

This example will be again discussed in the following
chapter(\textcolor{black}{Example 2.7}).

%.

\par
\noindent
%BBBBBBBBBBBBBBBBBB%SBSBSBS
\par
\noindent
{\small%%{\footnotesize
\vspace{0.1cm}
\begin{itemize}
\item[$\spadesuit$] \bf {{}}{Note }1.9{{}} \rm
It may be understandable to
consider
"observable" ="the partition of word".
%({\bf %\textcolor{black}{NZ}}
% and 
%{\bf %\textcolor{black}{NZ}}
% and {)}
For example,
\textcolor{black}{Fig. 1.2}
says that
$\big( f_{{c}}, f_{{h}} \big)$
is the partition
between
"cold" and "hot".
Also,
"{measuring instrument}"
is the instrument that choose a word among words.
In this sense,
we consider that
"observable"=
"measurement instrument".
Also,
John Locke's
famous
sayings
{\it
{\lq\lq}primary quality
{\rm
(e.g.,
length, weight, etc.)}{\rq\rq}}
and
{\it
{\lq\lq}secondary quality
{\rm (e.g.,
sweet, dark, cold, etc.)}{\rq\rq}}
urge us to associate the following correspondence:
$$
\text{state}\! \leftrightarrow \! \text{primary$\!$ quality},
\;
\qquad
\text{observable} \! \leftrightarrow \! \text{second$\!$ quality}
$$
These words form the basis of dualism.
%
%
%, %\textcolor{black}{NZ}${{\cdot}}$(1632--1704
%)%\textcolor{black}{NZ} and 
%%
%{Chap.{\;}} and {Chap.{\;}}
%(%\textcolor{black}{NZ}%\textcolor{black}{NZ}), %\textcolor{black}{NZ}
%{{measurement theory}}%\textcolor{black}{NZ}
%{{state}} and observable 
% and .
%, ${{\cdot}}$%\textcolor{black}{NZ}
%{Chap.{\;}}${{\cdot}}${Chap.{\;}}%\textcolor{black}{NZ},
%%\textcolor{black}{NZ}
%dualism%\textcolor{black}{NZ}{}
%%index{@${{\cdot}}$}
%%index{@{Chap.{\;}}, {Chap.{\;}}}
\end{itemize}
}
%%BBBBBBBBBBBBBBBBBB%SBSBSBSS
%
\subsubsection{Monism vs. dualism}
%\ssubsubsection{monism vs. dualism
%[I]}
\par
The readers may understand that
dualistic world-description is rather troublesome.
%, monism and ,
% and  and .
Here,
%\textcolor{black}{{{{}}}{Sec.1.2} }[
%monism
%]
recall
\textcolor{black}{Examples (({{}}R$_3$)
and
({{}}R$_4$))}.
That is,
\begin{itemize}
%[B%\textcolor{black}{NZ}%\textcolor{black}{NZ}] and , [300] and {{state}}.
\item[({{}}R$_3$)]
$\;\;$This flower is red
\\
$\Rightarrow$
[The object (i.e., this flower)] has a state such as [red]
%[%\textcolor{black}{NZ} and ], [] and {{state}}
\item[({{}}R$_4$)]
$\;\;$The water in this cup is cold
%$\;\;$%\textcolor{black}{NZ}{cup}%\textcolor{black}{NZ}%\textcolor{black}{NZ}{{{{c}}}}
\\
$\Rightarrow$
[The object (i.e., the water in this cup )] has a state such as [cold]
%,
%
\end{itemize}
The readers may feel
that
these
({{}}R$_3$)
and
({{}}R$_4$)
are unnatural.
Form the dualistic point of view,
"red"
and
"cold"
should be regarded as
"measured vales"
and not
"states".
This is a reason that,
in science
(i,e.,
the monastic world-description),
we may say
"This water is 55",
On the other hand,
in science,
we seldom say
"This water is hot".
The statement
"this water is hot"
is
not represented in
the monism
but
in the dualism.
as seen in
\textcolor{black}{
({{}}W$_4$)}
of
Example 1.1).
That is,
dualistic method
has a power to
clarify the difference between
"{{state}}{(}= the original property of {measuring object}{)}"
and
"measured value".
\begin{itemize}
\item[]
Compared with monism,
the dualistic view
has a possibility
to
extend the range of description.
\end{itemize}
Thus,
we push on to scientific dualism.
In spite that
we do not consider "the mind cannot be solved by the principle of a substance", we believe that
dualism (with a great power of expression) should be adopted.

\par

\par

\par
\noindent

\par
\noindent
\par
\noindent
%BBBBBBBBBBBBBBBBBB%SBSBSBS
\par
\noindent
{\small%%{\footnotesize
\vspace{0.1cm}
\begin{itemize}
\item[$\spadesuit$] \bf {{}}{Note }1.10{{}} \rm
Readers may want to
know
the conclusion of
"monism vs. dualism".
%%\textcolor{black}{NZ}.
In physics,
this is discussed in
Einstein--Bohr debate is characterized
as
"monism\textcolor{black}{{\cite{Eins}}}vs.dualism\textcolor{black}{{\cite{Bohr}}}"
in physics.
This is still unsolved.
%%\textcolor{black}{NZ},
%%index{@Einstein--}
%%index{@}
%,  and ,
%{the theory of relativity} and quantum mechanics%\textcolor{black}{NZ}
%
%{physics}%\textcolor{black}{NZ}{}
We think that
\begin{itemize}
\item[$(\sharp_1)$]
In the linguistic method,
dualism is superior to
monism.
That is because
the monism is not mature in the linguistic method
(\textcolor{black}{{Note }1.7}).
%
%monism vs. dualism
% and 
%dualism%\textcolor{black}{NZ}
%{{measurement theory}}
\end{itemize}
Also,
we think that
{statistics}
={dynamical system theory})
is
should be regarded as
the abbreviation of measurement theory
(cf. Chap.4).
Summing up the above arguments,
we have the following:
\par
\noindent
%\begin{table}[htbp] \small \caption{realistic and {linguistic world-description methods}
%$\diagdown$
%monism{$\cdot$}dualism
%}
\vskip0.5cm
\begin{center}
Table 1.1:
{realistic and {linguistic world-description methods}
$\diagdown$
monism{$\cdot$}dualism
}
%\BEGIN{tabular}{|l|l|*{2}{@{\quad\$}r|}}
%\BEGIN{tabular}{
\\
\begin{tabular}{
%@{\vrule width 1.8pt\ }c
%@{\vrule width 1.8pt\ }c|c
%@{\vrule width 1.8pt}
@{\vrule width 0.8pt\ }c
@{\vrule width 0.8pt\ }c|c
@{\vrule width 0.8pt}
}
\noalign{\hrule height 0.8pt}
%\noalign{\hrule height 1.8pt}
%$\diagdown$or
&$\quad$ monism  $\quad$ &$\quad$
$\quad$dualism $\quad$
$\quad$\\
\noalign{\hrule height 0.8pt}
%\noalign{\hrule height 1.8pt}
$\underset{\text{\scriptsize (realistic world-view)}}{\text{realistic world-description method}}$
& $\underset{\text{(
electromagnetism
,$\cdots$)}}{
\text{Newtonian mechanics }}$
&quantum mechanics  \\
\hline
$\underset{\text{(linguistic world-view)}}{\text{linguistic world-description method}}$
& $\underset{\text{({Note }1.7)}}{\text{state equation method}}$ & {
\text{measurement theory}}  \\
\noalign{\hrule height 0.8pt}
%\noalign{\hrule height 1.8pt}
\end{tabular}
\end{center}
%\end{table}
\par
\noindent
\par
\noindent
Therefore, {{measurement theory}}
has two kinds of "absurdness."
That is,
%idealism=linguistic world-view({Sec.8.1}) and ,
\begin{itemize}
\item[$(\sharp_2)$]
$
\text{the absurdness of {{measurement theory}}}
\cases
\text{idealism} &{\cdots} \text{linguistic world-view}
\\
\\
\text{dualism} &{\cdots}
\text{the Copenhagen interpretation}
\endcases
$
\end{itemize}
\end{itemize}
}
%%index{ and @}

%
%%BBBBBBBBBBBBBBBBBB%SBSBSBSS

\par

\par
\noindent
\par
\noindent
%BBBBBBBBBBBBBBBBBB%SBSBSBS
\par
\noindent
{\small%%{\footnotesize
\begin{itemize}
\item[$\spadesuit$] \bf {{}}{Note }1.11{{}} \rm
Readers may want to ask:
$$
\text{
What is science (other than physics) ?
}
$$
%\BEGIN{itemize}
%\item[]
The answer will be presented in {Sec.8.1}(m).
For the time being, think
%read, and advance.
%, %\textcolor{black}{NZ}\textcolor{black}{}2,
$$
\text{\bf various sciences (other than physics )=engineering}
$$
%,
%$$
%[]=[]+[]
%$$
% and ,
%%\textcolor{black}{NZ}
%%\textcolor{black}{NZ}.
We add some remark in what follows.
For example,
consider the question:
$$
\text{
a heart transplant operation
is medical or
engineering?
}
$$
A heart transplant operation
is,
of course,
most advanced science and technology.
However,
it is still unripe
if we regard a heart transplant operation
as engineering.
That is because
we can not yet create the robot
who has the technology of a heart transplant.
Such a robot will be realized in the development.
The heart transplant operation is still considered to be infancy
as one field of engineering.
\end{itemize}
}
%%BBBBBBBBBBBBBBBBBB%SBSBSBSS

\subsection{{Ordinary language}---
lawless area
---}%{Sec.1.3}
%BBBBBBBBBBBBBBBBBB%SBSBSBS
It is sure that
\begin{itemize}
\item[]
Ordinary language is human beings' greatest invention.
\end{itemize}
But,
ordinary language is elusive ambiguous and very strange monster
languages to take in, and it cannot be said that the framework is clear
Thus,
let us explain the relation among
ordinary language,
scientific language
and
mathematical language.

\par
For example,
clearly
mathematics
is included
in
the {ordinary language}
statements
(R$_1$),
(R$_2$)
and
(W$_1$).

The word Greek "logos" has a meaning of both "logic" and "language."
Thus,
considering that
"mathematics{ + }ordinary language"
$\subset$
wide ordinary language}",
we may assert that
"wide {ordinary language",
is the
Origin language
before
{world-description}(\textcolor{black}{{Chap.$\;$1}(O)}).

\renewcommand{\footnoterule}{
  \vspace{2mm}                      % %\textcolor{black}{NZ}
  \noindent\rule{\textwidth}{0.4pt}  
  \vspace{-3mm}
}

%%%%baseline
\baselineskip=18pt
\normalsize
\baselineskip=18pt
\par
If it be so,
we may see:
%{
%\ssmall
\begin{itemize}
\item[$\underset{}{\text{\normalsize (X$_1$)}}$]
the relation between
widely {ordinary language}
and{world-description method}:
{
%\ssmall
\\
$\overset{
}{\underset{\text{\scriptsize (before science)}}{
\text{
\fbox
{{\textcircled{\scriptsize 0}}
widely {ordinary language}}
}
}
}
$
$
\underset{\text{\scriptsize }}{\text{$\Longrightarrow$}}
$
$
\underset{\text{\scriptsize (Chap. 1(O))}}{\text{{world-description}}}
\cases
&
\!\!\!\!\!
\underset{\text{\scriptsize (Newtonian mechanics,etc.)}}{
{\textcircled{\scriptsize 1}}
{\text{realistic method} \qquad}
}
%}
\\
\\
&
\!\!\!\!\!
\underset{{\text{\scriptsize (measurement theory)}}}{
\text{\textcircled{\scriptsize 2}}
{\text{linguistic method}}
}
%{\text{\textcircled{\scriptsize 2}{linguistic method}(e.g., {{measurement theory}})}}
\endcases
$
}
\end{itemize}
\renewcommand{\footnoterule}{%
  \vspace{2mm}                      % %\textcolor{black}{NZ}
  \noindent\rule{\textwidth}{0.4pt}   % %\textcolor{black}{NZ}, 
  \vspace{-2mm}
}

%index{@{Chap.$\;$1}(X$_1$)}
And thus, we think:
\begin{itemize}
\item[(X$_2$)]
$\qquad$
every science holds
as a premise of
widely {ordinary language}\textcircled{\scriptsize 0}.
\end{itemize}
\renewcommand{\footnoterule}{%
  \vspace{2mm}                      % %\textcolor{black}{NZ}
  \noindent\rule{\textwidth}{0.4pt}   % %\textcolor{black}{NZ}, 
  \vspace{-2mm}
}

But, it is a matter of course
that
{ordinary language}
is not created for science,
and
the framework of
widely {ordinary language}
is not clear.
%%\textcolor{black}{NZ}.
Even if
there exists
the rule of {ordinary language},
it is too complicate
to write it.
Thus,
{ordinary language}
is so-called
"lawless area"
or
"disorderly language".

Although there may be many opinions,
it is sure that
the following is one of them:
\begin{itemize}
\item[(X$_3$)]
"\textcircled{\scriptsize 0} widely {ordinary language}"
includes
%%\textcolor{black}{NZ},
the statement
\textcolor{black}{(R$_1$),
(R$_2$),
(W$_1$) in Example 1.1}
and further,
%\BEGIN{itemize}
arithmetical word problems
(
{(\textcolor{black}{Example 7.6})}
or
{\textcolor{black}{(Chap.11 (A$_2$))}}
and so on),
\quad
%\underset{({Chap.{\;}}4, 7, 11{})}
{statistics
(=dynamical system theory)}
\end{itemize}
That is,
there is a reason to consider that
these somehow sink into ordinary language.

But,
{{measurement theory}}
is composed of two rules
(Axioms 1 and 2).
And thus,
it has the clear framework.
Therefore,
as
the statement (W$_1$)
in
\textcolor{black}{Example 1.1}
in
{ordinary language}
was written by
the measurement theoretical (W$_4$),
\begin{itemize}
\item[(X$_4$)]
As much as possible,
we start from
"\textcircled{\scriptsize 2}\ {{measurement theory}}"
and
not
"\textcircled{\scriptsize 0}\ widely {ordinary language}".
\end{itemize}

And
%%\textcolor{black}{NZ}
%,
\begin{itemize}
\item[(X$_5$)]
to establish metaphysics(called measurement theory)
as a discipline which forms the base of science,
or
as a basic language
by which
sciences are described.
Or equivalently,
\begin{itemize}
\item[]
\bf
{{measurement theory}}
is the special language by which
sciences are described.
And reversely,
sciences hold by the measurement theoretical description.
\end{itemize}
\rm
\end{itemize}
This is our purpose of this print.

\par
\noindent
\par
\noindent
%BBBBBBBBBBBBBBBBBB%SBSBSBS
\par
\noindent
{\small%%{\footnotesize
\vspace{0.1cm}
\begin{itemize}
\item[$\spadesuit$] \bf {{}}{Note }1.12{{}} \rm
The readers may ask:
$$
\text{
Why do we need to assert the
(X$_5$)?
}
$$
This answer was already presented in \textcolor{black}{{Note }1.11}.
That is,
\begin{itemize}
\item[]
%$\quad$
{{measurement theory}}
is the language for engineering
(or, sciences),
or
in other words,
for
quantification, mechanization, and automation.
%since it is useful to
\end{itemize}
%(, , ) and ,
%,
%(, )${{\cdot}}$
%
%%\textcolor{black}{NZ}
% and .
There may be several opinion for
the following disciplines:
\begin{itemize}
\item[]
Computational psychology, mathematical economics, financial engineering, cognitive science, management engineering, a medical engineering, educational technology, ergonomics, etc.
\end{itemize}
However,
in order to regard these
as sciences, there is no method besides describing these in terms of measurement theory.
Also, some may ask:
\begin{itemize}
\item[({}A$_1$)]
Why does measurement theory
hold?
$\;\;$
Is there another scientific language?
\end{itemize}
However, we have no answer to this question.
Let us add "two purposes" to
(X$_1$)
in what follows.
\small
\\
\\
$\underset{(Chap. 1)}{\text{(X$_1$)}}$
{\small
$\overset{
%({ordinary language})
}{\underset{\text{\scriptsize ({world-description}(=before science)}}{
\text{
\fbox
{{\textcircled{\scriptsize 0}}
widely {ordinary language}}
}
}
}
$
$
\underset{\text{\scriptsize }}{\text{$\Longrightarrow$}}
$
$
\underset{\text{\scriptsize (Chap. 1(O))}}{\text{{world-description}}}
\cases
&
\!\!\!\!\!
\overset{\text{\scriptsize (clarify God's principle)}}
{\underset{\text{\scriptsize (Newtonian mechanics,etc.)}}{
{\textcircled{\scriptsize 1}}
{\text{realistic method}}
}
}
%}
\\
\\
&
\!\!\!\!\!
\overset{\text{\scriptsize (create robots like human being)}}
{\underset{{\text{\scriptsize (measurement theory)}}}{
\text{\textcircled{\scriptsize 2}}
{\text{linguistic method}}
}
}
%{\text{\textcircled{\scriptsize 2}{linguistic method}(e.g., {{measurement theory}})}}
\endcases
$
}
\end{itemize}
}
%%BBBBBBBBBBBBBBBBBB%SBSBSBSS

\par
\noindent
Lastly,
we add the classification of
measurement theory
(as mentioned in the guide of this print).

\begin{itemize}
\item[$\underset{\text{\scriptsize }}{\text{(Y)}}$]
$
%\underset{(quantum mechanics)}
\underset{\text{\scriptsize (scientific language)}}{\text{
%\footnotesize
{{{measurement theory}}}}}
%{\footnotesize {{{measurement}}}}
\cases
\underset{}{\text{
%\footnotesize
quantum {{{measurement}}}}}:
\text{\footnotesize \textcolor{black}{{Chap. 3}}}
\\
\underset{}{\text{
%\footnotesize
classical {{{measurement}}}}}
%\footnotemark
\cases
\!\!\!
\text{\small continuous}
\cases
\!\!
\text{\footnotesize pure type}
%\longleftarrow
:
\text{\textcolor{black}{\footnotesize (Chaps. 2$\text{--}$9)}}\\
\!\!
\text{\footnotesize mixed type}
:
\text{\footnotesize
%\te
\textcolor{black}{{{{}}}{Sec.4.4}}
}
\\
\endcases
\\
\\
\!\!\!
\text{\small bounded}
\cases
\!\!
\text{\footnotesize pure type}
:
\text{\textcolor{black}{\footnotesize Chaps. 10, 11}}\\
\!\!
\text{\footnotesize mixed type}
:
\text{\footnotesize
\textcolor{black}{{Note }11.3}
}
\\
\endcases
\endcases
\endcases
$
\end{itemize}

%\footnotetext{
%,  and .
%, classical mechanical {world-view}({dynamical system theory}${{\cdot}}${statistics})
%, , {{measurement theory}}{},
%quantum mechanical {world-view}({{measurement theory}}\textcolor{black}{({}Y)})
%%\textcolor{black}{NZ}%\textcolor{black}{NZ}{}
%
%%\textcolor{black}{NZ} and .
%}
%\BEGIN{center}
%	0.2: {{measurement theory}}%\textcolor{black}{NZ}
%\END{center}
%\par
%{{{measurement theory}}}quantum mechanics%\textcolor{black}{NZ}%\textcolor{black}{NZ}
%{physics} and \footnote{{{measurement theory}}quantum mechanics, ,{\;}
%{\bf quantum mechanics} and . }. %\textcolor{black}{NZ}
%%\textcolor{black}{NZ}, {{measurement theory}} and 

%%%BBBBBBBBBBBBBBBBBequilibrium statistical mechanics
\par
\noindent

\renewcommand{\footnoterule}{%
  \vspace{2mm}                      % %\textcolor{black}{NZ}
  \noindent\rule{\textwidth}{0.4pt}   % %\textcolor{black}{NZ}, 
%  \vspace{-0.5pt}                     % %\textcolor{black}{NZ}
  \vspace{-5mm}
}

%========================================================
%\END{document}

%part I
%\part{classical {{{measurement theory}}}${{\cdot}}$}
%II II II II II

\baselineskip=18pt

\font\twvtt = cmtt10 scaled \magstep2
\font\fottt = cmtt10 scaled \magstep4
\par

\vskip3.5cm
\noindent
\par
\noindent
%222222222222222222222222222222222222222222222\tag{\tag
\section{Axiom${}_{\text{\scriptsize c}}^{\text{\scriptsize p}}$ 1
---
{{measurement}} \label{Chap2}}%{Chap.{\;}}2{}
%%\vspace{-0.8cm}
\noindent
{
\begin{itemize}
\item[{}]
{
\small
\par%[Abstract].
\rm
%\te
In Part II,
we study "classical (continuous pure type) measurement thory"
in the whole picture (Y) of measurement theory mentioned in
\textcolor{black}{Section1.3}.
This is characterized as follows:
\begin{align*}
\underset{\text{\scriptsize (scientific language)}}{\text{{} $\fbox{{{measurement theory}}}$}}
:=
{
%\overset{\text{\scriptsize [Axiom 1\textcolor{black}{(Sec. \REF{2secAxiom 1})}]}}
\overset{\text{\scriptsize [Axiom${}_{\text{\scriptsize c}}^{\text{\scriptsize p}}$ 1]}}
{
\underset{\text{\scriptsize
[probabilistic interpretation]}}{\text{{} $\fbox{{{measurement}}}$}}}
}
+
{
%\overset{\text{\scriptsize [Axiom 2\textcolor{black}{(SEc. \REF{6secAxiom 2})}]}}
\overset{\text{\scriptsize [Axiom${}_{\text{\scriptsize c}}^{\text{\scriptsize p}}$ 2]}}
{
\underset{\text{\scriptsize [{{the Heisenberg picture}}]}}
{\text{{}$\fbox{ causality }$}}
}
}
\end{align*}
}
{{{measurement theory}}} proclaims that
\small
\begin{itemize}
\item[$(\sharp)$]
After the example of the sentences of
\textcolor{black}{Axiom${}_{\text{\scriptsize c}}^{\text{\scriptsize p}}$ 1 and Axiom${}_{\text{\scriptsize c}}^{\text{\scriptsize p}}$ 2},
%the teacher,
every phenomenon should be described.
Or,
making a model of the sentences of
\textcolor{black}{Axioms 1 and 2},
describe every phenomenon.
\end{itemize}
In this chapter,
we are devote ourselves to
the mathematical formulation of Axiom${}_{\text{\scriptsize c}}^{\text{\scriptsize p}}$ 1
(measurement).
Axiom${}_{\text{\scriptsize c}}^{\text{\scriptsize p}}$ 2
(causality)
will be discussed in Chapter 6 later.
\vskip0.3cm
\end{itemize}
}
%\END{document}

%\cite{IIJ}

\normalsize
\baselineskip=18pt
\subsection{
Classical (continuous pure type) measurement theory}%2.1
%
%index{ and @locally compact space}
\rm
\subsubsection{{{State}} and Observable
---
the first quality and the secondary quality}%{Sec. 2.1.1}
%index{@{Chap.{\;}}, {Chap.{\;}}}
\par
If we expect that many scientists are interested in the philosophy of science
(i.e.,
the studies about the metaphysical aspects of science),
we can not do it without mathematics.
Therefore,
in this section
\textcolor{black}{2.1.1},
we prepare some elementary mathematical results.

%%\textcolor{black}{NZ}.
%
%({Sec. 2.1.2}),
%, {}
%

\par
Throughout this print,
put
${\mathbb N}=\{1,2,3,\ldots\}$
(i.e.,
the set of all natural numbers),
%%index{n@${\mathbb N}=\{1,2,3,\ldots\}$}
%%\textcolor{black}{NZ}
${\mathbb N}_0=\{0,1,2,\ldots\}$
(i.e.,
the set of all non-negative integers),
%%index{n0@${\mathbb N}_0=\{0.1,2,\ldots\}$}
${\mathbb Z}=\{0, \pm 1, \pm 2,\ldots\}$
(i.e.,
the set of all integers),
%%index{z@${\mathbb Z}$:}
%(%\textcolor{black}{NZ})
${\mathbb R}$
(i.e.,
the set of all real numbers).
%(\textcolor{black}{Appendix B.1)}.
%index{@${\mathbb R}$}
%index{r@${\mathbb R}$:}
Let $X$
be a set.
And let
${\cal P}(X)$
or
$2^X$
be a set of all subsets of $X$.
%%index{power@${\cal P}(X)$,$2^X$:}
%%index{@${\cal P}(X)$,$2^X$}
That is,
\begin{align*}
{\cal P}(X)=
2^X=\{ \Xi \;|\; \Xi \subseteq X \}
\end{align*}
%,
%$X${(}That is,
%{finite set}, ,
%$X=\{x_1,x_2,\ldots\}$
%%\textcolor{black}{NZ})%\textcolor{black}{NZ}, ${\cal P}(X)$
%$2^X$ and  and {red}{(Appendix B.1)}.
%%%index{@}}

\par
Let
$\Omega$
be
a
{\bf locally compact space},
%(\textcolor{black}{Appendix B.3(E)}),
for example,
$\Omega=$
${\mathbb N}$(
or,
the set of all natural numbers),
the real line
${\mathbb R}$,
the interval in
${\mathbb R}$,
the plane
(=2-dimensional space)
${\mathbb R}^2$
and so on.
%
%
%
%, 
%locally compact space$\Omega$
%.
% and ,
%%{finite set}$\{ \omega_1, \omega_2,\ldots, \omega_n \}$,
%%\textcolor{black}{NZ}${\mathbb N}$,
%%\textcolor{black}{NZ}${\mathbb R}$
%n
%$[a,b]$
%%
%%
%%$(a,b)$
%%.
%,
%2(={)}${\mathbb R}^2$
%%,
%%${\mathbb N}^2$
%locally compact space.
It is usually assumed that
a finite set
$\Omega$
has the
{discrete metric}$d_D$,
where
\begin{align*}
d_D(\omega,\omega')=
\cases
1
\quad
&
(\omega \not= \omega')
\\
0
&
(\omega = \omega')
\endcases
\tag{\color{black}{2.1}}
\end{align*}
% and ,
%$\Omega$
%%(, $(\Omega, d_D)$)
%
%locally compact space
% and .
This $( \Omega ,d_D )$
is called a
{\bf discrete metric space}.
%(\textcolor{black}{Appendix B.2(C)}).
%index{@discrete metric space}
%,
%$\Omega${finite set}%\textcolor{black}{NZ} and ,
%, discrete metric space and .
\par
Let
$C (\Omega)$
be the set of all complex valued bounded continuous function
$f: \Omega\to {\mathbb C}$,
that is,
%index{cOmega@$C(\Omega)$:}
%index{@$C(\Omega)$}
\begin{align*}
C (\Omega)
=\{ f \;|\;
f
\text{ is a complex valued continuous function on $\Omega$
such that}
\|f\|_{C (\Omega)}
%{
%
%\ssup}_{\omega \in \Omega} |f (\omega)|
< \infty
\}
%footnotmark
\end{align*}
where the norm
$\|f\|_{C (\Omega)} $
is defined by
%
%$\{ |f(\omega)| \in {\mathbb R} \;\;|\;\;
%\omega \in \Omega \}$
%%\textcolor{black}{NZ}(\textcolor{black}{Appendix B.2(B)}), That is,
\begin{align*}
\|f\|_{C (\Omega)}
=
{\sup_{\omega \in \Omega} |f (\omega)|
}
%\Big(
%{{=}}
%\text{
%
%$\{ a \in {\mathbb R} \; : \; |f (\omega)|{{\; \leqq \;}}a
%\;(\forall \omega \in \Omega )
%\}$
%%\textcolor{black}{NZ}
%}
%\Big)
%\footnotemark
\tag{\color{black}{2.2}}
\end{align*}
%index{sup@$\sup$:}
It is elementary that
the vector space
$C (\Omega)$
is a Banach space.
%(\textcolor{black}{Appendix B.6}).

\par
Our present purpose is to
read several examples in \textcolor{black}{{{{}}}Section 2.3}.
Thus, we have to add some mathematical preparations.
%
%%\textcolor{black}{NZ}
%%\textcolor{black}{NZ}{{measurement}}%\textcolor{black}{NZ} and {}
%,
%, ,
%, \textcolor{black}{{{{}}}2.3}
%.

%\par
%\noindent
%\vskip0.3cm
%BFBF
\par
\noindent
{\bf {Definition }2.1
[{}{Observable}, state space,
{{state}}{\rm{}}, measured value{\rm{}},
measured value space{\rm{}}]}$\;\;$%POPOPO
%index{@observable }
%index{@}
%index{@state space }
%index{@{{state}}}
%index{@measured value }
%index{@measured value }
%index{@probability }
\rm
A triplet ${\mathsf{O}} {=}
(X, {\cal F}, F)$
is called an {\it observable ({\rm or}, {measuring instrument})}
in $C(\Omega)$
if
it satisfies that
%index{measured value set@measured value setq{}$B%"[M{}(Babel set)}
%index{label set@label set (=measured value set)}
\rm
\begin{itemize}

\item[({}i)]
[{}Measurable space{}].
\rm %\ssl
$X$ is a set
({}called a
\it
{\lq\lq}measured value set{\rq\rq},
{\lq\lq}sample space{\rq\rq},
\rm %\ssl
or
\it
{\lq\lq}label set{\rq\rq}
\rm %\ssl),
),
and
${\cal F}$
(
$\subseteq$
${\cal P}(X)$
$({}\equiv  \{  \Xi{}: \Xi \subseteq X \}{})$
)
is the field.
%,
%(\textcolor{black}{Appendix B.5(A)}).
That is,
\begin{align*}
&
({\roman a}):
\emptyset
(=\text{empty set})\in {\cal F}, \;\; X \in {\cal F},
\quad
({\roman b}):
\Xi_i \in {\cal F} \quad
(i=1,2,\ldots,n)
%\;\;\; (n=1,2,\ldots)
\Longrightarrow {  \bigcup\limits_{i=1}^n
%\infty
}
\;\;
\Xi_i \in {\cal F}
%\Xi_1 \cap \Xi_2 \in {\cal F}
\\
&
({\roman c}):
\Xi \in {\cal F} \Longrightarrow X \setminus \Xi
%({{=}} \{ x \;| \; x \in X, x \notin \Xi \})
\in {\cal F}
\end{align*}
where
$X \setminus \Xi
(=\{ x \;| \; x \in X, x \notin \Xi \})$,
i.e.,
the compliment of
$\Xi$.
Also, the pair
$(X, {\cal F} )$
is called a (finitely) {measurable space}.
%{red}{(Appendix B.5(A))}.
%
%
%\item[(i)]
%[{}Field{}].
%\rm %\ssl
%$X$ is a set
%({}called a
%\it
%{\lq\lq}measured value set{\rq\rq},
%{\lq\lq}sample space{\rq\rq},
%\rm %\ssl
%or
%\it
%{\lq\lq}label set{\rq\rq}
%\rm %\ssl),
%and
%${\cal F}$
%is the subfield of
%the power set
%${\cal P}(X)$
%$({}\equiv  \{  \Xi{}: \Xi \subseteq X \}{})$,
\rm
\item[(ii)]
[Positivity].
\rm %\ssl
for every $\Xi$ $ \in $ $ {\cal F}     $,
$F({}\Xi)$
is an element in
${C(\Omega)} $
such that
$0 \le  F(\Xi)] \le 1$
(that is,
$0 \le [ F(\Xi)](\omega) \le 1
\;
(\forall \omega \in \Omega)
$
%\footnote{
%$"\forall \omega \in \Omega"$
%
%$(\text{for all }\omega \in \Omega)$
%(
%,
%
%$\Omega$%\textcolor{black}{NZ}%\textcolor{black}{NZ}$\omega$
%
%)
%%\textcolor{black}{NZ}.
%}
,
$F({}\emptyset{}) = 0 $
and
$F({}X{}) = 1 $
({}where
$0$ is the $0$-element in ${C(\Omega)}$),
\rm
\item[(iii)]
[Complete additivity].
\rm %\ssl
for any countable decomposition
$\{ \Xi_1 , \Xi_2, ..., \Xi_n , ... \}$
of $\Xi$,
$\Big($i.e.,
$ \Xi ,$
$ \Xi_n \in {\cal F} ,$
$
\cup_{n=1}^\infty \Xi_n = \Xi,$
$ \Xi_n \cap \Xi_m = \emptyset ({}\text{if }n \not= m   {})
\Big)$,
it holds that
\begin{align*}
 [{}F({}\Xi{}) ](\omega )
=
 \lim_{ N \to \infty }
 \sum_{ n=1}^N
[
{} F({}\Xi_n{}) ]
(\omega )
\; \; \quad ({}\forall \omega  \in \Omega {}).
\tag{2.3}
\end{align*}
%index{@}
\end{itemize}
\rm %\ssl
%The set $X$ is called a {\it measured value set}
\par
\rm
\renewcommand{\footnoterule}{%
  \vspace{2mm}                      % %\textcolor{black}{NZ}
  \noindent\rule{\textwidth}{0.4pt}   % %\textcolor{black}{NZ}, 
  \vspace{-3mm}
}
\rm
%\BEGIN{itemize}
%\item[({}i)]
\par
\noindent
\rm
%\par
%\noindent
%\par
%\noindent
\rm
\par
%\noindent
%%\rm
The $C(\Omega)$
is called a basic algebra.
%{\bf {basic algebra}}
% and .
%%index{@{basic algebra}$C(\Omega)$}
%,
Also,
the
$\Omega$
%{\bf state space }
and its element
$\omega (\in \Omega )$
is respectively
called
a state space
(or,
spectrum)
and
a state.
%{\bf {{state}}} and ,
And,
the
$X$
and
its element
$x (\in X)$
is respectively called a measured value space
and
a measured value.
%
%{\bf measured value }, %\textcolor{black}{NZ}
%{\bf measured value }
% and .
Let
$\omega \in \Omega$.
The triplet
$(X, {\cal F}, [F(\cdot )](\omega) )$
is called a
sample probability space.

% and .

%%BBBBBBBBBBBBBBBBBB%SBSBSBS{\mathsf O}
\par
\noindent
{\small%%{\footnotesize
\vspace{0.1cm}
\begin{itemize}
\item[$\spadesuit$] \bf {{}}{Note }2.1{{}} \rm
Some may think that
the function space $C(\Omega)$ is too simple.
%,  and .
However,
the $C(\Omega)$
or, the $L^\infty(\Omega,\nu)$
(\textcolor{black}{{Chap.{\;}10,11}
})
is an inevitable conclusion from dynamics
(
cf.
\textcolor{black}{\cite{Keio,
Saka}}
).
It is prohibited to consider the other function spaces.
If some intend to improve and extend
measurement theory,
they must start from the improvement of quantum mechanics.
In this sense,
the improvement of
measurement theory may be impossible
(\textcolor{black}{(F$'_3$) in Chap. 1}).
\end{itemize}
}
%%BBBBBBBBBBBBBBBBBB%SBSBSBSS
%\par
%\noindent
%
%
%
%
%
%\par
%\noindent
%\vskip0.3cm
%
\subsubsection{Examples of Observables}%{Sec. 2.1.2}
\par
In what follows, we shall mention several examples of
observables.
%, observable %\textcolor{black}{NZ}.
%\vskip0.1cm
\par
\noindent
%BFBF
{\bf Example 2.2}
{\bf [Existence observable {\rm{}}]}$\;\;$%POPOPO
%\rm
%$X${{measurement}} and  and ,
\par
\noindent
%\bf %BFBF
%\BEGIN{Exa} \label{Example 1.22}
%\rm
%\ssf
%[{}Existence observable
%in ${\cal A}$].
%index{existence observable@existence observable}
\rm
Let
${\mathsf O}^{} \equiv ({}X , {\cal F}, F^{{}})$
be any observable in
a $C(\Omega)$.
Define
the observable
${\mathsf O}^{{\rm (exi)}} \equiv ({}X , \{\emptyset , X\}, F^{{\rm (exi)}})$
in
a basic algebra
${C(\Omega)}$
such that:
\begin{align*}
F^{{\rm (exi)}}({} \emptyset {}) \equiv 0, \; \;
\quad
F^{{\rm (exi)}}({}X{}) \equiv I_{C(\Omega)}, \; \;
%\tag{1.67}
\end{align*}
which may be called the
\it
existence observable
\rm
({}or,
{\it
null observable}).
Consider
any
observable
${\mathsf O} =(X , {\cal F} , F{})$
in
$C(\Omega)$.
%\textcolor{black}{{Definition }2.1}
Note that
$\{ \emptyset , X \}$
$\subseteq$
${\cal F}$.
%(,
%$\{ \emptyset , X \}$
%
%${\cal F}$
%$\sigma -${{field}}
%)
%%
%%$\sigma -${{field}}
%, ,
And we see that
\begin{align*}
[F(\emptyset)](\omega)=0 ,
\quad
[F(X)](\omega)=1 \quad (\forall \omega \in \Omega )
\end{align*}
Thus, we see that
$(X , \{\emptyset , X\} , F^{\roman{(exi)}}{})$
$=$
$(X , \{\emptyset , X\} , F{})$,
and therefore,
we say that
%$C(\Omega)$
%%\textcolor{black}{NZ}
any observable
${\mathsf O} =(X , {\cal F} , F{})$
includes
the existence observable ${\mathsf O}^{\roman{(exi)}}$.
% and 
%.
%
%
% and {}
\par
\noindent
%\vskip0.3cm
%BFBF
\par
\noindent
{\bf Example 2.3}
{\bf [The resolution of the identity $I$
%{}${\bold 1}$%\textcolor{black}{NZ}{(}%\textcolor{black}{NZ}{)}
]}$\;\;$%POPOPO
%index{@$1$}
\rm
Also,
we may find the similarity between
an observable
${\mathsf O}$
and
\it
the resolution of the identity
$I$
\rm
%index{resolution of the identity@resolution of the identity}
in what follows.
Consider
an
observable
${\mathsf O}$
$\equiv$
$({}X , {\cal F} , F{}) $
in ${C(\Omega)}$
such that
$X$ is a countable
set
({}i.e.,
$X \equiv \{ x_1, x_2, ... \}$)
and
${\cal F}={\cal P}(X)$.
Then,
it is clear that
\begin{itemize}
\item[(i)]
$F(\{ x_k \}) \ge 0$
for all $k=1,2,...$
\item[(ii)]
$\sum_{k=1}^{\infty} F(\{ x_k \}) = I_{C(\Omega)}$
in the sense of weak topology
of ${C(\Omega)}$,
\end{itemize}
which imply that
the
$[{}F({}\{ x_k \}{}) \;{}: \; k=1,2,...,n{}]$
can be regarded as
\it
the resolution of the identity element
$I_{C(\Omega)}$.
\rm
Thus we say that
\begin{itemize}
\item[\textcolor{black}{}]
An observable ${\mathsf O}$
$\big($
$\equiv$
$({}X , {\cal F} , F{}) $
$\big)$
in ${C(\Omega)}$
can be regarded as
%index{fuzzy decomposition@fuzzy decomposition}
\begin{align*}
\text{
\it
{\lq\lq}
the resolution of the identity
$I_{C(\Omega)}$
}
%\tag{1.36}
\end{align*}
\rm
i.e.,
$[{}F({}\{ x_k \}{}) \;{}: \; k=1,2,...,n{}]$.
\end{itemize}
\par
\vskip0.3cm
\par
\noindent
%%%
\unitlength=0.40mm
%\vskip-0.4cm%%%
%%%%%%%%%%%%%%%%%
%\begin{figure*}[htbp]
%%%%%%%%%%%%%%%%%
%%%%%%%%%%%%%%%%%%
\begin{picture}(400,110)
\put(27,18){0}
\put(27,108){1}
%\put(80,127){ {\lq\lq}The figure of
%${\mathsf O} \equiv$
%$({}\{ x_1, x_2, x_3 \},$
%$ 2^{\{ x_1, x_2, x_3 \} },$
%$ F)$ in $C({}\Omega{})${\rq\rq} }
\put(44,93){\tiny $[F(\{x_1\})](\omega)$}
\put(170,88){\tiny $[F(\{x_2\})](\omega)$}
\put(300,90){\tiny $[F(\{x_3\})](\omega)$}
%%%%%%%%%
\put(350,18){$\Omega$}
\dottedline{3}(40,110)(340,110)
\put(40,110){\line(1,0){300}}
\put(40,20){\line(0,1){100}}
%\linethickness{0.15mm}
%%\thicklines
\put(40,20){\line(1,0){300}}
%%%%%%%%%%\linethickness{0.15mm}
\thicklines
\spline(40,110)(60,108)(80,102)(100,80)
(150,40)(200,30)(220,20)(240,20)
\spline(120,20)(130,20)(160,30)(250,50)
(270,80)(280,100)(300,105)(340,110)
\spline(40,20)(60,22)(80,28)(100,50)
(150,80)(250,70)
(270,50)(280,30)(300,25)(340,20)
\end{picture}
%\vskip-0.4cm%%%
%%%%%%%%%%%%%%%%%
%\BEGIN{FIGUre*}[htbp]
%%%%%%%%%%%%%%%%%
\vskip-0.5cm
%\caption{
\begin{center}{Figure 2.1:
${\mathsf O} \equiv$
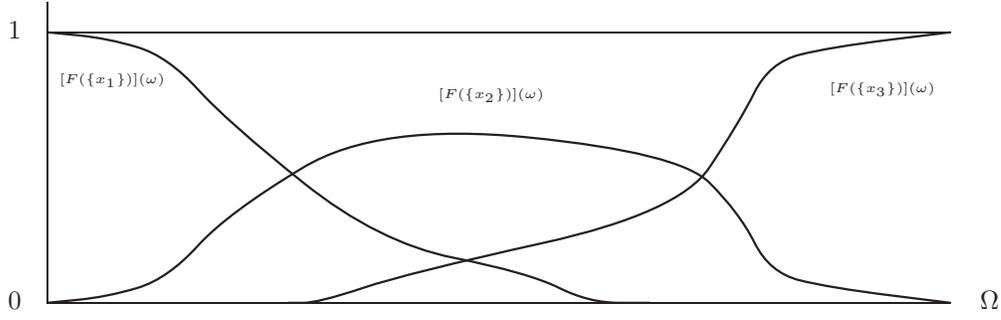
$({}\{ x_1, x_2, x_3 \},$
$ 2^{\{ x_1, x_2, x_3 \} },$
$ F)$ in $C({}\Omega{})$
}
\end{center}
\par
\noindent
In \textcolor{black}{Fig. 2.1},
assume that
$\Omega=[0,100]$
is the
axis of temperatures
($\SD$),
and
put
$X=\{ \text{C(="cold")}$,
$\text{L}$
$\text{(="lukewarm"}$
$=\text{"not hot}$
$\text{enough")}$,
$\text{H(="hot") }$
$
\}$.
And further,
put
$f_{x_1}=f_{\text{\scriptsize C}}$,
$f_{x_2}=f_{\text{\scriptsize L}}$,
$f_{x_3}=f_{\text{\scriptsize H}}$.
Then,
the resolution
$\{f_{x_1}, f_{x_2}, f_{x_3} \}$
can be regarded as
the
word's partition
$
\text{C(="cold")}$,
$\text{L(="lukewarm"="not hot}$
$\text{ enough")}$,
$\text{H(="hot") }
$.

\par
\noindent
Also, putting
\begin{align*}
{\cal F}
(=2^X)
 =\{ \emptyset,
\{x_1\},\{x_2\},\{x_3\},
\{x_1,x_2\},\{x_2,x_3\},\{x_1, x_3\},
X\}
\end{align*}
and
\begin{align*}
&
[F(\emptyset )](\omega ) = 0,
\;\;
[F(X)](\omega)=f_{x_1}(\omega)+f_{x_2}(\omega)+f_{x_3}(\omega)=1
\\
&
[F(\{x_1\})](\omega ) =f_{x_1}(\omega),
\;\;
[F(\{x_2\})](\omega ) =f_{x_2}(\omega),
\;\;
[F(\{x_3\})](\omega ) =f_{x_3}(\omega)
\\
&
[F(\{x_1,x_2\})](\omega ) =f_{x_1}(\omega)+f_{x_2}(\omega),
\;\;
[F(\{x_2,x_3 \})](\omega ) =f_{x_2}(\omega)+f_{x_3}(\omega)
\\
&
[F(\{x_1,x_3\})](\omega ) =f_{x_1}(\omega)+f_{x_3}(\omega)
%\;\;
%[F(\{x_2,x_3 \})](\omega ) =f_{x_2}(\omega)+f_{x_3}(\omega),
\end{align*}
then,
we have
the observable $(X, {\cal F}(=2^X), F)$
in $C([0,100])$.

\noindent
\par
\noindent
{\bf \vskip0.3cm
%BFBF
\par
\noindent
Example 2.4
[{{Triangle observable} {\rm}}{}]}$\;$%POPOPO
%index{@{{triangle observable} }}
\rm
Let testees drink water with various temperature
$\omega$$(0 {{\; \leqq \;}}\omega {{\; \leqq \;}}100)$.
And
you ask them
"How many degrees($\SD$) is roughly this water?
Gather the data,
( for example,
$h_{n}(\omega)$
persons say
$n$
${(}n=0,10,20,\ldots,90,100)$.
and
normalize them,
that is,
get
the polygonal lines.
For example,
define the state space
$\Omega$
by
the closed interval
$[0,100]$
$(
\subseteq {\mathbb R})$.
For each
$n  \in {\mathbb N}_{10}^{100}  = \{0,10,20,\ldots,100\}$,
define the (triangle)
continuous function
$g_{n}:\Omega \to [0,1]$
by
%(\textcolor{black}{FIGU 1.5}):
\begin{align*}
g_{n} (\omega)
=
\cases
0 & \quad (0 {{\; \leqq \;}}\omega {{\; \leqq \;}}n-10 ) \\
{\displaystyle \frac{\omega - n -10}{10} }
& \quad (n-10 {{\; \leqq \;}}\omega {{\; \leqq \;}}n ) \\
{\displaystyle
- \frac{\omega - n + 10}{10}
}
& \quad (n {{\; \leqq \;}}\omega {{\; \leqq \;}}n+10 ) \\
0 & \quad (n+10 {{\; \leqq \;}}\omega {{\; \leqq \;}}100 )
\endcases
\tag{\color{black}{2.4}}
%\tag{\color{black}{1.64}}
\end{align*}
%\omega\omegaxxxxxxxx
\par
\noindent
\par
\par
\noindent
\par
%\newpage
\par
\noindent
\par
\par
\noindent
%%%%eepic
%\vskip-0.4cm%%%
%\begin{figure}[htbp]
\unitlength=0.28mm
\begin{picture}(500,90)
\put(-80,0){{{
\put(98,2){\path(0,0)(50,70)(100,0)(150,70)(200,0)(250,70)(300,0)(350,70)
(400,0)(450,70)(500,0)
}
\put(48,2){\path(50,70)(100,0)(150,70)(200,0)(250,70)(300,0)(350,70)(400,0)
(450,70)(500,0)(550,70)}
\thicklines
\path(100,2)(600,2)
\path(100,2)(100,100)
\path(97,70)(104,70)
\put(87,70){1}
%\path(100,70)(240,70)(380,3)(600,4)
%\put(170,90){$f_{c}$}
%\path(100,4)(240,4)(380,70)(600,70)
%\put(445,90){$f_{h}$}
\multiput(98,0)(50,0){10}{
\dottedline{2}(0,0)(0,10)
}
\multiputlist(98,-10)(50,0)
{
%{\footnotesize BC700},
{\footnotesize 0},{\footnotesize 10},
{\footnotesize 20},{\footnotesize 30},
{\footnotesize 40},{\footnotesize 50},
{\footnotesize 60},{\footnotesize 70},{\footnotesize 80},{\footnotesize 90}
,{\footnotesize 100}
}
\multiputlist(105,80)(50,0)
{
%{\footnotesize BC700},
{\footnotesize $g_{0}$},{\footnotesize $g_{10}$},
{\footnotesize $g_{20}$},{\footnotesize $g_{30}$},
{\footnotesize $g_{40}$},{\footnotesize $g_{50}$},
{\footnotesize $g_{60}$},{\footnotesize $g_{70}$},{\footnotesize $g_{80}$},{\footnotesize $g_{90}$},{\footnotesize $g_{100}$}
}
%%%%%%%%%%%%%%%
}}}
\end{picture}
\vskip0.2cm
%\caption{
\begin{center}{Figure 2.2:
Triangle observable}
\end{center}
\par
\noindent
%\vskip1.0cm
%
\par
\noindent
Putting
$Y={\mathbb N}_{10}^{100}$
and
define the triangle observable
${\mathsf O}^{\triangle}= (Y , 2^Y, F^{\triangle} )$
such that
\begin{align*}
&
[F^{\triangle}(\emptyset )](\omega ) = 0,
\qquad
[F^{\triangle}(Y )](\omega ) = 1
\\
%[F(\emptyset )](\omega ) = 0,
%\quad
&
[F^{\triangle} (\Gamma )](\omega ) = \sum\limits_{n \in \Gamma } g_n (\omega )
\quad
(\forall \Gamma \in 2^{{\mathbb N}_{10}^{100}  })
%\tag{\color{black}{1.65}}
%%P%\TAG{3.5}
\end{align*}
Then,
we have the
triangle observable
${\mathsf O}^{\triangle}= (Y (=
{{\mathbb N}_{10}^{100}  }
), 2^Y, F^{\triangle} )$
in
$C([0,100])$.
%\omega(=47)]} )$
%is
%%\underset{[measurementy{}$B${{\cdot}}${}(B}
%{\text{about 40}}
%\\
%\text{about 50}
%\END{array}\right]
%$
%is given by
%$
%\left[\begin{array}{ll}
%{[F^{\triangle}( \{ 40 \})](47)=0.3}
%%{\text{about 40}}
%\\
%{[F^{\triangle}( \{ 50 \})](47)=0.7}
%\END{array}\right]
%$
%%
%%
%%$
%%\left[\begin{array}{ll}
%%[F^{\mbox{(tri)}}(\{ 40 \})](47)=
%%0.3
%%\\
%%[F^{\mbox{(tri)}}(\{ 50 \})](47)=
%%0.7
%%\END{array}\right]
%%$
%\END{itemize}
%\hfill{{$///$}}%%BFBFbfbf
%\END{Exa}  \rm
%
%eeeeeeeeeeeeeeeeeeeeeeeeeeeeee

%GGGGG
\par
\noindent
\vskip0.3cm
\par
\noindent
%BFBF
{\bf Example 2.5
[{Exact observable} {\rm}{}]}$\;\;$%POPOPO
%%index{@{exact observable} }
\rm
\sf
%[Exact observable in $C(\Omega)${}].
%%index{exact observable@exact observable}
\rm
Consider a commutative basic algebra
$C(\Omega)$.
%in \textcolor{black}{Example \REF{Example 1.2}}.
Let
${\cal B}_\Omega$
be the Borel field,
i.e.,
the smallest $\sigma$-field that contains
all open sets.
For each
$\Xi \in {\cal B}_\Omega$,
define the characteristic
function
$\chi_{{}_\Xi}: \Omega \to {\mathbb R}$
such that
\noindent
\begin{align*}
\chi_{{}_\Xi} ({}\omega{}) =
\cases
1 & \omega \in \Xi  \\
\\
0 & \omega \notin \Xi \quad
\endcases
%\tag{1.66}
\end{align*}
Put $[F^{{\rm (exa)}}(\Xi)](\omega)=\chi_\Xi(\omega)$
$(\Xi \in {\cal B}_\Omega , \omega \in \Omega )$.
The triplet
${\mathsf O}^{{\rm (exa)}}=
(\Omega, {\cal B}_\Omega ,  F^{{\rm (exa)}})$
may be called an {\it exact observable},
if
$\chi_{{}_\Xi} : \Omega \to {\mathbb R}$
is continuous
for all $\Xi (\in {\cal B}_\Omega )$.
For example,
when
$\Omega ={\mathbb N},
{\mathbb Z}$,
the
${\mathsf O}^{{\rm (exa)}}$
is an observable.
However,
when
when
$\Omega ={\mathbb R}$,
it is not so.
Of course,
we want to consider
the
${\mathsf O}^{{\rm (exa)}}$
is an observable.
%And we want to say
%\BEGIN{itemize}
%\item[(G$_1$)]
%the probability
%that
%a measured value obtained by
%the measurement
%$
%{\mathsf M}_{{C(\Omega)}} ({\mathsf O}^{{\rm (exa)}}
%:= ({}\Omega , {\cal B}_\Omega , F^{{\rm (exa)}}),  S_{[
%\delta_\omega]}{})
%$
%is given by
%$
%[F^{{\rm (exa)}}(\Xi)](\omega)
%$
%$(=
%\chi_\Xi(\omega)
%)$.
%\END{itemize}
%Or, equivalently,
%\BEGIN{itemize}
%\item[(G$_2$)]
%a measured value obtained by
%the measurement
%$
%{\mathsf M}_{{C(\Omega)}} ({\mathsf O}^{{\rm (exa)}}
%:= ({}\Omega , {\cal B}_\Omega , F^{{\rm (exa)}}),  S_{[
%\delta_\omega]}{})
%$
%is surely
%equal to
%$\omega$.
%\END{itemize}
However, it is not always true,
%possible
since the exact observable
${\mathsf O}^{{\rm (exa)}}$
does not always exist
(i.e.,
generally,
$\chi_{{}_\Xi}$
$\notin C(\Omega )$
)
in the basic algebraic formulation.
This is a weak point of
the basic algebraic formulation.
For this, we must prepare the bounded type formulation
as mentioned in \textcolor{black}{Chap. 10 and 11}.
However,
it is convenient to consider the
exact observable
${\mathsf O}^{{\rm (exa)}}$.
Thus,
\begin{itemize}
\item[(G$_3$)]
in spite of the wrong usage,
we sometimes use (G$_1$) or (G$_2$).
\end{itemize}
Of course,
when
$\Omega$
is finite,
or
$\Omega$
$=$
${\mathbb N}$,
any function
$f:\Omega \to {\mathbb R}$
is continuous,
and therefore,
we see that
${\mathsf O}^{\FIN}$
is an observable in
$C(\Omega)$.
%%\textcolor{black}{NZ}observable {}%\textcolor{black}{NZ}PPPPPPPFINPPPPP,
%${\mathsf O}^{\FIN} =
%({}\Omega , 2^{\Omega}, F^{\FIN}{})
%$$C(\Omega)$%\textcolor{black}{NZ}
%{\bf {exact observable} } and .
%%%\textcolor{black}{NZ} and ,
%%
%{{}}%\textcolor{black}{NZ}\RR$\!\!\!\!\; \;$
% and {}
\par
\par
%XXXXX\Gamma\Xi \Xi

%%BBBBBBBBBBBBBBBBBB%SBSBSBS{\mathsf O}
\par
\noindent
{\small%%{\footnotesize
\vspace{0.1cm}
\begin{itemize}
\item[$\spadesuit$] \bf {{}}{Note }2.2{{}} \rm
In usual case such as
$\Omega={\mathbb R}$,
the ${\mathsf O}^{\FIN}$
can not regarded as
the existence observable in
$C({\mathbb R})$.
This fact is a weak point in
{{{measurement theory}}}(continuous type).
This will be improved in
{{{measurement theory}}}(bounded type;
\text{\textcolor{black}{{Chap.{\;}10,11}}}
).
However,
in spite of the weak point,
{{{measurement theory}}}(continuous type)
is, of course,
fundamental.
\end{itemize}
}
%%BBBBBBBBBBBBBBBBBB%SBSBSBSS
%
%\ssubsecti

\par
\noindent
{\bf \vskip0.1cm
%BFBF
\par
\noindent
Example 2.6
[Round {observable} is not observable{}]}$\;\;$%POPOPO
%index{@observable }
\rm
%\textcolor{black}{Example 2.4}%\textcolor{black}{NZ}{(}{)}
%{--observable } and ,
% and , observable .
%.
Define the state space
$\Omega$
by
$\Omega = [0,100]$.
For each
$n  \in {\mathbb N}_{10}^{100}  {{=}} \{0,10,20,\ldots,100\}$,
define the discontinuous function
$g_{n}:\Omega \to [0,1]$
such that
\begin{align*}
g_{n} (\omega)
=
\cases
0 & \quad (0 {{\; \leqq \;}}\omega {{\; \leqq \;}}n-5 ) \\
1
& \quad (n-5 {{\; < \;}}\omega {{\; \leqq \;}}n +5) \\
0 & \quad (n+5 {{\; < \;}}\omega {{\; \leqq \;}}100 )
\endcases
%P%\TAG{6}
\end{align*}
%\omega\omegaxxxxxxxx
\par
\noindent
\par
\par
\noindent
\par
%\newpage
\par
\noindent
\par
%\vskip0.9cm
\par
\noindent
%%%%eepic
%\vskip-0.4cm%%%
%\begin{figure}[htbp]
\unitlength=0.20mm
%\unitlength=0.28mm
\begin{picture}(500,90)
\put(-80,0){{{
\put(98,2){\path(0,70)
(125,70)
}
\put(48,2){\path
(75,0)(75,70)
(125,70)(125,0)
(175,0)(175,70)
(175,0)
(275,0)
(275,70)(325,70)
(325,0)
(325,0)
(425,0)(425,70)
(475,70)(475,0)
(525,0)(525,70)
(550,70)(550,0)
(550,70)
(425,70)
}
\put(48,2){
\put(200,30){$\cdots \cdots$}
\put(350,30){$\cdots \cdots$}
}
\thicklines
\path(100,2)(600,2)
\path(100,2)(100,100)
\path(97,70)(104,70)
\put(87,70){1}
\multiput(98,0)(50,0){10}{
\dottedline{2}(0,0)(0,10)
}
\multiputlist(98,-10)(50,0)
{
{\footnotesize 0},{\footnotesize 10},
{\footnotesize 20},{\footnotesize 30},
{\footnotesize 40},{\footnotesize 50},
{\footnotesize 60},{\footnotesize 70},{\footnotesize 80},{\footnotesize 90}
,{\footnotesize 100}
}
\multiputlist(105,80)(50,0)
{
{\footnotesize $g_{0}$},{\footnotesize $g_{10}$},
{\footnotesize $g_{20}$},{\footnotesize ${}$},
{\footnotesize ${}$},{\footnotesize $g_{50}$},
{\footnotesize ${}$},{\footnotesize ${}$},{\footnotesize $g_{80}$},{\footnotesize $g_{90}$},{\footnotesize $g_{100}$}
}
%%%%%%%%%%%%%%%ffffffffffff
}}}
\end{picture}
\vskip0.2cm
\begin{center}{Figure 2.3:
Round observable
}
\end{center}
\par
\noindent
\def\RND{\scriptscriptstyle{\roman{RND}}}
%\def\RND{\roman{RND}}
%%\SW
\par
\noindent
%%\textcolor{black}{NZ} and ,
%$\{ g_n \}_{n \in {\mathbb N}_{10}^{100} }$
%$C([0,100])$%\textcolor{black}{NZ}{the resosution the unity}{}
Define
${\mathsf O}_{{\RND}}= (Y ({{=}} {\mathbb N}_{10}^{100})   , 2^Y, G_{{\RND}})$
such that
\begin{align*}
&
[G_{{\RND}}(\emptyset )](\omega ) = 0,
\quad
[G_{{\RND}}(Y )](\omega ) = 1
\\
%[G(\emptyset )](\omega ) = 0,
%\quad
&
[G_{{\RND}} (\Gamma )](\omega ) = \sum\limits_{n \in \Gamma } g_n (\omega )
\quad
(\forall \Gamma \in 2^Y =
2^{{\mathbb N}_{10}^{100}  })
%P%\TAG{7}
\end{align*}
Recall that
$g_n$
is not continuous.
Therefore,
the triplet
${\mathsf O}_{{\RND}}= (Y   , 2^Y, G_{{\RND}})$
is not an observable in
$C([0,100])$.
Of course,
we want to
regard
it as an observable.
For this,
we must prepare
\textcolor{black}{"bounded type {{{measurement theory}}}"}
(cf. \textcolor{black}{Chaps. 10,11}).

%POIUYTREWQ
\subsection{Axiom${}_{\text{\scriptsize c}}^{\text{\scriptsize p}}$ 1 %{Sec.2.2
---
There is no science without measurement
\label{2secAxiom 1}
}
%\ssubsection{{{measurement}} ({}\textcolor{black}{Axiom 1})
%-
%{{measurement}}
%---
%\label{2secAxiom 1}
%}
\par
As mentioned in \textcolor{black}{{Chap.$\;$1}},
{{{measurement theory}}}
is formulated as follows:.
That is,
\begin{align*}
\underset{\text{\scriptsize (scientific language)}}{\text{{} $\fbox{{{measurement theory}}}$}}
:=
{
\overset{\text{\scriptsize [Axiom${}_{\text{\scriptsize c}}^{\text{\scriptsize p}}$ 1]}}
%\overset{\text{\scriptsize [Axiom${}_{\text{\scriptsize c}}^{\text{\scriptsize p}}$ 1\textcolor{black}{(\REF{2secAxiom 1})}]}}
{
\underset{\text{\scriptsize
[probabilistic interpretation]}}{\text{{} $\fbox{{{measurement}}}$}}}
}
+
{
\overset{\text{\scriptsize [Axiom${}_{\text{\scriptsize c}}^{\text{\scriptsize p}}$ 2]}}
%\overset{\text{\scriptsize [Axiom${}_{\text{\scriptsize c}}^{\text{\scriptsize p}}$ 2\textcolor{black}{(\REF{6secAxiom 2})}]}}
{
\underset{\text{\scriptsize [{{the Heisenberg picture}}]}}
{\text{{}$\fbox{ causality }$}}
}
}
%%%%%%%%%%%%%CLAsSICAL
\end{align*}
In what follows,
we shall explain Axiom${}_{\text{\scriptsize c}}^{\text{\scriptsize p}}$ 1.
( For Axiom${}_{\text{\scriptsize c}}^{\text{\scriptsize p}}$ 2(causality {)}, see Chap. 6).
\par
%\vskip0.5cm
\par
%index{@(={measuring object})}

%\ssl
With any
\it
classical system
\rm
$S$,
a basic algebra
$ C(\Omega)$
can be associated in which
measurement theory of that system can be formulated.
A
\it
state
\rm
of the system $S$
is represented by
a
state
$\omega (\in \Omega $,
i.e.,
a
\it
state space
${})$.
\rm
%a
%\it
%quantity
%\rm
%is
%represented by
%a self-adjoint element of ${\cal A}$,
%or generally,
%an $n$-tuple of the
%$(${}commutative $)$ self-adjoint elements of ${\cal A}$.
Also,
an
\it
observable
\rm
%$($
%which is a kind of generalization of a quantity
%$)$
\rm
is represented by
${\bold O}$
$\equiv$
$({}X , {\cal F} , F{})$
in
the
$C(\Omega )$.

%
%
%$\!$\footnote{
%I like to image the following correspondence
%(measurement theory and philosophy):
%$$
%\text{\LL state\RR $\leftrightarrow$ \LL matter\RR}
%\qquad
%\text{\LL observable\RR $\leftrightarrow$ \LL idea\RR (= \LL form\RR )}
%$$
%}
%%index{idea@idea}
%%index{matter@matter}
%%index{form@form}
%%index{state@state}
%%index{observable@observable}
%The
%\it
%measurement%index{measurement@measurement}
%of an observable
%${\bold O}$
%for
%the system
%$S$
%with
%(or, in)
%the state
%$\rho^p$
%\ssl
%is represented by
%${\bold M}_{\cal A} \big({}{\bold O} , S_{[{}\rho^p] } \big)$
%%index{measure1@${\bold M}_{\cal A} \big({}{\bold O} , S_{[{}\rho^p] } \big)$}
%in the
%$C^*$-algebra
%${\cal A}$.
%Also, we can obtain a measured value $x$
%$({}\in X)$ by the measurement
%${\bold M}_{\cal A} \big({}{\bold O} , S_{[{}\rho^p] } \big)$.
%
%
%
%
%
%

%
%
%
%
%
%
%
%%\it
%{\bf (={measuring object})}
%\rm
%$S$,
%{basic algebra}
%$C (\Omega)$
%.
%$S$%\textcolor{black}{NZ}{\bf {{state}}}
%(,
%{\bf {{state}}}
%)
%,
%{\bf state space }$\Omega$%\textcolor{black}{NZ}
%$\omega$
%.
%\rm
%{\bf observable }
%\rm
%%$($
%%which is %\textcolor{black}{NZ} of a system quantity
%%$)$
%
%\rm
%$C (\Omega)$
%%\textcolor{black}{NZ}{observable }
%${\mathsf O}$
%${{=}}$
%$(X , {\cal F} , F{})$
%.
%,
\par
\vskip0.2cm
\par
\noindent
%%%index{\textcolor{black}{Axiom 1}@\textcolor{black}{Axiom 1} [{}{{measurement}}{}]}
\rm
\begin{itemize}
\item[({}a$_1$)]
{\bf {{An observer}}}
takes a measurement
$
\left[\begin{array}{ll}
{}
\text{of \bf observable }
[{\mathsf O}]
\\
%\text{
{}
\text{by \bf {measuring instrument}}
[{\mathsf O}]
%\text{
%{}[{\mathsf O}] \text{ and {measuring instrument}}
\end{array}\right]
$
for a {\bf {measuring object}}
with a {\bf {{state}}}.
%$\omega$ and {\bf {{state}}}
%{\bf {measuring object}},
%\\
%index{@{{measurement}}}
\end{itemize}
and thus, in short,
we write:
\begin{itemize}
\item[({}a$_2$)]
An observer take a
{\bf {{measurement}}}
${\mathsf M}_{C (\Omega)} \big({}{\mathsf O} , S_{[\omega] } \big)$
$\big($
or,
${\mathsf M}_{C (\Omega)} \big({}{\mathsf O} , S_{[\delta_\omega] } \big)$
$\Big)$.
%${\mathsf M}_{C(\Omega)}({\mathsf O}, S_{[\omega]})$
%index{measurement11@${\mathsf M}_{C (\Omega)} \big({}{\mathsf O} ,
%S_{[\omega] } \big)$:(continuous type ){{measurement}}}
%\footnote{
%$S_{[\omega]}$%\textcolor{black}{NZ}
%$S$
%System%\textcolor{black}{NZ}}
\end{itemize}
where
$\delta_\omega$
is a point measure at $\omega$.
Thus,
the value
$[F(\Xi)](\omega)$
is also
written by
%$\rho^p ({}F({}\Xi{}))$
%$\bigl($
%$\equiv$
$
{}_{{}_{{{\cal M}(\Omega)} } }
\Bigl\langle \delta_\omega , F ({}\Xi{})  \Bigl\rangle
{}_{{}_{{C(\Omega)} }}
$.

%\par
And further,
by
{{measurement}}
${\mathsf M}_{C (\Omega)} \big({}{\mathsf O} , S_{[{}\omega] } \big)$,
{\bf measured value } $x$
$({}\in X)$
is obtained.
%index{@measured value }
\par
\rm
{{}}
Axiom${}_{\text{\scriptsize c}}^{\text{\scriptsize p}}$ 1
is
the linguistic turn of
Born's quantum measurement
as follows
(cf.
\textcolor{black}{{}{{{}}}{Sec.3.2}}
).
\begin{align*}
\underset{\text{\scriptsize ({physical law})}}{
\text{\fbox{Born's quantum {{measurement}}
{}}}
}
\xrightarrow[\text{\scriptsize linguistic turn}]{\text{\scriptsize proverbalizing}}
\underset{\text{\scriptsize (linguistic rule)}}{
\text{\fbox{{{{measurement theory}}}{(}\textcolor{black}{Axiom${}_{\text{\scriptsize c}}^{\text{\scriptsize p}}$ 1}{)}}}
}
\end{align*}
\normalsize
\baselineskip=18pt
{}

Since
Axiom${}_{\text{\scriptsize c}}^{\text{\scriptsize p}}$ 1 below is
similar
to
[Axiom 1 Prototype]
(\pageref{proto}{ page}),
all readers can easily understand it.

%\BEGIN{itembox}[c]{
%\bf Axiom 1({{measurement}}) ${{\cdot}}$}
%\label{rule201}
%%index{1@Axiom 1[{}{{measurement}}{}(continuous type )]}
%\rm
%\BEGIN{itemize}
%%\item[
%%(i)]
%%{{measurement}}.
%\item[
%%(ii)
%]
%{basic algebra}
%$ C (\Omega)$
%{{measurement}}
%${\mathsf M}_{C (\Omega)} \big({}{\mathsf O}{{=}} ({}X, {\cal F} , F{}),
%S_{[{}\omega] } \big)$
%
%.
%{{measurement}}
%${\mathsf M}_{C (\Omega)}  \bigl({}{\mathsf O}  , S_{[{}\omega{}] } \bigl)$
%{{}}
%measured value
%$ x$
%$({}\in X  {})$
%,
%$ \Xi $
%$({}\in  {\cal F}{})$
%{\bf probability },
%${}[F({}\Xi{})](\omega)$
%.
%\END{itemize}
%\END{itembox}
%
%
%bbbbbbbbbbbbbbbbbbbbbbbbbbbbbbbbb

%\fbox{\parbox{160mm}{
{
%\large

%\BEGIN{itembox}[c]
\par
\noindent
\begin{center}
{\bf
Axiom${}_{\text{\scriptsize c}}^{\text{\scriptsize p}}$ 1
(measurement: continuous pure type)
%{ Axiom 1
%((continuous and pure type) classical measurement)
%%index{Axiom 1@Axiom 1 [{}Measurement axiom{}]}
}
%Linguistic world-description method in monism
\label{axiomcp1}
\end{center}
\par
\noindent
%\vskip0.1cm
\par
\noindent
\fbox{\parbox{155mm}{
\begin{itemize}
%\item[(i)]
%{{measurement}}
\item[]
{}\rm{}
Consider a measurement
${\mathsf M}_{C(\Omega)} \big({}{\mathsf O}:= ({}X, {\cal F} , F{})  , S_{[{}\omega] } \big)$
%%%index{ma1@${\mathsf M}_{C(\Omega)} \big({}{\mathsf O}, S_{[{}\rho^p] } \big)$}
formulated in a
basic algebra
${C(\Omega)}$.
Assume that
the measured value
$ x$
$({}\in X  {})$
is
obtained by the measurement
${\mathsf M}_{C(\Omega)}  \bigl({}{\mathsf O}  , S_{[{}\omega{}] } \bigl)$.
Then,
%it holds that
the probability
that
the
$ x$
$({}\in X  {})$
belongs to a set
$ \Xi $
$({}\in  {\cal F}{})$
is given by
$[{}F({}\Xi{})](\omega)$
$\bigl($
$\equiv$
$
{}_{{}_{{\cal M}(\Omega) } }
\Bigl\langle \delta_\omega , F ({}\Xi{})  \Bigl\rangle
{}_{{}_{{C(\Omega)} }}
$
$\bigl)$.
\end{itemize}
}
}
\par
\vskip0.5cm
\par
\noindent

%
%
%
%
%
%
%
%
%\par
%\vskip0.2cm
%\par
%\noindent
%\BEGIN{itembox}[c]{\bf Axiom 1
%((continuous and pure type) classical measurement)
%%index{Axiom 1@Axiom 1 [{}Measurement axiom{}]}
%}
%\label{axiomcp1}
%\label{axiomcp1}
%$\;\;\;$
%{}\rm{}
%Consider a measurement
%${\mathsf M}_{C(\Omega)} \big({}{\mathsf O}:= ({}X, {\cal F} , F{})  , S_{[{}\omega] } \big)$
%%%%index{ma1@${\mathsf M}_{C(\Omega)} \big({}{\mathsf O}, S_{[{}\rho^p] } \big)$}
%formulated in a
%basic algebra
%${C(\Omega)}$.
%Assume that
%the measured value
%$ x$
%$({}\in X  {})$
%is
%obtained by the measurement
%${\mathsf M}_{C(\Omega)}  \bigl({}{\mathsf O}  , S_{[{}\omega{}] } \bigl)$.
%Then,
%%it holds that
%the probability
%that
%the
%$ x$
%$({}\in X  {})$
%belongs to a set
%$ \Xi $
%$({}\in  {\cal F}{})$
%is given by
%$[{}F({}\Xi{})](\omega)$
%$\bigl($
%$\equiv$
%$
%{}_{{}_{{\cal M}(\Omega) } }
%\Bigl\langle \delta_\omega , F ({}\Xi{})  \Bigl\rangle
%{}_{{}_{{C(\Omega)} }}
%$
%$\bigl)$.
%%\hfill$(page PAio1})$
%\END{itembox}
%}
%%%%\TAG{PAio1} page}
%%$($
%%
\par
\par

%%eeeeeeeeeeeeeeeeeeeeeeeeeeeeeeeeeeeeeeeeeeeeeee

\def\BC{\left\{\begin{array}{ll}}
\def\EC{\end{array}\right.}
\rm
\par
%index{@probability }

\vskip0.5cm

If it writes without carrying out simple,
we see:

\begin{itemize}
\item[(b)]
When
an {\bf {{observer}}}
takes a measurement
of an {\bf observable
${\mathsf O}{{=}} ({}X, {\cal F} , F{})$
}
(or,
by a {\bf {measuring instrument}${\mathsf O}$}
)
for a {\bf {measuring object}}
with a
{\bf {{state}}}
$\omega$,
the probability
that
a {\bf measured value }
belongs to
$\Xi (\in {\cal F} )$
is given by
${}[F({}\Xi{})](\omega)$.
\end{itemize}
Measurement theory
says that
{\bf Describe every phenomenon
modeled on Axiom${}_{\text{\scriptsize c}}^{\text{\scriptsize p}}$ 1}.
That is,
the key-words
in
\textcolor{black}{Axiom${}_{\text{\scriptsize c}}^{\text{\scriptsize p}}$ 1}
is as follows.
\begin{itemize}
\item[(c)]
{{measurement}}, observer, system{(}{measuring object}{)}, {{state}}, observable , measured value
, probability
\end{itemize}
and
\begin{itemize}
\item[]
Use these key-words
modeled on Axiom${}_{\text{\scriptsize c}}^{\text{\scriptsize p}}$ 1.
\end{itemize}

%%\textcolor{black}{NZ}
%\BEGIN{itemize}
%\item[(d)]
%$\quad$
%{\bf [\textcolor{black}{Axiom 1}{(}{{measurement}})]=
%[observable {(}%\textcolor{black}{NZ})%\textcolor{black}{NZ}probabilistic interpretation]}
%\END{itemize}
% and .
%observable probability %\textcolor{black}{NZ},
%{{measurement}}probability .
%That is,
%\BEGIN{itemize}
%\item[(e)]
%$\quad
%$
%$
%\text{
%{{measurement}}, probability (\textcolor{black}{{Chap.$\;$1}%\textcolor{black}{NZ}the Copenhagen interpretation(U$_5$)})
%}
%$
%\END{itemize}
%{}

\par
The meanings of key-words
in (c)
are not explained.
Therefore,
the experimental verification is meaningless.
That is,
\begin{itemize}
\item[(d)]
Axiom${}_{\text{\scriptsize c}}^{\text{\scriptsize p}}$ 1({{measurement}})
is a metaphysical statement,
in the sense that
it can not be verified by experiment.
This is the remarkable feature of
linguistic method.
\end{itemize}
%\i
This should be compared to
physics
(i.e.,
realistic method).
Many readers may no be familiar to
linguistic method.
However,
several examinations in the following section
will promote reader's understanding.

%
% and , ,
%{world-description}(
%\textcolor{black}{{Chap.$\;$1}(O)}
%)%\textcolor{black}{NZ}2%\textcolor{black}{NZ}
%
%{realistic method} and {linguistic method}
%
% and .

%%BBBBBBBBBBBBBBBBBB%SBSBSBS{\mathsf O}
\par
\noindent
{\small%%{\footnotesize
\vspace{0.1cm}
\begin{itemize}
\item[$\spadesuit$] \bf {{}}{Note }2.3{{}} \rm
If metaphysics has history of failure,
this is due to
the serious trial to answer the following problem
\begin{itemize}
\item[$(\sharp)$]
What is the meaning of the key-words in (c)?
\end{itemize}
Although this $(\sharp)$ may be attractive,
however,
it is not productive.
What is important is
to know how to use the key-words in (c).
Of course,
this answer is mentioned in Axiom${}_{\text{\scriptsize c}}^{\text{\scriptsize p}}$ 1.
%
%
% and ,
%(b)%\textcolor{black}{NZ}%\textcolor{black}{NZ}(Axiom${}_{\text{\scriptsize c}}^{\text{\scriptsize p}}$ 1)
%{}
% and , ?,
%
% and 
%
% and %\textcolor{black}{NZ},
%{{measurement theory}} and {metaphysics}.
%{metaphysics}(=),
% and , ?
%(That is,  and , ?) and {Remark }{}
\end{itemize}
}
%%BBBBBBBBBBBBBBBBBB%SBSBSBSS
\par
\noindent

\par
\noindent
\vskip0.3cm

\par
The following example (under Axiom${}_{\text{\scriptsize c}}^{\text{\scriptsize p}}$ 1 )
is the same as \textcolor{black}{Example 1.1}.

\par
\noindent
\vskip0.1cm
%BFBF
\par
\noindent
{\bf Example 2.7
[Continued from \textcolor{black}{Example 1.1)]}$\;\;$%POPOPO
{\sf [The measurement of
"cold or hot"
for water in a cup]}$\;\;$%POPOPO
$\;$
\rm
In Example 1.1,
we consider
this [C-H]-thermometer
${\mathsf O} =$
$(f_{{c}},f_{{h}})$,
where
the state space
$\Omega=[0,100]$,
the measured value space
$X=\{
\text{c,h}\}$.
%\\
That is,
%
%Let testees drink water with various temperature
%$\omega$$(0 {{\; \leqq \;}}\omega {{\; \leqq \;}}100)$.
%And
%you ask them {\lq\lq}cold"
%or {\lq\lq}hot"
%alternatively.
%Gather the data,
%(
%for example,
%$g_{c}(\omega)$ persons say {\lq\lq}cold",
%$g_{h}(\omega)$ persons say {\lq\lq}hot")
%and
%normalize them,
%that is,
%get
%the polygonal lines
%such that
%\BEGIN{align*}
%&
%f_{c}(\omega)= \frac{g_{c}(\omega)}{\text{the numbers of testees}}
%\\
%&f_{h}(\omega)=\frac{g_{h}(\omega)}{\text{the numbers of testees}}
%\END{align*}
%And
\begin{itemize}
\item[]
$
\qquad
\qquad
f_{c} (\omega)
=
\cases
1 & \quad (0 {{\; \leqq \;}}\omega {{\; \leqq \;}}10 ) \\
\frac{70- \omega}{60}  & \quad (10 {{\; \leqq \;}}\omega {{\; \leqq \;}}70 ) \\
0 & \quad (70 {{\; \leqq \;}}\omega {{\; \leqq \;}}100 )
\endcases,
$
$
\qquad
f_{h} (\omega) = 1- f_{c} (\omega)
%%P%%P%\TAG{2.4}
$
\end{itemize}
\par
%\vskip5.9cm
\par
\noindent
%\noindent
%%%\vskip-0.4cm%%%
%\BEGIN{FIGUre}[htbp]
%%%%%eepic
%\unitlength=0.28mm
%\BEGIN{picture}(500,100)
%\put(-80,0){{{
%\thicklines
%\path(100,2)(597,2)
%\path(100,2)(100,100)
%\path(97,70)(104,70)%
%\put(87,70){1}
%\put(60,0){
%\path(40,70)(85,70)(382,3)(537,4)
%\path(40,4)(85,4)(382,70)(537,70)
%}
%\put(170,90){$f_{c}$}
%\put(445,90){$f_{h}$}
%\multiput(98,0)(50,0){11}{
%\dottedline{2}(0,0)(0,10)
%}
%\multiputlist(98,-10)(50,0)
%{
%%{\footnotesize BC700},
%{\footnotesize 0},{\footnotesize 10},
%{\footnotesize 20},{\footnotesize 30},
%{\footnotesize 40},{\footnotesize 50},
%{\footnotesize 60},{\footnotesize 70},
%{\footnotesize 80},{\footnotesize 90},
%{\footnotesize 100}
%%,{\footnotesize 70},
%}
%%%%%%%%%%%%%%%%
%}}}
%\END{picture}
%\vskip0.3cm
%\caption{Cold or hot?}
%\END{figure}
%%\par
%%\noindent
%Therefore, for example,
%\BEGIN{itemize}
%\item[({{}}A$_1$)]
%You choose one person from the testees,
%and
%you ask him/her {\lq\lq}cold"
%or {\lq\lq}hot"
%alternatively.
%Then the probability that
%he/she
%says
%$
%\left[\begin{array}{ll}
%{}
%\text{"cold"}
%%[{\mathsf O}]
%{}
%\text{"hot"}
%\END{array}\right]
%$
%is
%given by
%\rm
%$
%\left[\begin{array}{ll}
%{}f_{\text \rm c}(55)=0.25
%\\
%{}
%f_{\text \rm h}(55)=0.75
%\END{array}\right]
%$
%\END{itemize}
%This is described in terms of
%Axiom${}_{\text{\scriptsize c}}^{\text{\scriptsize p}}$ 1
%in what follows.
%%\BEGIN{itemize}
%Consider
%the state space
%${\cal M}_{+1}^p (\Omega )$
%$(\approx$
%$\Omega$
%$)=$ interval $[0, 100](\subset {\mathbb R})$
%and measured value space $X=\{c, h\}$.
Then,
we have
the (temperature) observable
${\mathsf O}_{ch}= (X , 2^X, F_{ch} )$
in
$C ( \Omega )$
such that
\begin{align*}
&
[F_{ch}(\emptyset )](\omega ) = 0,
\quad
&{}&
[F_{ch}(X )](\omega ) = 1
\\
%[F_{ch}(\emptyset )](\omega ) = 0,
%\quad
&
[F_{ch}(\{c\})](\omega ) = f_{c} (\omega ),
&{}&
[F_{ch}(\{h\})](\omega ) = f_{h} (\omega )
%\TAG{3.14}
%P%\TAG{3.11}
\end{align*}
Thus,
we get
a
measurement
${\mathsf M}_{C ( \Omega )} ( {\mathsf O}_{ch}, S_{[\delta_\omega]} )$
$($
$=$
${\mathsf M}_{C ( \Omega )} ( {\mathsf O}_{ch}, S_{[{\delta_\omega}]} )$.
Therefore,
for example,
putting
$\omega=55 \SD$,
we can,
by
\textcolor{black}{Axiom${}_{\text{\scriptsize c}}^{\text{\scriptsize p}}$ 1(page \pageref{axiomcp1})},
represent
the statement
\textcolor{black}{(W$_1$)}
in Example 1.1
as follows.
\begin{itemize}
\item[(e)]
the probability
that
a
{measured value}$x(\in X {{=}}  \{c, h\})$
obtained by
{measurement}
\\
\\
${\mathsf M}_{C ( \Omega )} ( {\mathsf O}_{ch}, S_{[
\omega(=55)]} )$
belongs to
%\\
%\\
set
$
\left[\begin{array}{ll}
{}
\emptyset
%(={\text })
\\
\{ \text{c}\}
%[{\mathsf O}]
\\
\{ {h} \}
\\
\{ {c} ,{h}\}
\end{array}\right]
$
is given by
\rm
$
\left[\begin{array}{ll}
{}
[F_{ch}( \emptyset )](55)= 0
\\
%[F_{ch}(
%\{
%{\text \rm c}
%\}
%)](55)=
%0.25
%\\
{}
[F_{ch}(
\{
{ c}
\}
)](55)=
0.25
\\
{}
[F_{ch}(
\{
{ h}
\}
)](55)=
0.75
\\
{}
[F_{ch}(
\{
{ c}
,
{ h}
\}
)](55)=
1
\end{array}\right]
$
\end{itemize}
If it writes without omitting, we see:
\begin{itemize}
\item[({}f)]
When
an {\bf {{observer}}}
takes a measurement
by
$
{\underset{{ \scriptsize
\text{{\bf {measuring instrument}}}
{
%{\mathsf O} =(f_{\text {{{{c}}}}},f_{\text {{{{h}}}}})
{\mathsf O}_{ch}= (X , 2^X, F_{ch} )
}
}}
%{AAA}
{\text{[[C-H]-{instrument}]}}
}
$
%by a {\bf {measuring instrument}${\mathsf O}$}
%)
\\
\\
for
$
\underset{\text{ \scriptsize  ({\bf system{(}{measuring object})})}}{\text{[{{{{water in cup}}}}]}}
$
with
$
{\underset{\text{ \scriptsize ({\bf {{state}}}$(=\omega \in \Omega)$
)}}{\text{[55 $\SD$]}}}
$,
the probability
that
{\bf measured}
{\bf value }
\\
\\
\\
$
\left[\begin{array}{ll}
{}
\text{{{{{c}}}}}
%[{\mathsf O}]
%
%\text{
\\
{}
\text{{{{{h}}}}}
\end{array}\right]
$
is obtained
is given
by
\rm
$
\left[\begin{array}{ll}
{}
f_{\text \rm {{{{c}}}}}(55)=0.25
\\
{}
f_{\text \rm {{{{h}}}}}(55)=0.75
\end{array}\right]
$
%%%
%\\
%\\
%$
%\underset{\text{}}{\text{{\bf observer}, }}
%{\underset{\text{ \scriptsize {\bf {{state}}}$\omega=$55}}{\text{[55]}}}
%$
%%\textcolor{black}{NZ}
%$
%\underset{\text{ \scriptsize \bf {measuring object}}}{\text{[{{{{Water in cup}}}}]}}
%$
%
%%\overset{\text{ \scriptsize  }}
%${\underset{\text{ \scriptsize 
%{\bf observable }${\mathsf O}_{{{{{c}}}}{{{{h}}}}}$}}{\text{[{{{{c}}}}{{{h}}}]}}}
%$
%
%\\
%\\
%$
%\underset{\text{ \scriptsize  {\bf {{measurement}}}${\mathsf M}_{C ( \Omega )} ( {\mathsf O}_{{{{{c}}}}{{{{h}}}}}, S_{[
%\omega(=55)]} )$}}{\text{[], }}
%$
%%\textcolor{black}{NZ}{\bf measured value }$x(\in X {{=}}  \{ {{{{c}}}} ,  {{{{h}}}} \})$
%,
%\par \noindent
%
%$
%\left[\begin{array}{ll}
%{}
%\emptyset
%(={\text {empty set}})
%\\
%\{ \text{{{{{c}}}}}\}
%{}
%\\
%\{ \text{{{{{h}}}}} \}
%\\
%\{ \text{{{{{c}}}}} ,\text{{{{{h}}}}}\}
%\END{array}\right]
%$
%%\\
%%\\
%%\rm
%
%{\bf probability }
%$
%\left[\begin{array}{ll}
%{}
%[F_{{{{{c}}}}{{{{h}}}}}( \emptyset )](55)= 0
%\\
%%[F_{{{{{c}}}}{{{{h}}}}}(
%%\{
%%{\text \rm {{{{c}}}}}
%%\}
%%)](55)=
%%0.25
%%\\
%{}
%[F_{{{{{c}}}}{{{{h}}}}}(
%\{
%{\text \rm {{{{c}}}}}
%\}
%)](55)=
%0.25
%\\
%{}
%[F_{{{{{c}}}}{{{{h}}}}}(
%\{
%{\text \rm {{{{h}}}}}
%\}
%)](55)=
%0.75
%\\
%{}
%[F_{{{{{c}}}}{{{{h}}}}}(
%\{
%{\text \rm {{{{c}}}}}
%,
%{\text \rm {{{{h}}}}}
%\}
%)](55)=
%1
%\END{array}\right]
%$
\end{itemize}

Here, note that
the key-words in (c)
---
observer, {{state}}, system {(={measuring object})}, observable (={measuring instrument}{)}, {{measurement}}, measured value , probability
---
are contained in the above (f).

\vskip2.0cm
\subsection{{{Simple examples of measurements}}}%2.3
\subsubsection{linguistic world-view
---
Wonder of man's linguistic competence
}%{Sec.2.3.1}
\par
The applied scope of physics
{physics}
(realistic world-description method)
is rather clear.
But
the applied scope of measurement theory
(as well as the proverb
"Even monkeys fall from trees"
)
is ambiguous.
%as
%{physics}
%is %\textcolor{black}{NZ},
%{{measurement theory}}({linguistic world-description method})
%%\textcolor{black}{NZ} and .
% and %\textcolor{black}{NZ} Even monkeys fall from trees%\textcolor{black}{NZ}
%%\textcolor{black}{NZ} and {}
%{realistic world-view},
% and .
%%\textcolor{black}{NZ}, 
%Axiom 1,
%%,

As mentioned in \textcolor{black}{{Note }2.3},
what we can do in measurement theory
is
\begin{itemize}
\item[(a)]
$
\cases
\text{(a$_1$):
Use the language defined by Axiom${}_{\text{\scriptsize c}}^{\text{\scriptsize p}}$ 1}
%Axiom 1%\textcolor{black}{NZ},  and }
\\
\\
\text{(a$_2$):
Trust in man's linguistic competence
(\textcolor{black}{{Chap.$\;$1}(M$_1$)})
}
\endcases
$
\end{itemize}
Thus,
some readers may doubt that
\begin{itemize}
\item[(b)]
$\qquad
\qquad
$
Is it science?
\end{itemize}
However,
the spirit of measurement theory
is different from
that of physics.
%linguistic world-view(language is before world),
%{realistic world-view}({world is before language})
% and ,
%{world-view }{}
%
%, .
%index{@linguistic world-view}
%index{@{realistic world-view}}
%5{}
%($\sharp_1$),
%($\sharp_2$)11.6.
%\END{itemize}
%}
%%%BBBBBBBBBBBBBBBBB
%
\subsubsection{Elementary examples
---
urn problem, etc.}%{Sec.2.3.2}
\par
Since {{measurement theory}}
is a language,
we can not master it
without examinations.
Thus,
we present simple examples
in what follows.
%,
%${{\cdot}}$, .
%\textcolor{black}{Axiom 1}({{measurement}})
%.
%, 
%.
%.
\par
\noindent
\vskip0.3cm
%BFBF
\par
\noindent
{\bf Example 2.8
[The number of balls in urn]}$\;\;$%POPOPO
\rm
In a certain urn $U$, some balls are contained.

\begin{itemize}
\item[({}a)]
Counting the number of the
balls contains $n$ balls,
we get "n".
\end{itemize}
Now we shall represent
the obvious statement
(a) in terms of measurement theory.
\par
Define the state $\omega_n$ of the urn $U$
such that
\begin{align*}
\omega_n  \quad \cdots \quad
\text{
%\LL{{}}
$n$ balls are contained in the urn $U$.
}
%\RR}
\qquad
(n=0,1,2,\ldots)
%\TAG{3.15}
%P%\TAG{13}
\end{align*}
Therefore,
the state space $\Omega$
is defined by
\begin{align*}
\Omega {{=}}
\{\omega_0,\omega_1,\omega_2,\ldots\}
\end{align*}
with the discrete metric.
\par
\noindent
\vskip0.3cm
\par
\noindent
%\vskip-0.4cm%%%
%\begin{figure}[htbp]
\unitlength=0.17mm
\begin{picture}(500,130)
\put(200,0){
\Thicklines
\put(-4,5){
\spline(45,100)(45,90)(0,40)(50,0)(75,0)
(100,0)(150,40)(105,90)(105,100)}
\put(165,5){
\spline(45,100)(45,90)(0,40)(50,0)(75,0)
(100,0)(150,40)(105,90)(105,100)}
\put(334,5){
\spline(45,100)(45,90)(0,40)(50,0)(75,0)
(100,0)(150,40)(105,90)(105,100)}
\thicklines
\put(57,116){{{state}}$\;\; \omega_0$}
\put(217,116){{{state}}$\;\;\omega_1$}
\put(372,116){{{state}}$\;\;\omega_2$}
\put(6,0){
%\filltype{white}\put(40,50){\circle*{10}}
%\filltype{white}\put(55,50){\circle*{10}}
%\filltype{white}\put(70,50){\circle*{10}}
%\filltype{white}\put(85,50){\circle*{10}}
%\filltype{black}\put(100,50){\circle*{10}}
%\filltype{white}\put(40,35){\circle*{10}}
%\filltype{white}\put(55,35){\circle*{10}}
%\filltype{white}\put(70,35){\circle*{10}}
%\filltype{white}\put(85,35){\circle*{10}}
%\filltype{black}\put(100,35){\circle*{10}}
}
\put(165,0){
\put(10,0){
%\filltype{white}\put(40,50){\circle*{10}}
%\filltype{white}\put(55,50){\circle*{10}}
%\filltype{black}\put(70,50){\circle*{10}}
%\filltype{black}\put(85,50){\circle*{10}}
%\filltype{black}\put(100,50){\circle*{10}}
%\filltype{white}\put(40,35){\circle*{10}}
%\filltype{white}\put(55,35){\circle*{10}}
%\filltype{black}\put(70,35){\circle*{10}}
%\filltype{black}\put(85,35){\circle*{10}}
%\filltype{black}\put(100,35){\circle*{10}}
\filltype{black}\put(70,30){\circle*{25}}
}
}
\put(334,0){
\put(10,0){
%\filltype{white}\put(40,50){\circle*{10}}
%\filltype{black}\put(55,50){\circle*{10}}
%\filltype{black}\put(70,50){\circle*{10}}
%\filltype{black}\put(85,50){\circle*{10}}
%\filltype{black}\put(100,50){\circle*{10}}
%\filltype{black}\put(40,35){\circle*{10}}
%\filltype{black}\put(55,35){\circle*{10}}
%\filltype{black}\put(70,35){\circle*{10}}
%\filltype{black}\put(85,35){\circle*{10}}
%\filltype{black}\put(100,35){\circle*{10}}
\filltype{black}\put(45,30){\circle*{25}}
\filltype{black}\put(95,30){\circle*{25}}
}
}
%(270,50)(280,30)(300,25)(340,20)
\put(530,50){$\cdots \cdots$}
}
\end{picture}
\vskip0.1cm
\begin{center}
{Figure 2.4:
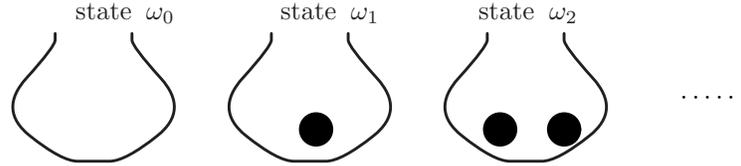
The number of balls in the urn}
\end{center}
\par
\noindent
%, ,
Define
the measured value space
by
${\mathbb N}_0$
$=\{0,1,2,\ldots\}$.
And
define
observable ${\mathsf O}_{}=({\mathbb N}_0, 2^{{\mathbb N}_0}, F)$
in
$C(\Omega)$
such that
\begin{align*}
[F( \Xi )](\omega_n )
=
\cases
1 & \quad ( n \in \Xi )
\\
0 & \quad (n \notin \Xi )
\endcases
\qquad
(\forall n \in {\mathbb N}_0 , \forall \Xi \subseteq {\mathbb N}_0)
\end{align*}
This observable ${\mathsf O}_{}=({\mathbb N}_0, 2^{{\mathbb N}_0}, F)$
is called a {\bf counting observable}.
%index{@observable }
Therefore,
the above statement \textcolor{black}{(a)}in {ordinary language}
can be translated to
the following (b) in terms of
\textcolor{black}{Axiom${}_{\text{\scriptsize c}}^{\text{\scriptsize p}}$ 1},
%%%%%%nkkkkkkkk
\begin{itemize}
\item[({}b)]
The probability the a measured value obtained by
{{measurement}}
${\mathsf M}_{C(\Omega)}({\mathsf O}_{}, S_{[\omega_n]})$
belong to $\Xi (\subseteq  {\mathbb N}_0)$
is given by
\begin{align*}
[F(\Xi)](\omega_n)
=
\left[\begin{array}{ll}
{}
1 & \quad ( n \in \Xi )
\\
0 & \quad (n \notin \Xi )
\end{array}\right]
\end{align*}
\end{itemize}
That is,
\begin{itemize}
\item[({}c)]
a measured value obtained by
{{measurement}}
${\mathsf M}_{C(\Omega)}({\mathsf O}_{}, S_{[\omega_n]})$
is surely $n$
%
%
%{{measurement}}
%${\mathsf M}_{C(\Omega)}({\mathsf O}_{}, S_{[\omega_n]})$measured value
%
%{(}probability 1)
%$\;
%n
%${}
\end{itemize}

%%%{\omega}_1{\omega}_2{\omega}_3%%%{\omega}_1{\omega}_2{\omega}_3
%%%%{\omega_1{\omega_2{\omega_3%%%{\omega_1{\omega_2{\omega_3
%\rm
%,
%{{measurement}}
%${\mathsf M}_{C(\Omega)}({\mathsf O}_{}, S_{[\omega_n]})$
%,
%,
%\BEGIN{itemize}
%\item[(d)]
%$
%%\text{observer}
%\underset{\text{}}{\text{observer, }}
%{\underset{\text{ \scriptsize {{state}}$\omega_n$}}{\text{[$n$]}}}
%\underset{\text{ \scriptsize  {measuring object}}}{\text{[{{Urn}}]}}
%,
%\;\;
%{\underset{\text{ \scriptsize 
%observable ${\mathsf O}_{}$}}{\text{[]}}}
%
%\underset{\text{ \scriptsize  {{measurement}}${\mathsf M}_{C ( \Omega )} ( {\mathsf O}_{}, S_{[
%\omega_n]} )$}}{\text{[], }}
%$
%%\\
%$
%\underset{\text{ \scriptsize  probability $1$}}{\text{[]}}
%\;\;
%\underset{\text{ \scriptsize  measured value }}{\text{[$n$]}}
%{}
%$
%\END{itemize}
% and .
%,
%{{Urn}}$U$, {{state}}$\omega_0$
% and  and {Remark }{},
%( and {)}{{state}}
%.
%%\qed

\normalsize \baselineskip=18pt
\par
\noindent
{\bf \vskip0.3cm
%BFBF
\par
\noindent
Example 2.9
%[{{about-observable} }
%
%%index{@{--observable }}
[Continued from \textcolor{black}{Example 2.4}
({{triangle observable} }{})]}$\;\;$%POPOPO
%Triangle observable]$\;$%POPOPO
%%index{rounding observable@{rounding observable}}
%{\sf [The rough measurement of the temperature of 
%water in a cup]}$\;\;$%POPOPO
$\;$
Let testees drink water with various temperature
$\omega \SD$
$(0 {{\; \leqq \;}}\omega {{\; \leqq \;}}100)$.
And
you ask them
"How many degrees($\SD$) is roughly this water?
Gather the data,
( for example,
$h_{n}(\omega)$
persons say
$n \SD$
${(}n=0,10,20,\ldots,90,100)$.
and
normalize them,
that is,
get
the polygonal lines.
For example,
define the state space
$\Omega$
by
the closed interval
$[0,100]$
$(
\subseteq {\mathbb R})$.
For each
$n  \in {\mathbb N}_{10}^{100}  = \{0,10,20,\ldots,100\}$,
define the (triangle)
continuous function
$g_{n}:\Omega \to [0,1]$
by
%(\textcolor{black}{FIGU 1.5}):
\begin{align*}
g_{n} (\omega)
=
\cases
0 & \quad (0 {{\; \leqq \;}}\omega {{\; \leqq \;}}n-10 ) \\
{\displaystyle \frac{\omega - n -10}{10} }
& \quad (n-10 {{\; \leqq \;}}\omega {{\; \leqq \;}}n ) \\
{\displaystyle
- \frac{\omega - n + 10}{10}
}
& \quad (n {{\; \leqq \;}}\omega {{\; \leqq \;}}n+10 ) \\
0 & \quad (n+10 {{\; \leqq \;}}\omega {{\; \leqq \;}}100 )
\endcases
%\tag{\color{black}{1.64}}
\end{align*}
%\omega\omegaxxxxxxxx
\par
\noindent
\par
%\par
%\noindent
%\par
%%\newpage
%\par
%\noindent
%\par
%\par
%\noindent
%%%%%eepic
%%\vskip-0.4cm%%%
%\BEGIN{FIGUre}[htbp]
%\unitlength=0.28mm
%\BEGIN{picture}(500,90)
%\put(-80,0){{{
%\put(98,2){\path(0,0)(50,70)(100,0)(150,70)(200,0)(250,70)(300,0)(350,70)
%(400,0)(450,70)(500,0)
%}
%\put(48,2){\path(50,70)(100,0)(150,70)(200,0)(250,70)(300,0)(350,70)(400,0)
%(450,70)(500,0)(550,70)}
%\thicklines
%\path(100,2)(600,2)
%\path(100,2)(100,100)
%\path(97,70)(104,70)
%\put(87,70){1}
%%\path(100,70)(240,70)(380,3)(600,4)
%%\put(170,90){$f_{c}$}
%%\path(100,4)(240,4)(380,70)(600,70)
%%\put(445,90){$f_{h}$}
%\multiput(98,0)(50,0){10}{
%\dottedline{2}(0,0)(0,10)
%}
%\multiputlist(98,-10)(50,0)
%{
%%{\footnotesize BC700},
%{\footnotesize 0},{\footnotesize 10},
%{\footnotesize 20},{\footnotesize 30},
%{\footnotesize 40},{\footnotesize 50},
%{\footnotesize 60},{\footnotesize 70},{\footnotesize 80},{\footnotesize 90}
%,{\footnotesize 100}
%}
%\multiputlist(105,80)(50,0)
%{
%%{\footnotesize BC700},
%{\footnotesize $g_{0}$},{\footnotesize $g_{10}$},
%{\footnotesize $g_{20}$},{\footnotesize $g_{30}$},
%{\footnotesize $g_{40}$},{\footnotesize $g_{50}$},
%{\footnotesize $g_{60}$},{\footnotesize $g_{70}$},{\footnotesize $g_{80}$},{\footnotesize $g_{90}$},{\footnotesize $g_{100}$}
%}
%%%%%%%%%%%%%%%%
%}}}
%\END{picture}
%\vskip0.2cm
%\caption{Triangle observable}
%\END{figure}
\par
\noindent
%\vskip1.0cm

\begin{itemize}
\item[(a)]
You choose one person from the testees,
and
you ask him/her
"How many degrees($\SD$) is roughly this water?".
Then the probability that
he/she
says
$
\left[\begin{array}{ll}
{}
\text{"about 40$\SD$"}
\\
%[{\mathsf O}]
{}
\text{"about 50$\SD$"}
\end{array}\right]
$
is
given by
\rm
$
\left[\begin{array}{ll}
{}g_{\text \rm 40}(47)=0.25
\\
{}
f_{\text \rm 50}(47)=0.75
\end{array}\right]
$
\end{itemize}

This is described in terms of
Axiom${}_{\text{\scriptsize c}}^{\text{\scriptsize p}}$ 1
in what follows.
%\BEGIN{itemize}

%
\par
\noindent
Putting
$Y={\mathbb N}_{10}^{100}$
and
define the triangle observable
${\mathsf O}^{\triangle}= (Y , 2^Y, F^{\triangle} )$
such that
\begin{align*}
&
[F^{\triangle}(\emptyset )](\omega ) = 0,
\qquad
[F^{\triangle}(Y )](\omega ) = 1
\\
%[F(\emptyset )](\omega ) = 0,
%\quad
&
[F^{\triangle} (\Gamma )](\omega ) = \sum\limits_{n \in \Gamma } g_n (\omega )
\quad
(\forall \Gamma \in 2^{{\mathbb N}_{10}^{100}  })
%\tag{\color{black}{1.65}}
%%P%\TAG{3.5}
\end{align*}
Then,
we have the
triangle observable
${\mathsf O}^{\triangle}= (Y (=
{{\mathbb N}_{10}^{100}  }
), 2^Y, F^{\triangle} )$
in
$C([0,100])$.
And we get a
measurement
${\mathsf M}_{C ( \Omega )} ( {\mathsf O}^{\triangle}, S_{[\delta_\omega]} )$.
For example,
putting
$\omega$=47$\SD$,
we see,
by
\textcolor{black}{Axiom${}_{\text{\scriptsize c}}^{\text{\scriptsize p}}$ 1(page \pageref{axiomcp1})},
that
\begin{itemize}
\item[(b)]
the probability
that
a measured value
obtained by the measurement
$
{\mathsf M}_{C ( \Omega )} ( {\mathsf O}^{\triangle},$
$S_{[
\omega(=47)]} )$
\\
is
$
\left[\begin{array}{ll}
%\underset{[measurementy{}$B${{\cdot}}${}(B}
{\text{about 40$\SD$}}
\\
\text{about 50$\SD$}
\end{array}\right]
$
is given by
$
\left[\begin{array}{ll}
{[F^{\triangle}( \{ 40 \})](47)=0.3}
%{\text{about 40}}
\\
{[F^{\triangle}( \{ 50 \})](47)=0.7}
\end{array}\right]
$
%
%
%$
%\left[\begin{array}{ll}
%[F^{\mbox{(tri)}}(\{ 40 \})](47)=
%0.3
%\\
%[F^{\mbox{(tri)}}(\{ 50 \})](47)=
%0.7
%\END{array}\right]
%$
\end{itemize}
\hfill{{$///$}}%%BFBFbfbf
%\END{Exa}  \rm
\par

{\bf \vskip0.2cm
%BFBF
\par
\noindent
Example 2.10}
\sf
[{}The urn problem].
\rm
There are two urns ${U}_1$
and
${U}_2$.
The urn ${U}_1$ [resp. ${U}_2$]
contains
8 white and 2 black balls
[resp.
4 white and 6 black balls]
%where
%$N$ is sufficiently large number.
({\it cf.} \textcolor{black}{Fig. 2.5}).

%TAB
%\BEGIN{align*}
\par
\noindent
%\vskip-0.4cm%%%
%\begin{table}[htbp] \small \caption{urn problem}
\begin{center}
Table 2.1: urn problem
\\
%\BEGIN{tabular}{|l|l|*{2}{@{\quad\$}r|}}
%\BEGIN{tabular}{
%||c
%||c|c
%||
%}
\begin{tabular}{
@{\vrule width 0.8pt\ }c
@{\vrule width 0.8pt\ }c|c
@{\vrule width 0.8pt}
}
\noalign{\hrule height 0.8pt}
Urn$\diagdown$ {{w}}${{\cdot}}${{b}} &$\quad$  white ball$\quad$ &$\quad$ black ball$\quad$\\
\noalign{\hrule height 0.8pt}
Urn $U_1$ & 8 & 2  \\
\hline
Urn $U_2$ & 4 & 6  \\
\noalign{\hrule height 0.8pt}
%Urn $U_3$  &  1 & 9  \\
%\hline
\end{tabular}
\end{center}
%\end{table}
\par
\noindent

\newpage

%Also,
%\BEGIN{itemize}
\par
\noindent
\unitlength=0.30mm
%%%%%%%%%%%%%%%%%
%\begin{figure*}[htbp]
%%%%%%%%%%%%%%%%%%
%\caption{
%${\mathsf O} \equiv$
%$({}\{ x_1, x_2, x_3 \},$
%$ 2^{\{ x_1, x_2, x_3 \} },$
%$ F)$ in $C({}\Omega{})$
%}
%\END{figure*}
%%%%%%%%%%%%%%%%%%
\begin{picture}(500,130)
\put(100,0){
\Thicklines
\put(0,5){
\spline(45,100)(45,90)(0,40)(50,0)(75,0)
(100,0)(150,40)(105,90)(105,100)}
\put(200,5){
\spline(45,100)(45,90)(0,40)(50,0)(75,0)
(100,0)(150,40)(105,90)(105,100)}
\thicklines
\put(73,116){$\omega_1$}
\put(273,116){$\omega_2$}
%%\put(373,116){$\omega_3$}
\put(10,0){
\filltype{white}\put(40,50){\circle*{10}}
\filltype{white}\put(55,50){\circle*{10}}
\filltype{white}\put(70,50){\circle*{10}}
\filltype{white}\put(85,50){\circle*{10}}
\filltype{black}\put(100,50){\circle*{10}}
\filltype{white}\put(40,35){\circle*{10}}
\filltype{white}\put(55,35){\circle*{10}}
\filltype{white}\put(70,35){\circle*{10}}
\filltype{white}\put(85,35){\circle*{10}}
\filltype{black}\put(100,35){\circle*{10}}
}
\put(200,0){
\put(10,0){
\filltype{white}\put(40,50){\circle*{10}}
\filltype{white}\put(55,50){\circle*{10}}
\filltype{black}\put(70,50){\circle*{10}}
\filltype{black}\put(85,50){\circle*{10}}
\filltype{black}\put(100,50){\circle*{10}}
\filltype{white}\put(40,35){\circle*{10}}
\filltype{white}\put(55,35){\circle*{10}}
\filltype{black}\put(70,35){\circle*{10}}
\filltype{black}\put(85,35){\circle*{10}}
\filltype{black}\put(100,35){\circle*{10}}
}
}
%\put(300,0){
%\put(10,0){
%\filltype{white}\put(40,50){\circle*{10}}
%\filltype{black}\put(55,50){\circle*{10}}
%\filltype{black}\put(70,50){\circle*{10}}
%\filltype{black}\put(85,50){\circle*{10}}
%\filltype{black}\put(100,50){\circle*{10}}
%\filltype{black}\put(40,35){\circle*{10}}
%\filltype{black}\put(55,35){\circle*{10}}
%\filltype{black}\put(70,35){\circle*{10}}
%\filltype{black}\put(85,35){\circle*{10}}
%\filltype{black}\put(100,35){\circle*{10}}
%}
%}
%\put(90,9){\footnotesize $\omega_1$}
%\put(300,9){\footnotesize $\psi_{1,2}(\omega_1)$}
}
\end{picture}
%%%%%%%%%%%%%%%%%
%\BEGIN{FIGUre*}[htbp]
%%%%%%%%%%%%%%%%%
\begin{center}
{Figure 2.5:
Urn problem}
\end{center}

%\caption{
%Urn problem
%}
%\end{figure*}
%%%%%%%%%%%%%%%%%%%
\par

%
%\BEGIN{center}
%%\BEGIN{tabular}{|l|l|*{2}{@{\quad\$}r|}}
%\BEGIN{tabular}{|c||c|c|l|r}
%\hline
%{} &$\quad$  white balls$\quad$ &$\quad$ black balls$\quad$\\
%\hline
%\hline
%urn $U_1$ & 8 & 2  \\
%\hline
%urn $U_2$ & 4 & 6  \\
%\hline
%urn $U_3$  &  1 & 9  \\
%\hline
%\END{tabular}
%\END{center}
%\vskip-1.0cm
%\BEGIN{align*}
%\TAG{1.44}
%\END{align*}
%\par

%%%%%%%%%%%%
\par
\noindent
Here,
consider the following {\lq\lq}statement
\textcolor{black}{(a)}":
\begin{itemize}
\item[(a)]
When one ball is picked up from the urn
$U_2$,
the probability that
the ball is white
is $0.4$.
%
%Pick out one ball at random
%from the
%urn $U_1$,
%and recognize the color (i.e.,
%"white" or {\lq\lq}black")
%of
%the ball
\end{itemize}
%(The term
%"at random" will be often omitted in this book.)
In measurement theory,
the {\lq\lq}measurement
\textcolor{black}{(a)}{\rq\rq}
is formulated as follows:
%%%{U}_1{\omega}_2{\omega}_3%%%{\omega}_1{\omega}_2{\omega}_3
%%%{\omega_1{\omega_2{\omega_3%%%{\omega_1{\omega_2{\omega_3
%\BEGIN{align*}
%\omega_1 = [{}8:2], \quad
%\omega_2 = [{}4:6], \quad
%\omega_3 = [{}1:9]. \quad
%\END{align*}
%Thus,
Assuming
\begin{align*}
U_1 \quad \cdots \quad
&
\text{{\lq\lq}the urn with the state $\omega_1
${\rq\rq}}
\\
U_2 \quad \cdots \quad
&
\text{{\lq\lq}the urn with the state $\omega_2
${\rq\rq}}
%\\
%U_3 \quad \cdots \quad
%&
%\text{{\lq\lq}the urn with the state $\omega_3${\rq\rq}}
%\tag{1.46}
\end{align*}
define
the state space
$\Omega$
by
$\Omega = \{ {\omega}_1 , {\omega}_2 \}$.
That is,
we assume the identification;
\begin{align*}
U_1 \approx \omega_1, \quad
U_2 \approx \omega_2, \quad
%U_3 \approx \omega_3. \quad
%\tag{1.47}
\end{align*}
\par
\noindent
%%TUBO
Put
{\lq\lq}$w${\rq\rq} = {\lq\lq white\rq\rq}$\!\!,\;$
{\lq\lq}$b${\rq\rq} = {\lq\lq black\rq\rq}$\!\!\;$,
and put
$X=\{w,b\}$.
%\BEGIN{itemize}
%\item[$(Q)$]
%\it
%Which is the chosen urn, ${\omega}_1$ or ${\omega}_2${}?
%\END{itemize}
%\rm
%
%%\rm
%
%\noindent
%[{\it Answer}].
%We regard $\Omega$
%$\big(\equiv$
%$\{ \omega_1 , \omega_2 \}
%\big)$ as the state space.
And define the observable
${\mathsf O}
\big(\equiv (X \equiv \{w,b\}, 2^{\{w,b\}}, F)
\big)$
in $C(\Omega)$
by
\begin{align*}
[F(\{w\})](\omega_1) = 0.8,& \qquad \qquad [F(\{b\})](\omega_1) = 0.2,
\\
{}[F(\{w\})](\omega_2) = 0.4,& \qquad \qquad [F(\{b\})](\omega_2) = 0.6.
%\tag{\color{black}{2.5}}
%
%\tag{1.48}
\end{align*}
Thus,
we get the measurement
$
{\mathsf M}_{C({}\Omega{})} ({}{\mathsf O} ,
S_{[{} \delta_{\omega_1}]})$.
Here,
\textcolor{black}{Axiom${}_{\text{\scriptsize c}}^{\text{\scriptsize p}}$ 1(page \pageref{axiomcp1})}
says that
\begin{itemize}
\item[(b)]
the probability
that
a measured value ${b}$
is obtained by
${\mathsf M}_{C({}\Omega{})} ({}{\mathsf O} ,
S_{[{} \delta_{\omega_1}]})$
is given
by
\begin{align*}
F(\{b\})(\omega_1) = 0.8
%\tag{1.50}
\end{align*}
\end{itemize}
\par

%
%
%we see that
%\BEGIN{align*}
%\text{\textcolor{black}{(B$_1$)}}
%=
%%{\mathsf M}_{C({}\Omega{})} ({}{\mathsf O} ,
%%S_{[{} \delta_{\omega_1}]})
%%\tag{1.49} \end{align*}
%Here,
%\textcolor{black}{Axiom${}_{\text{\scriptsize c}}^{\text{\scriptsize p}}$ 1(page PAio1})}
%says that
%\BEGIN{itemize}
%\item[(B$_2$)]
%the probability
%that
%a measured value ${b}$
%is obtained by
%${\mathsf M}_{C({}\Omega{})} ({}{\mathsf O} ,
%S_{[{} \delta_{\omega_1}]})$
%is given
%by
%\BEGIN{align*}
%F(\{b\})(\omega_1) = 0.8
%%\tag{1.50}
%\END{align*}
%\END{itemize}
\par
\noindent

%eeeeeeeeeeeeeeeeeeeeeeeeeeeee

%\qed
%%BBBBBBBBBBBBBBBBBB%SBSBSBS{\mathsf O}_{{{w}}{{b}}}F_{{{w}}{{b}}}FFFFFFF
%%BBBBBBBBBBBBBBBBBB%SBSBSBS{\mathsf O}FFFFFFFF
%%BBBBBBBBBBBBBBBBBB%SBSBSBS{\mathsf O}FFFFFFFF
\par
\noindent
{\small%%{\footnotesize
\vspace{0.1cm}
\begin{itemize}
\item[$\spadesuit$] \bf {{}}{Note }2.4{{}} \rm
Readers may fell that
\textcolor{black}{Example 2.7--Example 2.10} are too easy.
However, note that
\begin{itemize}
\item[$(\sharp)$]
Since how to write in addition to this was not learned, it wrote like this reluctantly.
\end{itemize}
As mentioned in (a) of \textcolor{black}{{Sec.2.3.1}},
what we can do
is
\begin{itemize}
\item[]
$\qquad$
to be faithful to Axiom${}_{\text{\scriptsize c}}^{\text{\scriptsize p}}$,
and
to
trust in Man's linguistic competence
\end{itemize}
If some find the other writing,
it will be praised as the greatest discovery on history of science.
That is because
it means the discovery beyond quantum mechanics.
%That is,
%
% and  and , {FIG.$\;$}(\textcolor{black}{{Chap.$\;$1}(X$_1$)}):
%\ssmall
%%index{@{Chap.$\;$1}(X$_1$)}
%%%{\textcircled{\scriptsize 0}}
%\\
%\\
%$\underset{({Chap.$\;$1})}{\text{(X$_1$)}}$
%$
%\!\!
%\overset{
%({ordinary language})
%}{\underset{({world-description}(=))}{
%\text{
%\fbox
%{{\textcircled{\scriptsize 0}}
%{ordinary language}}
%}
%}
%}
%$
%$
%\underset{\text{\scriptsize }}{\text{$\Longrightarrow$}}
%$
%$
%\!\!
%\underset{\text{\scriptsize ({Chap.$\;$1}(O))}}{\text{{world-description}}}
%\cases
%&
%\!\!\!\!\!\!
%{\text{\textcircled{\scriptsize 1}{realistic method}}}
%\\
%& {{\text (, )}}
%\\
%\\
%&
%\!\!\!\!\!\!
%{\text{\textcircled{\scriptsize 2}{linguistic method}}}
%\\
%& {{\text (language is before world)}}
%\ENDcases
%$
%\\
%\\
%, {ordinary language}\textcircled{\scriptsize 0}
%, monism and dualism and 
%,
%\textcolor{black}{{Chap.$\;$1}(F$_1$)}
% and .
\end{itemize}
}
%%BBBBBBBBBBBBBBBBBB%SBSBSBSS
\par
\noindent

%%BBBBBBBBBBBBBBBBBB%SBSBSBS{\mathsf O}
\par
\noindent
{\small%%{\footnotesize
\begin{itemize}
\item[$\spadesuit$] \bf {{}}{Note }2.5{{}} \rm
The statement
(a)
in \textcolor{black}{Example 2.10}
is not necessarily guaranteed:
\begin{itemize}
\item[]
When one ball is picked up from the urn
$U_2$,
the probability that
the ball is white
is $0.4$.
\end{itemize}
What we say is that
\begin{itemize}
\item[]
the statement (a) in ordinary language
should be written by
the measurement theoretical statement
%(a), 
(b)
\end{itemize}
It is a matter of course that
"probability"
can not be derived from mathematics itself.
For example,
the following
$(\sharp_1)$
and
$(\sharp_2)$
are not be guaranteed.
\begin{itemize}
\item[$(\sharp_1)$]
From the set $\{1,2,3,4,5\}$,
choose one number.
Then, the probability that the number is even
is
given by
$2/5$
\item[$(\sharp_2)$]
From the closed interval $[0,1]$,
choose one number $x$.
Then, the probability
that $x \in [a, b ]\subseteq [0,1]$
is
given by
$|b-a|$
\end{itemize}
The common sense
---
"probability"
can not be derived from mathematics, which is independent of our world
---
is well known as Bertrand's paradox
(cf. \textcolor{black}{\cite{
Keio}}).
Thus,
It is usual to add the term
"at random"
to
the above $(\sharp_1)$
and
$(\sharp_2)$.
In this print,
this term "at random"
is frequently omitted.
%,
% and  and .
%%index{@}
%%index{ and @}
%%, , ,
%
%% and ,
%.
% and {Remark }.
%\\
%(a):
%%BBBBBBBBBBBBBBBBB
\end{itemize}
}
%%BBBBBBBBBBBBBBBBBB%SBSBSBSS
%
\subsubsection{About the space in our world
---
Leibniz's relationalism}%{Sec. 2.3.3}
{
\bf
%BFBF
\par
\noindent
Example 2.11
[Approximate {{measurement}} of the position
of a particle]}$\;\;$%POPOPO
%\sf
%%otnote{[{}Normal observable]
%[How should the space be represented? $\;\;$Normal observable in $C({}{\mathbb R}{})${}].
\rm
Let $\Omega$ be an interval
of the one dimensional space
${\mathbb R}$.
Consider a particle P
with the position
$\omega_0 (\in \Omega = {\mathbb R})$.
%Let A and B be particles with the same masses $m$.
Consider the situation described in \textcolor{black}{Fig. 2.6}.
\par
%\noindent
%%%%%%%%%%%%%%%%%%
%\BEGIN{FIGUre*}[htbp]
%%%%%%%%%%%%%%%%%%
%\unitlength=0.8mm
%\BEGIN{picture}(150,20)
%\thicklines
%\put(-20,-10)
%{{
%%\put(65,20){\circle*{5}}%%%%%%%%
%%\put(60,20){\vector(-1,0){20}}%%%V
%%%\put(65,10){A}
%\put(-10,0){
%\put(112,23){\circle*{3}}%%%%%%%%%%%
%\put(60,20){\vector(1,0){110}}%%%V
%\put(110,13){$\omega_0$}
%\put(112,26){A}
%\put(160,12){$\Omega ={\mathbb R}$}
%}
%%\put(120,10){
%%%where {\lq\lq}the velocity of A{\rq\rq}
%%%$= -${\lq\lq}the velocity of B{\rq\rq}$\!\!\!\!.\; \;$
%%}
%}}
%\END{picture}
%%%%%%%%%%%%%%%%%%
%%\BEGIN{FIGUre*}[htbp]
%%%%%%%%%%%%%%%%%%
%%\vskip0.3cm
%\caption{
%A particle position as a state
%(How should space-time be represented?)
%}
%\END{figure*}
%%%%%%%%%%%%%%%%%%%
\par
%Here, note that
%%\newpage
%\BEGIN{itemize}
%\item[(E$_1$)]
%Space should be described as a kind of spectrum space
%%\hfill{\rm [FN]}
%\END{itemize}
\par
\noindent
Consider the following measurement \textcolor{black}{(a)}:
\begin{itemize}
\item[(a)]
measure
a particle's position roughly.
\end{itemize}
This \textcolor{black}{(a)} will be characterized as follows.
Let $\sigma$ be a fixed positive real.
Define the {\it normal observable} (or {\it Gaussian observable})
${\mathsf O}_{G^\sigma} \equiv ({\mathbb R}, {\cal B}_{\mathbb R}, G^{\sigma})$
in $C(\Omega)$
(where $Omega={\mathbb R}$) such that:
\begin{align*}
[G^{\sigma}(\Xi)] (\omega) = \frac{1}{\sqrt{2 \pi \sigma^2}}
\int_{\Xi} e^{- \frac{(x - \omega)^2}{2 \sigma^2}} dx
\quad (\forall \Xi \in {\cal B}_{\mathbb R}, \forall \omega \in \Omega
\equiv
[{}a, b{}]),
%%\TAG{3}
%\tag{1.62}
\end{align*}
which will be often used in this book.
See \textcolor{black}{Fig. 2.6}.
\par
\vskip0.5cm
\par
\noindent
\setlength{\unitlength}{0.7mm}
%%%%%%%%%%%%%%%%%
%\begin{figure*}[htbp]
%%%%%%%%%%%%%%%%%
%\caption{
%Urn problem
%}
%\END{figure*}
%%%%%%%%%%%%%%%%%%
%\BEGIN{picture}
\begin{picture}(100,70)
\thicklines
\put(115,10){
\put(-110,0){\vector(1,0){220}}
\put(110,-5){$x$}
\put(-5,55){$y$}
\put(0,0){\vector(0,1){50}}
\spline(-90,0.0)(-75,1.7)(-45,12.9)(-30,24.2)(-15,35.2)
(0,39.8)(15,35.2)(30,24.2)(45,12.9)(60,5.4)(75,1.7)
(90,0.0)
\put(-75,25){$y= \frac{1}{\sqrt{2 \pi \sigma^2}}
e^{- \frac{x^2}{2 \sigma^2}} $}
\dottedline{2}(30,0)(30,24)
\dottedline{2}(-30,0)(-30,24)
\dottedline{2}(60,0)(60,5.4)
\dottedline{2}(-60,0)(-60,5.4)
\put(30,-5){$\sigma$}
\put(-30,-5){$-\sigma$}
\put(57,-5){$2 \sigma$}
\put(-65,-5){$- 2\sigma$}
\allinethickness{0.07mm}
\spline(30,-7)(16,-10)(8,-11)
\spline(-30,-7)(-16,-10)(-8,-11)
\spline(60,-7)(32,-14)(8,-17)
\spline(-60,-7)(-32,-14)(-8,-17)
%\put(10,-10){\vector(1,0){20}}
%\put(10,-15){\vector(1,0){50}}
%\put(-10,-10){\vector(-1,0){20}}
%\put(-10,-15){\vector(-1,0){50}}
\put(-5,-12){\scriptsize 68.3\%}
\put(-5,-18){\scriptsize 95.4\%}
}
\end{picture}
%%%%%%%%%%%%%%%%%
%\BEGIN{FIGUre*}[htbp]
%%%%%%%%%%%%%%%%%
\vskip0.3cm
\begin{center}
{Figure 2.6:
Error function
}
\end{center}
%%%%%%%%%%%%%%%%%%
\par
\vskip1.0cm
\par
\noindent
Here, note that
$\frac{1}{\sqrt{2 \pi \sigma^2}}
\int_{-\infty}^{\infty}  e^{- \frac{x^2}{2 \sigma^2}} dx
=1
$
and
\begin{align*}
&
\frac{1}{\sqrt{2 \pi \sigma^2}}
\int_{-\sigma}^{\sigma}  e^{- \frac{x^2}{2 \sigma^2}} dx
=0.683...,
\quad
\frac{1}{\sqrt{2 \pi \sigma^2}}
\int_{-2 \sigma}^{2 \sigma}  e^{- \frac{x^2}{2 \sigma^2}} dx
=0.954...
\\
&
\frac{1}{\sqrt{2 \pi \sigma^2}}
\int_{-1.96 \sigma}^{1.96 \sigma}  e^{- \frac{x^2}{2 \sigma^2}} dx
\approx
0.95,
\quad
\frac{1}{\sqrt{2 \pi \sigma^2}}
\int_{- \infty }^{1.65 \sigma}  e^{- \frac{x^2}{2 \sigma^2}} dx
\approx
0.95
\tag{\color{black}{2.5}}
%\tag{\textcolor{black}{1.63}}
\end{align*}
Thus, the \textcolor{black}{(a)} is characterized as follows.
%that
\begin{itemize}
\item[(b)]
the probability that
a measured value
%$\vec{\omega}$
%$({}\in {\mathbb R}^d{})$
obtained
by the
measurement
${\mathsf M}_{{C} ({}{\mathbb R}{})}
({\mathsf O}_{G^\sigma} , S_{[{}\delta_{{\omega}_0}]})$
%%%%\newpage
belongs to
$\Xi$
$(\in {\cal B}_{{\mathbb R}}^{}{})$
is given by
$[G^{\sigma}(\Xi)] ({} {\omega}_0 {}) $.
\end{itemize}
\hfill{{$///$}}%%BFBFbfbf
%\END{Exa}
%
%%
%\par

\rm
\par \noindent {\bf Leibniz' opinion ({metaphysical space-time})}
%index{@}
\par
The problem
"What is space-time?"
is fundamental in any
{world-description} method.
For this problem,
measurement theory answers as follows
(for time, see \textcolor{black}{{Sec.6.4.2}}):
\begin{itemize}
\item[({}a)]
The position of
a matter is represented by
a state
(that is,
a state space ${mathbb R}^3$).
Therefore,
the position(\textcolor{black}{Example 2.11}),
the temperature
(\textcolor{black}{Example 2.7, Example 2.9}),
the number of balls (\textcolor{black}{Example 2.8})
and so on
are all
kinds of states.
That is,
\begin{itemize}
\item[]
{\bf
This world where we live
is regarded as a kind of state space,
and thus,
it is represented by
${\mathbb R}^3$}.
\end{itemize}
\end{itemize}
(For the general cases
(including quantum thheory), see \textcolor{black}{\cite{IQphi, ILing}})
As mentioned above,
space-time is not regarded as something special
in measurement theory.
This idea is considerably different from common sense
(i.e.,
the common sense
of the  theory of relativity).

\par
The above argument urges us to recall
Leibniz-Clarke correspondence
as follows.
\rm
%.
\par
\noindent
[{\bf Leibniz-Clarke Correspondence}]:
%%index{@}
%{\bf }
%,
Leibniz-Clarke Correspondence {(}1715--1716)
is important to know
both Leibniz's and Clarke's (=Newton's) ideas concerning space and time.
\par
\begin{itemize}
\item[({}b)]
Newton's absolutism says that
the space-time should be regarded as a receptacle of a "thing."
Therefore,
even if
"thing" does not exits,
the space-time exists.
On the other hand,
Leibniz's relationalism
says that
\begin{itemize}
\item[(b$_1$)]
Space is
a kind of state of "thing".
% and ${{\cdot}}${(}={{state}}{)}
%%,
%,
% and {{state}}
\item[(b$_2$)]
Time is an order of
occurring in succession
which changes one after another.
\end{itemize}
%,
%
%${{\cdot}}$, 
%{(}{)}{}
\end{itemize}
\par
\noindent

%BBBBBBBBBBBBBBBBBB%SBSBSBS
\par
\noindent
{\small%%{\footnotesize
\begin{itemize}
\item[$\spadesuit$] \bf {{}}{Note }2.6{{}} \rm
% and ,
%,
%{the theory of relativity}
% and .
%,
Many scientists may think that
\begin{itemize}
\item[]
{{Newton's}} assertion is understandable,
in fact,
his idea was inherited by Einstein.
On the other,
Leibniz's assertion is incomprehensible and literary.
Thus, his idea is not related to science.
\end{itemize}
However,
recall the classification
of the {world-description}:
\\
\\
%\BEGIN{itemize}
%\item[]
$\quad$
$
\underset{\text{\scriptsize (Chap. 1(O))}}{\text{{world-description}}}
\cases
%\textcircled{\scriptsize 1}:
\underset{\scriptsize
\text{}}{{\textcircled{\scriptsize 1}}\!:\!{
\text{realistic method (i.e., world is before language)}}}
%\cdots
%&
%{{Newton}}
%%(, , , , \cdots)
%\\
%&
%\xrightarrow[${{\cdot}}$]{}
%Einstein\xrightarrow[${{\cdot}}$]{}\cdots
%\\
\\
\\
%\textcircled{\scriptsize 2}:
\underset{\scriptsize
\text{}}{{\textcircled{\scriptsize 2}}\!:\!{
\text{linguistic method (i.e., language is before world)}}}
%\underset{\scriptsize
%\text{}}{{\textcircled{\scriptsize 2}}\!:\! {linguistic method}(language is before world)}
%%%\textcircled{\scriptsize 2}: realistic world-description method
%%\cdots
%&
%
%=
%{{measurement theory}}
\endcases
$
%\END{itemize}
\\
in which
Newton  and Leibniz respectively
devotes himself to
{\textcircled{\scriptsize 1}}
and
{\textcircled{\scriptsize 2}}.
Although Leibniz's assertion is not clear,
we believe that
\begin{itemize}
\item[]
Leibniz found the importance of
"linguistic space and time"
in science,
\end{itemize}
though he did not propose
his language.
\end{itemize}
}

As mentioned in (a),
{{{measurement theory}}}
adopts
Leibniz's relationalism.
For time, see \textcolor{black}{{}{{{}}}{Sec.6.4}}.

%,
%
%causality 
%, That is,
%causality  and .
%,
%{{{measurement theory}}},
%
%.

%BBBBBBBBBBBBBBBBBB%SBSBSBS
\par
\noindent
{\small%%{\footnotesize
\vspace{0.1cm}
\begin{itemize}
\item[$\spadesuit$] \bf {{}}{Note }2.7{{}} \rm
%index{@unsolved problem(
%${{\cdot}}$probability  and ?)}
Although
"What is space and time?"
is long time unsolved problem,
Newton's and Leibniz's ideas are characterized
in the following
classification
of the {world-description}:
\begin{itemize}
\item[$\underset{\text{\scriptsize (Chap.1)}}{\text{(O)}}$]
$
\cases
\textcircled{\scriptsize 1}:
\underset{\text{\scriptsize (realistic world view)}}{\text{Newotn, Clarke}}
&
\cdots
\overset{\text{\scriptsize (space-time in physics)}}{
\underset{\text{\scriptsize {\lq\lq}What is space-time?"}}{
\fbox{\text{realistic space-time}}}
}
\quad
\qquad
\text{(successors: Einstein, etc.)}
\\
\\
\textcircled{\scriptsize 2}:
\underset{\text{\scriptsize (linguistic world view)}}{\text{Leibniz}} 
&
\cdots
\overset{\text{\scriptsize (space-time in measurement theory)}}{
\underset{\text{\scriptsize {\lq\lq}How should space-time be represented?"}}{
\fbox{\text{linguistic space-time}}}
}
\;\;
\text{(i.e.,
spectrum, tree)}
\endcases
$
\end{itemize}
%({\it cf.} \textcolor{black}{Example 4.10} later
%).
Measurement theory is in Leibniz's side.
Thus, we consider that
%\textcolor{black}{Example \REF{Example 1.19}} says
\begin{itemize}
\item[($\sharp_2$)]
Space should be described as a kind of spectrum (=state space in classical situation)
\end{itemize}
That is, we think that
the Leibniz-Clarke debates
should be essentially regarded as
"linguistic world view
\textcircled{\scriptsize 2}"
v.s.
"realistic world view \textcircled{\scriptsize 1}".
%For time, see \textcolor{black}{Footnote \REF{Leibniz-Clarketime}
%(page \pageref{Leibniz-Clarketime}) in \textcolor{black}{Chapter 4}.
%}
%
%
%
%
%
%
%
%
%
%
%
%%\BEGIN{itemize}
%%\item[]
%$
%\underset{\text{\scriptsize ({Chap.$\;$1}(O))}}{\text{{world-description}}}
%\cases
%%\textcircled{\scriptsize 1}:
%&
%\underset{\scriptsize
%\text{( and realistic)}}{\textcircled{\scriptsize 1}: {realistic method}}
%\!\!\!\!
%{\cdots}
%{{Newton}}
%%(, , , , \cdots)
%\\
%&
%\qquad
%\quad
%\xrightarrow[${{\cdot}}$]{}
%Einstein\xrightarrow[${{\cdot}}$]{}\cdots
%\\
%\\
%%\textcircled{\scriptsize 2}:
%&
%\underset{\scriptsize
%\text{({metaphysics})}}{\textcircled{\scriptsize 2}: {linguistic method}}
%%\textcircled{\scriptsize 2}: realistic world-description method
%\!\!\!\!
%{\cdots}
%%
%\underset{\scriptsize \text{}}{}
%\\
%&
%\qquad
%\qquad
%\qquad
%\quad
%\xrightarrow[ and ]{}
%{{measurement theory}}
%\ENDcases
%$
%%\END{itemize}
\\
\\
Recall
the proverb
"Even monkeys fall from trees"
in
\textcolor{black}{{Sec.1.1}}
such that
%%index{monkey@ Even monkeys fall from trees}
%\BEGIN{align*}
%\overset{{{Newton}}}{\underset{}{\text{
%\fbox{ Even monkeys fall from trees}}}}
%\xrightarrow[proverbalizing]{}
%\overset{}{\underset{( and )}{\text{\fbox{
% Even monkeys fall from trees}}}}
%\END{align*}
%%\END{itemize}
% and ,
%,
\begin{itemize}
\item[($\sharp_1$)]
$\qquad
\qquad
$
$
\overset{\text{\scriptsize physical [space-time-probability]}}{\underset{\text{({realistic method})}}{\text{
\fbox{quantum mechanics}}}}
\xrightarrow[\text{proverbalizing}]{\text{linguistic turn}}
\overset{\text{\scriptsize linguistic [space-time-probability]
}}{\underset{\text{({linguistic method})}}{\text{\fbox{
{{measurement theory}}}}}}
$
\end{itemize}
% and .
Although quantum mechanics
will be explained in
\textcolor{black}{{Chap. 3}},
we can now conclude that
\begin{itemize}
\item[($\sharp_2$)]
It suffices to
use the terms
(space, time, probability)
in measurement theory
by an analogy of
quantum mechanics.
Moreover,
the usage is possible even if we do not quantum mechanics.
That is because these terms may be used
according to our common sense.
\end{itemize}
In linguistic method,
the question
"What is  space (time, probability)?"
is not important
(\textcolor{black}{{Note }2.3}).
What is important is
%
%{{measurement theory}}, ${{\cdot}}$${{\cdot}}$probability  and , ?,
%($\sharp_2$).
%%\textcolor{black}{{Note }2.3},
%% and ,
%{metaphysics} and ,
% and , ?,
\begin{itemize}
\item[($\sharp_3$)]
How should the term space (time, probability) be used ?
\end{itemize}
This is respectively answered in
(a) in this section,
Sec. 6.4.2,
Axiom${}_{\text{\scriptsize c}}^{\text{\scriptsize p}}$ 1.
%
%
%
%{}
%,
%${{\cdot}}$probability {
%Axiom${}_{\text{\scriptsize c}}^{\text{\scriptsize p}}$ 1\textcolor{black}{(\REF{2secAxiom 1})}}
%,
%
%{
%Axiom${}_{\text{\scriptsize c}}^{\text{\scriptsize p}}$ 2\textcolor{black}{(\REF{6secAxiom 2})}}
%.
% and , ?,
% and 
%
% and ,
%{{measurement theory}} and {metaphysics}.
%,
%{metaphysics},
%, (a)state space  and PART,
%%,
%
%%(
%(b$_1$)
%% and (b$_2$))
%,
%.
%%\\
%(iii):
%,
%{{Newton}}, Newtonian mechanics and {linguistic world-description method}
%, ,
%.
%\\
%(iv):
%.
%2(, {{Newton}})
%, (scientific language and {physics})
%${{\cdot}}$, .
\end{itemize}
}

%BBBBBBBBBBBBBBBBBB%SBSBSBS
\par
\noindent
{\small%%{\footnotesize
\begin{itemize}
\item[$\spadesuit$] \bf {{}}{Note }2.8{{}} \rm
The space-time in measuring object
is well discussed in the above.
However,
we have to say something about
"observer's time".
%
%{measuring object},
% and  and (
%, \textcolor{black}{{Sec.6.4.2}}),
%observer
%.
We conclude that
observer's time
is meaningless in measurement theory
as mentioned the Copenhagen interpretation(U$_2$)
in Chap. 1.
That is, the following question is nonsense
in measurement theory:
\begin{itemize}
\item[($\sharp_1$)]
When and where does an observer take a measurement
%observer{{measurement}} and 
%{{measurement}}{{state}} and 
\item[($\sharp_2$)]
Therefore,
the is no tense (present, past, future) in sciences.
\end{itemize}
%In fact,
%we know no science concerning tense.
%% and 
%%(
%\textcolor{black}{{Note }6.7}{)}.
%observer,  and 
%%
%%paradox
%%
%
%%
%.
%%,  and \footnote{
\end{itemize}
}

\par
\subsection{The age of engineering and various sciences }%2.4
\subsubsection{Measurement{Theory}
---
help the weak}
\par
Measurement {Theory} sided with the "Leibniz's relation theory" which may be called "endangered species"
 for the foregoing paragraph, and again with the John Locke's "the primary quality and the second quality" in \textcolor{black}{Note 1.9}. 
%index{@${{\cdot}}$}
%%index{@{Chap.{\;}}, {Chap.{\;}}}
%index{@{Chap.{\;}};{Chap.{\;}}}
%%In the preface, when seen from the viewpoint 
%of world description, it pointed out that 
%engineering and science were not remarkably so good compared with physics.
%%However, they are stood to the position of more than, such as 
%physics and a pair, by being based on Measurement {Theory}.
If it has a worldly way of speaking, measurement theory will be an ally of weak things
-
what cannot have confidence,
a thing currently considered to be suspicious,
a thing which is not trusted,
a thing like endangered species,
a thing which is not respected,
a cheap thing
-.
The concrete image is summarized as notes.
\par
\noindent
{\small%%{\footnotesize
\begin{itemize}
\item[$\spadesuit$] \bf {{}}Note 2.9{{}} \rm
We consider that
\begin{align*}
%\\
&
\;\;\;
\text{\bf The strong
{\rm}
}
&
&\text{\bf vs. }&
&
%\quad
\text{\bf The weak
{\rm }
}
\\
&
{\textcircled{\scriptsize 1}} \text{Newton's space-time}
&
&
\text{\ \ \ }
&
%\quad
&
\text{Leibniz's space-time}
\\
&
{\textcircled{\scriptsize 2}} \text{
materialism
}
&
&\text{\ \ \  }&
&
\text{idealism}
\\
&
{\textcircled{\scriptsize 3}} \text{physical science}
&
&\text{\ \ \  }&
&
\text{metaphysics}
\\
&
{\textcircled{\scriptsize 4}}
\text{monism}
&
&\text{\ \ \  }&
&
\text{dualism}
\\
&
{\textcircled{\scriptsize 5}}
\text{physics}
&
&\text{\ \ \ }&
&
\text{engineering(=various science)}
\quad
\text{(cf. Sec. 8.1(m))}
\\
&
{\textcircled{\scriptsize 6}}
\text{Einstein}
&
&\text{\ \ \ }&
&
\text{Fisher(or, von Neumann)}
\\
&
{\textcircled{\scriptsize 7}}
\text{theory of relativity}
&
&\text{\ \ \ }&
&
\text{quantum mechanics}
\\
&
{\textcircled{\scriptsize 8}}
\text{realistic world-view}
&
&\text{\ \ \ }&
&
\text{linguistic world-view}
\end{align*}
%%%%%PPPQQQ
%index{@${{\cdot}}$}
Supposing it decides "by majority", please understand the standard which classifies "a strong thing" and "a weak thing" to such an extent that it says that it will become like this.
Since it does not have the science view whose another side settled while one side has a "realistic science view", both are divided.
Namely,
\begin{itemize}
\item[]
"The strong side" has a common language called physics.
However, "the weak side" does not have a common language.
\end{itemize}
In this book,I claim that
\begin{itemize}
\item[]
a sight in which "the weak side" depends on the authority (mathematics, physics, application, culture) of the other place, and makes a living is caught from another's eyes, it is not a suitable good thing
\end{itemize}
, and needs to establish independent authority - linguistic science view - .
Namely,
\begin{itemize}
\item[$(\sharp_1)$]
\bf
If "the weak side" is bundled considering Measurement {Theory} as a common language, it can rank the strong side.
\end{itemize}
Although it is the same,
\begin{itemize}
\item[$(\sharp_2)$]
\bf
If the engineering theory - Measurement Theory - which bundles a weak side is made, it can rank the strong side.
\end{itemize}
This is an opinion of this book. (\textcolor{black}{{Chap.$\;$1}2}).
See Note 9.7 about upper {\textcircled{\scriptsize 6}}
, and see \textcolor{black}{Note 9.3} about {\textcircled{\scriptsize 7}}.
Moreover, {\textcircled{\scriptsize 8}} is a conclusion of this book.
%(in FIG. 1 of a "preface", and a positive 
%type, FIG. 1 of a "postscript." ).
%%)\textcolor{black}{9.3}.
\end{itemize}
}
%%BBBBBBBBBBBBBBBBBB%SBSBSBSS
%\par
%\noindent
In \textcolor{black}{Chapter 9}, equilibrium statistical mechanics is taken from the territory of physics, and it is regarded as one field of engineering and various sciences.
If it goes by this flow ,
\begin{itemize}
\item[(c)]
{\bf
Measurement theory is a good fellow who helps "a weak thing".
}
\end{itemize}
However, we should say that this view is superficial.
In fact, the position of "a strong side" and "a weak side"" is substantially reversed bordering on the second half of the 20th century
(as a symbolic incident, Apollo's landing on the moon (1969) occurs. ).
\renewcommand{\footnoterule}{
  \vspace{2mm}                      % 
  \noindent\rule{\textwidth}{0.4pt}  
  \vspace{-3mm}
}
It is what is called "the end of a big tale", and it is as follows, if the "ultimate purpose" is made to contrast and it writes.
\begin{align*}
\underset{\text{\scriptsize (The age of physics)}}{\text{The understanding of God's rule}}
\xrightarrow[\text{\scriptsize Apollo (1969)}]{}
\underset{\text{\scriptsize (The age of engineering)}}{
\text{
The manufacture of a robot like man }
}
\end{align*}
Now, physics is a storyteller of "the myth of space creation."
Although it may turn out that there is an irresistible charm in a myth lover and there may also be an opinion that it becomes a serious problem 1000 years after, I think that the meaning of "the seriousness of 1000-years-after" becomes delicate by lower (e).
All the things that are actually in personal appearance and are moving the world now are the products of engineering
- engineering works ${{\cdot}}$ construction, a car, an airplane, electrical machinery and apparatus, petroleum products, a computer, etc.  -
, and engineering is also bearing human beings' fate (or influences the fate of one country).
If this is taken into consideration ,
\begin{itemize}
\item[(d)]
By obtaining a formal language called Measurement theory, engineering and science make the scaffold steadfast, and sees adolescence - fast progressive era -.
\end{itemize}
This is the greatest message of this book.
(as mentioned in (I) of Chap. 13).
If it thinks simply that what is necessary is just to leave science to a robot after that if "about the same robot as a scientist" is made, the meaning of what "man contributes to development of engineering and science" will become doubtful.
It follows,
\begin{itemize}
\item[(e)]
The time of engineering ---
the best-before date of Measurement theory
(up to the day when "the time when a robot does science, and the time when a robot makes a robot" come)
--- will be at most four  hundreds of years from now on.
\end{itemize}

%BBBBBBBBBBBBBBBBBB%SBSBSBS
\par
\noindent
{\small%%{\footnotesize
\vspace{0.1cm}
\begin{itemize}
\item[$\spadesuit$] \bf {{}}Note 2.10{{}} \rm
About a scientific language, when a conclusion is written previously, it is as follows.
(\textcolor{black}{{Chap.$\;$1}2}).
\begin{itemize}
\item[($\sharp_1$)]
$
\cases
\textcircled{\scriptsize 0}:
\text{language in mathematics} &
%\text{\hspace{-4.0cm}}
\; \cdots \; \text{set theory}
\\
\quad
%\text{\footnotesize (${{\cdot}}$)}
\\
\textcircled{\scriptsize 1}:
\text{{language in physics}} &
%\text{\hspace{-4.0cm}}
\; \cdots \; 
\text{the theory of relativity, }......
\\
\quad
\text{\footnotesize (the realistic world-view:Clarify God's rule)}
\\
\textcircled{\scriptsize 2}:{\text{language in engineering}} &
%\text{\hspace{-4.0cm}}
\; \cdots \; \text{measurement theory}
\\
\quad
\text{\footnotesize (linguistic world-view:create a robot like a man)}
\endcases
$
\end{itemize}
%\text{PPP}QQQQQQQQQQQQ
\textcircled{\scriptsize 0}and\textcircled{\scriptsize 1}
will be common sense.
About \textcircled{\scriptsize 2}, it will argue about this book whole volume.
Although Gauss (1777 -- 1855) did not know set theory, the mathematical achievements were great, and although neither watt (1736 -- 1819) nor Edison (1847 - 1931) knew modern control theory, they did large invention.
Moreover, there are various ways to make "about the same robot as a scientist."
It is an extreme talk,:
\begin{itemize}
\item[($\sharp_2$)]
"Man" was made, when neglecting water, air, and a gravel without doing anything for about 4 billion years.
%
%\footnote{
%I have not told the joke.
%I think that it is a surprising thing that the wonderful machine
% ---man--- was made by evolution even if there were not
%"a language, mathematics, and scientific theories."
%Although it is quantum measurement {theory}($\approx$Quantum mechanics)
%to lead utilization of "the time crunch computer simulation of
%evolution", it is not concerned in this book.
%}
\end{itemize}
% and 
Therefore, large invention may be possible if there is even time (even if there is no control theory($\subset$Measurement {theory}; Chapter 7 )).
However, in order "to hurry", control theory is indispensable and this thought is spontaneous generation,
but I think that it was popularized by Norbert Wiener's (1894 year--1964 year) "cybernetics."
%index{@}
The reason which we must hurry
-
the reason the easy thing of "if there is even time" is not said.
-
is written in the following paragraph.
\end{itemize}
}
%
%%BBBBBBBBBBBBBBBBBB%SBSBSBSS
%
\subsubsection{Measurement Theory - For the further development of engineering and various science -}%2.4.2
\par
The reason into which engineering must be developed immediately is explained below.
Dr. Hawking (1942 --), a British theoretical physicist, is warning as follows on the website "Big Think."
:
\begin{itemize}
\item[(a)]
The resources of the earth are limited while population increases geometrically.
Furthermore, by the time progress of technology changed global environment also often and bad, it resulted.
We accomplished the development which should attract attention over the past 100 years.
But, it is not a thing which one chance merely remains behind on the earth in order to overcome 100 years of future, but is spread in the universe.
\end{itemize}
Even if Dr. Hawking does not say, everyone surely feel that "a big misfortune" will surely happen before long
if lives on this narrow earth.
% if it considers that "an earth escape thing and space settlement 
%things" are generally supported  
%,
%% and ,
%\BEGIN{itemize}
%\item[(b)]
%, 
%
%
%
%,
%${{\cdot}}$
%(=
%
%{Chap.{\;}})${{\cdot}}$,
%.
%%index{@${{\cdot}}$}
%, ${{\cdot}}$
% and ,
%, \footnote{
%
%,
%,
%
%${{\cdot}}$
%{}
%}.
%%
%\END{itemize}
% and .
%,
If Dr. Hawking's warning (a) is said at a word,
%,
\begin{itemize}
\item[(b)]
$\quad$
[
Human beings' continuation]
$\Longleftrightarrow$
[The construction of Space Colony]
\end{itemize}
The story of this book is developed by believing this.
%%%BBBBBBBBBBBBBBBBBB%SBSBSBSS

%BBBBBBBBBBBBBBBBBB%SBSBSBS
\par
\noindent
{\small%%{\footnotesize
\vspace{0.1cm}
\begin{itemize}
\item[$\spadesuit$] \bf {{}}Note 2.11 {{}} \rm
As mentioned above, I would like to consider
%,
\begin{itemize}
\item[$(\sharp)$]
{\bf
earth escape (construction and settlement of a space colony) hundreds of years after}
\end{itemize}
%index{@${{\cdot}}$}
as the present target of the human beings, us.
There may be some readers puzzled to the "space colony."
However, this is a device for fixing the purpose of "realization of a space colony" and keeping an argument from being spread.
For achievement of this $(\sharp)$, since human beings have to mobilize fully all the results (intellectual product) of the engineering and various science accumulated in the past,
I think that it is perfect as a target.
Moreover, if human beings reside permanently in the universe in 10,000 years, and if they look back upon history then, I will think that "the beginning of agriculture" and "earth escape" are commented as two major events of a history of man.
If that is right, we may call that the challenge$(\sharp)$ is "{\bf the greatest battle of human beings}."
\end{itemize}
}

%%BBBBBBBBBBBBBBBBBB%SBSBSBSS
\par
In this book, a "space colony" is often used as a metaphor.
This is because I think as follows.
\begin{itemize}
\item[(c)]
$
\quad
\text{
\bf
Since this book is writing of engineering, I should declare the clear purpose.
}
$
\end{itemize}
It is because "the standard of importance" will become diffuse and it will deviate from the meaning of (c) in the argument which is not shown the concrete target.
\begin{itemize}
\item[(d)]
In order to secure the firm foothold of engineering and to promote development of engineering further, Measurement theory was made as "a language of engineering."
\end{itemize}
The author's "spirit" is as follows.
\begin{itemize}
\item[(e)]
Measurement theory was made in order to defeat the greatest battle of human beings
-
Construction and settlement of the space colony of hundreds of years of after
-.
\end{itemize}
%\BEGIN{itemize}

%BBBBBBBBBBBBBBBBBB%SBSBSBS
\par
\noindent
{\small%%{\footnotesize
\vspace{0.1cm}
\begin{itemize}
\item[$\spadesuit$] \bf {{}}Note 2.12{{}} \rm
If you are a reader of a theoretical lover, you may ask as follows.
\begin{itemize}
\item[$(\sharp)$]
Although measurement theory includes quantum mechanics({Chap.$\;$1}(Y)),
why don't you propose the "measurement theory" include the theory of relativity?
\end{itemize}
However, this book is "writing of engineering" as mentioned above.
%%index{@${{\cdot}}$}
%${{\cdot}}$(=),
% and ,
% and  and ,
%
%,
%2.4.1(e)
%%
%%${{\cdot}}$(${{\cdot}}$)
%%
%%
% and .
Although I think that "refraining from doing too much" is an engineering sense, there is no mind which, of course, stops that the reader of a theoretical lover challenges the above-mentioned $(\sharp)$.
%${{\cdot}}$${{\cdot}}$
%% and ,
%,
\end{itemize}
}

\subsection{The Copenhagen interpretation
---
Only one measurement is permitted}%{Sec.2.5}
\par
In this section,
we examine
the Copenhagen interpretation(Chap. 1(U$_4$)),
i.e.,
"Only one measurement is permitted".
"Only one measurement"
implies
that
"only one observable"
and
"only one state".
That is,
we see:
\begin{itemize}
\item[]
$
\qquad
%\text
%\Leftrightarrow
\text{[{{only one measurement}}]}
\Longrightarrow
\cases
\text{only one observable {(}{=measuring instrument}{)}}
\\
\\
\text{{{only one state}}}
\endcases
$
\hfill{(2.6)}
\end{itemize}
%%PPPPPPPPPPPPPPPPP

%BBBBBBBBBBBBBBBBBB%SBSBSBS
\par
\noindent
{\small%%{\footnotesize
\begin{itemize}
\item[$\spadesuit$] \bf {{}}{Note }2.13{{}} \rm
%{{measurement theory}}, {{measurement}}({Chap.$\;$1}(U$_4$)).
%\item[] \bf  \rm %%%BBBBBBBBBBBBBBBBBBBB
%\\
%%,
Although
there may be several opinions,
the author believes that
the standard Copenhagen interpretation says that
only one measurement is permitted.
This spirit is inherited
to measurement theory.
%
%inheri
%
%When {measuring object}(\textcolor{black}{
%{Sec.1.2.3}{FIG.$\;$}1.1({{measurement}}{FIG.$\;$})}),
%{{state}}
%(,
%Schr\"odinger),
%{{measurement}},
%{{state}}{{measurement}} and .
%
%% and ,
%quantum mechanics, {{measurement}}
%\footnote{
% and ,
%quantum mechanics{}
%, quantum mechanics.
%{{measurement theory}}(=quantum mechanicslinguistic
%(cf. \textcolor{black}{\cite{INewi, ILing}}))
%{linguistic aspect} and quantum mechanics and 
%(\textcolor{black}{{Note }3.6, {Sec.9.3}}).
%}.
%{{{measurement theory}}}quantum mechanics{(}
%\textcolor{black}{{}{{{}}}{Sec.3.2}}{)},
%quantum mechanics{{measurement}},
%{{measurement theory}}.
%%,
%%quantum mechanics, {{measurement}}
%% and .
\end{itemize}
}
%%BBBBBBBBBBBBBBBBBB%SBSBSBSS
%
\subsubsection{"Observable is only one"
and simultaneous measurement}%{Sec. 2.5.1}
\par
%For example,
%consider the following measurement:
%%{{measurement}}
%%:
%\BEGIN{itemize}
%\item[({}a)]
%\baselineskip=18pt
\par
For example,
consider the following situation:
%{{measurement}}
%:
\begin{itemize}
\item[({}a)]
There is a cup in which water is filled.
Assume that the temperature
is
$\omega \SD$
$(0 {{\; \leqq \;}}\omega {{\; \leqq \;}}100)$.
%Consider
%
%
%There is water with temperature
%$\omega$$(0 {{\; \leqq \;}}\omega {{\; \leqq \;}}100)$
%in a cup.
Consider
two questions
"Is this water cold or hot?"
and
"How many degrees($\SD$) is roughly the water?".
This implies that
we take two
measurements
such that
\begin{itemize}
\item[]
$
\cases
\text{
$(\sharp_1)$:
${\mathsf M}_{C ( \Omega )} ( {\mathsf O}_{{{{{c}}}}{{{{h}}}}}
{{=}}
(\{{{{{c}}}},{{{{h}}}}\}
, 2^{\{{{{{c}}}},{{{{h}}}}\}}, F_{{{{{c}}}}{{{{h}}}}} ), S_{[\omega]} )$
in
\textcolor{black}{Example 2.7}
}
\\
\\
\text{
$(\sharp_2)$
:
${\mathsf M}_{C ( \Omega )}$
$ ({\mathsf O}_{\text{\scriptsize AB}}
$
$
{{=}}
({\mathbb N}_{10}^{100}   ,$
$
 2^{{\mathbb N}_{10}^{100} }, G_{\text{\scriptsize AB}} ),$
$
S_{[\omega]} )
$
in
\textcolor{black}{Example 2.9}
}
\endcases
$
\end{itemize}
\end{itemize}
However,
as mentioned in the above,
"only one observable"
must be demanded.
Thus, we have the following problem.
\par
\noindent
{\bf
\vskip0.3cm
%BFBF
\par
\noindent
{{Problem }}2.12}$\;\;$%POPOPO
\rm
%$\qquad$
Represent
two measurements
${\mathsf M}_{C ( \Omega )} ( {\mathsf O}_{{{{{c}}}}{{{{h}}}}}
{{=}}
(\{{{{{c}}}},{{{{h}}}}\}
, $
$2^{\{{{{{c}}}},{{{{h}}}}\}}, F_{{{{{c}}}}{{{{h}}}}} ), S_{[\omega]} )$
and
${\mathsf M}_{C ( \Omega )} ({\mathsf O}_{\text{\scriptsize AB}}
{{=}}
({\mathbb N}_{10}^{100}   ,$
$
 2^{{\mathbb N}_{10}^{100} }, G_{\text{\scriptsize AB}} ),$
$
S_{[\omega]} )
$
by
only one measurement.
\par
This will be answered in what follows.

\par
\noindent
{\bf
\vskip0.3cm
%BFBF
\par
\noindent
{Definition }2.13
[{}{\bf {Product measurable space}{\rm}}{}]}$\;\;$%POPOPO
%index{@{product measurable space}}
For each
$k =1,2,\ldots,n$,
consider
a measurable
$(X_k , $
${\cal F}_k )$.
The {{product space}}
%index{@{product space}}
$\bigtimes_{k=1}^n  X_k$
of
$X_k$
$(k=1,2,\ldots,n )$
is defined by
\begin{align*}
\bigtimes_{k=1}^n  X_k
=
\{ (x_1, x_2,\ldots, x_n ) \;|\; x_k \in X_k
\;\;
(k=1,2,\ldots,n )\}
%P%%\TAG{24}
\end{align*}
Similarly,
define
the
product
$\bigtimes_{k=1}^n  \Xi_k$
of
$\Xi_k
(
\in
{\cal F}_k
)$
$(k=1,2,\ldots,n )$
by
\begin{align*}
\bigtimes_{k=1}^n  \Xi_k
=
\{ (x_1, x_2,\ldots, x_n ) \;|\; x_k \in \Xi_k
\;\;
(k=1,2,\ldots,n )\}
\end{align*}
Further,
the
$\sigma$-{{field}}
$\bigstimes_{k=1}^n{\cal F}_k$
on
the {product space}
$\bigtimes_{k=1}^n  X_k$
by
\begin{itemize}
\item[]
$\bigstimes_{k=1}^n{\cal F}_k$
is
the smallest
%$\sigma$-
{{field}}
including
$ \{
\bigtimes_{k=1}^n  \Xi_k
\;|\;
\Xi_k \in {\cal F}_k
\;\;
(k=1,2,\ldots,n )
\}$
%(\textcolor{black}{Appendix B.5(A)})
\end{itemize}

$({}\bigtimes_{k=1}^n  X_k , \bigstimes_{k=1}^n{\cal F}_k)$
is called the {\{product measurable space}}.
%index{paa@$\bigstimes_{k=1}^n{\cal F}_k$:product $\sigma$-{field}}
%\par
%\vskip0.3cm
%\par
%%,
%%$$
%\{
%\bigtimes_{k=1}^n  \Xi_k
%\;|\;
%\Xi_k \in {\cal F}_k
%\;\;
%(k=1,2,\ldots,n )
%\}
%\ssubsetneqq
%\bigtimes_{k=1}^n{\cal F}_k
%$$
%.
%,
%{{product measurable space}}
%$({}\bigtimes_{k=1}^n  X_k , \bigtimes_{k=1}^n{\cal F}_k)$
%
%$({}\bigtimes_{k=1}^n  X_k ,\\ \bigotimes_{k=1}^n{\cal F}_k)$
% and 
%,
%(
%$\bigotimes_{k=1}^n$,
%\textcolor{black}{{Definition }2.17} and 
%)$({}\bigtimes_{k=1}^n  X_k , \bigtimes_{k=1}^n{\cal F}_k)$
% and .
Also,
in the case that
$(X,{\cal F})=(X_k,{\cal F}_k)$
$(k=1,2,\ldots,n)$,
the
{product space}
$\bigtimes_{k=1}^n  X_k$
is denoted by
$X^n$,
and
the
{product measurable space}
$({}\bigtimes_{k=1}^n  X_k ,$
$ \bigstimes_{k=1}^n{\cal F}_k)$
is denoted by
$(X^n,{\cal F}^n)$.

\par
\noindent
{\bf
\vskip0.3cm
%BFBF
\par
\noindent
{Definition }2.14
[{}simultaneous observable {}, simultaneous measurement{\rm
}{}]}$\;\;$%POPOPO
%index{@simultaneous observable }
\rm
For
$k =1,2,\ldots,n$
%$\in K {{=}}  \{ 1,2,\ldots |K| \}$,
consider
{observable } ${\mathsf O}_k$
$=$
$(X_k , $
${\cal F}_k , $
$F_k{})$
in
${C ( \Omega )}$.
Let
$({}\bigtimes_{k=1}^n  X_k , \bigstimes_{k=1}^n{\cal F}_k)$
be the {product measurable space}.
An {observable }
$\widehat{\mathsf O}$
$=$
$({}\bigtimes_{k\in K } X_k ,$
$ \bigstimes_{k=1}^n{\cal F}_k , $
$\widehat{F}{})$
in
${C ( \Omega )}$
is called the
simultaneous observable
of
$\{ {\mathsf O}_k \;: \; k=1,2,...,n \}$,
if it satisfies the following condition:
\begin{align*}
[{\widehat F}({}\Xi_1 \times \Xi_2 \times \cdots \times \Xi_{n}{})]
(\omega)
&
=
[F_1 ({}\Xi_1{})](\omega)
\cdot
[F_2 ({}\Xi_2{}) ](\omega)
%\cdot
%[F_3 ({}\Xi_3{}) ](\omega)
\cdots 
[F_n ({}\Xi_{n}{})]
(\omega)
%%%%2.24}
%\TAG{3.34}
%%\tag{\color{black}{2.7}}
%%%%%REDREDREDREDREDRE
\\
&
%\quad
(
\forall
\omega \in \Omega,
\;\;
\forall \Xi_k \in {\cal F}_k \;
(k=1,2,\ldots,n ))
\tag{\color{black}{2.7}}
\end{align*}
%%\it
% and ,
%observable
%$\widehat{\mathsf O}$
%$=$
%$({}\bigtimes_{k=1}^n  X_k , \bigstimes_{k=1}^n{\cal F}_k , \widehat{F}{})$
%,
%$\{{\mathsf O}_k\}_{k=1}^n$
%
%{\bf simultaneous observable } and .
%index{p2@$\bigtimes_{k=1}^n{\mathsf O}_k$:simultaneous observable }
\rm
$\widehat{\mathsf O}$
is also denoted by
$
\bigtimes_{k=1}^n{\mathsf O}_k$,
$\widehat{F}$
$=$
$
\bigtimes_{k=1}^n{F}_k$.
\rm
%imes_{k=1}^n{X}_k$
%$=$
%$\bigtimes_{k=1}^n{X}_k$,
%{}
Also,
the measurement
${\mathsf M}_{C(\Omega)} (\bigtimes_{k=1}^n{\mathsf O}_k, S_{[\omega]})$
is called the
{simultaneous measurement}.
\par
\vskip0.3cm
\par
%In {{{measurement theory}}}{(}
%\text{\textcolor{black}{{Chap.{\;}}IIPART}}
%{)},
%,
%
%($n=\infty$)
%,
%simultaneous observable $\bigtimes_{k=1}^\infty
%{\mathsf O}_k$
%%That is,
%%($n=\infty$
%%
%%)
%,
%bounded type {{{measurement theory}}}{(}\textcolor{black}{{Chap.{\;}}IVPART}{)}
%, ${{\cdot}}$
%\textcolor{black}{{(}10.14)}\footnote{
%,
%continuous type {{{measurement theory}}},
%
% and 
%\textcolor{black}{{(}
%{}3
%%%2%%
%)}.
%}.
%index{@simultaneous measurement}
%%index{papoduct@$\bigtimes_{k=1}^n$, $\quad$ $\bigtimes_{k=1}^n$}

%
%\END{align*}
%
%
\par
In what follows,
let us explain
the simultaneous measurement.
We want to take two
measurements
%{{state}}$\omega${measuring object},
%{{measurement}}
${\mathsf M}_{C(\Omega)}({\mathsf O}_1,$
$ S_{[\omega]})$
and
{{measurement}}
${\mathsf M}_{C(\Omega)}({\mathsf O}_2, S_{[\omega]})$.
That is, it suffices to image the following:
%\par
%\noindent
\vspace{-1.5cm}
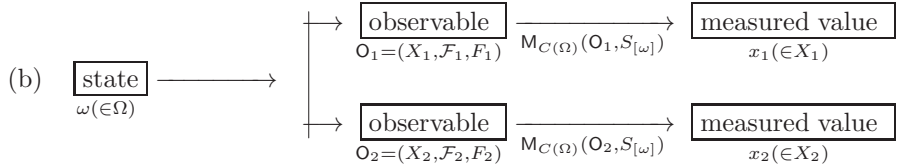
\begin{itemize}
\item[(b)]
$\text{
\unitlength=0.40mm
%\BEGIN{picture}(400,145)
\begin{picture}(400,70)
\put(-30,-70)
%\put(-30,-70)
%\put(0,-70)
{
\put(30,70){$\overset{{}}{\underset{\footnotesize \omega (\in \Omega )}
{\text{ \fbox{{{state}}}}}}
%}_{{(}{measuring object}{)}}
\xrightarrow[]{\qquad \qquad}$}
\path(111,53)(111,95)
\put(90,70)
{
$
\begin{array}{rr}
{}&{}
\longrightarrow
%\xrightarrow[{\mathsf M}_{C(\Omega)}({\mathsf O}_1, S_{[\omega]})]{}
%\underbrace{
\overset{{}}
%\underbrace{
{
\underset{\footnotesize {\mathsf O}_1{{=}} (X_1, {\cal F}_1, F_1 )}
{\text{ \fbox{observable }}}
}
%}_{}
\xrightarrow[{\mathsf M}_{C(\Omega)}({\mathsf O}_1, S_{[\omega]})]{}
%\underbrace{
\overset{{}}
{
\underset{\footnotesize x_1(\in X_1)}
{\text{ \fbox{measured value }}}
}
\\
{}&{}
\\
{}&{}
\longrightarrow
%\underbrace{
\overset{{}}
%\underbrace{
{
\underset{\footnotesize {\mathsf O}_2{{=}} (X_2 , {\cal F}_2, F_2 )}
{\text{ \fbox{observable }}}
}
%}_{}
\xrightarrow[{\mathsf M}_{C(\Omega)}({\mathsf O}_2, S_{[\omega]})]{}
%\underbrace{
\overset{{}}
{
\underset{\footnotesize x_2 (\in X_2)}
{\text{ \fbox{measured value }}}
}
\end{array}
%dddd
$
}
}
\end{picture}
}
$
\end{itemize}
%{the resosution the unity}$\{ f_{x_1}, f_{x_2}, f_{x_3} \}$
%
\par
\noindent
\vskip1.5cm
\par
\noindent
However,
the Copenhagen interpretation({Chap.$\;$1}(U$_4$))
says that
two measurements can not be taken,
Therefore,
combining
two observables ${\mathsf O}_1$
and
${\mathsf O}_2$,
we construct
the
simultaneous observable ${\mathsf O}_1\times {\mathsf O}_2$,
and
take
the simultaneous measurement
${\mathsf M}_{C(\Omega)} (
{\mathsf O}_1
\times
{\mathsf O}_2, S_{[\omega]})$
in what follows.
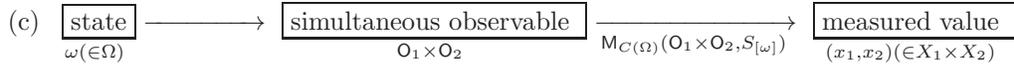
\begin{itemize}
\item[(c)]
$
%\overbrace{{
%\underbrace{
\overset{{}}{\underset{\footnotesize \omega (\in \Omega )}
{\text{ \fbox{{{state}}}}}}
%}_{{(}{measuring object}{)}}
\xrightarrow[]{\qquad \qquad}
%\underbrace{
\overset{{}}
%\underbrace{
{
\underset{\footnotesize {\mathsf O}_1 \times {\mathsf O}_2}
{\text{ \fbox{simultaneous observable }}}
}
%}_{}
\xrightarrow[
{\mathsf M}_{C(\Omega)} (
{\mathsf O}_1
\times
{\mathsf O}_2, S_{[\omega]})]{}
%\underbrace{
\overset{{}}
{
\underset{\footnotesize (x_1,x_2) (\in X_1 \times X_2 )}
{\text{ \fbox{measured value }}}
}
%}_
%}_{{(}observer{)}}
%}}^{{{measurement}}}
%\TAG{3.35}
$
\end{itemize}

%PPPPPPPPPPPPPPPPPPPPPPPPPPPPPPPPPPPPPPPPPPPP

\par
\noindent
\vskip0.3cm
%BFBF
\par
\noindent
{\bf Example 2.15
[The answer to {{Problem }}]}$\;\;$%POPOPO
%
%\textcolor{black}{(a)}.
%{{measurement}}
% and  and {}
%\END{itemize}
Consider the state space
$\Omega$
such that
$\Omega =$
$ [0,100]$,
the closed interval.
And consider two observables,
that is,
{{{{c}}}}{{{{h}}}}-observable
${\mathsf O}_{{{{{c}}}}{{{{h}}}}}= (X {{=}}  \{ {{{{c}}}} ,  {{{{h}}}} \}, 2^X, F_{{{{{c}}}}{{{{h}}}}} )$
(in \textcolor{black}{Example 2.7})
and
{{about-observable} }
${\mathsf O}_{\text{\scriptsize AB}}= (Y( {{=}} {\mathbb N}_{10}^{100})   , 2^Y, G_{\text{\scriptsize AB}} )$
(in \textcolor{black}{Example 2.9}).
Thus,
we get
the
simultaneous observable
${\mathsf O}_{{{{{c}}}}{{{{h}}}}} \times {\mathsf O}_{\text{\scriptsize AB}}$
$=$
$(\{ {{{{c}}}} ,  {{{{h}}}} \}\times {\mathbb N}_{10}^{100},
2^{\{ {{{{c}}}} ,  {{{{h}}}} \}\times {\mathbb N}_{10}^{100}},
F_{{{{{c}}}}{{{{h}}}}} \times G_{\text{\scriptsize AB}} )$,
take
the
simultaneous measurement
${\mathsf M}_{C(\Omega)}({\mathsf O}_{{{{{c}}}}{{{{h}}}}} \times {\mathsf O}_{\text{\scriptsize AB}}, S_{[\omega]})$.
For example,
putting
$\omega=55$,
we see
\begin{itemize}
\item[({}d)]
when
the simultaneous measurement
${\mathsf M}_{C(\Omega)}({\mathsf O}_{{{{{c}}}}{{{{h}}}}} \times {\mathsf O}_{\text{\scriptsize AB}}, S_{[55]})$
is taken, the probability
\begin{align*}
&
\text{that
}
\text{the measured value }
\left[\begin{array}{ll}
(\text{{{{{c}}}}}, \text{about 50$\SD$})
\\
(\text{{{{{c}}}}}, \text{about 60$\SD$})
\\
(\text{{{{{h}}}}}, \text{about 50$\SD$})
\\
(\text{{{{{h}}}}}, \text{about 60$\SD$})
\end{array}\right]
\text{is obtained is given by}
\left[\begin{array}{ll}
0.125
\\
0.125
\\
0.375
\\
0.375
\end{array}\right]
%\TAG{3.36}
\tag{\color{black}{2.8}}
%%%%%REDREDREDREDREDRE
%%\tag{\color{black}{2.7}}
\end{align*}
\end{itemize}
That is because
\begin{align*}
&
[(F_{{{{{c}}}}{{{{h}}}}} \times G_{\text{\scriptsize AB}})(
\{ (\text{{{{{c}}}}},
\text{about 50$\SD$}
) \}
)]
(55)
\\
=
&
[F_{{{{{c}}}}{{{{h}}}}}(\{ \text{{{{{c}}}}} \} )](55)
\cdot [G_{\text{\scriptsize AB}}(\{ \text{about 50$\SD$}\})](55)
=0.25 \cdot 0.5=0.125
\end{align*}
and similarly,
\begin{align*}
&
[(F_{{{{{c}}}}{{{{h}}}}} \times G_{\text{\scriptsize AB}})(
\{ (\text{{{{{c}}}}},
\text{about 60$\SD$}
) \}
)]
(55)
=0.25 \cdot 0.5=0.125
\\
&
[(F_{{{{{c}}}}{{{{h}}}}} \times G_{\text{\scriptsize AB}})(
\{ (\text{{{{{h}}}}},
\text{about 50$\SD$}
) \}
)]
(55)
=0.75 \cdot 0.5=0.375
\\
&
[(F_{{{{{c}}}}{{{{h}}}}} \times G_{\text{\scriptsize AB}})(
\{ (\text{{{{{h}}}}},
\text{about 60$\SD$}
) \}
)]
(55)
=0.75 \cdot 0.5=0.375
%\TAG{3.37}
\end{align*}

%_{{{{{c}}}}{{{{h}}}}}_{\text{\scriptsize AB}}_t_r\t_t_r

%%BBBBBBBBBBBBBBBBBB%SBSBSBS{\mathsf O}
\par
\noindent
{\small%%{\footnotesize
\vspace{0.1cm}
\begin{itemize}
\item[$\spadesuit$] \bf {{}}{Note }2.14{{}} \rm
%,
%,
% and , (,
%).
%%,
%,
%\textcolor{black}{{Note }2.5}.
%,
The above argument
does not have generality.
In
quantum mechanics,
a simultaneous observable ${\mathsf O}_1\times {\mathsf O}_2$
does not always exist
(\textcolor{black}{{Note }3.3}
and
Heisenberg's {uncertainty principle in Sec.3.4}).
\end{itemize}
}
%%BBBBBBBBBBBBBBBBBB%SBSBSBSS
%
\subsubsection{"State is only one" and parallel measurement}
\rm
\baselineskip=18pt
\par
\par
%For example,
%consider the following measurement:
%%{{measurement}}
%%:
%\BEGIN{itemize}
%\item[({}a)]
%\baselineskip=18pt
\par
For example,
consider the following situation:
%{{measurement}}
%:
\begin{itemize}
\item[({}a)]
There are two cups $A_1$ and $A_2$ in which water is filled.
Assume that the temperature of the water in the cup
$A_k$
$(k=1,2)$
is
$\omega_k \SD$
$(0 {{\; \leqq \;}}\omega_k {{\; \leqq \;}}100)$.
%Consider
%
%
%There is water with temperature
%$\omega$$(0 {{\; \leqq \;}}\omega {{\; \leqq \;}}100)$
%in a cup.
Consider
two questions
"Is the water in the cup $A_1$ cold or hot?"
and
"How many degrees($\SD$) is roughly the water in the cup $A_2$?".
This implies that
we take two
measurements
such that
\begin{itemize}
\item[]
$
\cases
\text{
$(\sharp_1)$:
${\mathsf M}_{C ( \Omega )} ( {\mathsf O}_{{{{{c}}}}{{{{h}}}}}
{{=}}
(\{{{{{c}}}},{{{{h}}}}\}
, 2^{\{{{{{c}}}},{{{{h}}}}\}}, F_{{{{{c}}}}{{{{h}}}}} ), S_{[\omega_1]} )$
in
\textcolor{black}{Example 2.7}
}
\\
\\
\text{
$(\sharp_2)$
:
${\mathsf M}_{C ( \Omega )}$
$ ({\mathsf O}_{\text{\scriptsize AB}}
$
$
{{=}}
({\mathbb N}_{10}^{100}   ,$
$
 2^{{\mathbb N}_{10}^{100} }, G_{\text{\scriptsize AB}} ),$
$
S_{[\omega_2]} )
$
in
\textcolor{black}{Example 2.9}
}
\endcases
$
\end{itemize}
\end{itemize}
However,
as mentioned in the above,
"only one state"
must be demanded.
Thus, we have the following problem.
\par
\noindent
{\bf
\vskip0.3cm
%BFBF
\par
\noindent
{{Problem }}2.16}$\;\;$%POPOPO
\rm
$\qquad$
Represent
two measurements
${\mathsf M}_{C ( \Omega )} ( {\mathsf O}_{{{{{c}}}}{{{{h}}}}}
{{=}}
(\{{{{{c}}}},{{{{h}}}}\}
, $
$2^{\{{{{{c}}}},{{{{h}}}}\}}, F_{{{{{c}}}}{{{{h}}}}} ), S_{[\omega_1]} )$
and
${\mathsf M}_{C ( \Omega )} ({\mathsf O}_{\text{\scriptsize AB}}$
$
{{=}}
({\mathbb N}_{10}^{100}   ,$
$
 2^{{\mathbb N}_{10}^{100} }, G_{\text{\scriptsize AB}} ),$
$
S_{[\omega_2]} )
$
by
only one measurement.
\par
This will be answered in what follows.
\par
\noindent

%\ssubsubsection{{}parallel measurement}{}}
\par
\noindent
{\bf
%\vskip0.3cm
%BFBF
\par
\noindent
{Definition }2.17
[{}{\bf Parallel observable {\rm
}}{}, parallel measurement{\rm
}]}$\;\;$%POPOPO
%index{@parallel {observable }}
\sf
[Parallel observable, parallel measurement].
\rm
For each
$k=1,2,$
$...,$
$n$,
consider
a measurement
${\mathsf M}_{{C(\Omega)}_k} \big({}{\mathsf O}_k \equiv
({}X , {\cal F} , F_k{}), S_{[{}\rho_k^p{}] } \big)$
in a $C^*$-algebra ${C(\Omega)}_k$.
Put
$\widehat{C(\Omega)}$
$=$
$\bigotimes_{k=1}^n {C(\Omega)}_k $,
i.e.,
the tensor product $C^*$-algebra of
$\{ {C(\Omega)}_k \;{}: $
$k=1,2,...,n \}$.
%And therefore, $\widehat{{\cal M}(\Omega)$ $=$ $\bigotimes_{k=1}^n {C(\Omega)}_k^* $.
Here,
consider
the tensor product $C^*$-observable
$\bigotimes_{k=1}^n {\mathsf O}_k $
$\equiv$
$(${}$X^n$,
$\bigstimes_{k=1}^n {\cal F}$,
$ {\widehat F} $
$\equiv \bigotimes_{k=1}^n F_k  $
$)$
in
$\widehat{C(\Omega)}$
$(${}$\equiv$
$\bigotimes_{k=1}^n {C(\Omega)}_k $
$)$
such that:
\begin{align*}
{\widehat F}
({}\Xi_1 \times \Xi_2 \times \cdots \times \Xi_n{})
=
F_1({}\Xi_1{}) \otimes
F_2({}\Xi_2{}) \otimes \cdots \otimes
F_n({}\Xi_n{})
\quad
({}\forall \Xi_k \in {\cal F}  , \; k=1,2,..., n{}) .
%\tag{1.77}
\tag{\color{black}{2.9}}
\end{align*}
Therefore,
we get
the measurement
${\mathsf M}_{\otimes {C(\Omega)}_k} ({}\bigotimes_{k=1}^n {\mathsf O}_k ,
S_{ [{}\bigotimes_{k=1}^n \rho_k^p{}] }{})$
in
$\bigotimes_{k=1}^n {C(\Omega)}_k $,
which is also denoted by
$\bigotimes_{k=1}^n {\mathsf M}_{{C(\Omega)}_k}
({}{\mathsf O}_k , S_{[{}\rho_k^p{}]}{})$
and
called the
{\it
parallel measurement}
%%index{repeated measurement@repeated measurement}
%index{parallel measurement@parallel measurement}
of
$\{
{\mathsf M}_{{C(\Omega)}_k}({}{\mathsf O}_k ,$
$ S_{[{}\rho_k^p{}]}{})\}_{k=1}^n $.

\par

%
%\END{align*}
%
%
\par
In what follows,
let us explain
the parallel measurement.
We want to take two
measurements
%{{state}}$\omega${measuring object},
%{{measurement}}
${\mathsf M}_{C(\Omega)}({\mathsf O}_1,$
$ S_{[\omega_1]})$
and
{{measurement}}
${\mathsf M}_{C(\Omega)}({\mathsf O}_2, S_{[\omega_2]})$.
That is, it suffices to image the following:
%\par
\begin{itemize}
\item[(b)]
$\quad$
$
%\overbrace{{
%\underbrace{
\cases
\overset{{}}{\underset{\footnotesize \omega_1 (\in \Omega_1 )}
{\text{ \fbox{{{state}}}}}}
%}_{{(}{measuring object}{)}}
\xrightarrow[]{\qquad \qquad}
%\underbrace{
\overset{{}}
%\underbrace{
{
\underset{\footnotesize {\mathsf O}_1}
{\text{ \fbox{observable }}}
}
%}_{}
\xrightarrow[{\mathsf M}_{C(\Omega_1)}({\mathsf O}_1, S_{[\omega_1]})]{\qquad \qquad}
%\underbrace{
\overset{{}}
{
\underset{\footnotesize x_1 (\in X_1 )}
{\text{ \fbox{measured value }}}
}
%}_{
%}_{{(}observer{)}}
%}}^{{{measurement}}}
\\
\\
\overset{{}}{\underset{\footnotesize \omega_2 (\in \Omega_2 )}
{\text{ \fbox{{{state}}}}}}
%}_{{(}{measuring object}{)}}
\xrightarrow[]{\qquad \qquad}
%\underbrace{
\overset{{}}
%\underbrace{
{
\underset{\footnotesize  {\mathsf O}_2}
{\text{ \fbox{observable }}}
}
%}_{}
\xrightarrow[{\mathsf M}_{C(\Omega_2)}({\mathsf O}_2, S_{[\omega_2]})]{\qquad \qquad}
%\underbrace{
\overset{{}}
{
\underset{\footnotesize x_2 (\in X_2 )}
{\text{ \fbox{measured value }}}
}
%}_}
%}_{{(}observer{)}}
\endcases
$
%\TAG{3.39}
\end{itemize}
\par
\noindent
However,
the Copenhagen interpretation({Chap.$\;$1}(U$_4$))
says that
two measurements can not be taken,
Let us
regard
two states $\omega_1$ and $\omega_2$
as
one state
$(\omega_1, \omega_2 )$
$( \in \Omega_1 \times \Omega_2 )$.
And further,
combining
two observables ${\mathsf O}_1$
and
${\mathsf O}_2$,
we construct
the
simultaneous observable ${\mathsf O}_1\times {\mathsf O}_2$,
and
take
the simultaneous measurement
${\mathsf M}_{C(\Omega_1 \times \Omega_2)} (
{\mathsf O}_1
\otimes
{\mathsf O}_2, S_{[(\omega_1, \omega_2 )]})$
in what follows.
%\BEGIN{itemize}
%\item[(c)]
%$
%
%
%
%
%
%
%
%,
%the Copenhagen interpretation({Chap.$\;$1}(U$_4$)),
%%Axiom${}_{\text{\scriptsize c}}^{\text{\scriptsize p}}$ 1(i),
%2{{measurement}} and .
%, 2{{state}}
%$\omega_1
%(\in \Omega_1)$
% and
%$\omega_2(\in \Omega_2)$
%{{state}}
%$(\omega_1,\omega_2)(\in \Omega_1\times \Omega_2)$
% and ,
%,
%2observable ${\mathsf O}_1$ and ${\mathsf O}_2$
%,
%parallel observable
%${\mathsf O}_1 \otimes {\mathsf O}_2$
%,
%parallel measurement
%${\mathsf M}_{C(\Omega_1 \times \Omega_2)}({\mathsf O}_1 \otimes
%{\mathsf O}_2
%, S_{[(\omega_1,\omega_2)]})$
%.
%{FIG.$\;$} and ,
\par
\begin{itemize}
\item[(c)]
%}
$
\overset{{}}{\underset{\footnotesize (\omega_1, \omega_2)  (\in \Omega_1
\times \Omega_2 )}
{\text{ \fbox{{{state}}}}}}
%}_{{(}{measuring object}{)}}
\xrightarrow[]{\;}
%\underbrace{
\overset{{}}
%\underbrace{
{
\underset{\footnotesize {\mathsf O}_1 \otimes {\mathsf O}_2}
{\text{ \fbox{parallel observable }}}
}
%}_{}
\xrightarrow[{\mathsf M}_{C(\Omega_1 \times \Omega_2)}({\mathsf O}_1 \otimes
{\mathsf O}_2
, S_{[(\omega_1,\omega_2)]})]{\qquad \qquad}
%\underbrace{
\overset{{}}
{
\underset{\footnotesize (x_1,x_2) (\in X_1 \times X_2 )}
{\text{ \fbox{measured value }}}
}
%}
%}_{{(}observer{)}}
$
\end{itemize}

%%%%PPPPPPPPPPPPPPPPPPPPPPPPP
%
%
%\par

\par
\noindent
\vskip0.3cm
%BFBF
\par
\noindent
{\bf Example 2.18
[{{Answer to Problem }}2.16]}$\;\;$%POPOPO
%
%{}
%\END{itemize}
Put
$\Omega_1 = \Omega_2  = [0,100]$,
and define
the state space
$\Omega_1 \times \Omega_2$.
And consider two observables,
that is,
{{{{c}}}}{{{{h}}}}-observable
${\mathsf O}_{{{{{c}}}}{{{{h}}}}}= (X {{=}}  \{ {{{{c}}}} ,  {{{{h}}}} \}, 2^X, F_{{{{{c}}}}{{{{h}}}}} )$
in $C(\Omega_1)$
(in \textcolor{black}{Example 2.7})
and
{{about-observable} }
${\mathsf O}_{\text{\scriptsize AB}}= (Y( {{=}} {\mathbb N}_{10}^{100})   , 2^Y, G_{\text{\scriptsize AB}} )$
in
$C(\Omega_2)$
(in \textcolor{black}{Example 2.9}).
Thus,
we get
the
parallel observable
${\mathsf O}_{{{{{c}}}}{{{{h}}}}} \times {\mathsf O}_{\text{\scriptsize AB}}$
$=$
$(\{ {{{{c}}}} ,  {{{{h}}}} \}\times {\mathbb N}_{10}^{100},
2^{\{ {{{{c}}}} ,  {{{{h}}}} \}\times {\mathbb N}_{10}^{100}},
F_{{{{{c}}}}{{{{h}}}}} \otimes G_{\text{\scriptsize AB}} )$
in $C(\Omega_1 \times \Omega_2 )$,
take
the
parallel measurement
${\mathsf M}_{C(\Omega_1 \times \Omega_2 )}({\mathsf O}_{{{{{c}}}}{{{{h}}}}} \otimes {\mathsf O}_{\text{\scriptsize AB}},
S_{[(\omega_1,\omega_2)]})$.
For example,
putting
$(\omega_1,\omega_2 ) =(25, 55)$,
we see the following.
% and , :

\begin{itemize}
\item[({}d)]
When the parallel measurement
${\mathsf M}_{C(\Omega_1 \times \Omega_2 )}({\mathsf O}_{{{{{c}}}}{{{{h}}}}} \otimes {\mathsf O}_{\text{\scriptsize AB}},
S_{[(25,55)]})$
is taken, the probability
\begin{align*}
&
\text{that
}
\text{the measured value }
\left[\begin{array}{ll}
(\text{{{{{c}}}}}, \text{about 50$\SD$})
\\
(\text{{{{{c}}}}}, \text{about 60$\SD$})
\\
(\text{{{{{h}}}}}, \text{about 50$\SD$})
\\
(\text{{{{{h}}}}}, \text{about 60$\SD$})
\end{array}\right]
\text{is obtained is given by}
\left[\begin{array}{ll}
0.375
\\
0.375
\\
0.125
\\
0.125
\end{array}\right]
%P%%\TAG{28}
\end{align*}
\end{itemize}
That is because
\begin{align*}
&
[(F_{{{{{c}}}}{{{{h}}}}} \otimes G_{\text{\scriptsize AB}})(
\{ (\text{{{{{c}}}}},
\text{about 50$\SD$}
) \}
)]
(25,55)
\\
=
&
[F_{{{{{c}}}}{{{{h}}}}}(\{ \text{{{{{c}}}}} \} )](25)
\cdot [G_{\text{\scriptsize AB}}(\{ \text{about 50$\SD$}\})](55)
=0.75 \cdot 0.5=0.375
\end{align*}
Thus, similarly,
\begin{align*}
&
[(F_{{{{{c}}}}{{{{h}}}}} \otimes G_{\text{\scriptsize AB}})(
\{ (\text{{{{{c}}}}},
\text{about 60$\SD$}
) \}
)]
(25,55)
=0.75 \cdot 0.5=0.375
%=0.25 \cdot 0.5=0.125
\\
&
[(F_{{{{{c}}}}{{{{h}}}}} \otimes G_{\text{\scriptsize AB}})(
\{ (\text{{{{{h}}}}},
\text{about 50$\SD$}
) \}
)]
(25,55)
=0.25 \cdot 0.5=0.125
\\
&
[(F_{{{{{c}}}}{{{{h}}}}} \otimes G_{\text{\scriptsize AB}})(
\{ (\text{{{{{h}}}}},
\text{about 60$\SD$}
) \}
)]
(25,55)
=0.25 \cdot 0.5=0.125
%\TAG{3.37}
\end{align*}

\par
\noindent
\vskip0.3cm
%BFBF
\par
\noindent
{\bf {Remark }2.19}$\;\;$%POPOPO
Also, for example,
putting
$(\omega_1,\omega_2 ) =(55, 55)$,
we see:
;
% and , :
\begin{itemize}
\item[({}e)]
parallel measurement
${\mathsf M}_{C(\Omega_1 \times \Omega_2 )}({\mathsf O}_{{{{{c}}}}{{{{h}}}}} \otimes {\mathsf O}_{\text{\scriptsize AB}},
S_{[(55,55)]})$,
Therefore,
\end{itemize}
\begin{align*}
&
\text{The probability that a measured value }
\left[\begin{array}{ll}
(\text{{{{{c}}}}}, \text{about 50$\SD$})
\\
(\text{{{{{c}}}}}, \text{about 60$\SD$})
\\
(\text{{{{{h}}}}}, \text{about 50$\SD$})
\\
(\text{{{{{h}}}}}, \text{about 60$\SD$})
\end{array}\right]
\text{is obtained is given by}
\left[\begin{array}{ll}
0.125
\\
0.125
\\
0.375
\\
0.375
\end{array}\right]
%\TAG{3.42}
\tag{\color{black}{2.10}}
%%%%%REDREDREDREDREDRE
\end{align*}
%\END{itemize}
That is because, we similarly, see
\begin{align*}
\cases
&
[F_{{{{{c}}}}{{{{h}}}}}(\{ \text{{{{{c}}}}} \} )](55)
\cdot [G_{\text{\scriptsize AB}}(\{ \text{about 50$\SD$}\})](55)
=0.25 \cdot 0.5=0.125
\\
&
[F_{{{{{c}}}}{{{{h}}}}}(\{ \text{{{{{c}}}}} \} )](55)
\cdot [G_{\text{\scriptsize AB}}(\{ \text{about 60$\SD$}\})](55)
=0.25 \cdot 0.5=0.125
\\
&
[F_{{{{{c}}}}{{{{h}}}}}(\{ \text{{{{{h}}}}} \} )](55)
\cdot [G_{\text{\scriptsize AB}}(\{ \text{about 50$\SD$}\})](55)
=0.75 \cdot 0.5=0.375
\\
&
[F_{{{{{c}}}}{{{{h}}}}}(\{ \text{{{{{c}}}}} \} )](55)
\cdot [G_{\text{\scriptsize AB}}(\{ \text{about 60$\SD$}\})](55)
=0.75 \cdot 0.5=0.375
\endcases
%\TAG{3.43}
\end{align*}
This is the same as
\textcolor{black}{Example 2.15}
{(}cf. \textcolor{black}{{Note }2.15} later).
\par
\noindent
\rm
\vskip0.3cm
%BFBF
\par
The follow is obvious, it is deep.
\par
\noindent
{\bf {{Theorem }}2.20}$\;\;$%POPOPO
%%index{@}
The sample probability space
of
a
simultaneous measurement
${\mathsf M}_{C(
%\bigtimes_{k=1}^n
\Omega)} (\bigtimes_{k=1}^n {\mathsf O}_k, S_{[\omega]})$
is the same as
that of
a
{parallel measurement}
${\mathsf M}_{C(
%\bigtimes_{k=1}^n
\Omega^n)}$
$ (
\bigotimes_{k=1}^n {\mathsf O}_k
, $
$S_{[
\otimes_{k=1}^n
\omega]})$.
\par
\noindent
\rm
{\it $\;\;\;\;${Proof.}}$\;\;$
$$
\text{
\textcolor{black}{(2.7)}
=
"\textcolor{black}{(2.9)}
in the case that
$\omega_k=\omega$
$(\forall k=1,2,\ldots,n)$".
}
$$
Thus, the proof is immediately follows.
\qed

%BBBBBBBBBBBBBBBBBB%SBSBSBS
\par
\noindent
{\small%%{\footnotesize
\vspace{0.1cm}
\begin{itemize}
\item[$\spadesuit$] \bf {{}}{Note }2.15{{}} \rm
%
%$\omega_1=\omega_2$,
%\textcolor{black}{(2.8)}
% and
%\textcolor{black}{(2.10)}{}
%,
%simultaneous measurement and parallel measurement and .
%%
%%\item[] \bf  \rm %%%BBBBBBBBBBBBBBBBBBBB
%%\\
%%%,
%%.
\textcolor{black}{{{Theorem }}2.20} is rather deep in the following sense.
For example,
"To toss a coin 10 times"
is
a simultaneous measurement.
On the other hand,
"To toss 10 coins once"
is
characterized as a parallel measurement.
The two have the same sample space.
This means that
the two are not distinguished by the sample space
and not the measurements
(i.e.,
a simultaneous measurement
and
a parallel measurement).
%
%{ordinary language} and  and  and 
%{{measurement theory}} and 
%,
%,
% and {}
%%{{measurement theory}}
%% and {}
%\textcolor{black}{{{Theorem }}2.20},
%\BEGIN{itemize}
%\item[$(\sharp_2)$]
%{ordinary language} and probability ,
%simultaneous measurement and parallel measurement and 
%, .
%%(\textcolor{black}{5.10}).
%\END{itemize}
However, this is peculiar to
classical pure measurements.
It does not hold
in
classical mixed measurements
and
quantum measurement.
%
%
%
%,
%{{measurement}}
%{}
%(
%\textcolor{black}{{Sec.1.3}{{measurement theory}}{FIG.$\;$}({}Y)}
%)
%{{measurement}}{{measurement}}, simultaneous measurement and parallel measurement,
% and , {{measurement}},
%simultaneous measurement and ,
%parallel measurement({Chap. 3}).
\end{itemize}
}

\subsubsection{{}The law of Large Numbers
---
How to find out the sample space}%{Sec. 2.6.1}
\par
Let
${\mathsf O} =(X, {\cal F}, F)$
be an observable in
$C(\Omega)$.
Consider its $n$-dimensional
parallel observable
$\widetilde{\mathsf O}
(=\bigotimes_{k=1}^n
{\mathsf O}
) =(X^n ,{\cal F}^n ,{\widetilde F}
({{=}} \bigotimes_{k=1}^n F ) )$
in $C(\Omega^n)$.
That is,
\begin{align*}
&
[{\widetilde F}({}\Xi_1 \times \Xi_2 \times \cdots \times \Xi_n{})]
(\omega_1,\omega_2,\ldots, \omega_n )
=
[F ({}\Xi_1{})](\omega_1 )
[F ({}\Xi_1{})](\omega_2 )
\cdots 
[F ({}\Xi_n{}) ](\omega_n )
%%%%2.24}
\\
&
\qquad \qquad \qquad
\qquad
%\qquad
(
\forall
(\omega_1, \omega_2,\ldots, \omega_n )\in \Omega^n,
\;\;
\forall \Xi_k \in {\cal F} \;
(k=1,2,\ldots,n ))
%P%\TAG{30}
\end{align*}
\par
Further,
put
$ {\cal M}_{+1} ({}X)$
$ =$
$ \{ \nu{}:$
$\nu$
is a probability measure on $X$
%(\textcolor{black}{Appendix B.5(F)})
$ \}$.
Define the map
$w: X^n \to {\cal M}_{+1} ({}X)$
such that,
%That is,
%$\Xi \in {\cal F}$,
%%%index{sharp@${{\ast}} [{}\; \cdot \;]$}
\begin{align*}
&
[
w({}x_1 , x_2 ,\ldots, x_n{}){}]
({}\Xi{})
=
\frac{ \sharp [{}\{ k{}: x_k \in \Xi \}{}] }
{n}
=
%[{}g ({}{\widetilde x }){}]({}\Xi{})=
\frac{ 1 }{ n } \sum\limits_{k=1}^n \chi_{{}_\Xi} ({}\pi_k ({}{\widetilde x} {}
{}) 
)
=
\frac{ 1 }{ n } \sum\limits_{k=1}^n \chi_{{}_\Xi} ({}x_k {}
{}) 
\\
&
\qquad
\qquad \qquad \qquad
\forall
\Xi \in {\cal F},
\;\;
{}\forall {\widetilde x}  = ({}x_1, x_2 ,\ldots, x_n{}) \in X^n{}{}
%%
%P%\TAG{31}
\end{align*}
where
$\sharp [{}A{}]$
$=$
{\lq\lq the number of the elements of a set $A$\rq\rq},
%index{@{characteristic function}$\chi_{{}_\Xi}$}
%index{characteristic@$\chi_{{}_\Xi}$:{characteristic function}}
%index{sharp@$\sharp [A]:$ $A$}
$\chi$ is a characteristic function
such that
$\chi_{{}_\Xi}(x)
=
1\;(x \in \Xi),$
$
=0\;(x \notin \Xi)$,
$\pi_k$
$: X^n  \to X $
is defined by
$\pi_k ({}{\widetilde x} {}) $
${{=}} \pi_k ({}x_1, x_2 ,\ldots, x_k,\ldots, x_n{})$
$= x_k $
\par
\vskip0.2cm
\par
Before
we present \textcolor{black}{{{Theorem }}2.21}(
the weal law of large numbers in measurement theory),
we add the following note.

%BBBBBBBBBBBBBBBBBB%SBSBSBS
\par
\noindent
{\small%%{\footnotesize
\vspace{0.1cm}
\begin{itemize}
\item[$\spadesuit$] \bf {{}}{Note }2.16{{}} \rm
The weak law of large numbers in probability theory
is as follows.
% and , ($\sharp$) and {}
\begin{itemize}
\item[$(\sharp)$]
Let
$(X, {\cal F}, P)$
be a probability space.
Consider its $n$-dimensional product probability
space
$(X^n, {\cal F}^n,  {P}^n)$.
%probability 
Let
$f: X \to {\mathbb R}$
be a measurable function
such that
$\int_X f(x) P(dx) = \mu$
and
$\int_X |f(x)-\mu |^2 P(dx) = \sigma^2 $.
Then it holds that
\begin{align*}
%\lim_{n \to \infty}
&
{P}^n
(\{ (x_1,x_2,...,x_n) \in X^n
\;|\;\;\;
|\frac{\sum_{k=1}^n f(x_k)}{n}
-\mu
|>{\varepsilon}
\}
)
{{\; \leqq \;}}
\frac{\sigma^2}{\varepsilon^2 n}
\\
&
\quad
\qquad \qquad \qquad
(\forall \varepsilon >0,
\forall n=1,2,...
)
\end{align*}
\end{itemize}
This theorem,
discovered by Jacob Bernoulli (1654-1705)
and was announced in 1713
(after his death),
is the most fundamental assertion in science.
Recall that
mathematics is independent of
our world.
Thus some may ask that
\begin{itemize}
\item[]
Why is a mathematical theorem
$(\sharp)$
useful?
\end{itemize}
This is due to
the fact that
the mathematical theorem $(\sharp)$
sinks into the
widely ordinary language
{\textcircled{\scriptsize 0}}
in the following diagram:
\\
\\
%index{@{Chap.$\;$1}(X$_1$)}
%%index{@{ordinary language}(X$_1$)}
$\underset{(Chap. 1)}{\text{(X$_1$)}} $
%{ordinary language} and {world-description method}
%:
%%\text
%%{\textcircled{\scriptsize 0}}
%\\
$
%\quad
%\qquad
\overset{
}{\underset{\text{(before science)}}{
\text{
\fbox
{
{\textcircled{\scriptsize 0}}
widely {ordinary language}}
}
}
}
$
$
\underset{\text{\scriptsize }}{\text{$\Longrightarrow$}}
$
$
\underset{\text{\scriptsize ({Chap.$\;$1}(O))}}{\text{{world-description}}}
\cases
&
\!\!\!\!\!\!
%\textcircled{\scriptsize 1}:
%\underset{\scriptsize
%\text{}}
{\text{\textcircled{\scriptsize 1}realistic scientific language}}
\\
&{\text{(Newtonian mechanics, etc.)}}
\\
\\
%\textcircled{\scriptsize 2}:
&
\!\!\!\!\!\!
%\underset{\scriptsize
%\text{}}
{\text{\textcircled{\scriptsize 2}{linguistic scientific language}}}
\\
&{\text{({{measurement theory}}, etc.)}}
\endcases
$
\\
\\
However,
the following theorem
2.21
(
the weak law of large numbers in measurement theory
)
is not in
{\textcircled{\scriptsize 0}}
but
{\textcircled{\scriptsize 2}}.
\end{itemize}
}
%%BBBBBBBBBBBBBBBBBB%SBSBSBSS
\par
\noindent

%_{{{{{c}}}}{{{{h}}}}}_t_t_t_{\text{\scriptsize AB}}_r_r
\par

% and

{\bf
%BFBF
\par
\noindent
{{Theorem }}2.21
[{}The weak law of large numbers in measurement theory{\rm
({\rm cf.$\;$}\textcolor{black}{\cite{IStat, Keio, INewi}})}]}$\;\;$%POPOPOPOIUYTREWQ
\rm
%Then we have the following proposition.
%\par
%\noindent
%%index{law of large numbers@law of large numbers}
%\bf  %2BFBF
%\BEGIN{Prp} \label{Proposition 1.25}
%\ssf
%[{}The law of large numbers, cf ${{{}}}$\cite{Kolm}{}].
%\ssl
Suppose the above parallel measurement
${\mathsf M}_{{C(\Omega^n)}} ({}\bigotimes_{k=1}^n {\mathsf O} ,
S_{ [(\omega,...,\omega)] })$
in
${C(\Omega^n)} $.
For
any
$\epsilon >0$
and
any
$\Xi $
$({}\in {\cal F}  {})$,
define
${\widehat D}_{\Xi , \epsilon }$
$({}\in \bigstimes_{k=1}^n {\cal F} $
$)$
by
%\END{document}
\begin{align*}
{\widehat D}_{\Xi , \epsilon }
=
\Bigl\{
{\widetilde x} = ({}x_1 , x_2 , ..., x_n{}) \in  X^n
\;{}: \;
\Bigl|
[{}w ({}{\widetilde x}{}){}] ({}\Xi{})
-
[F(\Xi)](\omega )
\Bigl|
< \epsilon
\Bigl\} .
\end{align*}
Then we see that
\begin{align*}
1 -
\frac{1}{ 4 \epsilon^2  n }
\;
\le
\;
[
{\widehat F}
({\widehat D}_{\Xi , \epsilon })
]
(\omega ,..., \omega)
\;
\le
\;
1,
\quad
(
\forall \omega
\in \Omega,
\forall \Xi
\in {\cal F}  ,
\forall \epsilon >0,
\forall n{}).
%%\TAG 2.3
\tag{\textcolor{black}{2.11}}
\end{align*}
\par
%\hfill{$///$}%%BFBFbfbf
%%\END{Pr}p}

\par
\noindent
\it
\par \noindent {\it $\;\;\;\;${Proof.}}
\rm
$\omega \in \Omega $,
$\Xi \in {\cal F}$.
Define
$\mu$
and
$\sigma$
such that
\begin{align*}
&
\mu
=
\int_X \chi_{{}_{\Xi}} (x ) [F(dx )](\omega)
=
[F(\Xi )](\omega)
\\
&
\sigma^2
=
\int_X |\chi_{{}_{\Xi}} (x ) -\mu|^2 [F(dx )](\omega)
=
[F(\Xi )](\omega) (1- [F(\Xi )](\omega) )
\end{align*}
Then,
the law of large numbers
(\textcolor{black}{Note 2.16})
says that
\begin{align*}
%({}\otimes_{k=1}^n  \rho_k^p{})
&
[{\widehat F}
(X^n \setminus {\widetilde D}_{\Xi , \varepsilon })
](\omega, \omega,\ldots, \omega )
{{\; \leqq \;}}
\frac{\sigma^2}{\varepsilon^2 n}
=
\frac{1}
{ \varepsilon^2 n }
%\max_{1 {{\; \leqq \;}}k {{\; \leqq \;}}n }
%\bigl[
[F(\Xi)](\omega)
(1-
[F(\Xi)](\omega)
%\rho_k^p ({}F_k ({}\Xi{})))
)
\\
=
&
\frac{1}
{\varepsilon^2  n }
\Big(\frac{1}{4} - ([F(\Xi)](\omega)-\frac{1}{2})^2 \Big)
{{\; \leqq \;}}
\frac{
1
}
{4 \varepsilon^2 n }
%P%\tag{3.35}
\end{align*}
Thus, we get
%which implies
\textcolor{black}{(2.11)}
\qed

%\par

\renewcommand{\footnoterule}{%
  \vspace{2mm}                      % 
  \noindent\rule{\textwidth}{0.4pt}   % ,                    
  \vspace{-3mm} 					% 
}
%%%%%%%%%%%%%%%%%%%%%%%%%%%%%%%%%%%%%%%%%%
\par
\noindent
\par
\noindent
%BBBBBBBBBBBBBBBBBB%SBSBSBS
\par
\noindent
{\small%%{\footnotesize
\vspace{0.1cm}
\begin{itemize}
\item[$\spadesuit$] \bf {{}}{Note }2.17{{}} \rm
%probability  and {{measurement theory}},
%probability , {{measurement theory}}
%(linguistic){world-description method}
%,  and ,
%.
As mentioned in
\textcolor{black}{{Note }1.1},
we believe that:
\begin{itemize}
\item[($\sharp_1$)]
Behind
a useful mathematical theory,
the powerful world view
is always hidden
\end{itemize}
Because
mathematic itself is independent of
our world.
In fact,
\begin{itemize}
\item[$(\sharp_2)$]
In the proof of
\textcolor{black}{{{Theorem }}2.21}({{The law of large numbers in measurement theory}}),
we can find the law of large numbers in probability theory
($(\sharp)$) in \textcolor{black}{{Note }2.16}).
\end{itemize}
Therefore, for example,

\begin{table}[h]%%b h(here) t p
\begin{center}
\begin{tabular}{l|ll}
       mathematics        &  & $\quad$  world-description method \\
\hline
%%%%
\text{differential geometry}
&
&$\quad$ 
\text{the theory of relativity}
\\
\text{differential equation}
&
&$\quad$ 
\text{{{Newton}} mechanics,
electromagnetism
}
\\
\text{Hilbert space}
&
&$\quad$
\text{quantum mechanics}
	\\
$\underset{\text{\scriptsize (Hilbert space)}}{\text{probability theory}}$
&
&$\quad$
{measurement theory}
	\\
\end{tabular}
%%\caption{Axiom}
%\BEGIN{center}
%	0.1: Axiom
%\END{center}
\end{center}
\end{table}
\end{itemize}
}
\par
\noindent
{\small%%{\footnotesize
\vspace{0.1cm}
\begin{itemize}
\item[$\spadesuit$] \bf {{}}{Note }2.18{{}} \rm
Now we can expect readers to believe in our
assertion
that
%
%{},
%,
%Axiom${}_{\text{\scriptsize c}}^{\text{\scriptsize p}}$ 1
%
%{metaphysics}
%%(\textcolor{black}{{Sec.2.2}(d)})
%
%.
%
%,
%,
%That is,
\begin{itemize}
\item[($\sharp_1$)]
{\bf
There is the metaphysics
(called measurement theory)
in the center of sciences.
}
\end{itemize}
% and  and .
% and ,
% and
% and .
%,
%{world-description}({ordinary language})
% and
%{world-description}:
%%index{@{Chap.$\;$1}(X$_1$)}
%\\
%\\
%$\underset{({Chap.$\;$1})}{\text{(X$_1$)}}$
%$\overset{
%({ordinary language})
%}{\underset{({world-description}(=))}{
%\text{
%\fbox
%{{\textcircled{\scriptsize 0}}
%{ordinary language}}
%}
%}
%}
%$
%$
%\underset{\text{\scriptsize }}{\text{$\Longrightarrow$}}
%$
%$\!\!
%\underset{\text{\scriptsize ({Chap.$\;$1}(O))}}{\text{{world-description}}}
%\!\!
%\cases
%&
%\!\!\!\!\!\!
%%\textcircled{\scriptsize 1}:
%%\underset{\scriptsize
%%\text{}}
%{\text{\textcircled{\scriptsize 1}{realistic method}}}
%\\
%&{\text{({world is before language})}}
%\\
%\\
%%\textcircled{\scriptsize 2}:
%&
%\!\!\!\!\!\!
%%\underset{\scriptsize
%%\text{}}
%{\text{\textcircled{\scriptsize 2}{linguistic method}}}
%\\
%&{\text{(language is before world)}}
%\ENDcases
%$
%\\
%\\
% and , ,
%\BEGIN{itemize}
%\item[($\sharp_2$)]
%realistic{world-description method}\textcircled{\scriptsize 1}, .
%,
%{ordinary language}\textcircled{\scriptsize 0}
%
%(\textcolor{black}{{Chap.$\;$1}(X$_4$)})
%\END{itemize}
%{}
%, ,
%(=={{measurement theory}})
%,
% and .
%%\
\end{itemize}
}

%%BBBBBBBBBBBBBBBBBB%SBSBSBSS

%
%%%%\omega^\omega^
%BBBBBBBBBBBBBBBBBB%SBSBSBS
\par
\noindent
{\small%%{\footnotesize
\begin{itemize}
\item[$\spadesuit$] \bf {{}}{Note }2.19{{}} \rm
As mentioned in Chap. 8,
measurement theory is deeply related to
traditional philosophies.
For example,
we see:
%{{measurement theory}}( and ){metaphysics},
% and ,
%{{measurement theory}}
% and  and .
%%.
%, 
%.
\\
\\
%index{@${{\cdot}}$}
$\cases
\text{{linguistic method}(language is before world)}
&{\cdots}
\text{
Saussure(\textcolor{black}{{Sec.8.1}})}
% and  and 
\\
\text{{{state}} and observable }
&{\cdots}
\text{Locke's
primary quantity
and secondary quantity
}
%{Chap.{\;}} and {Chap.{\;}}
%index{@{Chap.{\;}}, {Chap.{\;}}}
\\
\text{only one {{measurement}}}
&
\cdots
\text{only one state, no movement
}\text{ (Parmenides
%\textcolor{black}{{\cite{Hiro, Naga}}
%}
)}
%index{@}
\\
\text{observer's time(\textcolor{black}{{Note }2.8})}
&
\cdots
\text{Augustine's time, McTaggart's paradox}(\cite{McTa,IQphi})
\\
&\quad
\text{(the interpretation made dwarfish in {Note }6.6)}
%({\rm cf.} \textcolor{black}{{\cite{Hiro, Naga}}})
\\
\text{primary substance $\cdot$
secondary substance}
&{\cdots}
\text{The problem of universals}
(\textcolor{black}{{\cite{IUniversals}}})
\\
\text{observable is before {{state}}}
&{\cdots}
\text{Recognition constitutes the world}
\\
&\quad
\text{(Kant Copernican turn}
\text{ cf. {Sec.8.1})}
\\
&\quad \
\endcases
$
\\
\\
%{metaphysics}
%
%(That is,
%language is before world
%), 
% and (\textcolor{black}{{Sec.8.1}}),
%,
%,
%language is before world and 
%{metaphysics}
%
% and 
%. ,
%
%index{@}
%index{@}
%index{@}
\end{itemize}
}
\renewcommand{\footnoterule}{%
  \vspace{2mm}                      % 
  \noindent\rule{\textwidth}{0.4pt}   % , 
  \vspace{-5mm}
}

\vskip1.5cm
%====================================================
%33333333333333333333333333333333333333\tag\tag{
\section{From Quantum Mechanics to Measurement Theory
\label{Chap3}
}%{Chap.{\;}}3{}
%%\vspace{-0.8cm}
%\baselineskip=18pt\par
\noindent
\begin{itemize}
\item[{}]
{
\small
\par%[Abstract].
\rm
$\;\;\;\;$
As mention in Chap. 1,
we assert that
\begin{itemize}
\item[$(\sharp)$]
$\qquad
\qquad
$
$
\overset{\text{\scriptsize }}{\underset{\text{\scriptsize ({physics})}}{\text{
\fbox{quantum mechanics}}}}
\xrightarrow[\text{(linguistic turn)}]{\text{proverbalizing}}
\overset{\text{\scriptsize }}{\underset{\text{\scriptsize (scientific language)}}{\text{\fbox{
{{measurement theory}}}}}}
$
\end{itemize}
Therefore,
first we shall review
the elementary step of
quantum mechanics.
And further,
we derive
Axiom${}_{\text{\scriptsize c}}^{\text{\scriptsize p}}$ 1 in \textcolor{black}{{Sec.2.2}}
from
Born's quantum {{{measurement theory}}}.
Discussing EPR-paradox,
Schr\"odinger's cat, Heisenberg's {uncertainty principle}
we study the spirit of
quantum mechanics.
However,
our assertion
(i.e.,
the linguistic world-view)
is
the
reverse arrow of the $(\sharp)$,
that is,
"from measurement theory to quantum mechanics".
This will be discussed in
Sec. 9.3.
\vskip0.5cm
}
\end{itemize}

\subsection{The quick review on quantum mechanics}%{Sec.3.1}
\par
%index{@quantum mechanics}
\par
quantum mechanics is composed of two axioms
(i.e.,
%, 
%1 and  and ,
%quantum mechanicsPART,
%
%.
%, .
%\par
%\textcolor{black}{{Chap.$\;$1}},
%quantum mechanics and ,
%%, .
%quantum mechanics,
%2,
"Born's quantum measurement theory"
and
"{quantum kinetic equation}").
%{(}Heisenberg,
%,
% and Schr\"odinger{)}
%
%.
%%index{@{quantum kinetic equation}}
That is,
\begin{itemize}
\item[]
$
\underset{\text{\scriptsize ({physics})}}{\text{{} $\fbox{quantum mechanics}$}}
:=
{
\overset{\text{\scriptsize [quantum {{measurement}}]}}
{
\underset{\text{\scriptsize
[Born's probabilistic interpretation]}}{\text{{} $\fbox{{{measurement}}{}}$}}}
%[Bornprobabilistic interpretation(\textcolor{black}{{Sec. 3.1.1}})]}}{\text{{} $\fbox{{{measurement}}{}}$}}}
}
+
{
\overset{\text{\scriptsize [kinetic equation]}}
{
\underset{\text{\scriptsize [{quantum kinetic equation}
]}}
%(\textcolor{black}{{Note }10.2})]}}
{\text{{}$\fbox{causality }$}}
}
}
$
\end{itemize}

% and .
%

\par
%{},
%\textcolor{black}{{Sec.3.1}}
%quantum mechanics,

In \textcolor{black}{{Sec.3.2}},
we derive classical {{{measurement theory}}}(Axiom${}_{\text{\scriptsize c}}^{\text{\scriptsize p}}$ 1)
from
quantum mechanics(Born's quantum {{{measurement theory}}}).
This is rather concrete.
For the abstract argument, see
\textcolor{black}{\cite{IFuzz, Keio}}.
%.
%
\subsubsection{Born's quantum {{measurement theory}}}%{Sec. 3.1.1}
\par
Let
${\mathbb C}$
be the complex field
(i.e.,
the set of all complex numbers).
% and .
%%index{@${\mathbb C}$}
%%index{cc@${\mathbb C}$:}
Let
${\mathbb C}^n$
be the
$n$-dimensional complex space.
% and .
That is,
\begin{align*}
{\mathbb C}^n
=
\Big\{
\alpha
=
\bmatrix
        \; \alpha_1  \\
        \; \alpha_2 \\
        \vdots \\
        \; \alpha_n
\endbmatrix
\; \Big|\;
\alpha_1, \alpha_2, \ldots,\alpha_n
\text{is complex number}
\Big\}
%P%\TAG{1}
\end{align*}
%
%${\mathbb C}^n$,
The inner product
$\langle \cdot , \cdot \rangle$
and the norm
$|| \cdot ||$
is respectively defined by
\begin{align*}
\langle \alpha , \beta \rangle
=
\sum\limits_{k=1}^n
\overline{\alpha}_k \cdot \beta_k,
\qquad
||\alpha ||= |\langle \alpha , \alpha \rangle |^{1/2}
\quad
(\forall \alpha , \forall \beta \in {\mathbb C}^n )
%P%\TAG{2}
\end{align*}
(where $\overline{\alpha}_k$ is the conjugate complex number
of $\alpha_k$).
$H={\mathbb C}^n$
is called the $n$-dimensional Hilbert space.
Also,
an infinite dimensional Hilbert space
${\mathbb C}^\infty$
is defined by
$H=\{ \alpha \in {\mathbb C}^\infty  \;|\;
||\alpha||
=
$
$
(\sum\limits_{k=1}^\infty
\overline{\alpha}_k \cdot \alpha_k)^{1/2}
<
\infty
\}$.
% and .
%, {{measurement}}PART,
%$n= \infty$$n<\infty$
%{}
%% and ,
%%,
%%$n<\infty$
%%{}
%%index{ and @}
%}.
\par
Define the {state space}
${\widehat \Omega }$
such that
${\widehat \Omega } (\subset {\mathbb C}^n)$.
That is,
\begin{align*}
{\widehat \Omega}
=
\big\{
\alpha
\in {\mathbb C}^n
\;|\;
\;
|| \alpha || =1
\big\}
%P%\TAG{3}
\end{align*}
%%%%%%%|\alpha_1|^2 +|\alpha_2|^2 =1
where
$\alpha$ and $\beta$
are
identified if
$\alpha = e^{\theta \sqrt{-1} } \beta$
(fore some $\theta \in {\mathbb R}$).
% and , $\alpha =  \beta$ and .
%\text{ and }ffffffFFFFFFFFFFF
The
$\omega (\in {\widehat \Omega})$
is called the
{{{state}}}.

Let
$B({\mathbb C}^n)$
be
the set of all $n \times n$-complex matrices.
% and .
That is,
\begin{align*}
B({\mathbb C}^n)=
\Big\{
A=
\bmatrix
a_{11} & \cdots &a_{1n} \\
\vdots & \vdots & \vdots \\
a_{n1} & \cdots & a_{nn}
\endbmatrix
\; \Big|\;
a_{ij}
\text{
is complex numbers}
\Big\}
%}
\end{align*}
which is called the basic algebra.
%{\bf {basic algebra}} and .
The matrix $A (\in B({\mathbb C}^n))$
is said to be non-negative Hermitian matrix
if it satisfies the following
conditions
(i)
and
(ii):
\begin{itemize}
\item[({}i)]
$A$
is Hermitian,
that is,
$A=A^*$,
%
%$\bmatrix
%a_{11} & a_{12} \\
%a_{21} & a_{22}
%\ENDbmatrix
%$
%$=$
%$\bmatrix
%{\overline a}_{11} & {\overline a}_{12} \\
%{\overline a}_{21} & {\overline a}_{22}
%\ENDbmatrix
%$,
(where
$A^*$
is the conjugate transposed matrix)
%$A$)
%,
%%%%a}
\item[({}ii)]
$A {\; \geqq \;}0$,
That is,
$\langle \alpha , A \alpha \rangle {\; \geqq \;}0$
$\;\;$
$(\forall \alpha \in {\mathbb C}^n )$
%
%
%
%\\
%\BEGIN{align*}
%\bmatrix
%\overline{\alpha}_1 &
%\overline{\alpha}_2
%\ENDbmatrix
%%\overline{\alpha}_2]
%L
%%({}\{ \uparrow _{z}  \}{})
%\bmatrix
%        \; \alpha_1  \\
%        \; \alpha_2
%       \endbmatrix
%{\; \geqq \;}0
%\quad
%( \forall \alpha =
%\bmatrix
%        \; \alpha_1  \\
%        \; \alpha_2
%       \endbmatrix
%\in {\mathbb C}^2)
%\END{align*}LLLLLLLLLLLllllllllll
%LLFFFFFFF%\text{ and }afffffFFFFFFFFFFF
\end{itemize}

For simplicity,
assume that
the measured value space
$X$
is finite set.
The triplet
${\mathsf O} =(X, 2^X , F )$
is called an observable
in
the
{basic algebra}$B({\mathbb C}^n)$,
if it satisfies the following
(cf. \cite{Davi}):
%{\bf observable } and . That is,
\par
\noindent
\begin{itemize}
\item[({}i)]
The map
$F: 2^X \to B({\mathbb C}^n)$
satisfies that
(a):
$F(\emptyset )=0(=0-\text{matrix})$,
$F(X )=I (=\text{unit matrix})$
(b):
for each
$\Xi \in 2^X$,
$F(\Xi)$
is a non-negative Hermite matrix.
%
%$(\in B({\mathbb C}^2))$
%
%
%
%$\;\; (k=1,2,\ldots,K)$
%%\\
%{}
\item[({}ii)]
for each
$\Xi \in 2^X$,
it holds that
$F(\Xi ) = \sum\limits_{x\in \Xi } F(\{x\})
$
\end{itemize}
%observable ,
%{\cite{Davi}}
%
%(,
%observable  and 
%(\textcolor{black}{(3.2)})
%).
Also,
the
observable
${\mathsf O} =(X, 2^X , F )$
is said to be
a projective observable,
if
it satisfies
that
$F(\Xi)=(F(\Xi))^2$
$(\forall \Xi \in 2^X)$.
% and ,
%{\bf
%observable
%}
% and .
%%index{@observable }
\par
Here,
we have
the quantum
{{measurement}}
${\mathsf M}_{B({\mathbb C}^n)}({\mathsf O}=(X, 2^X , F ), S_{[\omega]})$
(where
$\omega \in {\widehat \Omega} $
).
That is,
\begin{itemize}
\item[]
$\;\;$
quantum {{measurement}}
${\mathsf M}_{B({\mathbb C}^n)}({\mathsf O}, S_{[\omega]})$
\\
\\
=
the quantum measurement
of
the observable
${\mathsf O}$
for
the {measuring object}
$S$
with a quantum
\\
$\;\;$
state
$\omega (\in {\widehat \Omega} )$
\end{itemize}

%index{@}
\par

Under the above preparation,
we can introduce
"Born's quantum {{{measurement theory}}}"
as follows.

\vskip0.5cm

%\BEGIN{itembox}[c]
\par
\noindent
\begin{center}
{\bf
Axiom(Q)
%${}_{\text{\scriptsize c}}^{\text{\scriptsize p}}$
1 [Born's quantum {{measurement theory}}]
}
\label{axiomq}
\end{center}
\par
\noindent
%\vskip0.1cm
\par
\noindent
\fbox{\parbox{155mm}{
\begin{itemize}
%\item[(i)]
%{{measurement}}
\item[]
{}\rm{}
Consider a measurement
${\mathsf M}_{B({\mathbb C}^n)}({\mathsf O}:= ({}X, 2^X, F{}), S_{[\omega]})$
formulated in a
basic algebra
$
B({\mathbb C}^n)
$.
Assume that
the measured value
$ x$
$({}\in X  {})$
is
obtained by the measurement
${\mathsf M}_{B({\mathbb C}^n)}({\mathsf O}, S_{[\omega]})$.
Then,
%it holds that
{{}}
the probability
that
a
measured value
$ x$
$({}\in X)$
is obtained
is
given by
$
\langle \omega , F(\{x\}) \omega \rangle
$.
%\END
\end{itemize}
}
}
\par
\vskip0.5cm
\par
\noindent

%
%
%
%
%
%\BEGIN{itembox}[c]{
%\bf
%\textcolor{black}{
%Axiom(Q)
%%${}_{\text{\scriptsize c}}^{\text{\scriptsize p}}$
%1 [Born's quantum {{measurement theory}}]
%%(quantum mechanicsprobabilistic interpretation)
%%Axiom${}_{\text{\scriptsize c}}^{\text{\scriptsize p}}$ 1(Born{{{measurement theory}}}:$B({\mathbb C}^n)$
%}
%}
%\label{axiomq}
%\label{Axiom(Q)}
%%index{1@Axiom${}_{\text{\scriptsize c}}^{\text{\scriptsize p}}$ 1[{{{measurement theory}}}]}
%%\BEGIN{itemize}
%%\item[]
%Consider a measurement
%${\mathsf M}_{B({\mathbb C}^n)}({\mathsf O}:= ({}X, 2^X, F{}), S_{[\omega]})$
%formulated in a
%basic algebra
%$
%B({\mathbb C}^n)
%$.
%Assume that
%the measured value
%$ x$
%$({}\in X  {})$
%is
%obtained by the measurement
%${\mathsf M}_{B({\mathbb C}^n)}({\mathsf O}, S_{[\omega]})$.
%Then,
%%it holds that
%{{}}
%the probability
%that
%a
%measured value
%$ x$
%$({}\in X)$
%is obtained
%is
%given by
%$
%\langle \omega , F(\{x\}) \omega \rangle
%$.
%%\END{itemize}
%\END{itembox}
%
%
%

Also,
the
parallel quantum measurement
$\bigotimes_{k=1}^N
{\mathsf M}_{B({\mathbb C}^n)}({\mathsf O}, S_{[\omega]})$
is possible,
and thus,
we can get
the
sample
probability space
$(X, {\cal F}, $
$\langle \omega ,F(\cdot ) \omega \rangle
)$.
%, quantum {{measurement theory}}{}

%
%%%%\omega^\omega^
%BBBBBBBBBBBBBBBBBB%SBSBSBS
\par
\noindent
{\small%%{\footnotesize
\vspace{0.1cm}
\begin{itemize}
\item[$\spadesuit$] \bf {{}}{Note }3.1{{}} \rm
%(i):
Since quantum mechanics is physics,
the following terms have reality:
\begin{itemize}
\item[({{}}$\sharp$)]
{{{measurement}}},
{{observer}}, {measuring object}, {{state}},
observable ($\approx${measuring instrument}), measured value , probability
\end{itemize}
\end{itemize}
}

%%BBBBBBBBBBBBBBBBBB%SBSBSBSS

%\vskip1.0cm

\par

% and .
\par

Now,
we shall
show that
an
Hermitian
matrix
$A (\in B({\mathbb C}^n ))$
can be
regarded
as
a
projective
observable.
For simplicity,
this is shown
in the case that
$n=3$.
We see
(for simplicity, assume that
$x_j \not= x_k $(if $j \not= k)$
)
%,
\begin{align*}
A
=
U^*
\bmatrix
x_1 & 0 & 0 \\
0 & x_2 & 0 \\
0 & 0 & x_3
\endbmatrix
U
\qquad
\tag{\color{black}{3.1}}
%%%%%REDREDREDREDREDRE
\end{align*}
where
$U$
$(\in B({\mathbb C}^3))$
is the unitary matrix,
%
%(That is,
%$U^* U=U U^*=I=$),
$x_k \in {\mathbb R}
$).
%
%
%$A$
%{}
%,
Putting
$X=\{ x_1, x_2, x_3 \}$,
\par
\noindent
{\small
\begin{align*}
&
F_A(\{x_1\})
=
U^*
\bmatrix
1 & 0 & 0 \\
0 & 0 & 0 \\
0 & 0 & 0
\endbmatrix
U,
\;
\quad
F_A(\{x_2\})
=
U^*
\bmatrix
0 & 0 & 0 \\
0 & 1 & 0 \\
0 & 0 & 0
\endbmatrix
U,
\;
\\
&
F_A(\{x_3\})
=
U^*
\bmatrix
0 & 0 & 0 \\
0 & 0 & 0 \\
0 & 0 & 1
\endbmatrix
U
\end{align*}
}
the,
we get
the projective observable ${\mathsf O}_A =(X,2^X, F_A)$
in
$B({\mathbb C}^3)$.
%
% and .
%,
%,
%observable ${\mathsf O}_A =(X,2^X, F_A)$
%,
%%\textcolor{black}{(3.1)}
%%,
%$A$
%$(=\sum\limits_{i=1}^3 x_i F_A(\{x_i\}) \;)$
%.
%,
%$A$
%
%observable ${\mathsf O}_A $
% and ,
Thus,
we have the following
identification:
\begin{align*}
\underset
{\text{\scriptsize ({}Hermitian matrix)}}{{A}}
\longleftrightarrow
\underset{\text{\scriptsize (projective {observable })}}
{{\mathsf O}_A =(X,2^X, F_A)
}
%%P%\TAG{44}
\tag{\color{black}{3.2}}
%%%%%]REDREDREDREDREDRE
\end{align*}
%index{@observable }
%%%%{2.32}
\par
\noindent
%\footnote{
%,
% and ,
%$x_1=x_2$
%
%
%$X=\{ x_1, x_3 \}$ and ,
%% and ,
%\BEGIN{align*}
%F_A(\{x_1\})
%=
%U^*
%\bmatrix
%1 & 0 & 0 \\
%0 & 1 & 0 \\
%0 & 0 & 0
%\ENDbmatrix
%U,
%\;\;
%F_A(\{x_3\})
%=
%U^*
%\bmatrix
%0 & 0 & 0 \\
%0 & 0 & 0 \\
%0 & 0 & 1
%\ENDbmatrix
%U
%\;\;
%%F(\[x_3\})
%%=
%%U^*
%%\bmatrix
%%0 & 0 & 0 \\
%%0 & 0 & 0 \\
%%0 & 0 & 1
%%\ENDbmatrix
%%U,
%%\;\;
%\END{align*}
% and ,
%observable ${\mathsf O}_A =(X,2^X, F_A)$
%.
%}.
\par
Let
$A
(\in B({\mathbb C}^n)
)$
be an Hermitian
matrix.
Under the identification
(3.2),
we have
the quantum {{measurement}}
${\mathsf M}_{B({\mathbb C}^n)}({\mathsf O}_A,$
$S_{[\omega]})$.
Born's quantum {{{measurement theory}}}
say that
\begin{itemize}
\item[]
The probability
that
a measured value $x(\in X)$
is
obtained by
the
quantum
{{measurement}}
${\mathsf M}_{B({\mathbb C}^n)}({\mathsf O}_A, S_{[\omega]})$
is given
by
$\langle \omega , F_A(\{x\}) \omega \rangle$.
\end{itemize}
Therefore,
the expectation of
a measured value
is given by
% ,
\begin{align*}
\int_X x
\langle \omega , F_A( dx ) \omega \rangle
=
\sum\limits_{x_i \in X } x_i \langle \omega , F_A(\{x_i \}) \omega \rangle
=
\langle \omega , A \omega \rangle
\end{align*}
Also,
its variance
$(\delta_A^\omega)^2$
is given by
\begin{align*}
(\delta_A^\omega)^2
=
%\frac{1}{3}
\sum\limits_{x_i \in X} (x_i -  \langle \omega , A \omega \rangle)^2
 \langle \omega , F_A(\{x_i \}) \omega \rangle
=
\langle A \omega , A \omega \rangle
-
|\langle \omega , A \omega \rangle|^2
%\right]^{1/2}
\tag{\color{black}{3.3}}
\end{align*}

\par

\vskip0.5cm
\par

\par \noindent {\bf Quantum {{{measurement theory}}}(
Stern--Gerlach experiment
Schtern--Gerlach
experiment (1922))}
\par

\rm
\par
Assume that
we examine the beam ({}of silver particles{})
after passing through the magnetic field.
Then,
as seen in the following figure,
we see that
all particles are deflected either equally upwards or equally downwards in a
50:50 ratio.
See \textcolor{black}{Fig. 3.1}.
%index{Stern--Gerlach experiment@Stern--Gerlach experiment}
\par
\noindent

%
%
%(That is,
%$N$ and
%$S$
%) and , \textcolor{black}{{FIG.$\;$}3.1},
%. 
%--.
%,
%\textcircled{\scriptsize }\textcircled{\scriptsize }
%?
%%index{@--}
\par
\noindent
%$\qquad \qquad$
%{\LL =(1922)\RR}
%\hfill{8)}%%%P%\TAG{8}
\par
\noindent
%\vskip-0.4cm%%%
%\begin{figure}[htbp]
\unitlength=0.25mm
%\unitlength=0.33mm
\begin{picture}(430,190)
\Thicklines
\put(90,180){\line(1,0){200}}
\put(185,150){S}
\put(185,50){N}
%\put(190,125){\line(5,2){200}}
\path(190,125)(290,180)
\path(190,125)(90,180)
\allinethickness{0.15mm}
\put(20,110){{ electron $e$}}
\put(25,80){\footnotesize {{state}}
$ \omega =
       \bmatrix
        \; \alpha_1  \\
        \; \alpha_2
       \endbmatrix
$
}
\put(50,100){\circle*{5}}
\put(50,100){\line(1,0){50}}
\put(100,100){\line(1,0){100}}
\put(200,100){\line(1,0){100}}
\put(100,100){\path(0,4)(6,0)(0,-4)}
\put(200,100){\path(0,4)(6,0)(0,-4)}
\put(280,100){\path(0,4)(6,0)(0,-4)}
\spline(295,100)(300,100)(350,120)(400,150)
\spline(295,100)(300,100)(350,80)(400,50)
\put(350,124){\path(0,1)(6,0)(0,-7)}
\put(350,76){\path(0,7)(6,0)(0,-1)}
\put(380,157){$[\uparrow]$}
\put(410,150){\textcircled{\scriptsize U}}
\put(380,44){$[\downarrow]$}
\put(410,50){\textcircled{\scriptsize D}}
\Thicklines
\put(140,75){\line(1,0){100}}
\path(90,20)(140,75)
\path(290,20)(240,75)
\put(90,20){\line(1,0){200}}
\path(400,175)(400,25)
\put(390,10){Screen}
\end{picture}
\vskip0.2cm
\begin{center}
{Figure 3.1:
Stern--Gerlach experiment
(1922)
}
\end{center}
\par
\noindent
%index{@observable }
\par

%index{spin observable@spin observable}
\par
\noindent
Consider the two dimensional Hilbert space
$V= {\mathbb C}^2$,
And therefore,
we get the non-commutative basic algebra
${C(\Omega)} = B(V)
(=B_c(V))$,
that is,
the algebra composed of all $2 \times 2$ matrices.
Note that
${C(\Omega)} = B(V)$
$= {B_c}({}V{})$
%=$
%$ {B_c}_I({}V{})$
%({}{\itrk 2.6 (i)})
since the dimension of $V$ is finite.

The spin state of
an electron $e$
is represented by
$\omega$
$\in$${\widehat \Omega}
(\subset {\mathbb C}^2 )$.
Put
$ \omega =
       \bmatrix
        \; \alpha_1  \\
        \; \alpha_2
       \endbmatrix
$
(
where,
$||\omega ||^2 = |\alpha_1|^2
+|\alpha_2|^2
=1$
).

Define
${\mathsf O}_z$
$\equiv$
$({}Z , 2^Z  , F_z  {})$,
the spin observable concerning the $z$-axis,
such that,
$Z= \{ \uparrow , \downarrow \}$
and
\begin{align*}
F_z ({}\{ \uparrow  \}{})
=
\bmatrix
1 & 0 \\
0 & 0
\endbmatrix
,
\quad
F_z ({}\{   \downarrow   \}{})
=
\bmatrix
0 & 0 \\
0 & 1
\endbmatrix ,
%\tag{1.68}
\end{align*}
\begin{align*}
F_z ({}\emptyset{})
=
\bmatrix
0 & 0 \\
0 & 0
\endbmatrix
,
\quad
F_z ({}\{   \uparrow ,\downarrow  \}{})
=
\bmatrix
1 & 0 \\
0 & 1
\endbmatrix .
%\tag{1.69} 
\end{align*}

Here,
Born's quantum {{{measurement theory}}}
{(}the probabilistic interpretation of quantum mechanics{)}
says that
\par
\noindent
\begin{itemize}
\item[]
%Born{{{measurement theory}}}
%\\
When
a quantum {{measurement}}${\mathsf M}_{B({\mathbb C}^2)}({\mathsf O}, S_{[\omega]})$
is taken,
the probability that
\begin{align*}
\text{}
\text{ a measured value }
\left[\begin{array}{ll}
\text{$\uparrow$}
\\
\text{$\downarrow$}
%a
%\\
%b
\end{array}\right]
\text{is obtained }
%\\
\text{is given by}
%&
%\cases
\left[\begin{array}{ll}
%\bmatrix
%\overline{\alpha}_1 &
%\overline{\alpha}_2
%\ENDbmatrix
%\overline{\alpha}_2]
\langle \omega, F^z(\{ \uparrow\})
\omega
\rangle =|\alpha_1|^2
\\
\\
%\bmatrix
%\overline{\alpha}_1 &
%\overline{\alpha}_2
%\ENDbmatrix
\langle \omega, F^z(\{ \downarrow\})
\omega
\rangle =
|\alpha_2|^2
%\bmatrix
%        \; \alpha_1  \\
%        \; \alpha_2
%       \endbmatrix
%       =|\alpha_2|^2
%\ENDcases
\end{array}\right]\text{{}}
\end{align*}
\end{itemize}
That is,
putting
$\omega$
$(=
\bmatrix
        \; \alpha_1  \\
        \; \alpha_2
\endbmatrix
\in {\widehat \Omega} )$,
we says that
\begin{itemize}
\item[]
\begin{align*}
&
\text{When the electron with a spin state {{state}} $\omega$
progresses in a magnetic field},
\\
&
\text{the probability that
the Geiger counter
}
\left[\begin{array}{ll}
\text{\textcircled{\scriptsize U}}
\\
\text{\textcircled{\scriptsize D}}
%a
%\\
%b
\end{array}\right]
\text{ sounds}
\\
&
%\\
\text{is give by}
%&
%\cases
\left[\begin{array}{ll}
\big[
\overline{\alpha}_1
\;\;\;
\overline{\alpha}_2
\big]
%\overline{\alpha}_2]
\bmatrix
1 & 0 \\
0 & 0
\endbmatrix
\bmatrix
        \; \alpha_1  \\
        \; \alpha_2
       \endbmatrix
       =|\alpha_1|^2
\\
\\
\big[
\overline{\alpha}_1
\;\;\;
\overline{\alpha}_2
\big]
%
%
%\bmatrix
%\overline{\alpha}_1 &
%\overline{\alpha}_2
%\ENDbmatrix
\bmatrix
0 & 0 \\
0 & 1
\endbmatrix
\bmatrix
        \; \alpha_1  \\
        \; \alpha_2
       \endbmatrix
       =|\alpha_2|^2
%\ENDcases
\end{array}\right]\text{}
\end{align*}
\end{itemize}
%\END

\par \noindent {\bf EPR-paradox}
%index{@EPR-paradox}
\par

Next,
let us explain
EPR-paradox
{(}Einstein--Podolsky--Rosen)
\textcolor{black}{{\cite{Eins, Sell}}}).
Consider
Two electrons $P_1$
and
$P_2$
and their spins.
%,
%index{ and @tensor Hilbert space
}
The
tensor Hilbert space
$H={\mathbb C}^2 \otimes {\mathbb C}^2$
is defined in what follows.
That is,
\begin{align*}
e_1=
\bmatrix
1 \\
0
\endbmatrix,
\quad
e_2
=
\bmatrix
0 \\
1
\endbmatrix
%%P%\TAG{7}
\end{align*}
(i.e.,
the complete orthonormal system $\{ e_1, e_2 \}$
in the ${\mathbb C}^2$),
\begin{align*}
{\mathbb C}^2 \otimes {\mathbb C}^2
=
\{
\sum\limits_{i,j=1,2} \alpha_{ij} e_i \otimes e_j
%\alpha_{11} e_1\otimes e_1
%+
%\alpha_{12} e_1\otimes e_2
%+
%\alpha_{21} e_2\otimes e_1
%+
%\alpha_{22} e_2\otimes e_2
\;|\;
\alpha_{ij}
\in {\mathbb C},
i,j=1,2
\}
\end{align*}
Put
%\{ e_i \otimes e_j\}_{ij=1,2} and 
%\END{align*}
$u=\sum\limits_{i,j=1,2} \alpha_{ij} e_i \otimes e_j$
and
$v=\sum\limits_{i,j=1,2} \beta_{ij} e_i \otimes e_j$.
And the inner product
$\langle u,v \rangle_{_{{\mathbb C}^2 \otimes {\mathbb C}^2}}$
is defined by
\begin{align*}
\langle u,v \rangle_{_{{\mathbb C}^2 \otimes {\mathbb C}^2}}
=
\sum\limits_{i,j=1,2} \overline{\alpha}_{i,j}\cdot \beta_{i,j}
%%P%\TAG{8}
\end{align*}

Therefore,
we have
the
tensor Hilbert space
$H={\mathbb C}^2 \otimes {\mathbb C}^2$
with
the complete orthonormal system
$\{ e_1 \otimes e_1,
e_1 \otimes e_2,
e_2 \otimes e_1,
e_2 \otimes e_2
\}$.

For each
$F\in B({\mathbb C}^2)$
and
$G\in B({\mathbb C}^2)$,
define
the
$F\otimes G \in B({\mathbb C}^2 \otimes {\mathbb C}^2)$
(i.e.,
linear operator $F\otimes G : {\mathbb C}^2 \otimes {\mathbb C}^2
\to
{\mathbb C}^2 \otimes {\mathbb C}^2
$
)
such that
\begin{align*}
(F \otimes G) ( u \otimes v) = Fu \otimes Gv
%P%\TAG{9}
\end{align*}
\par

Let us define
the singlet state
$s (\in {\mathbb C}^2 \otimes {\mathbb C}^2)$
of
two particles
$P_1$
and
$P_2$
by
%i
%{{state}}$s \in {\mathbb C}^2 \otimes {\mathbb C}^2$
%,
%{\bf {{state}}} and :
%%index{ and @{{state}}}
\begin{align*}
s=
\frac{1}{\sqrt 2} (e_1\otimes e_2 - e_2 \otimes e_1 )
%P%\TAG{10}
\end{align*}

Here, we see that
$\langle s,s \rangle_{_{{\mathbb C}^2 \otimes {\mathbb C}^2}}$
$=\frac{1}{2}\langle e_1\otimes e_2 - e_2 \otimes e_1 ,e_1\otimes e_2 - e_2 \otimes e_1  \rangle_{_{{\mathbb C}^2 \otimes {\mathbb C}^2}}$
$=\frac{1}{2}(1+1)=1$,
and thus,
$s$ is a state.
%${{state}}
Also,
assume that

%\END{document}

\begin{itemize}
\item[]
two particles $P_1$
and
$P_2$
are far.
\end{itemize}

\par

Let
${\mathsf O} =(X,2^X, F^z )$
in $B({\mathbb C}^2)$
(where
$X=\{\uparrow ,\downarrow \}$
)
be
the spin observable concerning the $z$-axis
such that
\begin{align*}
F^z( \{ \uparrow \})
=
\bmatrix
1 & 0 \\
0 & 0
\endbmatrix
,
\quad
F^z( \{ \downarrow \})
=
\bmatrix
0 & 0 \\
0 & 1
\endbmatrix
%%%%%2.40}
\end{align*}
The parallel observable
${\mathsf O}\otimes {\mathsf O} =(X^2,2^X\times 2^X, F^z \otimes F^z)$
in
$B({\mathbb C}^2\otimes {\mathbb C}^2)$
is defined by
\begin{align*}
&
(F^z \otimes F^z)(\{(\uparrow,\uparrow )\})=F^z(\{\uparrow \}) \otimes F^z(\{\uparrow \}),
\;\;\\
&
(F^z \otimes F^z)(\{(\downarrow,\uparrow )\})=F^z(\{\downarrow \}) \otimes F^z(\{\uparrow \})
\\
&
(F^z \otimes F^z)(\{(\uparrow,\downarrow )\})=F^z(\{\uparrow \}) \otimes F^z(\{\downarrow \}),
\;\;\\
&
(F^z \otimes F^z)(\{(\downarrow,\downarrow )\})=F^z(\{\downarrow \}) \otimes F^z(\{\downarrow \})
\end{align*}
Thus,
we get the
{{measurement}}
${\mathsf M}_{B({\mathbb C}^2\otimes {\mathbb C}^2)}({\mathsf O}\otimes {\mathsf O},S_{[s]})$
The,
Born's
quantum {{measurement theory}}
says that
\begin{itemize}
\item[]
When the parallel measurement{{measurement}}
${\mathsf M}_{B({\mathbb C}^2\otimes {\mathbb C}^2)}({\mathsf O}\otimes {\mathsf O},S_{[s]})$
is taken,
\\
the probability
that
the measured value
$
\left[\begin{array}{ll}
(\uparrow,\uparrow )
\\
(\downarrow,\uparrow )
%(\text{{{{{c}}}}}, \text{about 40})
\\
(\uparrow,\downarrow )
%(\text{{{{{h}}}}}, \text{about 30})
\\
(\downarrow,\downarrow )
%(\text{{{{{h}}}}}, \text{about 40})
\end{array}\right]
$
is obtained
\\
is given
by
$
\left[\begin{array}{ll}
\langle s,
(F^z \otimes F^z)(\{(\uparrow,\uparrow )\})s
\rangle_{_{{\mathbb C}^2 \otimes {\mathbb C}^2}}
=
0
\\
\langle s,
(F^z \otimes F^z)(\{(\downarrow,\uparrow )\})s
\rangle_{_{{\mathbb C}^2 \otimes {\mathbb C}^2}}
=
0.5
\\
\langle s,
(F^z \otimes F^z)(\{(\uparrow,\downarrow )\})s
\rangle_{_{{\mathbb C}^2 \otimes {\mathbb C}^2}}
=
0.5
\\
\langle s,
(F^z \otimes F^z)(\{(\downarrow,\downarrow )\})s
\rangle_{_{{\mathbb C}^2 \otimes {\mathbb C}^2}}
=
0
\end{array}\right]
%\TAG{4.56}
$
\hfill{{{\textcolor{black}{(3.4)}}
}}%%BFBFbfbf
\end{itemize}
That is because,
$F^z (\{\uparrow \})e_1=e_1$,
$F^z (\{\downarrow \})e_2=e_2,F^z (\{\uparrow \})e_2=
F^z (\{\downarrow \})e_1=0$
For example,
\begin{align*}
&
\langle s,
(F^z \otimes F^z)(\{(\uparrow,\downarrow )\})s
\rangle_{_{{\mathbb C}^2 \otimes {\mathbb C}^2}}
\\
=
&
\frac{1}{2}\langle
(e_1\otimes e_2 - e_2 \otimes e_1 )
,
(F^z(\{\uparrow \}) \otimes F^z(\{\downarrow \})
(e_1\otimes e_2 - e_2 \otimes e_1 )
\rangle_{_{{\mathbb C}^2 \otimes {\mathbb C}^2}}
\\
=
&
\frac{1}{2}\langle
(e_1\otimes e_2 - e_2 \otimes e_1 )
,
e_1\otimes e_2
\rangle_{_{{\mathbb C}^2 \otimes {\mathbb C}^2}}
%\\
%\qquad
%\qquad
%&
%\times
%\langle
%(e_1\otimes e_2 - e_2 \otimes e_1 )
%,
%F^z(\{\uparrow \}) \otimes F^z(\{\uparrow \})
%(e_1\otimes e_2 - e_2 \otimes e_1 )
%\rangle_{_{{\mathbb C}^2 \otimes {\mathbb C}^2}}
%\\
=
%&
\frac{1}{2}
\end{align*}

%\END{align*}
Here, it should be noted that
we can assume that
the $x_1$ and the $x_2$
({}in $({}x_1, x_2{})$
$ \in $
$\{$
$
(\uparrow _{z},\uparrow _{z}),$
$
(\uparrow _{z},\downarrow _{z}),$
$
(\downarrow _{z},\uparrow _{z}),
(\downarrow _{z},\downarrow _{z})
\}$)
are respectively obtained in Tokyo
and in New York
({}or,
in the earth and in the polar star{}).

\par
\noindent
\unitlength=0.8mm
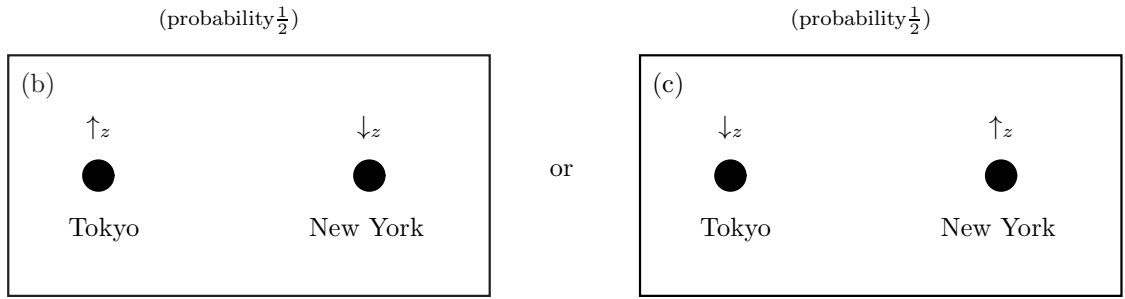
\begin{picture}(200,50)
\thicklines
\put(-20,0){{{{
\path(30,0)(110,0)(110,40)(30,40)(30,0)
\put(32,34){(b)}
\thicklines
\put(-20,0)
{{
\put(65,20){\circle*{5}}%%%%%%%%
\put(75,45){\footnotesize ({}probability$\frac{1}{2}$)}
\put(63,27){$\uparrow _{z}$}
%\put(60,20){\vector(-1,0){20}}%%%V
\put(60,10){Tokyo}
\put(0,0){
\put(110,20){\circle*{5}}%%%%%%%%%%%
\put(108,27){$\downarrow _{z}$}
%\put(115,20){\vector(1,0){20}}%%%V
\put(100,10){New York}
}
%\put(120,10){
%where \LL the velocity of A\RR
%$= -$\LL the velocity of B\RR$\!\!\!\!.\; \;$
%}
}}
}}}}
\thicklines
\put(100,20){or}
\put(85,0){{{{
\path(30,0)(110,0)(110,40)(30,40)(30,0)
\put(32,34){(c)}
\thicklines
\put(-20,0)
{{
\put(65,20){\circle*{5}}%%%%%%%%
\put(75,45){\footnotesize ({}probability$\frac{1}{2}$)}
\put(63,27){$\downarrow _{z}$}
%\put(60,20){\vector(-1,0){20}}%%%V
\put(60,10){Tokyo}
\put(0,0){
\put(110,20){\circle*{5}}%%%%%%%%%%%
\put(108,27){$\uparrow _{z}$}
%\put(115,20){\vector(1,0){20}}%%%V
\put(100,10){New York}
}
%\put(120,10){
%where \LL the velocity of A\RR
%$= -$\LL the velocity of B\RR$\!\!\!\!.\; \;$
%}
}}
}}}}
\end{picture}
\par
\noindent
This fact is,
figuratively speaking,
explained as follows:
\begin{itemize}
\item{}
Immediately after the particle in Tokyo
is measured and
the measured value $\uparrow _{z}$
[resp. $\downarrow _{z}${}]
is observed,
the particle in Tokyo informs
the particle in New York
\LL Your measured value has to be
$\downarrow _{z}$
[resp. $\uparrow _{z}${}]\RR$\!\!\!\!.\; \;$
\end{itemize}
Therefore,
the above fact implies that
quantum mechanics says that
%\BEGIN{itemize}
%\item{}
\it
there is something faster than light.
\rm
%%%%%%\END{itemize}
%index{paradox ({}de Broglie paradox{})@paradox ({}de Broglie paradox{})}
This is essentially the
same as
\it
the de Broglie paradox
\rm
({}{\it cf.} \cite{Sell}. Also see $\S9.3.3$).
That is,
\begin{itemize}
\item
if we admit quantum mechanics,
we must also admit the fact that
there is
\\
something faster than light
(i.e.,
so called "non-locality").
%In this section,
%we shall explain the above argument
%({}{\it cf.} ${{{}}}$\cite{DaBr, Sell}{}).
\hfill{\color{black}{(3.5)}}
%%%%\tag{2.76}
\end{itemize}
%Of course we admit PMT,
%and therefore,
%we believe that
%there is
%something faster than light.

%%{a} {b}
%\alpha \beta

%%%%\omega^\omega^
%BBBBBBBBBBBBBBBBBB%SBSBSBS
\par
\noindent
{\small%%{\footnotesize
\vspace{0.1cm}
\begin{itemize}
\item[$\spadesuit$] \bf {{}}{Note }3.2{{}} \rm
%,
%[EPR-paradox
%%\textcolor{black}{{\cite{Bohm}}}]
%,
%%$,
%%
%EPR\textcolor{black}{{\cite{Eins}}}
%.
%2
%$A_1$ and $A_2$ and ,
%2 and .
%,  and ,
%$t_0$, .
%% and .
%
As shown and emphasized in {\cite{ILing}},
quantum syllogism does not generally hold.
We believe that
this fact was, for the first time, discovered in
EPR-paradox
{{{}}}\textcolor{black}{{\cite{Eins}}}.
The reason that we think so is as follows.
Consider the two-particles system composed of
particles $P_1$
and $P_2$,
which is formulated in
a Hilbert space
$L^2({\mathbb R}^2)$.
Let
$\rho_{s} (\in {\frak S}^p (B_c( L^2({\mathbb R}^2)))$
be the EPR-state in EPR-paradox
(or, the singlet state in Bohm's situation).
Here, consider as follows:
\begin{itemize}
\item[{{{}}}{{(Z$_1$)}}]
Assume that $(x_1,  p_2)$ and $p'_2$
are obtained by the simultaneous measurement
of [ the position of $P_1$, the momentum of $P_2$]
and
[the momentum of $P_2$].
Since it is clear that
$p_2=p'_2$, thus,
we see that
\begin{align*}
\underset{\text{\scriptsize
[the position of $P_1$, the momentum of $P_2$]
)}}{(x_1,  p_2)} \;\;
\;\;
\Longrightarrow
\;\;
\underset{\text{\scriptsize [the momentum of $P_2$]}}{p_2}
\end{align*}
\end{itemize}
Here, for the definition of {\lq\lq}$\Longrightarrow$",
see ref. \textcolor{black}{{\cite{IFuzz}}}.
%\cite{Ishi5}}.
%the
\begin{itemize}
\item[{{{}}}{{(Z$_2$)}}]
Assume that $p_1$ and $p_2$
are obtained by the simultaneous measurement
of [the momentum of $P_1$]
and
[the momentum of $P_2$].
Since
the state
$\rho_{s} (\in {\frak S}^p (B_c( L^2({\mathbb R}^2)))$
is the EPR-state,
we see that $p_1=-p_2$,
that is, we see that
$$
\underset{\text{\scriptsize [
the momentum of $P_2$
]}}{p_2} \;\; \Longrightarrow
\;\;
\underset{\text{\scriptsize [the momentum of $P_1$]}}{-p_2}
$$
\item[{{{}}}{{(Z$_3$)}}]
Therefore,
if quantum syllogism holds,
{{{}}}{{(Z$_1$)}}
and
{{{}}}{{(Z$_2$)}}
imply that
\begin{align*}
\underset{\text{\scriptsize [the momentum of $P_1$]}}{- p_2}
\qquad
\end{align*}
that is,
the momentum of $P_1$ is equal to $-p_2$.
\end{itemize}
Since the above
{{{}}}{{(Z$_1$)}}-
{{{}}}{{(Z$_3$)}}
is not the approximately simultaneous measurement
({\it cf.} the definition {(N)}),
it is not related to
Heisenberg's uncertainty principle
{
(Theorem 3.2 later)}.
Thus,
the conclusion {{{}}}{{(Z$_3$)}}
is not contradictory to Heisenberg's uncertainty principle.
However,
now we can say that
the conclusion {{{}}}{{(Z$_3$)}}
is not true.
That is because
%
%In {{{}}}{\cite{Ishi1}},
%However,
%if we consider that
%the reason is not
%that
%th
the interpretation {{{}}}{({H$_2$})}
(i.e., only one measurement is permitted)
says,
as seen in \cite{INewi},
%as mentioned in {{{}}}{\cite{Ishi2}}
that
quantum syllogism
does not hold by the non-commutativity of the above three
observables,
i.e.,
\begin{itemize}
\item[]
$
\cases
\text{[the position of $P_1$, the momentum of $P_2$]}
\\
\text{[the momentum of $P_2$]}
\\
\text{[the momentum of $P_1$]}
\endcases
$
\end{itemize}
%Thus EPR-pardox is clarified
%in the linguistic interpretation of quantum mechanics.
Thus we see that
EPR-paradox is closely related to
the fact that
quantum syllogism does not hold in general.
%This should be compared with Bell's inequality
%\cite{Bell, Ishi2},
%which is believed to be closely connected to
%{\lq\lq}non-locality".
%
\end{itemize}
}

%%BBBBBBBBBBBBBBBBBB%SBSBSBSSPOIUYTREWQ
%
\subsubsection{Supplement
---
Bell's inequality}%{Sec. 3.1.2}
\baselineskip=18pt
\par
Let us have the argument in the previous section develop.
%,
%. supplement and .
%}.
\par
Put
$a=({a}_1,{a}_2) \in {\mathbb R}^2$,
$|a|=\sqrt{|{a}_1|^2+|{a}_2|^2}=1$.
Define the observable
${\mathsf O}_a$
$=$
$\bigl(X {{=}} \{ 1, -1 \}  ,$
$2^X  ,$
$ F_a \bigl)$
in
$B({}{\BBBC}^2{}) $
such that
\par
\noindent
{\small
\begin{eqnarray*}
F_{a}(\{1\})
&=&
\frac{1}{2}
\bmatrix
         1  & {a}_1 - {a}_2 {\sqrt{-1}}  \\
         {a}_1 + {a}_2 {\sqrt{-1}}     & 1
\endbmatrix,
\\
F_{a}(\{-1\})
&=&
\frac{1}{2}
\bmatrix
         1 & - {a}_1 + {a}_2 {\sqrt{-1}}  \\
         - {a}_1 - {a}_2 {\sqrt{-1}}     & 1
\endbmatrix.
\end{eqnarray*}
}
Further,
put
$b=({b}_1,{b}_2) \in {\mathbb R}^2$,
$|b|=\sqrt{|{b}_1|^2+|{b}_2|^2}=1$,
and,
by the same way,
define
the observable
${\mathsf O}_b$
$=$
$\bigl(X {{=}} \{ 1, -1 \}  ,2^X  , F_b \bigl)$
in
$B({}{\BBBC}^2{}) $.

%%%%\omega^\omega^
%BBBBBBBBBBBBBBBBBB%SBSBSBS
\par
\noindent
{\small%%{\footnotesize
\vspace{0.1cm}
\begin{itemize}
\item[$\spadesuit$] \bf {{}}{Note }3.3{{}} \rm
For example,
assume that
$a=(1,0),
b=(1/{\sqrt 2}, 1/{\sqrt 2})$.
Then,
the
simultaneous observable
${\mathsf O}_a \times {\mathsf O}_b$
in
$B({}{\BBBC}^2{}) $
does not exist.
The proof is easy,
thus,
it is omitted.
\end{itemize}
}

%%BBBBBBBBBBBBBBBBBB%SBSBSBSSPOIUYTREWQ

Of course,
we have the
parallel observable
$\widetilde{\mathsf O}_{ab} $
$({}{{=}}
{\mathsf O}_{a}
\otimes
{\mathsf O}_{b})$
$=$
$(X^2, 2^{X^2},F_a \otimes F_b )$
in
${ B }({\BBBC}^2 \otimes {\BBBC}^2{})$.
And further,
we have the{{measurement}}
${\mathsf M}_{B({}{\mathbb C}^2 \otimes {\mathbb C}^2{})}
(
\widetilde{\mathsf O}_{ab}
,S_{[{}s{}]}{})$,
where
$s$ is a singlet state.
Born's quantum {{{measurement theory}}}
says that
\begin{itemize}
\item[]
The probability that
a measured value
${\widetilde x}$
$({}= ({}x_1 , x_2{})${}$)$
$ \in X^2 $
$({{=}} \{ 1, -1 \}^2{}) $
is obtained by
the
{{measurement}}
${\mathsf M}_{B({}{\mathbb C}^2 \otimes {\mathbb C}^2{})}
({}\widetilde{\mathsf O}_{ab} ,S_{[{}s{}]}{})$
is given by
$\nu_{ab} ({}
\{({} x_1 , x_2) \}
)$,
where
\begin{align*}
\nu_{ab} ({}
\{({} x_1 , x_2) \}
)=
\langle
s
,
\bigl({}
%\otimes_{i,j=1,2}
({}F_{a}  \otimes F_{b}{})
({}\{({} x_1 , x_2) \}{})
\bigl)
s
\rangle_{_{{\mathbb C}^2 \otimes {\mathbb C}^2}}
\tag{\color{black}{3.6}}
\end{align*}
\end{itemize}
Now,
define the correlation function
$C_{ab}$
by
\begin{align*}
C_{ab}
&
=
\int_{X^2} x_1 \cdot x_2
\;\;
\nu_{ab}
({} dx_1 dx_2)
=
\int_{X^2} x_1 \cdot x_2
\;\;
\langle
s
,
\bigl({}
%\otimes_{i,j=1,2}
({}F_{a}  \otimes F_{b}{})
({} dx_1  dx_2)
\bigl)
s
\rangle_{_{{\mathbb C}^2 \otimes {\mathbb C}^2}}
\\
&
=
\sum\limits_{(x_1, x_2)\in X^2 }
%=-1,1}
x_1 \cdot x_2
\langle
s
,
\bigl({}
%\otimes_{i,j=1,2}
({}F_{a}  \otimes F_{b}{})
({}\{({} x_1 , x_2) \}{})
\bigl)
s
\rangle_{_{{\mathbb C}^2 \otimes {\mathbb C}^2}}
%\END{align*}
\intertext{A simple calculation shows(
{\rm cf.}
\textcolor{black}{{\cite{Sell}}}),
}
%\BEGIN{align*}
%P_{ab}
%\\
&
=
{a}_1 {b}_1 + {a}_2 {b}_2
%\tag{\color{black}{3.13}}
\end{align*}

\par
Here,
put
$a^1
(=({a}_1^1, {a}_2^1))$,
$a^2(=({a}_1^2, {a}_2^2))$,
$b^1(=({b}_1^1, {b}_2^1))$,
$b^2(=({b}_1^2, {b}_2^2))$.
And we have the following four measurements:
\begin{align*}
\begin{array}{l}
{\mathsf M}_{B({}{\mathbb C}^2 \otimes {\mathbb C}^2{})}
({}\widetilde{\mathsf O}_{a^1b^1} ,S_{[{}s{}]}{}),
\quad
{\mathsf M}_{B({}{\mathbb C}^2 \otimes {\mathbb C}^2{})}
({}\widetilde{\mathsf O}_{a^1b^2} ,S_{[{}s{}]}{}),
\\[1.5mm]
{\mathsf M}_{B({}{\mathbb C}^2 \otimes {\mathbb C}^2{})}
({}\widetilde{\mathsf O}_{a^2b^1} ,S_{[{}s{}]}{}),
\quad
{\mathsf M}_{B({}{\mathbb C}^2 \otimes {\mathbb C}^2{})}
({}\widetilde{\mathsf O}_{a^2b^2} ,S_{[{}s{}]}{})
\end{array}
%\tag{\color{black}{3.6}}
\end{align*}
Therefore,
we have the
parallel measurement
$\bigotimes_{i,j=1,2} {\mathsf M}_{ {B}({}{\BBBC}^2 \otimes {\BBBC}^2{})}
(\widetilde{\mathsf O}_{a^i b^j} , S_{[s{}]}{})$.
We
easily see that,
for each
$x\in \{-1,1\}$,
\begin{align*}
&
\nu_{a^1b^1}(\{x\} \times X)
=
\nu_{a^1b^2}(\{x\} \times X),
\;\;
&
&
\nu_{a^1b^1}(
X \times \{x\}
)
=
\nu_{a^2b^1}( X \times \{x\})
\;\;
\\
&
\nu_{a^2b^1}(\{x\} \times X)
=
\nu_{a^2b^2}(\{x\} \times X),
\;\;
&
&
\nu_{a^1b^2}(
X \times \{x\}
)
=
\nu_{a^2b^2}( X \times \{x\})
%\ENDcases
%P%\TAG{14}
\end{align*}
%
%
%
%
%$=$
%$-({}
%{a}^i_1 {b}^j_1 +
%{a}^i_2 {b}^j_2{})$.
%Thus,
%putting
Here, put
\begin{align*}
%%% \begin{align*}
a^1 = (0, 1),
\quad
b^1 = \Big({} \frac{1}{ \sqrt{2} },  \frac{1}{ \sqrt{2} }{}\Big),
\quad
a^2  = ({}1, 0),
\quad
\quad
b^2 = \Big({}\frac{1}{ \sqrt{2} }, - \frac{1}{ \sqrt{2} }{}\Big)
%P%\TAG{15}
\end{align*}
%%%\END{align*}
\par
\noindent
then,
we see
%\BEGIN{align*}
\begin{align*}
|C_{{}a^1 b^1{}} - C_{{}a^1  b^2{}}|
\;
+
\;
|C_{{}a^2  b^1{}} + C_{{}a^2  b^2 {}}|
=
2 \sqrt{2}
\tag{\color{black}{3.7}}
\end{align*}
%%%\BEGIN{align*}

%%{a} {b}
%\alpha \beta

%%%%\omega^\omega^
%BBBBBBBBBBBBBBBBBB%SBSBSBS
\par
\noindent
{\small%%{\footnotesize
\begin{itemize}
\item[$\spadesuit$] \bf {{}}{Note }3.4{{}} \rm
The theoretical conclusion (3.7)
is completely
verified by experiment
({\rm cf. }\textcolor{black}{{\cite{Sell}}}).
Also, the inequality such as
\begin{itemize}
\item[$(\sharp)$]
$
\qquad
\qquad
\qquad
\text{
"formula like the left-hand side of (3.7)"}
{\; \leqq \;}
2
$
\end{itemize}
is called
"Bell's inequality".
For example,
%(probability ), 
in mathematics,
Bell's inequality
is as follows.
%, :
Let $(Y,{\cal G}, \mu)$ be a probability space.
%probability  and .
Consider each
{measurable function}$f_k :Y \to \{-1,1\}$,
$(k=1,2,3,4)$.
And put
%,
$C_{13}=\int_Y f_1(y) \cdot f_3(y) \mu(dy)$,
$C_{14}=\int_Y f_1(y) \cdot f_4(y) \mu(dy)$,
$C_{23}=\int_Y f_2(y) \cdot f_3(y) \mu(dy)$,
$C_{24}=\int_Y f_2(y) \cdot f_4(y) \mu(dy)$.
Then,
adding the proof,
we can mention
{\bf Bell's inequality}
as follows.
\begin{align*}
&|C_{13}-C_{14}|+|C_{23}+C_{24}|
\\
=
&
|\int_Y f_1(y) \cdot f_3(y)- f_1(y) \cdot f_4(y) \mu(dy)|
+
|\int_Y f_2(y) \cdot f_3(y)- f_2(y) \cdot f_4(y) \mu(dy)|
{{\; \leqq \;}}2
\end{align*}
%(\textcolor{black}{(5.6)} and ).
%)}.
%\textcolor{black}{%index{@}}
%%index{ and @}
\end{itemize}
}

\vskip2.0cm

\subsection{
The derivation of
Axiom${}_{\text{\scriptsize c}}^{\text{\scriptsize p}}$ 1
from
Born's quantum {{measurement}}
}
%{Sec.3.2}
%\ssubsection{quantum mechanics{(}Born{{measurement}}
%{)}classical {{{measurement theory}}}{(}\textcolor{black}{Axiom${}_{\text{\scriptsize c}}^{\text{\scriptsize p}}$ 1}{)}}
\par
Let us
derive
Axiom${}_{\text{\scriptsize c}}^{\text{\scriptsize p}}$ 1
from
Born's quantum {{measurement}}.
That is,

\begin{align*}
\overset{}{\underset{\text{(quantum mechanics)}}{\text{
\fbox{
Born's quantum {{measurement}}
}}}}
\xrightarrow[{\text{Derivation}}]{}
\overset{}{\underset{\text{(measurement theory)}}{\text{\fbox{
{{Axiom${}_{\text{\scriptsize c}}^{\text{\scriptsize p}}$ 1
}}}}}}
\end{align*}
%\i

\par

\par
Let
$H$
be a Hilbert space
such that
$H=
{\mathbb C}^n$.
In this case,
we restrict
the state space
${ {\widetilde \Omega} } (
\subset
{\widehat \Omega}
\subset {\mathbb C}^n)$
such that
\begin{align*}
{ {\widetilde \Omega}} =\left\{
e_1,
e_2,
\ldots,
e_n
\right\}
=
\left\{
\bmatrix
        \; 1  \\
        \; 0 \\
        \; \vdots  \\
        \; 0
\endbmatrix,
\bmatrix
     \; 0  \\
        \; 1  \\
        \; \vdots  \\
        \; 0
\endbmatrix,
\ldots,
\bmatrix
     \; 0  \\
        \; 0  \\
        \; \vdots  \\
        \; 1
\endbmatrix
\right\}
\end{align*}
Thus,
$e (\in {\widetilde \Omega})$
is called a state.

Further,
define
$B_D({\mathbb C}^n)$
by
all $n \times n$-diagonal matrices,
which is called a basic structure.
% and ,
%
%{basic algebra} and .
\par
\noindent
\begin{itemize}
\item[(a)]
The above argument is not physical but mathematics.
%
%quantum mechanics, {basic algebra}
%$B({\mathbb C}^n)$ and ,
%,
%$B_D({\mathbb C}^n)$
%,
%{physics}.
%However,
we do not mind it
since our interest is the linguistic aspect of
quantum mechanics.
%
%,
%{linguistic aspect},  and .
\end{itemize}

Assume that
a measured value space $X$ is finite.
%{finite set} and .
The triple
${\widetilde{\mathsf O}} =(X,2^X,{\widetilde F})$
is called an observable
in
$B_D
({\mathbb C}^n)$,
if it satisfies the following (i) and (ii):
%observable  and . That is,
\par
\noindent
\begin{itemize}
\item[({}i)]
${\widetilde{\mathsf O}} =(X,2^X,
{\widetilde F}
)$
is an observable in $B
({\mathbb C}^n)$.
\item[({}ii)]
For each
$\Xi \in 2^X$,
$
{\widetilde F}
(\Xi)$
is a diagonal .
\end{itemize}
%, ${\mathsf O}=\{ F_1, F_2,\ldots F_K \}$
%{the resosution the unity}{}
Now we have
the
(diagonal matrices type{)} quantum {{measurement}}
${\mathsf M}_{B_D({\mathbb C}^n)}(\widetilde{\mathsf O}, S_{[e]})$
(where
$e \in { {\widetilde \Omega}}$
). That is,
\begin{align*}
&
{\mathsf M}_{B_D({\mathbb C}^n)}(\widetilde{\mathsf O}, S_{[e]})
\\
=
&
\text{
The measurement of
the observable
$\widetilde{\mathsf O}$
in
${B_D({\mathbb C}^n)}$
for the system with the
{{state}}$e (\in { {\widetilde \Omega}} )$
}
\end{align*}
\par
And we have
the
(diagonal matrices type{)} quantum {{measurement}}
theory
as follows.

\vskip0.5cm

%\BEGIN{itembox}[c]
\par
\noindent
\begin{center}
{\bf
\textcolor{black}{Axiom(D)}
1 quantum {{measurement}} diagonal matrix type
}
\label{rule302}
\end{center}
\par
\noindent
%\vskip0.1cm
\par
\noindent
\fbox{\parbox{155mm}{
\begin{itemize}
%\item[(i)]
%{{measurement}}
\item[]
{}\rm{}
Consider a measurement
${\mathsf M}_{B_D({\mathbb C}^n)}({\widetilde{\mathsf O}} {{=}} $
$(X,2^X,
{\widetilde F}
)$,
$
S_{[e]})$
%${\mathsf M}_{B({\mathbb C}^n)}({\mathsf O}:= ({}X, 2^X, F{}), S_{[\omega]})$
formulated in a
basic algebra
$B_D({\mathbb C}^n )$.
Assume that
the measured value
$ x$
$({}\in X  {})$
is
obtained by the measurement
${\mathsf M}_{B_D({\mathbb C}^n)}({\widetilde{\mathsf O}} $,
$
S_{[e]})$.
Then,
%it holds that
{{}}
the probability
that
a
measured value
$ x$
$({}\in X)$
is obtained
is
given by
$
\langle e , {\widetilde F}(\{x\}) e \rangle
$.
%\END
\end{itemize}
}
}
\par
\vskip0.5cm
\par
\noindent

%
%
%\vskip0.5cm
%
%\BEGIN{itembox}[c]{
%\bf
%%\textcolor{black}{Axiom${}_{\text{\scriptsize c}}^{\text{\scriptsize p}}$ 1({{{measurement theory}}}:${B_D({\mathbb C}^n)}$}
%\textcolor{black}{Axiom(D)
%%${}_{\text{\scriptsize c}}^{\text{\scriptsize p}}$
%1 quantum {{measurement}} diagonal matrix type}
%}
%\label{rule302}
%\label{axiom(D)}
%%%index{1@Axiom${}_{\text{\scriptsize c}}^{\text{\scriptsize p}}$ 1[{}{{measurement}}{}(continuous type )]}
%\rm
%%PPPPPPPPPPPP
%\BEGIN{itemize}
%%\item[
%%(i)]
%%{{measurement}}.
%\item[
%%(ii)
%]
%%{basic algebra}
%
%\END{itemize}
%\END{itembox}
%

\par
\vskip0.3cm

\par
Now,
let
$\widetilde{\mathsf O}=(X,2^X, {\widetilde F})$
be an observable
in
${B_D({\mathbb C}^n)}$.
Thus, for each
$x (\in X)$,
we see that
\begin{align*}
{\widetilde F}(\{x\})=
\bmatrix
f_{11}(\{x\}) & 0& \cdots &0 \\
0 & f_{22}(\{x\}) & \cdots &0 \\
\vdots & \vdots & \vdots &\vdots \\
0 & 0 & \cdots &f_{nn}(\{x\})\\
\endbmatrix
\in
B_D({\mathbb C}^n)
%P%\TAG{19}
\end{align*}
%
% and ,
%\END{align*}
%\langle e_k , \widetilde{F} (\{x\}) e_k \rangle
%=
%f_{kk}(\{x\})
%\END{align*}
%, %%%\Omega \omega \omega \Omega \omega \omega
Here, a discrete metric space
$\Omega=\{\omega_1, \omega_2,\ldots,
\omega_n \}$
is regarded as a space.
For each
$x\in X$,
define
$f_x: \Omega
%=\{\omega_1,\omega_2,\ldots,\omega_n \}
\to {\mathbb R}$
such that
\begin{align*}
f_x(\omega_k ) = f_{kk}(\{x\})
\bigl(=
\langle
e_k,
{\widetilde F}(\{x\}) e_k \rangle
\bigl)
\qquad
(\forall k =1,2,\ldots, n )
%%P%\TAG{20}
\end{align*}
Then,
we can regard
$\{ f_x \}_{x \in X}$
as
the resolution of the unity
on
$\Omega$.
Thus, putiing
%%%%%%%%%${\mathsf O}=(X, 2^X, F)\end{align*}{\mathsf O}=(X, 2^X, F)$
\begin{align*}
[F(\Xi )](\omega) = \sum\limits_{x\in \Xi } f_x (\omega)
\qquad
(\forall \omega \in \Omega )
\tag{\color{black}{3.8}}
\end{align*}
we get
the observable
${\mathsf O}=(X, 2^X, F)$
in
$C(\Omega )$.
%%%%%%%%%%%%%
Therefore, under the following identification:
\footnotesize
\begin{align*}
&
\left[\begin{array}{cc}
\text{
observable
$
\widetilde{\mathsf O}=(X,2^X,
{\widetilde F}
)
$
in
$
{B_D({\mathbb C}^n)}
$
}
\\
\\
{\widetilde \Omega }
=\{
e_1, e_2,...,e_n\}
\ni e_k
\end{array}\right]
\;\;
\mathop{\longleftrightarrow}_{\text{\footnotesize identification} }
\;\;
\left[\begin{array}{cc}
\text{
observable
$
{\mathsf O}=(X, 2^X, F)
$
in
$C(\Omega)$
}
\\
\\
\omega_k
\in
\{
\omega_1,\omega_2,...,\omega_n\}
=
\Omega
\end{array}\right]
\end{align*}
\normalsize \baselineskip=18pt
we see
\small
\begin{align*}
&
\text{
\fbox{
(diagonal matrix type)
quantum {{measurement}}
${\mathsf M}_{B_D({\mathbb C}^n)}(\widetilde{\mathsf O}, S_{[e_k]})$
}
}
\\
=
&
\text{
\fbox{
classical {{measurement}}
${\mathsf M}_{C(\Omega)}({\mathsf O}, S_{[\omega_k]})$
}
}
\end{align*}
\normalsize \baselineskip=18pt

\vskip0.5cm

\newpage

%\BEGIN{itembox}[c]
\par
\noindent
\begin{center}
{\bf
\textcolor{black}{Axiom${}_{\text{\scriptsize c}}^{\text{\scriptsize p}}$
1 ({{measurement}}: finite $\Omega$) }
}
%
%\textcolor{black}{Axiom(D)}
%1 quantum {{measurement}} diagonal matrix type
%}
\label{rule303}
\end{center}
\par
\noindent
%\vskip0.1cm
\par
\noindent
\fbox{\parbox{155mm}{
\begin{itemize}
%\item[(i)]
%{{measurement}}
\item[]
{}\rm{}
Consider a measurement
${\mathsf M}_{C(\Omega)}({{\mathsf O}} {{=}} $
$(X, 2^X, F),
S_{[\omega]})$
%${\mathsf M}_{B({\mathbb C}^n)}({\mathsf O}:= ({}X, 2^X, F{}), S_{[\omega]})$
formulated in a
basic algebra
$C(\Omega)$.
%Assume that
%the measured value
%$ x$
%$({}\in X  {})$
%is
%obtained by the measurement
%${\mathsf M}_{C(\Omega)}({{\mathsf O}} {{=}} $
%$(X, 2^X, F),
%S_{[\omega]})$.
Then,
%it holds that
{{}}
the probability
that
a
measured value
$ x$
$({}\in X)$
is obtained
by the measurement
${\mathsf M}_{C(\Omega)}({{\mathsf O}} {{=}} $
$(X, 2^X, F),
S_{[\omega]})$
is
given by
$[F(\{x\})](\omega)$.
\end{itemize}
}
}
\par
\vskip0.5cm
\par
\noindent

%
%
%Thus, we get:
%\BEGIN{itembox}[c]{
%\bf
%\textcolor{black}{Axiom${}_{\text{\scriptsize c}}^{\text{\scriptsize p}}$
%1 ({{measurement}}: finite $\Omega$) }
%%\textcolor{black}{Axiom${}_{\text{\scriptsize c}}^{\text{\scriptsize p}}$ 1({{measurement}}:finite $\Omega$) }
%}
%
%%index{1@Axiom${}_{\text{\scriptsize c}}^{\text{\scriptsize p}}$ 1[{{measurement}}:finite $\Omega$]}
%\BEGIN{itemize}
%%\item[
%%(i)]
%%{{measurement}}.
%\item[]
%Consider a measurement
%${\mathsf M}_{C(\Omega)}({{\mathsf O}} {{=}} $
%$(X, 2^X, F),
%S_{[\omega]})$
%%${\mathsf M}_{B({\mathbb C}^n)}({\mathsf O}:= ({}X, 2^X, F{}), S_{[\omega]})$
%formulated in a
%basic algebra
%$C(\Omega)$.
%Assume that
%the measured value
%$ x$
%$({}\in X  {})$
%is
%obtained by the measurement
%${\mathsf M}_{C(\Omega)}({{\mathsf O}} {{=}} $
%$(X, 2^X, F),
%S_{[\omega]})$.
%Then,
%%it holds that
%{{}}
%the probability
%that
%a
%measured value
%$ x$
%$({}\in X)$
%is obtained
%is
%given by
%$[F(\{x\})](\omega)$.
%\END{itemize}
%\END{itembox}

\def\BC{\left[\begin{array}{ll}}
\def\EC{\end{array}\right.}
\rm
\par

\vskip0.3cm
\par
If it writes without carrying out simple,
we see:

\begin{itemize}
\item[(b)]
When
an {\bf {{observer}}}
takes a measurement
of an {\bf observable
${\mathsf O}{{=}} ({}X, 2^X, F{})$
%${\mathsf O}{{=}} ({}X, {\cal F} , F{})$
}
(or,
by a {\bf {measuring instrument}${\mathsf O}$}
)
for a {\bf {measuring object}}
with a
{\bf {{state}}}
$\omega$,
the probability
that
a {\bf measured value }
belongs to
$\Xi (\in {2^X} )$
is given by
${}[F({}\Xi{})](\omega)$.
\end{itemize}

%
%%%%\omega^\omega^
%BBBBBBBBBBBBBBBBBB%SBSBSBS
\par
\noindent
{\small%%{\footnotesize
\begin{itemize}
\item[$\spadesuit$] \bf {{}}{Note }3.5{{}} \rm
Note that
A measurement
${\mathsf M}_{B({\mathbb C}^n)}({\mathsf O}:= ({}X, 2^X, F{}), S_{[\omega]})$
in
Axiom${}_{\text{\scriptsize c}}^{\text{\scriptsize p}}$ 1 (The probabilistic interpretation of quantum mechnaics)
is physics,
but
a measurement
${\mathsf M}_{C(\Omega)}({{\mathsf O}} {{=}} $
$(X, 2^X, F),
S_{[\omega]})$
in Axiom${}_{\text{\scriptsize c}}^{\text{\scriptsize p}}$ 1(Measurement; finite $\Omega$)
is language.
\end{itemize}
}

%%BBBBBBBBBBBBBBBBBB%SBSBSBSS
\baselineskip=18pt
\par
\vspace{0.3cm}

This is the mechanism of the derivation such that
\begin{itemize}
\item[(c)]
$
\qquad
\qquad
\overset{}{\underset{\text{(quantum mechanics)}}{\text{
\fbox{
Born's quantum {{measurement}}
}}}}
\xrightarrow[{\text{Derivation}}]{}
\overset{}{\underset{\text{(measurement theory)}}{\text{\fbox{
{{Axiom${}_{\text{\scriptsize c}}^{\text{\scriptsize p}}$ 1
}}}}}}
$
\end{itemize}

However,
the above is merely the starting point.
Our purpose is to asser as follows.
\begin{itemize}
\item[(d)]
What is important is
to
show the power of
classical {{{measurement theory}}},
and
to
decalre that it is a scientific language.
That is,
we have to show that
Axiom${}_{\text{\scriptsize c}}^{\text{\scriptsize p}}$ 1
is the most powerful proverb in sceince.
\end{itemize}

\subsection{Schr\"odinger's cat
}%{Sec.3.3}
%\ssubsection{Schr\"odinger's catquantum mechanics}
%
\par
In what follows,
we expain our opinion about the Copenhagen interpretation.

\subsubsection{The Copenhagen interpretation
---
measured value
is perceived by the brain
---}
\par
Schr\"odinger's cat
is the most famous paradox in quantum science.

%index{@Schr\"odinger's cat paradox}
\rm
[Schr\"odinger's cat paradox].
%%index{paradox ({}Schr\"odinger's cat paradox{})
%@paradox ({}Schr\"odinger's cat paradox{})}
%%index{Schr\"odinger's cat paradox@Schr\"odinger's cat paradox}
Note that
Schr\"odinger's cat
does not appear in the world of measurement theory.
Let us explain it as follows:
%({}according to the article ({}in Internet{}) by
%David M. Harrison, Department of Physics, University of Toronto{}).
In 1935
({}{\it cf.} \cite{Schr}{})
Schr\"odinger published an essay describing the conceptual problems
in quantum mechanics. A brief paragraph in this essay described
the cat paradox.

\begin{itemize}
\item[(a)]
Suppose we put a cat in a cage with a radioactive atom, a Geiger counter,
and a poison gas bottle; further suppose that the atom
in the cage has a half-life of one hour,
a fifty-fifty chance of decaying  within the hour.
If the atom decays, the Geiger counter will tick;
the triggering of the counter will
get the lid off the poison gas bottle, which will kill the cat.
If the atom does not decay, none of the above things happen,
and the cat will be alive.  Now the question:
\begin{itemize}
\item[(b)]
We then ask:
What is the state of the cat after one hour?
\end{itemize}
The answer according to quantum mechanics is that
%\BEGIN{itemize}
\item[(c)]
the cat is in a state which can be thought of as half-alive and half-dead,
that is,
the state such as
$\frac{\text{\LL \textcolor{black}{Fig.}3.2($\sharp_1$)\RR} + \text{\LL \textcolor{black}{Fig.}3.2($\sharp_2$))\RR}}{2}$
(or
more theoretocally,
$\frac{|\text{\LL \textcolor{black}{Fig.}3.2($\sharp_1$)\RR} \rangle + |\text{\LL \textcolor{black}{Fig.}3.2($\sharp_2$))\RR} \rangle }{{\sqrt 2}}$
).
\end{itemize}

%PPPPPPPPPPPPPP
%\END{document}
%EEEEEEEEEEEEEEEEEE

\baselineskip=18pt

\par
\noindent
%%\vskip-0.4cm%%%
%\begin{figure}[htbp]
\unitlength=0.62mm
%%\unitlength=0.72mm
%\BEGIN{picture}(1180,75)(10,0)
%%\unitlength=0.33mm
%%%%\put(25,12){A}
%\allinethickness{0.5mm}
%%\put(60,30){\oval(70,60)}
%\path(20,0)(100,0)(100,70)(20,70)(20,0)
%\put(25,64){{FIG.$\;$}3.2($\sharp_1$)}
%\put(14,-5){
%\path(30,50)(30,60)(40,60)(40,50)(30,50)
%\path(35,60)(35,65)(70,65)(70,45)
%\filltype{shade}\put(70,40){\circle*{10}}
%\filltype{black}\put(70,12){\circle*{13}}
%\put(55,32){\footnotesize }
%\put(60,20){\footnotesize }
%\put(32,53){$\cdots$}
%}
%\put(100,0){
%\put(14,-5){
%\path(30,50)(30,60)(40,60)(40,50)(35,45)
%\path(35,60)(35,65)(70,65)(70,45)
%\filltype{shade}\put(70,40){\circle*{10}}
%\filltype{black}\put(70,12){\circle*{13}}
%\put(52,44){\footnotesize !}
%\put(70,23){\vector(0,1){7}}
%\put(55,32){\footnotesize }
%\put(60,20){\footnotesize }
%}
%}
%%\put(160,30){\oval(70,60)}%%%%%%%%
%\put(100,0){
%\path(20,0)(100,0)(100,70)(20,70)(20,0)
%\put(25,64){{FIG.$\;$}3.2($\sharp_2$)}
%}
%%%%\put(160,30){\circle{60}}
%%\allinethickness{0.3mm}
%%%\filltype{white}
%\put(50,15){\circle{20}}
%\put(50,31){\circle{12}}
%\put(47,15){}
%\put(47,47){\footnotesize }
%\put(0,-11){
%\sspline(50,16)(40,17)(30,19)
%}
%\put(145,16){\ellipse{23}{17}}
%\put(162,14){\ellipse{13}{10}}
%\put(141,16){}
%%\put(151,47){\footnotesize }
%\put(10,0){
%\sspline(122,16)(122,6)(130,6)
%}
%\allinethickness{0.2mm}
%\put(-5,-20){
%\sspline(152,70)(152,60)(145,55)(145,50)
%\path(142,55)(145,50)(148,55)
%\put(136,57){\footnotesize }
%%{\footnotesize poison gas}
%}
%\END{picture}
%
%
%
%
%
%
%
%
%
%
%
%
%
%
%\par
%\noindent
%\unitlength=0.8mm
\begin{picture}(200,80)(10,0)
%%%\put(25,12){A}
%\put(40,16){\scriptsize $x_0$}
%\qbezier(40,20)(100,61)(157,42)
%\path(107,49)(115,48)(107,45)
%\put(40,20){\circle*{1}}
%\put(157,41){\circle*{1}}
%\put(155,45){\scriptsize $E({}x_0)$}
%%\put(157,40){\scriptsize $r$}
%%\put(76,42){\scriptsize $\beta$}
%\put(151,33){$ \; \rho^p$}
%\put(149,34){ \circle*{1} }
%\put(170,30){$ D_{x}^\gamma$}
%\put(153,63){ ${\frak S}^p ({{\cal M}(\Omega))$}
%\put(57,63){$X$}
%%\allinethickness{0.2mm}
%%\put(60,30){\line(1,0){30}}
%%\put(160,30){\line(1,0){30}}
%%\put(60,30){\line(5,3){26}}
%%\put(160,30){\line(1,4){5}}
%%\allinethickness{0.5mm}
%%\put(42,55){\line(4,-1){44}}
%%\put(142,55){\line(4,-1){44}}
\allinethickness{0.5mm}
%\put(60,30){\oval(70,60)}
\path(20,0)(100,0)(100,70)(20,70)(20,0)
\put(50,74){\textcolor{black}{Fig.}3.2($\sharp_1$)}
\put(14,-5){
\path(30,50)(30,60)(40,60)(40,50)(30,50)
\path(35,60)(35,65)(70,65)(70,45)
\filltype{shade}\put(70,40){\circle*{10}}
\filltype{black}\put(70,12){\circle*{13}}
\put(32,53){$\cdots$}
}
\put(100,0){
\put(14,-5){
\path(30,50)(30,60)(40,60)(40,50)(35,45)
\path(35,60)(35,65)(70,65)(70,45)
\put(70,20){\vector(0,1){10}}
\filltype{shade}\put(70,40){\circle*{10}}
\filltype{black}\put(70,12){\circle*{13}}
\put(52,40){tick !}
}
}
%\put(160,30){\oval(70,60)}
\put(100,0){
\path(20,0)(100,0)(100,70)(20,70)(20,0)
\put(50,74){\textcolor{black}{Fig.}3.2($\sharp_1$)}
}
%%%\put(160,30){\circle{60}}
%\allinethickness{0.3mm}
%%\filltype{white}
\put(50,15){\circle{20}}
\put(50,31){\circle{12}}
\put(47,15){cat}
\put(0,-11){
\spline(50,16)(40,17)(30,19)
}
\put(145,16){\ellipse{23}{17}}
\put(162,14){\ellipse{13}{10}}
\put(141,16){cat}
\put(10,0){
\spline(122,16)(122,6)(130,6)
}
\allinethickness{0.2mm}
\put(-5,-20){
\spline(152,70)(152,60)(145,55)(145,50)
\path(142,55)(145,50)(148,55)
\put(126,57){\footnotesize poison gas}
}
\end{picture}
\vskip0.2cm
\begin{center}
{Figure 3.2:
Schr\"odinger's cat
%%\label{Fig.4020}
}
\end{center}
\par
\noindent

\par
\noindent
Of course, this answer (A) is curious.
This is the so-called
Schr\"odinger's cat paradox.
This paradox is due to the fact that
micro mechanics and macro mechanics are mixed in
the above situation.
On the other hand,
as seen in (2.38),
micro mechanics ({}= quantum measurement theory{})
and
macro mechanics ({}= classical measurement theory{})
are always separated in measurement theory.
Therefore,
Schr\"odinger's cat
does not appear in the world of measurement theory,
though
this may be a surface solution
of Schr\"odinger's cat paradox.
\vskip0.2cm
\par

\rm
\begin{itemize}
\item[({}d)]
At the moment of the measurement
(i.e., at the moment of having opened the window of the box
and seeing inside),
"alive"
or
"dead"
is determined.
That is,

\begin{align*}
\text{half-alive and half-dead}
\xrightarrow[{\text{\scriptsize and seeing inside}}]{{\text{\scriptsize At the moment of having opened the window of the box}}}
\cases
\text{alive}
\\
\text{dead}
\endcases
\end{align*}
%}
\end{itemize}
This is famous Schr\"odinger cat paradox.
For author's opinion, see Note 3.6.
Here, note that
%paradox
%{.} ,{\;}
\par
\par
\begin{itemize}
\item[({}e$_1$)]
"At the moment of the measurement"
means
that
"At the moment that the signal reach observer's brain"
%$\qquad
%\qquad$
%{\bf , {{measurement}}{(}{)}}
\end{itemize}
The readers may doubt it.
However,
the dualism
is
to
partition
"brain"
and
"matter".

%
%({\rm cf.}
%\textcolor{black}{{\cite{Sell}}})
%
%\textcolor{black}{{Note }3.6($\sharp$)}
%.
%%
%%quantum mechanics
%%%
%%%the Copenhagen interpretation[\textcolor{black}{{Chap.$\;$1}(U$_1$)--(U$_7$))}]
%%%
%%
%%.
%\par
% and ,
%paradox
%, ,
%,
%%Einstein
%%({\rm cf.}
%%\textcolor{black}{{\cite{Sell}}}),
%.
%%index{@the Copenhagen interpretation}
%%, the Copenhagen interpretation
%%,
%PART,
%dualism and ,
%{(}={)} and  and
% and
%,  and  and .
%That is,
% and  and ,
%% and ,
\par
Thus, the dualism says that

\begin{itemize}
\item[({}e$_2$)]
$\qquad
\qquad$
{\bf There is no measurement without our brain.}
%, {{measurement}}{(}{)}}
\end{itemize}
Some may think,
from the point of the realistic view, that
the
({}e$_2$)
is absurd.
However, out proposal is linguistic.
We think that
this
(e$_2$)
is also one of the (linguistic) Copenhagen interpretation.
%%
%%n,
%%,
%%{Chap.$\;$1}the Copenhagen interpretation[\textcolor{black}{(U$_1$)--(U$_7$))}]
%---
%observer(U$_2$) and 
%---
%(, {FIG.$\;$}1.1({{measurement}}{FIG.$\;$}),  and ).
%
%{{measurement theory}},{\;},{\;}
%.{}
%%
%%%%%\omega^\omega^
%BBBBBBBBBBBBBBBBBB%SBSBSBS
\par
\noindent
{\small%%{\footnotesize
\vspace{0.1cm}
\begin{itemize}
\item[$\spadesuit$] \bf {{}}{Note }3.6{{}} \rm
%(i):
%%the Copenhagen interpretation
%%,
%%{physics}
%%.
%
%Schr\"odinger's cat
%(, (d) and ({\rm cf.}
%\textcolor{black}{\cite{Sell}})),
%observer{{measurement}}{{state}} and ,
%the Copenhagen interpretation[{Chap.$\;$1}\textcolor{black}{({{}}U$_1$)--({{}}U$_7$)}](\textcolor{black}{{Note }2.8}).
%
The cause of "confusion" is because it is caught by
realism.
According to
the linguistic world-view
(i.e.,
""The limits of my language mean the limits of my world."
due to Wittgenstein, Chap. 8 (m)),
\begin{itemize}
\item[($\sharp$)]
Schr\"odinger's cat
is out of the description of measurement theory,
and thus,
Schr\"odinger's cat does not exist
(cf .{}\textcolor{black}{\cite{IQphi, ILing}}).
\end{itemize}
\end{itemize}
}

\baselineskip=18pt
\subsection{Heisenberg's {uncertainty principle}
% and 
}%{Sec.3.4}
\subsubsection{
The thought experiment by
$\gamma$-rays
microscope}
%%index{
%1@Heisenberg's {uncertainty principle}
%(
%$\gamma$-rays
%microscope
%)}
\par
\rm
Heisenberg's {uncertainty principle}\textcolor{black}{{\cite{Heis}}}
is the following
{{Proposition }}3.1
[(i) and (ii)].
This is usually said to be the greatest scientific result in
the 20-th century.
However,
as mentioned is the following section,
it
is doubtful.
\par
\noindent
\vskip0.3cm
%\vskip0.3cm
%BFBF
\par
\noindent

\par
\noindent
{\bf {{Proposition }}3.1
}
%({\rm cf.}
%\textcolor{black}{{\cite{Heis}}})
\rm
\rm
\bf %BFBF
[Heisenberg's uncertainty relation]
\rm
({}{\it cf.} \cite{Heis}{}).
$\;\;$%POPOPO
\par
\rm
\begin{itemize}
\item[(i)]
{}\rm{}
The particle position q and momentum p can be measured
{\lq\lq simultaneously\rq\rq}$\!\!,\;$
if the {\lq\lq errors\rq\rq} $\Delta (q) $ and $\Delta (p)$
in determining the particle position and momentum
are permitted to be  non-zero.
\par
\rm
\item[(ii)]
{}\rm{}
Moreover, for any
$ \epsilon > 0 $ ,
we can take the above
{\lq\lq approximate simultaneous\rq\rq} measurement of the position q
and momentum p such that
$\Delta (q) <  \epsilon $
$(${}or $\Delta (p) < \epsilon $ $)$.
%\par
%\rm
%\item[(iii)]
%\ssl
However, the following Heisenberg's uncertainty relation
holds:
\begin{align*}
\Delta (q) \cdot  \Delta (p)
\ge \frac{ \hbar } {2},
\tag{3.9} 
\end{align*}
for all {\lq\lq approximate simultaneous\rq\rq} measurements of the particle position and momentum.
\end{itemize}
%{
%\footnotesize
%\ba

%
%
%\BEGIN{itemize}
%\item[({}i)]
%$x${{measurement}}. , $p${{measurement}}.
%, $x$ and $p${{measurement}} and ,
%
%$\Delta x$ and $\Delta p${{measurement}}.
%\item[({}ii)]
%, $\Delta x$ and $\Delta p$, Heisenberg's {uncertainty principle}
%. That is,
%\BEGIN{align*}
%\Delta x
%\cdot \Delta p {\doteqdot} \hbar (=
%\text{}/2\pi
%{\doteqdot}
%1.5547 \times 10^{-34} Js
%).
%%\TAG{4.59}
%\tag{\color{black}{3.9}}
%%%%%%REDREDREDREDREDRE
%\END{align*}
%%index{@}
%%index{hbar@$\hbar$:}
%, , ,
%.
%\END{itemize}
%%,

%

%\baselineskip=18pt

\par
\noindent
%%\vskip-0.4cm%%%
%\begin{figure}[htbp]
\unitlength=0.6mm
%\unitlength=0.8mm
\begin{picture}(200,80)(10,0)
%%%\put(25,12){A}
%\allinethickness{0.5mm}
%
\put(50,0){{{
\put(20,0){
\put(30,2){\vector(1,0){100}}
\put(30,2){\vector(0,1){70}}
\put(33,70){$z$}
\put(130,5){$x$}
\filltype{bla$ck}\put(70,5){\circle*{3}}
\put(69,18){$2 \varepsilon$}
\put(47,5){{\footnotesize electron $e$}}
\spline(65,15)(70,17)(75,15)
\put(100,7){\vector(-1,0){20}}
\put(100,10){$\hbar \nu$}
\put(60,30){{\scriptsize an object lens}}
\put(70,25){\ellipse{20}{5}}
\path(70,5)(80,25)(80,70)%(20,70)(20,0)
\path(70,5)(60,25)(60,70)(80,70)%(20,70)(20,0)
}
}}}
\end{picture}
%\vskip0.1cm
\begin{center}{Figure 3.3:
The thought experiment by
$\gamma$-rays
microscope
%\label{Fig.4030}
}
\end{center}
\par
\noindent
\par
\noindent
\baselineskip=18pt

\par

\par
Heisenberg's argument
(
the thought experiment by
$\gamma$-rays
)
is formed by the following
two approximate equalities
(3.10)
and
(3.11):
\begin{align*}
\Delta x = \frac{\lambda}{\sin \varepsilon }
\qquad
\text{(
the optical resolution
(concerning $x$-axixs),
$\lambda$: wavelength)}
\tag{\color{black}{3.10}}
%%%%%REDREDREDREDREDRE
\end{align*}
Also,
as seen in \textcolor{black}{{{Fig. 3.3}}},
the error of the momentum $p_x$
($x$-axis direction) is,
by
Compton recoil,
estimeted as
\begin{align*}
\Delta p_x = \frac{\hbar \nu}{c}\sin \varepsilon
\qquad
\text{
($c$: light speed, thus, $c= \lambda \cdot \nu$ ))}
%\
\tag{\color{black}{3.11}}
%%%%%REDREDREDREDREDRE
\end{align*}
Thus, Heisenberg's {uncertainty principle} \textcolor{black}{(3.9)}
is obtained by
$$
\Delta x \cdot \Delta p_x = (3.10) \times (3.11) \approx \hbar
$$
For precise argument, see \cite{Neum},
in which my favorite explanation is written.
%{{measurement}}
%$\Delta p $
% and .
%,
%\textcolor{black}{(3.10)}
% and
%\textcolor{black}{(3.11)}
%
%$\lambda = c/\nu
%$
%,
%Heisenberg's {uncertainty principle}\textcolor{black}{(3.9)}
%.
%\par
\par
%,
%\BEGIN{itemize}
%\item[]
%$\qquad \qquad \qquad \qquad \qquad $
%{{measurement}}
%
%\END{itemize}
% and .
%, , .
\par
However,
it should be noted that
the above is not argument in the framework of quantum mechanics.
% and 
%Heisenberg's {uncertainty principle}
% and ,
%quantum mechanics.
%Heisenberg,
%quantum mechanics
%%{(}, quantum mechanics{}{)}
%Heisenberg,
%quantum mechanics
%quantum mechanics
%.

%
%%%%\omega^\omega^
%BBBBBBBBBBBBBBBBBB%SBSBSBS
\par
\noindent
{\small%%{\footnotesize
\vspace{0.1cm}
\begin{itemize}
\item[$\spadesuit$] \bf {{}}{Note }3.7{{}} \rm
%(i):
We think that
Heisenberg's {uncertainty principle}({{Proposition }}3.1)
is meaningless.
That is because
%\BEGIN{itemize}
%\item[]
%%$\qquad
%%$
%Heisenberg's {uncertainty principle}({{Proposition }}3.1)
%,
% and , 
%\END{itemize}
% and . Heisenberg's {uncertainty principle}({{Proposition }}3.1)
%quantum mechanics and 
For example,
\begin{itemize}
\item[$(\sharp)$]
The approximate measurement
and "error" in {{Proposition }}3.1
are not defined.
%\item[$(\sharp_2)$]
%(3.9)
%$\doteqdot$
%(,
%$\geqq$
%),  and 
%\item[$(\sharp_3)$]
%EPR-paradox(\textcolor{black}{{Note }3.2}{{syllogism}}$(\sharp_3)$) and Heisenberg's {uncertainty principle}
%({{Proposition }}3.1)
%\textcolor{black}{(
%{\rm cf.}
%{\cite{QYuka}})}
%, {{Proposition }}3.1${{\cdot}}$
\end{itemize}
This will be improved
in
\textcolor{black}{{{Theorem }}3.4}
in the framwork of quantum mechanics.
That is,
Heisenber's thought experiment is an execellent
idea
before the discovery of quantum mechanics.
Some  may ask that
%({\rm cf.}
%\textcolor{black}{{\cite{Irep}}}).
\begin{itemize}
\item[]
If it be so,
why is Heisenberg's
{uncertainty principle}
({{Proposition }}3.1) famous?
%?
\end{itemize}
The author thinks that
\begin{itemize}
\item[]
 Heisenberg's
{uncertainty principle}
({{Proposition }}3.1)
was used as the slogan for advertisement
of quantum mechanics
in order to emphasize the difference
between
classical mechanics
and quantum mechanics.
\end{itemize}
\end{itemize}
}
%
%%BBBBBBBBBBBBBBBBBB%SBSBSBSS
%\vspace{-1.0cm}
%\textcolor{black}{%index{@{(}Heisenberg's {uncertainty principle})}}
\subsubsection{
The quantam mechanical formulation
of
Heisenberg's {uncertainty principle}}
\par
%,
%quantum mechanics and ,
%the Copenhagen interpretation.
%%index{@the Copenhagen interpretation}
%,
%quantum mechanics
%({Chap.$\;$1}the Copenhagen interpretation(U$_3$)),
%Heisenberg
%{{measurement}}{(}{)}, .

We think that
the standard Copenhagen interpretation says that
$$
\text{
\bf
{{
A measurement}} is not related to an interaction
}
$$
Therefore,
we do not trust Heisenberg's thought experiment by
$\gamma$-rays,
which clearly
is related to
a thing like an interaction.

Heisenberg's {uncertainty principle}
is often misunderstood
as
Robertson's uncertainty reltion
as follows.
%{uncertainty principle}{\rm ({\rm cf.}
%\textcolor{black}{{\cite{Neum, QYuka}}})} and
% and .
%%index{ and @{uncertainty principle}}
\par
\noindent
\vskip0.3cm
%BFBF
\renewcommand{\footnoterule}{%
  \vspace{2mm}                      % 
  \noindent\rule{\textwidth}{0.4pt}   % , 
  \vspace{-5mm}
}
\par
\noindent
{\bf {{Theorem }}3.2
[Robertson's {uncertainty principle}
%{\rm ({\rm cf.}
%\textcolor{black}{{\cite{Neum}}})}
\bf
]}$\;\;$%POPOPO
For simplicity,
Put $H={\mathbb C}^n$.
%
%\footnote{
%,
%.
%}
Let
$A$,
$B$
$(\in B({\mathbb C}^n))$
be
Hermitian,
thereffore,
by \textcolor{black}{(3.2)},
its observable representation
is ${\mathsf O}_A$,
${\mathsf O}_B$
respectively.
Here, consider the
{{measurement}}
${\mathsf M}_{B({\mathbb C}^n)}({\mathsf O}_A, S_{[\omega]})$
and
${\mathsf M}_{B({\mathbb C}^n)}({\mathsf O}_B, S_{[\omega]})$.
That is,
consider a
parallel measurement
${\mathsf M}_{B({\mathbb C}^n)\otimes B({\mathbb C}^n)}({\mathsf O}_A\otimes
{\mathsf O}_B, S_{[\omega \otimes \omega]})$.
Then,
it holds that:
\begin{align*}
\delta_A^\omega
\cdot
\delta_B^\omega
{\; \geqq \;}
\frac{1}{2}
|
\langle \omega , (AB-BA) \omega \rangle
|
\qquad
(\forall \omega \in {\widehat \Omega} (\subset {\mathbb C}^n))
%P%\TAG{25}
\end{align*}
where,
$\delta_A^\omega$
and
$\delta_B^\omega$
is defined by
\textcolor{black}{(3.3)}.
That is,
\begin{align*}
\cases
\delta_A^\omega
=
\left[
\langle A \omega , A \omega \rangle
-
|\langle \omega , A \omega \rangle|^2
\right]^{1/2}
\\
\delta_B^\omega
=
\left[
\langle B \omega , B \omega \rangle
-
|\langle \omega , B \omega \rangle|^2
\right]^{1/2}
\endcases
%P%\TAG{26}
\end{align*}
Of course,
the above holds in the case
of
infinite dimensional Hilbert space.
%,
%$H${(}That is, $n=\infty${)}
%.
For example,
when
$Q$
and
$P$
is respectively
the position
observable
and the momentum
observable
(i.e.,
when
$QP-PQ=\hbar {\sqrt{-1}}$
),
%index{@observable }
%index{@observable }
it holds that
$
\delta_Q^\omega
\cdot
\delta_P^\omega
{\; \geqq \;}
\frac{1}{2}
\hbar
$
%\qquad
%(\forall \omega \in {\widehat \Omega} (\subset H))
%%%P%\TAG{63}
%\END{align*}

\par
\vskip0.3cm

\par
%,
%quantum mechanics,
%( and , ),
%
% and ,
%.
% and ,
%$\delta_A^\omega$,
%observable $A$
%.
%$\delta_B^\omega$
%observable $B$
%{}
%{uncertainty principle},
%
%.
%{uncertainty principle},
%%,
%%
%Heisenberg's {uncertainty principle}
% and {}
%
%Heisenberg's {uncertainty principle}
%
%simultaneous measurement
%,
%{uncertainty principle}
% and .
% and ,
As pointed out
in
"mathematical foundatins of quantum mechanics(1932);\cite{Neum}",
Robertoson's uncertainty principle
is not the mathematical representation
of
Heisenberg's uncertainty principle.

\par
%\vskip0.3cm
\par
%,
%Heisenberg's {uncertainty principle}
%quantum mechanics
%
%.
%,
%,
%,
%Born{{{measurement theory}}}
%(\pageref{axiomq}{ page})
%{}
%POIUYTREWQ
Let us begin with
"approximately simultaneous observable".
\par
\noindent
\vskip0.1cm
%BFBF
\par
\noindent
{\bf {Definition }3.3
[Approximately simultaneous observable, error]}$\;\;$%POPOPOPO
For simplicity,
put
$H={\mathbb C}^n$.
And
let
$A$,
$B$
$(\in
B({\mathbb C}^n))$
be Ermite.
Let
$X$
and
$Y$
$( \subset {\mathbb R})$
be finite sets.
The observable
${\mathsf O}_{AB}=$
$(X \times Y, 2^{X \times Y}, F_{AB} )$
in
$B(
{\mathbb C}^n
)$
is called the
approximately simultaneous
observable
of
$A$
 and
$B$,
if it satisfies that
%simultaneous observable  and .
\begin{align*}
\cases
\langle \omega , A \omega \rangle =
\displaystyle \sum\limits_{x \in X } x \langle \omega , F_{AB}(\{x \} \times Y) \omega \rangle
\\
%\\
\langle \omega , B \omega \rangle =
\displaystyle \sum\limits_{y \in Y } y \langle \omega , F_{AB}(X \times \{y\}) \omega \rangle
\endcases
\qquad
(\forall \omega \in {\widehat \Omega} (\subset {\mathbb C}^n))
%P%\TAG{27}
\end{align*}
Further,
the errors
$\Delta_A^\omega $
and
$\Delta_B^\omega $
of
the
simultaneous measurement
${\mathsf M}_{B({\mathbb C}^n)}({\mathsf O}_{AB}, S_{[\omega]})$
is respectively defined by
\begin{align*}
\cases
\Delta_A^\omega
=
\Bigl[
\displaystyle \sum\limits_{x \in X } x^2 \langle \omega , F_{AB}(\{x \} \times Y) \omega \rangle
-
\langle A \omega , A \omega \rangle
\Bigr]^{1/2}
\\
%\\
\Delta_B^\omega
=
\Bigl[
\displaystyle \sum\limits_{y \in Y } y^2 \langle \omega , F_{AB}(X \times \{y \} ) \omega \rangle
-
\langle B \omega , B \omega \rangle
\Bigr]^{1/2}
\endcases
\quad
(\forall \omega \in {\widehat \Omega} (\subset {\mathbb C}^n))
%P%\TAG{28}
\end{align*}
%,
%$\Delta_A^\omega$
% and
%$\Delta_B^\omega$
%,
%simultaneous measurement,
%$\delta_A^\omega$
% and
%$\delta_B^\omega$
%\footnote{
%%,
%$\delta_A^\omega$
%${{\cdot}}$
%,
%$\Delta_A^\omega$
%{{measurement}}
% and .
%}.
\par
\vskip0.3cm
\par
Now,
we can present the
mathematical representation of
Heisenberg's {uncertainty principle}
as follows.
%quantum mechanics
%:
%index{2@Heisenberg's {uncertainty principle}(quantum mechanics)}
%\par
%\noindent
%
%\vskip0.3cm
%BFBF
\par
\noindent
{\bf {{Theorem }}3.4
[Heisenberg's {uncertainty principle}
{\rm
({\rm cf.}
{\textcolor{black}{{\cite{IshiU, Keio}}})
}]}$\;\;$%POPOPO
\rm
Put
$H={\mathbb C}^n$.
Let
$A$,
$B$
$(\in B({\mathbb C}^n))$
be Hermitian.
Then it holds that
\begin{itemize}
\item[({}i)]
A simultaneous measurement
${\mathsf M}_{B({\mathbb C}^n)}({\mathsf O}_{AB}, S_{[\omega]})$
of
$A$
and
$B$
exists.
\item[({}ii)]
Further,
Heisenberg's {uncertainty principle}
holds as follows.
\begin{align*}
\Delta_A^\omega
\cdot
\Delta_B^\omega
{\; \geqq \;}
\frac{1}{2}
|
\langle \omega , (AB-BA) \omega \rangle
|
\qquad
(\forall \omega \in {\widehat \Omega} (\subset {\mathbb C}^n))
%P%\TAG{29}
\end{align*}
\item[({}ii)$'$]
Of course,
the above holds in the case
of
infinite dimensional Hilbert space.
%,
%$H${(}That is, $n=\infty${)}
%.
For example,
when
$Q$
and
$P$
is respectively
the position
observable
and the momentum
observable
(i.e.,
when
$QP-PQ=\hbar {\sqrt{-1}}$
),
%index{@observable }
%index{@observable }
it holds that
\begin{align*}
\Delta_Q^\omega
\cdot
\Delta_P^\omega
{\; \geqq \;}
\frac{1}{2}
\hbar
\qquad
(\forall \omega \in {\widehat \Omega} (\subset H))
\tag{3.12}
\end{align*}
\end{itemize}

\par
%\textcolor{black}{{{Theorem }}3.4}quantum mechanics
%
%Heisenberg's {uncertainty principle}${{\cdot}}$
%,
%\textcolor{black}{
%{\cite{{Keio}}}{(}1991{)}
%}
%.

%
%%%%\omega^\omega^
%BBBBBBBBBBBBBBBBBB%SBSBSBS
\par
\noindent
\vspace{0.1cm}
{\small%%{\footnotesize
\begin{itemize}
\item[$\spadesuit$] \bf {{}}{Note }3.8{{}} \rm
%(i):
As mentioned in
\textcolor{black}{{Note }3.2},
quantum syllogism does not hold.
However,
it should be noted that
\begin{itemize}
\item[]
Heisenberg' {uncertainty principle}
and
the result concerning
{{syllogism}}$(\sharp_1)$--$(\sharp_3)$
in
(\textcolor{black}{{{Theorem }}3.1})
do not conradict.
That is because
Heisenberg' {uncertainty principle}
is related to
an approximate measurement.
\end{itemize}
\end{itemize}
}

%%BBBBBBBBBBBBBBBBBB%SBSBSBSS

%\par
%\noindent
%
\vskip0.5cm

\par
%BFBF
%In this section,
%
%(\textcolor{black}{{Sec.3.4}}), .
%That is,
What we did in this section
is
\begin{itemize}
\item[]
The theory
that is not described by quantum mechanics
quantum mechanics
(
i,e., Heisenberg's {uncertainty principle}\textcolor{black}{({{Proposition }}3.1)}
)
is described in quantum mechanics
such as
%, quantum mechanics( and )
%(,
Heisenberg's {uncertainty principle}\textcolor{black}{({{Theorem }}3.4)}.
\end{itemize}
Replacing
"quantum mechanics"
to
"measurement theory",
we have the spirit
such that
\par
\noindent
{\bf Our standing-point 3.5
{\bf [={Chap.$\;$1}(X$_4$)}]}$\;\;$%POPOPO
Thus, outstanding point is as follows.
\begin{itemize}
\item[$(\sharp_1)$]
\bf
The theory
described in ordinary language
should be described in measurement theory.
%
%,
%{{measurement theory}} and 
%
\end{itemize}
\rm
IN the following chapters,
the readers wii find that
\begin{itemize}
\item[$(\sharp_2)$]
Many ambiguous theories can be automatically sloved
if
they are descibed in measurement theory.
\end{itemize}

\par

%\vskip0.5cm
\par

% and ,
%\textcolor{black}{{Chap.{\;}}2{}},
%{}

%
%%%%\omega^\omega^
%BBBBBBBBBBBBBBBBBB%SBSBSBS
\par
\noindent
\vspace{0.1cm}
{\small%%{\footnotesize
\begin{itemize}
\item[$\spadesuit$] \bf {{}}{Note }3.9{{}} \rm
%(i):
The author proceeded
\begin{itemize}
\item[]
%$\qquad
%\quad$
from quantum mechanics
(cf.
\cite{IshiU}
)
to
classical {{measurement theory}}
(\textcolor{black}{\cite{IFuzz, IQfuz}})
\end{itemize}
Thus,
I was convinced at the time of the beginning as follows.
\begin{itemize}
\item[$(\sharp_)$]
Unless we know quantum mechanics,
we can not understand classical measurement theory.
\end{itemize}
However, it is not true.
This fact was taught by the students of my seminar.
That is because
%they can understand
%classical measurement theory
%without the knowlege of quantum mechanics.
%
%
%\footnote{
without the knowledge
of quantum mechanics,
we can understand und well use
measurement theory.
In fact,
the undergraduate students in my seminar
can understand measuremenmt theory
without the knowledge
of quantum mechanics.
Thus,
the author studied, from my students
the following fact
(= the main thema of this print= linguistic world view):
\begin{itemize}
\item[$(\sharp_2)$]
Even if we do not know
"monkey"
and "tree",
we can use the proverb
"Even monkeys fall from trees".
\end{itemize}
(Continued to \textcolor{black}{{Sec. 8.1.2}})
\end{itemize}
}
%BBBBBBBBBBBBBBBBBB%SBSBSBSS

%%%BBBBBBBBBBBBBBBBBB%SBSBSBSS

%%BFBF
%\par
%\noindent
%
%\par
%
\font\twvtt = cmtt10 scaled \magstep2
\font\fottt = cmtt10 scaled \magstep4
\par
%\noindent
%%BFBF%BFBF\item\item\item\item
%%%%%%%%4.00]:{{}}4.00]:{{}}4.00]:{{}}4.00]:{{}}4.00]:
%%%%%%%4.00]:{{}}4.00]:{{}}4.00]:{{}}4.00]:{{}}4.00]:
%44444444444444444444444444
%1.2.3.4.5.6.7.8.9.10.11......
\par
\noindent
\noindent
\par
\noindent
\vskip2.0cm
%4444444444444444444
\section{Fisher {statistics}\ I
\label{Chap4}
}%{Chap.{\;}}4{}
%\chapter[Fisher {statistics}(\textcolor{black}{Axiom${}_{\text{\scriptsize c}}^{\text{\scriptsize p}}$ 1})]{Fisher {statistics}
%\\
%{(}\textcolor{black}{Axiom${}_{\text{\scriptsize c}}^{\text{\scriptsize p}}$ 1}{)}}
%%\vspace{-0.8cm}
\rm
{\small%%{\footnotesize
\par\noindent
\rm\par
\par
\noindent
\begin{itemize}
\item[{}]
{
\small
\baselineskip=15pt
\par%[Abstract].
\rm
$\;\;\;\;$
{{{Measurement theory}}}
(continuous pure type) 
is formulated as follows.
%.
\begin{align*}
\underset{\text{\scriptsize (scientific language)}}{\text{{} $\fbox{{{measurement theory}}}$}}
:=
{
\overset{\text{\scriptsize
[Axiom${}_{\text{\scriptsize c}}^{\text{\scriptsize p}}$ 1]}}
%[Axiom${}_{\text{\scriptsize c}}^{\text{\scriptsize p}}$ 1\textcolor{black}{(\REF{2secAxiom 1})}]}}
{
\underset{\text{\scriptsize
[probabilistic interpretation]}}{\text{{} $\fbox{{{measurement}}}$}}}
}
+
{
\overset{\text{\scriptsize [Axiom${}_{\text{\scriptsize c}}^{\text{\scriptsize p}}$ 2]}}
%\overset{\text{\scriptsize [Axiom${}_{\text{\scriptsize c}}^{\text{\scriptsize p}}$ 2\textcolor{black}{(\REF{6secAxiom 2})}]}}
{
\underset{\text{\scriptsize [{{the Heisenberg picture}}]}}
{\text{{}$\fbox{ causality }$}}
}
}
\end{align*}
In \textcolor{black}{Chap. 2},
we explained Axiom${}_{\text{\scriptsize c}}^{\text{\scriptsize p}}$ 1.
%.
In this chaper, Fisher {statistics}
is described in terms of
Axiom${}_{\text{\scriptsize c}}^{\text{\scriptsize p}}$ 1.
%Axiom${}_{\text{\scriptsize c}}^{\text{\scriptsize p}}$ 1
%\footnote{
The term
"Fisher statistics"
is used in order to distiguish
Baysian statistics,
which
will be introduced as mixed measurement theory
in
\textcolor{black}{{Sec.4.4}}.
%
%.
%
%, probability ${{\cdot}}$ and ,  and :
%{
%\BEGIN{itemize}
%\item[]
%Newtonian mechanics, classical mechanics,
%probability ${{\cdot}}$, , 
%\footnote{{Chap.$\;$1}(F$_5$)}
%\END{itemize}
%}
%{
%\ssmall
%%\baselineskip=15pt
% and  and ,
%{}. , {}, {statistics},
%{statistics} and .
%}
}
\end{itemize}
}
\baselineskip=18pt
\def\BBbZ{{{\Bbb Z}}}

\par
\noindent
%\vskip1.0cm
\rm
\subsection{Why is {statistics} useful in science?%{Sec.4.1}
}
\subsubsection{Is the foundations of statistics firm?}%{Sec. 4.1.1}
\par
{Statistics}
is quite important decipline.
Statistics is indispesable for
life insurance,
DNA identification of a trial.
the determination of the economic policy of a country.,
etc.
Therefore,
{statistics} has to be
regarded as
"the disipline
with absolute authority".
However,
from the view-pint of
world-description,
statistics is
not firm.
That is because
the following question is not yet answered:
\begin{itemize}
\item[]
What kind of world-view
is statistics due?
%
%{statistics},{\;}{world-view}?
%
%{statistics},{\;}?
%.{}
%%${{\cdot}}${{measurement theory}}Axiom${}_{\text{\scriptsize c}}^{\text{\scriptsize p}}$ 1 and \ 2
%%{{measurement theory}}
%%,
%%{{measurement theory}}
%%
%%Axiom${}_{\text{\scriptsize c}}^{\text{\scriptsize p}}$ 1 and 2
%%
%%
\end{itemize}
That is,{\;}
Our \textcolor{black}{standing-point 3.5{(={Chap.$\;$1}(X$_4$))}}
says that
\begin{itemize}
\item[]
Every engineering
(or, science)
should be described by {{measurement theory}}
(i.e.,
Axiom${}_{\text{\scriptsize c}}^{\text{\scriptsize p}}$ 1 and \ 2).
\end{itemize}
\par
If it be so,
what we have to do
is
\begin{itemize}
\item[]
Statistical methods
---
Fisher maximim likelihood method,
confidence interval,
statistical hypothesis testing,
Bayes' metod,
etc.
---
are descibed in terms of
measurement theory
%{{{measurement theory}}} and
\end{itemize}
This will be done in this chapter.
Also,
this means
to
answer the problem:
\begin{itemize}
\item[]
Why is statistics useful in science?
\end{itemize}
% and  and \footnote{
%{statistics}
% and {statistics}.
%
% and .
%
%,
%\textcolor{black}{{Chap.$\;$1}(X$_1$)}{world-description}
%(\textcircled{\scriptsize 1}:{realistic world-view} and \textcircled{\scriptsize 2}:
%linguistic world-view)
%,
%\BEGIN{itemize}
%\item[$(\sharp)$]
%\textcircled{\scriptsize 1}Einstein and ,
%\textcircled{\scriptsize 2}, Fisher
%\END{itemize}
% and .
%%, , \textcolor{black}{{Chap.$\;$1}(F$_1$)},
%%{statistics},  and \textcircled{\scriptsize 0}{ordinary language},
%% and 
%%\textcircled{\scriptsize 2},  and .
%}.
%%
\subsubsection{Trial and {{measurement}}
\label{5sec{statistics}}
}
\par
The tem
"trial"
is studied in
mathematics
of
the high school.
However,
the following question is not easy.
\begin{itemize}
\item[]
Is the term "trial" a mathematical term?
\end{itemize}
Although it is not easy,
in what follows
we say something.

Let
$(X, {\cal F}, P)$
be a probability space.
%(\textcolor{black}{Appendix B.5(B)}).
Here,
$X$ is called a sample space,
and its element is said to be a sample.
Following common sense,
we define the
"trial"
as follows.
%,
\begin{itemize}
\item[]
the "trial"
is an experiment repeatable repeatedly
such as
"coin-tossing",
"throwing dice"
,
etc.
\end{itemize}
By the trial,
a sample
$x (\in X)$
is obtained.
%trial
When
a sample belongs to
$\Xi (\in {\cal F})$,
an event
$\Xi (\in {\cal F})$
is said to happen.

Now,
we think that
the following three sentences
\textcolor{black}{({}A$_1$)--({}A$_3$)}
asr same:
\begin{itemize}
\item[({}A$_1$)]
The probanility that
an event
$\Xi (\in {\cal F})$
happens is given by
$P(\Xi)$.
%\END{itemize}
%
%\BEGIN{itemize}
\item[({}A$_2$)]
When a trial is taken,
the probability that
a sample belongs to
an event
$\Xi (\in {\cal F})$
is given by
$P(\Xi)$.
\item[({}A$_3$)]
When a trial $(X, {\cal F}, P)$
is taken,
the probability that
a sample belongs to
an event
$\Xi (\in {\cal F})$
is given by
$P(\Xi)$.
\end{itemize}
%\textcolor{black}{({}A$_1$)--({}A$_3$)},
Since a trial is repeatable,
we can get a sample data,
and thus,
a sample probability space
$(X, {\cal F}, P)$.

In the statement
\textcolor{black}{({}A$_3$)},
a trial
$(X, {\cal F}, P)$
and a sample probability space
$(X, {\cal F}, P)$
overlap.
Thus,
the is not usual,
but
we adopt
often
the
\textcolor{black}{({}A$_3$)}
in this print.

%%{a} {b}
%\alpha \beta

%%%%\omega^\omega^
%BBBBBBBBBBBBBBBBBB%SBSBSBS
\par
\noindent
{\small%%{\footnotesize
\vspace{0.1cm}
\begin{itemize}
\item[$\spadesuit$] \bf {{}}{Note }4.1{{}} \rm
\textcolor{black}{({}A$_1$)}
may be mathematical,
on the other hand,
\textcolor{black}{({}A$_3$)}
may be linguistic.
However,
these can not be clarified without
the measurement theoretcal view-ponit.
%\textcolor{black}{({}A$_1$)--({}A$_3$)}
%PART.
%, {{measurement theory}},
% and {}
%%index{@}
\end{itemize}
}
%%BBBBBBBBBBBBBBBBBB%SBSBSBSS
\par
\noindent

\vskip0.3cm

Let $\Omega$
be a set, which is called a parameter space.
 and ,
 and .
For each parameter
$\omega (\in \Omega )$,
define a trial
$(X, {\cal F}, P_\omega)$
and consider a family of trials
$\{ (X, {\cal F}, P_\omega) \}_{\omega \in \Omega }$.
The,
we think that
the following three statements
\textcolor{black}{({}B$_1$)--({}B$_3$)}
are the same:.
\begin{itemize}
\item[({}B$_1$)]
Let $\omega_0 \in \Omega $.
A trial $(X, {\cal F}, P_{\omega_0})$ is taken,
the probability that
a sample belongs to
an event
$\Xi (\in {\cal F})$
is given by
$P_{\omega_0}(\Xi)$.
%
%$\Xi (\in {\cal F})$
%probability , $P_{\omega_0}(\Xi)$
\item[({}B$_2$)]
Let
$\omega_0 \in \Omega $.
When,
for a population
$S_{[\omega_0]}$,
a trial $(X, {\cal F}, P_{\omega_0})$ is taken,
the probability that
a sample belongs to
an event
$\Xi (\in {\cal F})$
is given by
$P_{\omega_0}(\Xi)$.
\item[({}B$_3$)]
When a trial
${\mathsf T}(\{ (X, {\cal F}, P_\omega) \}_{\omega \in \Omega },
S_{[\omega_0]})$
is taken,
the probability that
a sample belongs to
an event
$\Xi (\in {\cal F})$
is given by
$P_{\omega_0}(\Xi)$.
\end{itemize}
Similarly,,
a sample probability space
$(X, {\cal F}, P_{\omega_0})$
is obtained.

In the above
\textcolor{black}{({}B$_1$)--({}B$_3$)},
the term
"trial"
is used in confusion,
however,
we expect readers to read these such as
\begin{align*}
\text{
({}B$_1$)=({}B$_2$)=({}B$_3$)
}
\end{align*}
\par
Although
the statement
({}B$_3$)
may be familiar,
we adopt the \textcolor{black}{({}B$_3$)}.
And we present
Axiom${}_{\text{\scriptsize c}}^{\text{\scriptsize p}}$ 1( trial type)
as follows.

%index{ and @{statistics}}

\vskip0.5cm

%\BEGIN{itembox}[c]
\par
\noindent
\begin{center}
{\bf
\textcolor{black}{
Axiom(T)
1 (trial version)
}
}
\end{center}
\par
\noindent
%\vskip0.1cm
\par
\noindent
\fbox{\parbox{155mm}{
\begin{itemize}
%\item[(i)]
%{{measurement}}
\item[]
{}\rm{}
When a trial
${\mathsf T}(\{ (X, {\cal F}, P_\omega) \}_{\omega \in \Omega },
S_{[\omega_0]})$
is taken,
the probability that
a sample belongs to
an event
$\Xi (\in {\cal F})$
is given by
$P_{\omega_0}(\Xi)$.
\end{itemize}
}
}
\par
\vskip0.5cm
\par
\noindent

%\BEGIN{itembox}[c]{
%\bf
%\textcolor{black}{
%Axiom(T)
%1 (trial version)
%}
%}
%\label{rule401}
%\label{axiom(T)}
%\BEGIN{itemize}
%\item[]
%
%\END{itemize}
%\END{itembox}
%
%\vskip0.4cm
%\def\BC{\left[\begin{array}{ll}}
%\def\EC{\end{array}\right.}
%\rm
\par

This as well as
Axiom${}_{\text{\scriptsize c}}^{\text{\scriptsize p}}$ 1\textcolor{black}{(\pageref{axiomcp1}{ page})}
is linguistic.
Since we prepare
\textcolor{black}{({}A$_1$)--(B$_3$)},
readers are expected to understand
it
as "linguistic"
than
"mathematical".

\par

The following example make readers understand
the delicate diffrence between
"{{measurement}}"
and
"trial".
\par
\noindent
%BFBF
\par
\noindent
{\bf Example 4.1
[\textcolor{black}{Example 2.10}(urn problem):{{measurement}} and trial]}$\;\;$%POPOPO
Again
consider
\textcolor{black}{Example 2.10}(urn problem).
There are two urns ${U}_1$
and
${U}_2$.
The urn ${U}_1$ [resp. ${U}_2$]
contains
8 white and 2 black balls
[resp.
4 white and 6 black balls]
%where
%$N$ is sufficiently large number.
\textcolor{black}{({Fig.$\;$}4.1)}.
\par
\noindent
%%
%\vskip-0.4cm%%%
%\begin{figure}[htbp]
\unitlength=0.20mm
%\unitlength=0.30mm
\begin{picture}(500,130)
\put(80,0){{{
\put(100,0){
\Thicklines
\put(0,5){
\spline(45,100)(45,90)(0,40)(50,0)(75,0)
(100,0)(150,40)(105,90)(105,100)}
\put(200,5){
\spline(45,100)(45,90)(0,40)(50,0)(75,0)
(100,0)(150,40)(105,90)(105,100)}
\thicklines
\put(68,116){$U_1 {\approx} \omega_1$}
\put(268,116){$U_2 {\approx} \omega_2$}
%%\put(373,116){$U_3$}
\put(10,0){
\filltype{white}\put(40,50){\circle*{10}}
\filltype{white}\put(55,50){\circle*{10}}
\filltype{white}\put(70,50){\circle*{10}}
\filltype{white}\put(85,50){\circle*{10}}
\filltype{black}\put(100,50){\circle*{10}}
\filltype{white}\put(40,35){\circle*{10}}
\filltype{white}\put(55,35){\circle*{10}}
\filltype{white}\put(70,35){\circle*{10}}
\filltype{white}\put(85,35){\circle*{10}}
\filltype{black}\put(100,35){\circle*{10}}
}
\put(200,0){
\put(10,0){
\filltype{white}\put(40,50){\circle*{10}}
\filltype{white}\put(55,50){\circle*{10}}
\filltype{black}\put(70,50){\circle*{10}}
\filltype{black}\put(85,50){\circle*{10}}
\filltype{black}\put(100,50){\circle*{10}}
\filltype{white}\put(40,35){\circle*{10}}
\filltype{white}\put(55,35){\circle*{10}}
\filltype{black}\put(70,35){\circle*{10}}
\filltype{black}\put(85,35){\circle*{10}}
\filltype{black}\put(100,35){\circle*{10}}
}
}
%\put(300,0){
%\put(10,0){
%\filltype{white}\put(40,50){\circle*{10}}
%\filltype{black}\put(55,50){\circle*{10}}
%\filltype{black}\put(70,50){\circle*{10}}
%\filltype{black}\put(85,50){\circle*{10}}
%\filltype{black}\put(100,50){\circle*{10}}
%\filltype{black}\put(40,35){\circle*{10}}
%\filltype{black}\put(55,35){\circle*{10}}
%\filltype{black}\put(70,35){\circle*{10}}
%\filltype{black}\put(85,35){\circle*{10}}
%\filltype{black}\put(100,35){\circle*{10}}
%}
%}
%\put(90,9){\footnotesize $U_1$}
%\put(300,9){\footnotesize $\psi_{1,2}(U_1)$}
}
}}}
\end{picture}
\vskip0.1cm
\begin{center}{Figure 4.1:
Urn problem(={\textcolor{black}{Fig.}$\;$}2.5)}
\end{center}
\par
\noindent
\par
\noindent
\par
\noindent

\par
Like
\textcolor{black}{Example 2.10},
consider the following {\lq\lq}statement
\textcolor{black}{(a)}":
\begin{itemize}
\item[(a)]
When one ball is picked up from the urn
$U_2$,
the probability that
the ball is white
is $0.4$.
%
%Pick out one ball at random
%from the
%urn $U_1$,
%and recognize the color (i.e.,
%"white" or {\lq\lq}black")
%of
%the ball
\end{itemize}
%(The term
%"at random" will be often omitted in this book.)
Now,
let us describe the (a) interms of
"measurement"
and
"trial".
\par
\noindent
[I:\ Description by measurement]
$\;$
\par
This was already mentined in \textcolor{black}{Example 2.10}(urn problem).
\par
\noindent
[II:\ Description by trial]
\par
%\noindent
Let $\Omega=\{\omega_1, \omega_2 \}$
be a parameter space.
Consider the following identufication:
%:
\begin{align*}
{\text{Urn }}U_1 {\approx} \text{ parameter }\omega_1, \quad
{\text{Urn }}U_2 {\approx} \text{ parameter }\omega_2 \quad
%U_3 {\approx} \omega_3
\quad
%P%\TAG{19}
\end{align*}
%That is,
%
Define a trial
$\{({} \{ {{w}}, {{b}} \}, 2^{\{ {{w}}, {{b}} \}  }  ,$
$ P_\omega)
\}_{\omega \in \Omega }$
by
%\BEGIN{align*}
\begin{align*}
& P_{\omega_1}({}\{ {{w}} \}{})= 0.8,
& \quad & P_{\omega_1}({}\{ {{b}} \}{})= 0.2\; \\
& P_{\omega_2}({}\{ {{w}} \}{})= 0.4,
& \quad & P_{\omega_2}({}\{ {{b}} \}{})= 0.6
%\; \\
%& P_{\omega_3}({}\{ w \}{})= 0.1,
%& \quad & P_{\omega_3}({}\{ b \}{})= 0.9,
%%%%2.45} P_{\omega_1}P_{}Q_{}QP
\tag{\color{black}{4.1}}
\end{align*}
%%\BEGIN{align*}
%, {\lq $w$\rq} and {\lq $b$\rq}{{w}} and {{b}}
%.
%, :
%\BEGIN{align*}
%& F({}\{ w,b \}{})(\omega_1{})= 1, &\quad & F({}\{ w,b \}{})(\omega_1{})= 1\; \\
%& F({}\emptyset {})(\omega_2{})= 0, &\quad & F({}\emptyset {})(\omega_2{})= 0,
%%& F({}\{ w \}{})(\omega_3{})= 0.1, \qquad F({}\{ b \}{})(\omega_3{})= 0.9,
%%%%%2.45}
%\TAG{3.26}
%\END{align*}
%,
%trial${\mathsf T}($
%$\{({} \{ {{w}}, {{b}} \},$
%$ 2^{\{ {{w}}, {{b}} \}  }  , $
%$P_\omega)
%\}_{\omega \in \Omega },
%$
%$
%S_{[\omega_2]})$
%.
%%%%
%% and ,
%%\BEGIN{align*}
%%M_2^{} =
%%{\mathsf M}_{C (\Omega)} ({}{\mathsf O} ,
%%S_{ [{}{\omega_2}]}{}).
%%%%%%%2.46}
Therefore,
the statement (a)
is,
by Axiom${}_{\text{\scriptsize c}}^{\text{\scriptsize p}}$ 1(trial),
described as follows.
%Axiom${}_{\text{\scriptsize c}}^{\text{\scriptsize p}}$ 1(trial), 
% and :
\begin{itemize}
\item[({}b)]
When a trial
${\mathsf T}($
$\{({} \{ {{w}}, {{b}} \},$
$ 2^{\{ {{w}}, {{b}} \}  }  , $
$P_\omega)
\}_{\omega \in \Omega },
$
$
S_{[\omega_2]})$
is taken,
the probability
that
%{{}}probability  that
a sample
$
\left[\begin{array}{ll}
{{w}}
\\
{{b}}
\end{array}\right]
$
is obtained
%\\
is given by
$
\left[\begin{array}{ll}
P_{\omega_2}(\{{{w}}\})=
0.4
\\
%[F(\{ \text{{{{{c}}}}} \})]
P_{\omega_2}
(
\{{{b}} \})=
0.6
\end{array}\right]
$
\end{itemize}
%\qed

%$
\par
\vskip0.1cm
\par
The smilarity
between
[I:\ Description by {{measurement}}]
and
[II:\ Description by trial]
is due to
the followin theorem
(\textcolor{black}{{{Theorem }}4.2}).

{\bf
\par
\noindent
{{Theorem }}4.2}$\;\;$%POPOPO%POPOPO%
A trial
and a classical measurement
are mathematically
equavalent.
\par
\noindent
{\it $\;\;\;\;${Proof.}}$\;\;$
Consider
a trial
${\mathsf T}(\{(X,{\cal F}, P_\omega )\}_{\omega \in \Omega },
S_{[\omega_0]} )$
and a
{{measurement}}
${\mathsf M}_{C(\Omega)}({\mathsf O} {{=}} (X,{\cal F}, F), S_{[\omega_0]} )$
such that
\begin{align*}
 P_\omega (\Xi)
 =
[ F(\Xi)]
(\omega)
\qquad
(\forall \Xi \in {\cal F},
\omega \in \Omega )
\end{align*}
\renewcommand{\footnoterule}{
  \vspace{2mm}                      % 
  \noindent\rule{\textwidth}{0.4pt}  
  \vspace{-2mm}
}
This completes the proof.
\qed
\par
\vskip0.3cm
\par

%%%%\omega^\omega^
%BBBBBBBBBBBBBBBBBB%SBSBSBS
\par
\noindent
{\small%%{\footnotesize
\begin{itemize}
\item[$\spadesuit$] \bf {{}}{Note }4.2{{}} \rm
\textcolor{black}{{{Theorem }}4.2}
says that
"trial= classical {{measurement}}"
as a mathematical structure.
For example,
\begin{itemize}
\item[$(\sharp )$]
A trial
is repeatabe,
but
only one measurement is permitted.
\end{itemize}
Still,
the $(\sharp)$ is not problem if we introduce
a
parallel measurement(\textcolor{black}{{Sec. 2.5.2}}).
Further,
the spirit
of
Kolmogorov extension theorem
---
regarding many trials as one trial
---
is
caused by
the Copenhagen interpretation
(
only one measurement is permitted.
)
(cf \textcolor{black}{\cite{INewi, ILing}}).
However,
the spirit of {{measurement theory}}
(
linguistic world-view [\textcolor{black}{{Chap.$\;$1}(I)}]
and
the Copenhagen interpretation[\textcolor{black}{({{}}U$_1$)--({{}}U$_7$)}
)
is
omitted
in
a trial.
Thus we consider that
"trial"
is within
{ordinary language}.
Again consider
the (X$_1$)
(in (\textcolor{black}{{Chap.$\;$1}}
and
\textcolor{black}{{Note }2.4}
):
%%%%%%index{@{Chap.$\;$1}(X$_1$)}
\footnotesize
\begin{itemize}
\item[$\underset{(Chap. 1)}{\text{(X$_1$)}}$]
$\overset{
(\text{\scriptsize ordinary language})
}{\underset{\text{({before science})}}{
\text{
\fbox
{{\textcircled{\scriptsize 0}}
widely {ordinary language}}
}
}
}
$
$
\underset{\text{\scriptsize }}{\text{$\Longrightarrow$}}
$
$
\underset{\text{\scriptsize (Chap. 1 (O))}}{\text{{world-description}}}
\cases
&
\!\!\!\!\!\!
{\text{\textcircled{\scriptsize 1}{realistic method}}}
\\
\\
%\textcircled{\scriptsize 2}:
&
\!\!\!\!\!\!
%\underset{\scriptsize
%\text{}}
{\text{\textcircled{\scriptsize 2}{linguistic method}}}
\endcases
$
\end{itemize}
\normalsize
\footnotesize
where the trial is located in
\textcircled{\scriptsize 0}.
Also,
the following is important:
\begin{itemize}
\item[$(\sharp)$]
{{state}}
and
observable
are indespensable in measurement theory,
and thus,
it is connected to
quantum mechanics.
Since the trial is not connected to
quantum mechanics,
the problem
"monism or dualism?"
is neglected in the trial.
\end{itemize}
Therefore,
the overestimation of
\textcolor{black}{{{Theorem }}4.2}
(mathematical equivalence)
must be avoided.
\end{itemize}
}
%%BBBBBBBBBBBBBBBBBB%SBSBSBSS
\par
\noindent

%XXX

%%
%

\par

\par
%{\large
%\bf
%5.2.
\rm
\subsection{Fisher consider Born's reverse}%4.2
\par
As shown in
\textcolor{black}{{{Theorem }}4.2}%%%ABC,
meastement and {statistics}(trial)
are similar.
Therefore,
we can expect that
statistical methods
can be described in terms of
measurement theory.
In what follows,
this will be done.

\subsubsection{{Inference problem}
}
\par

%\textcolor{black}{{{Problem }}4.3}

%Let us explain
%Fisher maximim likelihood method.

%BFBF
\noindent
{\bf Problem 4.3}
\sf
[{}The urn problem by Fisher's maximum likelihood method].
% should be used.
%Coin-tossing and urn problem{}].
\rm
%index{urn problem@urn problem}
$\;$
There are two urns $U_1$ and $U_2$.
The urn $U_1$ [resp. $U_2$] contains $8$ white and $2$ black
balls
[resp. $4$ white and $6$ black balls].

\par
\noindent
%%Assume that the two urns can not be distinguished in appearance.
Here consider the following procedures
(i) and (2).
\begin{itemize}
\item[\textcolor{black}{(i)}]
One of the two (i.e., $U_1$ or $U_2$)
is chosen and is settled behind a curtain.
% by an unfair tossed-coin ($C_{p,1-p}$),
%%The chosen urn is denoted by $[*{}](\in \{U_1, U_2\})$.
Note,
for completeness, that you do not know whether it is
$U_1$ or $U_2$.
%since the two can not be distinguished in appearance.
\rm
\item[\textcolor{black}{(ii)}]
Pick up a ball out of
the urn chosen by the procedure (A$_1$).
And you find that the ball is white.
%\item[\textcolor{black}{(A}$_3$)]
%And, without returning the ball to the urn,
%pick out one ball from the urn.
\end{itemize}
\rm
\noindent
Here, we have the following problem:
\begin{itemize}
\item[\textcolor{black}{(iii)}]
Infer the probability
that
the ball obtained in the above (A$_3$)
is white
( or,
black)
?
\end{itemize}
%
%Now you sample, randomly, with
%replacement after each ball.
%%%\omega_w\omega_w%%\omega_b\omega_b\omega_w$ [

\noindent
\unitlength=0.27mm
%%%%%%%%%%%%%%%%%
%\begin{figure*}[htbp]
%%%%%%%%%%%%%%%%%ZZZZZZZZZZZzZZZZzZzZzZzZ
%\caption{
%Which is the hidden urn?
%}P$_
%\END{figure*}
%%%%%%%%%%%%%%%%%%
\begin{picture}(590,210)
%\p
%${\mathsf M}_{C(\Omega)}({\mathsf O}, S_{[{}\ast{}] }{})$}
\path(0,-20)(0,220)(590,220)(590,-20)(0,-20)
\Thicklines
\put(170,60){}
\put(155,50){\vector(1,0){30}}
\put(390,60){}
\put(410,50){\vector(-1,0){30}}
\put(200,5){
\path(-10,-10)(-10,120)(160,120)(160,-10)(-10,-10)
\allinethickness{0.1mm}
\multiput(-10,-10)(0,4){33}{\line(1,0){170}}
\Thicklines
\spline(70,70)(80,120)(130,140)
\path(120,143)(130,140)(120,130)
\filltype{white}\put(140,140){\circle*{10}}
\put(56,40){\Huge \bold [{}$\ast${}]}
\put(-180,190){\footnotesize
You do not know which the urn behind the curtain is,
$U_1$ or $U_2$.}
\put(-180,170){\footnotesize
Assume that you pick up a white ball from the urn.
%Don't return the ball to the urn.
}
%}
\put(-180,150){\footnotesize
The urn is $U_1$ or $U_2$?  $\quad$ Which do you think? }
%\put(-180,150){\footnotesize
%Further, pick out one ball from the urn.
%Infer the probability
%that
%the new picked ball
%is black.
%}
}
%
%
%
%
%
%\put(-180,190){\footnotesize
%You do not know which the urn behind the curtain is,
%$U_1$ or $U_2$, but the probability: $p$ and $1-p$.}
%\put(-180,170){\footnotesize
%Assume that you pick up a white ball from the urn.}
%%}
%\put(-180,150){\footnotesize
%The urn is $U_1$ or $U_2$?  $\quad$ Which do you think? }
%}
\put(0,5){
\spline(45,100)(45,90)(0,40)(50,0)(75,0)
(100,0)(150,40)(105,90)(105,100)}
\put(200,5){
\spline(45,100)(45,90)(0,40)(50,0)(75,0)
(100,0)(150,40)(105,90)(105,100)}
\put(400,5){
\spline(45,100)(45,90)(0,40)(50,0)(75,0)
(100,0)(150,40)(105,90)(105,100)}
\thicklines
\put(73,116){$U_1$}
\put(473,116){$U_2$}
%%\put(373,116){$U_3$}
\put(10,0){
\filltype{white}\put(40,50){\circle*{10}}
\filltype{white}\put(55,50){\circle*{10}}
\filltype{white}\put(70,50){\circle*{10}}
\filltype{white}\put(85,50){\circle*{10}}
\filltype{black}\put(100,50){\circle*{10}}
\filltype{white}\put(40,35){\circle*{10}}
\filltype{white}\put(55,35){\circle*{10}}
\filltype{white}\put(70,35){\circle*{10}}
\filltype{white}\put(85,35){\circle*{10}}
\filltype{black}\put(100,35){\circle*{10}}
}
\put(400,0){
\put(10,0){
\filltype{white}\put(40,50){\circle*{10}}
\filltype{white}\put(55,50){\circle*{10}}
\filltype{black}\put(70,50){\circle*{10}}
\filltype{black}\put(85,50){\circle*{10}}
\filltype{black}\put(100,50){\circle*{10}}
\filltype{white}\put(40,35){\circle*{10}}
\filltype{white}\put(55,35){\circle*{10}}
\filltype{black}\put(70,35){\circle*{10}}
\filltype{black}\put(85,35){\circle*{10}}
\filltype{black}\put(100,35){\circle*{10}}
}
}
\end{picture}
%%%%%%%%%%%%%%%%%
%\BEGIN{FIGUre*}[htbp]
%%%%%%%%%%%%%%%%%
\vskip0.3cm
\begin{center}{Figure 4.2:
Which is the hidden urn,
$U_1$ or $U_2$?
%(= FIG. 4.2)
}
\end{center}
%%%%%%%%%%%%%%%%%%

\par
%\vskip0.8cm
The answer is easy,
that is,
thr urn behind the curtain is
$U_1$.
That is because
the urn $U_1$
has more white balls than
$U_2$.
It is too easy,
but it includes the essence of
Fisher maximim likelihood method.

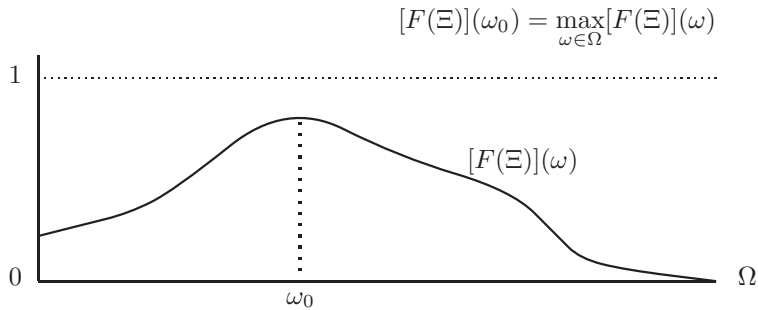
\subsubsection{Fisher maximim likelihood method
in {{{measurement theory}}}{}}%{Sec.4.2.2}
\par

\par
\noindent
\par
We begin with the following definition.
\par
\noindent
\bf %BFBF
{Notation 4.4}
\sf
[{}${\mathsf M}_{\cal A} ({\mathsf O}, S_{[*]})${}].
\rm
Consider a measurement
${\mathsf M}_{\cal A} ({}{\mathsf O} \equiv (X, {\cal F}, F),$
$ S_{[\rho^p]}{})$
formulated in a $C^*$-algebra ${\cal A}$.
In most measurements, it is usual to think that the state
$\rho^p \;(\in {\frak S}^p ({\cal A}^*))$ is unknown.
That is because the measurement
${\mathsf M}_{\cal A} ({\mathsf O}, S_{[\rho^p]})$
may be taken in order to know the state $\rho^p$.
Thus, when we want to stress that we do not know the state
$\rho^p$,
the measurement
${\mathsf M}_{\cal A} ({\mathsf O}, S_{[\rho^p]})$ is often denoted by
${\mathsf M}_{\cal A} ({\mathsf O}, S_{[*]})$.
%%index{ma2@
%index{measure3@${\mathsf M}_{\cal A} ({\mathsf O}, S_{[*]})$}
%${\mathsf M}_{\cal A} \big({}{\mathsf O}, S_{[{}\ast] } \big)$}
\hfill{{$///$}}%%BFBFbfbf
%\END{Not}
\par
\rm

%
%{basic algebra}
%$C(\Omega )$
%
%{{measurement}}
%${\mathsf M}_{C (\Omega)} $
%\newline
%$({}{\mathsf O} {{=}} (X, {\cal F}, F),$
%$ S_{[\omega]}{})$
%. 
%\BEGIN{itemize}
%\item[]
%{{measurement}},
%{{state}}
%$\omega \;(\in \Omega)$ and  and {}
%\END{itemize}
%{}
%\BEGIN{itemize}
%\item[]
%{{state}}$\omega${{measurement}}
%${\mathsf M}_{C (\Omega)} ({\mathsf O}, S_{[\omega]})$
%
%
%\END{itemize}
%{}
%, {{}}observer, {measuring object}{{state}}
%$\omega$ and  and 
%,
%%{{}}{{measurement}}
%${\mathsf M}_{C (\Omega)} ({\mathsf O}, S_{[\omega]})$
%
%${\mathsf M}_{C (\Omega)} ({\mathsf O}, S_{[*]})$ and .
\par
%\vskip0.5cm
\par
\rm
Using this notation,
we characterize our problem
(i.e.,
inference) as follows.
%${\mathsf M}_{C (\Omega)} ({\mathsf O}, S_{[*]})$,
%{}, :
	% as follows:
\begin{itemize}
\item[(a)]
%Infer{{}} {{state}}
%
%from{{}}{{measurement}} obtained by{{}}
Assume that a measured value obtained by
a
{{measurement}}
${\mathsf M}_{C (\Omega)}({\mathsf O} {{=}} (X, {\cal F}, F), S_{[*]})$
belongs to
$\Xi (\in {\cal F})$.
Then, infer the unknown {{state}}$[*] \;(\in \Omega)$
\end{itemize}

%Infer{{}} {{state}}
%
\par
Therefore,
the {{measurement}}
is "the view from the front",
that is,
%${\mathsf M}_{C(\Omega)}({\mathsf O}, S_{[\omega]})$
\begin{itemize}
\item[(b)]
$
\qquad
\qquad
(\text{observable} [{\mathsf O}], {\text{{state}}}[\omega(\in \Omega)])
\xrightarrow[{\mathsf M}_{C(\Omega)}({\mathsf O},
S_{[\omega]})]{
\quad {\text{\scriptsize measurement}} \quad}
\text{measured value} {[x (\in X)]}
%P%\tag{5.5}
$
\end{itemize}
On the other hand,
the {{inference}}
is "the view from the back",
that is,
\begin{itemize}
\item[(c)]
$
\qquad
\qquad
(\text{observable} [{\mathsf O}], \text{measured value} [x \in \Xi ( \in {\cal F})])
\xrightarrow[{\mathsf M}_{C(\Omega)}({\mathsf O},
S_{[\ast]})]{
\quad \text{\scriptsize inference}\quad}
%
%
%\xrightarrow[[{\mathsf M}_{C(\Omega}({\mathsf O},
%S_{[\ast]});\{x \} ] ]]{\quad inference \quad}
{{state}}{[\omega (\in \Omega)]}
%P%\tag{5.6}
$
\end{itemize}
In this sence,
the {{{inference problem}}
is the reverse problem of measurement.
Therefore,
it suffices to image
\textcolor{black}{{Fig.$\;$}4.3}.
%, {{inference problem}}{FIG.$\;$}{}
\rm
\par
\noindent
%%\vskip-0.4cm%%%
%\begin{figure}[htbp]
\par
\noindent
\unitlength=0.27mm
\begin{picture}(560,140)
\put(20,0){
\put(30,150){
%\Large
$
%\underbrace{\overset{{}}{\underset{\text{\footnotesize ({)}}}
%\;\;\;
%\underset{\|}{\text{ \fbox{{measuring object}}}}
%}}_{{(}{measuring object}{)}}
$
}
\put(30,100){
%\Large
$
%\underbrace{\overset{{}}{\underset{\text{\footnotesize (input{)}}}
\overset{\text{({measuring object})}}{\text{ \fbox{unknown {{state}}}}}
%}_{{(}{measuring object}{)}}
\xrightarrow[]{\qquad \qquad}
\underbrace{
%\underbrace{
\underset{\text{\footnotesize ({measuring instrument}{)}}}
{\text{ \fbox{observable }}}
%}_{}
\xrightarrow[\text{probabilistic}]{\qquad \qquad}
%\underbrace{
\underset{\text{\footnotesize (output{)}}}
{\text{ \fbox{measured value }}}
%}_{}
}_{{(}\text{observer}{)}}
%}}^{{{measurement}}}
$
}
\dashline{4}(160,0)(160,140)
\path(363,40)(363,10)(80,10)%(80,80)
\put(80,10){\vector(0,1){80}}
\put(200,15){inference}
%}
}
\end{picture}
\begin{center}{Figure 4.3:
The image of inference}
\end{center}
\par
\noindent
\baselineskip=18pt
\par
In order to answer {{the above problem }}\textcolor{black}{(a)},
we shall describe
Fisher maximim likelihood method
in terms of
{{{measurement theory}}}.

%%thich

\par
\noindent
{\bf
%\vskip0.3cm
%\vskip0.3cm
%BFBF
\par
\noindent
{{Theorem }}4.5
%index{@Fisher maximim likelihood method({{measurement theory}})}
[{}Fisher maximim likelihood method{(}{measurement theoretical representation}{)}
\rm
\textcolor{black}{
(
{\rm cf.\;\;}\cite{IStat, IWhat})}{\bf]}}$\;\;$%POPOPO
%Fisher maximim likelihood method.}
\rm
\rm
Consider a
{{measurement}}
${\mathsf M}_{C (\Omega)}({\mathsf O} $
${{=}} (X, {\cal F}, F),$
$ S_{[*]})$.
%When we know that{{}}

Assume that we know that
a {measured value }{{}}
obtained
by a {{measurement}}
${\mathsf M}_{C (\Omega)}({\mathsf O}, S_{[*]})$
belongs to
$\Xi \;(\in {\cal F})$.
Then,
there is a reason to infer that
the unknown state {{state}} $[*]$
is
$\omega_0 \;(\in \Omega)$
such that
% and inference and :
\begin{align*}
[F(\Xi)](\omega_0)
= \max_{\omega \in \Omega} [F(\Xi)](\omega)
%P%\tag{5.7}
\end{align*}
%That is,
%$[F(\Xi)](\omega ) {{\; \leqq \;}}[F(\Xi)](\omega_0)$
%$(\forall \omega \in \Omega)$
\par
\noindent
%%%
%%\vskip-0.4cm%%%
%\begin{figure}[htbp]
\unitlength=0.30mm
%\unitlength=0.35mm
\begin{picture}(400,110)
\put(27,18){0}
\put(27,108){1}
\put(350,18){$\Omega$}
\dottedline{3}(40,110)(340,110)
\put(150,10){$\omega_0$}
\put(40,20){\line(0,1){100}}
%\linethickness{0.15mm}
\thicklines
\put(40,20){\line(1,0){300}}
%\linethickness{0.15mm}
\thicklines
%\sspline(40,110)(60,108)(80,102)(100,80)
%(150,40)(200,30)(220,20)(240,20)
%\sspline(120,20)(130,20)(160,30)(250,50)
%(270,80)(280,100)(300,105)(340,110)
\spline(40,40)(60,45)(80,50)(100,60)
(150,100)(200,75)(250,60)
(270,40)(280,30)(300,25)(340,20)
\dottedline{5}(156,20)(156,90)
\put(230,70){$[{}F(\Xi)](\omega)$}
\end{picture}
\vskip-0.3cm
\begin{center}{Figure 4.4:
Fisher maximim likelihood method}
\end{center}
%\it
\par
\noindent
{\it $\;\;\;\;${Proof.}}$\;\;$
\rm
$\;$
Let
$\omega_1$
and $\omega_2$
be elements in $\Omega$
such that
% and ,
%
%$[{}\ast{}] = \omega_1 \; $
%
%$[{}\ast{}] = \omega_2 \;$
% and .
%
$[F(\Xi)](\omega_1) < [F(\Xi)](\omega_2)$.
Thus,
by
\textcolor{black}{Axiom${}_{\text{\scriptsize c}}^{\text{\scriptsize p}}$ 1}({{measurement}}),
\begin{itemize}
\item[(i)]
the probability that
a measured value obtained by
a
{{measurement}}${\mathsf M}_{C (\Omega)} ({\mathsf O}, S_{[\omega_1]})$
belongs to
$\Xi$
is equal to
$[F(\Xi)](\omega_1) $
\item[(ii)]
the probability that
a measured value obtained by
a
{{measurement}}${\mathsf M}_{C (\Omega)} ({\mathsf O}, S_{[\omega_2]})$
belongs to
$\Xi$
is equal to
$[F(\Xi)](\omega_2) $
\end{itemize}
Since we assume that
$[F(\Xi)](\omega_1) < [F(\Xi)](\omega_2)$,
we can conclude that
"(i) is more rare than (ii)".
Thus,
there is a reason to infer that
$[*]=\omega_2$.
\qed

\par
\noindent
%
%%BBBBBBBBBBBBBBBBBB%SBSBSBS

%%%BBBBBBBBBBBBBBBBBB%SBSBSBSS

\par
\noindent
{\small%%{\footnotesize
\vspace{0.1cm}
\begin{itemize}
\item[$\spadesuit$] \bf {{}}{Note }4.3{{}} \rm
%{statistics}
Fisher maximim likelihood method
in statistics
is easily obtained
if
a
{{measurement}}
${\mathsf M}_{C(\Omega)}({\mathsf O} {{=}}$
$ (X,{\cal F}, F), S_{[\omega_0]} )$
is replaced by
a
trial
${\mathsf T}(\{(X,{\cal F}, P_\omega )\}_{\omega \in \Omega },
S_{[\omega_0]} )$
in
\textcolor{black}{{{Theorem }}4.5}.
\end{itemize}
}
%%BBBBBBBBBBBBBBBBBB%SBSBSBSS

\par
\noindent
%index{@urn problem}

\par
\noindent
{\bf
Answer 4.6
[\textcolor{black}{{{The measurement theoretical answer to Problem }}4.3}{{{}}}]}$\;\;$%POPOPO
%$\;\;$
%\noi
Consider a
{{measurement}}
${\mathsf M}_{C (\Omega)} ({}{\mathsf O} {{=}}
({} \{ {{w}},$
$ {{b}} \}, 2^{\{ {{w}}, {{b}} \}  }  , F{})
,
S_{ [{}{\ast}]}{})$
in
\textcolor{black}{Example 2.10}.
The formula
\textcolor{black}{(2.5)}%%%%ABCNNN
says that
\begin{align*}
\max \{[F(\{{{w}}\})](\omega_1),
[F(\{{{w}}\})](\omega_2)
\}
=
\max \{0.8, 0.4\}
=
0.8
=
F(\{{{w}}\})](\omega_1)
%P%\tag{5.8}
\end{align*}
Therefore,
\textcolor{black}{{{Theorem }}4.5}
says that
%,
%{{state}}$\omega_1$inference,
%,
the urn behind the curtain
is
$U_1$.
\qed
%

%%{a} {b}
%\alpha \beta

%%%%\omega^\omega^
%BBBBBBBBBBBBBBBBBB%SBSBSBS
\par
\noindent
{\small%%{\footnotesize
\begin{itemize}
\item[$\spadesuit$] \bf {{}}{Note }4.4{{}} \rm
As seen in
\textcolor{black}{{Fig.$\;$}4.3},
inference (Fisher maximim likelihood method)
is the reverse of measurement.
%
%%,
%,
%Fisher,
%\BEGIN{itemize}
%\item[]
%Fisher,
%{{measurement}},
%%That is,
%{{measurement}},
%inference
%\END{itemize}
% and .
%,
% and , :
%%.
Here note that
\begin{itemize}
\item[]
Born's discovery
"the probabilitstic interpretation of quantum mechanics"
%probabilistic interpretation
(\pageref{axiomq}{ page})
:
\textcolor{black}{\cite{Born}}
(1926)
\\
Fisher's
great book
{\it
"Statistical Methods for Research Workers"
}{(}1925{)}
\end{itemize}
Thus,
it is surprising that
Fisher and Born
considered the same thing
in the different fields
in the same age.
\end{itemize}
}
%%BBBBBBBBBBBBBBBBBB%SBSBSBSS
%\par
%\noindent

\subsection{{{{Statistical methods in measurement theory}}}}%4.3
\subsubsection{Examples of Fisher maximim likelihood method}

%\END{document}

%index{@urn problem}
{\bf
\par
\noindent
Example 4.7
[{}Urn problem]}$\;\;$%POPOPO
Each urn $U_1$, $U_2$, $U_3$
contains
white balls and black ball such as:
% \textcolor{black}{Table} \textcolor{black}{\REF{5020}}.
\par
\noindent
%\vskip-0.4cm%%%
%\begin{table}[htbp] \small \caption{urn problem \label{5020}}
\begin{center}
Table 4.1:
urn problem
\\
%\BEGIN{tabular}{|l|l|*{2}{@{\quad\$}r|}}
\begin{tabular}{
@{\vrule width 0.8pt\ }c
@{\vrule width 0.8pt\ }c|c|c
@{\vrule width 0.8pt }}
\noalign{\hrule height 0.8pt}
{{w}}${{\cdot}}${{b}}$\diagdown$ {{Urn}} &$\quad$  {{Urn}} $U_1$ $\quad$ &$\quad$ {{Urn}} $U_2$ $\quad$
&$\quad$ {{Urn}} $U_3$ $\quad$ \\
\noalign{\hrule height 0.8pt}
white ball & 80\% & 40\%& 10\% \\
\hline
black ball & 20\% & 60\% & 90\%  \\
\noalign{\hrule height 0.8pt}
%{{Urn}} $U_3$  &  20\% & 20\% & 40\%& 20\%  \\
%\hline
\end{tabular}
\end{center}
%\end{table}
\par
\noindent
%\vskip-1.0cm
Here,
\begin{itemize}
\item[({}i)]
one of three urns is chosen,
but you do not knot it.
Pich up one ball from the unknown urn.
And you find that
its ball is white.
Then,
How do you infer the unknow urm,
i.e.,
$U_1$, $U_2$ or $U_3$?
\end{itemize}
Further,
\begin{itemize}
\item[({}ii)]
And further,
you pich up
another ball
from the unknown urn.
And you find that
its ball is black.
That i,
after all,
you have
one white ball and
one
one black ball.
Then,
How do you infer the unknow urm,
i.e.,
$U_1$, $U_2$ or $U_3$?
\end{itemize}
\par
\par
%%\End{document}
In what follows,
we shall answer the above problems
(i) and (ii)
in terms of measurement theory.
Put
$$
\omega_j
\longleftrightarrow
[
\text{the state such that urn $U_j$ is chosen}
]
\quad
(j=1,2,3)
$$
Thus,
we have the state space
%,
$\Omega$
$($
${{=}} \{ \omega_1 , \omega_2 , \omega_3 \}$
$)$
Further,
define the observable
${\mathsf O} = ({} \{ {{w}}, {{b}} \}, 2^{\{ {{w}}, {{b}} \}  }  , F{})$
in
$C({}\Omega{})$
such that
\begin{align*}
& F({}\{ {{w}} \}{})(\omega_1{})= 0.8,   & \;\;\;&
& F({}\{ {{w}} \}{})(\omega_2{})= 0.4,   & \;\;\;&
& F({}\{ {{w}} \}{})(\omega_3{})= 0.1   \\
& F({}\{ {{b}} \}{})(\omega_1{})= 0.2,   & \;\;\;&
& F({}\{ {{b}} \}{})(\omega_2{})= 0.6,   & \;\;\;&
& F({}\{ {{b}} \}{})(\omega_3{})= 0.9
%%%%2.45}
%P%\tag{5.10}
\end{align*}
%\BEGIN{align*}
%, {\lq ${{w}}$\rq} and {\lq $b$\rq}{{w}} and {{b}}
%.
%, :
%\BEGIN{align*}
%& F({}\{ {{w}},b \}{})(\omega_1{})= 1, &\quad & F({}\{ {{w}},b \}{})(\omega_1{})= 1\; \\
%& F({}\emptyset {})(\omega_2{})= 0, &\quad & F({}\emptyset {})(\omega_2{})= 0,
%%& F({}\{ w \}{})(\omega_3{})= 0.1, \qquad F({}\{ b \}{})(\omega_3{})= 0.9,
%%%%%2.45}
%\TAG{3.26}
%%\END{align*}
%,
%{{measurement}}
%${\mathsf M}_{C (\Omega)} ({}{\mathsf O} ,
%S_{ [{}{\ast}]}{})$
%.

%%%%

\par
\noindent
{\bf{Answer to ({}i)}:$\;\;$}
Consider the {{measurement}}
${\mathsf M}_{C (\Omega)}  ({}{\mathsf O}, S_{[{}\ast{}]})$,
by which
a measured value 
{\LL}{{w}}{\RR}
is obtained.
Therefore, we see
\begin{align*}
[F({ \{ {{w}} \} })] ({}\omega_1{}) =
0.8
=
\max_{ \omega \in \Omega }
[F({ \{ {{w}} \} })] (\omega)
=
\max
\{ 0.8, \; 0.4, \; 0.1 \}
%%%%%5.18}
%\Tag{5.12}
\end{align*}
\par
\noindent
Thus,
by
Fisher maximim likelihood method(\textcolor{black}{{{Theorem }}4.5}),
we see that
%({}That is,   5.6{})
\begin{align*}
[\ast] = \omega_1
%%%%%%%%%5.18}
%\Tag{5.11}
\end{align*}
Thus, we can infer that the unknown urn is
$U_1$.
%,
\par
\noindent
{\bf{Answer to ({}ii)}:$\;\;$}
$\;$
Next,
consider the
simultaneous measurement
${\mathsf M}_{C (\Omega)}  ({}\bigtimes_{k=1}^2 {\mathsf O} $
$ {{=}} $
$ ({}X^2 ,$
$ 2^{{}X^2} ,$
$ {\widehat F} {{=}} \bigtimes_{k=1}^2 F{}) ,$
$ S_{[{}\ast]})$,
by which
%.
%%,
%%$({}\bigtimes_{k=1}^2 F{})_{\Xi_1 \bigtimes \Xi_2 }
%%(\omega)$
%%$=$
%%$ F_{\Xi_1 }(\omega) \cdot F_{\Xi_2 }(\omega) $.
%%\BEGIN{itemize}
%%\ITEM[(c)]
%
%{{}}simultaneous measurement
%${\mathsf M}_{C (\Omega)}  ({}\bigtimes_{k=1}^2 {\mathsf O}, S_{[{}\ast]})$
%
%{{}}
a
measured value
$({} {{w}}, {{b}}{})$
is obtained.
Here, we see
\begin{align*}
[{\widehat F}(\{({{w}},{{b}})\})](\omega)=[F({\{{{w}}\}})]({}\omega)
\cdot
[F({\{{{b}}\}})]({}\omega)
\end{align*}
thus,
\begin{align*}
&[{\widehat F}(\{({{w}},{{b}})\})]({}\omega_1{})=0.16,
\;\;
[{\widehat F}(\{({{w}},{{b}})\})]({}\omega_2{})= 0.24,
\;\;
[{\widehat F}(\{({{w}},{{b}})\})]({}\omega_3{})= 0.09
%P%\tag{5.12}
\end{align*}
Thus,
by
Fisher maximim likelihood method(\textcolor{black}{{{Theorem }}4.5}),
we see that
%({}That is,   5.6{})
\begin{align*}
[\ast] = \omega_2
%%%%%%%%%5.18}
%\Tag{5.11}
\end{align*}
Thus, we can infer that the unknown urn is
$U_2$.
\qed
%Cf.{{}}next  5.
\par
\noindent
\par
\noindent

%%%%%%%%BBBBBBBBBBBB{{b}}bbbbbbbwwwwwwwwwwww
%

%
\par
\noindent
{\bf
%\vskip0.3cm
%\vskip0.3cm
%BFBF
\par
\noindent
Example 4.8
[{{Normal observable}}(i)]}$\;\;$%POPOPO
%index{@{{normal observable}}}
As mentioned in \textcolor{black}{Example 2.11},
consider
the {{normal observable}}
${\mathsf O}_{G_\sigma} $
${{=}}$
$({}{\mathbb R} , {\cal B}_{{\mathbb R}}^{} , {G_\sigma}
{})$
in
$C ({}{\mathbb R}{})$
(where
$\Omega={\mathbb R}$)
such that
\par
\noindent
\begin{align*}
&[{G_\sigma}(\Xi)] ({} {\mu} {}) =
\frac{1}{{\sqrt{2 \pi }  \sigma}}
\int_{\Xi} \exp[{}- \frac{1}{2 \sigma^2 }   ({}{x} - {\mu}  {})^2
] d{x}
\\
&
%\hspace{5cm}
\qquad
({}\forall  \Xi \in {\cal B}_{{\mathbb R}}^{},
\quad
\forall   {\mu}    \in \Omega= {\mathbb R}{})
%%%%%5.23}
\end{align*}
\par
\noindent
Thus,
%,
the
simultaneous observable
$\bigtimes_{k=1}^3 {\mathsf O}_{G_\sigma}$
({}in short,
${\mathsf O}_{G_\sigma}^3 $)
${{=}}$
$({}{\mathbb R}^3 , {\cal B}_{{\mathbb R}^3}^{} ,$
$ G_{\sigma}^3{})$
in $C ({}{\mathbb R}{})$
is defined by
\par
\noindent
\begin{align*}
&
[G_{\sigma}^3({\Xi_1 \times \Xi_2 \times \Xi_3 })] ({} {\mu} {})
=
[{G_\sigma}({\Xi_1})] ({}\mu{}) \cdot
[{G_\sigma}({\Xi_2})] ({}\mu{}) \cdot
[{G_\sigma}({\Xi_3})]({}\mu{})
\\
=
&
\frac{1}{({}{\sqrt{2 \pi }  \sigma)^3}}
\iiint_{\Xi_1 \times \Xi_2 \times \Xi_3}
\exp[{}- \frac{({}{x_1} - {\mu}  {})^2 +({}{x_2} - {\mu}  {})^2 +
({}{x_3} - {\mu}  {})^2     }{2 \sigma^2 }
] \\
&\hspace{5cm} \times d{x_1}  d{x_2} d{x_3}
\\
&
\qquad \qquad \qquad \qquad
({}\forall  \Xi_k \in {\cal B}_{{\mathbb R}}^{},k=1,2,3,
\quad
\forall   {\mu}    \in \Omega= {\mathbb R}{})
%%%%%5.24}
%P%\tag{5.14}
\end{align*}
Thus, we get the {{measurement}}
${\mathsf M}_{C({\mathbb R})} ({}{\mathsf O}_{G_\sigma}^3,
S_{[{}\ast{}] }{})$

Now we consider the following problem:
\begin{itemize}
\item[(a)]
By the {{measurement}}
${\mathsf M}_{C({\mathbb R})} ({}{\mathsf O}_{G_\sigma}^3,
S_{[{}\ast{}] }{})$
assume that
a {{}}measured value
$({}x^0_1, x^0_2 , x^0_3{})$
$(\in {\mathbb R}^3{})$
is obtained.
Then,
infer the unknown state
$[\ast]
(\in {\mathbb R})$.
\end{itemize}
\par
\noindent
{\bf{Answer(a)}$\;\;$}
%, $\Xi_i$
%
%
%{{}}measured value
%$({}x^0_1, x^0_2 , x^0_3{})$
%$({}\in {\mathbb R}^3{})$
%.
Put
\begin{align*}
\Xi_i = [ x_i^0 -\frac{1}{N},x_i^0 +\frac{1}{N}]
\qquad(i=1,2,3)
%P%\tag{5.15}
\end{align*}
Assume that
$N$
is sufficiently large.
Fisher maximim likelihood method(\textcolor{black}{{{Theorem }}4.5})
says that
the
unknown {{state}}$[{}\ast{}]$
$= \mu_0$
is found in what follows.
\begin{align*}
%&
[G_{\sigma}^3({\Xi_1 \times \Xi_2 \times \Xi_3 })] ({} {\mu_0} {})
%\\
=
%&
\max_{\mu \in {\mathbb R}}
[G_{\sigma}^3({\Xi_1 \times \Xi_2 \times \Xi_3 })] ({} {\mu} {})
\end{align*}
Since $N$ is sufficiently large,
we see
%.
\begin{align*}
&
\frac{1}{({}{\sqrt{2 \pi }  \sigma)^3}}
\exp[{}- \frac{({}{x^0_1} - {\mu_0}  {})^2 +({}{x^0_2} - {\mu_0}  {})^2 +
({}{x^0_3} - {\mu_0}  {})^2     }{2 \sigma^2 }
]
\\
=
&
\max_{\mu \in {\mathbb R}}
\Big[
\frac{1}{({}{\sqrt{2 \pi }  \sigma)^3}}
\exp[{}- \frac{({}{x^0_1} - {\mu}  {})^2 +({}{x^0_2} - {\mu}  {})^2 +
({}{x^0_3} - {\mu}  {})^2     }{2 \sigma^2 }
]
\Big]
%%%%%5.25}
%P%\tag{5.16}
\end{align*}
That is,
\begin{align*}
({}{x^0_1} - {\mu_0}  {})^2 +({}{x^0_2} - {\mu_0}  {})^2 +
({}{x^0_3} - {\mu_0}  {})^2
=
\min_{\mu \in {\mathbb R}}
\big\{
({}{x^0_1} - {\mu}  {})^2 +({}{x^0_2} - {\mu}  {})^2 +
({}{x^0_3} - {\mu}  {})^2
\big\}
%%%%%5.26}
%P%\tag{5.17}
\end{align*}
%$\mu_0$.
Therefore,
solving
$\frac{d}{d\mu} \{ \cdots \}=0$,
we conclude that
\begin{align*}
\mu_0
=
\frac{{x^0_1}+{x^0_2}+{x^0_3}}{3}
%%%%%5.27}
%P%\tag{5.18}
\end{align*}

\qed

\par
\noindent
{\bf
%\vskip0.3cm
%\vskip0.3cm
%BFBF
\par
\noindent
[{{Normal observable}}(ii)]}$\;\;$%POPOPO
Next, consider the case:
\begin{itemize}
\item[]
we know that
the length of the pencil $\mu$
is satisfied that
10cm $\mu$ $L$ cm $\le $30.
\end{itemize}
And we assume that
\begin{itemize}
\item[$(\sharp)$]
the length of the pencil $\mu$
and
the roughness $\sigma$ of
the ruler
are unknown.
%
%{measuring object}{{state}},
%$\mu$
% and
%$\sigma$
% and \footnote{$(\sharp)$
%
%%That is,
%%{measuring object}
%
% and ,
%{realistic world-view}
% and 
%%,
%(\textcolor{black}{{Note }6.14}
%($\sharp_2$)
%, 
%).
%observable dualism, ,
%{{state}} and ,
% and , {statistics}
%, ,
%, {statistics} and .
%}.
\end{itemize}
That is,
%${\mathbb R}_+$
%$=$
% and .
assume that
the state space
$\Omega$
$=$
$[10,30{}] \times {\mathbb R}_+ $
$
\big(
{{=}}
\{ \mu \in {\mathbb R} \; |\; 10 {{\; \leqq \;}}\mu {{\; \leqq \;}}30 \}
\times
\{ \sigma \in {\mathbb R}
\;|\;
\sigma > 0
\}
\big)
$

%{{normal observable}}
%${\mathsf O}_{G_\sigma} $
%${{=}}$
%$({}{\mathbb R} , {\cal B}_{{\mathbb R}}^{} , {G_\sigma}{})$
% and ,
Define the
observable ${\mathsf O}$
${{=}}$
$({}{\mathbb R} , {\cal B}_{{\mathbb R}}^{} , G)$
in
$C ({}[10,30 {}]\times {\mathbb R}_+{})$
such that
\par
\noindent
\begin{align*}
[G(\Xi)](\mu, \sigma )
=
[{G_\sigma}({\Xi})] (\mu)
\quad
({}\forall  \Xi \in {\cal B}_{{\mathbb R}}^{},
\;\;
\forall   ({\mu}, \sigma)
\in
\Omega=
[10,30] \times {\mathbb R}_+
)
%%%%%5.28}
%P%\tag{5.19}
\end{align*}
%\par
%\noindent
Therefore,
the
simultaneous observable
${\mathsf O}^3 $
${{=}}$
$({}{\mathbb R}^3 , {\cal B}_{{\mathbb R}^3}^{} ,$
$ G^3)$
in
$C ({}[10,30{}]\times {\mathbb R}_+{})$
is defined by
\par
\noindent
\begin{align*}
&
[G^3({\Xi_1 \times \Xi_2 \times \Xi_3 })] ({} {\mu},\sigma {})
=
[G(\Xi_1)](\mu, \sigma )
\cdot
[G(\Xi_2)](\mu, \sigma )
\cdot
[G(\Xi_3)](\mu, \sigma )
\\
=
&
\frac{1}{({}{\sqrt{2 \pi }  \sigma)^3}}
\int_{\Xi_1 \times \Xi_2 \times \Xi_3}
\exp[{}- \frac{({}{x_1} - {\mu}  {})^2 +({}{x_2} - {\mu}  {})^2 +
({}{x_3} - {\mu}  {})^2     }{2 \sigma^2 }
] d{x_1}  d{x_2} d{x_3}
\\
&
\qquad \qquad \qquad
({}\forall  \Xi_k \in {\cal B}_{{\mathbb R}}^{},k=1,2,3,
\quad
\forall   ({\mu},\sigma)    \in \Omega=
[10,30{}]\times {\mathbb R}_+{})
%%%%%5.29}
%\TAG{5.17}
\end{align*}
Thus,
we get the simultaneous measurement
${\mathsf M}_{C([10,30]\times {\mathbb R}_+)} ({}{\mathsf O}^3, S_{[{}\ast{}] }{})$.
Here,
we have the following problem:
\begin{itemize}
\item[(b)]
When
a {{}}measured value
$({}x^0_1, x^0_2 , x^0_3{})$
$({}\in {\mathbb R}^3{})$
is obtained by
the
{{measurement}}
${\mathsf M}_{C([10,30]\times {\mathbb R}_+)} ({}{\mathsf O}^3, S_{[{}\ast{}] }{})$,
infer the unknown state
$[\ast]
(
= (\mu_0,\sigma_0)
\in [10,30]\times {\mathbb R}_+)$,
i.e.,
the lenght $\mu_0$ of the pencile
and
the roughness $\sigma_0$
of the ruler.
\end{itemize}
\par
\noindent
{\bf{Answer (b)}$\;\;$}
By the same way of (a),
Fisher maximim likelihood method(\textcolor{black}{{{Theorem }}4.5})
says that
{{}}the unknown{{state}}
$[{}\ast{}]$
$= (\mu_0,\sigma_0)$
such that
%{{inference problem}} and  and :
% and ,
%, where
%$[{}\ast{}]$
%$= \mu_0$,  where
\begin{align*}
&
\frac{1}{({}{\sqrt{2 \pi }  \sigma_0)^3}}
\exp[{}- \frac{({}{x^0_1} - {\mu_0}  {})^2 +({}{x^0_2} - {\mu_0}  {})^2 +
({}{x^0_3} - {\mu_0}  {})^2     }{2 \sigma_0^2 }
]
\\
=
&
\max_{(\mu,\sigma) \in [10, 30] \times {\mathbb R}_+}
\Big\{
\frac{1}{({}{\sqrt{2 \pi }  \sigma)^3}}
\exp[{}- \frac{({}{x^0_1} - {\mu}  {})^2 +({}{x^0_2} - {\mu}  {})^2 +
({}{x^0_3} - {\mu}  {})^2     }{2 \sigma^2 }
]
\Big\}
%%%%%5.30}
%P%\tag{5.20}
\end{align*}
Thus,
solving
$\frac{\partial }{\partial \mu}\{\cdots \}=0$,
$\frac{\partial }{\partial \sigma}\{\cdots \}=0$
we see
\begin{align*}
&
\mu_0
=
\cases
10
\quad
\qquad
&
(\text{when }
\text{}
(x^0_1+x^0_2+x^0_3)/3< 10\; )
\\
\\
({x^0_1}+{x^0_2}+{x^0_3})/3
\quad
\qquad&
(\text{when }
\text{}
10 {{\; \leqq \;}}
(x^0_1+x^0_2+x^0_3)/3{{\; \leqq \;}}30 \; \text{})
\\
\\
30
\quad&
(\text{when }
\text{}
30
<
(x^0_1+x^0_2+x^0_3)/3 \; \text{})
\endcases
\\
&
\sigma_0
=
\sqrt{
\{
(x^0_1-{\widetilde \mu})^2
+
(x^0_2-{\widetilde \mu})^2
+
(x^0_3-{\widetilde \mu})^2
\}/3
}
\end{align*}
where
\begin{equation*}
{\widetilde \mu}=({x^0_1}+{x^0_2}+{x^0_3} )/3
\end{equation*}
\qed
%%%%

%
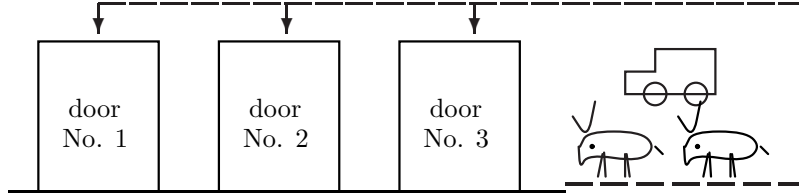
\subsubsection{{{Monty Hall problem}}
---
High school students' puzzle
}
\par
\par
The Monty Hall problem
is well-known and elementary .
Also it
is famous as the problem
in which even great mathematician P. Erd\"os
made a mistake
({\rm cf}.
\textcolor{black}{\cite{Hoff}}).
The Monty Hall problem
is as follows:

%
%{{Monty Hall problem}}, ${{\cdot}}$
%
%Let's make a dealprobability {}
%%1990, "Parade magazine" ${{\cdot}}$${{\cdot}}$
%%Ask Marilyn and  and , %,
%% and  and .
%
%(:,{\;}, 1998)
%
%%
%%%\bibitem{Hoff} Hoffman, P.
%%% \newblock {\em The Man Who Loved Only Numbers,
%%% The story of Paul Erd\"os and the search for mathematical truth,}
%%% \newblock {Hyperion, New York}
%%% \newblock {1998}
%%%\newblock{
%%%
%%% 
%%
%%\textcolor{black}{\cite{Hoff}})
%
% and ,
%, 
% and {}
%%,  and .
%,
%{{Monty Hall problem}} and .
%\par
%\vskip0.2cm
%\par
%{{}}{{Monty Hall problem}} and {}
%%\noindent
%%\noindent
{\bf
%\vskip0.1cm
%BFBF
\par
\noindent
{{Problem }}4.9{{}}
[{}{{Monty Hall problem}}
\rm
({\rm cf.}
\textcolor{black}{\cite{Keio,IMont}}){\bf]}}$\;\;$%POPOPO
\rm
%index{@{{Monty Hall problem}}}
{
\rm
You are on a game show and you are given the choice of three doors.
Behind one door is a car, and behind the other two are goats.
You choose, say, door 1, and the host, who knows where the car is,
opens another door,
behind which is a goat.
For example,
the host says that
\begin{itemize}
\rm
\item[($\flat$)]
the door 3 has a goat.
\end{itemize}
And further,
He now gives you the choice of
sticking with door 1 or switching to door 2?
What should you do?
}

%
%%%index{paradox({}{{Monty Hall problem}}{})
%%@paradox({}{{Monty Hall problem}}{})}
%\rm
%\par
%\noindent
%{
%%\footnotesize
%\BEGIN{itemize}
%\item[]
%.
%3door
%(That is, 1,
%2,
%3
%)
%
%1door{}, 2door
%{(}).
%, door.
%,
%.
%door and ?
%%\BEGIN{enumerate}
%\par\noindent
%, door and .
% and ,
%1door and .
% and ,
%
%, 3door
% and .
%, .
%1door,
%.
%2door
%?
% and .
%, ?
%%
%\END{itemize}
%}
%\par
%\noindent
%\vskip0.3cm
\par
\noindent
%%%%eepic
%%\vskip-0.4cm%%%
%\begin{figure}[htbp]
\unitlength=0.20mm
%\unitlength=0.27mm
\begin{picture}(500,140)
\put(60,0){{{
\thicklines
\put(430,55)
{{
\drawline[-15](-40,-30)(120,-30)(120,90)(-350,90)
\put(-350,90){\vector(0,-1){20}}
\put(-225,90){\vector(0,-1){20}}
\put(-100,90){\vector(0,-1){20}}
\path(0,30)(60,30)(60,60)(20,60)(20,45)(0,45)(0,30)
\put(20,30){\circle{15}}
\put(47,30){\circle{15}}
%\sspline(-5,50)(5,35)(10,60)
%\path(15,25)(12,10)
%\path(16,26)(17,10)
%\path(30,25)(30,10)
%\path(31,25)(33,10)
%\put(8,30){\circle*{2}}
%\path(50,30)(55,25)
}}
\put(400,20)
{{
\spline(0,30)(5,40)(40,40)(50,30)(40,20)(5,25)(0,15)(-1,30)
\spline(-5,50)(5,35)(10,60)
\path(15,25)(12,10)
\path(16,26)(17,10)
\path(30,25)(30,10)
\path(31,25)(33,10)
\put(8,30){\circle*{2}}
\path(50,30)(55,25)
}}
\put(470,20)
{{
\spline(0,30)(5,40)(40,40)(50,30)(40,20)(5,25)(0,15)(-1,30)
\spline(-5,50)(5,35)(10,60)
\path(15,25)(12,10)
\path(16,26)(17,10)
\path(30,25)(30,10)
\path(31,25)(33,10)
\put(8,30){\circle*{2}}
\path(50,30)(55,25)
}}
\thicklines
\put(20,20){\line(1,0){370}}
\put(40,20){
\path(0,0)(0,100)(80,100)(80,0)
\put(20,50){door}
}
\put(160,20){
\path(0,0)(0,100)(80,100)(80,0)
\put(20,50){door}
}
\put(280,20){
\path(0,0)(0,100)(80,100)(80,0)
\put(20,50){door}
}
\put(56,50){{No.} 1}
\put(176,50){{No.} 2}
\put(296,50){{No.} 3}
}}}
\end{picture}
\vskip-0.2cm
\begin{center}{Figure 4.5:
{{Monty Hall problem}}
}
\end{center}
\par
\noindent
{\bf{Answer}:$\;\;$}
Put
$\Omega = \{ \omega_1 , \omega_2 , \omega_3 \}$
with the discrete topology.
Assume that
each state
$\delta_{\omega_m}
(\in
{\frak S}^p (C(\Omega)^* ))$
means
\begin{align*}
\delta_{\omega_m}
\Leftrightarrow
\text{
the state that the car is }
\text{behind the door 1}
\quad
(m=1,2,3)
%\\
%\delta_{\omega_2}
%&
%\Leftrightarrow
%\text{
%the state that the car is}
%\text{behind the door 2}
%\\
%\delta_{\omega_3}
%&
%\Leftrightarrow
%\text{
%the state that the car is}
%\text{behind the door 3}
%%
%%
%%\\
%%\delta_{\omega_2}
%%\cdots \cdots
%%&
%%\text{
%%the state that the car is}
%\\
%&
%\text{
%behind the door number 2}
%\\
%\delta_{\omega_3}
%\cdots \cdots
%&
%\text{
%the state that the car is}
%\\
%&
%\text{behind the door number m}.
%\tag{3} 
\end{align*}
Define the observable
${\mathsf O}_1$
$\equiv$
$({}\{ 1, 2,3 \}, 2^{\{1, 2 ,3\}}, F_1)$
in $C({}\Omega{})$
such that
%\BEGIN{align*}
\begin{align*}
& [F_1({}\{ 1 \}{})](\omega_1{})= 0.0,\qquad
[F_1({}\{ 2 \}{})](\omega_1{})= 0.5,
\qquad
[F_1({}\{ 3 \}{})](\omega_1{})= 0.5,
%\footnotemark
\\
&
[F_1({}\{ 1 \}{})](\omega_2{})= 0.0,
\qquad
[F_1({}\{ 2 \}{})](\omega_2{})= 0.0, \qquad
[F_1({}\{ 3 \}{})](\omega_2{})= 1.0,
\\
& [F_1({}\{ 1 \}{})](\omega_3{})= 0.0,\qquad
[F_1({}\{ 2 \}{})](\omega_3{})= 1.0, \qquad
[F_1({}\{ 3 \}{})](\omega_3{})= 0.0,
%%
%\tag{4.2}
\tag{\color{black}{4.2}}
\end{align*}
%\footnotetext{
where
it is also possible to assume that
$F_1({}\{ 2 \}{})(\omega_1{})=\alpha$,
$F_1({}\{ 3 \}{})(\omega_1{}) =1- \alpha$
$
(0 < \alpha < 1)$.
%\BEGIN{align*}F_1FFFFFFFFFFFFF
The fact that
yuo say
"the door 1"
means
that
we have a measurement
${\mathsf M}_{C({}\Omega{})} ({}{\mathsf O}_1, S_{[{}\ast{}]})$.
%which should be regarded as
%the measurement theoretical representation of
%the measurement
%that
%\it
%you say "door 1".
\rm
Here, we assume that
\begin{itemize}
\item[a)]
{\lq\lq}a measured value $1$ is obtained by
the measurement
${\mathsf M}_{C({}\Omega{})} ({}{\mathsf O}_1, S_{[{}\ast{}]})${\rq\rq}
\\
$
\Leftrightarrow \text{The host says {\lq\lq}Door 1
% (number 1)
has a goat{\rq\rq}}$
\item[b)]
{\lq\lq}measured value $2$ is obtained
by
the measurement
${\mathsf M}_{C({}\Omega{})} ({}{\mathsf O}_1, S_{[{}\ast{}]})$
{\rq\rq}
\\
$
\Leftrightarrow \text{The host says {\lq\lq}Door 2
% (number 1)
has a goat{\rq\rq}}$
\item[c)]
{\lq\lq}measured value $3$ is obtained
by
the measurement
${\mathsf M}_{C({}\Omega{})} ({}{\mathsf O}_1, S_{[{}\ast{}]})$
{\rq\rq}
\\
$
\Leftrightarrow \text{The host says {\lq\lq}Door 3
% (number 1)
has a goat{\rq\rq}}$
%\\
\end{itemize}
\par
%$p_1 = p_2 = p_3 = 1/3$,
%then

%\BEGIN{align*}
\par
\noindent
Recall that,
in Problem 1, the host said
{\lq\lq}Door 3 has a goat{\rq\rq}$\!\!\!.\;$
This implies that
you get the measured value {\lq\lq}3{\rq\rq}
by the measurement
${\mathsf M}_{C({}\Omega{})} ({}{\mathsf O}_1, S_{[\ast]}{})$.
Therefore,
Theorem 1
(Fisher's maximum likelihood method)
says that
\it
you should pick
door number 2.
\rm
That is because
we see that
%{%\footnotesize
\begin{align*}
%&
[F_1({}\{3\}{})] ({}\omega_2{})
=
1.0
=
\max \{ 0.5, \; \; 1.0 , \; \; 0.0 \}
=
\max \{
[F_1({}\{3\}{})] ({}\omega_1{}),
[F_1({}\{3 \}{}){}]({}\omega_2{}),
[F_1({}\{3 \}{})] ({}\omega_3{})
\},
%\tag{5} 
\end{align*}
\par
\noindent
and thus,
there is a reason to infer that
$[\ast]$
$=$
$\delta_{\omega_2}$.
Thus,
you should
switch to door 2.
This is the first answer to
Problem 1
(Monty-Hall problem).

%2door{}
\qed

\renewcommand{\footnoterule}{%
  \vspace{2mm}                      % 
  \noindent\rule{\textwidth}{0.4pt}   % , 
  \vspace{-5mm}
}

\par
\noindent
{\small%%{\footnotesize
\vspace{0.1cm}
\begin{itemize}
\item[$\spadesuit$] \bf {{}}{Note }4.5{{}} \rm
The above answer is one of Answers of
Monty Hall problem.
Of course,
the answer
based on Bayes' theorem
%
%\footnote{
%[{\bf ({statistics})}]:
%%%index{@}
%$(X, {\cal F}, P)$probability  and .
%$C, D$ and ,
%{(}That is, $C, D \in {\Cal F}$ and ),
%$C$ and  and 
%$D$probability 
%$P_C (D)=\frac{P(C\cap D)}{P(C)}$
% and . ,
%$B$ and ,
%
%$A_1, A_2,\ldots ,A_n$.
%, $P(A_1)+P(A_2)+...+P(A_n)=1$
% and .  and , .
%\BEGIN{align*}
%P_B (A_k)=
%\frac{P(A_k)P_{A_k}(B)}{
%P(A_1)P_{A_1}(B)+
%P(A_2)P_{A_2}(B)+
%...
%+
%P(A_n)P_{A_n}(B)
%}
%\END{align*}
%,
%(\textcolor{black}{{Chap.$\;$1}(X$_1$)})
%{ordinary language}\textcircled{\scriptsize 0} and ,
%{statistics}{}
%{{measurement theory}}, {Remark }4.15.
%}
---
\textcolor{black}{{{Problem }}4.16}
---
is usual.
However, it is temporary.
Our final answer is presented in
{{Problem }}6.20 [{{Monty Hall problem}}]
in
\textcolor{black}{Sec. 6.4.5}.
\end{itemize}
}
%%BBBBBBBBBBBBBBBBBequilibrium statistical mechanicsPOIUYTREWQWERTY
\par
\noindent

%
%
%%BBBBBBBBBBBBBBBBBB%SBSBSBS
%\par
%\noindent
%{\footnotesize
%\BEGIN{itemize}
%\
%{{Monty Hall problem}}Answer,
%\footnote{
%[{\bf }]:
%%index{@}
%$(X, {\cal F}, P)$probability  and .
%$C, D$ and ,
%{(}That is, $C, D \in {\Cal F}$ and ),
%$C$ and  and 
%$D$probability 
%$P_C (D)=\frac{P(C\cap D)}{P(C)}$
% and . ,
%$B$ and ,
%
%$A_1, A_2,\ldots ,A_n$.
%, $P(A_1)+P(A_2)+...+P(A_n)=1$
% and .  and , .
%\BEGIN{align*}
%P_B (A_k)=
%\frac{P(A_k)P_{A_k}(B)}{
%P(A_1)P_{A_1}(B)+
%P(A_2)P_{A_2}(B)+
%...
%+
%P(A_n)P_{A_n}(B)
%}
%\END{align*}
%}Answer and .
%%index{@}
%\ITEM[] \bf  \rm %%%BBBBBBBBBBBBBBBBBBBB
%\\
%%.
%[{{{measurement theory}}}]
%{{Monty Hall problem}}
%Answer
%%
%\textcolor{black}{4.16}.
%\END{itemize}
%}
%%%BBBBBBBBBBBBBBBBBequilibrium statistical mechanics
%\par
%\noindent
%
%
%
%

\par
\noindent
%\bf
%{Remark }5.13.
%\rm
%%\par
%%\noindent
%%  2.27{})].
%%index{{{Monty Hall problem}}@{{Monty Hall problem}}}
%%index{paradox({}{{Monty Hall problem}}{})@paradox({}{{Monty Hall problem}}{})}
%, 
%4.12
%by{{}}.
%Since you get measured value  3, you get{{}}probability 
%$(\{1,2,3\}, 2^{\{1,2,3\}}, \nu_{s}{})$
%{}
%$\nu_s ({}\{1 \}{}) =0 $,
%$\nu_s ({}\{2 \}{}) =0 $
%{{ and }}
%$\nu_s ({}\{3 \}{}) =1 $.
%,
%{{}} $ \Delta$
%{}
%for  $\nu_1 , \nu_2 \in {\cal M}_{+1}({}\{ 1, 2,3\}{}){}$,
%\BEGIN{align*}
%{\Delta}   ({}\nu_1 , \nu_2{}) = | \nu_1 ({}\{1\}{}) - \nu_2 ({}\{1\}{}) |
%+
%| \nu_1 ({}\{2\}{}) - \nu_2 ({}\{ 2 \}{}) |
%+
%| \nu_1 ({}\{3\}{}) - \nu_2 ({}\{ 3 \}{}) |.
%%%%%%%%%%%%
%\
%Then, we see
%\BEGIN{align*}
%&
%{\Delta}   ({}\nu_s , [F({}\cdot{}){}]({}\omega_1{}){})=
%|0-0|+
%| 0 - 0.5| + | 1- 0.5| = 1,
%\\
%&
%{\Delta}   ({}\nu_s , [F({}\cdot{}){}]({}\omega_2{}){})
%=
%|0-0|+
%| 0 - 0| + | 1-1| = 0
%
%&
%{\Delta}   ({}\nu_s , [F({}\cdot{}){}]({}\omega_3{}){})
%=
%|0-0|+
%| 0 - 1| + | 1-0| = 2.
%%%%%\
%\
%,
%we can, by{{}},inferencethat
%$\omega_2$ is {{}}possible,
%That is,
%{{}}car is behind{{}}door {No.} 2.
%\par
%\qed
%
%

\par
%\vskip1.0cm
\par
\vskip2.0cm
\par
%%535353535
%\par
%\noindent
%{\bf \Large 5.3.
%
\rm
\subsubsection{Confidence interval}
\par
%%%%%%PPPPPPPPPPPPP

\par
\noindent
\par
%5353535353535353
Let
${\mathsf O} ({}\equiv ({}X, {\cal F} , F{}){})$
be an observable
%
%Consider a measurement
%${\mathsf M}_{C(\Omega)} \big({}{\mathsf O}:= ({}X, {\cal F} , F{})  ,$
%$ S_{[{}\ast{}] } \big)$
formulated in a
basic algebra
${C(\Omega)}$.
Assume that
$X$ has a metric $d_X$.
And assume
that
the state space
$\Omega$.
has the metric $d_\Omega$.
Let
$E:X \to \Omega$
be a continuous map,
which is called
\it
{\lq\lq} estimator{\rq\rq}$\!\!\!\!.\; \;$
\rm
Let
$\gamma$
be a real number such that
$0 \ll \gamma < 1$,
for example,
$\gamma = 0.95$.
For any
$ \omega({}\in\Omega)$,
define
the positive number
$\eta^\gamma_{\omega}$
$({}> 0)$
such that:
\begin{align*}
\eta^\gamma_{\omega}
=
\inf
\{
\eta > 0:
[ F ({}E^{-1} ({}
B(\omega ; \eta{})) ]
(\omega)
\ge \gamma
\}
%\tag{5.33}
\end{align*}
where
$B(\omega ; \eta)$
$=$
$\{ \omega_1
({}\in \Omega):
d_\Omega ({}\omega_1, \omega{}) \le \eta \}$.
For any
$x$
$({}\in X{})$,
put
\begin{align*}
D_x^{\gamma}
=
\{
{\omega}
(\in
\Omega)
:
d_\Omega ({}E(x),
\omega)
\le
\eta^\gamma_{\omega }
\}.
%\Big(%=
%B({}E(x) ; \eta^\gamma_{\omega })
%\Big)
%\tag{8}
\tag{\color{black}{4.3}}
\end{align*}

%%i.e.,
%$\rho_0^p = [{}\ast{}] $.
\par
\noindent
\unitlength=0.4mm
%%%%%%%%%%%%%%%%%
%\begin{figure*}[htbp]
%%%%%%%%%%%%%%%%%
%\vskip0.3cm
%\caption{
%Which is the hidden urn,
%$U_1$ or $U_2$?
%}
%\END{figure*}
%%%%%%%%%%%%%%%%%%
\begin{picture}(100,75)
%%\put(25,12){A}
\put(70,0){{
\put(40,16){\scriptsize $x_0$}
\qbezier(40,20)(100,61)(157,42)
\path(107,49)(115,48)(107,45)
\put(40,20){\circle*{1}}
\put(157,41){\circle*{1}}
\put(155,45){\scriptsize $E({}x_0)$}
%\put(157,40){\scriptsize $r$}
%\put(76,42){\scriptsize $\beta$}
\put(151,33){$ \; \omega_0$}
\put(149,34){ \circle*{1} }
\put(170,30){$ D_{x_0}^\gamma$}
\put(153,63){ $\Omega$}
\put(57,63){$X$}
%\allinethickness{0.2mm}
%\put(60,30){\line(1,0){30}}
%\put(160,30){\line(1,0){30}}
%\put(60,30){\line(5,3){26}}
%\put(160,30){\line(1,4){5}}
%\allinethickness{0.5mm}
%\put(42,55){\line(4,-1){44}}
%\put(142,55){\line(4,-1){44}}
\allinethickness{0.5mm}
\put(60,30){\oval(70,60)}
\put(160,30){\oval(70,60)}
%%\put(160,30){\circle{60}}
\allinethickness{0.3mm}
%\filltype{white}
\put(157,42){\ellipse{30}{30}}
}}
%\put(40,-20){\bf FIGU 3.
%\rm
%Inference interval
%$D^\gamma_{x_0}$
%}
\end{picture}
%%%%%%%%%%%%%%%%%
\begin{center}{Figure 4.6:
Inference interval
$D^\gamma_{x_0}$
}
\end{center}
%\BEGIN{FIGUre*}[htbp]
%%%%%%%%%%%%%%%%%
%\vskip0.3cm

%and further,COLOR
The $D_x^{\gamma}$ is called
\it
the $({}\gamma{})$-inference interval
of the
measured value
$x$.
%(%by
%measurement
%${\mathsf M}_{C(\Omega)} \big({}{\mathsf O}:= ({}X, {\cal F} , F{})  ,$
%$ S_{[{}\ast{}] } \big)$).
\rm
\par
The following is clear:
\begin{enumerate}
\item[]
\rm
for any
$\omega_0
({}\in
\Omega)$,
the probability,
that
the measured value $x$
obtained
by the measurement
${\mathsf M}_{C(\Omega)} \big({}{\mathsf O}:= ({}X, {\cal F} , F{})  ,$
$ S_{[\omega_0 {}] } \big)$
satisfies the following
condition $(\flat)$,
is larger than
$\gamma$
({}e.g., $\gamma= 0.95${}).
\rm
\begin{enumerate}
\item[$\text{(a)}$]
%$\qquad$
$ E(x) \in B({}\omega_0 ; {\eta }_{\omega_0}^\gamma{}) $
%$\qquad$
or equivalently,
%$\qquad$
$\quad  d(E(x),  \omega_0{}) \le  {\eta }^\gamma_{\omega_0}  $.
\end{enumerate}
\end{enumerate}
\par
\noindent
Assume that
%\BEGIN{enumerate}
%\item[(I)]
we get
a measured value
$x_0$
by
the measurement
${\mathsf M}_{C(\Omega)} \big({}{\mathsf O}:= ({}X, {\cal F} , F{})  ,$
$ S_{[\omega_0 {}] } \big)$.
%\END{enumerate}
%\par
%\noindent
Then,
%by the above $(\flat)$,
%we can assure,
%with the probability as more than $0.95$,
%that
we see the following equivalences:
\begin{itemize}
\item[(b)]
$
\qquad
\text{(a)} \; \Longleftrightarrow \;
d_\Omega ({}E({}x_0{}), \omega_0{}) \le \eta^\gamma_{\omega_0 }
\;
\Longleftrightarrow
\;
D_{x_0}^\gamma
\ni
\omega_0.
$
\end{itemize}
%where
%$\rho_0^p$
%is the unknown state,
%i.e.,
%$\rho_0^p = [{}\ast{}] $.

\par
\vskip1.0cm
\par

\par
Summing the above argument,
we have the following theorem.
\par
%%
%%%$ \omega({}\in\Omega)$,
%define
%the positive number
%$$
%$({}> 0)$
%%\noindent\omega_0
%{\bf
%BFBF
\par
\noindent
\bf
{{Theorem }}4.10
[{}Inference interval{}
{\rm
\textcolor{black}{
(
{\rm cf.}
\cite{IMeas, IWhat}
)}}]}$\;\;$%POPOPO
\rm
%\noindent
%\bf %BFBF
%Theorem 4.10
%\rm
%[{}Inference interval{}].
%%index{Inference interval@inference interval}
{}\rm{}
Let
${\mathsf O}:= ({}X, {\cal F} , F{}) $
be an observable in
${C(\Omega)}$.
Let
$\omega_0$
be any fixed state,
i.e.,
$\omega_0 \in
\Omega
$,
Consider
a measurement
${\mathsf M}_{C(\Omega)} \big({}{\mathsf O}:= ({}X, {\cal F} , F{})  ,$
$ S_{[\omega_0 {}] } \big)$.
Let
$E:X \to \Omega$
be an
estimator.
Let
$\gamma$ be such as $0 \ll \gamma < 1$
({}e.g., $\gamma = 0.95${}).
For any
$x
({}\in X{})$,
define
$D_x^{\gamma}$
as in (8).
Then, we see,
\begin{enumerate}
\rm
\item[(c)]
\it
the probability
that
the measured value
$x_0
({}\in X)$
obtained
by the measurement
${\mathsf M}_{C(\Omega)} \big({}{\mathsf O}:= ({}X, {\cal F} , F{})  ,$
$ S_{[\omega_0 {}] } \big)$
satisfies
the condition
that
\begin{align*}
\text{
$D_{x_0}^{\gamma} \ni \omega_0 $
},
%\tag{5.35}
\end{align*}
is larger than
$\gamma$.
\end{enumerate}
%\qed
\par
%\hfill$\blacksquare$

\vskip1.0cm

\rm
\par
\noindent
{\bf
%\vskip0.1cm
%BFBF
\par
\noindent
Example 4.11
[{}{{}}Urn problem{}]}$\;\;$%POPOPO
%
%
%%%index{urn problem@urn problem}
%\bf %BFBF
%Example 4.11
%\rm
%\rm
%[{}The urn problem{}].
Put
$\Omega$
$=$
$[{}0, 1{}]$,
i.e.,
the closed interval in ${\mathbb R}$.
We assume that
each
$\omega$
$({}\in \Omega \equiv [{}0, 1{}]{})$
represents an urn that contains a lot of
black balls
and white balls
such that:
\begin{align*}
&
\frac{
\text{ the number of white balls in the urn $\omega$}
}
{
\text{ the total number of balls in the urn $\omega$}
}
\\
\approx
&
\;\;
\omega
\quad
({}\forall \omega \in [0,1] \equiv \Omega{}).
%\tag{5.36}
\end{align*}
Define the observable
${\mathsf O} = ({} X \equiv \{ b, w \}, {\cal P}({\{ b, w \}  })  , F{})$
in
$C({}\Omega{})$
such that
%\BEGIN{align*}
\begin{align*}
&
F({}\emptyset{})(\omega)= 0, \quad
F({}\{ b \}{})(\omega)   = \omega, \quad
F({}\{ w \}{})(\omega{})
= 1- \omega ,\quad
F({}\{ b, w \}{})(\omega)= 1
\\
& \qquad
\qquad \qquad
({}\forall \omega \in [{}0, 1{}] \equiv \Omega{}).
%%\tag{5.37}
\end{align*}
%\BEGIN{align*}
\par
\noindent
Here,
consider the following measurement
$M_\omega$:
\begin{align*}
M_\omega
&
:=
\text
{
{\lq\lq} Pick out one ball from the
urn $\omega$,}
\text{
and recognize the color of the ball{\rq\rq}}
%%\tag{5.38}
\end{align*}
That is, we consider
%The measurement
%$M$
%is formulated as follows:
\begin{align*}
M_\omega =
{\mathsf M}_{C ({}\Omega{}) } ({}{\mathsf O} ,
S_{ [{}\delta_{\omega}]}{}).
%%\tag{5.39}
\end{align*}
Moreover we define the product observable
${\mathsf O}^N$
$\equiv$
$({}X^N , {\cal P}({}X^N{}) , F^N{})$,
such that:
\begin{align*}
&
[
F^N ({}\Xi_1 \bigtimes \Xi_2 \bigtimes \cdots \bigtimes \Xi_{N-1} \bigtimes \Xi_{N}{})
]
({}\omega{})
=
[F({}\Xi_1{})]({}\omega{}) \cdot
[F({}\Xi_2{})]({}\omega{}) \cdots
%[F({}\Xi_{N-1}{})]({}\omega{}) \cdot
[F({}\Xi_N{})]({}\omega{})
\\
&
\qquad
\qquad
(\forall
\omega
\in \Omega \equiv [0,1],
\quad
\forall
\Xi_1, \Xi_2, \cdots , \Xi_N \subseteq X \equiv \{ b,  w \}).
%%\tag{5.40}
\end{align*}
%For example,  put $N=800$.
Note that
\begin{align*}
&
\text{
{\lq\lq} take a measurement $M_\omega$ N times{\rq\rq}}
\Leftrightarrow
\text{
{\lq\lq} take a measurement
${\mathsf M}_{C(\Omega{})} ({}{\mathsf O}^{N} , S_{[\delta_{\omega}]}{})${\rq\rq}
}
%%\tag{5.41}
\end{align*}
Define the estimator
$E: X^{N} ({}\equiv \{ b,w \}^{N}{}) \to \Omega ({}\equiv [0,1]{})$
\par
\noindent
\begin{align*}
&
E({}x_1, x_2 , \cdots , x_{N-1} , x_{N}{})
=
\frac{
\sharp [{}
\{ n \in \{ 1,2, \cdots , N \} \; | \; x_n = b \}{}]
}
{N}
\\
&
\quad
\qquad
(\forall
x=
({}x_1, x_2 , \cdots , x_{N-1} , x_{N} {})
\in X^{N}
\equiv
\{b,w \}^{N}).
%\tag{9} 
\end{align*}
%For example, put$\gamma = 0.95$.
For each
$\omega ({}\in [0,1] \equiv \Omega{})$,
define the positive number
$\eta^\gamma_\omega$
such that:
\begin{align*}
\eta^\gamma_\omega
=
&
\inf
\Big\{
\eta > 0 \; \Big| \;
[
F^{N}
(\{ ({}x_1, x_2 , \cdots , x_{N}{})
\; | \;
\omega - \eta
\le
E({}x_1, x_2 , \cdots , x_{N}{})
\le
\omega+ \eta
\})]
({}\omega{})
>
0.95
\Big\}
\\
=
&
\displaystyle{
\mathop
{\text{\Large inf}}_{
[
F^{N}
(\{ ({}x_1, x_2 , \cdots , x_{N}{})
:
|
E({}x_1, x_2 , \cdots , x_{N}{})
-
\omega
|
\le
\eta
\})
]
({}\omega{})
>
0.95
}
}
\eta.
%%\tag{5.43}
\end{align*}
Put
\begin{align*}
D_x^\gamma
=
\{
\omega
({}\in \Omega{})
:
\;
| E({}x) - \omega | \le \eta_\omega^\gamma \}.
%%\tag{5.44}
\end{align*}
For example,
assume that
$N$ is sufficiently large and
$\gamma = 0.95$.
Then we see, from the property of binomial distribution, that
\begin{align*}
\eta_\omega^{0.95}
\approx
1.96
\sqrt{ \frac{\omega ({}1- \omega{})}{N} }
\intertext{and}
D_x^{0.95}
=
[{}E(x) -\eta_- , E(x) + \eta_+{}]
%\{
%\omega
%({}\in \Omega{})
%:
%\;
%| E({}x) - \omega | \le \eta_\omega^\gamma \}
%%\tag{5.45}
\end{align*}
where
\begin{align*}
\eta_-
=
\eta_{ E(x) - \eta_- }^{0.95},
\quad
\eta_+
=
\eta_{ E(x) + \eta_+ }^{0.95}.
%\{
%\omega
%({}\in \Omega{})
%:
%\;
%| E({}x) - \omega | \le \eta_\omega^\gamma \}
%%\tag{5.46}
\end{align*}
Under the assumption that
$N$ is sufficiently large,
%%and$\gamma = 0.95$
we can consider that
$$
\eta_-
\approx
\eta_+
\approx
\eta_{E(x)}^{0.95}
\approx
1.96
\sqrt{ \frac{E(x) ({}1- E(x){})}{N} }.
$$
Then we can conclude that
\begin{itemize}
\item[(d)]
\it
for any
urn
$\omega
({}\in
\Omega \equiv [0,1]))$,
the probability,
that
the measured value $x=$
$(x_1, x_2 , \cdots ,$
$ x_{N}{})$
obtained
by the measurement
${\mathsf M}_{C(\Omega)} \big({}{\mathsf O}^{N}   ,$
$ S_{[\delta_{\omega }{}] } \big)$
satisfies the following
condition $(\sharp{})$,
is larger than
$\gamma$
({}e.g., $\gamma= 0.95${}).
\begin{itemize}
\item[$(\sharp{})$]
%$\qquad$
$ | \omega - E(x) |
\le
1.96
\sqrt{ \frac{E(x) ({}1- E(x){})}{N} }
\le
\frac{0.98}{\sqrt{N}}
$.
\end{itemize}
\end{itemize}
%where
%$E$ is defined by (9).
%\END{enumerate}
%%\qed
%%\pa \hfill$\blacksquare$

%\vs\gamma\alpha \alphakip1.0cm
\par
\par

\par
%\noindent
%
\rm
\subsubsection{{{}}statistical hypothesis testing
---
St. Valentine's Day chocolate}%{Sec. 4.3.4}
\par
In what follows,
we shall describe
"statistical hypothesis testing"
in terms of measurement theory.
%{{{measurement theory}}}.
% and , .

%\BEGIN{itembox}[c]
\par
\noindent
\begin{center}
{\bf
\textcolor{black}{
St Valentine's Day chocolate
%Axiom${}_{\text{\scriptsize c}}^{\text{\scriptsize p}}$ 1(Born{{{measurement theory}}}:$B({\mathbb C}^n)$
}
}
\end{center}
\par
\noindent
%\vskip0.1cm
\par
\noindent
\fbox{\parbox{155mm}{
\begin{itemize}
\item[]
You (male) would like to verify the truth of the following hypothesis.
\begin{itemize}
\item[(a$'$)]
Hypothesis:
[Caroline is fond of you]
\end{itemize}
Here,
you wan to judge this hypothesis (a$'$).
by the result whether Caroline gives tomorrow's St Valentine's Day chocolate.
\begin{itemize}
\item[(b$'$)]
How do we consider the two cases;
\\
$
\cases
\text{
Case \textcircled{\scriptsize A}:

Caroline presents you a chocolate
}
\\
\text{
Case \textcircled{\scriptsize B}:
Caroline presents you a chocolate
}
\endcases
$
\end{itemize}
{}
\end{itemize}
}
}
\par
\vskip0.5cm
\par
\noindent

%
%
%Consider a measurement
%${\mathsf M}_{C(\Omega)}({{\mathsf O}} {{=}} $
%$(X, 2^X, F),
%S_{[\omega]})$
%%${\mathsf M}_{B({\mathbb C}^n)}({\mathsf O}:= ({}X, 2^X, F{}), S_{[\omega]})$
%formulated in a
%basic algebra
%$C(\Omega)$.
%Assume that
%the measured value
%$ x$
%$({}\in X  {})$
%is
%obtained by the measurement
%${\mathsf M}_{C(\Omega)}({{\mathsf O}} {{=}} $
%$(X, 2^X, F),
%S_{[\omega]})$.
%Then,
%%it holds that
%{{}}
%the probability
%that
%a
%measured value
%$ x$
%$({}\in X)$
%is obtained
%is
%given by
%$[F(\{x\})](\omega)$.
%
%
%%
%
%
%
%\BEGIN{itembox}[c]{
%
%%%index{1@Axiom${}_{\text{\scriptsize c}}^{\text{\scriptsize p}}$ 1[trial]}
%\label{rule402}
%\BEGIN{itemize}
%\item[]
%You (male) would like to verify the truth of the following hypothesis.
%\BEGIN{itemize}
%\item[(a$'$)]
%Hypothesis:
%[Caroline is fond of you]
%\END{itemize}
%Here,
%you wan to judge this hypothesis (a$'$).
%by the result whether Caroline gives tomorrow's St Valentine's Day chocolate.
%\BEGIN{itemize}
%\item[(b$'$)]
%How do we consider the two cases;
%\\
%$
%\cases
%\text{
%Case \textcircled{\scriptsize A}:
%
%Caroline presents you a chocolate
%}
%\\
%\text{
%Case \textcircled{\scriptsize B}:
%Caroline presents you a chocolate
%}
%\ENDcases
%$
%\END{itemize}
%{}
%\END{itemize}
%\END{itembox}

Eveyone guess as follows.
\begin{itemize}
\item[(c$'$)]
$
\cases
&\text{Case \textcircled{\scriptsize A}:
there is a great possibility that
Caroline is not fond of you.}
\\
&
\qquad
\qquad
\text{
Thus,
the hypothesis (a$'$) is should be rejected
}
\\
&
\text{Case \textcircled{\scriptsize B}:
This may be an obligatory-gift chocolate,
%thus,
%there is a poosibity that
%the hypothesis
%\text{(a$'$)}
%is not true,
}
\\
&
\qquad
\qquad
\text{
Thus,
the hypothesis (a$'$) is can not be rejected
}
%\\
%&
%\qquad
%\qquad
%\cdots
%\text{
%obligatory-gift chocolate
%}, 
%\text{(a$'$)}
%\\
%&
%\qquad
%\qquad
%\quad
%\quad
%\;\;
%.
\endcases
$
\end{itemize}

\rm
\rm

%TTTTTTTTTTTTTTTTTTTTTTTTTTTTT

\rm
\par
\noindent
\par
%\newpage
%\par
%\noindent
%For simplicity,
%we focus on
%classical measurements.
\vskip0.5cm
%~ Assertion
\par
\noindent
\par
In what follows
we shall study
{\lq\lq}statistical hypothesis testing{\rq\rq}$\!\!\!.\;$
\rm
Consider a measurement
%
%Assume that
%we know that the measured value $x \;(\in X )$
%is
%obtained by a measurement
${\mathsf M}_{C(\Omega)}({\mathsf O}\equiv(X, {\cal F}, F{}), S_{[*]}
)$
formulated in
${C(\Omega)}$.

Here, we assume that
$(X,
\tau{{}_X})$ is a topological space,
where
$\tau{{}_X}$
is the set of all open sets.
And assume that
$\overline{\cal F}={\cal B}_X$;
the Borel field,
i,e.,
the smallest $\sigma$-field that contains all
open sets in $X$.
Note that we can assume, without loss of generality,
that
$
F({\Xi})
\not=
0
$
for any open set $\Xi (\in \tau{{}_X} )$
such that $\Xi \not= \emptyset$.
That is because,
if
$
F({\Xi})
=
0
$,
it suffices to
redefine
$X$ by $X\setminus \Xi$.

\par
\noindent
\bf
{{Problem }}4.12
[{}Statistical hypothesis testing{}
{\rm
\textcolor{black}{
(
{\rm cf.}
\cite{IMeas, IWhat}
)}}
]
%}
$\;\;$%POPOPO
\rm
Assume the following hypothesis called {\lq\lq}{\it null hypothesis}":
\rm
\begin{itemize}
\item[\textcolor{black}{(a)}]
the unknown state
$[\ast]$
belongs to
a set ${N}_H$
$({}\subseteq
\Omega
)$.
\end{itemize}
Then, our problem is as follows.
\begin{itemize}
\item[\textcolor{black}{(b)}]
Define a proper
$[D] (\in {\cal F})$
such that
\begin{itemize}
\item[]
if
a measured valued
obtained by
the measurement
${\mathsf M}_{C(\Omega)}({\mathsf O}\equiv(X, {\cal F}, F{}), S_{[*]}
)$
belongs to
$[D]$,
then
we can deny
the null hypothesis
(a),
that is,
$[F([D])]( \omega )$
is sufficiently small
for any
$\omega \in {\mathcal N}_H$.
\end{itemize}
\end{itemize}

%\par
%\noindent

\newpage
\par
\noindent
%Problem 2.19
%TTTTTTTTTTTTTT
%Find a rejiction region ${\widehat R}_\alpha$ satisfying (C)
\unitlength=0.20mm
%\begin{figure*}[h]
\begin{picture}(400,110)
\put(150,0){{
\put(27,18){0}
%\put(27,50){$\alpha$}
\put(27,108){1}
\put(350,18){$ \Omega
%{\frak S}^p({C(\Omega)}^* )
$}
\dottedline{3}(40,110)(340,110)
%\put(150,10){$\omega_0$}
\put(40,20){\line(0,1){100}}
%\linethickness{0.15mm}
\thicklines
\put(40,20){\line(1,0){300}}
%\linethickness{0.15mm}
\thicklines
%\sspline(40,110)(60,108)(80,102)(100,80)
%(150,40)(200,30)(220,20)(240,20)
%\sspline(120,20)(130,20)(160,30)(250,50)
%(270,80)(280,100)(300,105)(340,110)
\spline(40,110)(60,109)(80,105)(110,100)
(150,30)(180,23)(200,27)(250,25)
(270,35)(280,70)(300,100)(340,110)
\dottedline{5}(40,40)(340,40)
\put(27,40){$\alpha$}
%\dottedline{5}(225,50)(225,20)
%\multiput(225,18)(3,0){40}{\line(0,2){4}}
\multiput(150,18)(3,0){40}{\line(0,2){4}}
\put(200,2){${{\mathcal N}_H} $}
\put(134,80){$
%{{}_{{C(\Omega)}^*}} \langle
%\omega,
[F(
%{\widehat R}^\alpha_
[D]
%{{\mathcal N}_H}
)]
(\omega)
%\rangle_{{}_{C(\Omega)}}
$}
}}
%\put(40,-20){\bf FIGU 4.
%\rm
%Null Hypothesis
%${{\mathcal N}_H}$
%}
\end{picture}
%%%%%%%%%%%%%%%%%
\begin{center}{Figure 4.7:
Null Hypothesis
${{\mathcal N}_H}$
%\label{Fig.047}
}
\end{center}
%\BEGIN{FIGUre*}[htbp]
%%%%%%%%%%%%%%%%%
%\vskip0.3cm

%\par
%\noindent
{\bf
%\vskip0.3cm
%\vskip0.3cm
%BFBF
\par
\noindent
{\bf{Answer}}:
$\;\;$
\rm
Define a function such that
$\Lambda_{{\mathcal N}_H}{}: X \to [0,1]$
\begin{align*}
\displaystyle
\Lambda_{{\mathcal N}_H}({}x)
=
\lim_{ \Xi \to \{ x \} }
\frac{
\displaystyle 
\sup_{ \omega \in {{\mathcal N}_H }} 
[{}F({}\Xi{}){}] ({}\omega {})
 }
{
\displaystyle 
\sup_{ \omega \in \Omega }
[{}F({}\Xi{}){}] ({}\omega {})
 }
\quad
({}\forall x \in X{})
%P%%\tag{5.36}
\end{align*}
\par
\noindent
Also,
for any
$\varepsilon \; ({}0 < \varepsilon {{\; \leqq \;}}1{})$,
define the
$[D]_{{\mathcal N}_H}^\varepsilon $
$({}\in {\cal F}{})$
such that
\begin{align*}
[D]_{{\mathcal N}_H}^\varepsilon
=
\{ x \in X \; | \;
\Lambda_{{\mathcal N}_H} ({}x{}) < \varepsilon \}
%P%%\tag{5.37}
\end{align*}
\par
\noindent
%%%
%\vskip-0.4cm%%%
%\begin{figure}[htbp]
\unitlength=0.30mm
%\unitlength=0.35mm
\begin{picture}(400,110)
\put(27,18){0}
\put(27,50){$\varepsilon$}
\put(27,108){1}
\put(350,18){$X$}
\dottedline{3}(40,110)(340,110)
%\put(150,10){$\omega_0$}
\put(40,20){\line(0,1){100}}
%\linethickness{0.15mm}
\thicklines
\put(40,20){\line(1,0){300}}
%\linethickness{0.15mm}
\thicklines
%\spline(40,110)(60,108)(80,102)(100,80)
%(150,40)(200,30)(220,20)(240,20)
%\spline(120,20)(130,20)(160,30)(250,50)
%(270,80)(280,100)(300,105)(340,110)
\spline(40,70)(60,75)(80,80)(100,90)
(150,100)(200,60)(250,40)
(270,35)(280,30)(300,25)(340,20)
\dottedline{5}(40,50)(340,50)
\dottedline{5}(225,50)(225,20)
\multiput(225,18)(3,0){40}{\line(0,2){4}}
\put(280,2){$[D]_{{\mathcal N}_H}^\varepsilon $}
\put(190,80){$\Lambda_{{\mathcal N}_H} (x) $}
\end{picture}
\vskip-0.1cm
\begin{center}{Figure 4.8:
$\Lambda_{{\mathcal N}_H} ({}x{}) $C$[D]_{{\mathcal N}_H}^\varepsilon $
}
\end{center}
\par
\noindent
\par
\noindent
And define
$\varepsilon_{\rm max}^{0.05}$
$(\in [0,1])$
(i.e.,
significant level=0.05)
%$(0 {\; \leqq \;} \varepsilon_{\rm max}^{0.05} {\; \leqq \;} 1)$
%%in
such that
\begin{align*}
\varepsilon_{\rm max}^{0.05}
=
\sup
\{
\varepsilon \; |
\;
\sup_{ \omega_0 \in {\mathcal N}_H }
[{}F({}[D]_{{\mathcal N}_H}^\varepsilon{}{}){}] ({}\omega_0 {})
{{\; \leqq \;}}0.05
\}
%P%%\tag{5.38}
\end{align*}
Thus, we can get the rejection region
$[D]_{{\mathcal N}_H}^{{\varepsilon_{\rm max}^{0.05}}} $
(
depending on
${\mathcal N}_H, \; {\mathsf O},\; \varepsilon (=0.05)$
).
%$
%'ð{\bf Šü‹pˆæ}'ƌĂԁD
%%%index{'«'«'á'­'¢'«@Šü‹pˆæ}
%'³'āC
%Ž–ŽÀ(b)'É'æ'èC
The following is obvious:
\begin{itemize}
\item[]
if
$[\ast]$
$\in {\mathcal N}_H$,
then
the probability that
a measured value
obtained by
the measurement
${\mathsf M}_{C (\Omega)} ({\mathsf O} $
${{=}}(X, {\cal F}, F{}) , S_{[\ast]})$
belongs to
$[D]_{{\mathcal N}_H}^{{\varepsilon_{\rm max}^{0.05}}} 
$
$(\in {\cal F} )$
is less than
$0.05$.
\end{itemize}
Since
it is quite rare that
"[measured value]
$\in
[D]_{{\mathcal N}_H}^{{\varepsilon_{\rm max}^{0.05}}}$",
we can deny
the hypothesis
(a).
%%%\large
\par
\vskip0.2cm
\par
\noindent
\noindent
%{\bf Œ‹˜_}
\par
\noindent
Therefore,
we can conclude that
\begin{itemize}
\item[(c)]
$
\cases
\text{
if
[measured value]
$\in
[D]_{{\mathcal N}_H}^{{\varepsilon_{\rm max}^{0.05}}}$,
then
we can deny
the hypothesis
(a).
}
\\
\\
\text{
if
[measured value]
$ \notin
[D]_{{\mathcal N}_H}^{{\varepsilon_{\rm max}^{0.05}}}$,
then
we can not deny
the hypothesis
(a).
}
\endcases
$
\end{itemize}
\par
\qed
\par

\par

%NNNNNNNNN{\cal N}{N}

\vskip0.5cm

\par
\par
\noindent
\par
\rm
\vskip2.0cm
\noindent
{\bf
Typical Examples in Classical Measurements
}

\rm

%}
%}
%

\par
\noindent
%\bf
%BFBF
\sf
%[{}Likelihood ratio test for the normal observable{}].
\rm
Our argument in the previous section
may be too abstract and general.
However, it is surely usual.
In this section,
this will be shown as easy examples
in classical measurements.

%%%%%%%\tag{\tag{
Put
$\Omega = {\mathbb R}$,
${C(\Omega)} = {C} ({}\Omega{})$.
Fix $\sigma >0$.
And consider the normal observable
${\mathsf O}_{\sigma} $
$\equiv$
$({}{\mathbb R}{} , {\cal B}_{{\mathbb R}{}}^{} , F_{\sigma}{})$
in
${C} ({}\Omega{})$ such that:
\par
\noindent
\begin{align*}
[F_{\sigma}({\Xi})] ({} {}{\omega} {}) =
\frac{1}{{\sqrt{2 \pi }\sigma{}}}
\int_{{\Xi}} \exp[{}- \frac{({}{}{x} - {}{\omega}  {})^2 }{2 \sigma^2}    {}] d {}{x}
\quad
\quad
({}\forall  {\Xi} \in {\cal B}_{{\mathbb R}{}}^{},
\quad
\forall   {}{\omega}    \in \Omega = {\mathbb R}{}).
%\tag{13}
\end{align*}
And further, consider the product observable
${\mathsf O}_{\sigma}^2 $
$\equiv$
$({}{\mathbb R}^2{} , {\cal B}_{{\mathbb R}^2{}}^{} ,
F_{\sigma}^2 )$
in
${C} ({}\Omega{})$.
That is,
\par
\noindent
\begin{align*}
&
[F_{\sigma}^2
(\Xi_1 \bigtimes \Xi_2)]
({}\omega{})
=
[F_{\sigma}(\Xi_1)](\omega) \cdot [F_{\sigma}(\Xi_2)]
({}\omega{})
=
\frac{1}{({{\sqrt{2 \pi }\sigma{}}})^2}
\iint_{\Xi_1 \times \Xi_2 }
\exp[{}- \frac{\sum_{k=1}^2 ({}{}{x_k} - {}{\omega}  {})^2
%+ ({}{}{x_2} - {}{\omega}  {})^2
}
{2 \sigma^2}    {}] d {}{x_1} d {}{x_2}
\\
&
\qquad
\qquad
\qquad
\qquad
({}\forall  \Xi_k \in {\cal B}_{{\mathbb R}{}}^{}
({}k=1,2),
\quad
\forall   {}{\omega}    \in \Omega = {\mathbb R}{}).
%\tag{14}
\end{align*}
In what follows, we consider the
measurement
${\mathsf M}_{{C(\Omega)}}({\mathsf O}_\sigma^2
=({}{\mathbb R}^2{} , {\cal B}_{{\mathbb R}^2{}}^{} ,
F_{\sigma}^2 )
,$
$ S_{[\ast]})$).
\par

\par
\noindent
{\bf
[Case(I)}; Two sided test, i.e., ${\mathcal N}_H = \{ \omega_0 \}${}].
Assume that
${\mathcal N}_H = \{ \omega_0
\}$,
$\omega_0 \in \Omega = {\mathbb R}$.
Note the identification
(1),
i.e.,
$\delta_{\omega_0} \approx \omega_0$.
Then,
we see
that,
for any
$ {({}x_1 , x_2{})} \in {\mathbb R}^2{}$,
\begin{align*}
&
\; \;
\Lambda_{{\mathcal N}_H}({}x_1, x_2{})
=
\sup_{\omega \in \{ \omega_0 \} }
L( (x_1. x_2 ), \delta_{\omega} )
=
\lim_{ {\Xi_1 \times \Xi_2} \to  ({}x_1 , x_2{})  }
\frac{
%\ssup_{ \omega \in { \{ \omega_0 \} } }
[F_{\sigma}^2
(\Xi_1 \bigtimes \Xi_2)]
({}\omega_0{})
}
{\sup_{ \omega \in \Omega }
[F_{\sigma}^2
(\Xi_1 \bigtimes \Xi_2)]
({}\omega{})
}
\\
&
=
\frac{
\exp[{}- \frac{({}{}{x_1} - {}{\omega_0}  {})^2 + ({}{}{x_2} - {}{\omega_0}  {})^2}
{2 \sigma^2}    {}]}
{
\exp[{}- \frac{({}{}{x_1} - {}{(x_1+x_2)/2}  {})^2 +
({}{}{x_2} - {}{(x_1+x_2)/2}  {})^2}
{2 \sigma^2}    {}]
}
=
\exp[{}- \frac{[{}({}x_1 + x_2{})- 2 \omega_0]^2}
{4 \sigma^2}    {}].
%\\
%&
%=
%\exp[{}- \frac{[{}({}x_1 + x_2{})/2-  \omega_0]^2}
%{2 ({}\sigma / \sqrt{2 })^2} {}]
%\tag{15}
\end{align*}
Also, for any
$\epsilon ({}> 0{})$,
define
${{D}}_{\{ \omega_0 \}}^\epsilon $
$({}\in {\cal B}_{{\mathbb R}^2} {})$
such that:
\begin{align*}
{{D}}_{\{ \omega_0 \} }^\epsilon
=
\{ {({}x_1 , x_2{})} \in {\mathbb R}^2 \; | \;
\Lambda_{\{ \omega_0 \}} ({} x_1 , x_2 {}) \le \epsilon \}.
%\tag{16} 
\end{align*}
Thus we can define
$\epsilon(\alpha)$
such that:
\begin{align*}
\epsilon(\alpha)
=
\sup
\{
\epsilon \; |
\;
\sup_{ \omega \in \{ \omega_0 \} }
[F_\sigma^2 ({{D}}_{\{ \omega_0 \}}^\epsilon{})](\omega)
\le \alpha
\}.
%\tag{17}
\end{align*}
Thus, putting $\alpha = 0.05$,
we see that
%%%%%%%%\Lambda_{\{ \omega_0 \}}
\begin{align*}
&
{\widehat R}^{0.05}_{\{ \omega_0 \}}
=
{{D}}^{\epsilon(0.05)}_{\{ \omega_0 \} }
\\
=
&
\{ ({}x_1 , x_2{})
\in {\mathbb R}^2 \; \; |\; \;
({}x_1 + x_2{})/ 2 \le \omega_0 - \frac{1.96 \sigma}{ {\sqrt 2}} \}
\bigcup
\{ ({}x_1 , x_2{})
\in {\mathbb R}^2  \; |\;
({}x_1 + x_2{})/ 2 \ge \omega_0 + \frac{1.96 \sigma}{ {\sqrt 2}} \}
\\
=
&
\text{{\lq\lq}Slash part in 
\textcolor{black}{{Fig. 4.9}}{\rq\rq}}
%\tag{18}
\end{align*}
\par
\noindent
%\newpage
%$\omega_0 = [{}\ast{}] $.
\par
\noindent
{${\:}$}
\par
\vskip2.0cm
\par
\noindent
\unitlength=0.4mm
%%%%%%%%%%%%%%%%%
%\begin{figure*}[htbp]
%%%%%%%%%%%%%%%%%
%%\vskip0.3cm
%\caption{
%u{}$BFY%=%flm{}(A%}
%\END{figure*}
%%%%%%%%%%twosided
\begin{picture}(300,170)
%%\put(25,12){A}
\put(100,50)
{{
%\path(-50,0)(100,0)
\put(120,-10){$x_1$}
\put(-10,90){$x_2$}
\put(-30,0){\vector(1,0){160}}
%\path(0,-30)(0,70)
\put(0,-20){\vector(0,1){120}}
%\put(-20,-20){\line(1,1){90}}
\multiput(-35,50)(3,-3){20}{\line(-1,-1){30}}
\multiput(-5,70)(3,-3){32}{\line(1,1){30}}
%\put(20,20){\circle*{2}}
%\sspline(0,0)(10,5)(20,20)
\path(15,-2)(15,2)
\path(40,-2)(40,2)
\path(-2,15)(2,15)
\path(65,-2)(65,2)
\path(-2,65)(2,65)
\path(-2,40)(2,40)
\put(2,37){$2 \omega_0$}
\put(35,-8){$2 \omega_0$}
%\put(11,6){$\omega_0$}
\put(17,3){$a$}
\put(62,-8){$b$}
\put(3,13){$a$}
\put(-6,62){$b$}
\put(40,94){ $a=2(\omega_0 - 1.96  \sigma/ {\sqrt 2})$}
\put(40,81){ $b=2(\omega_0 + 1.96  \sigma/ {\sqrt 2})$}
%\put(-30,13){\tiny $2(\omega_0 - 1.96  \sigma/ {\sqrt 2})$}
%\put(-30,63){\tiny $2(\omega_0 +1.96  \sigma/ {\sqrt 2})$}
\thicklines
\put(-55,70){\line(1,-1){90}}
\put(-15,80){\line(1,-1){105}}
}}
%
%\put(40,16){\scriptsize $x_0$}
%\qbezier(40,20)(100,61)(157,42)
%\path(107,49)(115,48)(107,45)
%\put(40,20){\circle*{1}}
%\put(157,41){\circle*{1}}
%\put(155,45){\scriptsize $L({}x_0)$}
%%\put(157,40){\scriptsize $r$}
%%\put(76,42){\scriptsize $\beta$}
%\put(151,33){$ \; \omega$}
\end{picture}
\vskip1.0cm
\begin{center}{Figure 4.9:
{Rejection region}
${\widehat{R}}^{0.05}_{\{ \omega_0 \}}$
%\label{Fig.4-9}
}
\end{center}

\par
\vskip1.0cm
\par
\noindent
{\bf [Case(II)}; One sided test, i.e., ${\mathcal N}_H = [\omega_0 , \infty{}) ${}].
Assume that
${\mathcal N}_H = [\omega_0 , \infty{})$,
$\omega_0 \in \Omega = {\mathbb R}$.
Then,
\begin{align*}
&
\; \;
\Lambda_{[{}\omega_0, \infty{}) }({} x_1 , x_2{})
=
\sup_{\omega \in [\omega_0, \infty) }
L( (x_1. x_2 ), \delta_{\omega} )
=
\underset{{ \omega \in [\omega_0,\infty) }}{\sup}
\lim_{ {\Xi_1 \times \Xi_2} \to  ({}x_1 , x_2{})  }
\frac{
[F_{\sigma}^2
(\Xi_1 \bigtimes \Xi_2)]
({}\omega{})
}
{
\underset{{ \omega \in \Omega }}{\sup}
[F_{\sigma}^2
(\Xi_1 \bigtimes \Xi_2)]
({}\omega{})}
\\
=
&
\underset{{ \omega \in [\omega_0,\infty) }}{\sup}
\exp[{}- \frac{[{}({}x_1 + x_2{})- 2 \omega]^2}
{4 \sigma^2}    {}]
%\\
%&
%=
%\lim_{ {\Xi_1 \times \Xi_2} \to \{ ({}x_1 , x_2{}) \} }
%\frac{
%%\ssup_{ \omega \in { \{ \omega_0 \} } }
%[F_{\sigma}^2
%(\Xi_1 \bigtimes \Xi_2)]
%({}\omega{})
%}
%{\sup_{ \omega \in \Omega }
%[F_{\sigma}^2
%(\Xi_1 \bigtimes \Xi_2)]
%({}\omega{})
%}
=
\cases
\exp[{}- \frac{[{}({}x_1 + x_2{})- 2 \omega_0]^2}
{4 \sigma^2}    {}]
\quad &
({}\frac{x_1 + x_2}{2} < \omega_0 {})
\\
1
\quad &
(\text{ otherwise }{})
\endcases
%\tag{19}
\end{align*}

\par
\noindent
Also, for any
$\epsilon ({}> 0{})$,
define
${{D}}_{ { [\omega_0, \infty{})  }}^\epsilon $
$({}\in {\cal B}_{{\mathbb R}^2}{})$
such that:
\begin{align*}
&
\;\;
{{D}}_{ { [{}\omega_0, \infty{})  }}^\epsilon
=
\{ ({}x_1 , x_2{}) \in {\mathbb R}^2 \; | \;
\Lambda_{{ [{}\omega_0, \infty{})  }} ({} x_1 , x_2 {}) \le \epsilon \}
=
\{ ({}x_1 , x_2{}) \in {\mathbb R}^2 \; | \;
\frac{ x_1 + x_2 }{2} - \omega_0< {\sqrt{ 4\sigma^2 \log \epsilon }} \}.
%\tag{20} 
\end{align*}
\par
\noindent
Thus we can define
$\epsilon(\alpha)$
such that:
\begin{align*}
\epsilon(\alpha)
=
\sup
\{
\epsilon \; |
\;
\sup_{ \omega \in { [{}\omega_0, \infty{})  } }
[F_{\sigma}^2
({{D}}_{ { [{}\omega_0, \infty{})  }}^\epsilon{})
]
({}\omega{})
\le
\alpha
\}.
%\tag{21} 
\end{align*}
\par
\noindent
Therefore, putting
$\alpha = 0.05$,
we see that
%%%%%%%%%\Lambda_{{\mathcal N}_H}
%{\lq\lq}the $0.05$-influential domain
%of
%$\nu_1$ for $\nu_2${\rq\rq}
%such that:
\begin{align*}
&
{\widehat{R}}^{0.05}_{[\omega_0, \infty)}
=
{{D}}^{\epsilon(0.05)}_{[\omega_0,\infty{}) }
=
\{ ({}x_1 , x_2{}) \in {\mathbb R}^2 \; \; |\; \;
\frac{({}x_1 + x_2{})}{ 2} \le \omega_0 - \frac{1.65  \sigma}{ {\sqrt 2}} \}
%\qquad
%(
%\text{
%{\rm cf.} \textcolor{black}{(1.63)}
%}).
=
\text{{\lq\lq}Slash part in \textcolor{black}{Fig. 4.10}{\rq\rq}}
%\tag{22}
\end{align*}
\par
\noindent
%\vskip0.7cm
%$\rho_0 = [{}\ast{}] $.
\par
\noindent
\noindent
%\unitlength=0.4mm

\par
\noindent
\unitlength=0.43mm
%%%%%%%%%%%%%%%%%
%\begin{figure*}[htbp]
%%%%%%%%%%%%%%%%%
%%\vskip0.3cm
%\caption{
%u{}$BFY%=%flm{}(A%}
%\END{figure*}
%%%%%%%%%%twosided
\begin{picture}(300,140)
%%\put(25,12){A}
\put(100,50)
{{
%\path(-50,0)(100,0)
\put(100,-10){$x_1$}
\put(-10,65){$x_2$}
\put(-30,0){\vector(1,0){130}}
%\path(0,-30)(0,70)
\put(0,-20){\vector(0,1){90}}
%\put(-20,-20){\line(1,1){90}}
\multiput(-35,50)(3,-3){20}{\line(-1,-1){30}}
%\multiput(-5,70)(3,-3){32}{\line(1,1){30}}
%\put(20,20){\circle*{2}}
%\sspline(0,0)(10,5)(20,20)
\path(15,-2)(15,2)
\path(40,-2)(40,2)
\path(-2,15)(2,15)
%\path(65,-2)(65,2)
%\path(-2,65)(2,65)
\path(-2,40)(2,40)
\put(2,37){$2 \omega_0$}
\put(35,-8){$2 \omega_0$}
%\put(11,6){$\omega_0$}
\put(17,3){$c$}
%\put(62,-8){$b$}
\put(3,13){$c$}
%\put(-6,62){$b$}
\put(25,50){ $c=2(\omega_0 - 1.65  \sigma/ {\sqrt 2})$}
%\put(40,56){ $b=2(\omega_0 + 1.96  \sigma/ {\sqrt 2})$}
%\put(-30,13){\tiny $2(\omega_0 - 1.96  \sigma/ {\sqrt 2})$}
%\put(-30,63){\tiny $2(\omega_0 +1.96  \sigma/ {\sqrt 2})$}
\thicklines
%%%PPPPPPPPPPPPPPP
%\put(-55,70){\line(1,-1){90}}
\put(-50,65){\line(1,-1){85}}
%\put(-15,80){\line(1,-1){105}}
}}
%\put(40,5){\bf FIGU 7.
%\rm
%Rejection region
%${\widehat{R}}^{0.05}_{[\omega_0, \infty)}$
%}
%
\end{picture}
\vskip-0.4cm
\begin{center}{Figure 4.10:
\rm
{Rejection region}
${\widehat{R}}^{0.05}_{[\omega_0, \infty)}$
%\label{Fig.4-10}
}
\end{center}

\vskip0.5cm

\par
\rm

\par
\noindent
%BBBBBBBBBBBBBBBBBB%SBSBSBS
{\small%%{\footnotesize
\begin{itemize}
\item[$\spadesuit$] \bf {{}}{Note }4.6{{}} \rm
Although
several statistical methods are described in terms of
measurements
in this section.
However,
it should be that
there are some gap
between
the statistical description
and
the measurement theoretical description.
Recall that measurement theory says that
\begin{itemize}
\item[$(\sharp_1)$]
the state after a measurement
is meaningless
\end{itemize}
Therefore,
\textcolor{black}{{{Theorem }}4.5}
[{}Fisher maximum likelihood method{(}{measurement theoretical representation}{)}
]
should be formally written as follows
\begin{itemize}
\item[$(\sharp_2)$]
{
%BFBF
\par
\noindent
{\bf {{Theorem }}4.5$'$}
%%index{@Fisher maximim likelihood method({{measurement theory}})}
[{}Fisher maximum likelihood method{(}{measurement theoretical representation}{)}
\rm
\textcolor{black}{
(
{\rm cf.}\cite{Keio, IWhat}
)}
\bf
]}$\;\;$%POPOPO
%Fisher maximim likelihood method.}
\rm
\rm
Let
${\mathsf O}\times  {\mathsf O}_1$
${{=}} (X \times Y , {\cal F}\times {\cal G} , F\times G)$
be
a simultaneous observable
in
$C(\Omega)$.
Consider
a
simultaneous measurement
${\mathsf M}_{C (\Omega)}({\mathsf O} \times {\mathsf O}_1,$
$ S_{[*]})$.
%.
%%When we know that{{}}
%{{measurement}}
%${\mathsf M}_{C (\Omega)}({\mathsf O}, S_{[*]})$
Assume that
a{measured value }{{}}
$(x, y )$
belongs to
$\Xi \times Y \;(\in {\cal F}\bigstimes {\cal G} )$.
Then,
there is a reason to infer that
\begin{itemize}
\item[]
the probability that
$y \in \Gamma$
is given by
$[G(\Gamma )](\omega_0 )$
($\forall \gamma \in {\cal G} )$
\end{itemize}
where
$
[F(\Xi)](\omega_0)
= \max_{\omega \in \Omega} [F(\Xi)](\omega)
$.
\end{itemize}
\textcolor{black}{{{Theorem }}4.5}
is the abbreviation
\textcolor{black}{{{{Theorem }}4.5$'$}}
Note that
$(\sharp_2)$
is applicable to quantum mechanics
(under the existence of
a simultaneous observable
${\mathsf O}\times  {\mathsf O}_1$
).
%).
%,
%$(\sharp_1)$
%
%{{measurement theory}}
%
%,
%statistical hypothesis testing(\textcolor{black}{{{Problem }}4.12}),
%{{measurement theory}}(Answer\textcolor{black}{{Note }5.6}).
%% and 
\par
\noindent
%%%%
%
%
%(a)?$\;\;$(b)?
%, ,  and {}
\end{itemize}
}
%%BBBBBBBBBBB Axiom${}_{\text{\scriptsize c}}^{\text{\scriptsize p}}$ 1 Axiom${}_{\text{\scriptsize c}}^{\text{\scriptsize p}}$ 2 Axiom${}_{\text{\scriptsize c}}^{\text{\scriptsize p}}$ 1

\subsection{{Bayesian }{statistics}}%{Sec.4.4}
%\ssubsection{supplement:{{{measurement theory}}}({Bayesian }{statistics})
\label{5secAxiom 1}
\subsubsection{What is mixed {{measurement}}?}%{Sec.4.4.1}
\baselineskip=18pt
\normalsize \baselineskip=18pt
\par
Recall the classification of{{{{measurement theory}}}}(\textcolor{black}{{Chap.$\;$1}(Y)}),
that is,
%index{@{{{measurement theory}}}}
\par
\noindent
%$\;\;$
%$\underset{\text{\scriptsize Chap. 1}}{\text{(Y)}}$
(a)
$
%\underset{(quantum mechanics)}
\underset{\text{\scriptsize (scientific language)}}{\text{
%\footnotesize
{{{measurement theory}}}}}
%{\footnotesize {{{measurement}}}}
\cases
\underset{}{\text{
%\footnotesize
quantum {{{measurement}}}}}{\textcircled{\scriptsize 6}}:
\text{\footnotesize \textcolor{black}{{Chap. 3}}}
\\
\underset{}{\text{
%\footnotesize
classical {{{measurement}}}}}
%\footnotemark
\cases
\!\!\!
\text{\small continuous}
\cases
\!\!
\text{\footnotesize pure type}{\textcircled{\scriptsize 1}}
%\longleftarrow
:
\text{\textcolor{black}{\footnotesize (Chaps. 2$\text{--}$9)}}\\
\!\!
\text{\footnotesize mixed type}
{\textcircled{\scriptsize 2}}:
\text{\footnotesize
\textcolor{black}{{{{}}}{Sec.4.4}}
}
\\
\endcases
\\
\\
\!\!\!
\text{\small bounded}
\cases
\!\!
\text{\footnotesize pure type}{\textcircled{\scriptsize 3}}
:
\text{\textcolor{black}{\footnotesize Chaps. 10, 11}}\\
\!\!
\text{\footnotesize mixed type}{\textcircled{\scriptsize 4}}
:
\text{\footnotesize
\textcolor{black}{{Notes }10.4, 11.3}
}
\\
\endcases
\endcases
\endcases
$
%\END{itemize}

In this preprint
we mainly devote ourselves to
classical pure {{measurement}}
({\textcircled{\scriptsize 1}} and {\textcircled{\scriptsize 3}})
and not
classical mixed {{measurement}}.
However,
in what follows,
only a few will describe this.

\par
\normalsize \baselineskip=18pt
\par
Let
$\Omega$
be a
locally compact space,
${\cal B}_{\Omega}$
be its Borel field.
${\cal M}(\Omega) $
%index{m1@${\cal M}(\Omega) $:}
and
${\cal M}_{+1}(\Omega)$
%index{m2@${\cal M}_{+1}(\Omega) $:state space }
are defined as follows.
%(\textcolor{black}{Appendix B.5(F)}).
\normalsize \baselineskip=18pt
{
%\ssmall
{
\begin{itemize}
\item[(b)]
$
\!\!\!
\cases
\text{\bf The space of complex valued measures}
:\;{\cal M}(\Omega)
\\
\qquad
=
\{ \nu (= \nu_1 -\nu_2 +{\sqrt{-1}} ( \nu_3 -\nu_4) )\;|\;
\text{$\nu_k$
$(k=1,2,3,4)$
is a finite measure on
$\Omega$
}
\}
\\
\text{\bf
The space of probability measures
}:
\;{\cal M}_{+1}(\Omega)
\\
\qquad
=
\{ \nu\;\;|\; \
\text{
$\nu$
is a finite measure on
$\Omega$
such that
$\nu(\Omega )=1$
}
\}
%P%\tag{5.40}
\endcases
$
\end{itemize}
}
}
\normalsize \baselineskip=18pt
In {{{measurement theory}}},
the
%,
%$\Omega$state space ,
${\cal M}_{+1}(\Omega)$
is called a
{\bf mixed state space }
(or, statistical state space,
compound state space),
$\nu (\in {\cal M}_{+1}(\Omega))$
is said to be
a {\bf {mixed state}}(or,
statistical state,
compound state).
%index{@{mixed state}}
%index{@state space }

\par
\vskip0.5cm
%\vskip1.0cm
\par
In what follows,
we shall explain
mixed {{measurement}}.
\par
\noindent
{\bf Example 4.13}
\bf
[Urn problem and coin tossing]
% and
%\textcolor{black}{Example 2.10}
%;urn problem{(}{{measurement}}{)}{}]
\rm
In \textcolor{black}{{{Problem }}4.3}[urn problem]
(and \textcolor{black}{Answer 4.6}),
putting
$\Omega=\{ \omega_1, \omega_2 \}$,
and consider
a
{{measurement}}
${\mathsf M}_{C (\Omega)} ({}{\mathsf O} {{=}}
({} \{ {{w}}, {{b}} \}, 2^{\{ {{w}}, {{b}} \}  }  , F{})
,
S_{ [{}{\ast}]}{})$,
where
${\mathsf O} = ({} \{ {{{w}},} {{b}} \},$
$ 2^{\{ {{w}}, {{b}} \}  }  ,$
$ F{})$
is defined by
the
\textcolor{black}{formula (2.5)}.
That is,
%\BEGIN{align*}
\begin{align*}
& F({}\{ {{w}} \}{})(\omega_1{})= 0.8, & \quad & F({}\{ {{b}} \}{})(\omega_1{})= 0.2
\\
& F({}\{ {{w}} \}{})(\omega_2{})= 0.4, & \quad & F({}\{ {{b}} \}{})(\omega_2{})= 0.6
%%%%%REDREDREDREDREDRE
\end{align*}
%,
Let us write \textcolor{black}{{Fig.$\;$}4.2(={Fig.$\;$}4.11)} again.
%.

\noindent
\unitlength=0.27mm
%%%%%%%%%%%%%%%%%
%\begin{figure*}[htbp]
%%%%%%%%%%%%%%%%%ZZZZZZZZZZZzZZZZzZzZzZzZ
%\caption{
%Which is the hidden urn?
%}P$_
%\END{figure*}
%%%%%%%%%%%%%%%%%%
\begin{picture}(590,230)
%\p
%${\mathsf M}_{C(\Omega)}({\mathsf O}, S_{[{}\ast{}] }{})$}
\path(0,-20)(0,220)(590,220)(590,-20)(0,-20)
\Thicklines
\put(170,60){}
\put(155,50){\vector(1,0){30}}
\put(390,60){}
\put(410,50){\vector(-1,0){30}}
\put(200,5){
\path(-10,-10)(-10,120)(160,120)(160,-10)(-10,-10)
\allinethickness{0.1mm}
\multiput(-10,-10)(0,4){33}{\line(1,0){170}}
\Thicklines
\spline(70,70)(80,120)(130,140)
\path(120,143)(130,140)(120,130)
\filltype{white}\put(140,140){\circle*{10}}
\put(56,40){\Huge \bold [{}$\ast${}]}
\put(-180,190){\footnotesize
You do not know which the urn behind the curtain is,
$U_1$ or $U_2$.}
\put(-180,170){\footnotesize
Assume that you pick up a white ball from the urn.
}
%}
\put(-180,150){\footnotesize
The urn is $U_1$ or $U_2$?  $\quad$ Which do you think? }
%\put(-180,150){\footnotesize
%Further, pick out one ball from the urn.
%Infer the probability
%that
%the new picked ball
%is black.
%}
}
%
%
%
%
%
%\put(-180,190){\footnotesize
%You do not know which the urn behind the curtain is,
%$U_1$ or $U_2$, but the probability: $p$ and $1-p$.}
%\put(-180,170){\footnotesize
%Assume that you pick up a white ball from the urn.}
%%}
%\put(-180,150){\footnotesize
%The urn is $U_1$ or $U_2$?  $\quad$ Which do you think? }
%}
\put(0,5){
\spline(45,100)(45,90)(0,40)(50,0)(75,0)
(100,0)(150,40)(105,90)(105,100)}
\put(200,5){
\spline(45,100)(45,90)(0,40)(50,0)(75,0)
(100,0)(150,40)(105,90)(105,100)}
\put(400,5){
\spline(45,100)(45,90)(0,40)(50,0)(75,0)
(100,0)(150,40)(105,90)(105,100)}
\thicklines
\put(73,116){$U_1$}
\put(473,116){$U_2$}
%%\put(373,116){$U_3$}
\put(10,0){
\filltype{white}\put(40,50){\circle*{10}}
\filltype{white}\put(55,50){\circle*{10}}
\filltype{white}\put(70,50){\circle*{10}}
\filltype{white}\put(85,50){\circle*{10}}
\filltype{black}\put(100,50){\circle*{10}}
\filltype{white}\put(40,35){\circle*{10}}
\filltype{white}\put(55,35){\circle*{10}}
\filltype{white}\put(70,35){\circle*{10}}
\filltype{white}\put(85,35){\circle*{10}}
\filltype{black}\put(100,35){\circle*{10}}
}
\put(400,0){
\put(10,0){
\filltype{white}\put(40,50){\circle*{10}}
\filltype{white}\put(55,50){\circle*{10}}
\filltype{black}\put(70,50){\circle*{10}}
\filltype{black}\put(85,50){\circle*{10}}
\filltype{black}\put(100,50){\circle*{10}}
\filltype{white}\put(40,35){\circle*{10}}
\filltype{white}\put(55,35){\circle*{10}}
\filltype{black}\put(70,35){\circle*{10}}
\filltype{black}\put(85,35){\circle*{10}}
\filltype{black}\put(100,35){\circle*{10}}
}
}
\end{picture}
%%%%%%%%%%%%%%%%%
%\BEGIN{FIGUre*}[htbp]
%%%%%%%%%%%%%%%%%
\vskip0.3cm
\begin{center}{Figure 4.11:
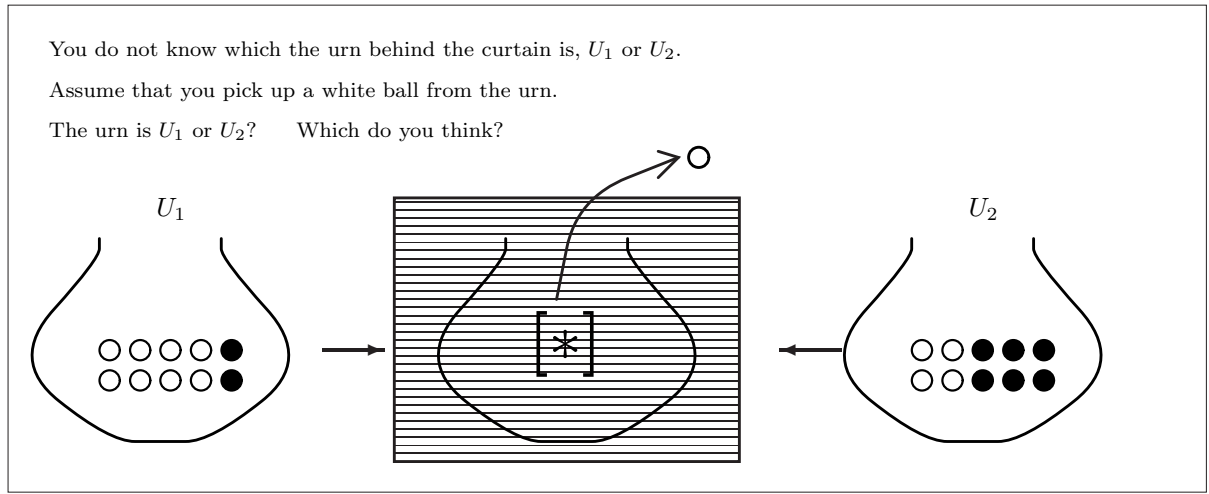
Inference interval
Which is the hidden urn,
$U_1$ or $U_2$?
(= \textcolor{black}{Fig.} 4.2)
}
\end{center}
%%%%%%%%%%%%%%%%%%

\par
\noindent
A mixed measurement is characterized
such as
"{{measurement}}
${\mathsf M}_{C (\Omega)} ({}{\mathsf O} {{=}}
({} \{ {{w}}, {{b}} \}, 2^{\{ {{w}}, {{b}} \}  }  , F{})
,
S_{ [{}{\ast}]}{})$"
+
"
probabilistic property of the unknown {{state}}$[\ast]$".
Let us explain it.
Consider the following two procedures
(c)
and
(d)(\textcolor{black}{{Fig.$\;$}4.12}):
\begin{itemize}
\item[(c)]
Consider an unfair coin-tossing
$(T_{p,1-p})$
such that
$( 0 {{\; \leqq \;}}p {{\; \leqq \;}}1 )$:
That is,
%\BEGIN{align*}
%$(T_{p,1-p})$,
\\
\\
$\quad$
$
\cases
\text{the possibility that "head" appears is}
%,
\text{ $100p \% $}
\\
\text{the possibility that "tail" appears is}
\text{ $100(1-p) \% $}
\endcases
$
\\
\\
If "head" [resp. "tail"]appears,
put
an urn $U_1({\approx} \omega_1)$
[resp. $U_2({\approx} \omega_2)$]
behind the curtain.
Assume that
you do not know
which urn is behind the curtain,
$U_1$
or $U_2$).
The unknown urn is denoted by
$[*{}](\in \{\omega_1, \omega_2\})$.
%Note,
%for completeness, that we do not know whether $[*]$ is
%$\omega_1$ or $\omega_2$,
%though we know the probabilistic rule 7).
%%since the two can not be distinguished in appearance.
%{{state}}
%$[\ast]$
%, {\bf {mixed state}} and ,

This situation is represented by
$$
\text{
$\nu_0 = p \delta_{\omega_1} + (1-p) \delta_{\omega_2}$
({}That is,
$\nu_0(\{\omega_1\}) = p$,  $\nu_0(\{\omega_2\}) = 1-p$)
}
$$
where $\delta_{\omega}$
is the point measure at $\omega$.
That is,
it suffices to consider that
$\nu_0$
is the distribution of $[\ast]$.
\rm
\rm
\item[(d)]
Consider the "measurement" such that
a ball is picked out from the unknown urn.
This "measurement"
is denoted by
${\mathsf M}_{C(\Omega)}({\mathsf O}, S_{[*]}(\nu_0))$,
and
called a mixed measurement.
\end{itemize}
\par

\par
\noindent
\unitlength=0.27mm
%%%%%%%%%%%%%%%%%
%\begin{figure*}[htbp]
%%%%%%%%%%%%%%%%%ZZZZZZZZZZZzZZZZzZzZzZzZ
%\caption{
%Which is the hidden urn?
%}P$_
%\END{figure*}
%%%%%%%%%%%%%%%%%%
\begin{picture}(590,220)
%\p
%${\mathsf M}_{C(\Omega)}({\mathsf O}, S_{[{}\ast{}] }{})$}
\path(0,-20)(0,220)(590,220)(590,-20)(0,-20)
\Thicklines
\put(170,60){p}
\put(155,50){\vector(1,0){30}}
\put(390,60){1-p}
\put(410,50){\vector(-1,0){30}}
\put(200,5){
\path(-10,-10)(-10,120)(160,120)(160,-10)(-10,-10)
\allinethickness{0.1mm}
\multiput(-10,-10)(0,4){33}{\line(1,0){170}}
\Thicklines
\spline(70,70)(80,120)(130,140)
\path(120,143)(130,140)(120,130)
\filltype{white}\put(140,140){\circle*{10}}
\put(56,40){\Huge \bold [{}$\ast${}]}
\put(-180,190){\footnotesize
You do not know which the urn behind the curtain is,
$U_1$ or $U_2$, but the probability: $p$ and $1-p$.}
\put(-180,170){\footnotesize
Assume that you pick up a white ball from the urn.}
%}
\put(-180,150){\footnotesize
The urn is $U_1$ or $U_2$?  $\quad$ Which do you think? }
}
\put(0,5){
\spline(45,100)(45,90)(0,40)(50,0)(75,0)
(100,0)(150,40)(105,90)(105,100)}
\put(200,5){
\spline(45,100)(45,90)(0,40)(50,0)(75,0)
(100,0)(150,40)(105,90)(105,100)}
\put(400,5){
\spline(45,100)(45,90)(0,40)(50,0)(75,0)
(100,0)(150,40)(105,90)(105,100)}
\thicklines
\put(73,116){$U_1$}
\put(473,116){$U_2$}
%%\put(373,116){$U_3$}
\put(10,0){
\filltype{white}\put(40,50){\circle*{10}}
\filltype{white}\put(55,50){\circle*{10}}
\filltype{white}\put(70,50){\circle*{10}}
\filltype{white}\put(85,50){\circle*{10}}
\filltype{black}\put(100,50){\circle*{10}}
\filltype{white}\put(40,35){\circle*{10}}
\filltype{white}\put(55,35){\circle*{10}}
\filltype{white}\put(70,35){\circle*{10}}
\filltype{white}\put(85,35){\circle*{10}}
\filltype{black}\put(100,35){\circle*{10}}
}
\put(400,0){
\put(10,0){
\filltype{white}\put(40,50){\circle*{10}}
\filltype{white}\put(55,50){\circle*{10}}
\filltype{black}\put(70,50){\circle*{10}}
\filltype{black}\put(85,50){\circle*{10}}
\filltype{black}\put(100,50){\circle*{10}}
\filltype{white}\put(40,35){\circle*{10}}
\filltype{white}\put(55,35){\circle*{10}}
\filltype{black}\put(70,35){\circle*{10}}
\filltype{black}\put(85,35){\circle*{10}}
\filltype{black}\put(100,35){\circle*{10}}
}
}
\end{picture}
%%%%%%%%%%%%%%%%%
%\BEGIN{FIGUre*}[htbp]
%%%%%%%%%%%%%%%%%
\vskip0.3cm
\begin{center}{Figure 4.12:
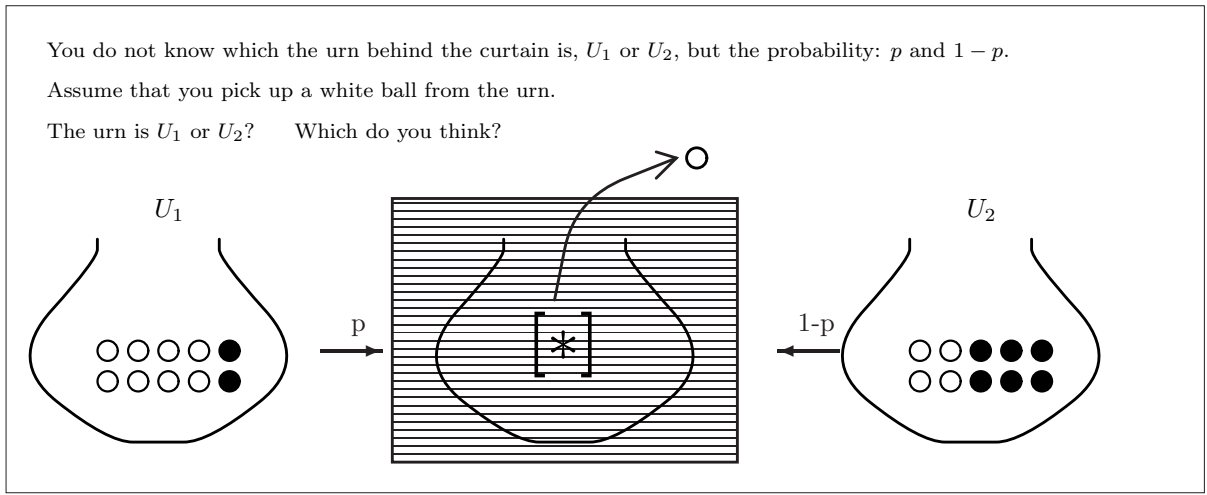
Which is the hidden urn,
$U_1$ or $U_2$?
( Mixed measurement)
}
\end{center}
%%%%%%%%%%%%%%%%%%

%\vskip0.5cm
\par
Now consider the following problem:
\begin{itemize}
\item[(e$_1$)]
Calculate the probability that
a white ball is picked out by
the mixed
{{measurement}}
${\mathsf M}_{C(\Omega)}({\mathsf O}, S_{[*]}(\nu_0))$!
%,
%white ballprobability .
\item[(e$_2$)]
And further,
when a white ball is picked out by
the mixed
{{measurement}}${\mathsf M}_{C(\Omega)}({\mathsf O}, S_{[*]}(\nu_0))$,
do you infer the unknown urn
$U_1$
or
$U_2$?
\end{itemize}
( See \textcolor{black}{({Fig.$\;$}4.12)}. )
%\END{itemize}

%%
\rm

\par
\noindent
\rm
{\bf Answer (e$_1$){$\;\;$}}
The following is clear:
\begin{itemize}
\item[(i)]
The possibility that "$[{}\ast{}] = {\omega_1} $"
is
$100p\%$.
Also,
The possibility that "$[{}\ast{}] = {\omega_2} $"
is
$100(1-p)\%$.
\end{itemize}
Further,
\begin{itemize}
\item[(ii)]
%
%$[{}\ast{}] = {\omega_1} $,
%That is,
the probability
that
a measured value
$x$
$({}\in \{ {{w}} , {{b}} \}{})$
is obtained by
a
{{measurement}}
${\mathsf M}_{C(\Omega)}({\mathsf O}, S_{[\omega_1]})$
is
\begin{align*}
&
[F(\{ x \}) ](
\omega_1
)
=
0.8
\; \;
(\text{when }x= {{w}}{}),
\quad
=
0.2
\; \;
(\text{when }x= {{b}}{})
\quad
\end{align*}
%,
%
%$[{}\ast{}] = {\omega_2} $,
%That is,
\\
the probability
that
a measured value
$x$
$({}\in \{ {{w}} , {{b}} \}{})$
is obtained by
a
{{measurement}}
${\mathsf M}_{C(\Omega)}({\mathsf O}, S_{[\omega_2]})$
is
\begin{align*}
&
[F(\{ x \}) ](
\omega_1
)
=
0.4
\; \;
(\text{when }x= {{w}}{}),
\quad
=
0.6
\; \;
(\text{when }x= {{b}}{})
\quad
\end{align*}
%,
%
\end{itemize}
\rm
%$\delta_{(\cdot)}$
%,
%$\nu_0 =p \delta_{\omega_1}
%+(1-p) \delta_{\omega_2}$
% and ,
%(i)+(ii)
%
%
%{{measurement}}
% and ,
%${\mathsf M}_{C(\Omega)}({\mathsf O}, S_{[{}\ast{}] }(\nu_0) )$
% and .
Therefore,
by
(i)
and
(ii),
the probability that
a measured value
$x$
$({}\in \{ {{w}} , {{b}} \}{})$
is obtained by
a mixed
{{measurement}}
${\mathsf M}_{C(\Omega)}({\mathsf O}, S_{[{}\ast{}] }(\nu_0) )$
is
\begin{align*}
P({}\{ x \}{})
&=
\int_\Omega
[F({}\{ x \}{})](
\omega)
\nu_0({}d \omega{})
=
p
[F({}\{ x \}{})](\omega_1)
+
(1-p)
[F({}\{ x \}{})](\omega_2)
\\
&=
\cases
0.8 p + 0.4 ({}1-p{})
\quad
&
(x={{w}}{}\; \text{ and })
\\
0.2 p + 0.6 ({}1-p{})
\quad
&
(x={{b}}{}\; \text{ and })
\endcases
%%P%\tag{5.42}
\tag{\color{black}{4.5}}
\end{align*}
This is the answer to
\textcolor{black}{{{Problem }}(e$_1$)}.
\qed
\par
\noindent
\rm
{\bf Answer(e$_2$){$\;\;$}}
\textcolor{black}{{{Problem }}(e$_2$)}
will be presented in \textcolor{black}{{Note }4.8}.
\qed

%BBBBBBBBBBBBBBBBBB%SBSBSBS
{\small%%{\footnotesize
\vspace{0.1cm}
\begin{itemize}
\item[$\spadesuit$] \bf {{}}{Note }4.7{{}} \rm
The following question is natural
That is,
\begin{itemize}
\item[]
In the above (i)
(or, (c)),
why  is "the possibility that $[{}\ast{}] = {\omega_1} $
is $100p\%$ $\cdots$"
replaced by
 "the probability that $[{}\ast{}] = {\omega_1} $
is $100p\%$ $\cdots$"
?
\end{itemize}
However,
the Copenhagen interpretation
says that
\begin{itemize}
\item[]
there is no probability without measurements.
\end{itemize}
This is the reason why the term
"probability" is not used.
%, {{measurement}}, probability 
\end{itemize}
}
%%BBBBBBBBBBBBBBBBBB%SBSBSBSS
\par
%\vskip1.0cm
\vskip1.0cm
\rm
\par
\subsubsection{Mixed {{measurement theory}}}%4.4.2
\par
\rm
As seen in the above,
we say that
%\textcolor{black}{Example 4.13},
\begin{itemize}
\item[(a)]
Pure {{measurement theory}}
is fundamental,
Adding the concept of
"mixed state",
we can construct
mixed {{measurement theory}}
as follows.
\begin{align*}
\underset{\text{\scriptsize ${\mathsf M}_{C({}\Omega{})} ({}{\mathsf O} , S_{[\ast ]}{(\nu)}) $}}{\text{{} $\fbox{
mixed {{measurement theory}}
}$}}
:=
{
\overset{\text{\scriptsize }}
{
\underset{\text{\scriptsize ${\mathsf M}_{C({}\Omega{})} ({}{\mathsf O} , S_{[\ast ]}{}) $}}{\text{{} $\fbox{
pure {{measurement theory}}
}$}}}
}
+
{
\overset{\text{\scriptsize }}
{
\underset{\text{\scriptsize $\nu$}}
{\text{{}$\fbox{ {mixed state}}$}}
}
}
%\footnotemark
%\TAG*{$\displaystyle{\mathop{1)}_{(=10))}}$}
%%%%%%%%%%%%%CLAsSICAL
%\TAG{3.0}
\end{align*}
{}
\end{itemize}

\rm
%\newpage

Thus we can present
the following
Axiom${}_{\text{\scriptsize c}}^{\text{\scriptsize m}}$ 1.
%%index{measurement12@${\mathsf M}_{C (\Omega)} \big({}{\mathsf O},
%$
%$
%S_{[{}\ast] }(\nu)
%\big)$:{(}continuous type {}){{measurement}}}
%index{@{{measurement}}}
%
%\BEGIN{itembox}[c]
\par
\noindent
\begin{center}
{\bf
Axiom${}_{\text{\scriptsize c}}^{\text{\scriptsize m}}$ 1 
(measurement: continuous mixed type)
%({{measurement}}) continuous$\cdot$mixed
}
\label{axiomcm1}
%\label{rule402}
\end{center}
%\END{document}
\par
\noindent
%\vskip0.1cm
\par
\noindent
\fbox{\parbox{155mm}{
\begin{itemize}
\item[]
%(ii)]
Consider a mixed{{measurement}}
${\mathsf M}_{C (\Omega)} \big({}{\mathsf O}{{=}}
$
$ ({}X, {\cal F} , F{}),
$
$
S_{[{}\ast] }(\nu)
\big)$
formulated in
$C(\Omega)$.
The,
the probability that
{{}}
a measured value
$ x$
$({}\in X  {})$
obtained by
the mixed {{measurement}}
${\mathsf M}_{C (\Omega)}  \bigl({}{\mathsf O}  , S_{[{}\ast{}] }(\nu) \bigl)$
belongs to
$ \Xi $
$({}\in  {\cal F}{})$
is given by
\begin{align*}
\int_\Omega
[F({}\Xi {})](
\omega)
\nu({}d \omega{})
%\
\end{align*}
%\hfill$22)$
%\END{
%\END{itemize}
\end{itemize}
}
}
\par
\vskip0.5cm
\par
\noindent

%
%
%
%
%
%\BEGIN{itembox}[c]{
%
%\label{axiomcm1}
%%index{1@Axiom${}_{\text{\scriptsize c}}^{\text{\scriptsize p}}$ 1[{}{{measurement}}{}(continuous type )]}
%%\label{rule501}
%%%index{ and @{statistics}probability }
%\BEGIN{itemize}
%%\item[(i)]
%%{{measurement}}.
%\item[]
%%(ii)]
%Consider a mixed{{measurement}}
%${\mathsf M}_{C (\Omega)} \big({}{\mathsf O}{{=}}
%$
%$ ({}X, {\cal F} , F{}),
%$
%$
%S_{[{}\ast] }(\nu)
%\big)$
%formulated in
%$C(\Omega)$.
%The,
%the probability that
%{{}}
%a measured value
%$ x$
%$({}\in X  {})$
%obtained by
%the mixed {{measurement}}
%${\mathsf M}_{C (\Omega)}  \bigl({}{\mathsf O}  , S_{[{}\ast{}] }(\nu) \bigl)$
%belongs to
%$ \Xi $
%$({}\in  {\cal F}{})$
%is given by
%\BEGIN{align*}
%\int_\Omega
%[F({}\Xi {})](
%\omega)
%\nu({}d \omega{})
%%\
%\END{align*}
%%\hfill$22)$
%%\END{align*}CRULE1
%%
%\END{itemize}
%\END{itembox}
%
%

\def\BC{\left[\begin{array}{ll}}
\def\EC{\end{array}\right.}
\rm
\par

%\vskip1.0cm

Thus we see:
\begin{itemize}
\item[(b)]
a mixed measurement is characterized as the following correspondence:
\begin{align*}
\text{
({mixed state}, observable )
}
\xrightarrow[\text{probabilistic}]{\quad \text{mixed {{measurement}}} \quad}
\text{measured value}
%P%\tag{5.44}
\end{align*}
\end{itemize}
\par
%,
%{FIG.$\;$}, {}
In {statistics},
both
"fluctuation "
and
"measurement error"
are represented by
random variable.
Thus,
we sometimes confuse the two.
On the other hand,
in {{measurement theory}},
the two have different mathematical structures,
that is,
\begin{itemize}
\item[(c)]
$
\cases
\text{(i)}:
\text{
[pure {{state}}]}
\xrightarrow[\text{\scriptsize fluctuation}]{}
[\text{{mixed state}}]
\\
\\
\text{(ii)}:
[\text{{exact observable}} ]
\xrightarrow[\text{\scriptsize measurement error}]{}
\text{[non-projective observable]}
\endcases
$
\end{itemize}
Thus,
in classical measurement theory,
we can avoid the confusions.
However,
as seen in
Heisenberg's {uncertainty principle}(\textcolor{black}{{{Theorem }}3.4}),
in quantum mechanics,
even pure state has a fluctuation.

\par
\noindent

\par
\vskip0.5cm

\par
The following example will promote the understanding of
mixed measurements.
%.
%\par
%\
%BFBF
\par
\noindent
{\bf
Example 4.14
[Coin-tossing]}
The coin-tossing
is  described in the following three
situations:
\par
\noindent
(i):
Let $\Omega$ be a state space of
states of a coin.
Let
$\omega_0$
$(\in \Omega )$
be a state of a fare coin.
Consider a {{measurement}}${\mathsf M}_{C(\{\Omega \})}({\mathsf O} {{=}} (\{h,t\},
2^{\{h,t\}}, F), S_{[\omega_0]})$
(where, "$h$"
and
$t$
respectively
means
"head"
and
"tail".
And
$F(\{h\})(\omega_0)=F(\{t\})(\omega_0)=1/2$).
Of course,
by the measurement,
the probability that
a measured value "$h$"
is obtained
is equal
to
$[F(\{h\})](\omega_0)=1/2$.
\par
\noindent
(ii):
Someone
(or, a robot)
tosses a fair coin.
And the measurement
is defined by
the check
of the result
(i.e.,
"$h$"
or
"$t$"
).
Define the state space
$\Omega$
such that
$\Omega=\{h,t\}$.
Then
the mixed exact measurement
is considered as
${\mathsf M}_{C(\{h,t\})}({\mathsf O}^{\FIN}, S_{[\ast]}
(\frac{1}{2}(\delta_h + \delta_t ) ))$
where
$\delta_{\omega}$($\in {\cal M}_{+1}(\Omega) $)
is the point measure at $\omega$.
Of course,
the probability that
a
measured value "$h$"
is obtained
is given by
$
\int_{\{h,t\}}
[F^{\FIN}(\{h\})](\omega)
\cdot
\frac{1}{2}(\delta_h + \delta_t )
(d \omega)
=1/2$.
%(\textcolor{black}{{Remark }6.27}).
\par
\noindent
(iii):
Someone
intentionally
grasped coin with the (right or left) hand,
and asked you "right or left?".
In this case,
you
may think
that
the right or the left is half-and-half.
The reason that you think so
is
due to
the principle of equal weight.
Here,
\begin{itemize}
\item[(d)]
the justification
of
the principle of equal weight
---
unless we have sufficient reason to regard one possible case
as more probable than another,
we treat them as equally probable
---
is the most famous problem in statistics.
\end{itemize}
%\footnote{
%{{Monty Hall problem}}{}
%}
%index{ and @}
This will be solved in
\textcolor{black}{{}{{{}}}Sec. 6.4.5}
({\rm cf.}
\textcolor{black}{\cite{Keio, IMont}}).

%\par
%\noindent
%\vskip0.3cm
\vskip0.3cm
\vskip0.3cm
%BFBF
\par
\noindent
{\bf
Remark 4.15
[Bayes' theorem in {{{measurement theory}}}]}
In this book,
we are not closely related to
mixed {{measurement}} theory.
But,
we have to mention only Bayes's theorem
in mixed measurement theory
as follows.
%, {{{measurement theory}}}
%, 
%({ordinary language} and ,
%\textcolor{black}{7}).
%%index{@({{{measurement theory}}})}
\begin{itemize}
\item[(e)]
[Bayes's theorem
in mixed measurement theory
(
{\rm cf.}
\textcolor{black}{{\cite{IMeas, IWhat}}})]
Consider a mixed
{{measurement}}
${\mathsf M}_{C (\Omega)} \big({}{\mathsf O}{{=}}
$
$ ({}X, {\cal F} , F{}),
$
$
S_{[{}\ast] }(\nu)
\big)$.
When we know that
a
{{}}measured value
%$ x$
%$({}\in X  {})$
belong to
$ \Xi $
$({}\in  {\cal F}{})$,
we can understand
that
the
{mixed state}$\nu$
changes to
a new {mixed state}$\nu_{\roman new} (\in {\cal M}_{+1}(\Omega ))$
such that
\begin{align*}
\nu_{\roman new}
(D)
=
\frac{\int_D
[F({}\Xi {})](
\omega)
\nu({}d \omega{})}{\int_\Omega
[F({}\Xi {})](
\omega)
\nu({}d \omega{})}
\qquad
(\forall D \in {\cal B}_\Omega )
%%%%P%\tag{5.45}
\tag{\color{black}{4.6}}
%%%%%REDREDREDREDREDRE
\end{align*}
\end{itemize}

\vskip0.3cm
\par
\noindent
%BBBBBBBBBBBBBBBBBB%SBSBSBS
{\small%%{\footnotesize
\begin{itemize}
\item[$\spadesuit$] \bf {{}}{Note }4.8{{}} \rm
Now we can answer \textcolor{black}{{{Problem }}(e$_2$)}
in\textcolor{black}{{Sec.$\;$}4.4.2}:
Since
"white ball"
is obtained by
a statical {{measurement}}\\
${\mathsf M}_{C(\Omega)}({\mathsf O}, S_{[*]}(\nu_0))$,
a new {mixed state}$\nu_{\roman new} (\in {\cal M}_{+1}(\Omega ))$
is given by
\begin{align*}
\nu_{\roman new}
(D)
&  % 
=
\frac{\int_D
[F({}\{ {{w}}\} {})](
\omega)
\nu_0 ({}d \omega{})}{\int_\Omega
[F({}
\{ {{w}}\}
{})](
\omega)
\nu_0({}d \omega{})}
=
\cases
\frac{\displaystyle 0.8 p
}{\displaystyle
0.8p+0.2(1-p)
}
\qquad
&
(\text{when }D=\{\omega_1 \})
\\
\\
\frac{\displaystyle 0.2(1- p)
}{\displaystyle
0.8p+0.2(1-p)
}
\quad
&
(\text{when }D=\{\omega_2 \})
\endcases
\end{align*}
\end{itemize}
}
%%BBBBBBBBBBBBBBBBBB%SBSBSBSS
\par

%\par
%\noindent
%%%%%
%%
%%
%%(a)?$\;\;$(b)?
%%, ,  and {}
%\END{itemize}
%}}.
%%%BBBBBBBBBBB

\par
\noindent
%BBBBBBBBBBBBBBBBBB%SBSBSBS
{\small%%{\footnotesize
\begin{itemize}
\item[$\spadesuit$] \bf {{}}{Note }4.9{{}} \rm
According to
the Copenhagen interpretation,
(cf. \textcolor{black}{{Note }4.6})
the (e)
in
\textcolor{black}{Remark 4.15}
should be described as the following
(e)$'$:
\textcolor{black}{
(
{\rm cf.}
\cite{IWhat}
)
}.
%\bf}$\;\;$%POPOPO.
\begin{itemize}
\item[(e)$'$]
Consider a simultaneous observable
${\mathsf O}\times  {\mathsf O}_1$
${{=}} (X \times Y , {\cal F}\times {\cal G} , F\times G)$
in
$C(\Omega)$,
and
a
simultaneous measurement
${\mathsf M}_{C (\Omega)}({\mathsf O} \times {\mathsf O}_1,$
$ S_{[*]}(\nu))$.

%.
%%When we know that{{}}
%{{measurement}}
%${\mathsf M}_{C
When we know that
a {measured value }{{}}
$(x, y )$
belongs to
$\Xi \times Y \;(\in {\cal F}\times {\cal G} )$,

the probability $P_\Gamma$
(i.e.,
the probability that
$y \in \Gamma (\in {\cal G} )$
)
is given by
\begin{align*}
P_\Gamma
=
\frac{\int_\Omega
[F({}\Xi {}) \cdot G({}\Gamma {})](
\omega)
\nu({}d \omega{})}{\int_\Omega
[F({}\Xi {})](
\omega)
\nu({}d \omega{})}
\qquad
\end{align*}
\end{itemize}
The \textcolor{black}{(e)}
should be regarded as the abbreviation
of the
{(e)$'$}.
\end{itemize}
}
%%BBBBBBBBBBBBBBBBBB%SBSBSBSSabbreviation;
\vskip0.3cm

\par
As an easy example,
we shall study {{Monty Hall problem}}
in {{{measurement theory}}}.
%\par
%\noindent
{\bf
%\vskip0.3cm
%\vskip0.3cm
%BFBF
\par
\noindent
{{Problem }}4.16{{}}
[{}{{Monty Hall problem}}
(Continued from \textcolor{black}{{{Problem }}4.9}){{}}
\rm
({\rm cf.}
\textcolor{black}{\cite{Keio,IMont}}
)
\bf
]}$\;\;$%POPOPO
\rm
%index{@{{Monty Hall problem}}}
%%index{paradox({}{{Monty Hall problem}}{})
%@paradox({}{{Monty Hall problem}}{})}
\rm
\par
\noindent

\rm
Suppose you are on a game show, and you are given
the choice of three doors
(i.e., {\lq\lq}number 1{\rq\rq}$\!\!\!,\;$ {\lq\lq}number 2{\rq\rq}$\!\!\!,\;$ {\lq\lq}number 3{\rq\rq}$\!\!)$.
Behind one door is a car, behind the others, goats.
You pick a door, say number 1.
Then,
the host,
who set a car behind a certain door,
says
\begin{itemize}
\rm
\item[\textcolor{black}{($\sharp_1$)}]
\rm
the car was set
behind the door
decided by the cast of the distorted dice.
That is,
the host set the car
behind the {\it $k$}-th door
({}i.e., {\lq\lq}number k{\rq\rq}$\!$)
with probability
$p_k$
(or,
weight
such that
$p_1 + p_2 + p_3 =1$, $ 0 \le p_1 , p_2 , p_3 \le 1 $
$)$.
%$p_1 = p_2 = p_3 = 1/3$
%$\Big)$
\end{itemize}
And further, the host says,
for example,
\begin{itemize}
\rm
\item[\textcolor{black}{($\flat$)}]
\rm
the door 3 has a goat.
\end{itemize}
He says to you,
{\lq\lq}Do you want to pick door number 2?{\rq\rq}
Is it to your advantage to switch your choice of doors?
\rm
\par

%%$F({}\{ 1 \}{})(\omega_1{})= 0.5$
%%and
%%$F({}\{ 2 \}{})(\omega_1{})= 0.5$
%%should be assumed in the
%%problem (P).
%}
%\BEGIN{align*}

\par
\noindent
{\bf Answer{$\;\;$}}
Recalling
\textcolor{black}{{{Problem }}4.9}
({{Monty Hall problem}}),
in what follows we study this problem.
Let
$\Omega$ and ${\mathsf O}_1$
be as in Section 3.1.
Under the hypothesis
($\sharp_1)$,
define the mixed state
$\nu_0$
$({}\in {\cal M}_{+1} ({}\Omega{}){})$
such that:
\begin{align*}
\nu_0 ({}\{ \omega_1 \}{}) = p_1,
\quad
\nu_0  ({}\{ \omega_2 \}{}) = p_2,
\quad
\nu_0  ({}\{ \omega_3 \}{}) = p_3
%%
%\tag{6}
\end{align*}
%where
%$p_1 + p_2 + p_3 =1$, $ 0 \le p_1 , p_2 , p_3 \le 1 $.
%%though
%it may suffice to assume that
%$p_1 = p_2 = p_3 = 1/3$.
Thus we have a mixed measurement
${\mathsf M}_{C({}\Omega{})} ({}{\mathsf O}_1, S_{[{}\ast{}]} ({}\{ \nu_0{}\}))$.
Note that
\begin{itemize}
\item[a)]
{\lq\lq}measured value $1$ is obtained
by the mixed measurement
${\mathsf M}_{C({}\Omega{})} ({}{\mathsf O}_1,
S_{[{}\ast{}]} ({}\{ \nu_0{}\}))${\rq\rq}
\\
$
\Leftrightarrow \text{the host says {\lq\lq}Door 1
has a goat{\rq\rq}}
$
%\\
%$
%%&
%\text{The probability is given by}
%\\
%\int_\Omega [F(\{1\})](\omega) \nu(d\omega)
%=
%0
%$
\item[b)]
{\lq\lq}measured value $2$ is obtained
by the mixed measurement
${\mathsf M}_{C({}\Omega{})} ({}{\mathsf O}_1,
S_{[{}\ast{}]} ({}\{ \nu_0{}\}))${\rq\rq}
\\
$
\Leftrightarrow \text{the host says {\lq\lq}Door 2
has a goat{\rq\rq}}
$
%\\
%$
%%&
%\text{The probability is given by}
%\\
%\int_\Omega [F(\{2\})](\omega) \nu(d\omega)
%=
%0.5 p_1 + 1.0 p_3
%$
\item[c)]
{\lq\lq}measured value $3$ is obtained
by the mixed measurement
${\mathsf M}_{C({}\Omega{})} ({}{\mathsf O}_1,
S_{[{}\ast{}]} ({}\{ \nu_0{}\}))${\rq\rq}
\\
$
\Leftrightarrow
$
the host says {\lq\lq}Door 3
has a goat{\rq\rq}
%&
%\text{The probability is given by}
%\\
%\int_\Omega [F(\{1\})](\omega) \nu(d\omega)
%=
%0.5 p_1 + 1.0 p_2
%$
\end{itemize}
\par
\noindent
Here, assume that,
by the
mixed
measurement
${\mathsf M}_{C({}\Omega{})} ({}{\mathsf O}_1, S_{[{}\ast{}]} ({}\nu_0{}))$,
you obtain a measured value $3$,
%that is, {\lq\lq}not 2{\rq\rq}$\!\!\!\!.\; \
which corresponds to
the fact that
the host said
{\lq\lq}Door 3 has a goat{\rq\rq}$\!\!\!.\;$
Then,
Theorem 3
(Bayes' theorem)
says that
the posterior state $\nu_{\rm post}$
$({}\in {\cal M}_{+1} ({}\Omega{}){})$
is given by
\begin{align*}
\nu_{\rm post}
= \frac{F_1(\{3\}) \times \nu_0}
%% {\sideset{}_{{}_{{\cal M}(\Omega)}} \and {}_{{}_{C(\Omega)}} \to
{\bigl\langle \nu_0, F_1(\{3\})
\bigr\rangle}.
%%%4
%%
%\tag{7}
\end{align*}
That is,
\begin{align*}
&
%\displaystyle{
%\mathop{
%\t
%{\lq\lq}\nu_0"
%}_{\s
%\longrightarrow
\nu_{\rm post} ({}\{ \omega_1 \}{})= \frac{\frac{p_1}{2}}{ \frac{p_1}{2} + p_2 },
\quad
\nu_{\rm post} ({}\{ \omega_2 \}{})= \frac{p_2}{ \frac{p_1}{2} + p_2 },
\quad
\nu_{\rm post} ({}\{ \omega_3 \}{}) =  0.
%%%4
%%
%\tag{8}
\end{align*}
Particularly,
we see that
\begin{itemize}
\item[($\sharp_2$)]
if
$p_1 = p_2 = p_3 = 1/3$,
then
it holds that
$\nu_{\rm post} ({}\{ \omega_1 \}{})=1/3$,
$\nu_{\rm post} ({}\{ \omega_2 \}{})=2/3$,
$\nu_{\rm post} ({}\{ \omega_3 \}{})=0$,
\it
and thus, you should pick
Door 2.
\rm
\end{itemize}

%\par
%\noindent
%{\it Remark 2}.
%The difference between
%Problem 1
%and
%Problem 2
%should be remarked.
%Since the ($\sharp_1$) in Problem 2
%%is the information from the host to you,
%Remark 2%Problem 1 and Problem 2 are completely different.
%%Although
%%the above (H) may be generally regarded as the
%proper answer of the Monty Hall
%problem,
%we do not admit that the (H) is proper.
%That is,
%we consider that
%the (H) is not the second answer to the Monty Hall problem.
%%as the answer to the Monty Hall
%%problem.
%%\hfill{$///$}

%BBBBBBBBBBBBBBBBBB%SBSBSBS
\par
\noindent
{\small%%{\footnotesize
\vspace{0.1cm}
\begin{itemize}
\item[$\spadesuit$] \bf {{}}{Note }4.10{{}} \rm
It is not natural to
assume
the rule ($\sharp_1$)
in
\textcolor{black}{{{Problem }}4.16}.
That is because
the host may intentionally set
the car behind a certain door.
Thus we think that
\textcolor{black}{{{Problem }}4.16}
is temporary.
For our formal proposal,
see
\textcolor{black}{{}{{{}}}Sec. 6.4.5}.
\end{itemize}
}
%%BBBBBBBBBBBBBBBBBequilibrium mixed mechanics
%
\baselineskip=18pt
\subsubsection{Entropy
---
The value of eyewitness information}%4.4.3
\normalsize \baselineskip=18pt
\par
\rm
\renewcommand{\footnoterule}{%
  \vspace{2mm}                      % 
  \noindent\rule{\textwidth}{0.4pt}   % , 
  \vspace{-3mm}
}
\rm
\par
As one of applications
({}of Bayes theorem{}),
we now study
the {\lq\lq entropy\rq\rq}
of the measurement(cf. \cite{Shan}).
Here we have the following definition.

%\footnote{
%,
%\textcolor{black}{{Note }9.3}.
%}.
\normalsize \baselineskip=18pt
\par
\noindent
\baselineskip=18pt
\par
\normalsize \baselineskip=18pt
\vskip0.2cm
\par
\noindent
\baselineskip=18pt
\par
\noindent
\bf %%BFBF
{Definition }4.17
[Entropy({\rm cf. }{\rm
\textcolor{black}{\cite{IStat}})}]$\;\;$%POPOPO
\rm
\baselineskip=18pt
Consider a mixed measurement
${\bold M}_{C({}\Omega{})} $
$\big({}{\bold O} \equiv ({}X , 2^X , F {}), $
$ S({}\rho_0{}) \big) $
in
a commutative basic algebra
$C({}\Omega{})$,
%$(${}%i.e.,
%$\rho_0 ({}\Omega{})=1$
%$)$,
where
the label set
$X$
is assumed to be at most countable,
i.e.,
$X$
$=$
$\{ x_1 , x_2 , ..., x_n , ... \}$.
%${\bold M}_{C_0 ({}\Omega{}){})} \big({}{\bold O} , S({}\rho_0{}) \big) $.
Then,
the
$H({}{\bold M}{})$,
the ({}fuzzy{}) entropy of
${\bold M}_{C({}\Omega{})} ({}{\bold O} , S({}\rho_0{}){})$,
{}\rm{}
is defined by

%\allowdisplaybreaks
\begin{align*}
&
\; \;
H
\Bigl({} {\bold M}_{C({}\Omega{})} ({}{\bold O} , S({}\rho_0{}){}) \Bigl)
\\
&
=
\sum_{n=1}^\infty
\Big(\int_{\Omega} [{}F({}\{ x_n \}{})] ({}\omega{})  \rho_0 ({}d \omega{})
\int_\Omega
      \frac{ [{}F({}\{ x_n \}{})] ({}\omega {}) }
             {{ \int_{\Omega}  [{}F({}\{ x_n \}{})] ({}\omega {})  \rho_0 ({}d \omega {}) } }
\log
      \frac{ [{}F({}\{ x_n \}{})] ({}\omega {}) }
             {{ \int_{\Omega}  [{}F({}\{ x_n \}{})] ({}\omega {})  \rho_0 ({}d \omega {}) } }
\rho_0 ({}d \omega{})
\Big)
%\tag{8.56}
\\
&
=
\sum_{n=1}^\infty
%\Big(
P({}\{ x_n \}{})
\cdot
I({}\{ x_n \}{})
\tag{4.7}
\end{align*}
\par
\noindent
\vskip-0.5cm
\par
\begin{align*}
\text{where, }
P({}\{ x_n \}{})
&
=
\int_{\Omega} [{}F({}\{ x_n \}{})] ({}\omega{})  \rho_0 ({}d \omega{})
\\
\Big(&
=
\text{
the probability that
a measured value
$x_n$ is obtained}
\Big)
\\
\\
I({}\{ x_n \}{})
&
=
\int_\Omega
      \frac{ [{}F({}\{ x_n \}{})] ({}\omega {}) }
             {{ \int_{\Omega}  [{}F({}\{ x_n \}{})] ({}\omega {})  \rho_0 ({}d \omega {}) } }
\log
      \frac{ [{}F({}\{ x_n \}{})] ({}\omega {}) }
             {{ \int_{\Omega}  [{}F({}\{ x_n \}{})] ({}\omega {})  \rho_0 ({}d \omega {}) } }
\rho_0 ({}d \omega{})
\\
&
=
\frac{1}{P({}\{ x_n \}{})}
\int_{\Omega}  [{}F({}\{ x_n \}{})] ({}\omega {})
\log
 [{}F({}\{ x_n \}{})] ({}\omega {})
\rho_0 ({}d \omega)
-
\log
P({}\{ x_n \}{})
\\
\Big(&
=
\text{ the information quantity
when
a measured value
$x_n$ is obtained}
\Big)
\end{align*}
%
%
%\int_\Omega
%      \frac{ [{}F({}\{ x_n \}{})] ({}\omega {}) }
%             {{ \int_{\Omega}  [{}F({}\{ x_n \}{})] ({}\omega {})  \rho_0 ({}d \omega {}) } }
%\log
%      \frac{ [{}F({}\{ x_n \}{})] ({}\omega {}) }
%             {{ \int_{\Omega}  [{}F({}\{ x_n \}{})] ({}\omega {})  \rho_0 ({}d \omega {}) } }
%\rho_0 ({}d \omega{})
%\tag{8.57}
%%%%%\eqno 5.6
%\END{align*}

The following is clear:
\begin{align*}
H({}{\bold M}{})
=
\sum_{n=1}^\infty \int_\Omega [{}F({}\{ x_n \}{})] ({}\omega{})
\log [{}F({}\{ x_n \}{})] ({}\omega{})   \rho_0 ({}d \omega{})
- \sum_{n=1}^\infty  P({}\{ x_n \}) \log  P(\{ x_n \}).
%%7
\tag{4.8}
\end{align*}
\par
\noindent
\par
\rm
\def\FND{\scriptscriptstyle{\roman{EXA}}}
\par
%\noindent

%%%%%%%%%%%%%%%%%%%%%%%%%%%%%%%%%%%%%

%%%BFBF
\par
\noindent
\par
\noindent
\bf
Example 4.18 [The offender is man or female?$\;$fast or slow?]
\rm
$\;\;$
Assume that
\begin{itemize}
\item[(a)]
There are
100 suspected persons
such as
$\{ s_1 , s_2 ,\ldots, s_{100} \}$,
in which
there is one criminal.
%$\{ \omega_1 , \omega_2 ,\ldots, \omega_{100} \}$
\end{itemize}
Define the state space
$\Omega$
$ = $
$\{ \omega_1 , \omega_2 ,\ldots, \omega_{100} \}$
such that
$$
{\text{state}} \omega_n
\cdots
\text{the state such that suspect } {s_n}
\text{ is a criminal}
%{{state}}
\qquad
(n=1,2,...,100)
$$

Define a male-observable
${\mathsf O}_{\roman m}$
$=$
$(X = \{ y_{\roman m} , n_{\roman m}  \}  , 2^{X} , M {})$
in
$C(\Omega)$
by
\begin{align*}
&
[M({\{ y_{\roman m} \} } )]({}\omega_n{})
=
m_{ y_{\roman m}  }({}\omega_n{})
=
\cases
0
\quad
&
({}n \; \text{is odd}{})
\\
1
&
({}n \; \text{is even}{})
\endcases
\quad
\\
&
[M({\{ n_{\roman m} \} } )]({}\omega_n{})
=
m_{ n_{\roman m}  }({}\omega_n{})
=
1-
[M({\{ y_{\roman m} \} } )]({}\omega_n{})
\end{align*}
For example,
\begin{itemize}
\item[]
Taking a {{measurement}}
${\mathsf M}_{C ({}\Omega {})} ({}{\mathsf O}_{\roman m}  ,
S_{[\omega_{17} ]}{})
$
---
the sex of the criminal $s_{17}$
---,
we get the measured value $n_{\roman m}$(=).
\end{itemize}
Also,
define
the fast-observable
${\mathsf O}_{\roman f}$
$ = $
$({}Y= \{ y_{\roman f} , n_{\roman f}  \}  , 2^{Y} , F {})$
in
$C(\Omega)$
by
\begin{align*}
&
[F({\{ y_{\roman f} \} } )]({}\omega_n{})
=
f_{ y_{\roman f} }({}\omega_n{})
=
{\displaystyle \frac{n-1}{99}
},
\qquad
\\
&
[F({\{ n_{\roman f} \} } )]({}\omega_n{})
=
f_{n_{\roman f}  }({}\omega_n{})
=
1-
[F({\{ y_{\roman f} \} } )]({}\omega_n{})
\end{align*}
%\END{document}
%
According to the principle of equal weight
(due to \textcolor{black}{{{Theorem }}6.21}
later),
there is a reason to consider that
a
{mixed state}${\nu_0}$
$( \in {\cal M}_{+1} ({}\Omega{})$
is
equal to
the {{state}}$\nu_e$
such that
${\nu_0} ({}\{ \omega_n \}{})
=\nu_e ({}\{ \omega_n \}{}) = 1/100$
$({}\forall n{})$.
Thus,
consider two
mixed {{measurement}}
${\mathsf M}_{C ({}\Omega {})} ({}{\mathsf O}_{\roman m}  , S_{[\ast ]}({}\nu_e{}){}) $
and
${\mathsf M}_{C ({}\Omega {})} ({}{\mathsf O}_{\roman f}  , S_{[\ast ]}({}\nu_e{}){}) $.
then,
by
\textcolor{black}{(4.8)},
we see:
\par
\noindent
\begin{eqnarray*}
H \big({\mathsf M}_{C ({}\Omega {})} ({}{\mathsf O}_{\roman m}  , S_{[\ast ]}({}\nu_e{})
&=& \int_\Omega m_{ y_{\roman m} } ({}\omega{})   \nu_e ({}d \omega{})
\cdot 
\log \int_{\Omega}
m_{ y_{\roman m}  } ({}\omega{})   \nu_e ({}d \omega{}) \\
& & -
\int_\Omega m_{\{ n_{\roman m}  \} } ({}\omega{})   \nu_e ({}d \omega{})
\cdot
\log \int_\Omega
m_{ n_{\roman m}  } ({}\omega{})   \nu_e ({}d \omega{})\\
&=&
- \frac{1}{2}  \log \frac{1}{2}
- \frac{1}{2}  \log \frac{1}{2}
=\log_2  2 = 1  \; \; \text{(bit)}
%\footnotemark{.}
\end{eqnarray*}

%---------------------------------------
%\BEGIN{align*}
%&
%\; \; \;
%H \big({\mathsf M}_{C ({}\Omega {})} ({}{\mathsf O}_{\roman m}  , S_{[\ast ]}({}\nu_e{}){})
%\big)
%\\
%&
%=
%{\small
%\text{
%$
%-
%\int_\Omega m_{ y_{\roman m} } ({}\omega{})   \nu_e ({}d \omega{})
%\cdot 
%\log \int_{\Omega}
%m_{ y_{\roman m}  } ({}\omega{})   \nu_e ({}d \omega{})
%-
%\int_\Omega m_{\{ n_{\roman m}  \} } ({}\omega{})   \nu_e ({}d \omega{})
%\cdot
%\log \int_\Omega
%m_{ n_{\roman m}  } ({}\omega{})   \nu_e ({}d \omega{})
%$
%}
%}
%\\
%&
%=
%- \frac{1}{2}  \log \frac{1}{2}
%- \frac{1}{2}  \log \frac{1}{2}
%=
%\log_2  2 = 1  \; \; \text{({}{})}\footnotemark{.}
%\END{align*}
%------------------------------------------------------------

{, }
\begin{eqnarray*}
H \big({\mathsf M}_{C ({}\Omega {})} ({}{\mathsf O}_{\roman f}  , S_{[\ast ]}({}\nu_e{}){})
\big) &=&
\int_\Omega f_{ y_{\roman f}  } ({}\omega{})     \log f_{ y_{\roman f}  } ({}\omega{})   \nu_e ({}d \omega{}) \\
 && \hspace{-4cm}+
\int_\Omega f_{ n_{\roman f}  } ({}\omega{})     \log f_{ n_{\roman f}  } ({}\omega{})   \nu_e ({}d \omega{})
 -
\int_\Omega f_{ y_{\roman f}  } ({}\omega{})   \nu_e ({}d \omega{})
\cdot 
\log \int_{\Omega}
f_{ y_{\roman f}  } ({}\omega{})   \nu_e ({}d \omega{})
\\
&& \hspace{-4cm}
-
\int_\Omega f_{ n_{\roman f}  } ({}\omega{})   \nu_e ({}d \omega{})
\cdot
\log \int_\Omega
f_{ n_{\roman f} } ({}\omega{})   \nu_e ({}d \omega{})
\\
&& \hspace{-4cm}{\doteqdot}
2 \int_0^1 \lambda  \log_2 \lambda  d \lambda  +1
=
- \frac{1}{ 2 \log_e 2} +1
=
0.278 \ldots \text{(bit)}
\end{eqnarray*}

%--------------------------------------------------------
%\BEGIN{align*}
%&H \big({\mathsf M}_{C ({}\Omega {})} ({}{\mathsf O}_{\roman f}  , S_{[\ast ]}({}\nu_e{}){})
%\big)
%\\
%=&
%\int_\Omega f_{ y_{\roman f}  } ({}\omega{})     \log f_{ y_{\roman f}  } ({}\omega{})   \nu_e ({}d \omega{})
%+
%\int_\Omega f_{ n_{\roman f}  } ({}\omega{})     \log f_{ n_{\roman f}  } ({}\omega{})   \nu_e ({}d \omega{})
%\\
%&
%-
%\int_\Omega f_{ y_{\roman f}  } ({}\omega{})   \nu_e ({}d \omega{})
%\cdot 
%\log \int_{\Omega}
%f_{ y_{\roman f}  } ({}\omega{})   \nu_e ({}d \omega{})
%\\
%&
%-
%\int_\Omega f_{ n_{\roman f}  } ({}\omega{})   \nu_e ({}d \omega{})
%\cdot
%\log \int_\Omega
%f_{ n_{\roman f} } ({}\omega{})   \nu_e ({}d \omega{})
%\\
%{\doteqdot}
%&
%2 \int_0^1 \lambda  \log_2 \lambda  d \lambda  +1
%=
%- \frac{1}{ 2 \log_e 2} +1
%=
%0.278 \cdots
%\; \; \text{({}{})}
%\qquad \qquad
%\qquad \qquad
%\END{align*}
%
%\footnotetext{
%$\log$$2$ and  and 
%,
%
% and .
%}
Therefore,
as
 eyewitness information,
"male of female" has more valuable than
"fast or slow".

%=================================================
\par
\noindent
\newpage %\vskip2.0cm
%55555555555555555555555555555555\tag{\tag{
\bf
\section{{{Practical Logic}}
\label{Chap5}
}%{Chap.{\;}}{}
%%\vspace{-0.8cm}
\pagestyle{headings}
\par
\noindent
\par
\noindent
\begin{itemize}
\item[{}]
{
\small
\baselineskip=15pt
\par%[Abstract].
\rm
$\;\;\;\;$
The term
"practical logic"
means
the logic in measurement theory.
It is certain that pure logic
is merely a kind of rule
in mathematics (or meta-mathematics).
If it is so,
the logic
is not guaranteed to be applicable to our world.
For instance,
mathematical logic does not assure the following famous statement:
\begin{itemize}
\item[{($\sharp_1 $)}]
\it
Since Socrates is a man and all men are mortal,
it follows that Socrates is mortal.
\end{itemize}
\rm
That is,
we think that
\begin{itemize}
\item[{($\sharp_2 $)}]
the above
($\sharp_1 $)
is not clarified yet.
%the unsolved proble.
\end{itemize}
{
\par
\noindent
In this chapter, we prove the $(\sharp_1)$ in classical systems.
As seen in Note 3.2,
it should  be recalled that
syllogism does not hold
in quantum systems.
}
}
\end{itemize}
%index{Aristotles@Aristotles}
{\footnotesize
\baselineskip=11pt
\begin{itemize}
\item[{}]
%{\footnotesize
%\baselineskip=11pt
\par
\rm
%({ does not
%necessarily hold in quantum systems.)
%We can assert,
%
%that
%this theorem (q.2)
%guarantees that
%the above (q.2) ({}or,
%the statement $[{}\sharp{}]${}) is {\lq\lq}theoretical true{\rq\rq}$\!\!\!\!.\; \;$
%Several variants may be interesting.
%For example,
%under the condition that
%{\lq\lq}$A \Rightarrow B$,
%$B \Rightarrow C${\rq\rq}$\!\!\!,\;$
%we can assert a kind of conclusion such as
%{\lq\lq}$C \Rightarrow A${\rq\rq}$\!\!\!\!.\; \;$
%For completeness,
%{\lq\lq}pure logic{\rq\rq} and
%{\lq\lq}practical logic{\rq\rq}
%must not be confused.
%The former is a basic rule on which mathematics is founded.
%On the other hand,
%the latter is a collection of
%theorems
%({}whose forms are similar to
%that of {\lq\lq}pure logic{\rq\rq}$\!$) in measurement theory.
%All results in this chapter are due to
%${{{}}}$\cite{Ifu1}.
%Also, this chapter can be skipped
%if readers want to study statistics
%in the framework of Smeasurement theory firstly ({\it cf.} chapters 8).
%}
%}
\end{itemize}
%
%}

%
%
%\END{itemize}
%}
%%%%\ENDtopmatter

\normalsize \baselineskip=18pt

\def\BBbZ{{{\Bbb Z}}}

\baselineskip=18pt
\subsection{Reconsider the logic of ordinary language}%{Sec.5.1}
\rm
\par
Logic has various aspects
such as the following
(A$_1$)--(A$_3$):.
\begin{itemize}
\item[(A$_1$)]
[Logic in mathematics or mathematical logic].
It is natural to
believe that
logic in mathematics
is most reliable.
But,
it should be noted that
mathematical logic is independent of our world.
Thus,
mathematical logic
can say nothing to our world,
unless some interpretation is added.
For example,
the following {{syllogism}} is obvious.
\begin{itemize}
\item[($\sharp$)]
"$A \Rightarrow B$,
$B \Rightarrow C$"
 then $"A \Rightarrow C"$
\end{itemize}
This is a rule in mathematical logic.
However,
the {{syllogism}}($\sharp$)
is closed in mathematics,
and thus,
it is not guaranteed to be related to
our world.

%
%it can say ,  and ,
%,
%,
%\footnote{
%,
%${{\cdot}}$
%{world-description}
%
% and .
% and  and 
% and  and 
%
%{world-description}.
%}.
%.
\item[(A$_2$)]
[{Ordinary language}].
There is logic
buried in {ordinary language}.
This logic is various,
for example,
"logic in marital dispute",
"logic in a court",
etc.
However,
the rule of ordinary language is not clear,
and therefore,
the following
($\sharp_1$)--($\sharp_4$)
have the room for reconsideration
(\textcolor{black}{{{Problem }}5.1}
and
\textcolor{black}{{Note }5.1--5.3}).
\begin{itemize}
\item[($\sharp_1$)]
Since Socrates is a man and all men are mortal,
it follows that Socrates is mortal.
%{Socrates},
%, .   , {Socrates}.
\item[($\sharp_2$)]
Flying arrow is not moving
(Zeno's paradoxes)
\item[($\sharp_3$)]
I think, therefore I am.
(Descartes)
\item[($\sharp_4$)]
"1 (kg) + 1 (kg) =2 (kg)"
(Edison; the master of invention)
(cf. \textcolor{black}{{Sec.11.5}})
\end{itemize}
%,
% and , \textcolor{black}{({}A$_1$)}{{syllogism}}($\sharp$)
% and ,
% and {ordinary language}{Socrates}($\sharp_1$) and 
%{}
%, \textcolor{black}{{{Problem }}5.1} and
%\textcolor{black}{{Note }5.1--5.3}
%.
\item[({}A$_3$)]
[Logic in the {world-descriptions}].
As mentioned frequently, we have to say that
\begin{itemize}
\item[$(\sharp_1)$]
mathematics,
independent of our world,
can not assert anything to the world
without the world-description
(e.g.,
Newtonian mechanics,
measurement theory,
etc.
).
%, ,
%${{\cdot}}$.
\end{itemize}
The spirit of
{world-description}
says that,
%,
\begin{itemize}
\item[$(\sharp_2)$]
First,
describe each phenomenon
by a language
(induced by {world-description}).
Next,
calculate its numerical representation.
\end{itemize}
%For example,
%consider Newtonian mechanics.
%mechanical phnomenon
%is frst described in terms of Netonian mechanics.
%And further,
%it suffices to
%the desription is calculated.
%$_1$)$(\sharp)${{syllogism}}
%% and
%,
%({}A$_2$)$(\sharp_1)${ordinary language}{{syllogism}}
%{{measurement theory}}
%
%,
%
%{} and .
%That is,
%{{measurement theory}} and {linguistic world-description method}{metaphysics}
%(, {\bf {{practical logic}}} and ) and .
%\textcolor{black}{{Sec.5.2}}.
\end{itemize}
\par
As in the above,
there is various logic
(A$_1$)--(A$_3$).
However,
recall
our standing point
\textcolor{black}{3.5{(={Chap.$\;$1}(X$_4$))}}
\begin{itemize}
\item[]
Describe any theory
(which is not yet described by measurement theory)
in terms of measurement theory.
\end{itemize}

%\END{document}

%BBBBBBBBBBBBBBBBBB%SBSBSBS
\par
\noindent
{\small%%{\footnotesize
\vspace{0.1cm}
\begin{itemize}
\item[$\spadesuit$] \bf {{}}{Note }5.1{{}} \rm
As seen in
\textcolor{black}{{Chap.$\;$1}(X$_1$)},
we have the following diagram;
%index{@{Chap.$\;$1}(X$_1$)}
\\
\\
$\underset{(Chap, 1)}{\text{(X$_1$)}}$
$\overset{
}{\underset{\text{\scriptsize (before science)}}{
\text{
\fbox
{{\textcircled{\scriptsize 0}}
widely {ordinary language}}
}
}
}
$
$
\underset{\text{\scriptsize }}{\text{$\Longrightarrow$}}
$
$
\underset{\text{\scriptsize (Chap. 1(O))}}{\text{{world-description}}}
\cases
&
\!\!\!\!\!
\underset{\text{\scriptsize (Newtonian mechanics,etc.)}}{
{\textcircled{\scriptsize 1}}
{\text{realistic method} \qquad}
}
%}
\\
\\
&
\!\!\!\!\!
\underset{{\text{\scriptsize (measurement theory)}}}{
\text{\textcircled{\scriptsize 2}}
{\text{linguistic method}}
}
%{\text{\textcircled{\scriptsize 2}{linguistic method}(e.g., {{measurement theory}})}}
\endcases
$
\\
\\
Here, it is natural to see that
\begin{itemize}
\item[$(\sharp_1)$]
"widely {ordinary language}\textcircled{\scriptsize 0}"
includes
\begin{itemize}
\item[$(\flat)$]
the statement $(\sharp_1)$
--
$(\sharp_4)$
in
$
\text{\textcolor{black}{({}A$_2$)}}$,
%
%
%%\quad
%$\underset{\text{\scriptsize ({{Theorem }}5.11)}}{{\text{syllogism}}}$,
%$\quad
%\underset{\text{\scriptsize (\textcolor{black}{Example 7.6})}}{\text{{elementari school word problem}}}$,
%\\
%$
%\underset{\text{\scriptsize ((A$_2$) in Chap. 11)}}{\text{motion function method}}$,
$\underset{\text{\scriptsize (Chaps. 4,7,11)}}{\text{{statistics}
(={dynamical system theory})}}
$
\end{itemize}
\end{itemize}
%\END{itemize}
%}
%
%\text{\scriptsize PPPPPPPPPPPPPPP
%\END{document}

However,
the framework
of
{ordinary language}
is not clear.
Therefore,
according to
\textcolor{black}{Our standing point 3.5}
{{(={Chap.$\;$1}(X$_4$))}},
our purpose is to
reconsider
this $(\flat)$
in
{{measurement theory}}\textcircled{\scriptsize 2}
\end{itemize}
}
%%BBBBBBBBBBBBBBBBBequilibrium mixed mechanics
\par
\noindent

About 2500 years ago,
Zeno (BC490-BC430)
pointed out the ambiguity of
the logic in ordinary language
as follows.
%
%{ordinary language} and 
%,
%{ordinary language}.
%paradox
%(Flying arrow is not moving){}
%2500, {(BC490-BC430)}:
%%index{@paradox}
\par
\noindent
%%BFBFBF
{\bf
{{Problem }}5.1
[Flying arrow is not moving]}$\;\;$%POPOPO
\begin{itemize}
\item[({}B$_1$)][{{Problem }}]:
Is
flying arrow is moving
or not?
%\footnote{
%{physics}${{\cdot}}$, .
%}
\item[({}B$_2$)][Zeno's answer]
Consider a flying arrow.
In any one instant of time, the arrow is not moving.
Therefore,
If the arrow is motionless at every instant, and time is entirely composed of instants, then
motion is impossible.
\qed
\end{itemize}

\vskip0.3cm
\par
If Zonon's logic
\textcolor{black}{({}B$_2$)}
is not true,
it is natural to consider that
the
$(\sharp_1)$
--
$(\sharp_4)$
in
\textcolor{black}{({}A$_2$)}
in ordinary language
cannot be believed easily.
Hence,
Zeno's paradoxes urge us to
ask the following question:
\begin{itemize}
\item[({}B$_3$)]
By what kind of world-description should
the flying arrow be described?
%{flying arrow}{world-description method}?
\end{itemize}
%index{@{{unsolved problem}}(paradox)}
This problem
\textcolor{black}{
({}B$_3$)
}
may be the most famous unsolved problem in science.
This problem
\textcolor{black}{({}B$_3$)}
will be solved in \textcolor{black}{Answer 11.11} in Chap. 11.
%.
\par
\vskip0.5cm
\par
As an answer to {{Problem }}
\textcolor{black}{({}B$_3$)},
some consider Newtonian mechanics.
However,
the "flying arrow"
is a symbol of
various motion$\cdot$change,
e.g.,
the growth of a tree,
economic growth of a country,
etc.
Thus,
Newtonian mechanics is not proper.
Of course,
some may assert
Laplace's demon.
However,
we think that
even a physical supremacist
hesitates to say
Laplace's demon.

%BBBBBBBBBBBBBBBBBB%SBSBSBS
\par
\noindent
{\small%%{\footnotesize
\vspace{0.1cm}
\begin{itemize}
\item[$\spadesuit$] \bf {{}}{Note }5.2{{}} \rm
If we know
the present state of the universe
and
the kinetic equation
(=the theory of everything),
and if we calculate it,
we can know everything
(from past to future).
There may be a reason to believe this idea.
This intellect is often referred to as Laplace's demon.
% (and sometimes Laplace's Superman, after Hans Reichenbach).
Laplace's demon is sometimes discussed as
the realistic-view
over which the degree passed.
Although the discussion about
\begin{itemize}
\item[]
$\qquad$
Laplace's Demon  vs.  measurement theory
\end{itemize}
is interesting,
we do not concerned with
Laplace demon in this book.
%\END{itemize}
%
%
%
% and {Remark }{}
% and ,
% and , proverbalizing and {metaphysics}
% and .
%, {{w}}.
\end{itemize}
}
%%BBBBBBBBBBBBBBBBBequilibrium mixed mechanics
\par
\noindent

%BBBBBBBBBBBBBBBBBB%SBSBSBS
\par
\noindent
{\small%%{\footnotesize
\begin{itemize}
\item[$\spadesuit$] \bf {{}}{Note }5.3{{}} \rm
%,
%.
%\textcolor{black}{{Chap.$\;$1}}
%{{{measurement theory}}}({}Y),
%That is,
Recall the classification of measurement theory
a follows.
%index{@quantum mechanics(={{measurement theory}})}
\begin{itemize}
\item[$\underset{(Chap. 1)}{\text{({}Y)}}$]
$
\quad
\overset{}{\underset{}{{{\text{measurement theory}}}}}
\cases
\underset{}{\text{quantum {{{measurement theory}}}}}
%\text{\textcolor{black}{{Chap. 3}}
%}
\\
\underset{}{\text{classical {{{measurement theory}}}}}
\endcases
%\\
%\ENDcases
%\TAG{0.4}
$
\end{itemize}
If we, from the quantum mechanical point of view,
discuss
$(\sharp_1)$
--
$(\sharp_4)$
in
({}A$_2$).
these meaning
will be clearer.
In what follows,
we mention them
as our final answers:
\begin{itemize}
\item[$(\sharp_1)$]
the {{syllogism}}
---
$(\sharp_1)$
in
$(A_2)$
---
does not hold
in quantum systems
(\textcolor{black}{{Note }3.8}),
%{{syllogism}}
%
%),
but
it is true in classical systems
(\textcolor{black}{{{Theorem }}5.11(i)}).
\item[$(\sharp_2)$]
The $(\sharp_2)$
(flying arrow)
in
$(A_2)$]
---
the trajectories of electron
---
is meaningless (\textcolor{black}{{Note }11.8})
in quantum systems,
but
it is true in classical systems(\textcolor{black}{{Sec.11.5}}).
\item[$(\sharp_3)$]
The statement "I think, $\cdots$"
($(\sharp_3)$ in $(A_2)$)
is nonsense
in both classical and quantum systems
(\textcolor{black}{{Sec.5.3.2}}).
\item[$(\sharp_4)$]
Edison's problem
$(\sharp_4)$
is true(\textcolor{black}{{Sec.11.5}}).
\end{itemize}
If it be so,
the readers may agree that
it is worth while studying
$(\sharp_1)$
--
$(\sharp_4)$
in
({}A$_2$).
\end{itemize}
}
%%BBBBBBBBBBBBBBBBBequilibrium mixed mechanics
%
\subsection{{}{Quasi-product }{observable } and marginal observable {}}%{Sec.5.2}
%\ssubsubsection{{}{quasi-product }{observable }{}}
\par
{}
\noindent
\par
As a generalization of
a
simultaneous observable
(\textcolor{black}{{Definition }2.14}),
we introduce the following
"quasi-product observable".

{\bf
%\vskip0.3cm
%\vskip0.3cm
%BFBF
\par
\noindent
{Definition }5.2
[{}{Quasi-product }{observable }{}]}$\;\;$%POPOPO
%index{@{quasi-product }{observable }}
\rm
For each
$k = 1,2,\ldots, n$,
consider
an observable
${\mathsf O}_k$
${{=}}$
$(X_k , $
${\cal F}_k , $
$F_k{})$
in
${C ( \Omega )}$.
%.
%
%%
%$({}\bigtimes_{k=1}^n  X_k , \bigtimes_{k=1}^n)$
%{product measurable space} and .
An {observable }
${\mathsf O}_{12...n}$
${{=}}$
$({}\bigtimes_{k=1}^n  X_k ,$
$ \bigstimes_{k=1}^n{\cal F}_k , $
${F}_{12...n}{})$
in
$C(\Omega)$
is called a
{\bf {quasi-product }{observable }}
of
$\{{\mathsf O}_k \}_{k=1}^n$,
and denoted by
$
{\mathop{\qp}_{k=1,2,\ldots,n}}{\mathsf O}_k
$
=
$
({}\bigtimes_{k=1}^n  X_k , \bigstimes_{k=1}^n{\cal F}_k,
{\mathop{\qp}_{k=1,2,\ldots,n}
}F_k{})$,
if it satisfied that
\begin{align*}
&
{F}_{12...n}({}X_1 \times \cdots \times X_{k-1}
\times \Xi_k \times
X_{k+1}
\times
\cdots
\times
X_n
)
% \times \Xi_2 \times \cdots \times \Xi_n{})
=
F_k ({}\Xi_k{})
\\
&
\quad  \qquad \qquad
(\forall \Xi_k \in {\cal F}_k , \forall k =1,2,\ldots,n )
%F_2 ({}\Xi_2{}) \cdots  F_n ({}\Xi_n{}) .
%%%%2.24}
%P%\tag{6.2}
\end{align*}
%\it
Of course,
a
simultaneous observable
is a kind of
{quasi-product }observable,
and thus,
a
{quasi-product }observable
is not generally determined uniquely.

%BBBBBBBBBBBBBBBBBB%SBSBSBS
\par
\noindent
{\small%%{\footnotesize
\begin{itemize}
\item[$\spadesuit$] \bf {{}}{Note }5.4{{}} \rm
Recall \textcolor{black}{(Example 2.10;
Urn problem)}.
Putting
$\Omega=\{\omega_1, \omega_2 \}$,
we,
by \textcolor{black}{(2.5)},
define
an observable
${\mathsf O}_{{{w}}{{b}}} = ({} \{ {{w}}, {{b}} \}, 2^{\{ {{w}}, {{b}} \}  }  , F{})$
in $C(\Omega)$.
Now we can
define the quasi-product
observable
$
{\mathsf O}\mathop{\qp}{\mathsf O}
$
$
=
{\mathsf O}_{12} = ({} \{ {{w}}, {{b}} \}\times
\{ {{w}}, {{b}} \}
, 2^{\{ {{w}}, {{b}} \} \times \{ {{w}}, {{b}} \} }  ,$
$ F_{12}{})$
in
$C({}\Omega{})$
%\BEGIN{align*}
\begin{align*}
& F_{12}({}\{ ({{w}}, {{w}}) \}{})(\omega_1{})= \frac{8 \times 7}{90},
& \quad & F_{12}({}\{ ({{w}},{{b}}) \}{})(\omega_1{})= \frac{8 \times 2}{90}
\\
& F_{12}({}\{ ({{b}},{{w}}) \}{})(\omega_1{})= \frac{2 \times 8}{90},
& \quad & F_{12}({}\{ ({{b}},{{b}}) \}{})(\omega_1{})= \frac{2 \times 1}{90}
%\END{align*}
%,
%\BEGIN{align*}
\\
& F_{12}({}\{ ({{w}}, {{w}}) \}{})(\omega_2{})= \frac{4 \times 3}{90},
& \quad & F_{12}({}\{ ({{w}},{{b}}) \}{})(\omega_2{})= \frac{4 \times 6}{90}
\\
& F_{12}({}\{ ({{b}},{{w}}) \}{})(\omega_2{})= \frac{6 \times 4}{90},
& \quad & F_{12}({}\{ ({{b}},{{b}}) \}{})(\omega_2{})= \frac{6 \times 5}{90}
\end{align*}
Here,
note that
${\mathsf O}_{12}$
is a quasi-product observable
and not a simultaneous observable.
\end{itemize}
}
%%BBBBBBBBBBBBBBBBBequilibrium mixed mechanics
\par
\noindent

\par
\par
\noindent
{\bf
\vskip0.3cm
\vskip0.3cm
%BFBF
\par
\noindent
{Definition }5.3
[{}Marginal {observable }{}]}$\;\;$%POPOPO
%%index{henkansokuryou@marginal {observable } ({}{})}
\rm
Consider an
{observable }
${\mathsf O}_{12...n}$
${{=}}$
$({}\bigtimes_{k=1}^n  X_k ,$
$ \bigstimes_{k=1}^n{\cal F}_k , $
${F}_{12...n}{})$
in
${C ( \Omega )}$.
For each
$1{{\; \leqq \;}}j {{\; \leqq \;}}n$, define
${F}^{(j)}_{12...n}$
by
\begin{align*}
{F}^{(j)}_{12...n}
(\Xi_j)
=
F_{12...n}(X_1 \times \cdots \times X_{j-1} \times \Xi_j \times X_{j+1}
\times \cdots \times X_{n})
\quad (
\forall
\Xi_j \in {\cal F}_j )
\end{align*}
Here,
${\mathsf O}^{(j)}_{12...n}$
${{=}}$
$(X_j ,$
$ {\cal F}_j , $
${F}^{(j)}_{12...n}{})$
is an observable in $C(\Omega)$.
The
${\mathsf O}^{(j)}_{12...n}$
is called a
{\bf marginal observable }
{(}precisely,
$(j)$-marginal observable
{)}
of
${\mathsf O}_{12...n}$.
%
%
%.
This can be generalized as follows.
For example, putting
${\mathsf O}_{12...n}^{(12)}$
${{=}}$
$({} X_1 \times X_2 ,$
$ {\cal F}_1 \times {\cal F}_2, $
${F}^{(12)}_{12...n}{})$,
\begin{align*}
{F}^{(12)}_{12...n}(\Xi_1 \times \Xi_2)
=
F^{(12)}_{12...n}(\Xi_1 \times \Xi_2 \times X_3 \times \cdots \times X_{n})
\quad (
\forall
\Xi_1 \in {\cal F}_1,
\forall
\Xi_2 \in {\cal F}_2 )
\end{align*}
we have the observable
${\mathsf O}_{12...n}^{(12)}$
${{=}}$
$({} X_1 \times X_2 ,$
$ {\cal F}_1 \times {\cal F}_2, $
${F}^{(12)}_{12...n}{})$.
Of course,
it holds that
${F}_{12...n}={F}^{(12...n)}_{12...n}{}$.

\par
\vskip0.5cm
\par

\par

An observable
${\mathsf O}$
${{=}}$
$(X, {\cal F} , F{})$
in $C(\Omega)$
has another representation as follows.
%,
%\par
\begin{align*}
{\roman{Rep}}^{\Xi}_{\omega}[{}{\mathsf O}]
%=
%{\roman{Rep}}[{}({}X, 2^{X}, F)]
=
\Big[
[{}F(\Xi)]
(\omega),
[{}F(\Xi^c)]
(\omega)
\Big]
%P%\tag{6.4}
\end{align*}
$\Xi^c$
$=$
$\{ x \in X \;|\; x \notin \Xi \}$,
i.e.,
the copmpliment of $\Xi$.
%\textcolor{black}{%index{@}}
%\textcolor{black}{%index{comliment@$\Xi^c$:}}
%\rm
Similarly,
an
observable
${\mathsf O}_{12}$
${{=}}$
$({}
X_1 \times X_2,
{\cal F}_1 \times {\cal F}_2,
%F {{=}}
F_{12}{})$
in
$C (\Omega)$
is represented by
\begin{align*}
{\roman{Rep}}_\omega^{\Xi_1\times \Xi_2}[{}{\mathsf O}_{12}]
=
\bmatrix
[{}F_{12} (\Xi_1 \times \Xi_2)] (\omega)
&
[{}F_{12}  (\Xi_1 \times \Xi_2^c)] (\omega)
\\
{}[{}F_{12}  (\Xi_1^c \times \Xi_2)] (\omega)
&
[{}F_{12}(\Xi_1^c \times \Xi_2^c)] (\omega)
\endbmatrix
%%\bmatrix
%%[{}F_{12} (\Xi_1 \times \Xi_2)]  (\omega)
%%&
%%[{}F_{12}  (\Xi_1 \times \Xi_2^c)]
%%(\omega)
%%\\
%%[{}F_{12}  (\Xi_1^c \times \Xi_2)]
%%(\omega)
%%&
%%[{}F_{12}  (\Xi_1^c \times \Xi_2^c)]
%%(\omega)
%%\ENDbmatrix
%P%\tag{6.5}
\end{align*}
%%%%EEEEEE
where,
it should be noted that
\par
\noindent
\begin{align*}
&
[{}F_{12}  (\Xi_1 \times \Xi_2{})  {}]  (\omega)
+
[{}F_{12}  (\Xi_1 \times \Xi_2^c{}) ]
(\omega)
=
[{}F^{(1)}_{12}(\Xi_1)]
(\omega)
\\
&
[{}F_{12} (\Xi_1^c \times \Xi_2^c{}) ]
(\omega)
+
[{}F_{12} (\Xi_1^c \times \Xi_2{}) ]
(\omega)
=
[{}F^{(1)}_{12}
(\Xi_1^c{})]
(\omega)
\\
&
[{}F_{12}  (\Xi_1 \times \Xi_2{})  {}]  (\omega)
+
[{}F_{12}  (\Xi_1^c \times \Xi_2{}) ]
(\omega)
=
[{}F^{(2)}_{12}(\Xi_2)]
(\omega)
\\
&
[{}F_{12} (\Xi_1 \times \Xi_2^c{}) ]
(\omega)
+
[{}F_{12} (\Xi_1^c \times \Xi_2^c{}) ]
(\omega)
=
[{}F^{(2)}_{12}
(\Xi_2^c{})]
(\omega)
%P%\tag{6.6}
\end{align*}

\vskip0.3cm
\vskip0.3cm
\rm
\rm
%Lemma ,
%{quasi-product }{observable }
%
%.

\par
\noindent
{\bf
%BFBF
\par
\noindent
Lemma 5.4
[The condition of {quasi-product }{observables }{\rm
(cf. \textcolor{black}{\cite{IFuzz}})
}]}$\;\;$%POPOPO
%\BEGIN{align*}
%X_1  = \{ y_1 , n_1  \}
%\quad
%\text{{ and }}
%\quad
%X_2  = \{ y_2 , n_2  \}.
%%%%% 3.29
%
%\END{align*}
Let
${\mathsf O}_{1}$
${{=}}$
$({}
X_1,
{\cal F}_1 ,
%F {{=}}
F_{1}{})$
and
${\mathsf O}_{2}$
${{=}}$
$({}
X_2,
{\cal F}_2 ,
%F {{=}}
F_{2}{} )$
be observables in
$C (\Omega)$.
Let
${\mathsf O}_{12}$
${{=}}$
$({}
X_1 \times X_2,
{\cal F}_1 \times {\cal F}_2,
%F {{=}}
F_{12}{}{{=}} F_1
\mathop{\qp}
F_2
)$
be a quasi-product observable
of
${\mathsf O}_{1}$
and
${\mathsf O}_{2}$.
That is, it holds that
\begin{align*}
F_1 = F_{12}^{(1)},
\qquad
F_2= F_{12}^{(2)}
\end{align*}
Then, putting
$\alpha_{_{\Xi_1 \times \Xi_2}} (\omega)
=
[{}F_{12} (\Xi_1 \times \Xi_2)]  (\omega)
$,
we see
\par
\noindent
%%%
%\allowdisplaybreaks
\begin{align*}
&
\; \;
\roman{Rep}_\omega^{\Xi_1\times \Xi_2}[{}{\mathsf O}_{12}]
=
\bmatrix
[{}F_{12} (\Xi_1 \times \Xi_2)]  (\omega)
&
[{}F_{12}  (\Xi_1 \times \Xi_2^c)]
(\omega)
\\
{}
[{}F_{12}  (\Xi_1^c \times \Xi_2)]
(\omega)
&
[{}F_{12}  (\Xi_1^c \times \Xi_2^c)]
(\omega)
\endbmatrix
\\
=
&
\bmatrix
\alpha_{_{\Xi_1 \times \Xi_2}} (\omega)
&
{}[{}F_1 (\Xi_1]  (\omega) -  \alpha_{_{\Xi_1 \times \Xi_2}} (\omega)
\\
{}
[{}F_2 (\Xi_2](\omega) - \alpha_{_{\Xi_1 \times \Xi_2}} (\omega)
&
{}
1+  \alpha_{_{\Xi_1 \times \Xi_2}} (\omega)-[{}F_1 (\Xi_1](\omega)-[{}F_2 (\Xi_2](\omega)
\endbmatrix
%\TAG{6.7}
\tag{\color{black}{5.1}}
%%%%%REDREDREDREDREDRE
\end{align*}
%%E
%That is,
%%%\BEGIN{tabular}{|l|l|*{2}{@{\quad\$}r|}}
%\BEGIN{center}
%\BEGIN{tabular}{|c||c|c|l|r}
%\hline
%{\text{marginal observable }} &$\quad  [{}F_2 (\Xi_2)  {}]  (\omega) \quad$
%&
%$\quad  [{}F_2 (\Xi_2^c)  {}]  (\omega)  \quad$\\
%\hline
%\hline
%$ [{}F_1 (\Xi_1)  {}]  (\omega)$
%& $  \alpha_{_{\Xi_1 \times \Xi_2}} (\omega) $ &
%$  [{}F_1 (\Xi_1)  {}]  (\omega) -  \alpha_{_{\Xi_1 \times \Xi_2} (\omega) $  \\
%\hline
%$ [{}F_1 (\Xi_1^c)  {}]  (\omega)$ &
%$[{}F_2  (\Xi_2)  {}]  (\omega) - \alpha_{_{\Xi_1 \times \Xi_2} (\omega)$
%& $1+  \alpha_{_{\Xi_1 \times \Xi_2} (\omega)
%-
%[{}F_1 (\Xi_1)  {}]  (\omega)
%-
%[{}F_2  (\Xi_2)  {}]  (\omega)$  \\
%\hline
%%$\omega_3$  &  1$\times N$ & 9$\times N$  \\
%%\hline
%\END{tabular}
%\END{center}
%\par
%\noindent
and
\begin{align*}
&
\max  \{
0,
[{}F_1 ({}  \Xi_{1} {}{})  {}]  (\omega)
+
[{}F_2  ({} \Xi_{2}  {}{})  {}]  (\omega)
-1
{}
\}
{{\; \leqq \;}}
\alpha_{_{\Xi_1 \times \Xi_2}} (\omega)
{{\; \leqq \;}}
%\\
%&
%\hspace{5cm}
\min
\{
[{}F_1 ({}  \Xi_{1} {}{})  {}]  (\omega)
,
\;
[{}F_2  ({} \Xi_{2}  {}{})  {}]  (\omega)
\}
\\
&
\hspace{4cm}
(\forall \Xi_1  \in {\cal F}_1,
\forall \Xi_2 \in {\cal F}_2,
\forall \omega \in \Omega{})
%\TAG{6.8}
\tag{\color{black}{5.2}}
%%%%%REDREDREDREDREDRE
%%%%% 3.2
\end{align*}
Reversely,
for any
$\alpha_{_{\Xi_1 \times \Xi_2} }$
$({}\in C({}\Omega{}){})$
such that
\textcolor{black}{(5.2)},
the observable ${\mathsf O}_{12}$
defined by
\textcolor{black}{(5.1)}
is a
{quasi-product }{observable}
of
${\mathsf O}_1$
 and
${\mathsf O}_2$.
Also,
it holds that
\par
\noindent
\begin{align*}
[{}F {}({} \Xi_{1} \times \Xi^c_{2}{}) {}{}  {}]  (\omega)
=0
\;
& \Longleftrightarrow
\;
\alpha_{_{\Xi_1 \times \Xi_2}} (\omega)=
[{}F_1 ({}  \Xi_{1} {}{})  {}]  (\omega)
%\\
%&
\Longrightarrow
\;
[{}F_1 ({}  \Xi_{1} {}{})  {}]  (\omega)
{{\; \leqq \;}}
[{}F_2 ({}  \Xi_{2} {}{})  {}]  (\omega)
%%%%%
\tag{\color{black}{5.3}}
%%%%%REDREDREDREDREDRE
\end{align*}
\par

Consider yes-no observables
${\mathsf O}_1$
$\equiv$
$({}X_1 , 2^{ X_1 }  , F_1{})$
and
${\mathsf O}_2$
$\equiv$
$({}X_2 ,$
$ 2^{ X_2 }  ,$
$ F_2{})$
in
%a commutative basic algebra
$C({}\Omega {})$
such that:
\begin{align*}
X_1  = \{ y_1 , n_1  \}
\quad
\text{ and }
\quad
X_2  = \{ y_2 , n_2  \}.
%\TAG 3.29
\end{align*}
Let
${\mathsf O}_{12}$
$\equiv$
$({}
X_1 \bigtimes X_2,
2^{
X_1 \bigtimes X_2 },
F
\equiv F_1 \bigtimes^{{\mathsf O}_{12}} F_2{})$
be a quasi-product observable
with the marginal observables
${\mathsf O}_1$
and
${\mathsf O}_2$.
\par
\noindent
Put
%%%
%\allowdisplaybreaks
\begin{align*}
&
\; \;
\roman{Rep}[{}{\mathsf O}_{12}]
=
\bmatrix
[{}F ({}\{({} y_{1} , y_{2}{}) \} {})  {}]  ({}\omega{})
&
[{}F ({}\{({} y_{1} , n_{2}{}) \} {})  {}]  ({}\omega{})
\\
{}[{}F ({}\{({} n_{1} , y_{2}{}) \} {})  {}]  ({}\omega{})
&
[{}F ({}\{({} n_{1} , n_{2}{}) \} {})  {}]  ({}\omega{})
\endbmatrix
\nonumber
\\
&
=
\bmatrix
\alpha ({}\omega{})
&
[{}F_1 ({}\{  y_{1} \} {})  {}]  ({}\omega{}) -  \alpha ({}\omega{})
\\
{}[{}F_2  ({}\{ y_{2}  \} {})  {}]  ({}\omega{}) - \alpha ({}\omega{})
&
1+  \alpha ({}\omega{})
-
[{}F_1 ({}\{  y_{1} \} {})  {}]  ({}\omega{})
-
[{}F_2  ({}\{ y_{2}  \} {})  {}]  ({}\omega{})
\endbmatrix,
%\label{3.3}
\textcolor{black}{\label{3.3}}
\end{align*}
%%%%
where
$\alpha \in C({}\Omega{})$.
$\Bigl($Note that
$ [{}F ({}\{({} y_{1} , y_{2}{}) \} {})  {}]  ({}\omega{})$
$+$
$ [{}F ({}\{({} y_{1} , n_{2}{}) \} {})  {}]  ({}\omega{})$
$=$
$ [{}F_1 ({}\{ y_{1} \} {})  {}]  ({}\omega{})$
and
$ [{}F ({}\{({} y_{1} , y_{2}{}) \} {})  {}]  ({}\omega{})$
$+$
$ [{}F ({}\{({} n_{1} , y_{2}{}) \} {})  {}]  ({}\omega{})$
$=$
$ [{}F_2 ({}\{ y_{2} \} {})  {}]  ({}\omega{})${}$\Bigl)$.
\par
\noindent
That is,

\begin{center}
%\BEGIN{tabular}{||c||c|c||l|r}
\begin{tabular}{
@{\vrule width 1.8pt\ }c
@{\vrule width 1.8pt\ }c|c
@{\vrule width 1.8pt }}
\noalign{\hrule height 1.8pt}
$F_1 \diagdown F_2$
{} &$\quad  [{}F_2 ({}\{ y_{2} \} {})  {}]  ({}\omega{}) \quad$
&
$\quad  [{}F_2 ({}\{ n_{2} \} {})  {}]  ({}\omega{})  \quad$
\\
\noalign{\hrule height 1.8pt}
$ [{}F_1 ({}\{ y_{1} \} {})  {}]  ({}\omega{})$
& $  \alpha ({}\omega{}) $ &
$  [{}F_1 ({}\{  y_{1} \} {})  {}]  ({}\omega{}) -  \alpha ({}\omega{}) $ 
\\
\hline
$ [{}F_1 ({}\{ n_{1} \} {})  {}]  ({}\omega{})$ &
$[{}F_2  ({}\{ y_{2}  \} {})  {}]  ({}\omega{}) - \alpha ({}\omega{})$
& $1+  \alpha ({}\omega{})
-
[{}F_1 ({}\{  y_{1} \} {})  {}]  ({}\omega{})
-
[{}F_2  ({}\{ y_{2}  \} {})  {}]  ({}\omega{})$ 
\\
%$\omega_3$  &  1$\times N$ & 9$\times N$  \\
%\hline
\noalign{\hrule height 1.8pt}
\end{tabular}
\end{center}
\par
\noindent
Then,
it holds that
\begin{align*}
\max \{ 0,
[{}F_1 ({}\{  y_{1} \} {})  {}]  ({}\omega{})
+
[{}F_2  ({}\{ y_{2}  \} {})  {}]  ({}\omega{})
-1
\}
\le
\alpha ({}\omega{})
&
\le
\min \{
[{}F_1 ({}\{  y_{1} \} {})  {}]  ({}\omega{})
,
\;
[{}F_2  ({}\{ y_{2}  \} {})  {}]  ({}\omega{})
\}
\nonumber
\\
&
(\forall \omega \in \Omega{}) .
%\label{3.4}
%%\label 3.2
\end{align*}
Conversely,
for any
$\alpha $
$({}\in C({}\Omega{}){})$
that satisfies \textcolor{black}{(5.2)},
the observable
${\mathsf O}_{12}$
defined by
%(7777.3)
\textcolor{black}{(5.1)}
is a quasi-product observable
with the marginal observables
${\mathsf O}_1$
and
${\mathsf O}_2$.
Also, note that
\par
\noindent
\begin{align*}
[{}F ({}\{({} y_{1} , n_{2}{}) \} {})  {}]  ({}\omega{})
=0
\;
\Leftrightarrow
\;
\alpha ({}\omega{})=
[{}F_1 ({}\{  y_{1} \} {})  {}]  ({}\omega{})
\;
\Rightarrow
\;
[{}F_1 ({}\{  y_{1} \} {})  {}]  ({}\omega{})
\le
[{}F_2 ({}\{  y_{2} \} {})  {}]  ({}\omega{}) .
%%\label 3.3
%\label{3.5}
\end{align*}
\par %\hf

\noindent
%\it
\par
\noindent
{\it $\;\;\;\;${Proof.}}{$\;\;$}
\rm
%, .
%$ 0 {{\; \leqq \;}}$
%
%\par
%\noindent
%\it
%\par {\it Proof.}
\rm
Though this lemma is easy, we add a brief proof for completeness.
Since
$ 0 \le $
\linebreak
$ [{}F ({}\{({} x_{1}^1 , x_{2}^2{}) \} {})  {}] $
$  ({}\omega{})$
$ \le 1$,
$({}\forall  x^1 , x^2 \in \{ y , n \}{})$,
we see, by
%(7777.3),
\textcolor{black}{(5.1)}
that
\begin{align*}
&
0 \le \alpha ({}\omega{})  \le 1,
\quad
0 \le
[{}F_1 ({}\{  y_{1} \} {})  {}]  ({}\omega{})
-  \alpha ({}\omega{})
\le 1,
\quad
0 \le
[{}F_2  ({}\{ y_{2}  \} {})  {}]  ({}\omega{})
- \alpha ({}\omega{})
\le 1,
\nonumber
\\
&
0
\le
1+  \alpha ({}\omega{})
-
[{}F_1 ({}\{  y_{1} \} {})  {}]  ({}\omega{})
-
[{}F_2  ({}\{ y_{2}  \} {})  {}]  ({}\omega{})
\le 1  ,
%%\label{3.6}
\end{align*}
%\nonumber \\
%\ENDaligned
%%%\label 3.4
%
%
which clearly implies \textcolor{black}{(5.2)}.
Conversely.
if
$\alpha$
satisfies \textcolor{black}{(5.2)},
then
we easily see \textcolor{black}{(5.1)}.
Also,
\textcolor{black}{(5.2)} is obvious.
This completes the proof.
\qed
\par
\noindent
{}
\par

\def\RD{\scriptscriptstyle{\roman{RD}}}
\def\RP{\scriptscriptstyle{\roman{RP}}}
\def\SW{\scriptscriptstyle{\roman{SW}}}
\par
\rm
%Next we provide several examples,
%which will promote a understanding of our theory.
%

\par

Let
${\mathsf O}_{12}$
${{=}}$
$({}
X_1 \times X_2,
{\cal F}_1 \times {\cal F}_2,
%F {{=}}
F_{12}{}{{=}} F_1
{\mathop{\qp}_{}}
F_2)$
be a quasi-product observable
(in $C(\omega )$)
of
${\mathsf O}_{1}$
${{=}}$
$({}
X_1,
{\cal F}_1 ,
%F {{=}}
F_{1}{})$
and
${\mathsf O}_{2}$
${{=}}$
$({}
X_2,
{\cal F}_2 ,
%F {{=}}
F_{2}{} )$.
Assume that
a measured value $(x_1, x_2)$
$(\in X_1 \times X_2 )$
is obtained by
a
{{measurement}}
${\mathsf M}_{C (\Omega)} ({\mathsf O}_{12} $
${{=}}(X_1 \times X_2,
{\cal F}_1 \times {\cal F}_2,
F_{12}{}{{=}} F_1
{\mathop{\qp}}
F_2), S_{[\omega]})$).
If we know that
$x_1 \in \Xi_1$,
then
we can calculate the probability
$P$
that
$x_2 \in \Xi_2$
(that is,
{\bf
the conditional probability
})
is given by
%\textcolor{black}{%index{@probability }}
\begin{align*}
P=
\frac{
[{}F_{12} (\Xi_1 \times \Xi_2{}){}]({}\omega{})
}
{
[{}F_{1} (\Xi_1 {}){}]({}\omega{})
%+
%[{}F_{12} (\Xi_1 \times \Xi_2^c){}]({}\omega{})
}
=
\frac{
[{}F_{12} (\Xi_1 \times \Xi_2{}){}]({}\omega{})
}
{
[{}F_{12} (\Xi_1 \times \Xi_2 {}){}]({}\omega{})
+
[{}F_{12} (\Xi_1 \times \Xi_2^c){}]({}\omega{})
}
%P%\tag{6.11}
%%%%% 3.2
\end{align*}
And it is,
by
\textcolor{black}{(5.2)},
estimated as follows.
\begin{align*}
&
\frac{\max  \{
0,
[{}F_1 ({}  \Xi_{1} {}{})  {}]  (\omega)
+
[{}F_2  ({} \Xi_{2}  {}{})  {}]  (\omega)
-1
{}
\}
}{{
[{}F_{12} (\Xi_1 \times \Xi_2){}]({}\omega{})
+
[{}F_{12} ({}\Xi_1 \times \Xi^c_2  {}){}]({}\omega{})
}}
%\\
{{\; \leqq \;}}
%&
P
%\frac{
%[{}F_{12} (\Xi_1 \times \Xi_2)  \}{}){}]({}\omega{})
%}
%{
%[{}F_{12} (\Xi_1 \times \Xi_2)  \}{}){}]({}\omega{})
%+
%[{}F_{12} (\Xi_1 \times \Xi_2^c){}]({}\omega{})
%}
%TAG{6.12}
%\\
{{\; \leqq \;}}
%\\
%&
%\hspace{4cm}
\frac{
\min
\{
[{}F_1 ({}  \Xi_{1} {}{})  {}]  (\omega)
,
\;
[{}F_2  ({} \Xi_{2}  {}{})  {}]  (\omega)
\}
}{{
[{}F_{12} (\Xi_1 \times \Xi_2{}){}]({}\omega{})
+
[{}F_{12} (\Xi_1 \times \Xi_2^c){}]({}\omega{})
}}
%P%\tag{6.12}
%%%%% 3.2
\end{align*}

%FFFFFFFFF\omega\omega_n

%\def\RD{\scriptscriptstyle{\roman{}}}
%\def\RP{\scriptscriptstyle{\roman{}}}
%\def\SW{\scriptscriptstyle{\roman{}}}
\par
\rm

\par
\noindent
{\bf
\vskip0.3cm
\vskip0.3cm
%BFBF
\par
\noindent
Example 5.5
[{}Tomatos{}]}$\;\;$%POPOPO
Let
$\Omega$
$   = $
$ \{ \omega_1 , \omega_2 , ...., \omega_N \}$
be a set of tomatoes,
which is regarded as
a compact Hausdorff space
with the
discrete topology.
%%Let $C({}\Omega{})$ be as in Example 2.2.
%Note that
%a tomato $\omega_n$
%is represented by a system $S_{[{}\delta_{\omega_n}]}$
%(%\it
%{\it cf.}
%\rm
%Example 2.3
%).
Consider yes-no observables
${\mathsf O}_{{\RD}}$
$\equiv$
$({}X_{\RD} , 2^{ X_{\RD} }  , F_{\RD}{})$
and
${\mathsf O}_{\SW}$
$\equiv$
$({}X_{\SW} , 2^{ X_{\SW} }  , F_{\SW}{})$
in
$C({}\Omega {})$
such that:
\begin{align*}
X_{\RD}
=
\{ y_{\RD} , n_{\RD} \}
\text{  and  }
X_{\SW}
=
\{ y_{\SW} , n_{\SW} \},
%\TAG 3.35
\end{align*}
where
we consider
that
{\lq\lq $y_{\RD}$\rq\rq}
and
{\lq\lq $n_{\RD}$\rq\rq}
respectively
mean
{\lq\lq RED\rq\rq}
and
{\lq\lq NOT RED\rq\rq}$\!.$
Similarly,
{\lq\lq $y_{\SW}$\rq\rq}
and
{\lq\lq $n_{\SW}$\rq\rq}
respectively mean
{\lq\lq SWEET\rq\rq}
and
{\lq\lq NOT SWEET\rq\rq}$\!.$
\par
\noindent
For example,
the $\omega_1$ is red and not sweet,
the $\omega_2$ is red and sweet,
etc. as follows.
\par
\vskip-0.5cm
\par

%$\omega_1$
%$\omega_2$, 
%.
\par
%\vskip-0.5cm
\par
\noindent
%%%
%\vskip-0.4cm%%%
%\begin{figure}[htbp]
\unitlength=0.25mm
%\unitlength=0.35mm
\begin{picture}(500,150)
%\put(27,18){0}
%\put(27,108){1}
%\put(350,18){$\Omega$}
%\dottedline{3}(40,110)(340,110)
%%\put(150,10){$\omega_0$}
%\put(20,90){\vector(1,0){295}}
%\put(20,20){\vector(0,1){150}}
\thicklines
\put(30,0)
{
{{
\put(0,80)
{
\put(0,0){\ellipse{50}{40}}
\path(-3,15)(10,30)
\spline(-10,13)(-3,10)(7,13)
\put(-8,-8){$\omega_1$}
}
\put(-13,45){$y_{{\RD}}$}
\put(-13,25){$n_{{\SW}}$}
}}
}
\put(130,0)
{
{{
\put(0,80)
{
\put(0,0){\ellipse{50}{40}}
\path(-3,15)(10,30)
\spline(-10,13)(-3,10)(7,13)
\put(-8,-8){$\omega_2$}
}
\put(-13,45){$y_{{\RD}}$}
\put(-13,25){$y_{{\SW}}$}
}}
}
\put(230,0)
{
{{
\put(0,80)
{
\put(0,0){\ellipse{50}{40}}
\path(-3,15)(10,30)
\spline(-10,13)(-3,10)(7,13)
\put(-8,-8){$\omega_3$}
}
\put(-13,45){$n_{{\RD}}$}
\put(-13,25){$y_{{\SW}}$}
}}
}
\put(310,0)
{
{{
\put(0,80)
{
\put(0,0){$\cdots$}
%{\ellipse{50}{40}}
%\path(0,20)(10,30)
%\put(-5,0){$\omega_1$}
}
\put(0,45){$\cdots$}%{$y_{\text{\scriptsize \RD}}$}
\put(0,25){$\cdots$}%{$n_{\text{\scriptsize \SW}}$}
}}
}
\put(400,0)
{
{{
\put(0,80)
{
\put(0,0){\ellipse{50}{40}}
\path(-3,15)(10,30)
\spline(-10,13)(-3,10)(7,13)
\put(-8,-8){$\omega_K$}
}
\put(-13,45){$n_{{\RD}}$}
\put(-13,25){$n_{{\SW}}$}
}}
}
%\put(10,170){${\mathbb R}^6$}
%\linethickness{0.15mm}
%\thicklines
%\put(20,90){\vector(1,0){295}}
%\put(300,70){T}
%\path(300,95)(300,85)
%\put(310,145){$w_1$}
%\put(310,135){$w_2$}
%\put(310,45){$w_{N}$}
%\dottedline{5}(320,120)(320,55)
%\linethickness{0.15mm}
\end{picture}
\vskip-0.3cm
\begin{center}{Figure 5.1:
Tomatos ( Red?, Sweet? )
}
\end{center}
\par
\noindent
\par
\noindent
\vskip-0.5cm
\par
\noindent

%That is,
%
%${\omega_k}_n$
%
%{\lq\lq \rq\rq}
%$\bigl[$ resp.  {\lq\lq \rq\rq}$\bigl]$
% and probability 
%$[{}F_{\RD} ({}\{ y_{\RD} \}{}){}]$
%$ ({}\omega_n{}) $
%$\bigl[$ resp.
%$[{}F_{\RD} ({}\{ n_{\RD} \}{}){}]$
%$ ({}\omega_n{}) $
%%$\bigl]$
%%.
%%Similarly, we consider the interpretation for the  {{measurement}}
%%${\bo\omega_n}]}{})$.
%$\;$
%({\it
%.
%\par \qed
%
%
%
%\par \noindent
%{\bf
%
%
%].
%}
\par
\noindent
Next,
consider the {quasi-product }{observable}
as follows.
\begin{align*}
{\mathsf O}_{12}
=
(X_{\RD} \times  X_{\SW} ,
2^{
X_{\RD} \times  X_{\SW} },
F
{{=}}
F_{\RD}
{\mathop{\qp}}
F_{\SW}{})
%%%% 3.39
%P%\tag{6.14}
\end{align*}
\par
\noindent
That is,
%%%
%\allowdisplaybreaks
{\small
\begin{align*}
&
\; \;
\roman{Rep}^{\{({} y_{{\RD}} , y_{{\SW}}{}) \}}_{\omega_k}
[{\mathsf O}_{12}]
=
\bmatrix
[{}F ({}\{({} y_{{\RD}} , y_{{\SW}}{}) \}{})  {}]  ({\omega_k})
&
[{}F  ({}\{({} y_{{\RD}} , n_{{\SW}}{}) \}{})  {}]  ({\omega_k})
\\
{}[{}F ({}\{({} n_{{\RD}} , y_{{\SW}}{}) \}{})  {}]  ({\omega_k})
&
[{}F ({}\{({} n_{{\RD}} , n_{{\SW}}{}) \}{})  {}]  ({\omega_k})
\\
\endbmatrix
\\
%\qquad
=
&
\bmatrix
\alpha_{_{\{({} y_{{\RD}} , y_{{\SW}}{}) \} }}
&
[{}F_{\RD} ({}\{  y_{{\RD}} \}{})  {}]  - 
\alpha_{_{\{({} y_{{\RD}} , y_{{\SW}}{}) \} }}
\\
{}[{}F_{\SW} ({}\{  y_{{\SW}} \}{})  {}]  - 
\alpha_{_{\{({} y_{{\RD}} , y_{{\SW}}{}) \} }} 
&
1+  \alpha_{_{\{({} y_{{\RD}} , y_{{\SW}}{}) \} }}
-
[{}F_{\RD} ({}\{  y_{{\RD}} \}{})  {}] 
-
[{}F_{\SW}  ({}\{ y_{{\SW}}  \}{})  {}] 
%
%\alpha_{_{\{({} y_{{\RD}} , y_{{\SW}}{}) \} }}
%({\omega_k})
%&
%[{}F_{\RD} ({}\{  y_{{\RD}} \}{})  {}]  ({\omega_k}) - 
%\alpha_{_{\{({} y_{{\RD}} , y_{{\SW}}{}) \} }}  ({\omega_k})
%\\
%{}[{}F_{\SW} ({}\{  y_{{\SW}} \}{})  {}]  ({\omega_k}) - 
%\alpha_{_{\{({} y_{{\RD}} , y_{{\SW}}{}) \} }}  ({\omega_k})
%&
%1+  \alpha_{_{\{({} y_{{\RD}} , y_{{\SW}}{}) \} }}  ({\omega_k})
%-
%[{}F_{\RD} ({}\{  y_{{\RD}} \}{})  {}]  ({\omega_k})
%-
%[{}F_{\SW}  ({}\{ y_{{\SW}}  \}{})  {}]  ({\omega_k})
%
\\
\endbmatrix
%P%\tag{6.15}
\end{align*}
}
\par
\noindent
where
$\alpha_{_{\{({} y_{{\RD}} , y_{{\SW}}{}) \} }}  ({\omega_k})$
satisfies
the \textcolor{black}{(5.2)}.
\rm
When
we know that
a tomato
${\omega_k} $
is red,
the probability
$P$
that
the tomato
${\omega_k} $
is sweet
is given by
\par
\noindent
\begin{align*}
P=
\frac{
[{}F ({}\{({}y_{\RD} , y_{\SW}{})  \}{}){}]({}{\omega_k}{})
}
{
[{}F ({}\{({}y_{\RD} , y_{\SW}{})  \}{}){}]({}{\omega_k}{})
+
[{}F ({}\{({}y_{\RD} , n_{\SW}{})  \}{}){}]({}{\omega_k}{})
}
=
\frac{
[{}F ({}\{({}y_{\RD} , y_{\SW}{})  \}{}){}]({}{\omega_k}{})
}
{
[{}F_{\RD} ({}\{  y_{{\RD}} \}{})  {}]  ({\omega_k})
}
%%%% 3.5
%P%\tag{6.16}[{}F_{\RD} ({}\{  y_{{\RD}} \}{})  {}]  ({\omega_k})
\end{align*}
Since
$[{}F ({}\{({}y_{\RD} , y_{\SW}{})  \}{}){}]({}{\omega_k}{})=
\alpha_{_{\{({}y_{\RD} , y_{\SW}{})  \}}
}
(\omega_k)$,
the conditional probability
$P$
is estimated by
{
%\small
\begin{align*}
&
\frac{\max  \{
0,
[{}F_1 ({}  \{  y_{{\RD}} \} {}{})  {}]  ({\omega_k})
+
[{}F_2  ({}  \{  y_{{\SW}} \} {}{})  {}]  ({\omega_k})
-1
{}
\}
}{{
[{}F_{\RD} ({}\{  y_{{\RD}} \}{})  {}]  ({\omega_k})
%[{}F ({}\{({}y_{\RD} , y_{\SW}{})  \}{}){}]({}{\omega_k}{})
%+
%[{}F ({}\{({}y_{\RD} , n_{\SW}{})  \}{}){}]({}{\omega_k}{})
}}
%\\
{{\; \leqq \;}}
%&
P
%\frac{
%[{}F ({}\{({}y_{\RD} , y_{\SW}{})  \}{}){}]({}{\omega_k}{})
%}
%{
%[{}F ({}\{({}y_{\RD} , y_{\SW}{})  \}{}){}]({}{\omega_k}{})
%+
%[{}F ({}\{({}y_{\RD} , n_{\SW}{})  \}{}){}]({}{\omega_k}{})
%}
%TAG{6.18}
%\\
{{\; \leqq \;}}
%\\
%&
%\hspace{4cm}
\frac{
\min
[{}F_1 ({}  \{  y_{{\SW}} \} {}{})  {}]  ({\omega_k})
,
\;
[{}F_2  ({}  \{  y_{{\SW}} \}  {}{})  {}]  ({\omega_k})
\}
}{{
[{}F_{\RD} ({}\{  y_{{\RD}} \}{})  {}]  ({\omega_k})
%[{}F ({}\{({}y_{\RD} , y_{\SW}{})  \}{}){}]({}{\omega_k}{})
%+
%[{}F ({}\{({}y_{\RD} , n_{\SW}{})  \}{}){}]({}{\omega_k}{})
}}
%P%\tag{6.17}
%%%%% 3.2{\omega_k}\omega\omega
\end{align*}
}

\subsection{Implication
---
The definition
of
"$\Rightarrow$"}%{Sec.5.3}
\subsubsection{Implication and contraposition}
\normalsize
\baselineskip=18pt
\par
In \textcolor{black}{Example 5.5},
consider the case that
$[{}F ({}\{({} y_{{\RD}} , n_{{\SW}}{}) \}{})  {}]
({}\omega{})=0$.
In this case, we see
\par
\noindent
\begin{align*}
\frac{
[{}F ({}\{({}y_{\RD} , y_{\SW}{})  \}{}){}]({}\omega{})
}
{
[{}F ({}\{({}y_{\RD} , y_{\SW}{})  \}{}){}]({}\omega{})
+
[{}F ({}\{({}y_{\RD} , n_{\SW}{})  \}{}){}]({}\omega{})
}
=1
%%%% 3.5
%P%\tag{6.18}
\end{align*}
Therefore,
when we know that
a tomato
$\omega $
is red,
the probability that
the tomato
$\omega $
is sweet
ig equal to
$1$.
That is,
\begin{align*}
\text{\LL $[{}F ({}\{({} y_{{\RD}} , n_{{\SW}}{}) \}{})  {}]
({}\omega{}) = 0$\RR}
\quad
\Longleftrightarrow
\quad
\Big[
\text{\LL Red\RR}
\Longrightarrow
\text{\LL Sweet\RR}
\Big]
%%%% 3.6
%P%\tag{6.19}
\end{align*}
{}
\par
%\newpage

%,
%{\lq\lq \rq\rq}
%$ \Longrightarrow$
%{\lq\lq \rq\rq}
% and {}
%\par \qed
\par
\vskip0.3cm
\vskip0.3cm
\par
Motivated by the above argument,
we have the following definition.

%%%%%%\par \qed
\par
\noindent
{\bf
%BFBF
\par
\noindent
{Definition }5.6
[{}\bf Inplication{}]}$\;\;$%POPOPO
%index{@}
\rm
%${\mathsf O}_1$
%${{=}}$
Let
${\mathsf O}_{12...n}$
${{=}}$
$({}\bigtimes_{k=1}^n  X_k ,$
$ \bigstimes_{k=1}^n{\cal F}_k , $
${F}_{12...n}
=
\underset{k=1,2,...,n}{\mathop{\qp}}
%\displaystyle{{\mathop{\qp}}_{k=1,2,...,n}}
F_k
{})$
be
a quasi-product observable
in $C(\Omega)$.
Let
$\Xi_1$
$ \in {\cal P} ({}X_1{}) $
and
$\Xi_2$
$ \in {\cal P} ({}X_2 {})$.
Then, the condition
\begin{align*}
{F}^{(ij)}_{12...n}
(\Xi_i \times ({} \Xi^c_j{}){})
]
(\omega)
=
0
%%%% 3.7
\end{align*}
is denoted by
\begin{align*}
[{\mathsf O}_{12...n}^{(i)};{\Xi_i}]
\underset{ {\mathsf M}_{C (\Omega)} ({}{\mathsf O}_{12...n} ,
S_{ [\omega]  }{}) }{ \Longrightarrow}
[{\mathsf O}_{12...n}^{(j)};{\Xi_j}]
\end{align*}
\par \hfill{$///$}
%%%\END{Def}
%$\blacksquare$
\par
\noindent
Of course, it should be read
as follows.
\begin{itemize}
\item[]
Assume that
a measured value
$(x_1, x_2)
(\in X_1 \times X_2)$
is obtained by
a
{{measurement}}${\mathsf M}_{C (\Omega)} ({}{\mathsf O}_{12} ,
S_{ [\omega]  }{})$.
When
we know that
$x_1 \in \Xi_1$,
then
we can assure that
$x_2 \in \Xi_2$.
{}
\end{itemize}
% and .
%.
%{basic algebra}${C ( \Omega )}$
%
%{observable }
%${\mathsf O}_{12...n}$
%${{=}}$
%$({}\bigtimes_{k=1}^n  X_k ,$
%$ \bigstimes_{k=1}^n{\cal F}_k , $
%${F}_{12...n}
%=
%\underset{k=1,2,...,n}{\mathop{\qp}}
%%\displaystyle{{\mathop{\qp}}_{k=1,2,...,n}}
%F_k
%{})$
%%
%.
%,
%$\omega \in \Omega $,
%$\Xi_i$
%$ \in {\cal F}_i $,
%$\Xi_j$
%$ \in {\cal F}_j$
% and ($1{{\; \leqq \;}}i , j {{\; \leqq \;}}n$),
%,
%\BEGIN{align*}
%{F}^{(ij)}_{12...n}
%(\Xi_i \times ({} \Xi^c_j{}){})
%]
%(\omega)
%=
%0
%%%%% 3.7
%%P%\tag{6.20}
%\END{align*}
% and ,
%\BEGIN{align*}
%[{\mathsf O}_{12...n}^{(i)};{\Xi_i}]
%\underset{ {\mathsf M}_{C (\Omega)} ({}{\mathsf O}_{12...n} ,
%S_{ [\omega]  }{}) }{ \Longrightarrow}
%[{\mathsf O}_{12...n}^{(j)};{\Xi_j}]
%%%%% 3.8
%%P%\tag{6.21}
%\END{align*}
% and .

\rm

\par
\noindent
{\bf  \vskip0.3cm
\vskip0.3cm
%BFBF
\par
\noindent
{{Theorem }}5.7
{\bf [{}Contraposition{}]}}$\;\;$%POPOPO
Let
${\mathsf O}_{12}$
$=$
$(X_1 \times X_2 ,$
$ {\cal F}_1 \times {\cal F}_2   ,$
$ F_{12}{}{{=}} F_1
{\mathop{\qp}}
F_2)$
be a quasi-product observable
in
${C (\Omega)}$.
Let
$\omega \in \Omega $.
Let
$\Xi_1$
$ \in {\cal F}_1 $
and
$\Xi_2$
$ \in {\cal F}_2$.
If it holds that
\begin{align*}
[{\mathsf O}_{12}^{(1)};{\Xi_1}]
\underset{ {\mathsf M}_{C (\Omega)} ({}{\mathsf O}_{12} ,
S_{ [\omega]  }{}) }{ \Longrightarrow}
[{\mathsf O}_{12}^{(2)};{\Xi_2}]
%%%% 3.8
%%P%\tag{6.22}
\tag{\color{black}{5.4}}
\end{align*}
then we see:
\begin{align*}
[{\mathsf O}_{12}^{(1)};{\Xi_1^c}]
\underset{ {\mathsf M}_{C (\Omega)} ({}{\mathsf O}_{12} ,
S_{ [\omega]  }{}) }{ \Longleftarrow}
[{\mathsf O}_{12}^{(2)};{\Xi_2^c}]
%%%% 3.8
%P%\tag{6.23}
\end{align*}
\par
\noindent
{\it $\;\;\;\;${Proof.}}$\;\;$
The proof is easy,
but
we add it.
Assume the condition
\textcolor{black}{(5.4)}.
%.
That is,
\begin{align*}
[F_{12}{}
(\Xi_1 \times ({}X_2 \setminus \Xi_2{}){})
]
(\omega)
=
0
%%%% 3.7
%P%\tag{6.24}
\end{align*}
Since
%\BEGIN{align*}
$\Xi_1 \times \Xi_2{}^c
=
(\Xi_1^c)^c \times \Xi_2^c
$
%%%% 3.7
%\TAG{6.22}
%\END{align*}
we see
\begin{align*}
[F_{12}{}
(
(\Xi_1^c)^c \times \Xi_2^c{})
%\Xi_1 \times ({}X_2 \setminus \Xi_2^c{}){})
]
(\omega)
=
0
%P%\tag{6.25}
\end{align*}
Therefore, we get
\begin{align*}
[{\mathsf O}_{12}^{(1)};{\Xi_1^c}]
\underset{ {\mathsf M}_{C (\Omega)} ({}{\mathsf O}_{12} ,
S_{ [\omega]  }{}) }{ \Longleftarrow}
[{\mathsf O}_{12}^{(2)};{\Xi_2^c}]
%%%% 3.8
%P%\tag{6.26}
\end{align*}
\qed

%BBBBBBBBBBBBBBBBBB%SBSBSBS
\par
\noindent
{\small%%{\footnotesize
\vspace{0.1cm}
\begin{itemize}
\item[$\spadesuit$] \bf {{}}{Note }5.5{{}} \rm
In what follows,
reconsider
the
statistical hypothesis testing
(\textcolor{black}{{{Sec. }}4.3.4}).
%(\textcolor{black}{{{Problem }}4.12}).
%
%\textcolor{black}{{Note }4.6}. 
%,
%\BEGIN{itemize}
%\item[]
%$[\ast ]\in N_H$,
%{{measurement}}
%${\mathsf M}_{C (\Omega)} ({\mathsf O} $
%${{=}}(X, {\cal F}, F{}) , S_{[\ast]})$
%
%measured value 
%$[D]_{N_H}^{{\varepsilon_{\rm max}^{0.05}}}
%$
%$(\in {\cal F} )$
%probability 
%$0.05$
%{}
%\END{itemize}
%.
%, ,
Consider
the observable
${\mathsf O}_{N_H} $
${=}(
Y(=\{0,1 \}), 2^{\{0,1 \}}, F_{N_H}{})$
in
$C (\Omega)$
such that
$$
[F_{N_H}(\{1\} )](\omega)=
\cases
1 \quad &( \omega \in N_H)
\\
0 \quad &( \omega \notin N_H)
\endcases
,
\qquad
[F_{N_H}(\{0\} )](\omega)=1-[F_{N_H}(\{1\} )](\omega)
$$
(
Even if
$F_{N_H}(\{ 1 \} ) \notin C(\Omega)$,
we do not mind it
(cf. Chap. 10)).
And consider the
simultaneous measurement
${\mathsf M}_{C (\Omega)} ({\mathsf O}_{N_H}\times {\mathsf O} $
${{=}}(Y \times X,  2^{\{0,1 \}} \times {\cal F}, F_{N_H} \times F{}) , S_{[\ast]})$.
If the
measured value
$(y,x)$
belongs to
$\{ 1 \}  \times X$,
the probability that
%
%$y=1$ and 
%(,
%%probability 1,
%$\omega \in N_H$ and ),
$x\notin
[D]_{N_H}^{{\varepsilon_{\rm max}^{0.05}}}
$
is
estimated by
% and probability ,
$$
\frac{
[(F_{N_H} \times F)
(\{1\} \times
(X \setminus [D]_{N_H}^{{\varepsilon_{\rm max}^{0.05}}})
)](\omega)}{[
(F_{N_H} \times F)
(\{1\} \times X)](\omega)}
=
%\frac{
[F
(X \setminus [D]_{N_H}^{{\varepsilon_{\rm max}^{0.05}}})
](\omega)
%
%}{[
%(F_{N_H} \times F)
%(\{1\} \times X)](\omega)}
\ge 0.95
$$
That is,
"$y=1$"
implies
%,
"mostly, $ x \notin [D]_{N_H}^{{\varepsilon_{\rm max}^{0.05}}} $".
This contraposition
(
i.e.,
if "$ x \in [D]_{N_H}^{{\varepsilon_{\rm max}^{0.05}}} $",
then
it is rare that
$y=1$)
is similar to
statistical hypothesis testing
%(\textcolor{black}{{{Sec. }}4.3.4})
(cf. \textcolor{black}{\cite{IWhat}}).
\end{itemize}
}
%%BBBBBBBBBBBBBBBBBequilibrium statistical mechanics
\par
\noindent

\par

\rm

\par
\noindent
{\bf
\subsubsection{"I think, therefore I am" is doubtful}%{Sec.5.3.2}
}
{
%BFBF
\par

The following example
is
somewhat
unnatural,
it may be dispensable for
the understanding
of
dualism.

% and .
\par
\noindent
{\bf Example 5.8}
[{}\bf {{Brain death}}{}]}$\;\;$%POPOPO
\rm
Let
$\omega_n$
$(\in \Omega=\{\omega_1,\omega_2,\ldots, \omega_N \}$)
be
the state
of Peter.
Let
${\mathsf O}_{12}$
$=$
$(X_1 \times X_2 ,$
$ 2^{ X_1   \times  X_2 }   ,$
$ F_{12}{}{{=}} F_1
{\mathop{\qp}}
F_2)$
be
the brain death observable
in
${C (\Omega)}$
such that
$X_1=\{ T, {\overline T}\}$
$X_2=\{ L, {\overline L}\}$,
where
$T$
$=$
$"think"$,
${\overline{T}}$
$=$
"not think",
$L$
$=$
$"live"$,
${\overline{L}}$
$=$
"not live".
For each
$\omega_n$
$(n=1,2,\ldots,N)$,
${\mathsf O}_{12}$
satisfies the condition in
\textcolor{black}{Table 5.1}.
\par
\noindent

%{{state}}$\omega_n$
%$(n=1,2,\ldots,N)$,
%\textcolor{black}{\REF{6010}} and .
\par
\noindent

%\begin{table}[htbp] \small \caption{Brain death observable
%${\mathsf O}_{12}$
%$=$
%$(X_1 \times X_2 ,$
%$ 2^{ X_1   \times  X_2 }   ,$
%$ F_{12}{})$
%\label{T6010}}
\begin{center}
Table 5.1:
Brain death observable
${\mathsf O}_{12}$
$=$
$(X_1 \times X_2 ,$
$ 2^{ X_1   \times  X_2 }   ,$
$ F_{12}{})$
\\
%\BEGIN{tabular}{||c||c|c||l|r}
\begin{tabular}{
@{\vrule width 1.8pt\ }c
@{\vrule width 1.8pt\ }c|c
@{\vrule width 1.8pt }}
\noalign{\hrule height 1.8pt}
$F_1 \diagdown F_2$ &$\quad  \overset{\quad}{[{}F_2 (\{ \text{\footnotesize L}\})  {}]  (\omega_n)} \quad$
&
$\quad
\overset{\quad}{
[{}F_2 (
\{
{\overline{\text{\footnotesize L}}}
\})  {}]  (\omega_n)} \quad$
\\
\noalign{\hrule height 1.8pt}
$ \overset{\quad}{[{}F_1 (\{ {\text{\footnotesize T}}\})  {}]  (\omega_n)}$
&
$
\underset{(=[F_{12}(
\{ {\text{\footnotesize T}}\}
\times
\{ {\text{\footnotesize L}}\}
)](\omega_n)
)}{(1+(-1)^n)/2}
$
%[{}F_1 (\{{\overline{\text{\footnotesize T}}}\})  {}]  (\omega)$
%$  \alpha_{\Xi_1 \times \Xi_2} (\omega) $
&
$\underset{(=[F_{12}(
\{ {\text{\footnotesize T}}\}
\times
\{ \overline{\text{\footnotesize L}}\}
)](\omega_n)
)}{0}
$
\\
\hline
$\overset{\quad}{[{}F_1 (\{{\overline{\text{\footnotesize T}}}\})  {}]  (\omega_n)}$
% [{}F_1 (\Xi_1^c)  {}]  (\omega_n)$ &
%$[{}F_2  (\Xi_2)  {}]  (\omega_n) -
%[{}F_1 (\Xi_1)  {}]  (\omega)$
&
$\underset{(=[F_{12}(
\{ \overline{\text{\footnotesize T}}\}
\times
\{ {\text{\footnotesize L}}\}
)](\omega_n))}{0}
$
&
$
\underset{(=[F_{12}(
\{ \overline{\text{\footnotesize T}}\}
\times
\{ \overline{\text{\footnotesize L}}\}
)](\omega_n))}{(1-(-1)^n)/2}
$
%-
%[{}F_2  (\Xi_2)  {}]  (\omega)$ 
\\
\noalign{\hrule height 1.8pt}
%$\omega_3$  &  1$\times N$ & 9$\times N$  \\
%\hline
\end{tabular}
\end{center}
%\end{table}
\par
\noindent

\par
\noindent
Since
$
[F_{12}(
\{ {\text{\footnotesize T}}\}
\times
\{ \overline{\text{\footnotesize L}}\}
)](\omega_n)
=0
$,
the following formula holds:
\begin{align*}
[{\mathsf O}_{12}^{(1)};{
\{
{\text{\footnotesize T}}
\}
}]
\underset{ {\mathsf M}_{C (\Omega)} ({}{\mathsf O}_{12} ,
S_{ [\omega_n]  }{}) }{ \Longrightarrow}
[{\mathsf O}_{12}^{(2)};{\{
{{\text{\footnotesize L}}}
\}}]
%%%% 3.8
%P%\TAG{6.27}
\end{align*}
Of course, this implies that
\begin{itemize}
\item[\textcolor{black}{}]
$\qquad$
Peter thinks,
therefore,
Peter lives.
%And
%Peter lives,
%therefore,
%Peter thinks,
%therefore,
%Peter lives.
\end{itemize}
This is the same as the statement
concerning brain death.
Note that
in the above example,
we see that
\begin{itemize}
\item[\textcolor{black}{}]
$\qquad$
observer$\longleftrightarrow$doctor,
$\qquad$
system$\longleftrightarrow$Peter,
\end{itemize}

This should not be confused with
the following famous Descartes' saying:

\begin{itemize}
\item[]
$\qquad$
{\it
\textcolor{black}{"I think, therefore I am".}
}
\end{itemize}
in which the following identification may be assumed:
\begin{itemize}
\item[]
$\qquad$
observer$\longleftrightarrow$I,
$\qquad$
system$\longleftrightarrow$I
\end{itemize}
And thus,
the above
is not a statement in dualism
(=measurement theory).
%Recall \textcolor{black}{Footnote \REF{FNIthink} in {Chap.$\;$1}}.
In order to propose \textcolor{black}{Fig. 1.1}
(i.e.,
dualism)
( that is, in order to establish the concept {\lq\lq}I" in science),
he started from the ambiguous statement
"I think, therefore I am".
Summing up,
we want to say the following irony:
\begin{itemize}
\item[]
$\qquad$
Descartes
proposed the dualism
(i.e.,
\textcolor{black}{Fig.} 1.1)
by the statement ($\sharp_1$) which is not understandable in dualism.
\end{itemize}

%\END{itemize}
%BFBF

Even in physic,
there is a case that
a meaningless statement,
which is useful to create the theory,
becomes famous.
This
(i.e.,
Heisenberg's {uncertainty principle}(\textcolor{black}{{{Proposition }}3.1})
)
is already pointed out
in
{\textcolor{black}{{}{{{}}}{Sec.3.4}}}.

\renewcommand{\footnoterule}{%
  \vspace{2mm}                      % 
  \noindent\rule{\textwidth}{0.4pt}   % , 
  \vspace{-5mm}
}
%BBBBBBBBBBBBBBBBBB%SBSBSBS
\par
\noindent
{\small%%{\footnotesize
\begin{itemize}
\item[$\spadesuit$] \bf {{}}{Note }5.6{{}} \rm
It is not true to consider that
every phenomena can be describe in terns of measurement theory.
%, {{measurement theory}} and 
%({Note }5.1).
%%%\textcolor{black}{{Note }2.8},
%%Heisenberg's {uncertainty principle}
%%\textcolor{black}{{{Theorem }}3.4},
%%{the theory of relativity}
%%{{measurement theory}}.
%%%{the theory of relativity},
readers may think that
the following can be described in measurement theory,
but I believe that it is impossible.
For example,
\begin{itemize}
\item[$(\sharp_1)$]
"observer's space-time",
"tense---past, present, future ---",
"Heidegger's saying:
{\lq\lq}In-der-Welt-sein{\rq\rq}",
"the measurement of a measurement",
"Only the present exists",
"subjective time (\textcolor{black}{{Note }6.7})",
\end{itemize}
Although these words can not be understood,
we think that
these are inconsistent
with
the Copenhagen interpretation
[\textcolor{black}{{Chap.$\;$1}(U$_1$)--(U$_7$))}].
\end{itemize}
}
%%BBBBBBBBBBBBBBBBBequilibrium statistical mechanics
\par
\noindent

\par

\rm

\subsection{Practical {{syllogism}}
---
Is {Socrates} mortal? }%{Sec.5.4}
\normalsize
\baselineskip=18pt
\par

The term:
"practical syllogism"
means
"syllogism in measurement theory.
And thus,
the syllogism should be proved in measurement theory.

%\vskip-1.0cm

\subsubsection{Combined observable
---
Observable is only one}%{Sec.5.4.1}
\normalsize
\baselineskip=18pt
\par
The Copenhagen interpretation says that
observable must be only one.
Thus,
many observables must be combined.

%index{@observable }
{\bf
%\vskip0.3cm
%\vskip0.3cm
%BFBF
\par
\noindent
{{Theorem }}5.9
{\bf [Combined observable{\rm
({\rm cf. }\textcolor{black}{${}$\cite{IFuzz}}})}]}$\;\;$%POPOPO
Let
${\mathsf O}_{12}{{=}} (X_1 \times X_2 , {\Cal F}_1 \times {\cal F}_2, F_{12})$
and
${\mathsf O}_{23}{{=}} $
$(X_2 \times X_3 ,$
$ {\Cal F}_2 \times {\cal F}_3, F_{23})$
be observables in
$C(\Omega)$.
Here, for simplicity,
assume that
$X_i{{=}} \{x^1_i, x^2_i,\ldots, x^{n_i}_i\}$
$(i=1,2,3)$
is finite,
Also,
assume that ${\cal F}_i = 2^{X_i}$.
Further assume that
\begin{align*}
{\mathsf O}_{12}^{(2)}
=
{\mathsf O}_{23}^{(2)}
\quad
(\text{That is, }\;\;
F_{12}(X_1 \times \Xi_2 )
=
F_{23}(\Xi_2 \times X_3 )
\quad(\forall \Xi_2 \in 2^{X_2}))
%P%\tag{6.28}
\end{align*}
Then,
we have
the observable
${\mathsf O}_{123}{{=}} (X_1 \times X_2 \times X_3,
{\Cal F}_1 \times {\cal F}_2 \times {\cal F}_3, F_{123})$
in $C(\Omega)$
such that
\begin{align*}
{\mathsf O}_{123}^{(12)}
=
{\mathsf O}_{12},
\quad
{\mathsf O}_{123}^{(23)}
=
{\mathsf O}_{23}
%P%\tag{6.29}
\end{align*}
That is,
\begin{align*}
F_{123}^{(12)}(\Xi_1 \times \Xi_2 \times X_3)
&
=
F_{12}(\Xi_1 \times \Xi_2 ),\;\;
F_{123}^{(23)}(X_1 \times \Xi_2 \times \Xi_3 )
=
F_{23}(\Xi_2 \times \Xi_3 )
\\
&
\quad(\forall \Xi_1 \in {\cal F}_1,
\forall \Xi_2 \in {\cal F}_2,
\forall \Xi_3 \in {\cal F}_3
))
\tag{\color{black}{5.5}}
\end{align*}
The
${\mathsf O}_{123}$
is called the combined observable of
${\mathsf O}_{12}$
and
${\mathsf O}_{23}$.
\par
\noindent
{\it $\;\;\;\;${Proof.}}$\;\;$
${\mathsf O}_{123}$
$=$
$(X_1 \times X_2 \times X_{3}  , $
${ {\cal F}_1  \times {\cal F}_{2} \times {\cal F}_3}, $
$F_{123}{})$
is, for example,
defined by
\par
\noindent
%%%
\begin{align*}
&
\; \;
[{}F_{123}
({}\{({}x_{1} , x_{2} , x_{3}^{}{}) \}{})
]
(\omega)
\\
&
=
\cases
{\displaystyle
\frac{
[{}F_{12} ({}\{({}x_{1} , x_{2}{}) \}{}){}]
(\omega)
\cdot
[{}F_{23} ({}\{({}x_{2} , x_{3}{}) \}{}){}]
(\omega)
}
{
[{}F_{12} ({}X_1 \times \{ x_{2}  \}{}){}]
(\omega)
}
}
\\
& \hspace{-2cm}
(
[{}F_{12} ({}X_1 \times \{ x_{2}  \}{}){}]
(\omega)
\;
\not= 0\text{ and })
\\
\\
0
\\
& \hspace{-2.0cm}
(
[{}F_{12} ({}X_1 \times \{ x_{2}  \}{}){}]
%[{}F_{2} ({}\{ x_{2}  \}{}){}]
(\omega)
= 0\text{ and })
\\
\endcases
\\
&
\qquad \qquad \qquad
(\forall \omega \in \Omega,
\forall
(x_1,x_2,x_3) \in X_1 \times X_2 \times X_3)
%\\
%&
%=
%\cases
%\frac{\displaystyle \frac{1}{3} \times 0.5}{\displaystyle \frac{1}{3} (0.5+ 1.0+0)}
%={\displaystyle \frac{1}{3}}
%\quad & (D=\{1\}\text{ and })
%\\
%\frac{\displaystyle \frac{1}{3} \times 1.0}{\displaystyle \frac{1}{3} (0.5+ 1.0+0)}
%=
%{\displaystyle \frac{2}{3}}
%\quad & (D=\{2\}\text{ and })
%\\
%\frac{\displaystyle \frac{1}{3} \times 0}{\displaystyle \frac{1}{3} (0.5+ 1.0+0)}
%=
%0
%\quad & (D=\{3\}\text{ and })
%\ENDcases
%%%\Txyyyyyyyyyy
\end{align*}
This clearly satisfies \textcolor{black}{(5.5)}.
\qed

\par
\vskip0.3cm
\par

{\bf
%BFBF
\par
\noindent
Remark 5.10
{\bf [Bell's inequality is useful]}}$\;\;$%POPOPO]
Put $X_1=X_2=X_3=X_4=\{-1, 1\}$.
Let
${\mathsf O}_{13}{{=}} (X_1 \times X_3 ,
2^{X_1} \times 2^{X_3} , F_{13})$,
${\mathsf O}_{14}{{=}} (X_1 \times X_4 ,
2^{X_1}  \times
2^{X_4}
, F_{14})$,
${\mathsf O}_{23}{{=}} $
$(X_2 \times X_3 ,$
$ 2^{X_2}  \times 2^{X_3} , F_{23})$
and
${\mathsf O}_{24}{{=}} $
$(X_2 \times X_3 ,$
$ 2^{X_2}  \times 2^{X_4} , F_{24})$
be observables in
$C(\Omega)$
such that
\begin{align*}
{\mathsf O}_{13}^{(1)}={\mathsf O}_{14}^{(1)},
\;\;
{\mathsf O}_{23}^{(2)}={\mathsf O}_{24}^{(2)},
\;\;
{\mathsf O}_{13}^{(3)}={\mathsf O}_{23}^{(3)},
\;\;
{\mathsf O}_{14}^{(4)}={\mathsf O}_{24}^{(4)}
\end{align*}
Define the probability measure $\nu_{ab}$
on
$\{-1, 1 \}^2$
by
the formula
\textcolor{black}{(3.6)}.
Assume that
there exists
a state
$\omega_0 \in \Omega $
such that
\begin{align*}
&
[F_{13}(\{(x_1,x_3)\})](\omega_0)=\nu_{a^1b^1}(\{(x_1,x_3)\},
\\
&
[F_{14}(\{(x_1,x_4)\})](\omega_0)=\nu_{a^1b^2}(\{(x_1,x_4)\}
\\
&
[F_{23}(\{(x_2,x_3)\})](\omega_0)=\nu_{a^2b^1}(\{(x_2,x_3)\},
\\
&
[F_{24}(\{(x_2,x_4)\})](\omega_0)=\nu_{a^2b^2}(\{(x_2,x_4)\}
\end{align*}
\par
Now we have the following problem:
\begin{itemize}
\item[(a)]
Does the observable
${\mathsf O}_{1234}{{=}} (\bigtimes_{k=1}^4
X_k ,
\bigtimes_{k=1}^4{\Cal F}_k , F_{1234})$
in $C(\Omega)$
satisfying the following
($\sharp$)?
\begin{itemize}
\item[($\sharp$)]
$
\displaystyle
{\mathsf O}_{1234}^{(13)}
=
{\mathsf O}_{13},
\;\;
{\mathsf O}_{1234}^{(14)}
=
{\mathsf O}_{14},
\;\;
{\mathsf O}_{1234}^{(23)}
=
{\mathsf O}_{23},
\;\;
{\mathsf O}_{1234}^{(24)}
=
{\mathsf O}_{24}
\;\;
$
\end{itemize}
\end{itemize}
In what follows,
we show that
the above observable ${\mathsf O}_{1234}$
does not exist.
\par
Assume that
the observable
${\mathsf O}_{1234}{{=}} (\bigtimes_{k=1}^4
X_k , $
$
\bigtimes_{k=1}^4{\Cal F}_k ,$
$ F_{1234})$
exists.
Then,
it suffices to show the contradiction.
Define
$C_{13}(\omega_0)$,
$C_{14}(\omega_0)$,
$C_{23}(\omega_0)$
and
$C_{24}(\omega_0)$
such that
%\newpage
\begin{align*}
\cases
\displaystyle
C_{13}(\omega_0)
=
\int_{\bigtimes_{k=1}^4 X_k} x_1 \cdot x_3
\;
[F_{1234}(\bigtimes_{k=1}^4 dx_k )](\omega_0)
\\
\hspace{1.1cm}
\bigl(
=
\int_{X_1 \times X_3} x_1 \cdot x_3
\;
\nu_{a^1 b^1}
(
dx_1 dx_3)
\bigl)
\\
\displaystyle
C_{14}(\omega_0)
=
\int_{\bigtimes_{k=1}^4 X_k} x_1 \cdot x_4
\;
[F_{1234}(\bigtimes_{k=1}^4 dx_k )](\omega_0)
\\
\hspace{1.1cm}
\bigl(=
\int_{X_1 \times X_4} x_1 \cdot x_4
\;
\nu_{a^1b^2}(
dx_1dx_4)
\bigl)
\\
\displaystyle
C_{23}(\omega_0)
=
\int_{\bigtimes_{k=1}^4 X_k} x_2 \cdot x_3
\;
[F_{1234}(\bigtimes_{k=1}^4 dx_k )](\omega_0)
\\
\hspace{1.1cm}
\bigl(=
\int_{X_2 \times X_3} x_2 \cdot x_3
\;
\nu_{a^2b^1}(
dx_2dx_3)
\bigl)
\\
\displaystyle
C_{24}(\omega_0)
=
\int_{\bigtimes_{k=1}^4 X_k} x_2 \cdot x_4
\;
[F_{1234}(\bigtimes_{k=1}^4 dx_k )](\omega_0)
\\
\hspace{1.1cm}
\bigl(=
\int_{X_2 \times X_4} x_2 \cdot x_4
\;
\nu_{a^2b^2}(
dx_2dx_4)
\bigl)
%\int_{X_2 \times x_4} x_2 \cdot x_4
%[F_{1234}(X_1 \times dx_2 \times X_3 \times dx_4 )](\omega_0)
\endcases
\end{align*}
Then,
we can easily get
the following Bell's inequality
(cf. Note 3.4).
\begin{align*}
&
|C_{13}(\omega_0)-C_{14}(\omega_0)|
+
|C_{23}(\omega_0)+C_{24}(\omega_0)|
\\
{{\; \leqq \;}}
&
\int_{\bigtimes_{k=1}^4 X_k}
\!\!\!
| x_1 | \cdot |x_3-x_4|
\;\;+
\!\!
|x_2 | \cdot |x_3+x_4|
\big[F_{1234}(\bigtimes_{k=1}^4 dx_k )\big](\omega_0)
%+
%\int_{\bigtimes_{k=1}^4 X_k}
%\!\!\!
%|x_2 | \cdot |x_3+x_4|
%\big[F_{1234}(\bigtimes_{k=1}^4 dx_k )\big](\omega_0)
\\
{{\; \leqq \;}}
&
2
\qquad
(\text{since } x_k \in \{-1, 1\} )
\tag{\color{black}5.6}
\end{align*}
\normalsize
\baselineskip=18pt
\par
\noindent
However,
the formula
\textcolor{black}{(3.7)}
says that
this (5.6) must be
$2{\sqrt 2}$.
Thus,
by contradiction,
we says that
${\mathsf O}_{1234}$ satisfying (a)
does not exist.
Thus we can not take a measurement
${\mathsf M}_{C(
\Omega)} (
{\mathsf O}_{1234},
S_{[\omega_0]}
)$.

However,
it should be noted that
\begin{itemize}
\item[(b)]
in stead of
${\mathsf M}_{C(
\Omega)} (
{\mathsf O}_{1234},
S_{[\omega_0]}
)$.
we can take a
parallel measurement
${\mathsf M}_{C(
\Omega^4)} (
{\mathsf O}_{13} \otimes
{\mathsf O}_{14} \otimes
{\mathsf O}_{23} \otimes
{\mathsf O}_{24}
,$
$
S_{[(\omega_0, \omega_0, \omega_0, \omega_0)]})$.
In this case,
we easily see that
(5.6)
=
$2{\sqrt 2}$
as the formula \textcolor{black}{(3.7)}.
\end{itemize}
That is,
\begin{itemize}
\item[(c)]
in the case of a parallel measurement,
Bell's inequality is broken
in both
quantum and classical systems.
\end{itemize}
%(\textcolor{black}{{}3.6.2}).
\par

%
%%%%\omega^\omega^
%BBBBBBBBBBBBBBBBBB%SBSBSBS
\par
\noindent
{\small%%{\footnotesize
\vspace{0.1cm}
\begin{itemize}
\item[$\spadesuit$] \bf {{}}{Note }5.7{{}} \rm
In the above argument,
Bell's inequality is used in the framework
of measurement theory.
This is of course true.
However,
since
mathematics is of course independent of
the world,
now we have the following question:
\begin{itemize}
\item[$(\sharp)$]
In order that
mathematical Bell's inequality
asserts something to
quantum mechanics,
what kind of idea
do we prepare?
\end{itemize}
We can not answer this question.
%.
%
% and .
%
%quantum mechanics(3.5)
%
%%
%({\rm cf.}
%\textcolor{black}{\cite{Sell}})
% and 
%.
%\textcolor{black}{%index{@()}}
%\textcolor{black}{3.5},
%\BEGIN{itemize}
%\item[$(\sharp_2)$]
%{{measurement theory}}
%,
%{{measurement theory}}
%
%\END{itemize}
%, (\textcolor{black}{4})
%{{Problem }}(a).
%.
\end{itemize}
}
\subsubsection{Practical {{syllogism}} and its variations}%5.4.2
\normalsize
\baselineskip=18pt
\par
%,
%,
%practical {{syllogism}}
%
%(\textcolor{black}{{Definition } 5.6}{})
%{{measurement theory}}
%
%.
%,
%\textcolor{black}{{{Theorem }}5.9}
%(observable )
%
%(
% and ,
%\textcolor{black}{{Note }3.2}$(\sharp_1)$ and
%$(\sharp_2)$),
%{{syllogism}}.
%%,
%,
%,
%\textcolor{black}{{{Theorem }}5.9}
%(observable )
%,
%,
%{{syllogism}}.
%%%index{ and @}
%%index{@({{syllogism}})}
%%%
%%\par
%
%
%bbbbbbbbbbbbbbbbbbbbbbbbbb
%
%
Now
we show several theorems
of practical syllogisms
({}i.e.,
theorems concerning {\lq\lq implication\rq\rq} in Definition 5.6{}).

%\noindent
{\bf
\vskip0.2cm
%\vskip0.3cm
%BFBF
\par
\noindent
{{Theorem }}5.11
[{}Practical {{syllogism}}
\rm
({\rm cf. }\textcolor{black}{${}$\cite{IFuzz}}){\bf]}}$\;\;$%POPOPO
%index{@practical {{syllogism}}}
\rm
$\;$

\rm
Let
${\mathsf O}_{123}$
$=$
$(X_1 \times X_2 \times X_3,$
$ {\cal F}_1 \times {\cal F}_2 \times {\cal F}_3 ,$
$ F_{123}{}{{=}} {\mathop{\qp}}_{k=1,2,3} F_k)$
be an observable in
${C (\Omega)}$
Fix
$\omega \in \Omega $,
$\Xi_1$
$ \in {\cal F}_1$,
$\Xi_2$
$ \in {\cal F}_2$,
$\Xi_3$
$ \in {\cal F}_3$
Then,
we see
the following
\rm
(i)
$\text{--}$
(iii).

\par
\noindent
\rm
(i).(practical {{syllogism}})
%.[{{Theorem }}5.7 and ]
\begin{align*}
[{\mathsf O}_{123}^{(1)};{\Xi_1}]
\underset{ {\mathsf M}_{C (\Omega)} ({}{\mathsf O}_{123} ,
S_{ [\omega]  }{}) }{ \Longrightarrow}
[{\mathsf O}_{123}^{(2)};{\Xi_2}] ,
\quad
[{\mathsf O}_{123}^{(2)};{\Xi_2}]
\underset{ {\mathsf M}_{C (\Omega)} ({}{\mathsf O}_{123} ,
S_{ [\omega]  }{}) }{ \Longrightarrow}
[{\mathsf O}_{123}^{(3)};{\Xi_3}]
%%%% 3.8
%P%\tag{6.32}
\end{align*}
implies
%\allowdisplaybreaks
\begin{align*}
&
\; \;
\roman{Rep}_\omega^{\Xi_1 \times \Xi_3}[{}{\mathsf O}^{(13)}_{123}]
=
\bmatrix
[{}F^{(13)}_{123} (\Xi_1 \times \Xi_3)]  (\omega)
&
[{}F^{(13)}_{123}  (\Xi_1 \times \Xi_3^c)]
(\omega)
\\
{}
[{}F^{(13)}_{123}  (\Xi_1^c \times \Xi_3)]
(\omega)
&
[{}F^{(13)}_{123}  (\Xi_1^c \times \Xi_3^c)]
(\omega)
\endbmatrix
\\
=
&
\bmatrix
[F^{(1)}_{123}(\Xi_1)](\omega)
&
0
\\
{}
[F^{(3)}_{123}(\Xi_3)](\omega)
-
[F^{(1)}_{123}(\Xi_1)](\omega)
&
1-
[F^{(3)}_{123}(\Xi_3)](\omega)
\endbmatrix
%P%\tag{6.33}
\end{align*}
%%%%
%
That is, it holds:
\begin{align*}
[{\mathsf O}_{123}^{(1)};{\Xi_1}]
%\underset{ {\mathsf M}_{C (\Omega)} ({}{\mathsf O}_{123} ,
%S_{ [\omega]  }{}) }{ \Longrightarrow}
%[{\mathsf O}_{123}^{(2)};{\Xi_2}] ,
%\quad
%[{\mathsf O}_{123}^{(2)};{\Xi_2}]
\underset{ {\mathsf M}_{C (\Omega)} ({}{\mathsf O}_{123} ,
S_{ [\omega]  }{}) }{ \Longrightarrow}
[{\mathsf O}_{123}^{(3)};{\Xi_3}]
%%%% 3.8
%%%%}
\tag{\color{black}{5.7}}
%%%%%REDREDREDREDREDRE
\end{align*}
\par
\noindent
\rm
(ii).
\rm
\begin{align*}
[{\mathsf O}_{123}^{(1)};{\Xi_1}]
\underset{ {\mathsf M}_{C (\Omega)} ({}{\mathsf O}_{123} ,
S_{ [\omega]  }{}) }{ \Longleftarrow}
[{\mathsf O}_{123}^{(2)};{\Xi_2}] ,
\quad
[{\mathsf O}_{123}^{(2)};{\Xi_2}]
\underset{ {\mathsf M}_{C (\Omega)} ({}{\mathsf O}_{123} ,
S_{ [\omega]  }{}) }{ \Longrightarrow}
[{\mathsf O}_{123}^{(3)};{\Xi_3}]
%%%% 3.8
%P%\tag{6.35}
\end{align*}
implies
\begin{align*}
&
\; \;
\roman{Rep}_\omega^{\Xi_1 \times \Xi_3}[{}{\mathsf O}^{(13)}_{123}]
=
\bmatrix
[{}F^{(13)}_{123} (\Xi_1 \times \Xi_3)]  (\omega)
&
[{}F^{(13)}_{123}  (\Xi_1 \times \Xi_3^c)]
(\omega)
\\
{}
[{}F^{(13)}_{123}  (\Xi_1^c \times \Xi_3)]
(\omega)
&
[{}F^{(13)}_{123}  (\Xi_1^c \times \Xi_3^c)]
(\omega)
\endbmatrix
\\
=
&
\bmatrix
\alpha_{_{\Xi_1 \times \Xi_3}}
&\;\;\;\;
[F^{(1)}_{123}(\Xi_1)](\omega)-\alpha_{_{\Xi_1 \times \Xi_3}}
\\
{}
[F^{(3)}_{123}(\Xi_3)](\omega)-\alpha_{_{\Xi_1 \times \Xi_3}}
&\;\;\;\;
1-\alpha_{_{\Xi_1 \times \Xi_3}}
-
[F^{(1)}_{123}(\Xi_1)]
-
[F^{(3)}_{123}(\Xi_3)]
\endbmatrix
\end{align*}
where
\begin{align*}
&
\hspace{-1cm}
\max \{
[F^{(2)}_{123}(\Xi_2)](\omega),
[F^{(1)}_{123}(\Xi_1)](\omega)+
[F^{(3)}_{123}(\Xi_3)](\omega) - 1 \}
\\
&
\hspace{0.5cm}{{\; \leqq \;}}
\alpha_{_{\Xi_1 \times \Xi_3}} ({}\omega{})
\tag{\color{black}{5.8}}
%\\
%&
%\hspace{2.5cm}
{{\; \leqq \;}}
\min \{ [F^{(1)}_{123}(\Xi_1)](\omega)  , [F^{(3)}_{123}(\Xi_3)](\omega)  \}
%\TAG{6.37}
%%%%%REDREDREDREDREDRE
\end{align*}
\par
\noindent
\rm
(iii).
\rm
\begin{align*}
[{\mathsf O}_{123}^{(1)};{\Xi_1}]
\underset{ {\mathsf M}_{C (\Omega)} ({}{\mathsf O}_{123} ,
S_{ [\omega]  }{}) }{ \Longrightarrow}
[{\mathsf O}_{123}^{(2)};{\Xi_2}] ,
\quad
[{\mathsf O}_{123}^{(2)};{\Xi_2}]
\underset{ {\mathsf M}_{C (\Omega)} ({}{\mathsf O}_{123} ,
S_{ [\omega]  }{}) }{ \Longleftarrow}
[{\mathsf O}_{123}^{(3)};{\Xi_3}]
%%%% 3.8
%P%\tag{6.38}
\end{align*}
implies
\small
%\allowdisplaybreaks
\begin{align*}
&
\; \;
\roman{Rep}_\omega^{\Xi_1 \times \Xi_3}[{}{\mathsf O}^{(13)}_{123}]
=
\bmatrix
[{}F^{(13)}_{123} (\Xi_1 \times \Xi_3)]  (\omega)
&
[{}F^{(13)}_{123}  (\Xi_1 \times \Xi_3^c)]
(\omega)
\\
{}
[{}F^{(13)}_{123}  (\Xi_1^c \times \Xi_3)]
(\omega)
&
[{}F^{(13)}_{123}  (\Xi_1^c \times \Xi_3^c)]
(\omega)
\endbmatrix
\\
=
&
\bmatrix
\alpha_{_{\Xi_1 \times \Xi_3}} (\omega)
&\;\;\;\;
[F^{(1)}_{123}(\Xi_1)](\omega)-\alpha_{_{\Xi_1 \times \Xi_3}} (\omega)
\\
{}
[F^{(3)}_{123}(\Xi_3)](\omega)-\alpha_{_{\Xi_1 \times \Xi_3}} (\omega)
&\;\;\;\;
1-\alpha_{_{\Xi_1 \times \Xi_3}} (\omega)
-
[F^{(1)}_{123}(\Xi_1)](\omega)
-
[F^{(3)}_{123}(\Xi_3)](\omega)
\endbmatrix
%P%\tag{6.39}
\end{align*}
\normalsize \baselineskip=18pt
where
\begin{align*}
&
\hspace{-1cm}
\max \{ 0 ,  [F^{(1)}_{123}(\Xi_1)](\omega)  + [F^{(3)}_{123}(\Xi_3)](\omega)
-  [F^{(2)}_{123}(\Xi_2)](\omega) \}
\\
& \hspace{1cm}
{{\; \leqq \;}}
\alpha_{_{\Xi_1 \times \Xi_3}} ({}\omega{})
%P%\tag{6.40}
%\\
{{\; \leqq \;}}
%&
\min \{[F^{(1)}_{123}(\Xi_1)](\omega)  ,
[F^{(3)}_{123}(\Xi_3)](\omega)
 \}
%%%% 4.20
\end{align*}
\def\RD{\scriptscriptstyle{\roman{RD}}}
\def\RP{\scriptscriptstyle{\roman{RP}}}
\def\SW{\scriptscriptstyle{\roman{SW}}}

\par
\noindent
\vskip0.3cm
\rm
\par
\noindent
{\it $\;\;\;\;${Proof.}}
$\;\;$
%\par
%\noindent
(i):
By the condition,
we see
$\;\;$
\begin{align*}
&
0=
[F^{(12)}_{123}(\Xi_1 \times \Xi_2^c )](\omega)
=
[F_{123}(\Xi_1 \times \Xi_2^c \times \Xi_3 )](\omega)
+
[F_{123}(\Xi_1 \times \Xi_2^c \times \Xi_3^c )](\omega)
\\
&
0=
[F^{(23)}_{123}(\Xi_2 \times \Xi_3^c )](\omega)
=
[F_{123}(\Xi_1 \times \Xi_2 \times \Xi^c_3 )](\omega)
+
[F_{123}(\Xi_1^c \times \Xi_2 \times \Xi_3^c )](\omega)
%P%\tag{6.41}
\end{align*}
Therefore,
\begin{align*}
&
0=
[F_{123}(\Xi_1 \times \Xi_2^c \times \Xi_3 )](\omega)
=
[F_{123}(\Xi_1 \times \Xi_2^c \times \Xi_3^c )](\omega)
\\
&
0=
[F_{123}(\Xi_1 \times \Xi_2 \times \Xi^c_3 )](\omega)
=
[F_{123}(\Xi_1^c \times \Xi_2 \times \Xi_3^c )](\omega)
%P%\tag{6.42}
\end{align*}
Hence,
\begin{align*}
&
[F^{(13)}_{123}(\Xi_1 \times \Xi_3^c )](\omega)
=
[F_{123}(\Xi_1 \times \Xi_2 \times \Xi_3^c )](\omega)
+
[F^{(13)}_{123}(\Xi_1 \times \Xi_2^c \times \Xi_3^c )](\omega)
=0
%P%\tag{6.43}
\end{align*}
Thus, we get,
\textcolor{black}{(5.7)}.

\rm
\par
\noindent
For the proof of (ii) and (iii),
see
\textcolor{black}{${}$\cite{IFuzz}}.
\qed

\def\RD{\scriptscriptstyle{\roman{RD}}}
\def\RP{\scriptscriptstyle{\roman{RP}}}
\def\SW{\scriptscriptstyle{\roman{SW}}}

\vskip 0.2cm
%\ssubsection
%%(
\par
\noindent
{\bf
\vskip0.3cm
\par
\noindent
Example 5.12
[Continued from {}\textcolor{black}{Example 5.5}]}$\;\;$%POPOPO
%}
\rm
%$\Omega$,
%$C({}\Omega{})$,
%
Let
${\mathsf O}_{{1}}$
${{=}}$
${\mathsf O}_{{\SW}}$
${{=}}$
$(X_{\SW} , $
$ 2^{ X_{\SW} }  ,$
$ F_{\SW}{})$
and
${\mathsf O}_{{3}}$
${{=}}$
${\mathsf O}_{\RD}$
${{=}}$
$(X_{\RD} ,$
$ 2^{ X_{\RD} }  ,$
$ F_{\RD}{})$
be as in
\textcolor{black}{Example 5.5}.
Putting
$X_{\RP}
=
\{ y_{\RP} , n_{\RP} \} $,
consider the new
{observable }
${\mathsf O}_{{2}}$
${{=}}$
${\mathsf O}_{{\RP}}$
${{=}}$
$(X_{\RP} , 2^{ X_{\RP} }  , F_{\RP}{})$.
%
%\BEGIN{align*}
%X_{\RP}
%=
%\{ y_{\RP} , n_{\RP} \} ,
%\TAG{6.44}
%\END{align*}
Here,
{\lq\lq $y_{\RP}$\rq\rq}
and
{\lq\lq $n_{\RP}$\rq\rq}
respectively means
"ripe"
and
"not ripe".
%%%
%\allowdisplaybreaks
Put
\begin{align*}
\roman{Rep}[{}{\mathsf O}_1{}]
&
=
\big[
[{}F_{\SW} ({}\{ y_{{\SW}} \}{})  {}]  ({\omega_k}),
[{}F_{\SW} ({}\{ n_{{\SW}} \}{})  {}]  ({\omega_k})
\big]
\\
\roman{Rep}[{}{\mathsf O}_2{}]
&
=
\big[
[{}F_{\RP} ({}\{ y_{{\RP}} \}{})  {}]  ({\omega_k}),
[{}F_{\RP} ({}\{ n_{{\RP}} \}{})  {}]  ({\omega_k})
\big]
\\
\roman{Rep}[{}{\mathsf O}_3{}]
&
=
\big[
[{}F_{\RD} ({}\{ y_{{\RD}} \}{})  {}]  ({\omega_k}),
[{}F_{\RD} ({}\{ n_{{\RD}} \}{})  {}]  ({\omega_k})
\big]
%P%\tag{6.45}
\end{align*}
Consider the following
{quasi-product }{observable:}
%%%
\begin{align*}
&
{\mathsf O}_{12}
=
(X_{\SW} \times  X_{\RP} ,
2^{
X_{\SW} \times  X_{\RP} },
F_{12}
{{=}}
F_{\SW} {\mathop{\qp}} F_{\RP}{})
%%P%\tag{6.46}
%\END{align*}
% and
%\BEGIN{align*}
\\
&
{\mathsf O}_{23}
=
(X_{\RP} \times  X_{\RD} ,
2^{
X_{\RP} \times  X_{\RD} },
F_{23}
{{=}}
F_{\RP}
{\mathop{\qp}} F_{\RD}{})
%P%\tag{6.47}
%
\end{align*}
Let
${{\omega_k}}$
$\in \Omega$.
And assume that
\par
\noindent
\begin{align*}
&
[{\mathsf O}_{123}^{(1)};{\{y_{\SW} \}}]
\underset{ {\mathsf M}_{C (\Omega)} ({}{\mathsf O}_{123} ,
S_{ [{\omega_k}]  }{}) }{ \Longrightarrow}
[{\mathsf O}_{123}^{(2)};{\{y_{\RP} \}}] ,
\\
&
[{\mathsf O}_{123}^{(2)};{\{y_{\RP} \}}]
\underset{ {\mathsf M}_{C (\Omega)} ({}{\mathsf O}_{123} ,
S_{ [{\omega_k}]  }{}) }{ \Longrightarrow}
[{\mathsf O}_{123}^{(3)};{\{y_{\RD} \}}]
%%%% 3.8
%\TAG{6.48}
\tag{\color{black}{5.9}}
%%%%%REDREDREDREDREDRE
\end{align*}
Then,
by
\textcolor{black}{{{Theorem }}5.11(i)},
we get:
%{{Theorem }}5.12[1]
\par
\noindent
%\BEGIN{align*}
%\allowdisplaybreaks
\begin{align*}
&
\; \;\;\;
\roman{Rep} [{}{\mathsf O}_{13}{}]
=
\bmatrix
[{}F_{13} ({} \{ y_{{\SW}}  \}\times  \{ y_{{\RD}} \}){}] ({}{{\omega_k}} {})
&
[{}F_{13} ({} \{ y_{{\SW}}  \}\times  \{ n_{{\RD}} \}){}] ({}{{\omega_k}} {})
\\
{}[{}F_{13} ({} \{ n_{{\SW}}  \}\times  \{ y_{{\RD}} \}){}] ({}{{\omega_k}} {})
&
[{}F_{13} ({} \{ n_{{\SW}}  \}\times  \{ n_{{\RD}} \}){}] ({}{{\omega_k}} {})
\\
\endbmatrix
%%%% 4.23
%P%\tag{6.49}
%\END{align*}
\\
&
=
\bmatrix
[{}F_{\SW} ({}\{ y_{\SW} \}{}){}]({}{{\omega_k}} {}) & 0 \\
{}[{}F_{\RD} ({}\{ y_{\RD} \}{}){}]({}{{\omega_k}} {}) - [{}F_{\SW} ({}\{ y_{\SW} \}{}){}]
({}{{\omega_k}} {})
&   1 -  [{}F_{\RD} ({}\{ y_{\RD} \}{}){}]({}{{\omega_k}} {}) \\
\endbmatrix
%P%\tag{6.50}
\end{align*}
\par
\noindent
Therefore,
when we know that
the tomato
${{\omega_k}} $ is sweet
by
{{measurement}}
${\mathsf M}_{C (\Omega)}  ({}  {\mathsf O}_{123} , S_{[{}{{{\omega_k}} }]}{})  $,
the probability that

${{\omega_k}} $
is red
is given by
\par
\noindent
\begin{align*}
\frac {
[{}F_{13} ({} \{ y_{{\SW}}  \}\times 
\{ y_{{\RD}} \}){}] ({}{{\omega_k}} {})
}
{
[{}F_{13} ({} \{ y_{{\SW}}  \}\times  \{ y_{{\RD}} \}){}] ({}{{\omega_k}} {})
+
[{}F_{13} ({} \{ y_{{\SW}}  \}\times  \{ n_{{\RD}} \}){}] ({}{{\omega_k}} {})
}
=
\frac{
[{}F_{\RD} (\{ y_{\RD} \}{}){}] ({}{{\omega_k}}{})
}
{
[{}F_{\RD} (\{ y_{\RD} \}{}){}] ({}{{\omega_k}}{})
}
=
1
%%%% 4.25
%P%\tag{6.51}
\tag{\color{black}{5.10}}
\end{align*}
Of course,
\textcolor{black}{(5.9)}
%respectively
means
%({})
\begin{align*}
\text{ {\lq\lq Sweet\rq\rq}  $\Longrightarrow$ {\lq\lq Ripe\rq\rq}  }
\qquad
\text{}
\qquad
\text{ {\lq\lq Ripe\rq\rq}  $\Longrightarrow$ {\lq\lq Red\rq\rq}  }
%%%% 4.26
%P%\tag{6.52}
\end{align*}
Therefore, by \textcolor{black}{(5.10)},
we get the following conclusion.
\begin{align*}
\text{ {\lq\lq Sweet\rq\rq}  $\Longrightarrow$ {\lq\lq Red\rq\rq}  }
%%%% 4.27
%P%\tag{6.53}
\end{align*}

\par
\noindent
However,
it is not useful in the market.
What we want to know
is such as
\begin{align*}
\text{ {\lq\lq Sweet\rq\rq}  $\Longrightarrow$ {\lq\lq Sweet\rq\rq}  }
%%%% 4.28
%P%\tag{6.54}
\end{align*}

This will be discussed in the following example.
%\qed

\par
\par
\noindent
\bf
%BFBF
{Example 5.13}
\rm
[{}Continued from Example 5.12, \cite{IFuzz}{}].
Instead of \textcolor{black}{(5.9)},
assume that
\begin{align*}
{\mathsf O}_1^{ \{y_1 \}}
\underset{ {\mathsf M}_{C({}\Omega{}) }({}{\mathsf O}_{12} ,S_{ [\delta_{\omega_n}{}] }{}) }
{ \Longleftarrow}
{\mathsf O}_2^{ \{y_2 \}} ,
\qquad
{\mathsf O}_2^{ \{y_2 \}}
\underset{ {\mathsf M}_{C({}\Omega{}) }({}{\mathsf O}_{23} ,S_{[{}\delta_{\omega_n}{}] }{}) }
{ \Longrightarrow}
{\mathsf O}_3^{ \{y_3 \}} .
%\ta
\tag{5.11} 
\end{align*}
%Assume the notation \textcolor{black}{(\REF{3.43})}.
When we observe that
the tomato
$\omega_n $
is {\lq\lq RED\rq\rq}$\!\!,\;$
we can infer,
by
the fuzzy inference
${\mathsf M}_{C({}\Omega)}({}  {\mathsf O}_{13} , S_{ [\delta_{\omega_n}{}] }{})  $,
the probability that
the tomato
$\omega_n $
is {\lq\lq SWEET\rq\rq}
is given by
\begin{align*}
Q
=
\frac {
[{}F_{13} ({} \{ y_{{\SW}}  \}\bigtimes  \{ y_{{\RD}} \}){}] ({}\omega_n {})
}
{
[{}F_{13} ({} \{ y_{{\SW}}  \}\bigtimes  \{ y_{{\RD}} \}){}] ({}\omega_n {})
+
[{}F_{13} ({} \{ n_{{\SW}}  \}\bigtimes  \{ y_{{\RD}} \}){}] ({}\omega_n {})
}
%\label4.30
%\label{3.50}
\end{align*}
which is, by \textcolor{black}{(5.8)}, estimated as follows:
%%%
%\allowdisplaybreaks
\begin{align*}
&
\; \; \;
\max \left\{
\frac{ [{}F_{\RP} ({}\{ y_{\RP} \}{})]  ({}\omega_n{})}
{ [{}F_{\RD} ({}\{ y_{\RD} \}{})]  ({}\omega_n{})} ,
\frac{
[{}F_{\SW} ({}\{ y_{\SW} \}{})]
+
[{}F_{\RD} ({}\{ y_{\RD} \}{})]
-1
}
{ [{}F_{\RD} ({}\{ y_{\RD} \}{})]  ({}\omega_n{})} \right\}
\nonumber
\le
Q
\le
\min
\{
\frac{ [{}F_{\SW} ({}\{ y_{\SW} \}{})]  ({}\omega_n{})  }
{[{}F_{\RD} ({}\{ y_{\RD} \}{})]  ({}\omega_n{})} ,
\;
1
\} .
%\label4.31
%\label{3.51}
\tag{9.12}
\end{align*}
\par
\noindent
Note that
\textcolor{black}{(5.11)}
%respectively
implies
({}and is implied by{})
\begin{align*}
\text{ {\lq\lq RIPE\rq\rq}  $\Longrightarrow$ {\lq\lq SWEET\rq\rq}  }
\qquad
\text{and}
\qquad
\text{ {\lq\lq RIPE\rq\rq}  $\Longrightarrow$ {\lq\lq RED\rq\rq}  } .
%\label4.32
%\label{3.52}
\end{align*}
And note that
the conclusion \textcolor{black}{(5.12)}
is somewhat like
\begin{align*}
\text{ {\lq\lq RED\rq\rq}  $\Longrightarrow$ {\lq\lq SWEET\rq\rq}  }.
%\label4.33
%\label{3.53}
\end{align*}
Therefore,
this conclusion
is peculiar to {\lq\lq fuzziness\rq\rq}$\!.$
\par \hfill{$///$}
%\END{Exa}
%BFBF

%BBBBBBBBBBBBBBBBBB%SBSBSBS
\par
\noindent
{\small%%{\footnotesize
\vspace{0.1cm}
\begin{itemize}
\item[$\spadesuit$] \bf {{}}{Note }5.8{{}} \rm
Recall
the
\textcolor{black}{{({}A$_2$) in Sec.5.1}},
that is,
\begin{itemize}
\item[($\sharp_1$)]
Since Socrates is a man and all men are mortal,
it follows that Socrates is mortal.
%{Socrates},
%, .   , {Socrates}.
\item[($\sharp_2$)]
Flying arrow is not moving.
\item[($\sharp_3$)]
I think, therefore I am.
\item[($\sharp_4$)]
Edison's "$1+1=2$"
\end{itemize}
In this chapter,
the above ($\sharp_1$)
and
($\sharp_3$)
are clarified.

%3,
% and .
If these four statements are
childish,
we can not explain the fact that
these tales have been transmitted from generation to generation,
and
these have been continuously
passed by
many persons with
sharp sensibility.
However,
measurement theory urges us to
understand ($\sharp_1$)--($\sharp_4$)
without sharp sensibility.
Recall \textcolor{black}{Chap. 1 (X$_1$)}:
%index{@{Chap.$\;$1}(X$_1$)}
\\
\\
$\underset{\text{(Chap. 1)}}{\text{(X$_1$)}}$
{\small
$\overset{
%({ordinary language})
}{\underset{\text{\scriptsize (before science)}}{
\text{
\fbox
{{\textcircled{\scriptsize 0}}
widely {ordinary language}}
}
}
}
$
$
\underset{\text{\scriptsize }}{\text{$\Longrightarrow$}}
$
$
\underset{\text{\scriptsize (Chap. 1(O))}}{\text{{world-description}}}
%\cases
%
%
%
%
%
%
%
%$\underset{({Chap.$\;$1})}{\text{(X$_1$)}}$
%$\overset{
%({ordinary language})
%}{\underset{({world-description}(=))}{
%\text{
%\fbox
%{{\textcircled{\scriptsize 0}}
%{ordinary language}}
%}
%}
%}
%$
%$
%\underset{\text{\scriptsize }}{\text{$\Longrightarrow$}}
%$
%$
%\underset{\text{\scriptsize ({Chap.$\;$1}(O))}}{\text{{world-description}}}
\cases
&
\!\!\!\!\!\!
%\textcircled{\scriptsize 1}:
%\underset{\scriptsize
%\text{}}
{\text{\textcircled{\scriptsize 1}{realistic method}}}
\\
&
{\text{({realistic world-view})}}
\\
\\
%\textcircled{\scriptsize 2}:
&
\!\!\!\!\!\!
%\underset{\scriptsize
%\text{}}
{\text{\textcircled{\scriptsize 2}{linguistic method}}}
\\
&
{\text{(linguistic world-view)}}
\endcases
$
}
\par
\noindent
If we believe in this,
our problem is as follows.
\begin{itemize}
\item[]
Should each
($\sharp_1$)--($\sharp_4$)
be
discussed in
%,
\textcircled{\scriptsize 0},
\textcircled{\scriptsize 1}
or
\textcircled{\scriptsize 2}
?
\end{itemize}
In this chapter,
($\sharp_1$)
and
%\textcircled{\scriptsize 2}
($\sharp_3$)
are explained.
The
($\sharp_2$)
and
($\sharp_4$)
will
be discussed in Chap. 11.
\end{itemize}
}
%%BBBBBBBBBBBBBBBBBequilibrium statistical mechanics
\par
\noindent

\par

%TAG{1.\tag2.%P%\TAG{%P%\tag{6.%P%\tag{5.%P%\TAG{%P%\tag{7.%P%\tag{8.%P%\tag{9.%P%\tag{10.
%TAG{1.\tag2.%P%\TAG{%P%\tag{6.%P%\tag{5.%P%\TAG{%P%\tag{7.%P%\tag{8.%P%\tag{9.%P%\tag{10.
%TAG{1.\tag2.%P%\TAG{%P%\tag{6.%P%\tag{5.%P%\TAG{%P%\tag{7.%P%\tag{8.%P%\tag{9.%P%\tag{10.
%\END{align*}\tag\tag)])])])]\iten\iten\item\item
%\END{align*}\tag\tagg)])])])]\iten\iten\item\item
%\END{align*}%%%%%\TAG
%\END{align*}\tag\tagg)])])])]\iten\iten\item\item
%\END{align*}\tag\tag
%TAG{1.\tag2.%P%\TAG{%P%\tag{6.%P%\tag{5.%P%\TAG{%P%\tag{7.%P%\tag{8.%P%\tag{9.%P%\tag{10.
%\END{align*}\tag\tag
%\END{align*}\tag\tag
%TAG{1.\tag2.%P%\TAG{%P%\tag{6.%P%\tag{5.%P%\TAG{%P%\tag{7.%P%\tag{8.%P%\tag{9.%P%\tag{10.
%TAG{1.\tag2.%P%\TAG{%P%\tag{6.%P%\tag{5.%P%\TAG{%P%\tag{7.%P%\tag{8.%P%\tag{9.%P%\tag{10.
%TAG{1.\tag2.%P%\TAG{%P%\tag{6.%P%\tag{5.%P%\TAG{%P%\tag{7.%P%\tag{8.%P%\tag{9.%P%\tag{10.
%%%%%%%2.00]:{{}}{Remark }3.00]:{{}}{Remark }4.00]:{{}}{Remark }5.00]:{{}}{Note }6.00]:
%%%%%%%7.00]:{{}}{Note }8.00]:{{}}{Note }9.00]:{{}}{Note }11.00]:{{}}{Remark }4.00]:
%TAG{1.\tag2.%P%\TAG{%P%\tag{6.%P%\tag{5.%P%\TAG{%P%\tag{7.%P%\tag{8.%P%\tag{9.%P%\tag{10.
%TAG{1.\tag2.%P%\TAG{%P%\tag{6.%P%\tag{5.%P%\TAG{%P%\tag{7.%P%\tag{8.%P%\tag{9.%P%\tag{10.
%TAG{1.\tag2.%P%\TAG{%P%\tag{6.%P%\tag{5.%P%\TAG{%P%\tag{7.%P%\tag{8.%P%\tag{9.%P%\tag{10. }

%666666666666666666666666666666666

\vskip2.0cm
\newpage
%666666666666666666666666666666666
\section{Axiom${}_{\text{\scriptsize c}}^{\text{\scriptsize pm}}$ 2 - causality
\label{Chap6}
}%{Chap.{\;}}{}
%%\vspace{-0.8cm}
%{(}\textcolor{black}{Axiom}2(POI){)}}
%\chapter[{}(\textcolor{black}{Axiom}2(POI))]{{}
%\\
%{(}\textcolor{black}{Axiom}2(POI){)}}
\pagestyle{headings}
\baselineskip=18pt
\rm
{\small%%{\footnotesize
\begin{itemize}
\item[{}]
{
\baselineskip=15pt
\par%[Abstract].
\rm
Measurement theory is formulated as follows:
\begin{align*}
\underset{\text{\scriptsize (scientific language)}}{\text{{} $\fbox{{{measurement theory}}}$}}
:=
{
%\overset{\text{\scriptsize [Axiom 1\textcolor{black}{(\REF{2secAxiom 1})}]}}
\overset{\text{\scriptsize [Axiom
%${}_{\text{\scriptsize c}}^{\text{\scriptsize p}}$
 1]}}
{
\underset{\text{\scriptsize
[probabilistic interpretation]}}{\text{{} $\fbox{{{measurement}}}$}}}
}
+
{
%\overset{\text{\scriptsize [Axiom 2\textcolor{black}{(\REF{6secAxiom 2})}]}}
\overset{\text{\scriptsize [Axiom
%${}_{\text{\scriptsize c}}^{\text{\scriptsize p}}$
 2]}}
{
\underset{\text{\scriptsize [{{the Heisenberg picture}}]}}
{\text{{}$\fbox{ causality }$}}
}
}
\end{align*}
Although the preceding chapter was introduction of the Axiom 1 about measurement,
from this chapter,
I explain the Axiom 2 about movement and change($\approx$causality).
If I say definitely roughly,
\begin{itemize}
\item[($\sharp$)]
Science is the learning of causality,
i.e., the learning about the phenomenon which can be expressed in the word "causality."
\end{itemize}
Therefore, in this chapter, we arrived at "causal relationship" in the main question at last.
However, in dualism, after understanding "measurement" enough, unless it comes out,
we cannot understand "movement and change."
In dualism, it is because we understand "movement and change"
as a debt of "measurement" and "causality."
}
\end{itemize}
}
\baselineskip=18pt
\def\BBbZ{{{\Bbb Z}}}

%------------------------------------------
\rm
\subsection{Outstanding-problem -What is causality? }%6.1
\subsubsection{Modern science started from the discovery of "causality."}%6.1.1
\par
When a certain thing happens, the cause exists.
This is called {\bf causality}.
You should just remember
the proverb of "{\bf smoke is not located on the place which does not have fire}."
%index{ and @ and , }
It is not so simple although you may think that it is natural.
For example, if you consider
\begin{itemize}
\item[]
Is it because that my feeling feels it refreshed this morning went to sleep well last night?
$\;\;$
,or is it because I go to favorite golf from now on?
\end{itemize}
you may be able to understand the difficulty of how to use the word "causality.
%Probably, especially about the difficulty of logic including time progress,
%you should just consider "The problem of the kinematic pair of [He does not study, if not scolded. ]" 
%of a high school student puzzle.
%(See the \textcolor{black}{footnote 1} of Chapter 8 about this answer. )
In daily conversation, it is used in many cases, mixing up "a cause (past)", "a reason (connotation)",
and "the purpose and a motive (future)."
\par
%(BC460--BC370)
It may be supposed that Heraclitus's{(}BC.540 -BC.480{)} "Everything changes."
and "Movement does not exist." of Parmenides {(}born around BC. 515{)}who is Zeno's teacher
are the beginning of research of movement and change.
%,
However, those meanings are not clear.
%({\rm cf. }\textcolor{black}{{\cite{Hiro, Naga}}}).
%index{ and @}
%index{@}
However, these two pioneers
-
Heraclitus and Parmenides
-
noticed first
that "movement and change" were the primary importance keywords in science(= "world description"{)} ,
i.e., it is
\begin{align*}
\text{
\bf
[World description ]=
[Description of movement and change ]
}
\end{align*}
%.
\par
However, Aristotle(BC384--BC322) further investigated about the essence of movement and change,
and he thought that all the movements had the "purpose."
%%index{ and @}
For example, supposing a stone falls,
that is because there is the purpose that the stone tries to go downward.
Supposing smoke rises,
that is because there is the purpose that smoke rises upwards.
Under the influence of Aristotle, "{\bf Purpose}" continued remaining
as a mainstream idea of "Movement"  for a long time of 1500 or more.
%index{@}
\par
Although "the further investigation" of Aristotle was what should be praised,
it was not able to be said  that "the purpose was to the point."
In order to free ourselves from Purpose and for human beings to discover
that the essence of movement and change is "causal relationship",
we had to wait for the appearance of Galileo, bacon, Descartes, Newton, etc.
$$
\text{
Revolution to "Causality" from "Purpose"
}
$$
is the greatest history-of-science top paradigm shift
-
It is not an overstatement even if we call it "{\bf birth of modern science}.
-
, and determined the "scientific revolution" after it.
%\footnote{
%In the present age, since it is supposed that "causality" is common sense,
%we may think that it is natural,
%but discovery of "causality" is the greatest history-of-science top paradigm shift
%on a par with the "Copenhagen theory" and the "theory of evolution."
%-
%All three smashed the outlook on nature since Aristotle.
%-
%In these three, "causality" is outstanding.
%It is because it is not an overstatement even if we say
%\BEGIN{itemize}
%\item[($\sharp$)]
%Science is the learning of causality, i.e., the learning about the phenomenon
%which can be expressed with the word "causality"
%-
%the proverb "smoke is not located on the place which does not have fire"
%- .
%\END{itemize}
%
%The science in the inside of ordinary language may have thousands of years or more of history.
%However, I thought that the scientific quality was different
%before discovery of causality and after that
%, and I say "modern science."
%}
\rm
\subsubsection{Four answers to "what is causality?"}%6.1.2
\par
As mentioned above, about "what is an essence of movement and change?",
it was once settled with the word "causality."
However, not all were solved now.
We do not yet understand "causality" fully.
In fact,
%:
%\BEGIN{align*}
%\text{
% and ?
%}
%\END{align*}
%.
% and ,
%, .
%,
\begin{itemize}
\item[]
\hspace{-0.5cm}
\bf
"What is causality?" is the most important outstanding problems in science.
\end{itemize}
\rm
%index{@( and ?)}
\par
\noindent
There may be a reader who is surprised with saying like this
although it is the outstanding problems in the present.
Below, I arrange the history of the answer to this problem.
%\footnote{
%
%{(},
%, , , \; \cdots \; 
%), 
% and ?
%
%.
%,
%, 
%, 
% and .
%,
%
% and ?
%
%{(}\textcolor{black}{7.5}{)}.
%}.
%\END{itemize}
%}
%

\par

%index{@}
\begin{itemize}
\item[
(a)
%\textcircled{\scriptsize 1}
]
{\bf
[Realistic causality]:
}
Newton advocated the realistic describing method of Newtonian mechanics as a final settlement of accounts of ideas, such as Galileo, bacon, and Descartes, and he thought as follows.
:
\begin{itemize}
\item[]
"Causality" actually exists in the world.
The equation of motion of Newton described faithfully this "causality"
that exists in fact by the differential equation
-
the equation of a causal chain
-.
\end{itemize}
\end{itemize}
This realistic causality may be a very natural idea,
and you may think that you cannot think in addition to this.
In fact, probably, we may say that the current of the realistic causal relationship
which continues like
"Newtonian mechanics$\longrightarrow$
Electricity and magnetism$\longrightarrow$
Theory of relativity$\longrightarrow$
$\; \cdots \;$"
is a scientific flower.

%,
However, there is also another idea and there is three "nonexistent causalities" as follows.
\begin{itemize}
\item[
(b)
%\textcircled{\scriptsize 2}
]
{\bf
[Cognitive causality]:
}
Hume, Kant, etc. who are philosophers thought as follows. :
\begin{itemize}
\item[]
We can not say that "Causality" actually exists in the world, or that it do not exist in the world.
And when we think that "something" in the world is "causality",
we should just believe that the it has "causality".
%${{\cdot}}$.
%${{\cdot}}$ and  and , ${{\cdot}}$ and .
% and .
\end{itemize}

\end{itemize}
Several readers may regard this as it being "a kind of rhetoric", 
moreover, several readers may be convinced in "That is right if you say so."
Surely, since you are looking through the prejudice "causality", you may look such.
It is Kant's famous "Copernican revolution"(that is, "recognition constitutes the world." ) which is considered
that the recognition circuit of causality is installed in the brain,
and when it is stimulated by "something" and reacts, "there is causal relationship."
(Refer to later \textcolor{black}{Section 8.1}.)
Probably, many readers doubt about the substantial influence which this (b) had on the science after it.
However, in this book(Refer to later \textcolor{black}{Section 8.1}.) , I adopted the friendly story to the utmost to Kant.
% and  and .
%(\textcolor{black}{8.1}).
%index{@}
\par
\noindent
\begin{itemize}
\item[
(c)
%\textcircled{\scriptsize 3}
]
{\bf
[Mathematical causality{(}Dynamical system theory{)}]:
}
Since dynamical system theory has developed as the mathematical technique in engineering,
they have not investigated "What is causality?" thoroughly.
However,
\begin{itemize}
\item[]
In dynamical system theory, we think that there is mathematics of an equation of state previously
(i.e., the time first degree alliance differential equation of the variable \textcolor{black}{(1.1)})
, and the phenomenon described with the equation has "causality."
(Refer to \textcolor{black}{{Chap.$\;$1}(E$_1$)}).
\end{itemize}

\end{itemize}
With the ordinary feeling of science, you may tend to understand this (c)
because you thinks somehow "=."
However, you should be cautious of it being a typical example of the form of the mathematics buried into ordinary language.
However, for the purpose of "Being helpful", I think that (c) should be evaluated more.
\par
\noindent
\begin{itemize}
\item[
(d)
%\textcircled{\scriptsize 4}
]
{\bf
[Linguistic causal relationship {(}Measurement{}Theory{)}]:
}
The causal relationship of measurement theory is decided by the Axiom 2 of this chapter.
If I say in detail,:
\begin{itemize}
\item[]
Although measurement theory consists of the two \textcolor{black}{Axioms} 1 and 2,
it is the \textcolor{black}{Axiom} 2 that is concerned with causal relationship.
When describing a certain phenomenon in a language called measurement theory and using the \textcolor{black}{Axiom 2},
we think that the phenomenon has causality.
\end{itemize}
%.
\end{itemize}
Although it is above, it is the next difference when (a)--(d) is summarized.
\begin{itemize}
\item[]
(a)
%\textcircled{\scriptsize 1}
World is first$\;\;$
(b)
%\textcircled{\scriptsize 2}
Recognition is first$\;\;$
\\
(c)
%\textcircled{\scriptsize 3}
Mathematics(buried into ordinary language)  is first
$\;\;$
\\
(d)
%\textcircled{\scriptsize 4}
Language (Measurement Theory) is first
\end{itemize}
\par
Now, in measurement theory, we assert the next as said repeatedly. :
\begin{itemize}
\item[]
$\qquad
\qquad$
Measurement theory is a basic language which describes various sciences.
\end{itemize}
Supposing this is recognized, we can assert the next.
Namely,
\begin{itemize}
\item[]
$\quad
\quad$
{\bf
In science, causality is claimed in upper (d).
}
\end{itemize}
This is an answer of measurement theory to "What is causality?",
and I explain these details after the following paragraph.
\par\noindent
%BBBBBBBBBBBBBBBBBB%SBSBSBS
{\small%%{\footnotesize
\vspace{0.1cm}
\begin{itemize}
\item[$\spadesuit$] \bf {{}}Note 6.1{{}} \rm
For the question
"What is space-time?",
there are two answers as follows.
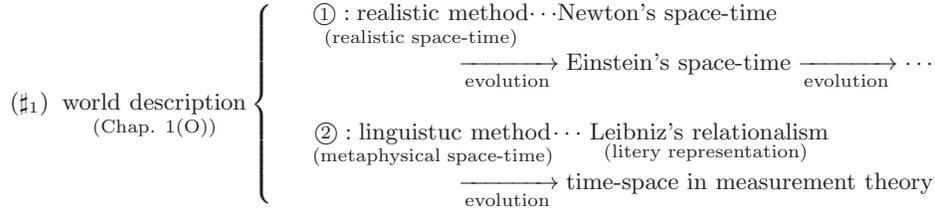
\begin{itemize}
\item[($\sharp_1$)]
%\BEGIN{itemize}
%\item[]
$
\underset{\text{\scriptsize (Chap. 1(O))}}{\text{world description}}
\cases
%\textcircled{\scriptsize 1}:
&
\underset{\scriptsize
\text{(realistic space-time)}}{\textcircled{\scriptsize 1}: \text{realistic method}}
{\cdots}
\text{Newton's space-time}
\\
&
\quad
\qquad
\qquad
\quad
\xrightarrow[\text{\scriptsize evolution}]{}
\text{Einstein's space-time}
\xrightarrow[\text{\scriptsize evolution}]{}
\cdots
\\
\\
%\textcircled{\scriptsize 2}:
&
\underset{\scriptsize
\text{(metaphysical space-time)}}{\textcircled{\scriptsize 2}:
\text{linguistuc method}}
%\textcircled{\scriptsize 2}: \text{realistic method}
{\cdots}
\underset{\text{\scriptsize (litery representation)}}{\text{
Leibniz's relationalism}}
\\
&
\quad
\qquad
\qquad
\quad
\xrightarrow[\text{\scriptsize evolution}]{}
\text{time-space in measurement theory}
\endcases
$
\end{itemize}
Concerning
"What is causality?",
a similar argument is possible.
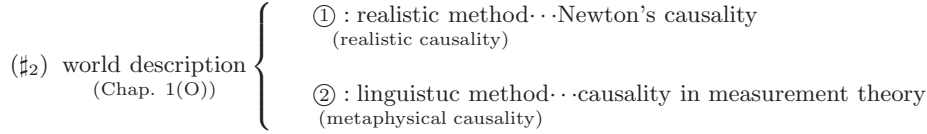
\begin{itemize}
\item[($\sharp_2$)]
$
\underset{\text{\scriptsize (Chap. 1(O))}}{\text{world description}}
\cases
%\textcircled{\scriptsize 1}:
&
\underset{\scriptsize
\text{(realistic causality)}}{\textcircled{\scriptsize 1}: \text{realistic method}}
{\cdots}
\text{Newton's causality}
%
%\\
%&
%\quad
%\qquad
%\qquad
%\quad
%\xrightarrow[\text{\scriptsize evolution}]{}
%\text{Einstein's space-time}
%\xrightarrow[\text{\scriptsize evolution}]{}
%\cdots
\\
\\
%\textcircled{\scriptsize 2}:
&
\underset{\scriptsize
\text{(metaphysical causality)}}{\textcircled{\scriptsize 2}:
\text{linguistuc method}}
%\textcircled{\scriptsize 2}: \text{realistic method}
{\cdots}
%
%\underset{\text{\scriptsize (litery representation)}}{\text{
%Leibniz's relationalism}}
%\\
%&
%\quad
%\qquad
%\qquad
%\quad
%\xrightarrow[\text{\scriptsize evolution}]{}
\text{causality in measurement theory}
\endcases
$
%
%
%
%
%
%
%
%
%
%
%
%
%$
%\underset{\text{\scriptsize ({Chap.{\;}}1{}(O))}}{\text{}}
%\cases
%\textcircled{\scriptsize 1}:
%\underset{\scriptsize
%\text{}}{}
%\; \cdots \; 
%&
%\text{(a)}
%\\
%\\
%\textcircled{\scriptsize 2}:
%\underset{\scriptsize
%\text{}}{}
%%\textcircled{\scriptsize 2}: 
%\; \cdots \; 
%&
%\underset{(POI)}{\text{(d)}}
%\ENDcases
%$
\end{itemize}
Also,
\begin{itemize}
\item[$(\sharp_3)$]
In \textcolor{black}{Sec. 8.1},
we discuss
the relation
among
(b), (c) and (d).
\end{itemize}
\end{itemize}
}
\par
\noindent
%BBBBBBBBBBBBBBBBBB%SBSBSBS
{\small%%{\footnotesize
\begin{itemize}
\item[$\spadesuit$] \bf {{}}Note 6.2{{}} \rm
As one of the by-products of measurement theory,
I can reply to the outstanding problems
\begin{itemize}
\item[($\sharp_1$)]
What are time, space, causality, and probability?
\end{itemize}
in a metaphysical position(linguistic position of an anti physics supreme principle).
In metaphysics, answering to "What is "
is defining how to use the language .
(
\textcolor{black}{Note 2.3},
\textcolor{black}{Note 5.6}
).
%%index{@(
%${{\cdot}}$ and ?)}
%%index{@(
%${{\cdot}}$ and ?)}
Therefore, this ($\sharp_1$) is equivalent to the following ($\sharp_2$).
\begin{itemize}
\item[($\sharp_2$)]
To propose the linguistic universe describing method containing the word "time, space, causality, and probability."
% and
\end{itemize}

Of course, in this book, measurement theory (i.e., establishment of the linguistic method)
is proposed as this answer.
\end{itemize}
}

\rm

\rm
\subsection{Causality
---
No smoke without fire}
%index{ and @No smoke without fire}
\subsubsection{{{The Heisenberg picture}}
and {{the Schr\"odinger{ picture}}}}
\par
%,
%{{measurement theory}}causality .
%\textcolor{black}{{Chap.{\;}}2{}} and {},
%${{\cdot}}$, ,
%causality , .

\def\PAR{\roman{par}}
\noindent
\par
Let
$\Omega$
be a state space.
state space.
Let
$C(\Omega)$
be a space of all
real continuous valued functions
on $\Omega$.
Also, recall
${\cal M}(\Omega)$
and
${\cal M}_{+1}(\Omega)$
in
\textcolor{black}{{Sec.4.4.1}(b)}.

\par
\noindent
{\bf {Definition }6.2
[{{Causal operator}}{\rm{(causal operator)}}]}$\;\;$%POPOPO
%index{@{{causal operator}}}
Let
$\Omega_1$
and
$\Omega_2$
be state spaces.
a continuous linear operator
$\Phi_{1,2}:C(\Omega_2)
\to C(\Omega_1) $
is called a causal operator
(or, Markov causal operator)
if it
satisfies
(i)$\text{---}$(iii):
%index{@causality {{the Heisenberg picture}}}
\begin{itemize}
\item[(i)]
$f_2 \in C(\Omega_2), \;\; f_2 {\; \geqq \;}0$
$\Longrightarrow$
$\Phi_{12}f_2 {\; \geqq \;}0$
\item[(ii)]
$\Phi_{12} 1_2 = 1_1$
where,
$1_k (\omega_k ) = 1$
$(
\forall \omega_k \in \Omega_k, k=1,2)$
\item[(iii)]
There exists a continuous linear operator${\Phi}^*_{1,2}:{\cal M}_{}(\Omega_1)
\to {\cal M}_{}(\Omega_2) $
such that
%\footnote{
%${\Phi}^*_{1,2}:{\cal M}_{+1}(\Omega_1)
%\to {\cal M}_{+1}(\Omega_2) $
%.
%{,
%$\Omega$
%,
%(i) and (ii)(iii),
%(iii) and .
%}
%}
:
\begin{align*}
&
\int_{\Omega_1}  [\Phi_{1,2} f_2](\omega_1) \;\;\rho_1 (d \omega_1 )
=
\int_{\Omega_2}   f_2(\omega_2)
\;\;
({\Phi}^*_{1,2} \rho_1) (d \omega_2 )
%\phi_{1,2}:{\cal M}_{+1}(\Omega_1)
\\
& \hspace{4cm}
(\forall \rho_1 \in {\cal M}(\Omega_1),
\forall f_2 \in C(\Omega_2))
%%P%\tag{7.6}
\end{align*}
If $\Omega$ is compact,
this condition is
a consequence of the above
(i) and (ii).
\end{itemize}

The {{causal operator}}
$\Phi_{1,2}:C(\Omega_2)
\to C(\Omega_1) $
is regarded as
{{the Heisenberg picture}}
representation of
the
"causality".
Also, the dual causal operator
${\Phi}^*_{1,2}$
${\Phi}_{1,2}$
is
called
{{the Schr\"odinger picture}}
representation of
the
"causality".
%index{@dual {{causal operator}}}
%index{faiover@${\Phi}^*_{1,2}$:dual {{causal operator}}}
%\END{itemize}
In addition,
the {{causal operator}}${\Phi}_{1,2}$
is called a
{\bf
{{deterministic causal operator}}},
if
there exists
a
{continuous map }$\phi_{1,2}:\Omega_1 \to \Omega_2$
such that
%index{@{{deterministic causal operator}}}
\begin{align*}
&
[\Phi_{1,2}f_2](\omega_1)
=
f_2(\phi_{1,2}(\omega_1))
\quad
(\forall f_2 \in C(\Omega_2 ), \forall \omega_1
\in \Omega_1 )
\tag{\color{black}{6.1}}
\end{align*}
\rm
This {continuous map }$\phi_{1,2}:\Omega_1 \to \Omega_2$
is said to be a
{\bf {{deterministic causal map}}}.

%index{@{{deterministic causal map}}}
\par
\noindent
%\begin{figure}[htbp]
\par
\noindent
\unitlength=0.28mm
%\unitlength=0.35mm
\begin{picture}(500,130)
\put(90,9){\footnotesize $\omega_1$}
\put(300,9){\footnotesize $\phi_{1,2}(\omega_1)$}
\qbezier(103,9)(200,0)(290,9)
\put(290,9){\line(-4,1){11}}
\put(290,9){\line(-2,-1){11}}
\put(390,16){$\Omega_2$}
\put(0,16){$\Omega_1$}
%\dottedline{3}(40,110)(340,110)
\put(94,84){\line(1,0){200}}
\path(104,86)(94,84)(104,82)
\path(94,84)(94,20)
\put(94,20){\circle*{5}}
\path(294,84)(294,20)
\put(294,20){\circle*{5}}
\path(292,74)(294,84)(296,74)
\put(350,80){$f_2$}
\put(30,80){$\Phi_{1,2} f_2$}
%\put(20,20){\line(0,1){100}}
%\linethickness{0.15mm}
\thicklines
\put(20,20){\line(1,0){160}}
\put(220,20){\line(1,0){160}}
%\linethickness{0.15mm}
\thicklines
\spline(20,50)(40,40)(80,100)(100,78)
(120,63)(140,70)(160,80)(180,110)
\spline(220,95)(240,60)(284,68)(300,94)
(320,100)(340,80)(360,70)(380,50)
%\sspline(120,20)(130,20)(160,30)(250,50)
%%(270,80)(280,100)(300,105)(340,110)
%\sspline(40,40)(60,45)(80,50)(100,60)
%(150,90)(250,80)
%(270,50)(280,30)(300,25)(340,20)
\end{picture}
\begin{center}{Figure 6.1:
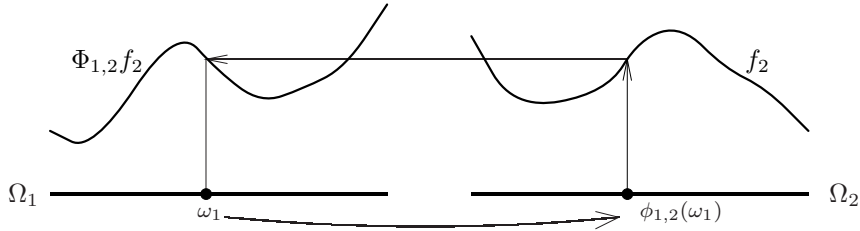
{{deterministic causal map}}$\phi_{1,2}$ and {{deterministic causal operator}}$\Phi_{1,2}$
}
\end{center}
\rm
\par

%
%\par
%\noindent
%{
%\par
%\noindent
%%%%%eepic
%%\vskip-0.4cm%%%
%\BEGIN{FIGUre}[htbp]
%\unitlength=0.35mm
%\BEGIN{picture}(500,130)
%\put(-50,0){
%%\put(0,0)%%%{AAAAA}
%\put(100,0){
%\put(50,85){\vector(1,0){158}}
%%\sspline(40,90)(80,110)(120,90)
%}
%%%%%%%%%%%%%%%%%%%%%%%%%%%%%
%%\put(200,0){
%%\put(50,85){\vector(1,0){98}}
%%}
%%%%%%%%%%%%%%%%%%%%%%%%%%%%%
%\thicklines
%\put(100,0){
%\put(40,70){\circle{100}}
%%\put(50,85){\vector(1,0){98}}
%\put(50,85){\circle*{4}}
%\put(47,10){$\Omega_1$}
%\put(47,75){$\omega_1$}
%\put(35,60){\scriptsize (POI){{state}}}
%\put(105,72){{{deterministic causal map}}$\phi_{1,2}$}
%%\put(105,80){$\phi_{1,2}$}
%%\sspline(40,90)(80,110)(120,90)
%}
%%%%%%%%%%%%%%%%%%%%%%%%%%%%%
%\put(260,0){
%\put(60,70){\circle{100}}
%%\put(50,85){\vector(1,0){98}}
%\put(50,85){\circle*{4}}
%\put(47,10){$\Omega_2$}
%\put(44,73){$\omega_2$}
%\put(35,60){\scriptsize (POI){{state}}}
%%\sspline(40,90)(80,110)(120,90)
%}
%%%%%%%%%%%%%%%%%%%%%%%%%%%%%
%%\put(300,0){
%%\put(50,70){\circle{70}}
%%\put(50,85){\circle*{4}}
%%
%%\put(47,14){$\Omega_2$}
%%%\put(-90,0){\spline(40,90)(80,110)(120,90)}
%%}
%}
%\END{picture}
%%\vskip-0.3cm
%\caption{
%{{deterministic causal map}}
%}
%\END{figure}
%}
%\par
%\noindent
%\par
%\noindent

\par
%\qed
\noindent
%{{causal operator}}$[\Phi_{1,2}{{}} \rm C(\Omega_2 )
%\to C (\Omega_1 )$
%{\bf causality {{the Heisenberg picture}}}
% and .
%,
%$\phi$
%
%{{deterministic causal map}} and ,
%$\Phi$
%{\bf {{deterministic causal operator}}} and .

%
%\par
%\par
%\noindent
%%%
%\unitlength=0.35mm
%\BEGIN{picture}(500,130)
%\put(90,9){\footnotesize $\omega_1$}
%\put(300,9){\footnotesize $\phi_{1,2}(\omega_1)$}
%\qbezier(103,9)(200,0)(290,9)
%\put(290,9){\line(-4,1){11}}
%\put(290,9){\line(-2,-1){11}}
%\put(390,16){$\Omega_2$}
%\put(0,16){$\Omega_1$}
%%\dottedline{3}(40,110)(340,110)
%\put(94,84){\line(1,0){200}}
%\path(104,86)(94,84)(104,82)
%\path(94,84)(94,20)
%\put(94,20){\circle*{5}}
%\path(294,84)(294,20)
%\put(294,20){\circle*{5}}
%\path(292,74)(294,84)(296,74)
%\put(350,80){$f_2$}
%\put(30,80){$\Phi_{1,2} f_2$}
%%\put(20,20){\line(0,1){100}}
%%\linethickness{0.15mm}
%\thicklines
%\put(20,20){\line(1,0){160}}
%\put(220,20){\line(1,0){160}}
%%\linethickness{0.15mm}
%\thicklines
%\sspline(20,50)(40,40)(80,100)(100,78)
%(120,63)(140,70)(160,80)(180,110)
%\sspline(220,95)(240,60)(284,68)(300,94)
%(320,100)(340,80)(360,70)(380,50)
%%\sspline(120,20)(130,20)(160,30)(250,50)
%%%(270,80)(280,100)(300,105)(340,110)
%%\sspline(40,40)(60,45)(80,50)(100,60)
%%(150,90)(250,80)
%%(270,50)(280,30)(300,25)(340,20)
%\END{picture}

%QQQQQQQQQQQQQQQQQQQQQQQQQQQQ
\par
\noindent
{\bf {{Theorem }}6.3
[{{Causal operator}} and observable ]}$\;\;$%POPOPO
For any observable $
{\mathsf O}_2$
$=$
$(X , {\cal F} , F_2{})$
in $C(\Omega_2)$,
the
{{}}$(X , {\cal F} , \Phi_{1,2} F_2{})$
is an observable in ${C (\Omega_1)}$.
We denote that
$\Phi_{1,2} {\mathsf O}_2$
$=$
$(X , {\cal F} , \Phi_{1,2} F_2{})$.
\par
\noindent
{\it $\;\;\;\;${Proof.}}{$\;\;$}
For any $\Xi$
$(\in {\cal F} )
$,
consider the countable decomposition
$\{\Xi_1, \Xi_2, \ldots, \Xi_n, \ldots\}$
$\Big($
that is,
$\Xi = \bigcup\limits_{n=1}^\infty \Xi_n $, 
$\Xi_n \in {\cal F}, (n = 1, 2, \ldots)$,
$\Xi_m \cap \Xi_n =\emptyset
\;\;
(m \not= n )$
$\Big)$.
Recalling the condition:
\textcolor{black}{(2.3)},
we see,
for any
$\rho_1 (\in {\cal M}(\Omega_1))$,
\begin{align*}
&
\int_{\Omega_1}
\Big[
[{}\Phi_{1,2} F_2{}]({}
\bigcup\limits_{n=1}^\infty \Xi_n
{})
\Big](\omega_1)
\;\;\rho_1 (d \omega_1 )
=
\int_{\Omega_2}
[ F_2{}({}
\bigcup\limits_{n=1}^\infty \Xi_n
{})]
(\omega_2)
\;\;{}{\Phi}^*_{1,2}\rho_1 (d \omega_2 )
\\
=
&
\sum\limits_{n=1}^\infty \int_{\Omega_2}
[ F_2{}({}
%\bigcup\limits_{n=1}^\infty
\Xi_n
{})]
(\omega_2)
\;\;{}{\Phi}^*_{1,2}\rho_1 (d \omega_2 )
=
\sum\limits_{n=1}^\infty
\int_{\Omega_1}
\Big[
[{}\Phi_{1,2} F_2{}]({}
%\bigcup\limits_{n=1}^\infty
\Xi_n
{})
\Big](\omega_1)
\;\;\rho_1 (d \omega_1 )
%
%
%
%\\
%=
%&
%
%
%
%
%
%
%
%
%\Phi_{1,2} ({}F_2 ({}
%\bigcup\limits_{n=1}^\infty \Xi_n
%{}))
%=
%\ssum\limits_{n=1}^\infty \Phi_{1,2} ({}F_2 ({}\Xi_n)
%%+ F_2 ({} \Xi'{}))
%\\
%=&
%[\Phi_{1,2} ({}F_2{})]({}\Xi)
%+
%[\Phi_{1,2} (F_2)]({}\Xi'{}).
\quad
%%%%3.11}
%%P%\tag{7.2}
\end{align*}
%$\rho_1$
% and ,
Thus, $\Phi_{1,2} {\mathsf O}_2$
$=$
$(X , {\cal F} , \Phi_{1,2} F_2{})$
is an observable in
$C(\Omega_1)$.
\qed

\par
\noindent
\vskip0.3cm
\par
\noindent
{\bf {{Theorem }}6.4}$\;\;$%POPOPO
Consider a {continuous map }$\phi_{1,2}:\Omega_1 \to \Omega_2$.
The
operator${\Phi}_{1,2}
:C(\Omega_2) \to C(\Omega_1)$
is defined by
the formula
\textcolor{black}{(6.1)},
that is,
\begin{align*}
&
[\Phi_{1,2}f_2](\omega_1)
=
f_2(\phi_{1,2}(\omega_1))
\quad
(\forall f_2 \in C(\Omega_2 ), \forall \omega_1
\in \Omega_1 )
\end{align*}
Then,
the
operator ${\Phi}_{1,2}
:C(\Omega_2) \to C(\Omega_1)$
is a
{{deterministic causal operator}}.
This means that
"{continuous map }={{deterministic causal map}}".
\par
\noindent
{\it $\;\;\;\;${Proof.}}$\;\;$
It suffices to show the existence of
the dual {{causal operator}}${\Phi}^*_{1,2}:{\cal M}_{}(\Omega_1)
\to {\cal M}_{}(\Omega_2) $.
This is shown as follows.
\begin{align*}
[{\Phi}^*_{1,2}
\rho_1]
(D_2)
=
\rho_1 ( \phi_{12}^{-1}(D_2))
\qquad
(\forall D_2 \in {\cal B}_{\Omega_2},
\forall \rho_1 \in {\cal M}(\Omega_1) )
%%P%\tag{7.3}
\tag{\color{black}{6.2}}
%%%%%REDREDREDREDREDRE
\end{align*}
Thus,
we get
the dual {{causal operator}} ${\Phi}^*_{1,2}:{\cal M}_{}(\Omega_1)
\to {\cal M}_{}(\Omega_2) $.
\qed
\par
\noindent
\vskip0.3cm
\vskip0.3cm
%BFBF
\par
\noindent
{\bf {{Theorem }}6.5}$\;\;$%POPOPO
Let
${\Phi}_{1,2}:C(\Omega_2)
\to C(\Omega_1) $
be a
{{{deterministic causal operator}}}.
Then, it holds:
\begin{align*}
{\Phi}_{1,2} (f_2 \cdot g_2 )
=
{\Phi}_{1,2} (f_2  )
\cdot
{\Phi}_{1,2} (g_2 )
\qquad
(\forall f_2, \forall g_2 \in C(\Omega_2 ))
\end{align*}
\par
\noindent
{\it $\;\;\;\;${Proof.}}$\;\;$
Let
$f_2$,
$
g_2$
be elements in
$C(\Omega_2)$
Let
$\phi_{1,2}:\Omega_1 \to \Omega_2$
be
a {{deterministic causal operator}}
of a {{deterministic causal operator}}
${\Phi}_{1,2}:C(\Omega_2)
\to C(\Omega_1) $.
Then,
we see:
\begin{align*}
&
[{\Phi}_{1,2} (f_2 \cdot g_2 )]
(\omega_1)
=
(f_2 \cdot g_2 )(\phi_{1,2}{(}\omega_1))
=
f_2(\phi_{1,2}{(}\omega_1))
\cdot
g_2(\phi_{1,2}{(}\omega_1))
\\
=
&
[{\Phi}_{1,2} (f_2 )]
(\omega_1)
\cdot
[{\Phi}_{1,2} (g_2 )]
(\omega_1)
=
[{\Phi}_{1,2} (f_2  )
\cdot
{\Phi}_{1,2} (g_2 )]
(\omega_1)
%\\
%&
\qquad
(
\forall \omega_1 \in \Omega_1)
%\forall f_2, \forall g_2 \in C(\Omega_2 ))
\end{align*}
This completes the proof.
\qed

\subsubsection{A simple example
---
A matrix representation in the case of finite state space}
%, }%{Sec.6.2.2}
\par
\noindent
{\bf Example 6.6
[{{Deterministic causal operator}}, deterministic dual {{causal operator}}, {{deterministic causal map}}]}$\;\;$%POPOPO
Define the two states space
$\Omega_1$
and
$\Omega_2$
such that
$\Omega_1=\Omega_2={\mathbb R}$.
Define the {{deterministic causal map}}
$\phi_{1,2} : \Omega_1 \to \Omega_2 $
such that
\begin{align*}
\omega_2 = \phi_{1,2}(\omega_1 )= 3 \omega_1 +2
\qquad
(\forall \omega_1 \in \Omega_1 ={\mathbb R})
\end{align*}
Then,
by \textcolor{black}{(6.2)},
we get
the deterministic dual
{{causal operator}}
${\Phi}^*_{1,2}:{\cal M}_{}(\Omega_1)
\to {\cal M}_{}(\Omega_2) $
such that
\begin{align*}
{\Phi}^*_{1,2} \delta_{\omega_1} = \delta_{3 \omega_1 +2}
\qquad
(\forall \omega_1 \in \Omega_1 )
\end{align*}
where
$\delta_{(\cdot)}$
is the point measure.
%(\in
%{\cal M}_{+1}(\Omega_1)
Also,
the
{{deterministic causal operator}}$\Phi_{1,2}:C(\Omega_2)
\to C(\Omega_1) $
is defined by
\begin{align*}
[\Phi_{1,2}
(f_2)](\omega_1 )
=
f_2(3 \omega_1 +2)
\qquad
(\forall f_2 \in C(\Omega_2), \forall \omega_1 \in \Omega_1 )
\end{align*}

\par
\noindent
\vskip0.3cm
\vskip0.3cm
%BFBF
\par
\noindent
{\bf Example 6.7
[Dual {{causal operator}}${{\cdot}}${{causal operator}}]}$\;\;$%POPOPO
Put
$\Omega_1=\{ \omega_1^1, \omega_1^2, \omega_1^3 \}$
and
$\Omega_2=\{ \omega_2^1 , \omega_2^2\}$.
And define
$\rho_1 (\in {\cal M}_{+1}(\Omega_1))$
such that
\begin{align*}
\rho_1= a_1 \delta_{\omega_1^1} +a_2 \delta_{\omega_1^2} +a_3 \delta_{\omega_1^3}
\quad
(0{{\; \leqq \;}}a_1,a_2 ,a_3 {{\; \leqq \;}}1,
a_1+a_2 +a_3=1)
\end{align*}
Then,
the dual {{causal operator}}
${\Phi}^*_{1,2}: {\cal M}_{+1}(\Omega_1) \to  {\cal M}_{+1}(\Omega_2)$
is represented by
\begin{align*}
{\Phi}^*_{1,2}(\rho_1 ) = &
(c_{11}a_1+c_{12}a_2+c_{13}a_3)\delta_{\omega_2^1}
+
(c_{21}a_1+c_{22}a_2+c_{23}a_3)\delta_{\omega_2^2}
\\
\;
&
(0 {{\; \leqq \;}}c_{ij} {{\; \leqq \;}}1,
\sum\limits_{i=1}^2 c_{ij} =1)
\end{align*}
 and .
Consider the identification:${\cal M}(\Omega_1) {\; \approx \;} {\mathbb R}^3$,
${\cal M}(\Omega_2) {\; \approx \;} {\mathbb R}^2$,
That is,
\begin{align*}
&
{\cal M}(\Omega_1)
\ni
\alpha_1 \delta_{\omega_1^1} +\alpha_2 \delta_{\omega_1^2} +
\alpha_3 \delta_{\omega_1^3}
\underset{\text{\scriptsize (identification)}}{\longleftrightarrow}
\bmatrix
\alpha_1
\\
\alpha_2
\\
\alpha_3
\\
\endbmatrix
\in {\mathbb R}^3
\\
&
{\cal M}(\Omega_2)
\ni
\beta_1 \delta_{\omega_2^1} +\beta_2 \delta_{\omega_2^2}
\underset{\text{\scriptsize (identification)}}{\longleftrightarrow}
\bmatrix
\beta_1
\\
\beta_2
\\
%\beta_3
%\\
\endbmatrix
\in {\mathbb R}^2
\end{align*}
%%%%%%%POIUYTREWQ
Then,
putting
\begin{align*}
&
{\Phi}^*_{1,2}(\rho_1 ) =
\beta_1\delta_{\omega_2^1}
+
\beta_2\delta_{\omega_2^1}
=
\bmatrix
\beta_1
\\
\beta_2
\\
%\beta_3
%\\
\endbmatrix,
\\
&
\rho_1
=
\alpha_1 \delta_{\omega_1^1} +\alpha_2 \delta_{\omega_1^2} +
\alpha_3 \delta_{\omega_1^3}
=
 \bmatrix
\alpha_1
\\
\alpha_2
\\
\alpha_3
\\
\endbmatrix
\end{align*}
write, by matrix representation,
as follows.
\begin{align*}
{\Phi}^*_{1,2} (\rho_1 ) =
\bmatrix
\beta_1
\\
\beta_2
\\
%\beta_3
%\\
\endbmatrix
=
\bmatrix
c_{11} & c_{12} & c_{13}
\\
c_{21} & c_{22} & c_{23}
%\\
%c_{11} & c_{12} & c_{13}
\\
\endbmatrix
\bmatrix
\alpha_1
\\
\alpha_2
\\
\alpha_3
\\
\endbmatrix
\end{align*}
%POIUYTREWQ\alpha_a_\beta_b_
Next, from this dual {{causal operator}}
${\Phi}^*_{1,2}: {\cal M}_{}(\Omega_1) \to  {\cal M}_{}(\Omega_2)$,
we shall construct
a
{{causal operator}}
$\Phi_{1,2}: C(\Omega_2) \to  C(\Omega_1)$.
%,
Consider the identification:${C}(\Omega_1)
{\; \approx \;}
{\mathbb R}^3$,
${C}(\Omega_2) {\; \approx \;} {\mathbb R}^2$,
that is,
\begin{align*}
{C}(\Omega_1)
\ni
f_1
\underset{\text{(identification)}}{\longleftrightarrow}
\bmatrix
f_1 ( \omega_1^1)
\\
f_1 ( \omega_1^2)
\\
f_1 ( \omega_1^3)
\\
\endbmatrix
\in {\mathbb R}^3,
\qquad
{C}(\Omega_2)
\ni
f_2
\underset{
\text{(identification)}
}{\longleftrightarrow}
\bmatrix
f_2 ( \omega_2^1)
\\
f_2 ( \omega_2^2)
\\
%f_1 ( \omega_1^3)
%\\
\endbmatrix
\in {\mathbb R}^2
\end{align*}
Let
$f_2 \in C(\Omega_2)$,
$f_1 = \Phi_{1,2} f_2$.
Then,
we see
\begin{align*}
\bmatrix
f_1(\omega_1^1)
\\
f_1(\omega_1^2)
\\
f_1(\omega_1^3)
\\
%b_3
%\\
\endbmatrix
=
f_1
=
\Phi_{1,2} (f_2 ) =
\bmatrix
c_{11} & c_{21}
\\
c_{12} & c_{22}
\\
c_{13} & c_{23} &
\\
\endbmatrix
\bmatrix
f_2(\omega_2^1)
\\
f_2(\omega_2^2)
\\
\endbmatrix
\end{align*}
Therefore,
the relation between
the dual {{causal operator}}${\Phi}^*_{1,2}$
and
{{causal operator}}$\Phi_{1,2}$
is represented as the
the transposed matrix.
\par
\noindent
\vskip0.3cm
%\vskip0.3cm
%BFBF
\par
\noindent
{\bf Example 6.8
[Deterministic dual {{causal operator}}, {{deterministic causal map}}, {{deterministic causal operator}}]}$\;\;$%POPOPO
Consider the case that
dual {{causal operator}}
${\Phi}^*_{1,2}: {\cal M}(\Omega_1)
({\approx} {\mathbb R}^3)
\to  {\cal M}(\Omega_2)
({\approx} {\mathbb R}^2)
$
ha s the matrix representation such that
\begin{align*}
{\Phi}^*_{1,2}(\rho_1 ) =
\bmatrix
b_1
\\
b_2
\\
%b_3
%\\
\endbmatrix
=
\bmatrix
0 & 1& 1
\\
1 & 0& 0
%\\
%c_{11} & c_{12} & c_{13}
\\
\endbmatrix
\bmatrix
a_1
\\
a_2
\\
a_3
\\
\endbmatrix
\end{align*}
In this case,
it is the deterministic dual {{causal operator}}.
%{\Phi}^*_{1,2}: {\cal M}_{+1}(\Omega_1) \to  {\cal M}_{+1}(\Omega_2)$
%, dual {{causal operator}}
%$,
This
{{deterministic causal operator}}
$\Phi_{1,2}: C(\Omega_2) \to  C(\Omega_1)$
is represented by
\begin{align*}
\bmatrix
f_1(\omega_1^1)
\\
f_1(\omega_1^2)
\\
f_1(\omega_1^3)
\\
%b_3
%\\
\endbmatrix
=
f_1
=
\Phi_{1,2} (f_2 ) =
\bmatrix
0 & 1
\\
1& 0
\\
1 & 0 &
\\
\endbmatrix
\bmatrix
f_2(\omega_2^1)
\\
f_2(\omega_2^2)
\\
\endbmatrix
\end{align*}
with the {{deterministic causal map}}
$\phi_{1,2}: \Omega_1 \to \Omega_2$
such that
\begin{align*}
\phi_{1,2}(\omega_1^1)= \omega_2^2,
\quad
\phi_{1,2}(\omega_1^2)= \omega_2^1,
\quad
\phi_{1,2}(\omega_1^3)= \omega_2^1
\quad
\end{align*}

%\phi\phi\phi
%\noindent
%%%%%%
%\unitlength=0.40mm
%\BEGIN{picture}(400,95)
%\put(27,70){$\Omega_1 \cdots \cdots$}
%\put(100,77){$\omega_1^1$}
%\put(100,70){$\cdot$}
%\put(100,65){\vector(2,-1){85}}
%\put(150,77){$\omega_1^2$}
%\put(150,70){$\cdot$}
%\put(150,65){\vector(-1,-1){40}}
%\put(200,77){$\omega_1^3$}
%\put(200,70){$\cdot$}
%\put(200,65){\vector(0,-1){40}}
%\put(27,20){$\Omega_2 \cdots \cdots$}
%\put(100,20){$\cdot$}
%\put(100,13){$\omega_2^1$}
%\put(150,20){$\cdot$}
%\put(150,13){$\omega_2^2$}
%\END{picture}
%\par
%
%
%
\rm
\subsubsection{{{Sequential causal operator}}
---
A chain of causalities}%6.2.3
\par
%index{@semi-ordered tree}

Let $(T,\le)$ be a tree, i.e., a partial ordered
finite set such that {\lq\lq$t_1 \le t_3$ and $t_2 \le t_3$\rq\rq} implies {\lq\lq$t_1 \le t_2$ or $t_2 \le t_1$\rq\rq}\!.
Assume that
there exists an element $t_0 \in T$,
called the {\it root} of $T$,
such that
$t_0 \le t$ ($\forall t \in T$) holds.

Put $T^2_\le = \{ (t_1,t_2) \in T^2{}: t_1 \le t_2 \}$.
An element $t_0 \in T$ is called a {\it root} if
$t_0 \le t$ ($\forall t \in T$) holds.
%Note that the sub-tree
%$T_{t_0} \equiv$
%$\{ t \in T \; | \; t \ge t_0 \}$
%has the root $t_0$.
%index{root@root}
Since we usually consider the subtree $T_{t_0}$
$({}\subseteq T{})$
with the root
$t_0$,
we assume that the tree-like ordered set
%$(T,\le)$
has a root.
In this chapter,
assume,
for simplicity,
that $T$ is finite
(though it is sometimes infinite in applications).
\par
For simplicity, assume that
$T$ is finite, or a finite subtree of a whole tree.
Let
$T$
$({}= \{ 0, 1, ..., N \}{})$
be a tree with the root $0$.
Define the
%index{parent map@parent map}
\it
parent map
\rm
$\pi{}: T \setminus \{ 0 \} \to T$
such that
$\pi ({}t{}) = \max \{ s \in T{}: s < t \}$.
It is clear that
the tree
$({}T\equiv \{ 0,1,..., N\} , \le{})$
can be identified with
the pair
$({}T\equiv \{ 0,1,..., N\} , \pi:
T \setminus \{ 0 \} \to T)$.
Also,
note that,
for any $t \in T \setminus \{ 0 \}$,
there uniquely exists a natural number $h(t)$
$(${}called the
\it
height
\rm
of $t$ $)$
such that
$\pi^{h(t)} ({}t{}) = 0 $.
Here,
$\pi^2 ({}t{}) = \pi ({}\pi (t))$,
$\pi^3 ({}t{}) = \pi ({}\pi^2 (t))$,
etc.
Also, put
$\{ 0,1,..., N \}^2_{{}_{\le}} $
$=$
$\{ ({}m,n) \; | \; 0 \le m \le n \le N \} $.
%Thus,
%the general sy
In \textcolor{black}{{Fig.$\;$}6.2},
see
the root
$t_0$,
the parent map:
$\pi(t_3)=\pi(t_4)=t_2$,
$\pi(t_2)=\pi(t_5)=t_1$,
$\pi(t_1)=\pi(t_6)=\pi(t_7)=t_0$
\par
\noindent
%%\vskip-0.4cm%%%
%\begin{figure}[htbp]
\setlength{\unitlength}{0.7mm}
\begin{picture}(100,60)(40,0)
\put(60,0){
\put(33,18){\makebox(10,10)[r]{${t_0}$}}
\put(65,30){\makebox(10,10)[r]{${t_1}$}}
\put(91,45){\makebox(10,10)[r]{${t_2}$}}
\put(131,50){\makebox(10,10)[r]{${t_3}$}}
\put(131,35){\makebox(10,10)[r]{${t_4}$}}
\put(91,14){\makebox(10,10)[r]{${t_5}$}}
\put(60,10){\makebox(10,10)[r]{${t_6}$}}
\put(60,0){\makebox(10,10)[r]{${t_7}$}}
\put(60,33){\vector(-3,-1){15}} % 0,1
\put(60,20){\vector(-3,1){15}} % 0,6
\put(60,7){\vector(-3,2){15}} % 0,7
\put(85,48){\vector(-3,-2){13}} % 1,2
\put(90,20){\vector(-3,2){13}} % 1,5
\put(128,55){\vector(-3,-1){13}} % 2,3
\put(128,40){\vector(-3,2){13}} % 2,4
\put(50,16){$\pi$}
\put(47,32){$\pi$}
\put(47,5){$\pi$}
\put(73,45){$\pi$}
\put(83,25){$\pi$}
\put(118,57){$\pi$}
\put(118,38){$\pi$}
}
\end{picture}
%\vskip-0.3cm
\begin{center}{Figure 6.2:
Tree
}
\end{center}
%BFBF
\hfill{$///$}
\par
\noindent
{\bf
{Definition }6.9
[{{Sequential causal operator}}]}$\;\;$%POPOPO
$\;$
%%index{ippansisutemu@ ({}{})}
%index{@sequential observable}
%\rm
%{{}}
%${\bold S}_{[\omega_{t_0}]} {{=}} $
%$[{}S_{[\omega_{t_0}]}, \{ \Phi_{t_1,t_2}{}: $
%${C (\Omega_{t_2})} \to {C (\Omega_{t_1})} \}_{(t_1,t_2) \in T^2_{\leqq}}{}]$
%{\rm (i)$\text{--}$(iii)} and ,
%{{state}}
%$\omega_{t_0}$ and 
%%atisfies s .
The family

$\{ \Phi_{t_1,t_2}{}: $
${C (\Omega_{t_2})} \to {C (\Omega_{t_1})} \}_{(t_1,t_2) \in T^2_{\leqq}}$
$\Big($
or,
$\{$
% \Phi_{t_1,t_2}{}: $
${C (\Omega_{t_2})} \overset{\Phi_{t_1,t_2}}\to {C (\Omega_{t_1})} \}_{(t_1,t_2) \in T^2_{\leqq}}$
%%{{}}{{sequential causal operator}}
%%$\{ \Phi_{t_1,t_2}{}: $
%%${C (\Omega_{t_2})} \to {C (\Omega_{t_1})} \}_{(t_1,t_2) \in T^2_{\leqq}}$
%%{{}}
%$\{ {C (\Omega_{t})} \overset{\Phi_{\pi(t), t}}\to
%{C (\Omega_{\pi(t)})} \}_{(t_1,t_2) \in T^2_{\leqq}}
%$
$\Big)$
is called a
{\bf {{sequential causal operator}}},
if it satisfies that
(\textcolor{black}{{Fig.$\;$}6.3}):
\begin{itemize}
\item[{\rm (i)}]
For each$t \;(\in T)$,
a basic algebra ${C (\Omega_{t})}$
is determined.
% {{}}ssociated.
%
%$t_0 \;(\in T)$ {{}}$T$
% and .
%%.
% $S${{state}} $\omega_{t_0} \;(\in
%\Omega_{t_0})$, That is, {{}}{{state}}$\omega_{t_0}$.
\item[{\rm (ii)}]
For each $(t_1,t_2) \in T^2_{\leqq}$, a {{causal operator}}
$\Phi_{t_1,t_2}{}: {C (\Omega_{t_2})} \to {C (\Omega_{t_1})}$
is defined such as
$\Phi_{t_1,t_2} \Phi_{t_2,t_3} = \Phi_{t_1,t_3}$
$(\forall (t_1,t_2)$,
$\forall (t_2,t_3) \in T^2_{\leqq})$.
Here,
$\Phi_{t,t} : C(\Omega_t) \to C(\Omega_t)$
is the identity operator.
%operator and .
\end{itemize}
%, $T$ and .
\par
\noindent

{
%\ssmall
\par
\noindent
%%\vskip-0.4cm%%%
%\begin{figure}[htbp]
%\ssetlength{\unitlength}{0.57mm}
\setlength{\unitlength}{0.45mm}
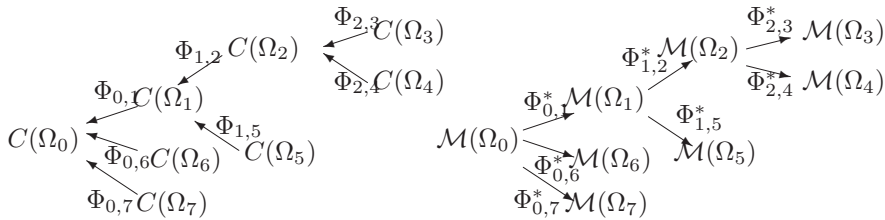
\begin{picture}(100,60)(40,0)
\put(20,0){
\put(33,18){\makebox(10,10)[r]{${C (\Omega_0)}$}}
\put(70,30){\makebox(10,10)[r]{${C (\Omega_1)}$}}
\put(98,45){\makebox(10,10)[r]{${C (\Omega_2)}$}}
\put(141,50){\makebox(10,10)[r]{${C (\Omega_3)}$}}
\put(141,35){\makebox(10,10)[r]{${C (\Omega_4)}$}}
\put(103,14){\makebox(10,10)[r]{${C (\Omega_5)}$}}
\put(75,12){\makebox(10,10)[r]{${C (\Omega_6)}$}}
\put(71,-1){\makebox(10,10)[r]{${C (\Omega_7)}$}}
\put(60,33){\vector(-3,-1){15}} % 0,1
\put(60,20){\vector(-3,1){15}} % 0,6
\put(60,7){\vector(-3,2){15}} % 0,7
\put(85,48){\vector(-3,-2){13}} % 1,2
\put(90,20){\vector(-3,2){13}} % 1,5
\put(128,55){\vector(-3,-1){13}} % 2,3
\put(128,40){\vector(-3,2){13}} % 2,4
\put(50,16){$\Phi_{0,6}$}
\put(47,35){$\Phi_{0,1}$}
\put(45,3){$\Phi_{0,7}$}
\put(71,47){$\Phi_{1,2}$}
\put(83,25){$\Phi_{1,5}$}
\put(118,57){$\Phi_{2,3}$}
\put(118,38){$\Phi_{2,4}$}
}
\put(150,0){
\put(33,18){\makebox(10,10)[r]{${{\cal M} (\Omega_0)}$}}
\put(70,30){\makebox(10,10)[r]{${{\cal M} (\Omega_1)}$}}
\put(98,45){\makebox(10,10)[r]{${{\cal M} (\Omega_2)}$}}
\put(141,50){\makebox(10,10)[r]{${{\cal M} (\Omega_3)}$}}
\put(141,35){\makebox(10,10)[r]{${{\cal M} (\Omega_4)}$}}
\put(103,14){\makebox(10,10)[r]{${{\cal M} (\Omega_5)}$}}
\put(72,12){\makebox(10,10)[r]{${{\cal M} (\Omega_6)}$}}
\put(71,-1){\makebox(10,10)[r]{${{\cal M} (\Omega_7)}$}}
\put(44,26){\vector(3,1){15}} % 0,1
\put(44,23){\vector(3,-1){15}} % 0,6
\put(44,15){\vector(3,-2){15}} % 0,7
\put(81,38){\vector(3,2){13}} % 1,2
\put(81,30){\vector(3,-2){13}} % 1,5
\put(110,50){\vector(4,1){13}} % 2,3
\put(110,45){\vector(4,-1){13}} % 2,4
\put(47,13){$\Phi^*_{0,6}$}
\put(44,32){$\Phi^*_{0,1}$}
\put(41,3){$\Phi^*_{0,7}$}
\put(73,45){$\Phi^*_{1,2}$}
\put(89,28){$\Phi^*_{1,5}$}
\put(110,57){$\Phi^*_{2,3}$}
\put(110,38){$\Phi^*_{2,4}$}
}
%\END{picture}
%\END{picture}
\end{picture}
%\END{picture}
%\vskip-0.3cm
\begin{center}{Figure 6.3:
{{Sequential causal operator}}
and
dual {{sequential causal operator}}
in the case
of
\textcolor{black}{{Fig.$\;$}6.2}
}
\end{center}
}

%{\cal M}CCCCCCCCCCCCC

\par
\noindent
The dual family
$\{ {\Phi}^*_{t_1,t_2}{}: $
${{\cal M} (\Omega_{t_1})} \to {{\cal M} (\Omega_{t_2})} \}_{(t_1,t_2) \in T^2_{\leqq}}$
is called
the {\bf dual {{sequential causal operator}}}
%(,
%{\bf
%causality {{the Schr\"odinger{ picture}}}}{)}
% and 
of a
{{sequential causal operator}}
$\{$
% \Phi_{t_1,t_2}{}: $
${C (\Omega_{t_2})} \overset{\Phi_{t_1,t_2}}\to {C (\Omega_{t_1})} \}_{(t_1,t_2) \in T^2_{\leqq}}$.
%
%\BEGIN{itemize}
%\ITEM[{\rm (i)}]
%$t \;(\in T)$,
%state space $\Omega_{t}$
%.
%% {{}}ssociated.
%%\ITEM[{\rm (ii)}]
%%$t_0 \;(\in T)$ {{}}$T$
%% and .
%%%.
%% $S${{state}} $\omega_{t_0} \;(\in
%%\Omega_{t_0})$, That is, {{}}{{state}}$\omega_{t_0}$.
%\ITEM[{\rm (i)}]
%$(t_1,t_2) \in T^2_{\leqq}$, 
%$\phi_{t_1,t_2}{}: {\Omega_{t_1}} \to {\Omega_{t_2}}$
%,
%$\phi_{t_2,t_3} \phi_{t_1,t_2}  = \phi_{t_1,t_3}$
%$(\forall (t_1,t_2)$,  $(t_2,t_3) \in T^2_{\leqq})$
%%.
%%\END{itemize}
%%, $T$ and .
%\par
%\noindent
%
%
%
%\BEGIN{align*}
%\ssetlength{\unitlength}{0.7mm}
%\BEGIN{picture}(100,60)(40,0)
%\put(33,18){\makebox(10,10)[r]{${C (\Omega_0)}$}}
%\put(80,30){\makebox(10,10)[r]{${C (\Omega_1)}$}}
%\put(103,45){\makebox(10,10)[r]{${C (\Omega_2)}$}}
%\put(151,50){\makebox(10,10)[r]{${C (\Omega_3)}$}}
%\put(151,35){\makebox(10,10)[r]{${C (\Omega_4)}$}}
%\put(113,14){\makebox(10,10)[r]{${C (\Omega_5)}$}}
%\put(80,10){\makebox(10,10)[r]{${C (\Omega_6)}$}}
%\put(80,0){\makebox(10,10)[r]{${C (\Omega_7)}$}}
%%
%\put(60,33){\vector(-3,-1){15}} % 0,1
%\put(60,20){\vector(-3,1){15}} % 0,6
%\put(60,7){\vector(-3,2){15}} % 0,7
%\put(85,48){\vector(-3,-2){13}} % 1,2
%\put(90,20){\vector(-3,2){13}} % 1,5
%\put(128,55){\vector(-3,-1){13}} % 2,3
%\put(128,40){\vector(-3,2){13}} % 2,4
%%
%\put(50,16){$\Phi_{0,6}$}
%\put(47,35){$\Phi_{0,1}$}
%\put(47,5){$\Phi_{0,7}$}
%\put(71,47){$\Phi_{1,2}$}
%\put(83,25){$\Phi_{1,5}$}
%\put(118,57){$\Phi_{2,3}$}
%\put(118,38){$\Phi_{2,4}$}
%\END{picture}
%%%%%%3.23}
%\TAG{7.29}
%\END{align*}
%
%
%

\subsubsection{{{Sequential causal operator}}
---
Simultaneous differential equation of the first order
}%{Sec.6.2.4}
\rm
\par

\noindent
%\vskip0.3cm
%\vskip0.3cm
%BFBF
\par
\noindent
{\bf
Example 6.10
[Pheasants and rabbits problem]}$\;\;$%POPOPO
Consider the following situation:
\begin{itemize}
\item[(a)]
[Pheasants and rabbits problem]
A number of $m$ pheasants and $n$ rabbits are placed
together in the same cage.
Then
there are
$m+n$ heads and $2m+4n$ legs.
% and $m$ and $n$.
%, $m+n$,
%$2m+4n${.}
 \end{itemize}
%\END{QA}
\par
\noindent
In what follows,
this statement in ordinary language
will be changed to
the measurement theoretical statement.
%{ordinary language}(a),
%{{measurement theory}}
%
%{{sequential causal operator}}
%
%.
Putting
$\mathbb{N}_0=\{0, 1, 2, \ldots\}$,
define
% and ,
state space
$\Omega_0$,
$\Omega_1$,
$\Omega_2$
such that
\begin{align*}
\Omega_0 = \mathbb{N}_0 \times \mathbb{N}_0,
\quad
\Omega_1 = \mathbb{N}_0,
\quad
\Omega_2 = \mathbb{N}_0
\end{align*}
Put
$T(0)=\{0,1,2\}$
with the
{parent map}$\pi: \{1,2\} \to \{0,1,2\}$
such that
\begin{align*}
\pi (1)=0,
\quad
\pi(2)=0
\end{align*}
Then,
the
{{deterministic causal map}}
$\phi_{0,1}:\Omega_0 \to \Omega_1$
and
$\phi_{0,2}:\Omega_0 \to \Omega_2$
is respectively represented by
\begin{align*}
\phi_{0,1}(m,n)=m+n,
\quad
\phi_{0,2}(m,n)=2m+4n
\end{align*}
Therefore, by \textcolor{black}{{{Theorem }}6.4},
the {{deterministic causal operator}}
$\Phi_{0, 1}: C (\Omega_1) \to
C (\Omega_0)$
and
$\Phi_{0, 2} : C (\Omega_2)
\to C (\Omega_0)$
are defined as follows.
\begin{align*}
\cases
[\Phi_{0, 1} (f_1)] (m,n)
= f_1 (m + n)
\qquad
&
(\forall f_1 \in C (\Omega_1),
\forall (m,n) \in \Omega_0)
\\
\\
{}
[\Phi_{0, 2} (f_2)] (m,n)
= f_2 (2 m+ 4 n)
&
(\forall f_2 \in C (\Omega_2),
\forall (m,n) \in \Omega_0)
\endcases
\end{align*}
\par
\noindent
Thus, we get
a deterministic {{sequential causal operator}}
$\{ {C (\Omega_{t})} \overset{\Phi_{0, t}}\to {C (\Omega_{0})} \}_{ t
=1,2}$.
%

%\par
%%index{measure2@${\frak M}({}\{ {\mathsf O }_t \}_{t \in T},
%{\bold S}_{[{}\omega_{t_0}{}] }  {})$}

%%\BEGIN{align*}
%%%\label{eq:3-04}
%%%\xymatrix@R=5mm@C=20mm
%%%{
%%& \ar[ld]_{\Phi_{0, 1}} C (\Omega_1)\\
%%C (\Omega_0) & \\
%%& \ar[lu]^{\Phi_{0, 2}} C (\Omega_2)
%%%}#############{3.19}
%%%\raisebox{-10mm}{.}
%%
%%%
%%
%\par
%\BEGIN{align*}
%\mbox{
%\ssetlength{\unitlength}{0.7mm}
%\BEGIN{picture}(80,60)(0,20)
%%%\BEGIN{picture}(80,80)(0,20)
%\put(80,50){\makebox(10,10)[r]{${}$}}
%\put(80,70){\makebox(10,10)[r]{${C (\Omega_1)}$}}
%\put(30,45){\makebox(10,10)[r]{${C (\Omega_0)}$}}
%\put(85,15){\makebox(10,10)[r]{${C (\Omega_2)}$}}
%%\put(60,55){\vector(-3,-1){13}}
%\put(60,70){\vector(-3,-2){15}}
%\put(60,22){\vector(-3,2){15}}
%%\put(87,48){\vector(-3,-2){13}}
%%\put(87,22){\vector(-3,2){13}}
%%\put(112,30){\vector(-3,-2){13}}
%%\put(112,7){\vector(-3,2){13}}
%%
%%\put(50,56){$\Phi_{0,2}$}
%%\put(50,43){$\cdots \cdots$}
%%\put(50,38){$\cdots \cdots$}
%\put(47,72){$\Phi_{0,1}$}
%\put(47,20){$\Phi_{0,2}$}
%%\put(73,46){$\Phi_{1,5}$}
%%\put(80,28){$\Phi_{1,2}$}
%%\put(98,28){$\Phi_{2,4}$}
%%\put(97,7){$\Phi_{2,3}$}
%\END{picture}
%}
%\qquad \qquad \qquad \qquad
%\TAG{7.5}
%\END{align*}
%\par
%\par
%\noindent
%%,
%%{{state}}$(m_0,n_0)$
%%${\bold S}_{[(m_0,n_0)]}$
%%$=$
%%$[(m_0,n_0) \; :\;$
%%$\{ \Phi_{0,1}{}:
%%{C (\Omega_{1})} \to {C (\Omega_{0})} ,
%%\Phi_{0,2}{}: $
%%${C (\Omega_{2})} \to {C (\Omega_{0})}
%%\}
%%]$.
%
%
%
%

%\rm
%\par
%\noindent
%\vskip0.3cm
%\vskip0.3cm
%BFBF
\par
\noindent
{\bf
Example 6.11
[State equation}]}$\;\;$%POPOPO
\rm
Let $T={\mathbb R}$
be the time axis.
%$T${finite set},
%{)}
For each $t ( \in T)$,
consider the state space
$\Omega_t = {\mathbb R }^n$
($n$-dimensional real space).
And consider
simultaneous differential equation of the first order
\begin{align*}
%\overset{\scriptsize{
%
%}}{\underset{\scriptsize{}}{\text{\fbox{{state equation}}}}}
%\cdots
&
\cases
\frac{d\omega_1}{dt}{} (t)=v_1(\omega_1(t),\omega_2(t),\ldots,\omega_n(t), t)
\\
\frac{d\omega_2}{dt}{} (t)=v_2(\omega_1(t),\omega_2(t),\ldots,\omega_n(t), t)
\\
\cdots \cdots
\\
\frac{d\omega_n}{dt}{} (t)=v_n (\omega_1(t),\omega_2(t),\ldots,\omega_n(t), t)
\endcases
%\tag*{{\color{black}$\displaystyle{\mathop{(6.2)}}$}}
\tag*{{\color{black}$\displaystyle{\mathop{(6.3)}_{(=(1.1))}}$}}
\end{align*}
which is called a
{\bf
{state equation}
}(\textcolor{black}{{Chap.$\;$1}(1.1)}).
Let
$\phi_{t_1,t_2}: \Omega_{t_1} \to \Omega_{t_2}$,
$(t_1 {\; \leqq \;} t_2)$
be
a
{{deterministic causal map}}
induced by the state equation (6.3).
It is clear that
$\phi_{t_2,t_3} (\phi_{t_1,t_2} (\omega_{t_1}))
=
\phi_{t_1,t_3} (\omega_{t_1})
$
$(\omega_{t_1} \in \Omega_{t_1} , t_1 {{\; \leqq \;}}t_2 {{\; \leqq \;}}t_3)$.
Therefore,
by
\textcolor{black}{{{Theorem }}6.4},
we have the
deterministic {{sequential causal operator}}
$\{ \Phi_{t_1,t_2}{}: $
${C (\Omega_{t_2})} \to {C (\Omega_{t_1})} \}_{(t_1,t_2) \in T^2_{\leqq}}$.
%
%\BEGIN{align*}
%&
%\qquad
%(\Phi_{t_1,t_2} f_2)
%(\omega^{t_1})
%=
%f_2
%(\phi_{t_1,t_2}(\omega^{t_1})),
%\\
%&
%(\forall f_2 \in C(\Omega_{t_2}),
%\forall \omega^{t_1}
%=
%(
%\omega^{t_1}_1,
%\omega^{t_1}_2,
%...,
%\omega^{t_1}_n
%)
%\in \Omega_{t_1}
%)
%\END{align*}
%

\rm
%\par
%\noindent
%\vskip0.3cm
%\vskip0.3cm
%BFBF
\par
\noindent
{\bf
Example 6.12
[Difference equation
of the second order
]}$\;\;$%POPOPO
Consider the discrete time $T=\{0, 1 ,2,\ldots \}$
with the parent map
$\pi: T\setminus\{0\} \to T$
such that
$\pi(t )=t-1
\;
(\forall t =1,2,...)$.
For each
$t ( \in T)$,
consider a state space
$\Omega_t$
such  that
$\Omega_t = {\mathbb R }$.
For example,
consider the following difference equation:
That is,
$\phi: \Omega_{t}\times \Omega_{t+1} \to \Omega_{t+2}$
satisfies as follows.
\begin{align*}
\omega_{t+2} =\phi( \omega_t , \omega_{t+1} ) = \omega_t + \omega_{t+1} +2
\qquad
(\forall t \in T )
\end{align*}
Here, note that
the {{state}}
$\omega_{t+2}$
depends on
both
$\omega_{t+1}$
and
$\omega_{t}$
(i.e.,
multiple markov property).
This must be modified as follows.
For each $t ( \in T )$
consider a new state space
${\widetilde \Omega}_t =$
$\Omega_t \times \Omega_{t+1}
= {\mathbb R }\times {\mathbb R }$.
And define the
{{deterministic causal map}}
$\widetilde{\phi}_{t,t+1} :
{\widetilde \Omega}_t \to
{\widetilde \Omega}_{t+1}$
as follows.
\begin{align*}
&
(\omega_{t+1}, \omega_{t+2} )
=
\widetilde{\phi}_{t,t+1}
(\omega_t, \omega_{t+1} )
=
(
\omega_{t+1} ,
\omega_t + \omega_{t+1} +2
)
\\
& \hspace{5cm}
(\forall
(\omega_t , \omega_{t+1})
\in
{\widetilde \Omega}_t,
\forall t \in T )
\end{align*}
Therefore,
%\textcolor{black}{Example 6.11} and ,
by \textcolor{black}{{{Theorem }}6.4},
the {{deterministic causal operator}}
$\widetilde{\Phi}_{t,t+1} :
C({\widetilde \Omega}_{t+1}) \to
C({\widetilde \Omega}_{t})$
is defined by
\begin{align*}
&
\quad
[\widetilde{\Phi}_{t,t+1}
{\tilde f}_t](
\omega_t , \omega_{t+1}
)
=
{\tilde f}_t(\omega_{t+1} ,
\omega_t + \omega_{t+1} +2
)
\\
&
%\qquad
%\qquad
(\forall
(\omega_t , \omega_{t+1})
\in
{\widetilde \Omega}_t,
\forall {\tilde f}_t \in C({\widetilde \Omega}_{t+1}) ,
\forall t \in T\setminus \{0\})
)
\end{align*}
Thus, we get the
%{\tilde f}ffffffff
deterministic {{sequential causal operator}}
$\{ {\widetilde{\Phi}}_{t,t+1}{}: $
$C (\widetilde{\Omega}_{t+1}) \to C (\widetilde{\Omega}_{t}) \}_{t \in
T \setminus \{0\}
}$.

\par
\noindent
%It is  to note that
\par
\noindent
%BBBBBBBBBBBBBBBBBB%SBSBSBS
{\small%%{\footnotesize
\vspace{0.1cm}
\begin{itemize}
\item[$\spadesuit$] \bf {{}}{Note }6.3{{}} \rm
%?
%%?
%\item[] \bf  \rm %%%BBBBBBBBBBBBBBBBBBBB
%\\
In measurement theory,
multiple markov process
and
time-lag process
are prohibited.
%
%${{\cdot}}$, ,
%{{state}}, {{state}}
% and  and  and , 
%.
%mutuple markov proscee
%,
%state space ,
%{{sequential causal operator}}, {{{measurement theory}}}{}
%{state equation}\textcolor{black}{(6.3)}
%{differential equation}.
%,
%${{\cdot}}$ and ,
%${{\cdot}}$.
\end{itemize}
}
%%BBBBBBBBBBBBBBBBBB%SBSBSBSS
\par
\noindent
{\bf Example 6.13
[Random walk]}$\;\;$%POPOPO
%index{@}
Put
${\mathbb Z}=\{0,\pm1,\pm2,\ldots\}$.
Define the dual {{causal operator}}${\Phi}^*: {\cal M}_{+1}({\mathbb Z}) \to {\cal M}_{+1}({\mathbb Z})$
such that
\begin{align*}
{\Phi^*}( \delta_i ) = \frac{\delta_{i-1} + \delta_{i+1}}{2}
\qquad
(i \in {\mathbb Z})
\end{align*}
where $\delta_{(\cdot )}
(\in
{\cal M}_{+1}({\mathbb Z})
)$ is a point measure.
Therefore,
the
{{causal operator}}$\Phi: C({\mathbb Z}) \to C({\mathbb Z})$
is defined by
\begin{align*}
[\Phi ( f ) ](i)= \frac{f({i-1}) + f({i+1})}{2}
\qquad
(\forall f \in C({\mathbb Z}),
\forall i \in {\mathbb Z})
\end{align*}
Now,
consider the discrete time
$T=\{0,1,2,\ldots,N\}$.
For each
$t (\in T )$,
a state space $\Omega_t$
is define by
$\Omega_t= {\mathbb Z}$.
Then,
we have
the
{{sequential causal operator}}
%$[{}\{ {\mathsf O}_t \}_{ t \in T} ,$
$\{  \Phi_{\pi(t), t }(=\Phi ){}: $
${C (\Omega_t)} \to {C (\Omega_{\pi(t)})} \}_{ t \in T\setminus \{0\} }$
%$]$

\par
\par
\rm
\subsection{{{Realized causal}} observable
---
{{Only one measurement}}}%{Sec.6.3}
Let
$(T(t_0), {{\; \leqq \;}})$
(or,
$T(t_0)
= \{ t_0, t_1,\ldots , t_N \}$
)
be a semi-ordered tree
with the root
$t_0$.
Mainly
consider
the parent map representation
$(T{{=}} \{ t_0,t_1,\ldots, t_N\} , \pi:
T \setminus \{ t_0 \} \to T)$.
\par
\noindent
\bf
{Definition }6.14
[Sequential observable{}]$\;\;$%POPOPO
\rm
$\;$
%%index{@ ({}{})}
%index{@sequential observable}
%\rm
%{{}}
%${\bold S}_{[\omega_{t_0}]} {{=}} $
%$[{}S_{[\omega_{t_0}]}, \{ \Phi_{t_1,t_2}{}: $
%${C (\Omega_{t_2})} \to {C (\Omega_{t_1})} \}_{(t_1,t_2) \in T^2_{\leqq}}{}]$
%{\rm (i)$\text{--}$(iii)} and ,
%{{state}}
%$\omega_{t_0}$ and 
%satisfies s .
%$(T, {{\; \leqq \;}})$
%semi-ordered tree and .
Consider a {{sequential causal operator}}
$\{ \Phi_{t_1,t_2}{}: $
${C (\Omega_{t_2})} \to {C (\Omega_{t_1})} \}_{(t_1,t_2) \in T^2_{\leqq}}$.
Assume that
for each
$t \in T$,
an {observable }${\mathsf O }_t {{=}} ({}X_t, {\cal F}_t , F_t{})$
in ${C (\Omega_{t})}$
is determined.
The,
the pair
$[{}\{ {\mathsf O}_t \}_{ t \in T} ,
\{  \Phi_{t_1,t_2}{}: $
${C (\Omega_{t_2})} \to {C (\Omega_{t_1})} \}_{(t_1,t_2) \in T^2_{\leqq}}$
$]$
is called
a
{\bf
sequential observable},
\rm
and denoted by
$[{}{\mathsf O}_T{}]$
or
$[{}{\mathsf O}_{T(t_0)}{}]$.
That is,
$[{}{\mathsf O}_T{}]$
$=$
$[{}\{ {\mathsf O}_t \}_{ t \in T} ,
\{  \Phi_{t_1,t_2}{}: $
${C (\Omega_{t_2})} \to {C (\Omega_{t_1})} \}_{(t_1,t_2) \in T^2_{\leqq}}$
$]$.
Using the {parent map}$\pi: T\setminus \{0\} \to T$,
we also write
that
$[{}{\mathsf O}_T{}]$
$=$
$[{}\{ {\mathsf O}_t \}_{ t \in T} ,
$
$
\{
{C (\Omega_{t})} $
$\xrightarrow[]{{\Phi_{ \pi({}t{}), t } }}$
$
C (\Omega_{\pi({}t{})})
\}_{t \in T \setminus \{ 0 \}{}) {}}]$.

\rm
%$\omega_{t_0}$
%$\in \Omega_{t_0}$
% and .
%{{}}
%${\bold S}_{[\omega_{t_0}]} {{=}} $
%$[{}S_{[\omega_{t_0}]}, \{ \Phi_{t_1,t_2}{}: $
%${C (\Omega_{t_2})} \to {C (\Omega_{t_1})} \}_{(t_1,t_2) \in T^2_{\leqq}}{}]$
%
%{\bf {{state}}
%$\omega_{t_0}$
%}
% and .
\rm
\par
%\qed

%%we see as follows:
%$(T {{=}} \{ 0, 1,\ldots N \}, \pi{}: T \setminus \{0\} \to T{})$
%{{}} $0$,
%% with   and let
%${\bold S}_{[\omega_0]}
%{{=}} [{}S_{[\omega_0]}, $
%${C (\Omega_{t})} \overset{\Phi_{\pi(t),t}}\to
% {C (\Omega_{\pi(t)})} \; (t \in T \setminus \{0\}){}]$
% and .
\par

\vskip0.3cm
\par
\vskip0.3cm
\par
%\textcolor{black}{{Sec. 2.5.1}},
According to the Copenhagen interpretation
(i.e.,
only one observable is permitted
),
we must
regard
many observables
$\{ {\mathsf O}_t \}_{ t \in T}$
in
%
%{{measurement theory}}
%
%,
%{{measurement}}
%,
%,
%observable 
%.
%,
a sequential observable
$[{}{\mathsf O}_T{}]$
${{=}}$
$[{}\{ {\mathsf O}_t \}_{ t \in T} ,
\{  \Phi_{t_1,t_2}{}: $
${C (\Omega_{t_2})} \to {C (\Omega_{t_1})} \}_{(t_1,t_2) \in T^2_{\leqq}}$
$]$
as only one observable.
This is realized as follows.

%BFBF
\par
\noindent
{\bf
{Definition }6.15
[{{{Realized causal}} observable }{}{\rm{}}]}$\;\;$%POPOPO
\rm
%index{@{{realized causal}} observable }
Let
$[{\mathsf O}_{T(t_0)}]$
$=$
$[{}\{ {\mathsf O}_t \}_{ t \in T},\\
\{  \Phi_{\pi(t), t }{}: $
${C (\Omega_t)} \to {C (\Omega_{\pi(t)})} \}_{ t \in T\setminus \{t_0\} }$
$]$
be a
sequential observable.
For each
 $s$
$({}\in T{})$,
put
$T_s = \{ t \in T \;|\; t {\; \geqq \;}s \}$.
And define
the
{{}}{observable }
$\widehat{\mathsf O}_s {{=}} (\bigtimes_{t \in T_s } X_t, $
$\bigstimes_{t \in T_s } {\cal F}_t, {\widehat F}_s)$
in
$C (\Omega_s)$
as follows.
%such that:
\par
\noindent
\begin{align*}
\widehat{\mathsf O}_s
=
\cases
{\mathsf O}_s
\quad
&
\text{$ ( s \in T \setminus \pi (T) \; \text{ and }${})}
\\
\\
{\mathsf O}_s
{\times}
({}\bigtimes_{t \in \pi^{-1} ({}\{ s \}{})} \Phi_{ \pi(t), t} \widehat {\mathsf O}_t{})
\quad
&
\text{($ s \in \pi (T) \; \text{ and }${})}
\endcases
\tag{6.4}
%%%%%3.25}
%P%\tag{7.10}
\end{align*}
using this iteratively,
after all we get
the observable
$\widehat{\mathsf O}_{t_0{}}$
$=$
$(\bigtimes_{t \in T } X_t, $
$\bigstimes_{t \in T } {\cal F}_t,$
${\widehat F}_{t_0})$
in
$C(\Omega_{t_0})$.
Put
$\widehat{\mathsf O}_{t_0{}}$
$=$
$\widehat{\mathsf O}_{T(t_0){}}$.
The
$\widehat{\mathsf O}_{T(t_0){}}$
$=$
$(\bigtimes_{t \in T } X_t, $
$\bigstimes_{t \in T } {\cal F}_t,$
${\widehat F}_{t_0})$
is called the
{\bf {{realized causal}} observable}
of the
sequential observable$[{}{\mathsf O}_{T(t_0)}{}]$
$=$
$[{}\{ {\mathsf O}_t \}_{ t \in T} ,
\{  \Phi_{\pi(t), t }{}: $
${C (\Omega_t)} \to {C (\Omega_{\pi(t)})} \}_{ t \in T\setminus \{t_0\} }$
$]$.

\vskip0.5cm
\par
\noindent

\rm

\par
\noindent
%\vskip0.3cm
%\vskip0.3cm
%BFBF
\par
\noindent
{\bf
Example 6.16
[{}Simple example{\rm
(Continued from Fig. \textcolor{black}{6.3})}]}$\;\;$%POPOPO
%\bf
% 7.8.
%\rm
%\rm
Suppose that a tree $(T \equiv \{ 0, 1, ..., 6, 7 \}, \pi)$
has an ordered structure such that
$\pi(1) = \pi(6) = \pi(7) = 0$,
$\pi(2) = \pi(5) = 1$,
$\pi(3) = \pi(4) = 2$.
$\Big($See Fig.
%%EQ
\textcolor{black}{(6.3)}.$\Big)$%\TAG{28}
$\;$
%Consider
%
%
%
%
%
%%a causal relation $ [
%%\{ {\cal A}_t \overset{\Phi_{\pi(t), t}}\to
%%{\cal A}_{\pi(t)} \}_{ t \in T \setminus \{0\} }{}]$
%%with the initial system $S_{[\rho_0^p]}$.
%
%\BEGIN{align*}
%\ssetlength{\unitlength}{0.7mm}
%\BEGIN{picture}(100,60)(40,0)
%\put(33,18){\makebox(10,10)[r]{${\cal A}_0$}}
%\put(60,30){\makebox(10,10)[r]{$C(\Omega_1)$}}
%\put(83,45){\makebox(10,10)[r]{$C(\Omega_2)$}}
%\put(106,50){\makebox(10,10)[r]{${\cal A}_3$}}
%\put(106,35){\makebox(10,10)[r]{${\cal A}_4$}}
%\put(83,18){\makebox(10,10)[r]{${\cal A}_5$}}
%\put(60,15){\makebox(10,10)[r]{${\cal A}_6$}}
%\put(60,0){\makebox(10,10)[r]{${\cal A}_7$}}
%%
%\put(60,33){\vector(-3,-1){15}} % 0,1
%\put(60,20){\vector(-3,1){15}} % 0,6
%\put(60,7){\vector(-3,2){15}} % 0,7
%\put(85,48){\vector(-3,-2){13}} % 1,2
%\put(85,25){\vector(-3,2){13}} % 1,5
%\put(108,55){\vector(-3,-1){13}} % 2,3
%\put(108,40){\vector(-3,2){13}} % 2,4
%%
%\put(50,16){$\Phi_{0,6}$}
%\put(47,35){$\Phi_{0,1}$}
%\put(47,5){$\Phi_{0,7}$}
%\put(71,47){$\Phi_{1,2}$}
%\put(78,30){$\Phi_{1,5}$}
%\put(98,57){$\Phi_{2,3}$}
%\put(98,38){$\Phi_{2,4}$}
%\END{picture}
%\tag{\textcolor{black}{4.28}}
%\END{align*}
%

%
%\noindent
%Also, for each $t \in \{ 0, 1, ..., 6, 7 \}$, consider an observable
%${\mathsf O}_t \equiv $
%$(X_t, {\cal F}_t, F_t)$ in a $C^*$-algebra ${\cal A}_t$.
%Thus,
%we have
Consider a
sequential observable
$[{}{\mathsf O}_T{}]$
$=$
$[{}\{ {\mathsf O}_t \}_{ t \in T} ,
$
$
\{
{C (\Omega_{t})} {{\Phi_{ \pi({}t{}), t } }\atop{\rightarrow}} $
$
C (\Omega_{\pi({}t{})})
\}_{t \in T \setminus \{ 0 \}{}) {}}]$.
Now,
we shall construct its
{{realized causal}} observable $\widehat{\mathsf O}_{T(t_0){}}$
$=$
$(\bigtimes_{t \in T } X_t, $
$\bigstimes_{t \in T } {\cal F}_t,$
${\widehat F}_{t_0})$
in what follows.

Put
\begin{align*}
\widehat{\mathsf O}_t
=
{\mathsf O}_t
\quad
\text{
and thus}
\quad
\widehat{F}_t
=
{F}_t
\quad
({}t = 3,4,5,6,7).
%%\tag{4.29} 
\end{align*}
\rm
First we construct the product observable
$\widehat{\mathsf O}_2$ in ${C(\Omega_2)}$
such as
\begin{align*}
\widehat{\mathsf O}_2
= (X_2 \times X_3 \times X_4,
{\cal F}_2 \bigstimes {\cal F}_3 \bigstimes {\cal F}_4,
{\widehat F}_2 {})
\quad
\text{ where }
{\widehat F}_2
=
F_2 \bigtimes ({}\bigtimes_{t=3,4} \Phi_{2,t} {\widehat F}_t{}) ,
%
%\TAG{30}
\end{align*}
Iteratively, we construct the following:
\begin{align*}
\CD
 {C(\Omega_0)}
  @<\Phi_{0,1}<<
 {C(\Omega_1)}P
  @<\Phi_{1,2}<<
 {C(\Omega_2)} \\
 F_0 \bigtimes \Phi_{0,6} {\widehat F}_6 \bigtimes \Phi_{0,7} {\widehat F}_7
  @.
 F_1 \bigtimes \Phi_{1,5} {\widehat F}_5
  @. \\
  @VVV
  @VVV \\
 \underset{(F_0 \times \Phi_{0,6} {\widehat F}_6 \times \Phi_{0,7}
{\widehat F}_7 \times \Phi_{0,1} {\widehat F}_1)}
{{\widehat F}_0}
  @<{\Phi_{0,1}}<<
 \underset{(F_1 \times \Phi_{1,5} {\widehat F}_5 \times
 \Phi_{1,2} {\widehat F}_2{})} {{\widehat F}_1}
  @<{\Phi_{1,2} }<<
 \underset{(F_2 \times \Phi_{2,3} {\widehat F}_3 \times
\Phi_{2,4} {\widehat F}_4)}
 {{\widehat F}_2} .
\endCD
%%\tag{4.30}
\end{align*}
%\bigtimes
That is, we get the product observable
$\widehat{\mathsf O}_1 \equiv ({\bigtimes}_{t=1}^5 X_t,
{{\bigstimes}_{t=1}^5 {\cal F}_t},
{\widehat F}_1)$
of ${\mathsf O}_1$,
$\Phi_{1,2} \widehat{\mathsf O}_2$ and
$\Phi_{1,5} \widehat{\mathsf O}_5$, and finally, the product observable
\begin{align*}
\widehat{\mathsf O}_0 \equiv ({\bigtimes}_{t=0}^7 X_t,
{{\bigstimes}_{t=0}^7 {\cal F}_t},
{\widehat F}_0
(=
F_0 \times ({}\bigtimes_{t=1,6,7} \Phi_{0,t} {\widehat F}_t{})
)
%\tag{\textcolor{black}{4.31}}
\end{align*}
of ${\mathsf O}_0$,
$\Phi_{0,1} \widehat{\mathsf O}_1$,
$\Phi_{0,6} \widehat{\mathsf O}_6$ and
$\Phi_{0,7} \widehat{\mathsf O}_7$.
Then,
we get the realization of
%$\widehat{\mathsf O}_0$
%is called
%\it
%the realization
%\rm
%(or,
%{\it
%the Heisenberg picture representation})
%\rm
%of
%\it
a sequential observable
\rm
%index{realization@realization of sequential observable}
%
$[{}\{{\mathsf O}_t\}_{t \in T },$
$\{ {C(\Omega_t)} \overset{\Phi_{\pi(t), t}}\to
{C(\Omega_{\pi(t)})} \}_{ t \in T \setminus \{0\} }{}]$.
For completeness,
${\widehat F}_0
$
is represented by
\par
\noindent
\begin{align*}
&
\widehat{F}_0 (\Xi_0 \times \Xi_1 \times \Xi_2 \times \Xi_3 \times \Xi_4 \times \Xi_5 \times \Xi_6 \times \Xi_7)]
\\
=
&
F_0(\Xi_0)
\times
\Phi_{0,1} \biggl(
F_1(\Xi_1)
\times
\Phi_{1,5}F_5(\Xi_5)
\times
\Phi_{1,2}
\Big(
F_2(\Xi_2) \times \Phi_{2,3}F_3(\Xi_3)
\times \Phi_{2,4}F_4(\Xi_4)
\Big)
\biggl)
\\
&
\qquad \qquad
\qquad \qquad
\times
\Phi_{0,6}(
F_6(\Xi_6))
\times
\Phi_{0,7}(
F_7(\Xi_7))
\tag{6.5}
\end{align*}
\rm
%

%\noindent
%\vskip0.3cm
%\vskip0.3cm
%BFBF
\par
\noindent
%^{\roman{(exi)}}
{\bf Remark 6.17}$\;$%POPOPO
In the above example,
consider the case that
${\mathsf O}_t$
($t=2,6,7$)
is not determined.
In this case,it suffices to
define
${\mathsf O}_t$
by
the {{}}{existence observable }
${\mathsf O}^{\roman{(exi)}}_t {{=}} (X_t ,
\{ \emptyset, X_t \}, F^{\roman{(exi)}}_t)$.
Then,
we see that
\begin{align*}
&
\widehat{F}_0 (\Xi_0  \times \Xi_1 \times
X_2 \times \Xi_3 \times \Xi_4 \times \Xi_5 \times X_6
\times X_7
)
\\
=&
F_0(\Xi_0)
\times
\Phi_{0,1} \biggl(
F_1(\Xi_1)
\times
\Phi_{1,5}F_5(\Xi_5)
\times
\Phi_{1,2}
\Big(
\Phi_{2,3}F_3(\Xi_3)
\times \Phi_{2,4}F_4(\Xi_4)
\Big)
\biggl)
\tag{6.6}
%\\
%&
%\qquad
%\times
%\Phi_{0,6}(
%F_6(\Xi_6)
\end{align*}
This is true.
However,
the following is not wrong.
Putting
$T'$
$=$
$\{0,1,3,4,5 \}$,
consider the
$[{}{\mathsf O}_{T'}{}]$
$=$
$[{}\{ {\mathsf O}_t \}_{ t \in {T'}} ,
\{  \Phi_{t_1,t_2}{}: $
${C (\Omega_{t_2})} \to {C (\Omega_{t_1})} \}_{(t_1,t_2) \in (T')^2_{{\; \leqq \;}}}$
$]$.
Then,
the {{realized causal}} observable
$\widehat{\mathsf O}_{T'(0){}}$
$=$
$(\bigtimes_{t \in T' } X_t, $
$\bigstimes_{t \in T' } {\cal F}_t,$
${\widehat F}'_{0})$
is defined by
%POIUYTREWQ
\begin{align*}
&
\widehat{F}'_0 (\Xi_0  \times \Xi_1 \times
%X_2 \times
\Xi_3 \times \Xi_4 \times \Xi_5
%\times X_6
%\times X_7
)
=
F_0(\Xi_0)\\
&
\times
\Phi_{0,1} \Big(
F_1(\Xi_1)
\times
\Phi_{1,5}F_5(\Xi_5)
\times
%\Phi_{1,2}(
\Phi_{1,4}F_4(\Xi_4)\times \Phi_{1,3}F_3(\Xi_3)
\times \Phi_{1,4}F_4(\Xi_4)
\Big)
\tag{6.7}
%\\
%&
%\qquad
%\times
%\Phi_{0,6}(
%F_6(\Xi_6)
\end{align*}
which is different from the true
\textcolor{black}{(6.6)}.
W e may sometimes omit
"existence observable".
However,
if we do so,
we omit it on the basis of careful cautions.

\par
\noindent
\vskip0.3cm
\vskip0.3cm
%BFBF
\par
\noindent
{\bf {{Theorem }} 6.18
[Deterministic sequential observable{{realized causal}} observable
]} %POPOPO
\rm
Let
$(T(t_0), {{\; \leqq \;}})$
be a tree.
%$=$
%$\{ T=\{0,1,2,\ldots,N\},
%\pi: T \setminus \{0\} \to T\}$
Let
$[{}{\mathbb O}_T{}]$
$=$
$[{}\{ {\mathsf O}_t \}_{ t \in T} ,
\{  \Phi_{t_1,t_2}{}: $
${C (\Omega_{t_2})} \to {C (\Omega_{t_1})} \}_{(t_1,t_2) \in T^2_{\leqq}}$
$]$
be deterministic causal observable.
Then,
the realization
$\widehat{\mathsf O}_{{t_0}{}} $
$
\equiv ({\bigtimes}_{t \in T} X_t,{{\bigstimes}_{t \in T} {\cal F}_t},
{\widehat F}_{t_0})
$
is represented by
\begin{align*}
\widehat{\mathsf O}_{{t_0}{}} = \bigtimes_{t\in T} \Phi_{{t_0},t} {\mathsf O}_t
%P%\TAG{7.13}
%%\tag{4.33} 
\end{align*}
That is, it holds that
\begin{align*}
&
[\widehat{F}_{t_0} (
\bigtimes_{t\in T} \Xi_t \
%times \Xi_1 \times \cdots \times \Xi_N
)]
(\omega_{t_0} ) = \bigtimes_{t\in T} [\Phi_{{t_0},t} {F}_t (\Xi_t )](\omega_{t_0} )
= \bigtimes_{t\in T} [{F}_t (\Xi_t )](\phi_{{t_0},t} \omega_{t_0} )
%P%\TAG{7.14}
\\
&
\quad \qquad \quad \qquad \quad \qquad \quad \qquad
%\qqquad
(\forall \omega_{t_0} \in \Omega_{t_0}, \forall \Xi_t \in {\cal F}_t )
%%\tag{4.34} 
\end{align*}
%{t_0}{t_0}00000000000
\par
\noindent
%BFBF
%%%%%%%%%%%%\END{The}
\par
\noindent
{\it $\;\;\;\;${Proof.}}$\;\;$
It suffices to prove
the classical case of
\textcolor{black}{Example 4.15}.
%(i.e., FIGU 6.3)}.
Using
\textcolor{black}{Theorem 4.2}
repeatedly,
we see that
\allowdisplaybreaks
\begin{align*}
%%%%%%%%%%%POIUYTREWQ
&
\textcolor{black}{(4.31)}
=
{\widehat F}_0
=
F_0 \times ({}\bigtimes_{t=1,6,7} \Phi_{0,t} {\widehat F}_t{})
\\
=
&
F_0 \times (
%{}\bigtimes_{t=1,6,7}
\Phi_{0,1} {\widehat F}_1{}
\times
\Phi_{0,6} {\widehat F}_6{}
\times
\Phi_{0,7} {\widehat F}_7{}
)
=
F_0 \times (
%{}\bigtimes_{t=1,6,7}
\Phi_{0,1} {\widehat F}_1{}
\times
\Phi_{0,6} {F}_6{}
\times
\Phi_{0,7} { F}_7{}
)
\\
=
&
\Big(
\bigtimes_{t=0,6,7}
\Phi_{0,t} F_t
\Big)
\times (
%{}\bigtimes_{t=1,6,7}
\Phi_{0,1} {\widehat F}_1{}
)
=
\Big(\bigtimes_{t=0,6,7}
\Phi_{0,t} F_t
\Big)
\times
%{}\bigtimes_{t=1,6,7}
\Phi_{0,1}
(
F_1 \times ({}\bigtimes_{t=2,5} \Phi_{1,t} {\widehat F}_t{})
{}
)
\\
=
&
\Big(
\bigtimes_{t=0,1,6,7}\Phi_{0,t} F_t
\Big)
\times
%{}\bigtimes_{t=1,6,7}
\Phi_{0,1}({}\bigtimes_{t=2,5} \Phi_{1,t} {\widehat F}_t{})
=
\Big(
\bigtimes_{t=0,1,6,7}\Phi_{0,t} F_t
\Big)
\times
%{}\bigtimes_{t=1,6,7}
\Phi_{0,1}({}
\Phi_{1,2} {\widehat F}_2{}
\times
\Phi_{1,5} {\widehat F}_5{})
\\
=
&
\Big(
\bigtimes_{t=0,1,5,6,7}\Phi_{0,t} F_t
\Big)
\times
%{}\bigtimes_{t=1,6,7}
\Phi_{0,1}({}
\Phi_{1,2} {\widehat F}_2{}
)
=
\Big(
\bigtimes_{t=0,1,5,6,7}\Phi_{0,t} F_t
\Big)
\times
%{}\bigtimes_{t=1,6,7}
\Phi_{0,1}({}
\Phi_{1,2}
(F_2 \times ({}\bigtimes_{t=3,4} \Phi_{2,t} {\widehat F}_t{}))
{}
)
\\
=
&
\bigtimes_{t=0}^7 \Phi_{0,t} F_t
%
%P%\TAG{7.15}
%%\tag{4.35} 
\end{align*}
This completes the proof.
\qed
\par
\noindent
{}

%BBBBBBBBBBBBBBBBBB%SBSBSBS
\par
\noindent
{\small%%{\footnotesize
\begin{itemize}
\item[$\spadesuit$] \bf {{}}{Note }6.4{{}} \rm
Note that
a simultaneous observable
(and
a parallel observable)
can be regarded as a kind of
{{realized causal}} observable.
%
%
%  and  and {}
%,
%parallel observable .
%
%$k = 1,2$
%,
%{basic algebra}${C ( \Omega_k )}$
%
%{observable } ${\mathsf O}_k$
%${{=}}$
%$(X_k , {\cal F}_k , F_k{})$
%.
%,
%state space
%$\Omega_0$
%
%$\Omega_1 \times \Omega_2$
%.
%{{deterministic causal map}}
%$\phi_{0,1}: \Omega_0 \to \Omega_1$
% and
%$\phi_{0,2}: \Omega_0 \to \Omega_2$
%
%\BEGIN{align*}
%\Omega_0 =
%\Omega_1 \times \Omega_2
%\ni (\omega_1, \omega_2) \overset{\phi_{0,k}}{\to}
%\omega_k \in \Omega_k
%\qquad
%(k=1,2)
%\END{align*}
%.
%{{deterministic causal map}}$\phi_{0,k}$
%
%{{deterministic causal operator}}
%$\Phi_{0,k}: C(\Omega_k) \to C(\Omega_0)$
% and .
%{parent map}$\pi (1)=0$,
%$\pi(2)=0$
% and ,
%
%$[ \{{\mathsf O}_k\}_{k=1,2} ;
%\{ \Phi_{0,k}: C(\Omega_k) \to C(\Omega_0)\}_{k=1,2}
%]$
%.
%{{realized causal}} observable
%%
%%$C({\Omega_0})$parallel observable
%$\widetilde{\mathsf O}$
%${{=}}$
%$(X_1 \times X_2  , {\cal F}_1 \boxtimes {\cal F}_2 , \widetilde{F}{})$
%,
%%EEEEEE
%\BEGIN{align*}
%[{\widetilde F}({}\Xi_1 \times \Xi_2 )]
%(\omega_1,\omega_2 )
%=
%[F_1 ({}\Xi_1{})](\omega_1 )
%\cdot
%[F_2 ({}\Xi_1{})](\omega_2 )
%%%%%2.24}
%%\TAG{7.25}
%\END{align*}
%,
%parallel observable
%${\mathsf O}_1
%\otimes
%{\mathsf O}_2$
%.
%That is,
%parallel observable
%
%{{realized causal}} observable  and  and .
% and {Remark }\textcolor{black}{{Note }6.15}.
%%(
%%\textco
%%olor{red}{{Sec. 7.3.3}(f$_2$) and (f$_3$)})
%%).
%% and . ,
%%{{measurement theory}}{metaphysics},
%%
%% and .
\end{itemize}
}

\subsection{\textcolor{black}{Axiom${}_{\text{\scriptsize c}}^{\text{\scriptsize pm}}$ 2}
%)
---"No smoke without fire"
%{Sec.6.4}--
\label{6secAxiom 2}
}
%\ssubsection{causality No smoke without fire}
%index{No smoke without fire.
@No smoke without fire}

\rm
\subsubsection{{{The Heisenberg picture}}}%{Sec.6.4.1}
\rm
\par

Summing up the arguments in the previous section,
we assert Axiom${}_{\text{\scriptsize c}}^{\text{\scriptsize pm}}$ 2 as follows.
,
%Axiom${}_{\text{\scriptsize c}}^{\text{\scriptsize pm}}$ 2.
%
%causality
%---
%No smoke without fire and 
%---
%
%{{measurement theory}} and .

\vskip0.5cm
\rm
%\newpage
%%index{measurement12@${\mathsf M}_{C (\Omega)} \big({}{\mathsf O},
%$
%$
%S_{[{}\ast] }(\nu)
%\big)$:{(}continuous type {}){{measurement}}}
%index{@{{measurement}}}
%
%\BEGIN{itembox}[c]
\par
\noindent
\begin{center}
{\bf
\textcolor{black}{Axiom${}_{\text{\scriptsize c}}^{\text{\scriptsize pm}}$ 2} (causality : continuous type)
}
\label{axiomcpm2}
\end{center}
%\END{document}
\par
\noindent
%\vskip0.1cm
\par
\noindent
\fbox{\parbox{155mm}{
\begin{itemize}
\item[(i)]
A chain of causalities
\\
A chain of causalities
is represented by
{{sequential causal operator}}
$ \{ \Phi_{t_1,t_2}{}: $
${C (\Omega_{t_2})} \to {C (\Omega_{t_1})} \}_{(t_1,t_2) \in T^2_{\leqq}}$.
\\
\item[(ii)]{{realized causal}} observable
\\
A sequential observable
$[{}{\mathsf O}_{T(t_0)}{}]$
${{=}}$
$[{}\{ {\mathsf O}_t \}_{ t \in T} ,
\{  \Phi_{t_1,t_2}{}: $
${C (\Omega_{t_2})} \to {C (\Omega_{t_1})} \}_{(t_1,t_2) \in T^2_{\leqq}}$
$]$
is realized by
its
{{realized causal}} observable
%$
%\widehat{\mathsf O}_{T{}}
%$
$\widehat{\mathsf O}_{T(t_0){}}$
$=$
$(\bigtimes_{t \in T } X_t, $
$\bigstimes_{t \in T } {\cal F}_t, {\widehat F}_{t_0})$.
%\hfill
\end{itemize}
}
}
\par
\vskip0.5cm
\par
\noindent

%
%
%
%
%
%
%\BEGIN{itembox}[c]{
%\bf
%\textcolor{black}{Axiom${}_{\text{\scriptsize c}}^{\text{\scriptsize pm}}$ 2} (causality ) continuous$\cdot$pure
%}
%\label{rule601}
%\label{axiomcpm2}
%%index{2@Axiom${}_{\text{\scriptsize c}}^{\text{\scriptsize pm}}$ 2[{}causality {}(continuous type )]}
%%\BEGIN{itemize}
%%\item[(i)]
%(i):
%A chain of causalities
%\\
%A chain of causalities
%is represented by
%{{sequential causal operator}}
%$ \{ \Phi_{t_1,t_2}{}: $
%${C (\Omega_{t_2})} \to {C (\Omega_{t_1})} \}_{(t_1,t_2) \in T^2_{\leqq}}$.
%\\
%(ii):{{realized causal}} observable
%\\
%A sequential observable
%$[{}{\mathsf O}_{T(t_0)}{}]$
%${{=}}$
%$[{}\{ {\mathsf O}_t \}_{ t \in T} ,
%\{  \Phi_{t_1,t_2}{}: $
%${C (\Omega_{t_2})} \to {C (\Omega_{t_1})} \}_{(t_1,t_2) \in T^2_{\leqq}}$
%$]$
%is realized by
%its
%{{realized causal}} observable
%%$
%%\widehat{\mathsf O}_{T{}}
%%$
%$\widehat{\mathsf O}_{T(t_0){}}$
%$=$
%$(\bigtimes_{t \in T } X_t, $
%$\bigstimes_{t \in T } {\cal F}_t, {\widehat F}_{t_0})$.
%%\hfill$13)$
%%\END{itemize}
%\END{itembox}
%

\normalsize
\baselineskip=18pt

\par
Therefore,
we get
a
continuous$\cdot$pure type classical {{{measurement theory}}}
as follows.
%:
%(POI){{{measurement theory}}}, quantum mechanicsverbalizing,
\begin{align*}
%\dashbox{5}
\underset{\text{\scriptsize (scientific language)}}{\text{{}
$\fbox{pure {{{measurement theory}}}}$}}
:=
{
\overset{\text{\scriptsize [Axiom${}_{\text{\scriptsize c}}^{\text{\scriptsize p}}$ 1]}}
{
\underset{\text{\scriptsize
[probabilistic interpretation]}}{\text{{} $\fbox{(pure) {{measurement}}}$}}}
}
+
{
\overset{\text{\scriptsize [Axiom${}_{\text{\scriptsize c}}^{\text{\scriptsize pm}}$ 2]}}
{
\underset{\text{\scriptsize [{{the Heisenberg picture}}]}}
{\text{{}$\fbox{ causality }$}}
}
}
%\TAG*{$\displaystyle{\mathop{1)}_{(=10))}}$}
%%%%%%%%%%%%%CLAsSICAL
\end{align*}

\par
%{{realized causal}} observable
%$\widehat{\mathsf O}_{T{}}$
%,
%\textcolor{black}{Axiom${}_{\text{\scriptsize c}}^{\text{\scriptsize p}}$ 1}
%{(}{{measurement}}{)},
Thus, we say that
\begin{itemize}
\item[(a)]
The probability that a measured value
$(x_t)_{t \in T}$
obtained by
the
{{measurement}}
${\mathsf M}_{C(\Omega_{t_0} )}(\widehat{\mathsf O}_{T{}},
S_{[\omega_{t_0}]})$
belongs to
${\widehat \Xi}( \in \bigstimes_{t\in T}{\cal F}_t )$
is given by
$[{\widehat F}_{t_0}(
%\bigtimes_{t\in T} \Xi_t
{\widehat \Xi}
)](\omega_{t_0})$.
\end{itemize}
For completeness,
note that
%bounded type {{{measurement theory}}},
\begin{itemize}
\item[(b)]
A state
$\omega_0$
is fixed,
and thus,
it does not change.
%index{@{{state}}}
\end{itemize}

%BBBBBBBBBBBBBBBBBB%SBSBSBS
\par
\noindent
{\small%%{\footnotesize
\vspace{0.1cm}
\begin{itemize}
\item[$\spadesuit$] \bf {{}}{Note }6.5{{}} \rm
The (i) in Axiom${}_{\text{\scriptsize c}}^{\text{\scriptsize pm}}$ 2
is fundamental.
However,
the (ii)
may be regarded as
is the consequence of
"Only one observable is permitted"
(i.e.,
the Copenhagen interpretation).
%
%,
%(ii), (i) and ,
%the Copenhagen interpretation
%(That is,
%{{measurement}}) and .
\end{itemize}
}

\baselineskip=18pt
\par
\noindent
\vskip0.3cm
\vskip0.3cm
%BFBF
\par
\noindent
{\bf {Remark }6.19
[Mixed {{measurement theory}}
]}$\;\;$%POPOPO
In mixed {{{measurement theory}}},
Axiom${}_{\text{\scriptsize c}}^{\text{\scriptsize pm}}$ 2
is valid.
That is,
it is common
in pure and mixed measurement theories.
Thus,
we get mixed measurement theory
as follows.
\begin{align*}
\underset{\text{\scriptsize (scientific language)}}{\text{{}
$\fbox{mixed {{measurement theory}}}$}}
:=
{
\overset{\text{\scriptsize [Axiom${}_{\text{\scriptsize c}}^{\text{\scriptsize p}}$ 1]}}
{
\underset{\text{\scriptsize
[probabilistic interpretation]}}{\text{{} $\fbox{mixed {{measurement}}}$}}}
}
+
{
\overset{\text{\scriptsize [Axiom${}_{\text{\scriptsize c}}^{\text{\scriptsize pm}}$ 2]}}
{
\underset{\text{\scriptsize [{{the Heisenberg picture}}]}}
{\text{{}$\fbox{ causality }$}}
}
}
%\TAG*{$\displaystyle{\mathop{1)}_{(=10))}}$}
%%%%%%%%%%%%%CLAsSICAL
\end{align*}
That is,
\begin{itemize}
\item[(c)]
The probability that a measured value
$(x_t)_{t \in T}$
obtained by
the
{{measurement}}
${\mathsf M}_{C(\Omega_{t_0} )}(\widehat{\mathsf O}_{T{}},
S_{[\ast]}(\nu))$
belongs to
${\widehat \Xi}( \in \bigstimes_{t\in T}{\cal F}_t )$
is given by
$\int_{\Omega_{t_0}}
[{\widehat F}_{t_0}(
{\widehat \Xi}
)](\omega_{t_0})
\nu_{t_0} (
d \omega_{t_0}
)$
{}
\end{itemize}
%.
%,
%{{{measurement theory}}},
%\BEGIN{itemize}
%\ITEM{}
%\END{itemize}
Also, a mixed state
$\nu_{t_0}$
is fixed,
and thus,
it does not change.
\rm
\renewcommand{\footnoterule}{
  \vspace{2mm}                      % 
  \noindent\rule{\textwidth}{0.4pt}  
  \vspace{-3mm}
}
\subsubsection{How should time be represented?
---
Leibniz's relationalism}%6.4.2
%index{@}
\baselineskip=18pt
\par
In \textcolor{black}{{Sec. 2.3.3}},
we conclude that
\begin{itemize}
\item[(d)]
The space of our world is described as a kind of
state space
(or precisely,
spectrum).
\end{itemize}
In ,
Leibniz-Clarke Correspondence
(\textcolor{black}{{Sec. 2.3.3}}),
%
%
%
% {(}1715--1716)
%is importat to know
%both Leibniz's and Clarke's (=Newton's) ideas concerning space and time.
%\par
%\BEGIN{itemize}
%\item[({}b)]
%Newton's absolutism says that
%the space-time should be regarded as a receptacle of a "thing."
%Therefore,
%even if
%"thing" does not exits,
%the space-time exists.
%On the other hand,
%Leibniz's relationalism
%says that
%\BEGIN{itemize}
%\item[(b$_1$)]
%Space is
%a kind of state of "thing".
%% and ${{\cdot}}${(}={{state}}{)}
%%%,
%%,
%% and {{state}}
Leibniz says
"Time is an order of
occurring in succession
which changes one after another".
Measurement theory
agrees to
Leibniz's opinion
as follows.
\begin{itemize}
\item[(e)]
Time axis ${\mathbb R}$
(
or,
${\mathbb Z}$)
is described as a kind of
the semi-ordered tree $(T, \leqq )$.
\end{itemize}
\baselineskip=18pt
% and \footnote{
%{}(\textcolor{black}{{{Chap.{\;}}II}}), ,
%$T${finite set} and ,
%\textcolor{black}{{{Chap.{\;}}IV}}( and , $T={\mathbb R}, 
%{\mathbb Z}$)). }.
%
%,
%{{sequential causal operator}}
%$ \{ \Phi_{t_1,t_2}{}: $
%${C (\Omega_{t_2})} \to {C (\Omega_{t_1})} \}_{(t_1,t_2) \in T^2_{\leqq}}$
%
%{\bf
%{{sequential causal operator}}
%}
% and .
%%(\textcolor{black}{Example 6.23}).
%%index{@{{sequential causal operator}}}

%, ${\mathbb R}$
%(,
%${\mathbb Z}$)
%causality  and , causality  and ,
%causality causality ,
%causality causality 
% and
%
%.
%, ,
% and ,
%causality 
% and .
%
%BBBBBBBBBBBBBBBBBB%SBSBSBS
\par
\noindent
{\small%%{\footnotesize
\vspace{0.1cm}
\begin{itemize}
\item[$\spadesuit$] \bf {{}}{Note }6.6{{}} \rm
After Newton,
physical space-time
may be superior to
{metaphysical }space-time.
The reason may be as follows.
\begin{itemize}
\item[($\sharp_1$)]
The classification of
the {world-description}(\textcolor{black}{Chap. 1 (O)})
---
{realistic method} and {linguistic method}
---
is not firm.
%(That is,
%{metaphysics} and )
\item[($\sharp_2$)]
Many people investigate
"What is space-time?"
and not
"How should space-time be represented?"
\end{itemize}
For example,
the following two definitions
are famous:
\begin{itemize}
\item[($\sharp_3$)]
[Leibniz]:
"Time is an order of
occurring in succession
which changes one after another".
%index{@}
\item[($\sharp_4$)]
[Augustinus(354--430)]:
Only present exists.
Past is in memory.
Future is in presentiment.
\end{itemize}
However,
these are rather literary.
than scientific.
%, ,
%{{measurement theory}},
%
%
%$(\sharp_3)$
% and
%$(\sharp_4)$
%,
%(e) and ,
%
% and \footnote{
%, ($\sharp_4$)
%, 
% and ,
%{{measurement theory}} and 
%(That is, {{measurement theory}}),
%\BEGIN{itemize}
%\item[$(\sharp)$]
%[=$t_0$],
%[={{state}}],
%[={{realized causal}} observable ]
%\END{itemize}
% and .
%,
%, (e), .
%, $(\sharp)$,
%{measuring object}{}
%{{measurement theory}}, (observer)
%(\textcolor{black}{{Note }6.7})
%}.
\end{itemize}
}
%%BBB

\par
\noindent
%It is  to note that
\par
\noindent
%BBBBBBBBBBBBBBBBBB%SBSBSBS
{\small%%{\footnotesize
\begin{itemize}
\item[$\spadesuit$] \bf {{}}{Note }6.7{{}} \rm
In this section,
the space-time in {measuring object}
is explained.
However,
measurement theory assert that
observer's space-time
does not exist.
Thus,
there is no
tense
---
past, present, future
---
in science.
%
%
%{(}${{\cdot}}${)}
%(\textcolor{black}{{Note }2.8}).
%,
%,  and .
%
%,
%, observer
%,
%% and ,
%%observer
%%
%% and .
%
%,
%{ordinary language}, ,
%observer
%,
% and .
%,
%observer.
% and ,
For example,
note that
\begin{itemize}
\item[$(\sharp)$]
In Axiom${}_{\text{\scriptsize c}}^{\text{\scriptsize p}}$ 1,
"the probability that a measured value was obtained"
and
"the probability that measured value will be obtained"
are confused.
\end{itemize}
That is,
measurement theory
is not concerned with
observer's time
(or,
subjective time).
\end{itemize}
}
\rm
\subsubsection{{{Why does measurement theory hold?}}}%6.4.3
\par
Now we have several key-words in measurement theory.
%, {{measurement theory}}.
%\item[] \bf  \rm %%%BBBBBBBBBBBBBBBBBBBB
%\\
%.
\begin{itemize}
\item[(f$_1$)][\textcolor{black}{Axiom${}_{\text{\scriptsize c}}^{\text{\scriptsize p}}$ 1}]:
{{measurement}}(observer, {measuring object}, {{measurement}}, observable , {{state}}, measured value , probability )
\item[(f$_2$)][\textcolor{black}{Axiom${}_{\text{\scriptsize c}}^{\text{\scriptsize pm}}$ 2}]:
causality (semi-ordered tree, {{sequential causal operator}},
{{realized causal}} observable
\item[(f$_3$)][The Copenhagen interpretation]:
space as a kind of state space (or, precisely,
spectrum) ,
time as a kind of semi-ordered tree.
%semi-ordered tree and 
%(, (f$_3$), state space  and semi-ordered tree and 
% and ,
%(f$_1$) and (f$_2$) and )
%%({}\textcolor{black}10
%%%7%%
%})
%)
\end{itemize}
Measurement theory says that
these words should be used according to
modeled on Axiom${}_{\text{\scriptsize c}}^{\text{\scriptsize p}}$s 1 and 2.
%{},
%
%,  and {{{measurement theory}}}{}
%,
%{{{measurement theory}}}{(}Axiom${}_{\text{\scriptsize c}}^{\text{\scriptsize p}}$ 1 and 2)${{\cdot}}$,
% and ,
% and (\textcolor{black}{3.5}).
Thus, the following question is natural:,
\begin{itemize}
\item[(g)]
Why can various sciences be described by the only two axioms?
% (POI)2
%[{{measurement}} and causality ]
%,  and  and ,
%{\bf ${{\cdot}}$}{}
\end{itemize}
% and .
%${{\cdot}}${{{{c}}}},
%${{\cdot}}$ and 
% and
%.
%{physics}?, ({{Newton}})
%, (=[Simple is best])
%.
%,
%?, ,
% and  and {}
%%,
%% and ,
%%
%%(, ).

\par
\noindent
%BBBBBBBBBBBBBBBBBB%SBSBSBS
{\small%%{\footnotesize
\vspace{0.1cm}
\begin{itemize}
\item[$\spadesuit$] \bf {{}}{Note }6.8{{}} \rm
%.
% and ,
%${{\cdot}}$,
% and {}
%,
The above question (g)
may be deeper than
the question such that
\begin{itemize}
\item[$\underset{(Chap. 1)}{\text{(F$_5$)}}$]
Why are two mathematical theories
({differential equation}
and probability theory)
useful in science?
\end{itemize}
That is because
this is solved if the (g) is answered.
Thus, what is important
is the following.
\begin{itemize}
\item[$(\sharp)$]
Why is measurement theory
{{measurement theory}}
---
the language of quantum mechanics
% mechanics
---
applicable to various sciences?
$\;\;$
Why is the absurd theory
(i.e.,
dualism and the Copenhagen interpretation)
necessary?
\end{itemize}
We have no answer to this problem.
\end{itemize}
}
%%BBBBBBBBBBBBBBBBBB%SBSBSBSS
%
\baselineskip=18pt
\subsubsection{{{State}} change ---{{the Schr\"odinger{ picture}}}---}%{Sec.6.4.4}
\par
%%%%%%%%%
The Copenhagen interpretation
---
{Chap.$\;$1}(U$_4$)
%\textcolor{black}{\REF{2secAxiom 1}}
%\textcolor{black}{Axiom${}_{\text{\scriptsize c}}^{\text{\scriptsize p}}$ 1}(i
---
says that
"only one measurement is permitted",
which
implies
"Only one state and only one observable"
Thus,
we construct
the {{realized causal}} observable
%$
%\widehat{\mathsf O}_{T{}}
%$
$\widehat{\mathsf O}_{T{}}$
$=$
$(\bigtimes_{t \in T } X_t, $
$\bigstimes_{t \in T } {\cal F}_t, {\widehat F}_{t_0})$
from the
sequential observable
$[{}{\mathsf O}_T{}]$
${{=}}$
$[{}\{ {\mathsf O}_t \}_{ t \in T} ,
\{  \Phi_{t_1,t_2}{}: $
${C (\Omega_{t_2})} \to {C (\Omega_{t_1})} \}_{(t_1,t_2) \in T^2_{\leqq}}$
$]$.
And we take a
{{measurement}}${\mathsf M}_{C(\Omega_0)}(
\widehat{\mathsf O}_{T{}}
, S_{[\omega_0]})$.
%
%.
%,
%\textcolor{black}{{}{{{}}}{Sec.6.4.1}(b)}
%,
%\BEGIN{align*}
%{{state}}.
%%\par
%\END{align*}
% and .
%{{{measurement theory}}}{}
\par
\noindent
%BBBBBBBBBBBBBBBBBB%SBSBSBS
{\small%%{\footnotesize
\vspace{0.1cm}
\begin{itemize}
\item[$\spadesuit$] \bf {{}}{Note }6.9{{}} \rm
Summing up
the above argument as follows.
classical mechanical {world-view}causality ,
%{differential equation}\textcolor{black}{(1.1)
\begin{align*}
\underset{\text{\scriptsize (the Copenhagen interpretation)}}{[{\text{
only one measurement is permitted}}]}
\Longrightarrow
\underset{}{\text{[Only one observable]}}
\Longrightarrow
\underset{\text{\scriptsize (Axiom${}_{\text{\scriptsize c}}^{\text{\scriptsize pm}}$ 2(ii))}}{\text{[{{realized causal}} observable ]}}
\end{align*}
{}
\end{itemize}
}
%%BBBBBBBBBBBBBBBBBB%SBSBSBSS

%{}

\par
However,
as a convenient method,
we sometimes use
%,  and ,
the state change
due to
{{the Schr\"odinger{ picture}}}.
This is not general
but it may be understandable.
Thus,
in what follows,
we explain this view-point
(i.e.,
state change).

\par
We begin with the simplest example.
Put $T=\{0,1\}$,
Consider a
{{deterministic causal operator}}
$\Phi_{0,1}{}: $
${C (\Omega_{1})} \to {C (\Omega_{0})}$
with  a
{{deterministic causal map}}
$\phi_{0,1}:\Omega_0 \to \Omega_1$.
Let
${\mathsf O}_1=(X_1, {\cal F}_1, F_1)$
be an observable in
${C (\Omega_{1})} $.

be
 and .
${\mathsf O}_1=(X_1, {\cal F}_1, F_1)$
${C (\Omega_{1})} ${{deterministic causal operator}}
 and .
${\mathsf O}_1=(X_1, {\cal F}_1, F_1)$
in ${C (\Omega_{1})} $
Consider a
{{measurement}}
${\mathsf M}_{C(\Omega_0)}(
\Phi_{0,1} {\mathsf O}_1, S_{[\omega_0]})$.
Axiom${}_{\text{\scriptsize c}}^{\text{\scriptsize p}}$ 1({{measurement}}
says that
\begin{itemize}
\item[(a)]
The probability that
a measured value
obtained by a
{{measurement}}
${\mathsf M}_{C(\Omega_0)}(
\Phi_{0,1} {\mathsf O}_1, S_{[\omega_0]})$
belongs to
$\;$
$\Xi_1
(\in {\cal F}_1 )$
is given by
$\;$
$
[\Phi_{0,1} F_1(\Xi_1)](\omega_0)
$.
\end{itemize}
Next, consider a
{{measurement}}
${\mathsf M}_{C(\Omega_1)}(
{\mathsf O}_1, S_{[\phi_{0,1}(\omega_0)]})$
The, Axiom${}_{\text{\scriptsize c}}^{\text{\scriptsize p}}$ 1({{measurement}}
%(\textcolor{black}{\REF{2secAxiom 1}})
)
also says that
%,
\begin{itemize}
\item[(b)]
The probability that
a measured value
obtained by a
{{measurement}}
${\mathsf M}_{C(\Omega_1)}(
{\mathsf O}_1, S_{[\phi_{0,1}(\omega_0)]})$
%${\mathsf M}_{C(\Omega_0)}(
%\Phi_{0,1} {\mathsf O}_1, S_{[\omega_0]})$
belongs to
$\;$
$\Xi_1
(\in {\cal F}_1 )$
is given by
$\;$
$
[F_1(\Xi_1)](
\phi_{0,1}(\omega_0)
)
$
\end{itemize}
Here, note that
$
[\Phi_{0,1} F_1(\Xi_1)](\omega_0)
$
$=$
$
[F_1(\Xi_1)](
\phi_{0,1}(\omega_0)
)
$.
% and
%${\mathsf O}_1$
%,
Thus,
(a) and (b)
imply that
\begin{itemize}
\item[(c)]
${\mathsf M}_{C(\Omega_0)}(
\Phi_{0,1} {\mathsf O}_1, S_{[\omega_0]})$
and
${\mathsf M}_{C(\Omega_1)}(
{\mathsf O}_1, S_{[\phi_{0,1}(\omega_0)]})$
can be identified.
\end{itemize}
Also, we may consider that
\begin{itemize}
\item[(d)]
$\omega_0$
is a state at time
$t=0$,
and
$\phi_{0,1}(\omega_0)$
is a state at time
$t=1$.
Writing it in diagram  as follows.
\begin{align*}
\Omega_0
\ni
\underset{\text{(time 0)}}{\omega_0}
\xrightarrow[\quad \text{\scriptsize (change)} \quad]{}
\underset{\text{(time 1)}}{\phi_{0,1}(\omega_0)}
\in
\Omega_1
\end{align*}
\end{itemize}
{}

\par
Now we shall generalize the above argument.
Consider
a sequential {{{causal operator}}}
$
\{  \Phi_{t_1,t_2}{}: $
${C (\Omega_{t_2})} \to {C (\Omega_{t_1})} \}_{(t_1,t_2) \in T^2_{\leqq}}$
Its dual sequential causal operator
is
$
\{  {\Phi}^*_{t_1,t_2}{}: $
${{\cal M}(\Omega_{t_1})} \to {{\cal M}(\Omega_{t_2})} \}_{(t_1,t_2) \in T^2_{\leqq}}$.
Further,
define the
sequential observable
$[{}{\mathsf O}_{T(t_0)}{}]$
${{=}}$
$[{}\{ {\mathsf O}_t \}_{ t \in T} ,
\{  \Phi_{t_1,t_2}{}: $
${C (\Omega_{t_2})} \to {C (\Omega_{t_1})} \}_{(t_1,t_2) \in T^2_{\leqq}}$
$]$.
Let
$\omega_0 \in \Omega_{t_0}$
be a state at time $t$.
%
%{{state}} and ,
%%index{@{{state}}}
%$\delta_{(\cdot)}$(\textcolor{black}{Appendix B.5(C)})
%,
The
$\{ {\Phi}^*_{0, t} \delta_{\omega_0} \}_{t \in T}$
is called a state change due to
{{the Schr\"odinger{ picture}}}.
%%index{@{
%{state}}{{the Schr\"odinger{ picture}}}}
Therefore,
\begin{itemize}
\item[(e)]
{{the Schr\"odinger{ picture}}}
urges us to image that a state changes
as follows.
\begin{align*}
{\cal M}_{+1} (\Omega_0)
\ni
\underset{\text{\scriptsize (time 0)}}{\delta_{\omega_0}}
\xrightarrow[\quad  \quad]{}
\underset{\text{\scriptsize (time 1)}}{\Phi_{0,t}^* (\delta_{\omega_0})}
\in
{\cal M}_{+1}(\Omega_t)
\end{align*}
\end{itemize}
Here it should be noted that,
$ {\Phi}^*_{0, t} \delta_{\omega_0} $
is not generally a point measure
but
a mixed state
(i,e.,
$ {\Phi}^*_{0, t} \delta_{\omega_0} $
$\in$
${\cal M}_{+1} (\Omega_t )$).
\par
\renewcommand{\footnoterule}{%
  \vspace{2mm}                      % 
  \noindent\rule{\textwidth}{0.4pt}   % , 
  \vspace{-5mm}
}

%For each $t \in T $,
%,
%For a {mixed state}$
%{\Phi}^*_{0, t} \delta_{\omega_0}
%( \in {\cal M}_{+1} (\Omega_t ))$,
%
%observable
%${\mathsf O}_t=(X_t, {\cal F}_t, F_t)$

Consider a
mixed {{measurement}}
${\mathsf M}_{C(\Omega_t)}({\mathsf O}_t, S( {\Phi}^*_{0, t}  \delta_{\omega_0}))$.
%
% and \footnote{${\mathsf M}_{C(\Omega_t)}({\mathsf O}_t, S( {\Phi}^*_{0, t}  \delta_{\omega_0}))$,
%$S_{[\ast]}$$[\ast]$,
%.
%{{the Schr\"odinger{ picture}}}, 
%.
%}.
%%
Then,
by
\textcolor{black}{Axiom${}_{\text{\scriptsize c}}^{\text{\scriptsize m}}$ 1}(mixed {{measurement}})
in {Sec.$\;$4.4},
we see:
%,
\begin{itemize}
\item[(f)]
The probability that
a measured value
obtained by a
mixed {{measurement}}
${\mathsf M}_{C(\Omega_t)}({\mathsf O}_t, S( {\Phi}^*_{0, t}  \delta_{\omega_0}))$
%
%${\mathsf M}_{C(\Omega_1)}(
%{\mathsf O}_1, S_{[\phi_{0,1}(\omega_0)]})$
%%${\mathsf M}_{C(\Omega_0)}(
%\Phi_{0,1} {\mathsf O}_1, S_{[\omega_0]})$
belongs to
$\;$
$\Xi_t
(\in {\cal F}_t )$
is given by
\begin{align*}
\int_{\Omega_t} F_t (\Xi_t ) [{\Phi}^*_{0, t}  \delta_{\omega_0}]
(d \omega_t )
\;
\bigl(=
[
{\Phi}_{0, t}
F_t(\Xi_t)
]
(\omega_0)
\bigl)
\end{align*}
\end{itemize}
Thus,
for each
$t (\in T)$,
we have the following identification:
\begin{equation*}
{\mathsf M}_{C(\Omega_{t_0})}(
\Phi_{t_0,t} {\mathsf O}_t, S_{[\omega_0]})
=
{\mathsf M}_{C(\Omega_t)}(
{\mathsf O}_t, S_{[\ast]}({\Phi}^*_{t_0,t}\delta_{\omega_0}))
\end{equation*}
However, it should be noted that
\begin{align*}
\bigtimes_{t \in T}
\int_{\Omega_t} F_t (\Xi_t ) [{\Phi}^*_{0, t}  \delta_{\omega_0}]
(d \omega_t )
\not=
[{\widehat F}_{t_0}
(\bigtimes_{t \in T} \Xi_t)
]
(\omega_0)
\tag{\color{black}{6.8}}
%%%%%REDREDREDREDREDRE
\end{align*}
As seen in
\textcolor{black}{{{Theorem }}6.18},
in the particular case of
deterministic {{sequential causal operator}},
$\not=$
can be replaced by
$=$
in
\textcolor{black}{(6.8)}.

\subsubsection{The principle of equal weight
---
Famous {{unsolved problem}}}%6.4.5
\par
Reconsidering

{{Monty Hall problem}}
(\textcolor{black}{{{Problem }}4.9, {{Problem }}4.16}),
we
present
the final answer of
{{Monty Hall problem}}.

%

%
%%
%
%\par
%\noindent
{\bf
%\vskip0.3cm
%\vskip0.3cm
%BFBF
\par
\noindent
{{Problem }}6.20{{}}
[{}{{Monty Hall problem}}
(Continued from \textcolor{black}{{{Problem }}4.9, {{Problem }}4.16}){{}}
\rm
({\rm cf.}
\textcolor{black}{\cite{Keio,IMont}}){\bf]}}$\;\;$%POPOPO
\rm
%index{@{{Monty Hall problem}}}
%%index{paradox({}{{Monty Hall problem}}{})
%@paradox({}{{Monty Hall problem}}{})}
\rm
\par
\noindent
\it
Suppose you are on a game show, and you are given
the choice of three doors
(i.e., {\lq\lq}number 1{\rq\rq}$\!\!\!,\;$ {\lq\lq}number 2{\rq\rq}$\!\!\!,\;$ {\lq\lq}number 3{\rq\rq}$\!\!)$.
Behind one door is a car, behind the others, goats.

\begin{enumerate}
\rm
\item[\textcolor{black}{($\sharp_2$)}]
\it
You choose a door
by the cast of the fair dice,
i.e.,
with probability
$1/3$.
\end{enumerate}
According to the rule
{\rm \textcolor{black}{($\sharp_2$)}},
you pick a door, say number 1, and the host,
who knows where the car is,
opens another door,
behind which is a goat.
For example,
the host says that
\begin{itemize}
\rm
\item[($\flat$)]
\it
the door 3 has a goat.
\end{itemize}
He says to you,
{\lq\lq}Do you want to pick door number 2?{\rq\rq}
Is it to your advantage to switch your choice of doors?
\rm
\par

%%$F({}\{ 1 \}{})(\omega_1{})= 0.5$
%%and
%%$F({}\{ 2 \}{})(\omega_1{})= 0.5$
%%should be assumed in the
%%problem (P).
%}
%\BEGIN{align*}

%In what follows we study this problem.

\par
\noindent
%\ssubsection{Partial mixed measurements}
%Measurement theory ( basic algebraic formulation)}
\par
\rm
\par
\noindent
{\bf Answer{$\;\;$}}
As \textcolor{black}{Problem 4.9}
(Monty Hall problem),
consider a state space
$\Omega = \{ \omega_1 , \omega_2 , \omega_3 \}$.
And the observable
${\mathsf O}=(X, {\cal F}, F)$
is defined by
the formula
\textcolor{black}{(4.2)}.
The map
$\phi:\Omega \to \Omega $
is defined by
\begin{align*}
\phi(\omega_1) =\omega_2,
\quad
\phi(\omega_2) =\omega_3,
\quad
\phi(\omega_3) =\omega_1
\quad
\end{align*}
we get a causal operator
$\Phi:C(\Omega) \to C(\Omega)$
by
$[\Phi(f)](\omega)
=
f(\phi(\omega))$
$(\forall f \in C(\Omega), \;\forall \omega \in \Omega )$.
Assume that
a car is behind the door $k$
$(k=1,2,3)$.
Then, we say that
\begin{itemize}
\item[(a)]
By the dice-throwing,
you get
$
%\text{}
\left[\begin{array}{ll}
1,2
\\
3,4
%(\text{{{{{c}}}}}, \text{about 40})
\\
5,6
%(\text{{{{{h}}}}}, \text{about 30})
%(\text{{{{{h}}}}}, \text{about 40})
\end{array}\right],
\text{then, take a measurement}
\left[\begin{array}{ll}
{\mathsf M}_{C (\Omega)} ({}{\mathsf O}, S_{[{}\omega_k{}]})
\\
{\mathsf M}_{C (\Omega)} ({}\Phi{\mathsf O}, S_{[{}\omega_k{}]})
\\
{\mathsf M}_{C (\Omega)} ({}\Phi^2{\mathsf O}, S_{[{}\omega_k{}]})
\end{array}\right]
%\TAG{4.56}
%\tag{\color{black}{3.11}}
$
%%%%%REDREDREDREDREDRE
\end{itemize}
%\END{itemize}.
%\textcolor{black}{4.9},
% and ,
%3 and ,
%
%${\mathsf M}_{C (\Omega)} ({}\Phi {\mathsf O}, S_{[{}\ast{}]})$
%
%{{}}3
% and ,
%
%
% and .
%,
We,
by
\textcolor{black}{{}{{{}}}Sec.6.4.4(c)},
see the following identifications:
${\mathsf M}_{C (\Omega)} ({}\Phi{\mathsf O}, S_{[{}\omega_k{}]})$
$=$
${\mathsf M}_{C (\Omega)} ({}{\mathsf O}, S_{[{}\phi(\omega_k){}]})$,
${\mathsf M}_{C (\Omega)} ({}\Phi^2{\mathsf O},$
$ S_{[{}\omega_k{}]})$
$=$
${\mathsf M}_{C (\Omega)} ({}{\mathsf O},$
$ S_{[{}\phi^2 (\omega_k){}]})$.
Thus,
the above (a)
is equal to
\begin{itemize}
\item[(b)]
By the dice-throwing,
you get
$
%\text{}
\left[\begin{array}{ll}
1,2
\\
3,4
%(\text{{{{{c}}}}}, \text{about 40})
\\
5,6
%(\text{{{{{h}}}}}, \text{about 30})
%(\text{{{{{h}}}}}, \text{about 40})
\end{array}\right]
\text{then, take a measurement}
\left[\begin{array}{ll}
{\mathsf M}_{C (\Omega)} ({}{\mathsf O}, S_{[{}\omega_k{}]})
\\
{\mathsf M}_{C (\Omega)} ({}{\mathsf O}, S_{[{}\phi (\omega_k){}]})
\\
{\mathsf M}_{C (\Omega)} ({}{\mathsf O}, S_{[{}\phi^2(\omega_k){}]})
\end{array}\right]
%\TAG{4.56}
%\tag{\color{black}{3.11}}
%%%%%REDREDREDREDREDRE
$
\end{itemize}
Here, note that
$\frac{1}{3}(\delta_{\omega_k}+\delta_{\phi(\omega_k)}
+\delta_{\phi^2 (\omega_k)})$
$=$
$\frac{1}{3}(\delta_{\omega_1}+\delta_{\omega_2}
+\delta_{\omega_3})$
$(\forall k=1,2,3)$.
Thus,
the (b)
is identified with
the mixed measurement
${\mathsf M}_{C (\Omega)} ({}{\mathsf O}, S_{[{}\ast{}]}(\nu_e ))$
where
$\nu_e =\frac{1}{3}(\delta_{\omega_1}+\delta_{\omega_2}
+\delta_{\omega_3})$.
Therefore,
\textcolor{black}{Problem 6.20}
is the same as
\textcolor{black}{Problem 4.16}.
Hence,
you should choose
the  door 2.
\qed

\par
\noindent
%It is  to note that
\par
\noindent
%BBBBBBBBBBBBBBBBBB%SBSBSBS
{\small%%{\footnotesize
\begin{itemize}
\item[$\spadesuit$] \bf {{}}{Note }6.10{{}} \rm
The above argument is easy.
That is,
since you have no information,
we choose the door by a fair dice throwing.
In this sense,
the principle of equal weight
---
unless we have sufficient reason to regard one possible case
as more probable than another,
we treat them as equally probable
---
is clear in measurement theory.
However,
it should be noted that
the above argument is based on dualism.
%
%,
%dualism(observer and {measuring object}) and {Remark }{}
\end{itemize}
}
%%BBBBBBBBBBBBBBBBBB%SBSBSBSS

%
%In this section,
%we explain the (c).
\par
\vskip0.2cm
%\par
%, 
%
% and ,
%probability  and  and 
%
%
%, \textcolor{black}{{{Theorem }}6.21}.
%%index{ and @}
%\textcolor{black}{Example 4.14(d)},
%\BEGIN{itemize}
%\item[]
%,
%{statistics}
%
%That is,
%{ordinary language} and probability 
%
%
%{{unsolved problem}}(\textcolor{black}{\cite{Nipp}})
%\END{itemize}
% and ,
%.
%%\footnote{
%%{{Monty Hall problem}}{}
%%}
%%index{ and @}
%%, \textcolor{black}{{}{{{}}}6.4.5}
%%({\rm cf.}
%%\textcolor{black}{\cite{Keio, Ieas10}})
%%.
\par
%formation
%84848484
%%BFBF
The following is the measurement theoretical formulation of
"the principle of equal weight".
\noindent
{\bf {{Theorem }}6.21
[The principle of equal weight]}$\;\;$%POPOPO
Consider a finite state space $\Omega$,
that is,
$\Omega=\{\omega_1,\omega_2,\ldots,\omega_n\}$.
Let
${\mathsf O}=(X, {\cal F}, F)$
be an observable in
$C(\Omega)$.
%notations.
Consider a {{measurement}}$
{\mathsf M}_{C(\Omega )}({\mathsf O} ,
S_{[
\ast]}
)
$.
If the observer has no information for the state
$[\ast]$,
there is a reason to
that
this measurement is identified with
the mixed measurement
$
{\mathsf M}_{C(\Omega )}({\mathsf O} ,
S_{[
\ast]}(\nu_e)
)
$,
where
$\nu_e =\frac{1}{n}
\sum\limits_{i=1}^n \delta_{\omega_i}
(
\in
{\cal M}_{+1} (\Omega)
)
$.
\par
\noindent
{\it $\;\;\;\;${Proof.}}$\;\;$
The proof is a easy consequence of the above
Monty Hall problem
(or,
see
\textcolor{black}{\cite{Keio, IWhat}}.
\qed

\subsection{Examples in "Axiom${}_{\text{\scriptsize c}}^{\text{\scriptsize p}}$ 1({{measurement}}){ + }Axiom${}_{\text{\scriptsize c}}^{\text{\scriptsize pm}}$ 2(causality )"
}%{Sec.6.5}
In
dualism({{measurement theory}}),
Axiom${}_{\text{\scriptsize c}}^{\text{\scriptsize pm}}$ 2(causality)
is not independently used,
but
it is used with Axiom${}_{\text{\scriptsize c}}^{\text{\scriptsize p}}$ 1({{measurement}}).
% and .

%\ssubsection{causality (Axiom${}_{\text{\scriptsize c}}^{\text{\scriptsize pm}}$ 2){ + }{{measurement}}(Axiom${}_{\text{\scriptsize c}}^{\text{\scriptsize p}}$ 1)}%{Sec.6.5}
\subsubsection{Parallel structure}%{Sec.6.5.1}
\par
%index{@}
\par
Consider a semi-ordered tree
$(T(0) {{=}} \{ 0,1,\ldots,N \},
\pi{}: T\setminus \{0\} \to T )$
with {\bf parallel structure}
such that
$\pi({}t) = 0$
$({}\forall t \in T \setminus \{ 0 \}{})$.

\par
\noindent
%\vskip-0.4cm%%%
%\begin{figure}[htbp]
{
\setlength{\unitlength}{0.7mm}
\begin{picture}(80,55)(0,20)
\put(40,0){
%%\BEGIN{picture}(80,80)(0,20)
\put(70,60){\makebox(10,10)[r]{${C (\Omega_1)}$}}
\put(70,45){\makebox(10,10)[r]{${C (\Omega_2)}$}}
\put(70,22){\makebox(10,10)[r]{${C (\Omega_N)}$}}
\put(30,41){\makebox(10,10)[r]{${C (\Omega_0)}$}}
\put(60,65){\vector(-3,-2){15}}
\put(60,50){\vector(-3,-1){13}}
\put(60,27){\vector(-3,2){15}}
%\put(87,48){\vector(-3,-2){13}}
%\put(87,22){\vector(-3,2){13}}
%\put(112,30){\vector(-3,-2){13}}
%\put(112,7){\vector(-3,2){13}}
%
\put(47,67){$\Phi_{0,1}$}
\put(50,51){$\Phi_{0,2}$}
\put(50,43){$\cdots \cdots$}
\put(50,38){$\cdots \cdots$}
\put(47,25){$\Phi_{0,N}$}
}
\end{picture}
}
%\END{picture}
%\vskip-0.3cm
\begin{center}{Figure 6.4:
Parallel structure
}
\end{center}
%}
%\qquad \qquad \qquad \qquad
%%%%%%3.20}
%%P%\tag{7.18}
%\END{align*}
\par
\noindent
% $t \in T$,
%${C (\Omega_{t})}${observable }
%${\mathsf O}_t$
%${{=}}$
%$(X_t , {\cal F}_t , F_t{})$
%.
%%, our is to represent
%
Consider
a
sequential observable$[{}{\mathsf O}_T{}]$
${{=}}$
$[{}\{ {\mathsf O}_t \}_{ t \in T} ,
\{  \Phi_{\pi(t), t }{}: $
${C (\Omega_t)} \to {C (\Omega_{\pi(t)})} \}_{ t \in T\setminus \{0\} }$
$]$
and its
{{realized causal}} observable $\widehat{\mathsf O}_{T{}}$
${{=}}$
$(\bigtimes_{t=0}^N   X_t ,$
$ \bigstimes_{t=0}^N   {\cal F}_t  ,$
$ {\widehat F}_{{0}}{})$,
That is,
\begin{align*}
{\widehat F}_{0}({}\Xi_0 \times   \Xi_{1} \times \Xi_{2} \times \cdots \times \Xi_{{N}}{})
=
\bigtimes_{t \in T} \Phi_{0,t}F_t (\Xi_t )
\end{align*}
Thus, we have the
{{measurement}}
$
%{\frak M}({}\{ {\mathsf O }_t \}_{t \in T} ,
%{\bold S}_{[{}\omega_0{}] }   {})
%=
{\mathsf M}_{C (\Omega)}
(\widehat{\mathsf O}_{T{}}
{{=}}
(\bigtimes_{t \in T}  X_t , \bigstimes_{t \in T}  {\cal F}_t ,
{\widehat F}_0{}),
S_{[\omega_0]} {}).
%%%%% 2.5
%%%%%3.19}
$
Therefore,
in the case of
parallel structure,
note that
the equality
"="
holds
in
\textcolor{black}{(6.8)}.

That is, we see
\begin{itemize}
\item[]
the probability
that
a measured value
obtained by a parallel
{{measurement}}
${\mathsf M}_{C (\Omega)}
(\widehat{\mathsf O}_{T{}}
{{=}}
$
$
(\bigtimes_{t \in T}  X_t , $
$\bigstimes_{t \in T}  {\cal F}_t ,
{\widehat F}_0{}),
S_{[\omega_0]} {})$
belongs to
$\Xi_0 \times \Xi_1 \times \cdots \times \Xi_N$
is given by
\begin{align*}
[{\widehat F}_0{}(
\bigtimes_{t \in T}
\Xi_t
)](\omega_0)
&=
\bigtimes_{t \in T} [\Phi_{0,t}F_t (\Xi_t )](\omega_0)
%\\
%&=
%\bigtimes_{t \in T} [F_t (\Xi_t )](\phi_{0,t} (\omega_0))
%P%\tag{7.20}
\end{align*}
\end{itemize}

%\qed

\par
\noindent
\vskip0.3cm
\vskip0.3cm
%BFBF
\par
\noindent
{\bf
Example 6.22
[Before pheasants and rabbits problem{\rm
([\textcolor{black}{Example 6.10}]+[{{measurement}}])}]}$\;\;$%POPOPO
Like \textcolor{black}{Example 6.10},
consider the following problem:
\begin{itemize}
\item[(a)]
[Pheasants and rabbits problem]
A number of $m$ pheasants and $n$ rabbits are placed
together in the same cage.
Then
$m+n$ heads and $2m+4n$ legs are counted.
Find the number of pheasants and rabbits.
% and $m$ and
%$n$.
% and 5,
%14{}
%
%
%
%
% and $m$ and $n$.
%, $m+n$,
%$2m+4n${.}
\end{itemize}
%\END{QA}
\par
\noindent
{ordinary language}

In
\textcolor{black}{Example 6.10},
the statement
(a)
in ordinary language
is
understood as
\begin{itemize}
\item[]
"$m$", "$n$", "$m+n$" and "$2m+4n$"
are states
\end{itemize}
However,
here
we understand
as follows.
\begin{itemize}
\item[]
"$m$" and "$n$"
are states,
but
"$m+n$" and "$2m+4n$"
are measured values
\end{itemize}
As
mentioned in
\textcolor{black}{Example 6.10},
put
\begin{align*}
\Omega_0 = \mathbb{N}_0 \times \mathbb{N}_0,
\quad
\Omega_1 = \mathbb{N}_0,
\quad
\Omega_2 = \mathbb{N}_0
%P%\tag{7.21}
\end{align*}
Putting
$T=\{0,1,2\}$,
$\pi(1)=0$,
$\pi(2)=0$.
then
we get
a {{sequential causal operator}}
$\{ {C (\Omega_{t})} \overset{\Phi_{\pi(t), t}}\to {C (\Omega_{\pi(t)})} \}_{ t \in T \setminus \{0\} }$.
For each
$t \in \{1, 2\}$,
consider
{\it {exact observable} }${\mathsf O}^{\FIN}_t $
$=(\mathbb{N}_0, 2^{\mathbb{N}_0}, F_t^{\FIN})$
in
$C (\Omega_t)$.
Thus, we get
the
sequential observable$[{}{\mathsf O}_T{}]$
${{=}}$
$[{}\{ {\mathsf O}_t \}_{ t=1,2} ,
\{  \Phi_{\pi(t), t }{}: $
${C (\Omega_t)} \to {C (\Omega_{\pi(t)})} \}_{ t=1,2 }$
$]$
and its
{{realized causal}} observable
$\widehat{\mathsf O}_0 {=}(\mathbb{N}_0 \times \mathbb{N}_0,
2^{\mathbb{N}_0 \times \mathbb{N}_0}, \widehat{F}_0)$
such that
\begin{align*}
 [\widehat{F}_0 (\Xi_1 \times \Xi_2)] (m,n)
&= [\Phi_{0, 1} F^{\FIN}_1(\Xi_1)] (m,n)
\cdot [\Phi_{0, 2} F^{\FIN}_2(\Xi_2)] (m,n)\\
&= [F^{\FIN}_1 (\Xi_1)] (m + n)
\cdot [F^{\FIN}_2 (\Xi_2)]
(2 m + 4 n)
\\
&
\qquad
\qquad
(\forall
\Xi_1, \forall \Xi_2 \in 2^{\mathbb{N}_0},
\forall (m,n) \in \Omega_0)
%P%\tag{7.22}
\end{align*}
%(\forall
%\Xi_1, \Xi_2 \in 2^{\mathbb{N}_0},
%(m,n) \in \Omega_0)
\par
\noindent
Hence, we get
the {{measurement}}
${\mathsf M}_{C(\Omega_0)}
(\widehat{\mathsf O}_0, S_{[(m,n)]})$.
It is clear that
\begin{itemize}
\item[(b)]
By
{{measurement}}
${\mathsf M}_{C(\Omega_0)}
(\widehat{\mathsf O}_0, S_{[(m,n)]})$,
we get a
measured value
$(m+n, 2m+4n)$
with probability $1$.
\end{itemize}
Here,
it should be noted
that
there are many interpretations in
dualism
(measurement theory).

%}
%
\subsubsection{Series structure
---
{{Measurement}}
of time}%{Sec.6.5.2}
\par
\rm
Assume that the semi-ordered tree
$(T {{=}} \{ 0,1,\ldots,N \},
\pi{})$
has the series structure
such that
That is,
$\pi({}t) = t-1$
$({}\forall t \in T \setminus \{ 0 \}{})$.
Consider a
{{sequential causal operator}}
$\{ {C (\Omega_{t})} \overset{\Phi_{t-1, t}}\to {C (\Omega_{t-1})}
\}_{ t \in T \setminus \{0\} }$
such that
\begin{align*}
{
{\text{$ {C (\Omega_0)} $}}
}
%{{\footnotesize \Phi_{0, 1 } }\atop{\longleftarrow}}
\mathop{\longleftarrow}^{\Phi_{0, 1 } }
{\text{$ {C (\Omega_{1})} $}}
%{{ \Phi_{ 1, 2 } }\atop{\longleftarrow}}
\mathop{\longleftarrow}^{\Phi_{1,2 } }
{\text{$ {C (\Omega_{2})} $}}
%{{ \Phi_{ 2, 3 } }\atop{\longleftarrow}}
\mathop{\longleftarrow}^{\Phi_{2,3 } }
\cdots
\cdots
\cdots
%{{ \Phi_{ N-2, N-1 } }\atop{\longleftarrow}}
\mathop{\longleftarrow}^{\Phi_{ N-2, N-1 } }
{\text{$  C (\Omega_{{N-1}}) $}}
%{{ \Phi_{ N-1, N } }\atop{\longleftarrow}}
\mathop{\longleftarrow}^{\Phi_{N-1, N } }
{\text{$  C (\Omega_{N}) $}}
%%%%3.15}
%P%\tag{7.23}
\end{align*}
Now,
consider a sequential observable$[{}{\mathsf O}_T{}]=$
$[{}\{ {\mathsf O}_t \}_{ t \in T} ,
\{  \Phi_{\pi(t), t }{}: $
${C (\Omega_t)} \to {C (\Omega_{\pi(t)})} \}_{ t \in T\setminus \{0\} }$
$]$.
And let
us
construct
the
{{realized causal}} observable
$\widehat{\mathsf O}_{T}$
%${{=}}$
%${\mathsf O}_{0}
%{{\times}}
%\Phi_{0,1} \widehat{\mathsf O}_{1}$
${{=}}$
$(\bigtimes_{t=0}^N   X_t ,$
$ \bigstimes_{t=0}^N   {\cal F}_t  ,$
$ {\widehat F}_{{0}}{})$.
%6.14 and ,
%
%.

%\begi

Firstly,
put,
$\widehat{\mathsf O}_{N}$
$({}{{=}}  $ $(X_N , {\cal F}_N , {\widehat F}_N{})${}$)$
$=$
${\mathsf O}_{N}$
$({}{{=}}  $ $(X_N , {\cal F}_N , F_N{}){})$.
We have the
simultaneous observable
$\widehat{\mathsf O}_{N-1}$
${{=}}$
${\mathsf O}_{N-1}
{\times}
\Phi_{N-1,N} {\mathsf O}_N$
${{=}}$
$(X_{N-1} \times X_{N} , {\cal F}_{N-1} \boxtimes {\cal F}_n  , {\widehat F}_{{N-1}}{})$
in
$C (\Omega_{{N-1}})$.
That is,
\begin{align*}
{\widehat F}_{{N-1}}({}\Xi_{{N-1}}  \times \Xi_{{N}}{})
=
\bigl(F_{{N-1} }  {{\times}}
%%{\widehat{\mathsf O}}_{N-1} }
\bigl({}\Phi_{{{N-1} } , {{N} } } F_{{N} })
\bigl)
({}\Xi_{{N-1}}  \times \Xi_{{N}}{})
%%%%3.16}
%P%\tag{7.24}
\end{align*}
%({}though{{}}  and {{}}uniqueness are {}).
Similarly,
we get the
simultaneous observable
$\widehat{\mathsf O}_{N-2}$
${{=}}$
${\mathsf O}_{N-2}
{{\times}}
\Phi_{N-2,N-1} \widehat{\mathsf O}_{N-1}$
${{=}}$
$(X_{N-2} \times X_{N-1} \times X_{N} ,$
$ {\cal F}_{N-2} \boxtimes   {\cal F}_{N-1} \boxtimes {\cal F}_n  ,$
$ {\widehat F}_{{N-2}}{})$
in
$C (\Omega_{{N-2}})$.
That is,
\begin{align*}
%\align &
{\widehat F}_{{N-2}} ({}\Xi_{{N-2}}  \times \Xi_{{N-1}}
\times \Xi_{{N}}{})
%\\
=
%&
\bigl(F_{{N-2} }
{{\times}}
%qpqp
(\Phi_{{{N-2} } , {{N-1} } }
{\widehat F}_{{N-1}})
\bigl)
(\Xi_{N-2} \times  \Xi_{{N-1}}  \times \Xi_{{N}}{})
%\ENDalign
%%%%3.17}
%P%\tag{7.25}
\end{align*}
Iteratively,
%%%%%%%%%%%%%POIUYTREWQ
{
\footnotesize
\begin{align*}
%\minCDarrowwidth{0.3cm}
\CD
                   [{}{{C (\Omega_0)}}]
   @<{\Phi}_{0,1}<<   [{}{{C (\Omega_1)}}]
   @<{\Phi}_{1,2}<< \cdots %\cdots
%@<{\Phi}<< [{}{C (\Omega_{N-2})}]
   @<{\Phi}_{N-2,N-1}<< [{}{C (\Omega_{N-1})}]
   @<{\Phi}_{N-1,N}<< [{}{{C (\Omega_N)}}]  \\
           { F_0 }
   @.  F_1
   @. \cdots  %\cdots
%   @.  F_{N-2}
   @.  F_{N-1}
   @.  F_N \\
   @VVV @VVV @. @VVV @VVV  \\
                      {{({}F_{0} {\times}{  \Phi {\widehat F}_{1}}}{}) }\atop{{=\widehat F}_{0}}
   @<{\Phi}_{0,1}<<  {{({}F_{1} {\times}  \Phi {\widehat F}_{2}{}) }\atop{{=\widehat F}_{1}} }
   @<{\Phi}_{1,2}<< \cdots %\cdots
%@<{\Phi}<<
%{{({}F_{N-2} {\times}  \Phi {\widehat F}_{N-1}{}) }\atop{={\widehat F}_{N-2}}}
   @<{\Phi}_{N-2,N-1}<< {{({}F_{N-1} {\times}  \Phi {\widehat F}_N{}) }\atop{={\widehat F}_{N-1}}}
   @<{\Phi}_{N-1,N}<< {{(F_N)}\atop{={\widehat F}_{N}}}
\endCD
   %%%%% 2.4
%%P%\tag{7.26}
\end{align*}
}
%
%
%
%
%
%
%
%,
%%%%%%%%%%%%%%POIUYTREWQ
%{\scriptsize
%\BEGIN{align*}
%%\minCDarrowwidth{0.3cm}
%\CD
%                   [{}{{C (\Omega_0)}}]
%   @<{\Phi}<<   [{}{{C (\Omega_1)}}]
%   @<{\Phi}<< \cdots
%   @<{\Phi}<< [{}{C (\Omega_{N-2})}]
%   @<{\Phi}<< [{}{C (\Omega_{N-1})}]
%   @<{\Phi}<< [{}{{C (\Omega_N)}}]  \\
%           { F_0 }
%   @.  F_1
%   @. \cdots
%   @.  F_{N-2}
%   @.  F_{N-1}
%   @.  F_N \\
%   @VVV @VVV @. @VVV @VVV  @ VVV \\
%                      {{({}F_{0} {\times}{  \Phi {\widehat F}_{1}}}{}) }\atop{{=\widehat F}_{0}}
%   @<{\Phi}<<  {{({}F_{1} {\times}  \Phi {\widehat F}_{2}{}) }\atop{{=\widehat F}_{1}} }
%   @<{\Phi}<< \cdots
%   @<{\Phi}<<
%{{({}F_{N-2} {\times}  \Phi {\widehat F}_{N-1}{}) }\atop{={\widehat F}_{N-2}}}
%   @<{\Phi}<< {{({}F_{N-1} {\times}  \Phi {\widehat F}_N{}) }\atop{={\widehat F}_{N-1}}}
%   @<{\Phi}<< {{(F_N)}\atop{={\widehat F}_{N}}}
%\ENDCD
%   %%%%% 2.4
%%%P%\tag{7.26}
%\END{align*}
%}
\par
\noindent
%\BEGIN{align*}
And finally,
we get the simultaneous observable
$\widehat{\mathsf O}_{0}$
${{=}}$
${\mathsf O}_{0}
{{\times}}
\Phi_{0,1} \widehat{\mathsf O}_{1}$
${{=}}$
$(\bigtimes_{t=0}^N   X_t ,$
$ \bigstimes_{t=0}^N   {\cal F}_t  ,$
$ {\widehat F}_{{0}}{})$
in
$C (\Omega_{{0}})$.
%\BEGIN{align*}
%AAAA
That is,
\begin{align*}
{\widehat F}_{0}({}\Xi_0 \times   \Xi_{1} \times \Xi_{2}
\times \cdots \times \Xi_{{N}}{})
=
\bigl(F_0
{{\times}}
(\Phi_{{0}  , {{1} } }
{\widehat F}_{1})
\bigl)
({}\Xi_0 \times   \Xi_{1} \times \Xi_{2} \times \cdots \times \Xi_{{N}})
%%%%%3.18}
%P%\tag{7.26}
\end{align*}
Here,
$\widehat{\mathsf O}_{0}$
%%index{@ of sequential observable}
%%index{@ of sequential observable}
is
the
{{realized causal}} observable $\widehat{\mathsf O}_{T}${}
of
the
{{}}sequential observable
$[{}\{ {\mathsf O}_t \}_{ t \in T} ,
\{  \Phi_{\pi(t), t }{}: $
${C (\Omega_t)} \to {C (\Omega_{\pi(t)})} \}_{ t \in T\setminus \{0\} }$
$]$.
%{{state}}$\omega_0 (\in \Omega_0)$
%
Thus,
we get
{{measurement}}:
%${\frak M}({}\{ {\mathsf O }_t \}_{t \in T} ,$
%$ {\bold S}_{[{}\omega_0{}] } {})$
%.
%That is,
\begin{align*}
%{\frak M}({}\{ {\mathsf O }_t \}_{t \in T} ,
%{\bold S}_{[{}\omega_0{}] }   {})
%=
{\mathsf M}_{C (\Omega_0)}
(\widehat{\mathsf O}_{T{}}
{{=}}
(\bigtimes_{t \in T}  X_t , \bigstimes_{t \in T}  {\cal F}_t ,
{\widehat F}_0{}),
S_{[\omega_0]} {})
%%%%% 2.5
%%%%%3.19}
%P%\tag{7.27}
\end{align*}
\par
%\noindent

\rm
\par
\noindent
\vskip0.3cm
\vskip0.3cm
%BFBF
\par
\noindent
{\bf Example 6.23
[{{Measurement of discrete time}}]}$\;\;$%POPOPO
Considering
a state space ${\mathbb Z}=\{0,\pm1,\pm2,\ldots\}$,
define the
{{deterministic causal map}}$\phi: {\mathbb Z} \to {\mathbb Z}$
by
\begin{align*}
{\mathbb Z} \ni i \mapsto i+1 \in {\mathbb Z}
%%P%\tag{7.28}
\end{align*}
and thus,
define
the
{{deterministic causal operator}}
$\Phi:C({\mathbb Z})\to C({\mathbb Z})$
by
\begin{align*}
[\Phi(f)](i)= f(i-1)
\qquad
(\forall i \in {\mathbb Z}, \forall f \in C({\mathbb Z}))
\end{align*}
Consider
the {exact observable}
${\mathsf O}^{\FIN} = ( {\mathbb Z}, 2^{{\mathbb Z}}. F^{\FIN} )$
in
$C({\mathbb Z})$.
Let
$T=\{0,1,2,$
$\ldots,$
$N\}$
be
the discrete time,
that is,
$(T(0), {{\; \leqq \;}})$.
For
$t (\in T)$,
consider a state space $\Omega_t = {\mathbb Z}$.
Define the
{{deterministic causal operator}}
$\Phi_{m,n} : C(\Omega_n) \to C(\Omega_m)$
($0 {{\; \leqq \;}}m {{\; \leqq \;}}n {{\; \leqq \;}}N$)
such that
\begin{align*}
\Phi_{m,n} =\overbrace{\Phi \cdot \Phi \cdots \cdots \Phi}^{n-m}
=\Phi^{n-m}
\end{align*}
Putting
${\mathsf O}^{\FIN}_t = {\mathsf O}^{\FIN}$
$(\forall t \in T)$,
we have
the sequential deterministic causal
{exact observable}
$[\{ {\mathsf O}^{\FIN}_t\}_{t\in T},
\{ \Phi: C(\Omega_{m}) \to
C(\Omega_{m-1})\}_{m \in T \setminus \{0\}} \}]$
Thus,
by
\textcolor{black}{{{Theorem }}6.18},
we get
the
{{realized causal}} observable
$\widehat{\mathsf O}
=(\bigtimes_{m \in T}{\mathbb Z},$
$
\bigstimes_{m \in T} 2^{\mathbb Z},
{\widehat F}^{})$
such that
\begin{align*}
{\widehat F}^{}( \bigtimes_{m \in T} \Xi_m )
=
\bigtimes_{m \in T} \Phi^m F^{\FIN}(\Xi_m )
%\quad
%(\forall \Xi_m \in \Omega_m ({{=}} {\mathbb Z} ))
\end{align*}
Therefore,
for
the initial state
$\omega_0 (\in \Omega_0 )$,
we get
the
"time {{measurement}}"
${\mathsf M}_{C(\Omega_0)}(\widehat{\mathsf O}, S_{[\omega_0]})$.
The measured value
is clearly
\begin{align*}
(\omega_0, \omega_0 +1, \omega_0 +2,\ldots, \omega_0 +N)
\end{align*}
{}That is,
if
the initial state space
$\Omega_0$
is assumed to be
at time
$\omega_0$,
then
the clock
(at time $t$)
shows
time
$\omega_0 +t$.
\par
\noindent
%BBBBBBBBBBBBBBBBBB%SBSBSBS
{\small%%{\footnotesize
\vspace{0.1cm}
\begin{itemize}
\item[$\spadesuit$] \bf {{}}{Note }6.11{{}} \rm
The above example says that
each time $t$ $(\in T=\{0,1,\ldots,N\})$
is not a state,
but
a state of the clock
is a state,
i.e.,
$\omega_t$.
And thus it can be measured.
\end{itemize}
}
%{Chap.{\;}}2{}, .
% and 
%%BBBBBBBBBBBBBBBBB
%\BEGIN{itemize}
%\ITEM[({}B$_1$)]
%$
%\c

\def\FIN{{\roman{(exa)}}}
\def\EXI{{\roman{(exi)}}}
\rm
\par
\noindent
\vskip0.3cm
\vskip0.3cm
%BFBF
\par
\noindent
{\bf Example 6.24
[Random walk(Continued from Example 6.13)]}$\;\;$%POPOPO
%index{@}
${\mathbb Z}=\{0,\pm1,\pm2,\ldots\}$
 and .
In \textcolor{black}{Example 4.10},
put
%Suppose that
%a tree
%$(T \equiv \{ 0,1,..., N \},
%\pi{})$
%has a {\lq\lq series\rq\rq} structure,
%i.e.,
%$\pi({}t) = t-1$
%$({}\forall t \in T \setminus \{ 0 \}{})$.
%Consider a
%general system
%${\mathbb S}_{[{}\delta_0{}] }  $
%$\equiv$
%$[{} S_{[\delta_0]}  $,
%$ {\cal A}_t {{\Phi_{ \pi({}t{}), t } }\atop{\rightarrow}}
%{\cal A}_{\pi({}t{})}$
%$({}$
%$t \in T \setminus \{ 0 \}{}) {}]$
%with the initial system $ S_{[\delta_0]}  $,
%that is,
%\BEGIN{align*}
%{
%{\text{$ {\cal A}_0 $}}
%}
%%{{\footnotesize \Phi_{0, 1 } }\atop{\longleftarrow}}
%\mathop{\longleftarrow}^{\Phi_{0, 1 } }
%{\text{$ {\cal A}_{1} $}}
%%{{ \Phi_{ 1, 2 } }\atop{\longleftarrow}}
%\mathop{\longleftarrow}^{\Phi_{1,2 } }
%{\text{$ {\cal A}_{2} $}}
%%{{ \Phi_{ 2, 3 } }\atop{\longleftarrow}}
%\mathop{\longleftarrow}^{\Phi_{2,3 } }
%\cdots
%\cdots
%\cdots
%%{{ \Phi_{ N-2, N-1 } }\atop{\longleftarrow}}
%\mathop{\longleftarrow}^{\Phi_{ N-2, N-1 } }
%{\text{$  {\cal A}_{{N-1}} $}}
%%{{ \Phi_{ N-1, N } }\atop{\longleftarrow}}
%\mathop{\longleftarrow}^{\Phi_{N-1, N } }
%{\text{$  {\cal A}_{N} $}}.
%\TAG{4.35} \end{align*}
%\par
%\noindent
%\par
%%\noindent
%Let
%
%Consider a commutative $C^*$-algebra
%${C} ({\Bbb Z})$.
%
%
%Here, put
\begin{align*}
C(\Omega_t) = {C}({}{\Bbb Z}{})
\quad
(\forall  t \in T = \{ 0,1,..., N \}{})
%%\tag{4.23} 
\end{align*}
where
${\Bbb Z}$
is the set of all integers,
i.e.,
${\Bbb Z}=\{0, \pm 1, \pm 2 , ... \}$.
And define
a causal operator
$\Phi_{ t-1, t}
({}\equiv \Phi{}):
C(\Omega_t) ({}\equiv {C}({}{\Bbb Z}{}){})
\to
{C(\Omega_{t-1})} ({}\equiv {C}({}{\Bbb Z}{}){})$
such that:
\begin{align*}
({}\Phi f{})
(n)
=
({}\Phi_{ t-1, t} f{})
(n)
=
\frac{ f({}n+1) + f({}n-1) }{2}
\qquad
\qquad
(\forall f \in C(\Omega_t) ({}\equiv {C}({}{\Bbb Z}{}){}),
\forall n \in {\Bbb Z}{}).
%%\tag{4.24} 
\end{align*}
Further, define the sequence observable
$[{}\{ {\mathsf O}_t \}_{ t \in T} ,$
$\{  \Phi_{\pi(t), t }(=\Phi ){}: $
${C (\Omega_t)} \to {C (\Omega_{\pi(t)})} \}_{ t \in T\setminus \{0\} }$
$]$
as follows:
Putting
$X_t=\Omega_t={\mathbb Z}$,
for example,
define
\begin{align*}
{\mathsf O}_t =
\cases
\text{exact observable:}{\mathsf O}_t^{\roman{(exa)}}
=({\mathbb Z}(\;=X_t), 2^{{\mathbb Z}}, F_t^{\roman{(exa)}})
\quad
&
(t=2,4)
\\
\\
\text{existence observable:}{\mathsf O}_t^{\text{\scriptsize \roman{(exi)}}}
=({\mathbb Z}(\;=X_t), \{\emptyset, {{\mathbb Z}}\}, F_t^{\roman{(exi)}})
\quad
&
(otherwise)
\endcases
\end{align*}
%,
%
%$[{}\{ {\mathsf O}_t \}_{ t \in T} ,
%\{  \Phi_{\pi(t), t }{}: $
%${C (\Omega_t)} \to {C (\Omega_{\pi(t)})} \}_{ t \in T\setminus \{0\} }$
%$]$
%.
Since
existence observables can be ignored,
we have the realized observable
$\widehat{\mathsf O}_0$
$=$
$({\mathbb Z}^2
(=X_2 \times X_4 ), 2^{\mathbb Z} \times 2^{\mathbb Z}, \widehat{F})$
such that
\begin{align*}
{\widehat F}^{}(\Xi_2 \times \Xi_4 )
=
\Phi^2 \Big( F^{\roman (exa)}( \Xi_2 )
\times \Phi^2(F^{\roman (exa)} ( \Xi_4 ))
\Big)
\quad
(\forall \Xi_2 \in 2^{X_2 },
\forall \Xi_4 \in 2^{X_4})
\end{align*}
Let $0$
$(\in \Omega_0 ={\mathbb Z})$
be a state.
Consider the measurement
measurement${\mathsf M}_{C(\Omega_0)}(\widehat{\mathsf O}_0,
S_{[0]})$.
For example,
we shall calculate
\begin{itemize}
\item[\textcolor{black}{(F)}]
the probability
that
the measured value belongs to
$\{0,1\}\times \{-1,0\}$
${{=}} \Xi_2 \times \Xi_4 (\subseteq X_2 \times X_4)$is equal to
$\frac{1}{4}$.
\end{itemize}
This is easily shown as follows.
Using the characteristic function
$\chi_{\{ \cdot \}}$,
we see
%%%%%\chi_\chi_
\begin{align*}
&
[
F^{\roman (exa)}_2
(\Xi_2)](\omega) =
\chi_{_{\Xi_2}}(\omega)
=\chi_{_{\{0 \}}}(\omega) + \chi_{_{\{1 \}}}(\omega)
&\;\;&
( \forall \omega \in \Omega_2 ={\mathbb Z})
\\
&
[F^{\roman (exa)}_4 (\Xi_4)](\omega) =
\chi_{_{\Xi_4}}(\omega)
=
\chi_{_{\{0 \}}}
(\omega) +
\chi_{_{\{-1 \}}}(\omega)
&\;\;& ( \forall \omega \in \Omega_4 ={\mathbb Z})
\end{align*}
And therefore,
$$
(\Phi
\chi_{_{\{m\}}})(\omega)
=
\frac{1}{2}(\chi_{_{\{m\}}}(\omega+1)
+
\chi_{_{\{m\}}}(\omega-1))
=
\frac{
\chi_{_{\{m-1\}}}(\omega)
+
\chi_{_{\{m+1\}}}(\omega)}{2}
$$
thus,
\begin{align*}
\Phi^2 (F^{\FIN} (\Xi_4)) =\Phi
\Big(
\frac{\chi_{_{\{-1\}}} +\chi_{_{\{1\}}}}{2}
+
\frac{\chi_{_{\{-2\}}}+ \chi_{_{\{0\}}}}{2}
\Big)
=
%&
\frac{1}{4}
\Big(
\chi_{_{\{ -3 \} }}
+
\chi_{_{\{ -2 \} }}
+
2\chi_{_{\{ -1 \} }}
+
2\chi_{_{\{ 0 \} }}
+
\chi_{_{\{ 1 \} }}
+
\chi_{_{\{ 2 \} }}
\Big).
%+
%\chi_{_{\{ 3 \} }}
\end{align*}
{further}
\begin{align*}
F^{\roman (exa)}(\Xi_2) \times \Phi^2 (F^{\roman (exa)}(\Xi_4))
=
&
(\chi_{_{\{0 \}}}
 + \chi_{_{\{1 \}}}
)\times
\frac{1}{4}
\Big(
\chi_{_{\{ -3 \} }}
+
\chi_{_{\{ -2 \} }}
+
2\chi_{_{\{ -1 \} }}
+
2\chi_{_{\{ 0 \} }}
+
\chi_{_{\{ 1 \} }}
+
\chi_{_{\{ 2 \} }}
\Big)
\\
=
&
\frac{1}{4}(2\chi_{_{\{ 0 \} }}
+
\chi_{_{\{ 1 \} }}
).
\end{align*}
{From this,}
\begin{align*}
&
\Phi^2 \Big(
F^{\roman (exa)}(\Xi_2) \times \Phi^2 (F^{\roman (exa)}(\Xi_4))
\Big)
=
\frac{1}{8}
\Phi
\Big(
2\chi_{_{\{ -1 \} }}
+2\chi_{_{\{ 1 \} }}+
\chi_{_{\{ 0 \} }}
+
\chi_{_{\{ 2 \} }}
\Big)
\\
=
&
\frac{1}{16}
\Big(
2\chi_{_{\{ -2 \} }}
+
2\chi_{_{\{ 0 \} }}
+
2\chi_{_{\{ 0 \} }}
+
2\chi_{_{\{ 2 \} }}
+
\chi_{_{\{ -1\} }}
+
\chi_{_{\{ 1\} }}
+
\chi_{_{\{ 1 \} }}
+
\chi_{_{\{ 3 \} }}
\Big)
\\
=
&
\frac{1}{16}
\Big(
2\chi_{_{\{ -2 \} }}
+
\chi_{_{\{ -1\} }}
+
4\chi_{_{\{ 0 \} }}
+
2\chi_{_{\{ 1\} }}
+
2\chi_{_{\{ 2 \} }}
+
\chi_{_{\{ 3 \} }}
\Big)
%P%\TAG{7.28}
\end{align*}

\par
\par
\noindent
Thus,
we conclude that
\begin{align*}
\text{(F)}
=
&
{\widehat F}(\{0,1\} \times \{-1,0\})](0)
\\
=
&
\frac{1}{16}
\Big(
2\chi_{_{\{ -2 \} }}
(0)
+
\chi_{_{\{ -1\} }}(0)
+
4\chi_{_{\{ 0 \} }}(0)
+
2\chi_{_{\{ 1\} }}(0)
+
2\chi_{_{\{ 2 \} }}(0)
+
\chi_{_{\{ 3 \} }}(0)
\Big)
=\frac{4}{16}
=
\frac{1}{4}
\end{align*}
\hfill{$///$}
%BF

\subsection{Two kinds of absurdness
---
idealism and dualism}%{Sec.6.6}
%\ssubsection{linguistic world-view}
\par
As mentioned in
{Note }1.10, {{measurement theory}}
has two kinds of absurdness.
%.
That is,
%idealism=linguistic world-view({Sec.8.1}) and ,
\begin{itemize}
\item[$(\sharp_2)$]
$
{\text{
Two kinds of absurdness
}}
\cases
\text{idealism} &{\cdots}\underset{\text{(Fit feet to shoes)}\qquad }{
\text{linguistic world-view}}
\\
\\
\text{dualism} &{\cdots}\underset{\text{(A spectator does not go up to the stage)}}{\text{the Copenhagen interpretation}}
\endcases
$
\end{itemize}
%2%index{ and @}
In what follows,
we explain these.
%, {Remark }.
%
\subsubsection{The Copenhagen interpretation
---
A spectator does not go up to the stage
}%{Sec.6.6.1}
\par
\noindent
{\bf Remark 6.25[
A spectator does not go up to the stage
]}$\;\;$%POPOPO
Consider the elementary problem
with
two steps
(a)
and
(b):
\begin{itemize}
\item[(a)]
Consider an urn, in which 3 white balls and 2 black balls
are.
Consider the following trial:
\begin{itemize}
\item[]
Pick out one ball from the urn.
If it is black,
you return it in the urn
If it is white,
you do not return it and have it.
Assume that you take three trials.
\end{itemize}.
\item[(b)]
Then,
calculate the probability that
you have
2 white ball
after (a)(i.e., three trials).
\end{itemize}
\par
\noindent
{\bf{Answer}$\;\;$}
Put ${\mathbb N}_0$
$=\{0,1,2,\ldots\}$.
Assume that
there are
$m$
white balls
and
$n$ black balls
in the urn.
This situation is represented by a state
$
(m,n) \in {\mathbb N}_0^2
$.
We can define the dual causal operator
${\Phi^*}: {\cal M}_{+1}({\mathbb N}_0^2)$
$
\to
{\cal M}_{+1}({\mathbb N}_0^2)$
such that
\par
\noindent
\begin{align*}
{\Phi^*}(\delta_{(m,n)}) =
\cases
\frac{m}{m+n} \delta_{(m-1,n)}+\frac{n}{m+n} \delta_{(m,n)}
& \quad
(
{ \text{when} \;\; m \not= 0 \; )}
\\
\delta_{(0,n)}
& \quad
{(\text{when }  m = 0 \; )}.
%%\\
%%\delta_{(0,n)}
%%&  m \not= 0)
\endcases
\end{align*}
where
$\delta_{(\cdot)}$
is the point measure.

Let
$T=\{0,1,2,3\}$
be discrete time.
For each
$t$
$\in T$,
put
$\Omega_t = {\mathbb N}_0^2$.
Thus, we see:
\par
\noindent
\begin{align*}
&
{[\Phi^*]}^3 (\delta_{(3,2)}) =
{[\Phi^*]}^2
\left(
\frac{3}{5}\delta_{(2,2)}
+
\frac{2}{5}\delta_{(3,2)}
\right)
%\\
=
%{\small
%\text{
%$
{\Phi^*}
\left(
(\frac{3}{5} (\frac{2}{4} \delta_{(1,2)}
+\frac{2}{4} \delta_{(2,2)} )
+
\frac{2}{5}{(}  \frac{3}{5}  \delta_{(2,2)}+\frac{2}{5}  \delta_{(3,2)}
)
\right)
\\
=
%&
%{\Phi^*}
%\left(
%\frac{3}{10} \delta_{(1,2)}
%+\frac{27}{50} \delta_{(2,2)}
%+
%\frac{4}{25}  \delta_{(3,2)}
%\right)
%%$
%%}
%%}
%\\
%%\\
%=
&
%{\overline \Phi}
%\left(
\frac{3}{10} (
\frac{1}{3}  \delta_{(0,2)}+\frac{2}{3}  \delta_{(1,2)}
)
+\frac{27}{50}
(
\frac{2}{4}  \delta_{(1,2)}+\frac{2}{4}  \delta_{(2,2)}
)
+
\frac{4}{25}
(
\frac{3}{5}  \delta_{(2,2)}+\frac{2}{5}  \delta_{(3,2)}
)
\\
=
&
%{\overline \Phi}
%\left(
\frac{1}{10}  \delta_{(0,2)}+\frac{47}{100}  \delta_{(1,2)}
+\frac{183}{500}
\delta_{(2,2)}
+
\frac{8}{125}  \delta_{(3,2)}
%\right)
%\right)
%P%\tag{7.29}
\end{align*}
\par
\noindent
Define the observable
${\mathsf O} =({\mathbb N}_0,2^{{\mathbb N}_0}, F^{})$
in
$C(\Omega_3)$
such that
\begin{align*}
[F^{}(\Xi)](m,n)
=
\cases
1 & \qquad (m,n ) \in \Xi \times {\mathbb N}_0
\subseteq \Omega_3
\\
0 & \qquad (m,n ) \notin \Xi \times {\mathbb N}_0
\subseteq \Omega_3
\endcases
\end{align*}
Therefore,
the probability
that
a measured value "$2$"
is obtained by
the
{{measurement}}
${\mathsf M}_{C({\mathbb N}_0^2)}(\Phi^3{\mathsf O}, S_{[(3,2)]})$
is
given by
%, measured value $2$probability ,
%That is,
%white ball$2$probability ,
\begin{align*}
[\Phi^3 (F (\{2\}))](3,2)
=
\int_{\Omega_3}
[F(\{2\})](\omega)
({[\Phi^*]}^3 (\delta_{(3,2)}) )(d \omega)
=
\frac{183}{500}
\end{align*}
{}
\qed
%,
\par
The above may be easy,
but
we should note that
%, \textcolor{black}{(c)}{Remark }{}
%{}
%,
\begin{itemize}
\item[(c)]
the part \textcolor{black}{(a)}
is related to
causality,
and
the part \textcolor{black}{(b)}
is related to
{{measurement}}.
\end{itemize}
%{Remark }.
Thus, the observer
is not in the (a).
%, \textcolor{black}{(a)},
%observer,
%{{{measurement theory}}}, the Copenhagen
%interpretation({Chap.$\;$1}(U$_1$),
%{measuring object}, observer and .
Figuratively speaking,
we say:
$$
\text{
A spectator does not go up to the stage
}
$$
%, {{{measurement theory}}} and {}
Thus,
someone in the (a)
should be regard as
"robot".
%
%BBBBBBBBBBBBBBBBBB%SBSBSBS
\par
\noindent
{\small%%{\footnotesize
\vspace{0.1cm}
\begin{itemize}
\item[$\spadesuit$] \bf {{}}{Note }6.12{{}} \rm
% and ,
%Descartes{} and .
%,
The part (a) is not related to
"probability".
That is because
The spirit of
measurement theory
says that
\begin{itemize}
\item[]
there is no probability without measurements.
\end{itemize}
although
something like "probability"
in the (a)
is called
"Markov probability".
\end{itemize}
}
\subsubsection{linguistic world-view
---
Fit feet to shoes
}%6.6.2I find the shoes fit for feet.
\par
{Ordinary language}
has everything,
i,e.,
monism,
dualism,
tense
and so on.
Also,
in ordinary language,
there is no clear rule
how to use
the terms:
"measurement"
and
"causality".

%is
%a monster  language
%"causality"
%and
%"mea and 
%%
%{}
%,
%monism and dualism and ,
% and  and ,
%.
%, ,
%{{measurement}} and causality 
% and ,
%{{measurement theory}} and {ordinary language}
%.
% and ,
%{ordinary language},
%{{measurement}} and causality  and 
%.
%%
%%dualism and ,
%%,
%%{{measurement}} and causality 
%%
%% and .
%, {Remark }.

\par
\noindent
%\vskip0.3cm
%\vskip0.3cm
%BFBF
\par
\noindent
{\bf Remark 6.26
[Confusion of {{Measurement}} and causality (Continued from
\textcolor{black}{Example 2.7})]}$\;\;$%POPOPO
Recall \textcolor{black}{Example 2.7}
[The measurement of
"cold or hot"
for water].
Consider the
{{measurement}}
${\mathsf M}_{C ( \Omega )} ( {\mathsf O}_{{{{{c}}}}{{{{h}}}}},$
$ S_{[\omega]} )$
where
$\omega=5 (\SD)$.
Then
we say
that
\begin{itemize}
\item[(a)]
By
the {\bf {{measurement}}}
${\mathsf M}_{C ( \Omega )} ( {\mathsf O}_{{{{{c}}}}{{{{h}}}}}, S_{[
\omega(=5)]} )$,
the probability that
a
{\bf measured value }
\\
\\
$x(\in X
=\{{{{{c}}}}, {{{{h}}}}\})$
belongs to
a set
$
\left[\begin{array}{cc}
{}
\emptyset
(={\text {empty set}})
\\
\{ \text{{{{{c}}}}}\}
%[{\mathsf O}]
%
%\text{
{}
\\
\{ \text{{{{{h}}}}} \}
\\
\{ \text{{{{{c}}}}} ,\text{{{{{h}}}}}\}
\end{array}\right]
$
is equal to
\rm
$
\left[\begin{array}{cc}
{}
%[F(\emptyset)]
%(5)=
0
\\
{}
[F(\{ {{{{c}}}} \})]
(5)=1
\\
{}
[F(\{ {{{{h}}}} \})](5)
=0
\\
{}
%[F(\{ {{{{h}}}}, {{{{c}}}} \})]
%(5)=
1
\end{array}\right]
$
{}
\end{itemize}
Here,
we should not think:
\begin{itemize}
\item[]
$\qquad$
"5$\SD$"
is the cause and
"cold"
is a result.
\end{itemize}
That is,
we {\bf never} consider that
\begin{itemize}
\item[(b)]
$\qquad
\qquad
$
$
\underset{\text{{(}cause{)}}}{\fbox{5 $\SD$}}
\longrightarrow
\underset{\text{{(}result{)}}}{\fbox{cold}}
$
\end{itemize}
%, {{{measurement theory}}}{}
The reason is that
\textcolor{black}{Axiom${}_{\text{\scriptsize c}}^{\text{\scriptsize pm}}$ 2}
is not used in (a),
though
the (a) may be sometimes regarded as
the causality
(b)
in ordinary language.
%
%BBBBBBBBBBBBBBBBBB%SBSBSBS
\par
\noindent
{\small%%{\footnotesize
\vspace{0.1cm}
\begin{itemize}
\item[$\spadesuit$] \bf {{}}{Note }6.13{{}} \rm
However,
from the different point of view,
the above (b)
can be justified as follows.
Define the
dual {{causal operator}}
$
{\Phi^*}
:
{\cal M}([0, 100])
\to
{\cal M}(\{{{{{c}}}}, {{{{h}}}}\})$
by
\begin{align*}
&
[{\Phi^*}
\delta_\omega
](D)
=
f_{ {{{{c}}}} }(\omega)
\cdot
\delta_{\text{\scriptsize C}}
(D)
+
f_{{{{{h}}}} }(\omega)
\cdot
\delta_{\text{\scriptsize H}}(D)
\qquad
(\forall \omega \in [0,100],\;\;
\forall D \subseteq \{{{{{c}}}}, {{{{h}}}}\})
\end{align*}
Then,
the (b) can be regarded as "causality".
That is,
\begin{itemize}
\item[$(\sharp)$]
$
\text{
\bf
"measurement or causality"
depends on
how to describe a phenomenon.
}
$
\end{itemize}
This is the
{linguistic world-description method}.
%(\textcolor{black}{{Note }6.12$(\sharp_2)$}).
%(, $F(\{h})(\omega_0)=F(\{t})(\omega_0)=1/2$)
\end{itemize}
}
%%BBBBBBBBBBBBBBBBBB%SBSBSBSSM}M}M}
\par
\noindent

\noindent
\vskip0.2cm
\par
\noindent
%%BFBF
{\bf Remark 6.27
[Confusion of {{mixed measurement}}
and Markov causality
%(Continued from
%Example 4.13(urn problem: mixed {{measurement}}))
]}$\;\;$%POPOPO
Reconsider
\textcolor{black}{Example 4.13}(urn problem:mixed {{measurement}}).
Consider
a
state space $\Omega=\{\omega_1, \omega_2 \}$,
and
define
the
{observable }
${\mathsf O} = ({} \{ {{w}}, {{b}} \}, 2^{\{ {{w}}, {{b}} \}  }  , F{})$
in
$C({}\Omega{})$
by
the formula
\textcolor{black}{(2.5)}.
Define the
{mixed state}
by
$\nu_0 =p \delta_{\omega_1}
+(1-p) \delta_{\omega_2}$.
Then
the probability
that
a measured value
$x$
$({}\in \{ {{w}} , {{b}} \}{})$
is obtained by
the mixed {{measurement}}
${\mathsf M}_{C(\Omega)}({\mathsf O}, S_{[{}\ast{}] }(\nu_0) )$
is,
by
\textcolor{black}{(4.5)},
given
by
\begin{align*}
P({}\{ x \}{})
&=
\int_\Omega
[F({}\{ x \}{})](
\omega)
\nu_0({}d \omega{})
=
p
[F({}\{ x \}{})](\omega_1)
+
(1-p)
[F({}\{ x \}{})](\omega_2)
\\
&=
\cases
0.8 p + 0.4 ({}1-p{})
\quad
&
(\text{when }x={{w}}{}\; )
\\
0.2 p + 0.6 ({}1-p{}))
\quad
&
(\text{when }x={{b}}{}\; )
\endcases
\tag{\color{black}{6.9}}
%%%%%%%%%%%\TAG{4.51}
\end{align*}
Now,
define a new state space $\Omega_0$
by
$\Omega_0=\{\omega_0\}$.
And define the
dual {{Markov causal operator}}
${\Phi^*}: {\cal M}_{+1}(\Omega_0)$
$
\to
{\cal M}_{+1}(\Omega)$
by
%
%$\delta_{(\cdot)}$
%,
${\Phi^*}(\delta_{\omega_0})$
$
=p \delta_{\omega_1}
+(1-p) \delta_{\omega_2}$.
Thus,
we have the
{{Markov causal operator}}
${\Phi}: C(\Omega)$
$
\to
C(\Omega_0)$.
Here,
consider
a pure {{measurement}}
${\mathsf M}_{C(\Omega_0)}(\Phi{\mathsf O}, S_{[\omega_0]})$.
Then,
the probability that
a measured value
$x$
$({}\in \{ {{w}} , {{b}} \}{})$
is obtained by
the measurement
is given by
\begin{align*}
P({}\{ x \}{})
&=
[\Phi (F (\{ x \}))](\omega_0)
=
\int_\Omega
[F({}\{ x \}{})](
\omega)
\nu_0({}d \omega{})
%=
%p
%[F({}\{ x \}{})](\omega_1)
%+
%(1-p)
%[F({}\{ x \}{})](\omega_2)
\\
&=
\cases
0.8 p + 0.4 ({}1-p{})
\quad
&
(\text{when }x={{w}}{}\; )
\\
0.2 p + 0.6 ({}1-p{}))
\quad
&
(\text{when }x={{b}}{}\; )
\endcases
%%%%%%%%%%%\TAG{4.51}
\end{align*}
which is equal to the \textcolor{black}{(6.9)}.
Therefore,
the
mixed {{measurement}}
${\mathsf M}_{C(\Omega)}({\mathsf O}, S_{[{}\ast{}] }(\nu_0) )$
can be regarded as
the pure {{measurement}}
${\mathsf M}_{C(\Omega_0)}(\Phi{\mathsf O}, S_{[\omega_0]})$.
%
%BBBBBBBBBBBBBBBBBB%SBSBSBS
\par
\noindent
{\small%%{\footnotesize
\vspace{0.1cm}
\begin{itemize}
\item[$\spadesuit$] \bf {{}}{Note }6.14{{}} \rm
%\textcolor{black}{Example 4.13}
%(\textcolor{black}{Example 4.14(ii)}
%
%\textcolor{black}{{Note }4.7})
%probability (={mixed state})
%,
%%%%%%POIUYTREWQ
%\textcolor{black}{{Remark }6.27}, probability 
% and {Remark }{}
%%index{${{\cdot}}$@probability ${{\cdot}}$probability }
%That is,
In the above arguments,
we see that
\begin{itemize}
\item[$(\sharp)$]
$
\qquad
\qquad
\text{
\normalsize \baselineskip=18pt
\bf
Concept depends on the description
%how to , 
}
$
\end{itemize}
This is the
{linguistic world-description method}{}.
As mentioned in Note 2.3,
we are not concerned with
the question
"what is $\bigcirc \bigcirc$?".
The reason is due to
the $(\sharp)$.
\end{itemize}
}
%
%BBBBBBBBBBBBBBBBBB%SBSBSBS
\par
\noindent
{\small%%{\footnotesize
\begin{itemize}
\item[$\spadesuit$] \bf {{}}{Note }6.15{{}} \rm
%simultaneous measurement\textcolor{black}{{Definition }2.14}
%, causality 
% and ,
%\textcolor{black}{{Note }6.4}, causality 
%.
%,
As mentioned in
\textcolor{black}{{Note }6.13},
"Measurement or Causality"
depends on
the description.
Some may recall Nietzsche's famous saying:
%(1844--1900):
%%index{@}
\begin{itemize}
\item[]
$\qquad$
{\bf
There are no facts, only interpretations.
}
\end{itemize}
This is just
the {linguistic world-description method}{}
with the spirit:
"Fit feet (=world) to shoes (language)".
%
%\\
%\\
%$\underset{(Chap. 1)}{\text{(X$_1$)}}$
%$\overset{
%}{\underset{\text{(before science)}}{
%\text{
%\fbox
%{{
%\textcircled{\scriptsize 0}
%}
%widely {ordinary language}}
%}
%}
%}
%$
%$
%\underset{\text{\scriptsize }}{\text{$\Longrightarrow$}}
%$
%%\BEGIN{itemize}
%%\item[]
%%$\qquad \qquad$
%$
%\underset{\text{\scriptsize (Chap. 1(O))}}{\text{{world-description}}}
%\cases
%%\textcircled{\scriptsize 1}:
%&
%\!\!\!\!\!\!
%\underset{\scriptsize
%\text{}}{{\textcircled{\scriptsize 1}}{\text{realistic method}}}
%\\
%&
%\text{({world is before language})}
%%\cdots
%%\underset{(POI)}{}\underset{(POI)}{}
%\\
%&
%%\quad
%\text{\footnotesize
%$\underset{(POI)}{feet}\underset{(POI)}{}$
%}
%\\
%\\
%&
%\!\!\!\!\!\!
%%\textcircled{\scriptsize 2}:
%\underset{\scriptsize
%\text{}}{{\textcircled{\scriptsize 2}}{linguistic method}}
%\\
%&
%\text{(language is before world)}
%\\
%&
%%\quad
%\text{\footnotesize
%$
%\underset{(POI)}{}\underset{(POI)}{}$
%}
%\ENDcases
%$
%%\END{itemize}
%\\
%.
%,
%%
%% and  and ,
%%%
%%{linguistic world-description method}{\textcircled{\scriptsize 2}},
%% and  and ,
%%{ordinary language}{\textcircled{\scriptsize 0}},
%%
%%{}
%%, {ordinary language},
%%=
%% and , 
%%.
%(\textcolor{black}{{Note }1.12}) and ,
%\BEGIN{itemize}
%\item[$(\sharp_1)$]
%, , , , ,
%, , , .....
%\END{itemize}
%
%,
%,
% and . ,
% and  and 
%,
%\BEGIN{itemize}
%\item[$(\sharp_2)$]
%{{measurement theory}},
%{Chap.{\;}} and
%,
%${{\cdot}}$
%\END{itemize}
% and .
%,
%({Sec.8.1}(m)).
%%, , ${{\cdot}}$(2.4.2)
%%{}
%%%2.4.2[${{\cdot}}$],
%%%{{measurement theory}}{}
%%%index{@${{\cdot}}$}
%%

\end{itemize}
}
%%BBBBBBBBBBBBBBBBBB%SBSBSBSS
\par

%%%${\omega_r$ or ${\omega_b$

%%%%%%% %P%\tag{6 %P%\tag{4 \end{align*}\end{align*}\end{align*}
%%%

%\END{align*}\tag\tag\bf\bf\bf\bf
%\END{align*}\tag\tag
%\END{align*}%%%%%\TAG
%\END{align*}\tag\tag
%\END{align*}\tag\tag
%\END{align*}\tag\tag
%\END{align*}\tag\tag
%\END{align*}\tag\tag)])])])]\iten\iten\item\item
%\END{align*}\tag\tagg)])])])]\iten\iten\item\item
%\END{align*}%%%%%\TAG
%\END{align*}\tag\tagg)])])])]\iten\iten\item\item
%\END{align*}\tag\tag
\par
%\vskip0.5cm
%\noindent
%\vskip2.0cm
\par
\noindent
%77777777777777777777777777777777777777777777
%%\ssection{Fisher {statistics}\ }%{Chap.{\;}}4{}
\section{Fisher {statistics}\ II
\label{Chap7}
}%{Chap.{\;}}{}
%\chapter[
%Fisher {statistics}({}II)(\textcolor{black}{Axiom${}_{\text{\scriptsize c}}^{\text{\scriptsize p}}$ 1} and 2)
%]{
%Fisher {statistics}({}II)
%\\
%{
%(\textcolor{black}{Axiom${}_{\text{\scriptsize c}}^{\text{\scriptsize p}}$ 1} and 2)
%}}
%
%%\vspace{-0.8cm}
\par
\noindent
{\small%%{\footnotesize
\par
\par
\rm
\pagestyle{headings}
%\markright{6. Fisher {statistics} II (related to \textcolor{black}{Axiom${}_{\text{\scriptsize c}}^{\text{\scriptsize pm}}$ 2})}
\noindent
\begin{itemize}
\item[{}]
{
\small
\baselineskip=15pt
\par%[Abstract].
\rm
%%%%%we see that
$\;\;\;\;$
As mentioned before,
{{{measurement theory}}} is formulted as follows.
%:
That is,
\begin{align*}
%\dashbox{5}
\underset{\text{\scriptsize (scientific language)}}{\text{{} $\fbox{{{{measurement theory}}}}$}}
:=
{
\overset{\text{\scriptsize [Axiom${}_{\text{\scriptsize c}}^{\text{\scriptsize p}}$ 1]}}
{
\underset{\text{\scriptsize
[probabilistic interpretation]}}{\text{{} $\fbox{{{measurement}}}$}}}
}
+
{
\overset{\text{\scriptsize [Axiom${}_{\text{\scriptsize c}}^{\text{\scriptsize pm}}$ 2]}}
{
\underset{\text{\scriptsize [{{the Heisenberg picture}}]}}
{\text{{}$\fbox{ causality }$}}
}
}
%\footnotemark
%\TAG*{$\displaystyle{\mathop{1)}_{(=10))}}$}
%%%%%%%%%%%%%CLAsSICAL
%\TAG{3.0}
\end{align*}
\par
\noindent
In \textcolor{black}{Chap. 5},
we studied
Fisher {statistics}
in
\textcolor{black}{Axiom${}_{\text{\scriptsize c}}^{\text{\scriptsize p}}$ 1}.
In this chapter,

Fisher {statistics} will be discussed in
%{(}, {{regression analysis}}{)}
%{{measurement theory}}
%(
\textcolor{black}{Axiom${}_{\text{\scriptsize c}}^{\text{\scriptsize p}}$ 1} and 2.
}
\end{itemize}
}
\baselineskip=18pt

\font\fottt= cmtt10 scaled \magstep3
\font\fott= cmtt10 scaled \magstep2
\par
\noindent

\baselineskip=18pt
\subsection{{{Measurement
(= the view from the front)}},
Inference$\cdot$Control
(= the view from the back)}%7.1
\par
%,
%{statistics}={dynamical system theory}
%
%{differential equation} and probability  and 
% and 
%
% and 
%(\textcolor{black}{{Sec. 1.1.1}}\textcolor{black}{2}),
%{{inference problem}}{statistics},
%{{control problem}}{dynamical system theory}
% and  and , .
%,
%2
%{}
% and ,
%{statistics}={dynamical system theory}.
%
\subsubsection{{{inference problem}}{(}{statistics}{)}}%{Sec. 7.1.1}
\par
\noindent
{\bf
{{Problem }}7.1
[{}{{}}{{Inference problem}} and {{regression analysis}}]}$\;\;$%POPOPO
\rm
Let $\Omega $
$\equiv$
$\{ \omega_1 , \omega_2 , ... , \omega_{100} \}$
be a set of all students of a certain high school.
Define $h : \Omega \to [{}0, 200{}]$
and
$w : \Omega \to [{}0, 200{}]${}
such that:
\begin{align*}
&
h ({}\omega_n{}) =
\text{ {\lq\lq}the height of a student $\omega_n${\rq\rq} }
\quad
({}n= 1,2,..., 100{})
\\
&
w ({}\omega_n{}) =
\text{ {\lq\lq}the weight of a student $\omega_n${\rq\rq} }
\quad
({}n= 1,2,..., 100{})
\tag{7.1} 
\end{align*}
\par
\noindent
For simplicity,
put,
$N=5$.
For example,
see
Table 
\textcolor{black}{7.1}.
%.
%%\begin{table}[htbp] \small \caption{
%Height and weight
%}
\begin{center}
Table 7.1:
Height and weight
\\
%\BEGIN{tabular}{|l|l|*{2}{@{\quad\$}r|}}
%\BEGIN{tabular}{||c||c|c|c|c|c||l|r}
\begin{tabular}{
@{\vrule width 0.8pt\ }c
@{\vrule width 0.8pt\ }c|c|c|c|c
@{\vrule width 0.8pt }}
\noalign{\hrule height 0.8pt}
Height$\cdot$ Weight
 $\diagdown$ Student & $\omega_1$ & $\omega_2$ &
$\omega_3$ &$\omega_4$ & $\omega_5$
%& $\omega_1$
%& $\omega_2$ & $\omega_1$ & $\omega_2$ & $\omega_1$ & $\omega_2$
\\
\noalign{\hrule height 0.8pt}
Height ($h(\omega)$) & 150 & 160 & 165 & 170 & 175
%& 160  & 155 & 160 & 155 & 160
\\
\hline
%\hline
Weight($w(\omega)$) & 65 & 55 & 75 & 60& 65
%& 60 & 50 & 60& 50 & 60
\\
\noalign{\hrule height 0.8pt}
%  &  10 & 90  \\
%\hline
\end{tabular}
\end{center}
%\end{table}
%\par
%%\noindent
%
%
%
%
%
%
%
%
%
%
%
%\unitlength=0.6mm
%%%%%%%%%%%%%%%%%%
%\BEGIN{FIGUre*}[htbp]
%%%%%%%%%%%%%%%%%%
%%%\vskip-1.0cm
%%\caption{
%%Deterministic map
%%}
%%\END{figure*}
%%%%%%%
%\BEGIN{picture}(200,75)
%%%\put(25,12){A}
%\put(40,16){\scriptsize $\omega$}
%\put(147,11){\circle*{2}}
%\put(40,20){\circle*{2}}
%\put(187,51){\circle*{2}}
%\qbezier(40,20)(100,41)(147,11)
%\put(180,55){$h ({}\omega{})$}
%\qbezier(40,20)(100,81)(187,51)
%\put(144,15){$w ({}\omega{})$}
%\put(0,12){
%\path(107,49)(115,48)(107,45)
%}
%\put(0,12){
%\path(107,49)(115,48)(107,45)
%}
%\put(0,-23){
%\path(107,52)(115,48)(107,45)
%}
%\put(57,63){$\Omega$}
%\put(120,10){\line(1,0){80}}
%\put(120,0){0}
%\put(155,0){100}
%\put(195,0){200}
%\put(120,8){\line(0,1){3}}
%\put(160,8){\line(0,1){3}}
%\put(200,8){\line(0,1){3}}
%\put(0,40){
%\put(120,10){\line(1,0){80}}
%\put(120,0){0}
%\put(155,0){100}
%\put(195,0){200}
%\put(120,8){\line(0,1){3}}
%\put(160,8){\line(0,1){3}}
%\put(200,8){\line(0,1){3}}
%}
%\allinethickness{0.5mm}
%\put(60,30){\oval(70,60)}
%%%\put(160,30){\oval(70,60)}
%%%\put(160,30){\circle{60}}
%\allinethickness{0.3mm}
%%\filltype{white}
%%%\put(157,42){\ellipse{40}{23}}
%\END{picture}
%%%%%%%%%%%%%%%%%%
%%\BEGIN{FIGUre*}[htbp]
%%%%%%%%%%%%%%%%%%
%%\vskip-1.0cm
%\caption{
%Student's height and weight
%}
%\END{figure*}
%%%%%%%
\par
\noindent
%$\Big($Note that this is a special case of
%\textcolor{black}{FIG. (q.20)}.$\Big) \; $
Assume that:
\begin{itemize}
\item[\textcolor{black}{(a}$_1$)]
The principal of this high school knows
the both functions $h$ and $w$.
That is, he knows the exact data of the height and weight
concerning all students.
\end{itemize}
Also, assume that:
\begin{itemize}
\item[\textcolor{black}{(a}$_2$)]
Some day, a certain student helped a drowned girl.
But,
he left without reporting the name.
Thus,
all information that the principal knows is as follows:
\begin{enumerate}
\item[(i)]
he is a student of his high school.
\item[(ii)]
his height [resp. weight{}] is about 170 cm
[resp. about 80 kg{}].
\end{enumerate}
\end{itemize}
Now we have the following question:
\begin{itemize}
\item[\textcolor{black}{(b)}]
Under the above assumption (a$_1$) and (a$_2$),
how does the principal infer who is he?
\end{itemize}
This will be answered in \textcolor{black}{Answer 5.4}.
\par

\subsubsection{{{control problem}}{(}{dynamical system theory}{)}}%{Sec. 7.1.2}
\par
\rm
Adding
{{measurement}}
equation
$g: {\mathbb R}^3 \to {\mathbb R}$
to
{{}}{state equation}\textcolor{black}{(6.3)},
we get
{dynamical system theory}\textcolor{black}{(7.2)}.
That is,
%:
%index{@{dynamical system theory}}
%index{@{state equation}}
%index{@{{measurement}}}
%%%%\fboxsep=5mm
%%%
%\BEGIN{itemize}
%\ITEM[({}B$_1$)]
\begin{align*}
\fbox
{
\roman{
{dynamical system theory}
}
}
=
\cases
{\roman{(i)}}:
\underset{
(\text{initial} \omega(0)=\alpha)}{\frac{d \omega (t)}{dt} =
v({}\omega(t),  t{}, e_1(t), \beta)}
\; & \cdots \text{({ {state equation}})}
%%}
\\
\\
{\roman{(ii)}}:
x(t) = g({}\omega(t), t{}, e_2 (t) ) \;
&  \cdots
\mbox{({}
{{{measurement}}})}
\endcases
%%%%%%{1.17}%%TAG
%%%%%10.12}
%
\tag{\color{black}{7.2}}
\end{align*}
where
%$x_0$
%,
$\alpha, \beta  $
%$\lambda_2$
are parameters,,
$e_1 (t)$
is noise,
$e_2 (t)$ is measurement error.
%\footnote{
%{dynamical system theory},
%$e_1(t)$ and {{measurement}}$e_2(t)$
%,  and .
%, {dynamical system theory} and  and .
%}.
%$u_1(t), u_2(t)${\bf (, {)}}.
%\END{itemize}
\par
%,
%\textcolor{black}{(i)}probability {differential equation},
%{{{measurement theory}}}\textcolor{black}{Axiom${}_{\text{\scriptsize c}}^{\text{\scriptsize pm}}$ 2}.
%\textcolor{black}{(ii)}, {statistics}probability {{{measurement theory}}}\textcolor{black}{Axiom${}_{\text{\scriptsize c}}^{\text{\scriptsize p}}$ 1}
%.
%,
%{dynamical system theory}{{{measurement theory}}}.
\par

\vskip0.3cm
\par
The following examole
is the simplest problem
concerning
inference.

\par
\noindent
{\bf \vskip0.3cm
\vskip0.3cm
%BFBF
\par
\noindent
{{Problem }}7.2
[{}{{}}{{Control problem}} and {{regression analysis}}]}$\;\;$%POPOPO
\rm
We have a rectangular water tank filled with water.
%Consider the phenomenon that a fixed,
%but unknown, quantity of water flows into the tank.
%And, assume that we do not know the quantity of
%water which has filled up the tank.
%
Assume that
the height of water at time $t$ is given
by the following function $h(t)$:
%with unknown parameters
%$\alpha_0$ and $\beta_0$
\begin{align*}
\frac{dh}{dt}=\beta_0,
\text{ then }
h(t) = \alpha_0 + \beta_0 t,
\tag{7.3}
\end{align*}
where
$\alpha_0$
and
$\beta_0$
are unknown fixed parameters
such that
$\alpha_0$ is the height of water filling the tank at the beginning and $\beta_0$ is the increasing height of water per unit time.
The measured height $h_m(t)$ of water at time $t$
is assumed to be represented by
\begin{align*}
h_m(t) = \alpha_0 + \beta_0 t + e(t),
%\tag{5.4}
\end{align*}
where $e(t)$ represents a noise
(or more precisely,
a measurement error) with some suitable conditions.
And assume that
we obtained the measured data of the heights of water
at $t=1,2,3$ as follows:
\begin{align*}
h_m(1)=0.5, \quad h_m(2)=1.6, \quad h_m(3)=3.3.
%\label{1data}
\tag{7.4}
\end{align*}
\par
\noindent
%%%
\unitlength=0.35mm
%%%%%%%%%%%%%%%%%
%\begin{figure*}[htbp]
%%%%%%%%%%%%%%%%%
%%\vskip-1.0cm
%\caption{
%Deterministic map
%}
%\END{figure*}
%%%%%%
\begin{picture}(400,130)
\put(100,0){
\path(130,112)(130,70)
\path(133,112)(133,70)
\path(136,112)(136,70)
\path(139,112)(139,70)
\put(165,30){$h(t)$}
\put(168,25){\vector(0,-1){15}}
\put(168,35){\vector(0,1){22}}
\thicklines
%\path(10,75)(160,75)%%%%y=1
%%\put(110,0){$\Omega$}
%\put(0,75){1}
%\put(0,10){0}
\put(10,10){\line(0,1){100}}%%% y axis
\put(10,10){\line(1,0){150}}%%%% x axis
\path(160,10)(160,110)
%%\put(70,55){$I$}
\path(180,120)(140,120)(140,110)
\path(129,110)(129,130)(180,130)
\multiput(10,10)(0,3){17}{\line(1,0){150}}
}
\end{picture}
%%%%%%%%%%%%%%%%%
%\BEGIN{FIGUre*}[htbp]
%%%%%%%%%%%%%%%%%
%\vskip-1.0cm
\begin{center}{Figure 7.1:
Water tank
}
\end{center}
%%%%%%
%\vskip-0.5cm

\par
\noindent
Under this setting, we consider the following problem:
\begin{itemize}
\item[(c$_1$)]
[Control]:
Settle the state $(\alpha_0, \beta_0)$
such that
measured data
(7.4)
will be obtained.
\end{itemize}
or, equivalently,
\begin{itemize}
\item[(c$_2$)]
[Inference]:
when measured data
(7.4)
is obtained,
infer the unknown state $(\alpha_0, \beta_0)$.
\end{itemize}
This will be answered in \textcolor{black}{Answer 5.4}.
\par
%%%%\hfill{$///$}%%BFBFbfbf
%%\END{Pr}b}
%

%
%
%
%
%
%eeeeeeeeeeeeeeeeeeeeeeeeeeeeeeeeeeeeeeeeeee
%
%
%\par
%, 
%$({\roman c}_1)${{inference problem}}$({\roman c}_2)$ and {}
%%
%\BEGIN{itemize}
%\item[(${\rm c}_2$)]
%%
%[{\bf {{inference problem}}}]:
%$t=1,2,3$
%{{measurement}}
%%% of{{}}s of water
%%at $t=1,2,3$ as follows:
%\BEGIN{align*}
%x(1)=1.9, \quad x(2)=3.0, \quad x(3)=4.7
%%\label{1data}
%%P%\tag{8.7}
%\END{align*}
%%.
%%{{measurement}}
%%%%EQ
%%5)
% and .
% and ,
%$\alpha$
% and
%$\beta$
%inference.
%%
%% of{{}}water  at $t=2$
%\END{itemize}
%,
Note that
(${\rm c}_1$)=(${\rm c}_2$)
from the theoretical point of view.
Thus we consider that
\begin{itemize}
\item[(d)]
{\bf
{{Inference problem}}
and
{{control problem}}
are
the same problem.
And these are characterized as the reverse problem of measurements.
}
\end{itemize}
 and {Remark }(\textcolor{black}{{Sec.4.2.2}(c)}).
%2.
%index{@{{control problem}}}
%index{@{{inference problem}}}

\subsection{{{Regression analysis}}
---
causality{ + }Fisher maximum likelihood method}%7.2
\par
Combining
Axiom${}_{\text{\scriptsize c}}^{\text{\scriptsize pm}}$ 2(causality)
and
Fisher maximum likelihood method(\textcolor{black}{{{Theorem }}4.5}){)},
we can easily prove the following.

\par
\noindent
{\bf
%\vskip0.3cm
%\vskip0.3cm
%BFBF
\par
\noindent
{{Theorem }}7.3
[{{regression analysis}}{\rm{}}
\rm
({\rm cf.}
\textcolor{black}{\cite{IMeas}})
\bf
]}$\;\;$%POPOPO
\rm
%index{@{{regression analysis}}}
Let
$(T{{=}} \{ t_0,t_1,$
$\ldots,$
$ t_N\} , \pi:
T \setminus \{ t_0 \} \to T)$
be semi-ordered tree.
Let
$\widehat{\mathsf O}_{T{}}$
${{=}} (
\bigtimes_{t \in T} X_t, $
${\bigstimes_{t \in T} {\cal F}_t}, $
$
{\widehat F}_{t_0})$
be the realized causal observable
of
a
{{}}sequential observable
$[{}\{ {\mathsf O}_t \}_{ t \in T} ,
\{  \Phi_{\pi(t), t }{}: $
$C({\Omega}_{t}{})\to C({\Omega }_{\pi(t)}{}) \}_{ t \in T\setminus \{t_0\} }$
$]$.
Consider a
{{measurement}}
%an {observable } $\widehat{\mathsf O}_{T{}}$ in
%
%we have{{}} to represent{{}}{{measurement}}
\begin{align*}
{\mathsf M}_{C (\Omega_{t_0})}  ({}\widehat{\mathsf O}_{T{}} {{=}} (
\bigtimes_{t \in T} X_t, {\bigstimes_{t \in T} {\cal F}_t},
{\widehat F}_{t_0}), S_{[\ast]}{})
\qquad
%(\text{{\rm cf.}~{{Theorem }}2.7}).
%%
%P%\tag{8.9}
\end{align*}
Assume that
a measured value by the measurement
belongs to
${\widehat \Xi}
\;(\in
{\bigstimes_{t \in T} {\cal F}_t}
)$.
Then,
there is a reason to infer that
\begin{align*}
[{}\ast{}]
=
{\omega_{t_0}}
%P%\tag{8.10}
\end{align*}
where
{{}}${\omega_{t_0}} \;(\in \Omega_{t_0})$
is defined by
\begin{align*}
[\widehat F_{t_0} (
{\widehat \Xi}
)](\omega_{t_0}) =
\max_{\omega \in \Omega_{t_0}} [\widehat F_{t_0} (
{\widehat \Xi}
){}](\omega)
%P%\tag{8.11}
\end{align*}
\par
\noindent
The poof is a direct consequence of
Axiom${}_{\text{\scriptsize c}}^{\text{\scriptsize pm}}$ 2(causality)
and
Fisher maximum likelihood method(\textcolor{black}{{{Theorem }}4.5}){)}.
Thus, we omit it.
\qed

\par
\rm
\par
\par
\vskip0.3cm
\par
Now we can present the answer to
\textcolor{black}{{{Problem }}7.1}.

%{{measurement theory}}
%That is, {{regression analysis}}(\textcolor{black}{{{Theorem }}7.3})).
%{{regression analysis}} I.
\par
\noindent
%BFBF
\par
\noindent
{\bf
Answer 7.4
[(Continued from {}\textcolor{black}{{{Problem }}7.1}({{inference problem}})){{regression analysis}}]}$\;\;$%POPOPO
%\par
%\par
%\noindent
%%BFBF
%%\BEGIN{Ans} \label{Answer 5.4}
%%\sf
%[Answer to \textcolor{black}{Problem 5.1}].
\rm
For each
$t=1,2$,
let
${\mathsf O}_{G_{\sigma_t}} {{=}} ({\mathbb R}, {\cal B}_{\mathbb R}, G_{\sigma_t})$
be the normal observable
with
a
standard deviation
$\sigma_t >0$
in
$C (\Omega_t)$.
That is,
%({\rm cf.}\textcolor{black}{ 2.17}).:
\begin{align*}
[G_{\sigma_t}(\Xi)] (\omega) = \frac{1}{\sqrt{2 \pi \sigma_t^2}}
\int_{\Xi} e^{- \frac{(x - \omega)^2}{2 \sigma_t^2}} dx
\quad (\forall \Xi \in {\cal B}_{\mathbb R}, \forall \omega \in
\Omega_t
)
%%%%%%3}
%\tag{5.9}
\end{align*}
% $t \in \{ 1,2,3 \}$, $\Omega_t =\Omega$
% and .
%
\par
\noindent
%,
Thus,
we have
a
deterministic sequence observable
%observable
$[ \{{\mathsf O}_{G_{\sigma_t}}\}_{t=1,2}
,
\{
\Phi_{0, t}
:
C(\Omega_t) \to C(\Omega_0)\}_{t=1,2}
]$.
Its realization
$\widehat{\mathbb O}_{T{}}$
${{=}}$
$({\mathbb R}^2, {\cal F}_{{\mathbb R}^2}, {\widehat F}_0)$
is defined by
{\small
\begin{align*}
&
%{\small
%\text{
%$
[{\widehat F}_0({}\Xi_1 \times \Xi_2 {})]
(\omega)
=
[\Phi_{0,1} G_{\sigma_1} ] (\omega)
\cdot
[\Phi_{0,2} G_{\sigma_2}] (\omega)
%\cdot
%[\Phi_{0,3} G_{\sigma}] (\omega)
=
[G_{\sigma_1} ({\Xi_1})] ({}\phi_{0,1}(\omega){})
\cdot
[G_{\sigma_2} ({\Xi_2})] ({}\phi_{0,2}(\omega){})
%$
%}
%}
%\cdot
%[G_{\sigma}({\Xi_3})] ({}\phi_{0,3}(\omega){})
%
%({}\Xi_{1} \times \Xi_{2} \times \Xi_{{3}}{})]
\\
\\
&
\qquad \qquad
\qquad \qquad
({}\forall \Xi_1, \Xi_2 \in
{\cal B}_{\mathbb R}
,
\;
\forall \omega \in \Omega_0
=
\{\omega_1, \omega_2,\ldots,\omega_5\}
)
%P%\TAG{8.15}
\end{align*}
}
%,
%{{}}measurement
%${\mathsf M}_{C (\Omega)}  ($
%$\widehat{\mathbb O}_{T{}}, $
%%${{=}}$
%%$({\mathbb R}^3,$
%%$ {\CAL B}_{{\mathbb R}^3 },$
%%$ {\widehat F}_{{0}}{}),$
%$S_{[\ast]}{}).$
%%%EQ
%.
Let $N$
be sufficiently large.
Define intervals
$\Xi_1, \Xi_2 \subset {\mathbb R}$
by
\begin{align*}
\Xi_1 =\left[{}165 - \frac1{N}, 165 + \frac1{N}\right],
\qquad
\Xi_2 =\left[{}65 - \frac1{N}, 65+ \frac1{N} \right]
%P%\TAG{8.16}
\end{align*}
The measured data
obtained by
a measurement
${\mathsf M}_{C (\Omega_0)}  (\widehat{\mathbb O}_{T{}},
%${{=}}$
%$({\mathbb R}^3,$
%$ {\CAL B}_{{\mathbb R}^3 },$
%$ {\widehat F}_{{0}}{}),$
S_{[\ast]}{})$
is
%\par
\begin{align*}
(165,65) \; (\in {\mathbb R}^2)
%P%\TAG{8.17}
\end{align*}
Thus,
measured value
belongs to
$\Xi_1 \times \Xi_2$.
Using
Regression analysis
(
\textcolor{black}{Theorem 7.3}),
\textcolor{black}{Problem 7.1(b)}
is characterized
as follows:
\begin{itemize}
\item[(${{\sharp}}$)]
Find
$\omega_0$
$(\in \Omega_0)$
such as
$$
[{\widehat F}_0(\{\Xi_1 \times \Xi_2)]
%\Xi_{1}^\delta \times \Xi_{2}^\delta \times \Xi_{{3}}^\delta)]
({}\omega_0
)
=
\max_{\omega \in \Omega }
[{\widehat F}_0(\{\Xi_1 \times \Xi_2)]
%\Xi_{1}^\delta \times \Xi_{2}^\delta \times \Xi_{{3}}^\delta)]
({}\omega
)
$$
%%P%\TAG{8.18}
\end{itemize}
%\par
%\noindent
%\BEGIN{align*}
%\Xi_1 =[{}170 - \frac1{N}, 170 + \frac1{N}],
%\Xi_2 =[{}80 - \frac1{N}, 80+ \frac1{N}],
%\TAG{8.19}
%\END{align*}
% and .
%
%we see, under{{}}assumption that $N$ is
Since $N$ is sufficiently large,
\begin{align*}
%%EQ
({{\sharp}} )
\Longrightarrow
&
\max_{ \omega \in \Omega_0 }
\frac1{{\sqrt{(2 \pi)^2 \sigma_1^2 \sigma_2^2}{}}}
\mathop{
\int\int}_{
\Xi_1 \times \Xi_2  }
%
%[{}6.2 - \frac{1}{N}, 6.2 + \frac{1}{N}] \times
%[{}9.9 - \frac{1}{N}, 9.9 + \frac{1}{N}] \times
%[{}14.1 - \frac{}{N}, 14.1 + \frac{1}{N}]
%}
\exp
{
[
{}- \frac{
({}{}{x_1} - h(\omega){})^2
}{2 \sigma_1^2}
- \frac{
({}{}{x_2} - {}w(\omega){})^2
}{2 \sigma_2^2}
{}]
}
d {}{x_1} d {}{x_2}
%P%\TAG{8.18}
%
%xuu
%
%
%
%
%
%
%
%
%
%
%
%
%[{\widehat F}_{0}(%\{ 6.2 \} \times \{ 9.9\} \times \{14.1 \}
%%\Xi_{1}^\delta \times \Xi_{2}^\delta \times \Xi_{{3}}^\delta
%)]
%({}\alpha, \beta{})
%\quad
%\text{({}for  $N${})}
\\
\Longrightarrow
&
\max_{ \omega \in \Omega_0 }
\exp
{
[
{}- \frac{
({}{}{165} - h(\omega){})^2
}{2 \sigma_1^2}
- \frac{
({}{}{65} - {}w(\omega){})^2
}{2 \sigma_2^2}
{}]
}
\\
%&
%\qquad \qquad \qquad \qquad \qquad \qquad \qquad \qquad
%\qquad \qquad \qquad
%\text{({}mean square min {})}
%\\
\Longrightarrow
&
\min_{ \omega \in \Omega_0 }
{
[
\frac{
({}{}{165} - h(\omega){})^2
}{2 \sigma_1^2}
+ \frac{
({}{}{65} - {}w(\omega){})^2
}{2 \sigma_2^2}
{}]
}
%\\
%&
%\qquad \qquad \qquad \qquad \qquad \qquad \qquad \qquad
\qquad
\text{(
for simplicity,
assume that
$\sigma_1=\sigma_2$)}
\\
\Longrightarrow
&
\text{When $\omega_4$, minimum value }\frac{
({}{}{165} - 170{})^2+({}{}{65} - 60{})^2
}{2 \sigma_1^2}
\text{is obtained}
\\
\Longrightarrow
&
\text{The student is $\omega_4$}
%&
%\BEGIN{cases}
%\\
%%({}1.9 -({}\alpha + \beta{}){}) +
%%({}3.0 - ({}\alpha + 2 \beta{}){}) +
%%({}4.7 - ({}\alpha + 3 \beta{}){}) = 0
%%\\
%%({}1.9 -({}\alpha + \beta{}){}) +
%%2 ({}3.0 - ({}\alpha + 2 \beta{}){}) +
%%3 ({}4.7 - ({}\alpha + 3 \beta{}){}) = 0
%\END{cases}
%\\
%&
%\Rightarrow
%\text{
%\textcolor{black}{}
%}
%&
%P%\TAG{8.19}
\end{align*}
Therefore,
we can infer that
the student who helps the girl
is
$\omega_4$.
\qed
%BFBF
%\END{Ans}

%

\par
%\vskip1.5cm
\par

Next we shall present the answer to
\textcolor{black}{{{Problem }}7.2}.
%{{regression analysis}} I.
\par
\noindent
\vskip0.2cm
%\vskip0.3cm
%BFBF
\par
\noindent
{\bf
Answer 7.5
[(Continued from {}\textcolor{black}{{{Problem }}7.2}({{control problem}})){{regression analysis}}]}$\;\;$%POPOPO

In what follows,
from the measurement theoretical point of view,
we shall answer Problem \textcolor{black}{(7.2)}.
Let
$T=\{0,1,2\}$
be
a series ordered set
such that
the parent map $\pi :T\setminus\{0\} \to T$
is defined by
$\pi ( t ) = t-1$
$\;(t=0,1,2)$.
Put
$\Omega_0
= [0,\; 2] \times [0,\; 2]$,
$\Omega_1 =[0,\; 4] \times [0,\; 2]$,
$\Omega_2 = [0,\; 6] \times [0,\; 2]$.
For
each
$t=1,2$,
consider
a continuous map
$\phi_{\pi(t),t}{}: \Omega_{\pi(t)} \to \Omega_t $
such that
\begin{align*}
&
%\phi_{0,1}(\omega_0  )=
\phi_{0,1}(\alpha , \beta ) = (\alpha + \beta, \beta)
&\;&
(\forall \omega_0 =(\alpha, \beta ) \in \Omega_0 )
\\
&
%\phi_{1,2 }(\omega_1 ) =
\phi_{1,2}(\alpha , \beta ) = ( \alpha + \beta, \beta)
&\;&
(\forall \omega_1 =
(\alpha, \beta )
\in \Omega_1
%\forall t=1,2,3
).
%\\
%&
%\phi_{2,3 }(\omega_2 ) = \omega_2+ \beta
%&\quad&
%(\forall \omega_2 \in \Omega_2
%=[0,\; 6]
%%\forall t=1,2,3
%)
%\tag{27} 
\end{align*}
Then,
we get
the deterministic causal operators
thus,
$\{\Phi_{\pi(t),t}{}: C(\Omega_{t}) \to C(\Omega_{\pi(t)}
)
\}_{t \in \{1,2\}}$
such that
\begin{align*}
(\Phi_{0,1}f_1)(\omega_0) \!=\! f_1( \phi_{0,1}(\omega_0))
&
\quad
\;
(\forall f_1 \in C(\Omega_1),
\forall \omega_0 \in \Omega_0)
\\
(\Phi_{1,2}f_2)(\omega_1) \!=\! f_2( \phi_{1,2}(\omega_1))
&
\quad
\;
(
\forall f_2 \in C(\Omega_2),
\forall \omega_1 \in \Omega_1).
%%\tag{428}
\end{align*}
Thus,
we have the causal relation as follows.
%%%
\begin{align*}
{
{\text{$ {C (\Omega_0)} $}}
}
%{{\footnotesize \Phi_{0, 1 } }\atop{\longleftarrow}}
\mathop{\longleftarrow}^{\Phi_{0, 1 } }
{\text{$ {C (\Omega_{1})} $}}
%{{ \Phi_{ 1, 2 } }\atop{\longleftarrow}}
\mathop{\longleftarrow}^{\Phi_{1,2 } }
{\text{$ {C (\Omega_{2})} $}}.
\end{align*}
Put
$\phi_{0,2}(\omega_0)=\phi_{1,2}(\phi_{0,1}(\omega_0))$,
$\Phi_{0,2}=\Phi_{0,1}\cdot \Phi_{1,2}$.

Let ${\mathbb R}$ be the set of real numbers.
Fix
$\sigma>0$.
For each
$t=0,1,2$,
define
the {\it
normal observable}
${\mathsf O}_{t} {{\equiv}} ({\mathbb R}, {\cal B}_{\mathbb R}, G^n_{\sigma})$
in
$C (\Omega_t)$
such that
%({\rm cf.$\;$}{}{|{}$B!#!{}(B.17}).qe
\begin{align*}
&
[G^n_{\sigma}(\Xi)] (\omega_t ) = \frac{1}{\sqrt{2 \pi \sigma^2}}
\int_{\Xi} \exp({- \frac{(x - \alpha)^2}{2 \sigma^2}}) dx
\\
&
(\forall \Xi \in {\cal B}_{\mathbb R}, \forall \omega_t
=(\alpha, \beta) \in \Omega_t
{{=}}
[{}0, \; 2t+2{}]\times [0, \;2]).
%%%%%%3}
%P%\TAG{8.23}
%%\tag{429}
\end{align*}
Thus,
we get
the
sequential deterministic causal observable
$[{\mathbb O}_{T{}}]$
${{=}}$
$[ \{{\mathsf O}_{t}\}_{t=0,1,2}
,
\{\Phi_{\pi(t),t}{}: C(\Omega_{t}) \to C(\Omega_{\pi(t)}
)
\}_{t=1,2}
]$.
Then,
from Theorem 6.12,
the
realized causal observable
$\widehat{\mathsf O}_{0{}}$
${{\equiv}}$
$({\mathbb R}^3, {\cal B}_{{\mathbb R}^3}, {\widehat F}_0)$
in
${C(\Omega_0{})}$
is
obtained
as follows:
\begin{align*}
&
[{\widehat F}_0({}\Xi_0 \times \Xi_1 \times \Xi_2{})]
(\omega_0)
=
\big[
%\Phi_{0,1}
\big(G^n_{\sigma} ({\Xi_0})
\Phi_{0,1} (G^n_{\sigma} ({\Xi_1})
\Phi_{1,2} (G^n_{\sigma} ({\Xi_2})
))
\big)
\big]
(\omega_0)
\\
=
&
[G^n_{\sigma} ({\Xi_0})] ({}\omega_0{})
\cdot
[G^n_{\sigma} ({\Xi_1})] ({}\phi_{0,1}(\omega_0){})
%\\
%&
%\qquad \qquad
\cdot
[G^n_{\sigma}({\Xi_2})] ({}\phi_{0,2}(\omega_0){})
%
%({}\Xi_{1} \times \Xi_{2} \times \Xi_{{3}}{})]
\\
&
%%%\qquad \qquad
({}\forall \Xi_0, \Xi_1, \Xi_2 \in {\cal B}_{\mathbb R},
\;
\forall \omega_0 =({}\alpha, \beta{}) \in \Omega_0
%= [0, \; 1] \times [0, \; 2]
{}).
%P%\TAG{8.24}
%%\tag{430} 
\end{align*}
We have the measurement
${\mathsf M}_{C({}\Omega_0{})} (\widehat{\mathsf O}_0,$
$
S_{[\ast]}{}\;)$.
We see that
the measured value
$(x_0, x_1, x_2 )$
obtained by
the measurement
${\mathsf M}_{C({}\Omega_0{})} (\widehat{\mathsf O}_0,$
$
S_{[\ast]}{}\;)$
is equal to
\par
\begin{align*}
(0.5, \; 1.6, \; 3.3) \; (\in {\mathbb R}^3).
%%\tag{431}
\end{align*}
\par
\noindent
Define the closed interval
$\Xi_t$
$(t=0,2,3)$
such that
\begin{align*}
\Xi_0 =[{}0.5 - \frac1{2N}, 0.5 + \frac1{2N}],
\;\;
\Xi_1 =[{}1.6 - \frac1{2N}, 1.6 + \frac1{2N}],
\;\;
\Xi_2 =[{}3.3 - \frac1{2N}, 3.3 + \frac1{2N}],
\end{align*}
for sufficiently large $N$.
Here,
Fisher's method
says that
it suffices to
solve the problem.
\begin{itemize}
\item[($\sharp$)]
%{\lq\lq}
Find
$({}\alpha_0, \beta_0{})$
such as
\begin{align*}
\max_{ ({}\alpha, \beta{}) \in \Omega_0 }
[{\widehat F}_0(
\Xi_0
\times
\Xi_1
\times
\Xi_2
]
({}\alpha, \beta{})
%%\tag{432}
\end{align*}
%{\rq\rq}
%\text{ for sufficiently large $N$ }{\lq\lq.
%\hfill{(13)}%%%%\tag{4
\end{itemize}
\par
\noindent
%Putting

Putting
\begin{align*}
&
U(x_0, x_1,x_2, \alpha, \beta)
=
\sum_{k=0}^2
({}{}{x_k} - ({}\alpha + k \beta{}){})^2
\end{align*}
we have the following problem
that is equivalent to
\textcolor{black}{($\sharp$)}:
\begin{itemize}
\item[]
Calculate
\begin{align*}
\frac{\partial}{\partial \alpha}{U(0.5,1.6,3.3, \alpha,\beta) }
%=
=0,
\;\;
\frac{\partial}{\partial \beta}{U(0.5,1.6, 3.3, \alpha,\beta)}
%=
%-\sum_{k=0}^2
%k
%({}{}{x_k} - ({}\alpha + k \beta{}){})
=0,
\end{align*}
%%{\lq\lq}
%Find
%$({}\alpha_0, \beta_0{})$
%such as
%\BEGIN{align*}
%%&
%\min_{ ({}\alpha, \beta{}) \in \Omega_0 }
%\exp \Big(- \frac{U(x_0, x_1,x_2, \alpha, \beta)}{2 \sigma^2}
%\Big)
%\Leftrightarrow
%\max_{ ({}\alpha, \beta{}) \in \Omega_0 }
%U(x_0, x_1,x_2, \alpha, \beta).
%\END{align*}
\end{itemize}
Then, we get
\begin{align*}
%\BEGIN{cases}
%({}0.5 -({}\alpha + \beta{}){}) +
({}\alpha, \beta{}) =
(0.4, 1.4)
%P%\TAG{8.27}
%%\tag{433}
\end{align*}
%Thus, we see, by the statement \textcolor{black}{(O$_2$)}, that
%\BEGIN{itemize}

Therefore,
in order to get the
measured value
$(1.9, \; 3.0, \; 4.7)$,
the control state
$(\alpha, \beta)$
should be defined by
$(0.4, \; 1.4)$.

Here, again note
the equivalence of
%\textcolor{black}{{Sec. 7.1.2}}
%(d)
\textcolor{black}{{{control problem}}$(c_1)$}
and
\textcolor{black}{{{inference problem}}$(c_2)$}.
\qed
\par

\par
%\vskip1.6cm
\par
\par
\noindent
{\bf %NFBFBF
Example 7.6
[Pheasants and rabbits problem
\rm
%({\rm cf.}
%\textcolor{black}{\cite{IElem}})
{\bf]}
}$\;\;$%POPOPO
\rm
Consider the following situation:
\begin{itemize}
\item[(a)]
[Pheasants and rabbits problem]
A number of pheasants and rabbits are placed
together in the same cage. 7 heads and 22 feet are counted.
Find the number$m$ of pheasants and the number $n$ of rabbits.
% and $m$ and
%$n$.
% and 5,
%14{}
\end{itemize}
% and ,  $(m,n)$.
%\END{QA}
\par
\noindent
\par
\noindent
{\bf{Answer}$\;\;$}
This problem
---
Pheasants and rabbits problem
---
has various aspects.
Usually, we consider that,
\begin{itemize}
\item[(b)]
The statement (a) is in ordinary language.
%, {ordinary language}
\end{itemize}
This aspect (b)
may assert that
"5", "14", "$(m,n)$"
should be states.
%
%Answer.
%,
%{world-description},
%monism{world-description}(\textcolor{black}{1.2.1}),
%,
%5, 14,
%$(m,n)${{state}} and 
% and (\textcolor{black}{Example 6.10}).
%Answer.
However,
in what follows,
we regard the problem
(a) as the inference problem.
That is,
\begin{itemize}
\item[(c)]
Regarding $(5,15)$
as an exact measured value,
infer the state $(m,n)$.
\end{itemize}

Put
$\mathbb{N}_0=\{0, 1, 2, \ldots\}$,
$\Omega_0 = \mathbb{N}_0 \times \mathbb{N}_0$, 
$\Omega_1 = \mathbb{N}_0${ and }$\Omega_2 = \mathbb{N}_0$.
Define the
{{causal operator}}
$\Phi_{0, 1}: C (\Omega_1) \to
C (\Omega_0)$
and
$\Phi_{0, 2} : C (\Omega_2)
\to C (\Omega_0)$
such that
\begin{align*}
&
[\Phi_{0, 1} (f_1)] (m,n)
{=}f_1 (m + n),
\quad
[\Phi_{0, 2} (f_2)] (m,n)
= f_2 (2 m+ 4 n)
\\
&
\qquad
(
\forall
f_i \in C (\Omega_i),
i = 1, 2,
\quad
\forall (m,n) \in \Omega_0)
%P%%%\tag{48.28}
\end{align*}
\par
\noindent
%,{.}
%%\BEGIN{align*}
%%%\label{eq:3-04}
%%%\xymatrix@R=5mm@C=20mm
%%%{
%%& \ar[ld]_{\Phi_{0, 1}} C (\Omega_1)\\
%%C (\Omega_0) & \\
%%& \ar[lu]^{\Phi_{0, 2}} C (\Omega_2)
%%%}#############{3.19}
%%%\raisebox{-10mm}{.}
%%
%%%
%%
%\par
%\BEGIN{align*}
%\mbox{
%\ssetlength{\unitlength}{0.7mm}
%\BEGIN{picture}(80,60)(0,20)
%%%\BEGIN{picture}(80,80)(0,20)
%\put(80,50){\makebox(10,10)[r]{${}$}}
%\put(80,70){\makebox(10,10)[r]{${C (\Omega_1)}$}}
%\put(30,45){\makebox(10,10)[r]{${C (\Omega_0)}$}}
%\put(85,15){\makebox(10,10)[r]{${C (\Omega_2)}$}}
%%\put(60,55){\vector(-3,-1){13}}
%\put(60,70){\vector(-3,-2){15}}
%\put(60,22){\vector(-3,2){15}}
%%\put(87,48){\vector(-3,-2){13}}
%%\put(87,22){\vector(-3,2){13}}
%%\put(112,30){\vector(-3,-2){13}}
%%\put(112,7){\vector(-3,2){13}}
%%
%%\put(50,56){$\Phi_{0,2}$}
%%\put(50,43){$\cdots \cdots$}
%%\put(50,38){$\cdots \cdots$}
%\put(47,72){$\Phi_{0,1}$}
%\put(47,20){$\Phi_{0,2}$}
%%\put(73,46){$\Phi_{1,5}$}
%%\put(80,28){$\Phi_{1,2}$}
%%\put(98,28){$\Phi_{2,4}$}
%%\put(97,7){$\Phi_{2,3}$}
%\END{picture}
%}
%\qquad \qquad \qquad \qquad
%\TAG{8.29}
%\END{align*}
%\par
\par
\noindent
For each
$t \in \{1, 2\}$,
consider the
{\it {exact observable} }${\mathsf O}^{\FIN}_t $
in
$C (\Omega_t)$.
That is,  ${\mathsf O}^{\FIN}_t
= (\mathbb{N}_0, 2^{\mathbb{N}_0}, \allowbreak
F^{\FIN})$
satisfies
\begin{align*}
%\label{eq:3-05}
[F^{\FIN}](\Xi)] (n)
=
\cases
1 \quad (n \in \Xi)\\
0 \quad (n \notin \Xi)
\endcases
%P%%%\tag{48.30}
\end{align*}
%,
%\BEGIN{align*}
%[G(\{k\})] (n)
%=
%\cases
%0 \quad & (k {{\; \leqq \;}}n-2 )
%\\
%\alpha \quad & (k =n-1 )
%\\
%%1-2 \alpha \quad & (k =n )
%%\\
%1-2 \alpha \quad & (k =n )
%\\
%\alpha \quad & (k =n+1 )
%\\
%0 \quad &(k {\; \geqq \;}n+1)
%\ENDcases
%\TAG{3.21}
%
Hence,
we get
the sequential deterministic causal
{exact observable}
$[ \{
{\mathsf O}^{\FIN}_t
\}_{t=1,2}
,
\{\Phi_{0,t}{}: C(\Omega_{t}) \to$
$ C(\Omega_{0}
)
$
$
\}_{t \in \{1,2\}}
]$.
Then,
the {{realized causal}} observable
${\widehat {\mathsf O}}_0 = (\mathbb{N}_0 \times \mathbb{N}_0,
2^{\mathbb{N}_0 \times \mathbb{N}_0}, \widehat{F})$
in
$C (\Omega_0)$
is defined by
\begin{align*}
 [\widehat{F} (\Xi_1 \times \Xi_2)] (m,n)
&= [\Phi_{0, 1} F^{\FIN}] (m,n)
\cdot [\Phi_{0, 2} F^{\FIN}] (m,n)
= [F^{\FIN} (\Xi_1)] (m + n)
\cdot [F^{\FIN} (\Xi_2)]
(2 m + 4 n)
\\
&
(\forall \Xi_1, \Xi_2 \in 2^{\mathbb{N}_0},
\forall (m,n) \in \Omega_0
)
%P%%%\tag{48.31}
\end{align*}
\par
\noindent

Assume that
a measured value
\[
 (5, 14) \in \mathbb{N}_0 \times \mathbb{N}_0
\]
is
obtained by
the
{{measurement}}
${\mathsf M}_{C (\Omega_0)}
({\widehat{\mathsf O}}_0, S_{[*]})$.

Therefore,
Fisher maximum likelihood method(\textcolor{black}{{{Theorem }}4.5})
says that
\begin{itemize}
\item[($\sharp$)]
find
$(m, n) (\in \mathbb{N}_0 \times
\mathbb{N}_0)
$
such that
$$
[\widehat{F} (\{5\} \times \{14\})] (m, n)
=
\max_{(m, n) (\in \mathbb{N}_0 \times
\mathbb{N}_0)}[\widehat{F} (\{5\} \times \{14\})] (m, n)
$$
\end{itemize}
{}
Therefore,
\begin{align*}
\text{($\sharp$)}
&
\Longrightarrow
\max_{(m, n) \in \mathbb{N}_0 \times \mathbb{N}_0}
\bigl([F^{\FIN} (\{5\})] (m + n)
\cdot
[F^{\FIN} (\{14\})] (2 m + 4 n)
\bigr)\\
&\Longrightarrow
%\mbox{``}
[F^{\FIN} (\{5\})] (m + n) = 1
%\mbox{''}
\quad
\mbox{and} \quad
%\mbox{``}
[F^{\FIN} (\{14\})] (2 m + 4 n) = 1
%\mbox{''}
\\
&\Longrightarrow
%\mbox{``}
m + n = 5
%\mbox{''}
\quad \mbox{} \quad
%\mbox{``}
2 m + 4 n = 14
%\mbox{''}
\\
&\Longrightarrow
m = 3, n = 2
%P%%%\tag{48.32}
\end{align*}
Thus,
there is a reason to infer that
$(m,n)$
$=$
$(3,2)$.
\qed
\vskip1.0cm
\par
\noindent
{\small%%{\footnotesize
\vspace{0.1cm}
\begin{itemize}
\item[$\spadesuit$] \bf {{}}{Note }7.1{{}} \rm
Since {{measurement theory}}
is based on dualism,
it has
several interpretations of
"Pheasants and rabbits problem"
as follows.
\par
\noindent
%% \vskip-0.3cm%%%
%\begin{table}[htbp] \small \caption{Several interpretations of "Pheasants and rabbits problem"}
\begin{center}
Table 7.2:
Several interpretations of "Pheasants and rabbits problem"
\\
%\BEGIN{tabular}{|l|l|*{2}{@{\quad\$}r|}}
%\BEGIN{tabular}{||c||c|c|c|c|c||l|r}
\begin{tabular}{
@{\vrule width 0.8pt\ }c
@{\vrule width 0.8pt\ }c|c|c
@{\vrule width 0.8pt }}
\noalign{\hrule height 0.8pt}
Variations $\diagdown$ $(m,n,(5,14))$ & $m$ & $n$ &
$(5,14)$
%& $\omega_1$
%& $\omega_2$ & $\omega_1$ & $\omega_2$ & $\omega_1$ & $\omega_2$
\\
\noalign{\hrule height 0.8pt}
$\underset{
\text{(i.e., without mrasurement)}
}{\text{usual interpretation in monism}}$
& {\text{state}} & {\text{state}} & {\text{state}}
%& 160  & 155 & 160 & 155 & 160
\\
\hline
%\hline
$\underset{\quad}{\text{dualism(\textcolor{black}{Example 7.6})}}$& {{state}} & {{state}} & measured value 
%& 60 & 50 & 60& 50 & 60
\\
\hline
%\hline
$\underset{\text{(cf. Example 11.11)}}{\text{dualism}
%({}\textcolor{black}{1}}
%%%8%%
%)
}$
&
measured value  & measured value
& measured value 
%& 60 & 50 & 60& 50 & 60
\\
\noalign{\hrule height 0.8pt}
%  &  10 & 90  \\
%\hline
\end{tabular}
\end{center}
%\end{table}
\par
\noindent
%%\vskip-0.5cmPOIUYTREWQ
Recall \textcolor{black}{Chap. 1(X$_1$)},
that is,
\begin{itemize}
%\item[(X$_1$)]
\item[$\underset{(Chap. 1)}{\text{(X$_1$)}}$]
%$\underset{({Chap.$\;$1})}{\text{(X$_1$)}}$
%{ordinary language} and {world-description method}
%:
%%\text
%%{\textcircled{\scriptsize 0}}
%\\
%%{\textcircled{\scriptsize 0}}
%\\
$
%\quad
%\qquad
\overset{
}{\underset{\text{(before science)}}{
\text{
\fbox
{
{\textcircled{\scriptsize 0}}
widely {ordinary language}}
}
}
}
$
$
\underset{\text{\scriptsize }}{\text{$\Longrightarrow$}}
$
$
\underset{\text{\scriptsize ({Chap.$\;$1}(O))}}{\text{{world-description}}}
\cases
&
\!\!\!\!\!\!
%\textcircled{\scriptsize 1}:
%\underset{\scriptsize
%\text{}}
{\text{\textcircled{\scriptsize 1}realistic scientific language}}
\\
&{\text{(Newtonian mechanics, etc.)}}
\\
\\
%\textcircled{\scriptsize 2}:
&
\!\!\!\!\!\!
%\underset{\scriptsize
%\text{}}
{\text{\textcircled{\scriptsize 2}{linguistic scientific language}}}
\\
&{\text{({{measurement theory}}, etc.)}}
\endcases
$
\end{itemize}
And recall \textcolor{black}{Chap. 1(X$_3$)},
that is,
\begin{itemize}
\item[(X$_3$)]
"\textcircled{\scriptsize 0} widely {ordinary language}"
includes
\begin{itemize}
\item[$(\sharp)$]
the
arithmetical word problems
(
{(\textcolor{black}{Example 7.6})}
or
{\textcolor{black}{(Chap.11 (A$_2$))}}
and so on),
\quad
%\underset{({Chap.{\;}}4, 7, 11{})}
{statistics
(=dynamical system theory)}
\end{itemize}
\end{itemize}
And our standing point is
to reconsider
each in the $(\sharp)$
from the measurement theoretical point of view.
If fact we can see several
"pheasants and rabbits problems
as in \textcolor{black}{Table} 7.2.
\end{itemize}
}
%%BBBBBBBBBBBBBBBBBB%SBSBSBSS

\subsection{{{Measurement theory}}
is valueless if not used  }%7.3
\baselineskip=18pt
\par
Two world-views
(i.e.,
the {realistic world-view}
and
the linguistic world-view)
are composed of
different principles.
That is,
\begin{itemize}
\item[(a)]
The
{realistic world-view}
needs official guarantee of the experts.
The number of specialists may be about 100.
On the other hand,
The
{linguistic world-view}
does not need
official guarantee of the specialists
but
popular support.
\end{itemize}
In other words,
\begin{itemize}
\item[(b)]
The linguistic world-view
is worthless
if
it is used by
many persons.
\end{itemize}
In chapters 4,7,
we assert that
\begin{itemize}
\item[(c)]
{statistics}${{\cdot}}${dynamical system theory}
is characterized as
the abbreviation of
{{measurement theory}}
\end{itemize}
If it is so,
measurement theory
satisfies the condition
(b).

%
%{statistics}${{\cdot}}${dynamical system theory},
%(c),
%(a) and .
%%.
%
%, , {{measurement theory}} and ,
%,
%.
%,
%{{measurement theory}},
%%,
%(c).
%% and
%%.
%%%%{}
%%%,
%%%\BEGIN{itemize}
%%%\item[(d)]
%%%,
%%{{measurement theory}}
%%(=[{{measurement theory}}+quantum {{measurement theory}}]) and ,
%%
%%%{statistics}${{\cdot}}${dynamical system theory}, {{measurement theory}}
%%\END{itemize}
%%.
%%
%%BBBBBBBBBBBBBBBBBB%SBSBSBS\omega\omega_n
%\par
%\noindent
%{\small%%{\footnotesize
%\vspace{0.1cm}
%\BEGIN{itemize}
%\item[$\spadesuit$] \bf {{}}{Note }7.2{{}} \rm
%%Pheasants and rabbits problem, dualism
%% and ,
%,
%{{measurement theory}} and .
% and ,
%{statistics}${{\cdot}}${dynamical system theory}
%{{measurement theory}}.
%,
%{{w}} and {{w}} and
%(That is,  and  and )
% and .
%,
% and , ,
%
%
%{{measurement theory}},
%
%
%.
%,
%, ,
%\BEGIN{itemize}
%\item[$(\sharp)$]
%{differential equation} and probability  and 
%,
% and {{measurement theory}}, )
% and (\textcolor{black}{{Note }6.8}).
%, quantum mechanics,
%.
%\END{itemize}
% and  and .
%,
%${{\cdot}}$.
%%{}
%\END{itemize}
%}
%%%BBBBBBBBBBBBBBBBBB%SBSBSBSS
%\par
%%TAG
%
%

%%\part{{{measurement theory}}(idealism and )}
%\part{ and {{measurement theory}}}
%88888888888888888888888888888888888888888888
%\vskip3.0cm
%\newpage
%88888888888888888888888888888888888888888888
%\ssection{In the inside of the genealogy of dualism idealism }%{Chap.{\;}}{}
\section{Reconsideration of traditional philosophies
in measurement theory
\label{Chap8}
%In the inside of the genealogy of dualism idealism
}%{Chap.{\;}}{}
%%\vspace{-0.8cm}
\baselineskip=18pt\par
\noindent
\begin{itemize}
\item[{}]
{
\small
\baselineskip=15pt
\par%[Abstract].
\rm
$\;\;\;\;$
If we investigate "language" , a philosophical domain must be trodden in somewhat inevitably.
This chapter explains the genealogy of the idealism shown in \textcolor{black}{Fig. 8.2} later:
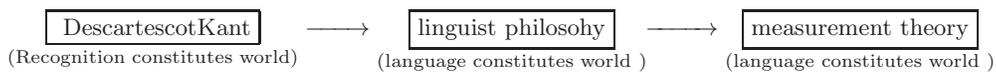
\begin{itemize}
\item[]
$
\overset{}{\underset{\text{\scriptsize (Recognition constitutes world)}}{\fbox{\text{
Descartes$\cot$Kant}}}
}
\xrightarrow[\qquad \text{\bf
\scriptsize }]{\text{\scriptsize
}}
\overset{}{\underset{\text{\scriptsize ({language constitutes world}
)}}{\fbox{\text{{linguist philosohy}}}}}
\!\!
\xrightarrow[\qquad\text{
\scriptsize}]{\text{\scriptsize
}}
\overset{}{\underset{\text{\scriptsize ({language constitutes world}
)}}{\fbox{\text{{{measurement theory}}}}}}
$
\end{itemize}
Although this was the scenery seen from measurement {theory}, I wrote that the reader of science also understood.
Of course, there is no classification of liberal arts and science, and
\begin{itemize}
\item[]
if you do not understand measurement {theory}, I think that you do not idealism.%\footnotemark
\end{itemize}
%\footnotetext{
%As stated in "not studying unless he is scolded" of Section 6.1.1, the kinematic pair of a proposition including time progress is fairly puzzling.
%In this case, it becomes
%$$
%being because measurement theory being known before it, supposing you understand idealism].
%$$
%If it tells repeatedly, it is not "you understand measurement theory is known if you understand idealism."
%}
}
\end{itemize}
\normalsize \baselineskip=18pt
\rm
\par
%\vskip1.5cm
\par
\subsection{Genealogy of dualism idealism }%8.1
\baselineskip=18pt
\subsubsection{Descartes and Kant}%8.1.1
\par
Although measurement {theory} is metaphysics
(i.e.,
a
learning which cannot decide whether right or wrong by experiments
), if you only master measurement theory, it is not necessarily indispensable to understand measurement {theory} in relation with philosophy.

However, the consideration of the relation of measurement theory and philosophy is indispensable
to check a position of measurement theory in the science(i.e., to verify the opinion of this book:
\begin{itemize}
\item[]
Science is describing by measurement theory.
\end{itemize}
).{}
The head family of metaphysics is philosophy and philosophers consider many things truly.
And measurement {theory} is subject to influences of some from their work.
Although the philosopher has not argued in ordinary language and the author could not necessarily fully understand their ideas,
I thought that what was suggested from philosophy should have been written.
\begin{itemize}
\item[(a)]
The spirit of measurement {theory} (Copenhagen interpretation) resembles the philosophy made into the main stream to the extent that it will be called surreptitious use of originality, if I do not write.
\end{itemize}
This chapter explains this.
Now, I will put up the image (\textcolor{black}{Fig. 1.1 of {Chap.$\;$1}}) of measurement again.
\par
\vskip0.5cm

\noindent
%\vskip-0.4cm%%%
%\begin{figure}[htbp]
\noindent
\unitlength=0.7mm
\begin{picture}(200,80)(10,0)
%\put(0,-40)
{{{
\allinethickness{0.5mm}
\drawline[-40](80,0)(80,52)(30,52)(30,0)
\drawline[-40](130,0)(130,52)(170,52)(170,0)
%\drawline[-20](10,-13)(240,-13)(240,80)(10,80)(10,-13)
\path(20,0)(200,0)%y{}$B<J{}(B(100,0)(100,70)(20,70)(20,0)
\put(20,70){\bold{}}
\put(14,-5){
\put(37,50){$\bullet$}
}
\put(100,0){
\put(14,-5){
}
}
\put(50,25){\ellipse{17}{25}}%zV
\put(50,44){\ellipse{10}{13}}%zI
\put(0,34){\put(43,25){\bf \footnotesize{observer}}
\put(42,20){\scriptsize{(I(=mind))}}
}
%\put(48,25){\bf \footnotesize{u{}$B%!!+{}(B}
%\put(42,20){\scriptsize{(measurement{}$B<T{}(B)}}
\put(7,7){\path(46,27)(55,20)(58,20)}
\path(48,13)(47,0)(49,0)(50,13)%yJ
\path(51,13)(52,0)(54,0)(53,13)%yJ
%\put(0.20){
%\put(147,33){\bf \footnotesize {G}
%\put(121,28){\scriptsize (measurementymxzqbrurwr{}$B%%%b{}(B)}
%%\put(140,23){\scriptsize {(}measurementymxz{)}}
%}
\put(0,26){
\put(147,33){\bf \footnotesize system}
\put(148,28){\scriptsize (matter)}
%\put(140,23){\scriptsize {(}measurementymxz{)}}
}
\path(152,0)(152,20)(165,20)(150,50)(135,20)(148,20)(148,0)%{{}$B<X{}(B
%\put(130,55){\bf \footnotesize [$(\sharp_1)$\textcolor{black}{v{}$B<>{}(Bv{}$B~U{}(B*qzr*}2(t{}$B.X{}(Biu{}$B<V{}(Bv{}$B<W{}(B)]}
\put(10,0){
}
\allinethickness{0.2mm}
\put(0,-5){
\put(130,39){\vector(-1,0){60}}
\put(70,43){\vector(1,0){60}}
\put(92,51){\bf \scriptsize [observable]}
\put(58,50){\bf \scriptsize }
\put(57,48){\bf \scriptsize [measured value]}
\put(91,44){\scriptsize \textcircled{\scriptsize a}interfere}
\put(91,33){\scriptsize \textcircled{\scriptsize b}perceive a reaction}
\put(130,48){\bf \scriptsize [state]}
}
\put(-30,0){
%\put(75,60){\bf \footnotesize [\textcolor{black}{v{}$B<>{}(Bv{}$B~U{}(B*qzr*}1{(}measurement)]}
%
}
}}}
\end{picture}
%\hfill{\rm [FN]}\footnotemark
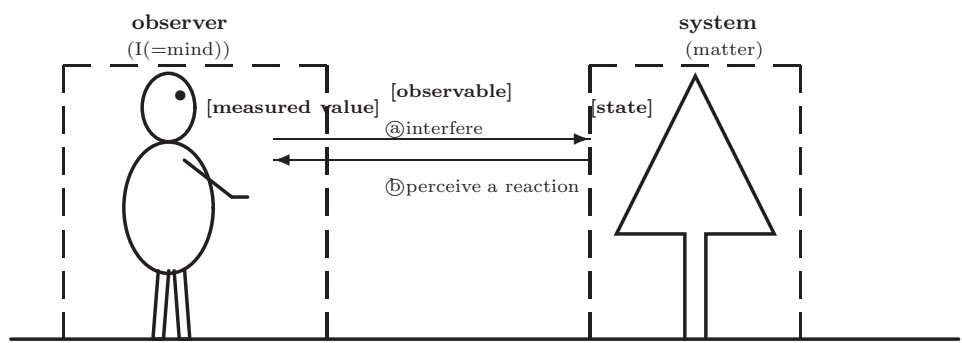
\begin{center}{Figure 8.1 (=\textcolor{black}{Fig.} 1.1):
The image of {\lq\lq}measurement(=\textcircled{a}+\textcircled{b})"
in dualism
}
\end{center}
\par
\noindent

%
%
%
%
%
%\par
%\noindent
%\BEGIN{FIGUre}[htbp]
%\noindent
%\unitlength=0.7mm
%\BEGIN{picture}(200,62)(10,0)
%{{{
%\allinethickness{0.5mm}
%\drawline[-40](70,0)(70,52)(30,52)(30,0)%
%\drawline[-40](115,0)(115,52)(155,52)(155,0)%
%\path(20,0)(160,0)%(100,0)(100,70)(20,70)(20,0)
%\put(20,70){\bold{}}
%\put(14,-5){
%\put(32,50){$\bullet$}
%}%
%\put(100,0){
%\put(14,-5){
%}
%}
%\put(45,25){\ellipse{17}{25}}%
%\put(45,44){\ellipse{10}{13}}%
%\put(2,34){\put(43,25){\bf \footnotesize{}}
%\put(39,20){\scriptsize{(POI)}}
%}
%\put(7,7){\path(41,27)(50,20)(53,20)}%
%\path(43,13)(42,0)(44,0)(45,13)%
%\path(46,13)(47,0)(49,0)(48,13)%
%\put(0,26){
%\put(131,33){\bf \footnotesize }
%\put(117,28){\scriptsize (, )}
%}
%\path(135,0)(135,20)(148,20)(133,50)(118,20)(131,20)(131,0)%
%\put(10,0){
%}
%\allinethickness{0.2mm}
%\put(0,-5){
%\put(112,39){\vector(-1,0){43}}%
%\put(70,43){\vector(1,0){43}}%
%\put(75,51){\bf \scriptsize [${{\cdot}}$]}
%\put(52,50){\bf \scriptsize []}
%\put(51,46){\scriptsize (POI)}
%\put(75,46){\scriptsize \textcircled{\scriptsize a} }
%\put(75,33){\scriptsize \textcircled{\scriptsize b} }
%\put(117,48){\bf \scriptsize []}
%}
%\put(-30,0){
%}
%}}}
%\END{picture}
%\caption{{FIG.$\;$}({Chap.{\;}}1{}{FIG.$\;$}1.1)}
%\END{figure}
%
%\par
%
%

\par
\noindent
Moreover, as shown in \textcolor{black}{Fig.} 8.2, we consider that
\begin{itemize}
\item[(b)]
{
 Measurement {} theory is also the linguistic version of not only quantum mechanics but Descartes-Kant philosophy."}
\end{itemize}
If written by diagram,
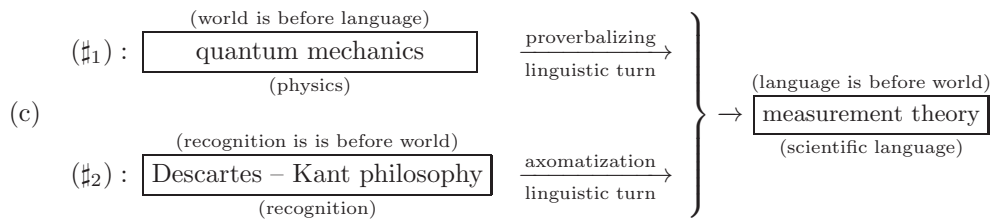
\begin{itemize}
\item[(c)]
$
\left.\begin{array}{ll}
(\sharp_1):
\;
\overset{\text{\scriptsize ({world is before language})}}{\underset{\text{\scriptsize ({physics})}}{\fbox{\;\;\;\;\;\;\text{quantum mechanics}\;\;\;\;\;\;}}}
&
\xrightarrow[\text{\scriptsize linguistic turn
\scriptsize }]{\text{\scriptsize proverbalizing
}}
\\
\\
(\sharp_2):
\;
\overset{\text{\scriptsize ({recognition is is before world})}
}{\underset{\text{\scriptsize (recognition)}}{\fbox{\text{{Descartes -- Kant
philosophy}}}}
}
&
\xrightarrow[\text{\scriptsize
linguistic turn
}]{\text{\scriptsize
axomatization
}}
\end{array}\right\}
\xrightarrow[\text{
\scriptsize }]{\text{\scriptsize
}}
\overset{\text{\scriptsize (language is before world)}}{\underset{\text{\scriptsize (scientific language)}}{\fbox{\text{\small{{{{measurement theory}}}}}}}}
$
\end{itemize}
%where
%the $(\sharp_1)$
%was explained in \textcolor{black}{{Chap. 3}}.
%In this section, we explain the $(\sharp_2)$.
%
%
%
%
%
%
%
%
%
%
%
%
%
%
%\BEGIN{itemize}
%\item[(c)]
%$
%\left.\begin{array}{ll}
%(\sharp_1):
%\;
%\overset{(, )}{\underset{(POI)}{\fbox{\;\;\;\;\;\;\text{}\;\;\;\;\;\;}}}
%&
%\xrightarrow[\text{\scriptsize 
%\sscriptsize }]{\text{\scriptsize
%
%}}
%\\
%\\
%(\sharp_2):
%\;
%\overset{(, )}{\underset{(POI)}{\fbox{\text{\footnotesize{--}}}}
%}
%&
%\xrightarrow[\text{\scriptsize 
%\sscriptsize }]{\text{\scriptsize
%
%}}
%\END{array}\right\}
%\xrightarrow[\text{
%\sscriptsize }]{\text{\scriptsize
%
%}}
%\overset{(, )}{\underset{(POI)}{\fbox{\text{\small{{}}}}}}
%$
%\END{itemize}
\textcolor{black}{Chapter 3} explained the portion of this quantum mechanics of (c)$(\sharp_1)$.
This section explains a lower portion$(\sharp_2)$.
\par
%\vskip1.0cm

\par
If a conclusion is described previously, correspondence of the keyword of Descartes-Kant philosophy and measurement {} theory (Copenhagen interpretation) will become as it is shown in \textcolor{black}{Table 8.1}.
%%%index{@{Chap.{\;}}, {Chap.{\;}}}
%%index{@{Chap.{\;}};{Chap.{\;}}}
%
%
%
%
%
%CCCCCCCCCCCCCCCCCCCCCCCCCCCCC
%,
%Descartes--Kant and {{{measurement theory}}}
%(the Copenhagen interpretation),
%\textcolor{black}{8.1}.
%index{@{Chap.{\;}}, {Chap.{\;}}}
%\begin{table}[htbp] \small \caption{
%Descartes--Kant philosophy and {{{measurement theory}}}
%(cf. \textcolor{black}{\cite{IQphi, ILing}})
%}
\vskip-1.5cm
\begin{center}
Table 8.1:
Descartes--Kant philosophy and {{{measurement theory}}}
(cf. \textcolor{black}{\cite{IQphi, ILing}})
\par
%\vskip-0.5cm
\noindent
\begin{align*}
\begin{array}{@{\vrule width 0.8pt\ }c|c|c@{\vrule width 0.8pt }}
%\BEGIN{tabular}{
%@{\vrule width 0.8pt\ }c
%@{\vrule width 0.8pt\ }c|c|c
%@{\vrule width 0.8pt }}
\noalign{\hrule height 0.8pt}
\underset{\text{\scriptsize (recognition)}}{\text{Descartes--Kant}}
&
\multicolumn{1}{@{}c@{}|}
{\begin{array}{c|c}
\multicolumn{2}{c}{\quad{\textcircled{\scriptsize 1}:\text{I}}\quad}
\\
\hline
\underset{\text{\scriptsize (sense organ, secondary quantity)}}{\quad \textcircled{\scriptsize 2}
:\text{body} \quad}&\underset{\quad \qquad \text{\scriptsize (brain)}\qquad \quad}{
\textcircled{\scriptsize 3}
\text{:mind}}
\end{array}
}
%{||c|}{Descartes} & \multicolumn{2}{c||}{{{{measurement theory}}}}
&
\underset{\text{\scriptsize (
primary quantity{)}}}{
\textcircled{\scriptsize 4}
:\text{matter}}
%\multicolumn{2}{||c|}{Descartes} & \multicolumn{2}{c||}{{{{measurement theory}}}}
\\
\noalign{\hrule height 0.8pt}
\underset{\text{\scriptsize (language)}}{\text{measurement}}
&
\multicolumn{1}{@{}c@{}|}
{\begin{array}{c|c}
\multicolumn{2}{c}{
\textcircled{\scriptsize 1}
:{\text{observer}}\quad}
\\
\hline
%\hline
\quad \quad
\underset{\text{\scriptsize ({measuring instrument})}}{
\textcircled{\scriptsize 2}
:\text{observable} }&\underset{\text{\scriptsize (perceive by brain)}}{
\textcircled{\scriptsize 3}
:\text{measured value } }
\end{array}
}
%{||c|}{Descartes} & \multicolumn{2}{c||}{{{{measurement theory}}}}
&\underset{
\text{\scriptsize
{(}{{state}}=property of {measuring object}{)}}}{
\textcircled{\scriptsize 4}
:\text{measuring object}}
%\multicolumn{2}{||c|}{Descartes} & \multicolumn{2}{c||}{{{{measurement theory}}}}
\\
\noalign{\hrule height 0.8pt}
\end{array}
\end{align*}
\end{center}
%\end{table}
%\par
%\noindent
%{}
%index{@{Chap.{\;}}, {Chap.{\;}}}
\par

%
%
%
%, ,
% and 
%
%
%
%
\par
If you see this table,
you can guess  a correspondence-related meaning generally, but I will add some notes.
\begin{itemize}
\item[\textcircled{\scriptsize 1}]
["I"$\leftrightarrow$"observer"]$\;\;$:I think that this does not need to explain.
\item[\textcircled{\scriptsize 2}]
["body"$\leftrightarrow$"observable"]:
Probably, this will also be good since
Body$\leftrightarrow$Sense organ
$\leftrightarrow$Observable
$\leftrightarrow$
Observable.
%First of all, in the Descartes philosophy, I think that this can be agreed since "consciousness" was made an issue of when the most although the "body" is made an issue of.
%However, the supplement of the "Body$\leftrightarrow$Observable" is written by \textcolor{black}{Note 8.1}.
\item[
\textcircled{\scriptsize 3}
]
["mind"$\leftrightarrow$"measured value"]:$\;\;$This may be unexpected.
However, measurement is perceiving a signal with an observer's brain.
And the "perceived value" is then called "measured value."
For example, if an observer does not look at the value even if the scale of the voltmeter has pointed out 1.5V (namely, if it does not reach to the observer's
brain ), we can not call it measurement.
The phenomenon "the scale of a voltmeter points out 1.5V" is a perfect physical phenomenon, and if it becomes so much, it can be said that the world of monism (only "thing") is enough.
Therefore, if it says in motto, we can say "measured value does not exist without a brain. "{(}\textcolor{black}{{}Chapter 3.3}{)},
and this is "the standard interpretation of quantum mechanics."
%,
%,
%{(}\textcolor{black}{{}3.7}{)}
% and .
I think that dualism is to consider 
that "self {(}Brain${{\cdot}}$Heart{)}" is a special existence.
%, ,
%${{\cdot}}$,
%.
\item[
\textcircled{\scriptsize 4}
]
["matter"$\leftrightarrow$"measuring object"]$\;\;$
This is also natural and it will not be necessary to explain it.
%
%,
%% and 
%%{(} and {)}
%%,
%%.
% and  and ,
% and .
%, {},
%,
%{}, =.
%,
%, {},
% and , 
% and .
%%\textcircled{\scriptsize 1}$\text{--}$\textcircled{\scriptsize 3} and , \textcircled{\scriptsize 4}
%%, .
\end{itemize}
As mentioned above, there is correspondence of \textcolor{black}{Table} 8.1 about the basic keywords of Descartes-Kant philosophy and measurement theory.
Therefore, about the basic keywords, we may 
consider that "Descartes Kantianism = measurement theory."
Of course, since both are dualism, the similarity may be natural one.

\renewcommand{\footnoterule}{
  \vspace{2mm}                      % 
  \noindent\rule{\textwidth}{0.4pt}  
  \vspace{-3mm}
}
%
%BBBBBBBBBBBBBBBBBB%SBSBSBS\omega\omega_n
\par
\noindent
{\small%%{\footnotesize
\vspace{0.1cm}
\begin{itemize}
\item[$\spadesuit$] \bf {{}}Note 8.1{{}} \rm
Since there is the "\textcircled{\scriptsize 2}:body(= Observable)" in the middle of "self" and a "thing" so that it may understand,
if \textcolor{black}{Fig.} 1.1 (image figure of measurement) of {Chap.$\;$1} is seen, it is good though it belongs to the direction of a "thing."
When thinking so, it may be called "mind-body dualism",
but "to which it belonging" and "how to call" are not important.
% because of the reason written to the \textcolor{black}
%{footnote} 6 of {Chap.$\;$1}.
Moreover, if I think "Body$\leftrightarrow$Observable", 
there may be quite big "Body."
I think that it is so much comfortable even if it considers that "glasses" is a part of body.
However, as an example of "the measuring instrument to a measuring object",
if it considers "the scale of the voltmeter to voltage should shake", "the jet stream to an airplane",
"the Polesta r( for checking a direction)" may be sufficient.
Since "Body(=Observable)" is "what exists in the middle of a brain and a measuring object ",
we can consider that a jet stream is the body(=Observable).
However,
\begin{itemize}
\item[]
it should be the same as arguing about "What is a monkey? $\;\;$What is a tree?
" in "Even monkeys fall from trees" of an idiom to have such a discussion
---
namely, "What is the body? " etc.
---
, and we should notice that a productive argument is not expectable.
%%index{@}
\end{itemize}
If too earnest to a question of "What is $\bigcirc \bigcirc$?" as the \textcolor{black}{Note 2.3} described, it will fit into a dead end.
It is because
\begin{itemize}
\item[]
a concept is decided in the context in a linguistic science view.
(\textcolor{black}{Note 6.14})
\end{itemize}
%(Also check the \textcolor{black}{footnote 5} of Chapter 4 again. )
\end{itemize}
}
%%BBBBBBBBBBBBBBBBBB%SBSBSBSS
\par
\noindent

\noindent
%BBBBBBBBBBBBBBBBBB%SBSBSBS\omega\omega_n
\par
\noindent
{\small%%{\footnotesize
\begin{itemize}
\item[$\spadesuit$] \bf {{}}Note 8.2{{}} \rm
Although monism and materialism won a great success in physics.
On the other hand,
in philosophy, they adhered to dualism and idealism.
%index{@}
%index{@}
%\item[] \bf  \rm %%%BBBBBBBBBBBBBBBBBBBB
%\\
%,
If it carries out from the common sense feeling of science, we think that 
the following question is natural.
\begin{itemize}
\item[$(\sharp_1)$]
Why has "strange theories", such as idealism and dualism, clung to many wise people with the talent which is equal to Newton or Einstein?
\end{itemize}
This question will not be canceled if there is no diagram of the following world description in mind.
%index{@{Chap.{\;}}1{}(X$_1$)}
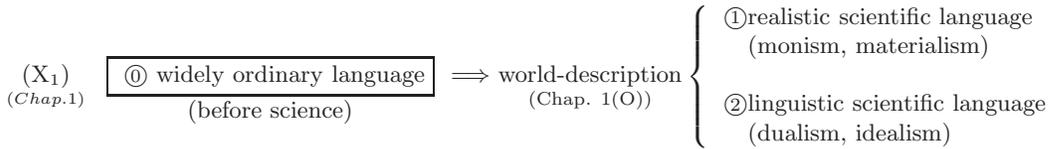
\begin{itemize}
%\item[(X$_1$)]
\item[$\underset{(Chap. 1)}{\text{(X$_1$)}}$]
%$\underset{({Chap.$\;$1})}{\text{(X$_1$)}}$
%{ordinary language} and {world-description method}
%:
%%\text
%%{\textcircled{\scriptsize 0}}
%\\
%%{\textcircled{\scriptsize 0}}
%\\
$
%\quad
%\qquad
\overset{
}{\underset{\text{(before science)}}{
\text{
\fbox
{
{\textcircled{\scriptsize 0}}
widely {ordinary language}}
}
}
}
$
$
\underset{\text{\scriptsize }}{\text{$\Longrightarrow$}}
$
$
\underset{\text{\scriptsize (Chap. 1(O))}}{\text{{world-description}}}
\cases
&
\!\!\!\!\!\!
%\textcircled{\scriptsize 1}:
%\underset{\scriptsize
%\text{}}
{\text{\textcircled{\scriptsize 1}realistic scientific language}}
\\
&{\text{(monism, materialism)}}
\\
\\
%\textcircled{\scriptsize 2}:
&
\!\!\!\!\!\!
%\underset{\scriptsize
%\text{}}
{\text{\textcircled{\scriptsize 2}{linguistic scientific language}}}
\\
&{\text{(dualism, idealism)}}
\endcases
$
\end{itemize}
Supposing philosophy has a difficult portion, it is to
use ambiguous ordinary language{\scriptsize 0}
(Of course,there are also philosophical fields (ethical philosophy etc.) which must be done so plentifully.),
but I would like to think that it is true about the strong will which refused the realistic world view, and sharp intuition.
%\vskip-1.0cm
%.
%
% and .
%{(}{}), 
% and .
%%index{@}
\end{itemize}
}
%%BBBBBBBBBBBBBBBBBB%SBSBSBSS

\baselineskip=18pt
\par

\vskip1.0cm
{\bf Immanuel Kant}(1724--1804) is a philosopher with the biggest influence in modernization,
and advocated what is called "Copernican revolution"
(that is, "recognition constitutes the world." ) in epistemology.
%index{@}
%index{ and @}
%.
% and ,
% and ,
%.
%\par
%{}1{}, .
As general explanation of "pure reason criticism" of Kant
\cite{Kant}, explanation of MSN (Encarta encyclopedia ) is quoted below.
%index{@}
\par
\noindent
{\small
\begin{itemize}
\item[(d)]
{\bf [Pure reason criticism ]}
It is "pure reason criticism" that makes the basis of Kant's critical philosophy,
and the target suited seeing and reaching to an extreme of man's cognitive ability.
As a result, it is clarified that
man's cognitive ability is merely passively struck with things of the world, and it does not only take,
and that it is working actively in the world rather and completes the object of the recognition itself.
Although built, the world is not necessarily completed from nothing like God.
The world is a certain form and there is already it,
and in order to materialize recognition, the information from this world acquired by pushing in feeling is required as a material.
However, this information is only the disorderly confused thing as it is.
Man's cognitive ability must be pushed in a fixed form with which he is originally endowed, and must give orderly order to the information on this confused feeling.
Moreover, it is since the object of the first recognition to unify by it is summarized.
According to Kant, the form with which man is endowed is as follows.
\par
\noindent
%(d)
$
\cases
{\roman (i)}:&
\text{
Form of sensitivity{(}intuition{)}(Space-time
(=${\mathbb R}\times {\mathbb R}^3$))}
\\
{\roman (ii)}:&
\text{Form of understanding{(}thinking{)}
(For example, the concept of a quantity,}
\\
&
\text{whether it is single or a large number, the concept of a relation like causality, etc.)}
\endcases
$
%$\hfill$34)$%tag{
%\END{align*}
\par
\noindent
If that is right, in spite of being unable to prove the proposition "all the thing is among time and space",
and "all follow causal relationship", they will be unconditionally applied to the object of all the experiences experientially.
It is because the object will not be constituted without space, time, and the form of causal relationship.
It is like it being considered that the utterance "the world is green" is right for all human beings,
when all human beings see the world for example, having covered green sunglasses.
(MSN -- {(} the Encarta encyclopedia. 2009 DVD Japanese version(translated by the author))
).
\end{itemize}
}
%%%BBBBBBBBBBBBBBB

\par
Probably, we may consider the following correspondence compared with measurement theory
because Kant has said that it related to space, time, or causal relationship in upper (i) of (d), and the portion of (ii).
\begin{itemize}
\item[(e)]
$
\qquad
\left\{\begin{array}{ll}
{}
\text{Sensitivity}
\\
\text{Understanding}
%[{\mathsf O}]
%
%%\text{
%{}
%\\
%\{ \text{{{{{h}}}}} \}
%\\
%\{ \text{{{{{c}}}}} ,\text{{{{{h}}}}}\}
\end{array}\right\}
$
$
\longleftrightarrow
$
$
\left\{\begin{array}{ll}
{}
%\textcolor{black}
{\text{Axiom${}_{\text{\scriptsize c}}^{\text{\scriptsize p}}$ 1 (measurement)}}
\\
{}
%\textcolor{black}
{\text{Axiom${}_{\text{\scriptsize c}}^{\text{\scriptsize pm}}$ 2 (causality)}}
\end{array}\right\}
$
\end{itemize}
\par
%\vskip1.5cm
\par
\subsubsection{Linguistic revolution and measurement {theory}
---
Idealism which a monkey can not understand}%8.1.2
\par
Now, Descartes-Kant philosophy
% must have advocated "philosophy is the queen of 10,000 study."
develops for the purpose of "theory of basing of science", and, probably, may conclude that the compilation was made by Kant.
Possibly the intention suffered a setback.
%index{@}
However, if we may regard it as
\begin{align*}
\underset{\text{\scriptsize (Kantianism)}}{\text{"Theory of basing of science"}
}
=
\underset{\text{\scriptsize (Measurement {theory} )}}{\text{"The basic language which describes science" }}
%=
\end{align*}
the purpose of Descartes-Kant philosophy and measurement theory will become the same.
\par
However,
\begin{itemize}
\item[(f$_1$)]
Although Descartes-Kant philosophy and measurement theory are dualism with the same purpose and the correspondence
(\textcolor{black}{Table} 8.1 and (e){) mentioned above} is among both, even if both are alike, why have not they resembled it closely?
\end{itemize}
I think that the reason
---
Although it is having stated repeatedly since \textcolor{black}{Note 2.3 }
---
is the next
((f$_2$),(f$_3$)).
\begin{itemize}
\item[(f$_2$)]
Descartes Kant
philosophy
% {(}= -- " -- philosophy " to tell -- {)} which 
investigates in detail about the basic keywords "I" {(} "body" , "mind",
etc.{)}.
\end{itemize}
On the other hand,
\begin{itemize}
\item[(f$_3$)]
Measurement theory tells the world as directions of the \textcolor{black}{Axioms} 1 and 2.
and not investigates the  keywords "observer"
( "observable", "measured value" , etc.  {)}.
%\footnote{
%\textcolor{black}{{}{{{}}}6.7},
%}.
\end{itemize}
That is, I think that it is the difference between "Philosophy told about" and "Philosophy told by".
In this sense, Newtonian mechanics is also "Philosophy told by".

Although this difference is decisive, Descartes-Kant philosophy and measurement {} theory (Copenhagen interpretation) are considerably alike.
For example, there is the next resemblance. :
\begin{itemize}
\item[(g$_1$)]
The importance of "space-time", "causal relationship", and "measurement ($\approx$ recognition)" was observed.
(Therefore, it means that Kant was sure of "the miracle of \textcolor{black}{Section 6.4.3 (g)}".
)
\item[(g$_2$)]
$
\underset{\text{\scriptsize (Copernican turn)}}{\text{[Recognition constitutes world]}}
\Longleftrightarrow
\underset{{\text{\scriptsize
(the Copenhagen interpretation(Chap. 1(${\roman U}_6)$)}}}{\text{[
Observable is before state
]}}$
\end{itemize}
Moreover, it is as follows, if explanation of the Encarta encyclopedia of (d) is imitated and measurement {} theory is described.
\par
\noindent
{\small
\begin{itemize}
\item[(h)]
{\bf [Measurement {} Theory ]}
Measurement {} Theory is the linguistic describing method about an everyday phenomenon.
%,
% and .
%,
The description by measurement theory does not simply describe things of the world as it is passively.
It is working actively in the world rather, completes the object as a fiction and describes it.
Although completed as a fiction, it does not necessarily complete from nothing.
Since the world is a certain form and is already there, in order to materialize description, the information from this world acquired by pushing in feeling is required as a material.
However, this information is only the disorderly confused thing as it is.
The description by measurement theory must be pushed in a fixed form with which measurement theory is originally equipped,
must give orderly order to this confused information, and must summarize the first description(fiction) to unify by it.
The form with which measurement theory is equipped is as follows.
\par
\noindent
%(h)
$
\cases
{\roman (i)}:&
\text{
\textcolor{black}{Axiom${}_{\text{\scriptsize c}}^{\text{\scriptsize p}}$ }1(
Measurement{)}
}
\\
{\roman (ii)}:
&
\text{
\textcolor{black}{Axiom${}_{\text{\scriptsize c}}^{\text{\scriptsize p}}$ }2(Causal relationship{)}
}
\endcases
$
%\hfill$35)$%%%tag{
%\END{align*}
\par
\noindent
If that is right, in spite of being unable to prove "all the thing is among time and space",
and the proposition "all follow causal relationship", they will be unconditionally applied to the object of all the experiences experientially.
Space, time, and causal relationship are because the object will not be described without the form {(}\textcolor{black}Axiom${}_{\text{\scriptsize c}}^{\text{\scriptsize p}}$ 1 and 2 ).
Supposing it has the rule that only the word "green" can be used as a color, it is like what we can only describe "the color in the world is green."
%(, MSN{(})).
\end{itemize}
}
%%%BBBBBBBBBBBBBBB
%,
%(d) and (h),
% and  and
%.

\baselineskip=18pt
\par
From the above thing, we understand the similarity of a Kantianism and measurement theory.

The differences among both are a "recognition version" and a "language version."
Supposing that is right,
you will think that you want the proposition which is unconditionally applied to the object of all the experiences to correspond with measurement theory{(}\textcolor{black}{Axiom${}_{\text{\scriptsize c}}^{\text{\scriptsize p}}$} 1 and 2 )
in spite of {\bf a priori overall judgment} of Kant
(That is, a proposition which is unconditionally applied to the object of all the experiences in spite of the ability not to prove experientially (empirical validation cannot be carried out)).
 I would like to think so, since both aim at establishment of metaphysics.
%
%BBBBBBBBBBBBBBBBBB%SBSBSBS\omega\omega_n
\par
\noindent
{\small%%{\footnotesize
\vspace{0.1cm}
\begin{itemize}
\item[$\spadesuit$] \bf {{}}Note 8.3{{}} \rm
Measurement {theory} is materialized from the following two beliefs(\textcolor{black}{Section 2.3.1 (a)}).
\begin{itemize}
\item[]
%$\qquad$
$
\cases
\text{
Faithful to Axiom 1 and 2
}
\\
\text{
Reliance to man's linguistic competence and cognitive ability
}
%\textcolor{black}{({Chap.{\;}}1{}({}M$_1$))}
\endcases
$
\end{itemize}

In measurement theory, about the portions of "the linguistic competence and cognitive ability of man", we only merely wonder and we do not do investigation beyond it.
However, it may be thought that the direction of "wonder of the linguistic competence and cognitive ability of man" was trodden in in the Kantianism.
Even if the learning which tells "recognition" was inherited in a modern style to science (= material study (Psychology, cognitive science, brain science, artificial intelligence, etc. )),
the direction which Kant aimed at must be metaphysical world description.
As analogy of "atomism (Demokritos) to atomism (theory of elementary particles)",
there may be some some readers who think "the unripe state of science (= material study) is philosophy".
However, if the Kantianism (pure reason criticism) is considered so, Kant does not rest in peace.
Although the metaphysical opinion is carried out, if it is misunderstood and criticized in case of material study, Kant may be embarrassed.
%will not be able to bear, either.
Of course, there is no philosopher who is doing confusion of metaphysics and material study.
%
%, \textcolor{black}{{Chap.{\;}}1{}({}A)}:
%\BEGIN{align*}
%{}, ?
%\BEGIN{align*}
% and 
% and 
%.
%, , 
%,
%{} and ,  and , .
\end{itemize}
}
%%BBBBBBBBBBBBBBBBBB%SBSBSBSS
%

%BBBBBBBBBBBBBBBBBB%SBSBSBS\omega\omega_n
\par
\noindent
{\small%%{\footnotesize
\begin{itemize}
\item[$\spadesuit$] \bf {{}}Note 8.4{{}} \rm
I have heard the opinion said "It is because that Gauss(1777--1855) refrained from the official announcement of non-Euclidean geometry wanted to avoid friction with the Kantists who claim "Space-time ($={\mathbb R}\times {\mathbb R}^3$)
is the sensitivity with which man is endowed." ."
Although this truthfulness is not certain,
I think that it is a fact that the influence of the Kantianism of those days was so greatest that it was not amusing even if there was a talk said like this.
If it considers from now on,
it is
natural to doubt
%ncomprehensible
"why generally it was supported?",
but I would like to make as a fiction the plot in which the Kantianism 
greatly affected modern science, in the evolution
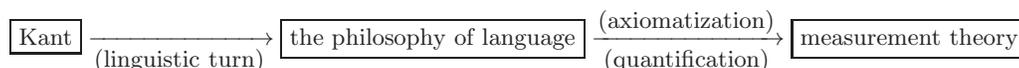
\begin{align*}
\text{\fbox{Kant}}
\xrightarrow[\text{(linguistic turn)}]{}
\text{\fbox{the philosophy of language}}
\xrightarrow[\text{(quantification)}]{\text{(axiomatization)}}
\text{\fbox{measurement theory}}
\end{align*} 
It is because only a negative answer will be contemporarily thought of to a problem :
\begin{itemize}
\item[]
What on earth was the dualism idealism (Plato, Descartes, Kant) made into a philosophical main stream?
\end{itemize}
if our fiction does not exist.
%\textcolor{black}{6.2.1(a)},
% and  and ,
% and .
\end{itemize}
}
%%BBBBBBBBBBBBBBBBBB%SBSBSBSS
\par
\noindent

\vskip1.0cm

%index{ and @}
%index{@}
As mentioned above,
though it is [(d):Pure reason criticism (epistemology) ]$\doteqdot$[(h):Measurement {theory} (language) ],
"[Recognition]$\not=$[Language]" is also worried too.
When becoming it so, "{\bf the linguistic turn}"
---
namely, revolution to "linguistic philosophy" from "epistemology"
---
carried out by philosophers,
such as Saussure(1857--1913) and Wittgenstein(1889-1951), after Kant has a meaning important for measurement theory.
That is,
\begin{align*}
\text{
from "recognition" to "language" }
\end{align*}
If it writes diagrammatically,
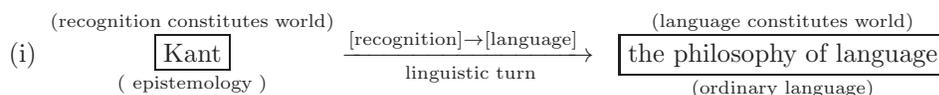
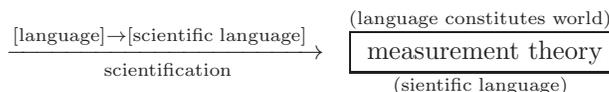
\begin{itemize}
\item[(i)]
$
\overset{\text{\scriptsize (recognition constitutes world)}}{\underset{
\text{\scriptsize (
epistemology
)}}{\fbox{\text{Kant}}}
}
%\END{array}\right]
\xrightarrow[\text{
\scriptsize linguistic turn}]{\text{\scriptsize [recognition]$\rightarrow$[language]
}}
\!\!
\quad
\overset{
\text{\scriptsize (language constitutes world)}
}{\underset{\text{\scriptsize (ordinary language)}}{\fbox{\text{the philosophy of language}}}}
\!\!
$
\\
\\
\\
$\qquad$
$\qquad$
$\qquad$
$
\xrightarrow[\text{
\scriptsize scientification}]{\text{\scriptsize [language]$\rightarrow$[scientific language]
}}
\!\!
\quad
\overset{
\text{\scriptsize (language constitutes world)}
}{\underset{\text{\scriptsize (sientific language)}}{\fbox{\text{
measurement theory}}}}
$
\end{itemize}
%\renewcommand{\footnoterule}{%
%  \vspace{2mm}                      % 
%  \noindent\rule{\textwidth}{0.4pt}   % , 
%  \vspace{-5mm}
%}
%\footnote{
Note that "Recognition constitutes the world. " is used in two meanings((g$_2$),(i)).
Although (g$_2$) is still used in the Copenhagen interpretation of measurement theory({Chap.$\;$1}(U$_5$)),
it was revolved in (i) to "Language constitutes the world."
\par
\vskip1.0cm
\par
That is, the phrases:
%({\rm cf. }\textcolor{black}{{\cite{Hiro, Naga}}}):
\begin{itemize}
\item[]
"Language is before the world"
\quad
"The limits of my language mean the limits of my world"
\quad
"Language constitutes the world"
\quad
"Language game"
\end{itemize}
of the linguistic philosophy which philosophers like and use is borrowed, and it is only a cut about these at the basic spirit of measurement theory.

We are impressed by the sharp sensitivity of the philosophers
who arrived at such a phrase in ordinary language,
without having an easy concrete model called measurement theory
(i.e.,
Axioms 1 and 2).
Even though that was right, the author
was taught the following fact
(j)
from the students of my seminar
and not from Saussure (\textcolor{black}{Refer to Note 3.9}):
\begin{itemize}
\item[(j)]
Even if we do not know a "monkey" and a "tree", we can use the proverb that 
"Even monkeys fall from trees".
\end{itemize}

%index{@}
\vskip1.0cm

%\par
%,
%,
%\BEGIN{itemize}
%\ITEM[$\bullet$]
%{} and  and ,
%
%
%%\END{itemize}
%,
%,
%, 
%
% and .
\par
Ordinary language is a monster language taken in vaguely -even in case of a realistic science view and linguistic scientific view.
In ordinary language, no clear things can be said in a strict meaning.
For example,
"Which came first, the world or the language?
also has the side like "which came first, the chicken or the egg?"
and it is not that "which came first" is so clear.
On the other hand, in measurement theory, it can be said completely that "language is first"
That is because Axioms 1 and 2 are "perfect mystic words",
and it does not exist from the starts, such as the world of corresponding (any fields other than quantum mechanics).

%index{@}
It follows,
\begin{itemize}
\item[(k)]
Measurement theory presented the meaning of idealism
(the spirit of "language is before world") in the form which everyone can understand.
%index{@}
%,
%{},
%
%
\end{itemize}
If that is right, idealism should be known also by whom, but regrettably a monkey does not understand it.
That is because
\begin{itemize}
\item[(l)]
$\qquad
\quad$
measurement theory is based on man's linguistic competence.
\end{itemize}
But a monkey can count some apples,
and there is also computer software comparable as the world champion of chess.
However, it will be thought at least for about 50 years from now on that idealism belongs only to man.
\par
%,
%{}, 
% and ,
% and .
Wittgenstein does not have a worldly way of speaking like "fitting feet with shoes(\textcolor{black}{Note 6.15})."
He said lucidly as follows more smartly.
\begin{itemize}
\item[(m)]
{\bf
Language constitutes the world. }.
Therefore, since we decided that measurement theory described science, what it described is the world of many science and engineering.
(\textcolor{black}{Note 6.15}$(\sharp_2)$).
That is, I understand that all the following (m$_1$)--(m$_3$) is the same meaning.
\begin{itemize}
\item[(m$_1$)]
\bf
Engineering and science are the worlds described by measurement theory.
\rm
\item[(m$_2$)]
\bf
The limit of a language called measurement theory is the limit of the world of engineering and science.
\rm
\item[(m$_3$)]
\bf
Measurement theory = The language of engineering and science
\rm
$\;\;$
(Note 1.2)
% and ,
%
\end{itemize}
\rm
\end{itemize}
I think that there are very beneficial to measurement theory.
If we trust Wittgenstein, it means that he had answered by upper (m) about
\begin{itemize}
\item[(n)]
{\bf What is various science? }
\end{itemize}

\par
%,
%,
% and ,
%, .
%%\footnote{
%%,  and ,
% and  and .
%,
%.
%%,
%%.
%%,
%%
%%,
%%, , 3
%% and .
%%,
%%{} and .
%%%,
%%% and ,
%%%,
%%%(
%%%{Chap.{\;}}11{}1
%%).
%%{} and 
%% and ,
%%, .
%%%index{@}
%}.
%
%BBBBBBBBBBBBBBBBBB%SBSBSBS\omega\omega_n
\par
\noindent
{\small%%{\footnotesize
\vspace{0.1cm}
\begin{itemize}
\item[$\spadesuit$] \bf {{}}Note 8.5{{}} \rm
The above (m) is not "an eternal definition" of "science."
It is because there is a best-before date in measurement theory as Section 2.4.2 [space colony] described.
Moreover, it is because you should propose another linguistic science language and constitute "the world of another science"
, if you are not pleased with this "world of science (measurement theory constitutes)".
%index{@${{\cdot}}$}
%,
%,
%,
%,
%% and .
%%, =(8.1) and ,
%\BEGIN{itemize}
%\item[$(\sharp)$]
%$
%
%\cases
% &\; \cdots \; 
%\\
% &\; \cdots \; 
%\ENDcases
%$
%\END{itemize}
% and .
%%index{ and @}
\end{itemize}
}
%%BBBBBBBBBBBBBBBBBB%SBSBSBSS

%\newpage

\subsection{Where is measurement theory in traditional philosophies}%8.2
\baselineskip=18pt
\par
The argument to this chapter is summarized
as the following
\textcolor{black}{Fig. 8.2}.
%
%it means that the problem \textcolor{black}{(F$_1$)--(F$_5$)} of \textcolor{black}{{Chap.$\;$1}}
%was answered like \textcolor{black}{(F$'_1$)--(F$'_5$)}.
%Therefore, it means that "\textcolor{black}{FIG. 1} of
% a preface" is provisional and we had got  of the following positive type.
%

%---------------------------  and {FIG.$\;$}    ------------------
%BBBBBBBBBBBBBBBBBB%SBSBSBS

%\par
\noindent
%\begin{figure}[htbp]
%%
%%---------------------------  and {FIG.$\;$}    ------------------
%%BBBBBBBBBBBBBBBBBB%SBSBSBS
%{\small%%{\footnotesize
%\vskip0.5cm
%\caption{${{\cdot}}${FIG.$\;$}}
%\END{figure}
%\par
\noindent
%\BEGIN{FIGUre}[htbp]
%%
\unitlength=0.25mm
%\unitlength=0.33mm
{\small
\begin{picture}(500,256)
\put(0,95){$
%\xrightarrow[]{}
\;\;\;\;
%\;\;
%\underset{({world-description})}{} %
\cases
\overset{{{{}}}}{\underset{}{\text{\fbox{Kant}}}}
\xrightarrow[{Sec.8.1}]{\qquad \quad \quad \qquad}
\overset{{{{}}}}{\underset{}{\text{\fbox{{linguistic philosohy}}}}}
\xrightarrow[{Sec.8.1}]{\qquad \quad \qquad \quad \;\;\;\;\;}
%\xrightarrow[  \quad \;\;\;\;\;\;\;]{}
\\
\\
\qquad
\qquad
\quad
%\qquad
\overset{\text{(classical mechanical world-view)}}{\text{
\fbox{{statistics}:({\bf trial})}
}}
\xleftarrow[\;\; \quad \text{\scriptsize abbreviation} \quad ]{}
\!\!\!\!
\overset{\text{\scriptsize (language)}}{\underset{\text{\scriptsize (idealism)}}{\text{\fbox{
{\bf measurement theory}}}}}
\\
\\
\\
\underset{}{\text{\fbox{Newton}}}
\xrightarrow[]{}
\!\!
\cases
\underset{\text{\scriptsize ({the causal world-view})}}{\text{\fbox{{state equation method}(Chap. 1(E$_1$))}}}
\\
\\
\underset{{Sec.3.1}}{\text{\fbox{quantum mechanics}}}
\xrightarrow[]{}
\cases
\underset{\text{\scriptsize (linguistic)}}{(1)}\xrightarrow[\;\;\; \; \;\;{Sec.3.2} \;\;\;\;\;\;]{\;\;
\;\;}
\\
\\
\\
%\\
\underset{\text{\scriptsize (realistic)}}{(2)}
\xrightarrow[\;\;\; \quad \quad]{}
\endcases
\\
\\
\underset{\text{\scriptsize (Einstein)}}{\text{\fbox{{the theory of relativity}}}}
\;\;
\xrightarrow[]{\qquad \quad \quad \quad \;\;\;\;\;\;\;\;\;\;\;}
\endcases
\endcases
$
}
%%%%%%%%%%
\put(420,69){\vector(0,1){72}}%%%%%%
\put(396,246){\vector(0,-1){25}}
\put(190,130){\dashline{3}(0,0)(0,35)}
\put(192,145){\footnotesize ({missing link})}
%%%%%%%%%%%%%%%%%%%%%%%%
%%%%%%%%%%%%%%%%%%%%%%%%%%%%%%
\put(45,-20)
{
{
\put(295,73)
{
$
{
\left.\begin{array}{ll}
{
\left.\begin{array}{ll}
\;\;
\\
\;\;
\\
\;\;
\\
\;\;
\\
\;\;
\\
\;\;
\end{array}\right.
}
\\
\\
\\
\\
{
\left.\begin{array}{ll}
\;\;
\\
\;\;
\\
\;\;
\quad
\end{array}\right\}
}
\xrightarrow[]{}
\!\!\!
\overset{\text{\scriptsize ({physics})}}{\underset{\text{\scriptsize (realism)}}{\text{\fbox{
{the theory of everything}}}}}
\end{array}\right.
}
$}
\put(325,-10)
%\put(315,-5)
{\dashline{3}(10,65)(0,65)(0,-35)(222,-35)(222,65)(200,65)
%{PHYSICS
\put(15,59){\footnotesize [{\bf {realistic world-view}}(unsolved)]}
}
%%%%%%%%%
%%%%%%%%%%%%%%%%%%%%%%%%%%
\put(325,150)
{\dashline{3}(35,88)(0,88)(0,10)(222,10)(222,88)(200,88)
%LINGUISTIC
\put(45,85){\footnotesize [{\bf linguistic world-view}]}
}
%%%%%%%%
}
}
\end{picture}
}
\vskip1.5cm
\begin{center}{Figure 8.2:
The  development of the {world-descriptions}}
%\caption{{world-description}${{\cdot}}${FIG.$\;$}}
\end{center}
\rm
\par
\noindent
\baselineskip=18pt
\normalsize \baselineskip=18pt
Here, although the branch
$\Big($
i.e.,
$
\underset{}{\text{\fbox{quantum mechanics}}}
\xrightarrow[\; {{{}}}\;]{}
\cases
(1)
\xrightarrow[]{}
\\
(2)
\xrightarrow[]{}
\endcases
$
$\Big)$
of quantum mechanics was described in the \textcolor{black}{Note 3.6},
it argues in \textcolor{black}{Section 9.3}.
%
%(\textcolor{black}{{{{}}}9.3}).}
%
%BBBBBBBBBBBBBBBBBB%SBSBSBS
\par
\noindent
{\small%%{\footnotesize
\vspace{0.1cm}
\begin{itemize}
\item[$\spadesuit$] \bf {{}}Note 8.6{{}} \rm
%(i):
The Note 3.9 also described,

The author progressed in the order
\begin{itemize}
\item[]
%$\qquad
%\quad$
from "Quantum mechanics(Theorem 3.4 (Formulation within the quantum mechanics of the uncertainty principle of Heisenberg ))"
to "Classic measurement theory \textcolor{black}{\cite{IFuzz, IQfuz}}"
\end{itemize}
I expected that classic measurement theory became a completely different thing from dynamical system theory and statistics at the beginning.
In this meaning,
 $$
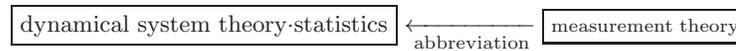
\fbox{\text{dynamical system theory$\cdot$statistics}}
\xleftarrow[\text{\scriptsize abbreviation}]{}
\fbox{\text{\scriptsize measurement theory}}
$$
of \textcolor{black}{Fig.} 8.2 was too appropriate and a disappointment.
Conversely, I got the firm belief "there is only measurement theory for world description."

%{\textcolor{black}{FIG.}$\;$}8.2 and {\textcolor{black}{FIG.}$\;$} and 
%.
\end{itemize}
}
%BBBBBBBBBBBBBBBBBB%SBSBSBSS

%

\subsection{Supplement: About ordinary language }

\baselineskip=18pt
\par

It is how to use the word "ordinary language" that I strayed most while writing this book, and I do not still define it clearly.
Therefore, I am getting confused and using the word "ordinary language" in whole this book.
That is, since I am getting confused and using "Case 1" and "Case 2" like
\begin{itemize}
\item[]
$
\cases
\text{Case 1 : $\;\;$[ordinary language] $\cap$ [measurement theory] = $\emptyset$ }
\\
\text{Case 2 : $\;\;$[ordinary language] $\supset$ [measurement theory] }
\endcases
$
\end{itemize}
, I would like to add some "supplement:ordinary language" here.
However, since it does not necessarily become clear in particular in this section, you may skip this section.

%
%\vskip0.5cm
%,  and ,
%(e)
%(,
%{Chap.{\;}}1{}(X$_1$))
% and ${{\cdot}}$
%% and
%%${{\cdot}}$${{\cdot}}$
%,
%
%, .
%%
%%BBBBBBBBBBBBBBBBBB%SBSBSBS\omega\omega_n
%\par
%\noindent
%%{
%%%\ssmall%%{\footnotesize
%%\vspace{0.1cm}
%%\BEGIN{itemize}
%%\item[$\spadesuit$] \bf {{}}12.2{{}}
%

\rm
Although we do not know in the time of when human beings invented language, language is continuing developing continuously after it.
Therefore, we may consider that "development of language" is almost synonymous( that is, proportional) with "development of civilization."

In this sense, we may think that
%\ssmall
\begin{itemize}
\item[$(\sharp_1)$]
the theme of this book is one of the trials which make power of expression of ordinary language rich.
\end{itemize}
I explain this below.
I persisted in the following diagram in this book.

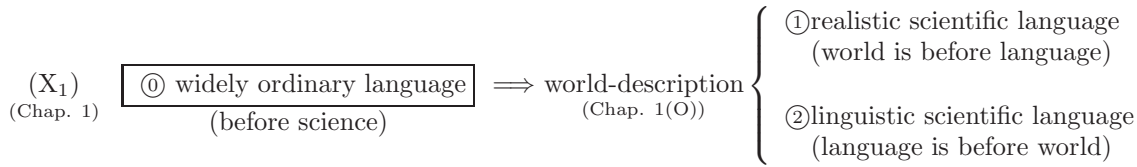
\begin{itemize}
%\item[(X$_1$)]
\item[$\underset{\text{\scriptsize (Chap. 1)}}{\text{(X$_1$)}}$]
%$\underset{({Chap.$\;$1})}{\text{(X$_1$)}}$
%{ordinary language} and {world-description method}
%:
%%\text
%%{\textcircled{\scriptsize 0}}
%\\
%%{\textcircled{\scriptsize 0}}
%\\
$
%\quad
%\qquad
\overset{
}{\underset{\text{(before science)}}{
\text{
\fbox
{
{\textcircled{\scriptsize 0}}
widely {ordinary language}}
}
}
}
$
$
\underset{\text{\scriptsize }}{\text{$\Longrightarrow$}}
$
$
\underset{\text{\scriptsize (Chap. 1(O))}}{\text{{world-description}}}
\cases
&
\!\!\!\!\!\!
%\textcircled{\scriptsize 1}:
%\underset{\scriptsize
%\text{}}
{\text{\textcircled{\scriptsize 1}realistic scientific language}}
\\
&{\text{(world is before language)}}
\\
\\
%\textcircled{\scriptsize 2}:
&
\!\!\!\!\!\!
%\underset{\scriptsize
%\text{}}
{\text{\textcircled{\scriptsize 2}{linguistic scientific language}}}
\\
&{\text{(language is before world)}}
\endcases
$
\end{itemize}

%
%
%
%
%
%
%
%
%%index{@{Chap.{\;}}1{}(X$_1$)}
%\\
%\\
%$\underset{({Chap.{\;}}1{})}{\text{(X$_1$)}}$
%$
%\!\!
%\overset{
%(POI)
%}{\underset{((=))}{
%\text{
%\fbox
%{{\textcircled{\scriptsize 0}}
%}
%}
%}
%}
%$
%$
%\underset{\text{\scriptsize }}{\text{$\Rightarrow$}}
%$
%$
%\underset{\text{\scriptsize ({Chap.{\;}}1{}(O))}}{\text{}}
%\cases
%&
%\!\!\!\!\!\!
%{\text{\textcircled{\scriptsize 1}}}
%\\
%& (, )
%\\
%\\
%&
%\!\!\!\!\!\!
%{\text{\textcircled{\scriptsize 2}}}
%\\
%& (, )
%%{\text{\textcircled{\scriptsize 2}(, )}}
%\ENDcases
%$
%\\
%\\
And since it could not say that the framework of ordinary language\textcircled{\scriptsize 0} in a broad sense was clear,
I chose to start from measurement theory\textcircled{\scriptsize 2}.
However, if it further gives a broad interpretation of ordinary language, it is reasonable also for thinking as follows.
\begin{itemize}
\item[$(\sharp_2)$]
$
\qquad
\qquad
(\textcircled{\scriptsize 0}
\cup
\textcircled{\scriptsize 1}
\cup
\textcircled{\scriptsize 2}
)
\subset
\text{
"more widely ordinary language"
}
$
\end{itemize}
Rather, it is more natural to think like this.

Mathematics and physics may also be "ordinary language" for a mathematician or a physicist.
Moreover, I will not deny the opinion that it is an expedient diagram for this (X$_1$) to make contrast with physics and measurement theory conspicuous.
%(refer to \textcolor{black}{the footnote 6 of {Chap.$\;$1}})
That is, I will not argue about
"Ordinary language of (X$_1$) vs. Ordinary language of $(\sharp_1)$".
It is because avoiding the argument in connection with ordinary language as much as possible must have been the plan which cohered.

If that is right, it is the same as $(\sharp_1)$ to have carried out in this book.
If I repeat,
\begin{itemize}
\item[$(\sharp_3)$]
Into a lawless area called ordinary language, a steadfast small dualism language area called measurement theory was found out (adding),
and power of expression of ordinary language was made somewhat rich.
\end{itemize}
%\textcircled{\scriptsize 3}, \textcircled{\scriptsize 2}
%,
Of course, it is not only measurement theory to make ordinary language rich.
\begin{itemize}
\item[$(\sharp_4)$]
"Mathematics", "physics", "the good", "justice", "love", "freedom", "art", "the theory of evolution", "\text{DNA}","the Internet", "democracy", "economy", "environment"
$\; \cdots$ 
\\
That is, all the things learned in the education of primary schools and junior and senior high schools,
and the language and the concept and theory of the special field of study of the extension also make ordinary language rich.
\end{itemize}
if that is right, we would like to come to compare each of $(\sharp_4)$ and measurement theory

It is possible for a certain grade to carry out this comparison.
As mentioned above, it is because it will be quite fair to consider
\begin{itemize}
\item[$(\sharp_5)$]
all the researches of all learning are estimated by the measure "which made language powerful richly."
\end{itemize}
if "Development of language
$\underset{\text{\tiny proportionality}}{\varpropto}$
Development of civilization" is right

Measurement theory made language rich.
However, in addition to it, the language of
 $$
\text{
Kant, Fischer, Wittgenstein, von Neumann
}
$$
was swung, for example, and language was made still more powerful.

Moreover, I think that this measure$(\sharp_5)$ is almost the same as "the standard of the importance of this book"
-
"Is it how much helpful for construction and settlement of a space colony?"
, namely,
"Is it how much helpful in order that human beings may survive?"
-.
%index{@${{\cdot}}$}
%2.11,
%
%%
%%${{\cdot}}$
%%

\par
\vskip1.2cm
\par

%%%\TAG(A)(B)(C)(D)(E)(F)(G)(H)(I)(J)(K)(L)(M)

%%%\TAG(A)(B)(C)(D)(E)(F)(G)(H)(I)(J)(K)(L)(M)
%%%\TAG(A)(B)(C)(D)(E)(F)n(G)(H)n(I)(J)(K)(L)(M)
\def\RND{\scriptscriptstyle{\roman{RND}}}
%\par
%\noindent
%\part{}
%999999999999999999
%\vskip3.0cm
%\newpage
\section{Equilibrium statistical mechanics
\label{Chap9}
}%{Chap.{\;}}{}
%%\vspace{-0.8cm}%\pi\pi\piXXXXXXXYYYYYYYYYDDDDDDRRRRRR
\baselineskip=18pt\par
\noindent
\begin{itemize}
\item[{}]
{
\small
\baselineskip=15pt
\par%[Abstract].
\rm
$\;\;\;\;$
Our purpose is
to establish the following spirit:
\begin{itemize}
\item[$(\sharp_1)$]
Describing ordinary phenomena by
a metaphysical language
(i.e.,
measurement theory),
we make engineering
(or,
science).
\end{itemize}
Following this spirit,
\begin{itemize}
\item[$(\sharp_2)$]
we study equilibrium statistical mechanics
in measurement theory
\end{itemize}
And therefore,
we conclude that
equilibrium statistical mechanics
is not physics.
In addition,
we assert that
%{{unsolved problem}}{}
%, {}, equilibrium statistical mechanics.
%{Sec.9.3}:
\begin{itemize}
\item[$(\sharp_3)$]
quantum mechanics is
the greatest examples of
the applications of measurement theory.
%?
\end{itemize}
}
\end{itemize}

\subsection{equilibrium statistical mechanics}%9.1
\par
\baselineskip=18pt
It is usual to
consider that
equilibrium statistical mechanics
is constructed on the base of dynamical system theory.
On the other hand,
we construct
equilibrium statistical mechanics
in
measurement theory.

%
%
% and , {},
% and .
%{state equation}
%$$
%PV = nRT
%$$
% and {}
%, $P$, $V$, $n$,
%$R$, $T${}
%%Newtonian mechanics,
%%{},  and  and .
%
%
%\par
%%index{@}
%equilibrium statistical mechanics
%
%
%{dynamical system theory}${{\cdot}}${statistics}(\textcolor{black}{{{{}}}9.2(b)})
%
%,
%
%,
%probability \footnote{probability ,
% and  and ,
% and \textcolor{black}{{{Theorem }}6.21}
% and .
%,
%{},  and , probability 
% and .
%}
%
% and 
%({\rm cf.}
%\textcolor{black}{\cite{Ruel}}),
%
%{{{measurement theory}}}.
%%index{ and @probability }
%%\textcolor{black}{%index{ and @{(}probability {)}}}
%%equilibrium statistical mechanics
%%
%%{}
%,
%\textcolor{black}{3.5},
%That is,
%\BEGIN{itemize}
%\item[]
%{{measurement theory}}
%%
%%(, Heisenberg's {uncertainty principle}\textcolor{black}{(3.9))}
%,
%{{measurement theory}}
%
%\END{itemize}
%.
%
%
%{}, equilibrium statistical mechanics
%{{{measurement theory}}}
%\textcolor{black}{({\rm cf.}
%\cite{Keio, IErgo})}
%,
%,
%\BEGIN{itemize}
%\item[]
Thus, we have the following problem:
\begin{align*}
\underset{\text{\scriptsize (new method due to {{measurement theory}})}}{\fbox{equilibrium statistical mechanics}}
\;\;\;\;
\text{vs.}
\;\;
\underset{
\text{\scriptsize(the conventional method due to {dynamical system theory})}}{\fbox{equilibrium statistical mechanics}}
\end{align*}
%\END{itemize}
\par
\noindent
This
"vs."
must
be settled
in future.
%%%%%%%%%
%\par
%\noindent
\subsubsection{The dynamical aspect of equilibrium statistical mechanics
---
Ergodic hypothesis}%9.1.1
\par

\par
%A state of a system composed of
Assume that about $N ({}{{\approx}} 10^{24}{})$ particles
{(}for example,
hydrogen molecules)
move in a box.
It is natural
to
assume the following
phenomena
\textcircled{\scriptsize 1}
$
\text{--}
$
\textcircled{\scriptsize 4}:
%%%$B0l1~G'$a$h$&!%%%(B
\begin{itemize}
\item[\textcircled{\scriptsize 1}]
Every particle obeys Newtonian mechanics.
\item[\textcircled{\scriptsize 2}]
Every particle moves uniformly in the box.
For example,
a particle does not halt in a corner.
\item[\textcircled{\scriptsize 3}]
Every particle moves with the same statistical behavior
concerning time.
\item[\textcircled{\scriptsize 4}]
The motions of particles are $(${}approximately{}$)$ independent of each other.
\end{itemize}

In what follows
we shall
devote ourselves to the problem:
\begin{itemize}
\item[(a)]
how to describe
the above
equilibrium statistical mechanical
phenomena
\textcircled{\scriptsize 1}
$
\text{--}
$
\textcircled{\scriptsize 4}
in terms of measurement theory.
\end{itemize}
For completeness,
again note that
measurement theory is a kind of language.

\par
In this preprint,
the knowledge of
statistical mechanics
is not
required.
Thus,
we add the allegory
as follows.

%
%BBBBBBBBBBBBBBBBBB%SBSBSBS
\par
\noindent
{\small%%{\footnotesize
\vspace{0.1cm}
\begin{itemize}
\item[$\spadesuit$] \bf {{}}{Note }9.1{{}} \rm
The original idea may be due to
L. Boltzmann:
({\it
Vorlesungen \"uber Gastheorie},
1895,
J Ambrosius Barth, 1923).
Let us explain
\textcircled{\scriptsize 2}
$
\text{--}
$
\textcircled{\scriptsize 4}
as allegory
as follows.
%%%index{@{}}
%\item[] \bf  \rm %%%NOTENOTENOTE
%\\
100 kindergarteners are carrying out a swing [SW], a sliding way [SL], and sand play [SN] to the lunch break of 1 hour in the yard of the kindergarten.
Then,
\textcircled{\scriptsize 2}
$
\text{--}
$
\textcircled{\scriptsize 4}
can be understood as the following allegory:.
\begin{itemize}
%\ITEM[\textcircled{\scriptsize 1}]
%{{Newton}}
%.
\item[\textcircled{\scriptsize 2}]
Every kindergartner is fickle and changes play one after another.
For example,
a kindergartner plays as follows.
\begin{itemize}
\item[
$
({{\sharp}})
$
]
$
\underset{\scriptsize (5\text{ min.})}{\fbox{\text{SW}}}
{ \! \rightarrow \!}
\underset{\scriptsize (3\text{ min.})}{\fbox{\text{SL}}}
{ \! \rightarrow \!}
\underset{\scriptsize (6\text{ min.})}{\fbox{\text{SN}}}
{ \! \rightarrow \!}
\underset{\scriptsize (7\text{ min.})}{\fbox{\text{SL}}}
{ \! \rightarrow \!}
\underset{\scriptsize (9\text{ min.})}{\fbox{\text{SW}}}
{ \! \rightarrow \!}
\underset{\scriptsize (8\text{ min.})}{\fbox{\text{SL}}}
{ \! \rightarrow \!}
\underset{\scriptsize (9\text{ min.})}{\fbox{\text{SW}}}
{ \! \rightarrow \!}
\underset{\scriptsize (6\text{ min.})}{\fbox{\text{SN}}}
{ \! \rightarrow \!}
\underset{\scriptsize (7\text{ min.})}{\fbox{\text{SW}}}
%{ \! \rightarrow \!}
%\underset{\scriptsize (3)}{\fbox{}}
%%\rightarrow
%\underset{\scriptsize (4)}{\fbox{}}
$
\end{itemize}
\item[\textcircled{\scriptsize 3}]
Every kindergartner has the same palatability.
Therefore, the sum total time of each three play is the same.
For example,
for every kindergartner,
we see that
\begin{align*}
\cases
\text{the time which played the swing is}
&
\quad
\text{30 minutes}
\\
\text{the time which played the sliding way is}
&
\quad
\text{18 minutes}
\\
\text{the time which played the sand play is}
&
\quad
\text{12 minutes}
\\
\endcases
\end{align*}
{}
%That is,
% and 
%.
\item[\textcircled{\scriptsize 4}]
Every kindergartner is independently
playing
% the spirit of "being independence and self-respect mostly."
It is hardly influenced by other kindergarteners' play.
They do not do group action.
\end{itemize}
Imaging
the above
\textcircled{\scriptsize 2}--\textcircled{\scriptsize 4},
readers may read as follows.
\end{itemize}
}
%%BBBBBBBBBBBBBBBBBB%SBSBSBSS
\par
\noindent

\vskip0.3cm%%%%%%
%%%%%%%%
\par
\noindent
{\bf
%\large
About \textcircled{\scriptsize 1}}
\par
\noindent
\par
In Newtonian mechanics, any state of a system composed of
$N ({}\approx 10^{24}{})$ particles is represented by a point
$({}q , p{})$
$\bigl(\equiv$
(position, momentum)
$=
$
$({} q_{1n}, q_{2n}, q_{3n} ,$
$  p_{1n} , p_{2n} ,$
$ p_{3n}{})_{n=1}^N $
$\bigl)$
in
a phase
({}or state{}) space
${\mathbb R}^{6N}$.
%({\it cf.} the formula {}{(2.}8)).
Let ${\cal H}: {\mathbb R}^{6N} \to {\mathbb R}$
be a Hamiltonian
such that
\begin{align*}
&
{\cal H}
\big(
({} q_{1n}, q_{2n}, q_{3n} ,
p_{1n} , p_{2n} ,
p_{3n}{})_{n=1}^N
\big)
%=
%\text{monentum energy}+\text{potential enagy}
%\\
=
%&
[\sum\limits_{n=1}^N
\sum\limits_{k=1,2,3}
\frac{(p_{kn})^2}{2 \times \text{particle's mass}}
]
%\\
%&
%\qquad
\! + \!
U(
({} q_{1n}, q_{2n}, q_{3n} )_{n=1}^N
).
\tag{2}
\end{align*}
%a posi
Let
$\{ {{{}} \psi}^{{}_E}_t \}_{ - \infty  < t < \infty }$
be the flow
on
the energy surface
${{\Omega}}_{{}_E}$
%(%$=$
%$\{ ({}q, p{}) \in {\mathbb R}^{6N}{}:
%{\cal H}({}q,p{}) = E \}$
%)
induced by
the
Newton equation
with the Hamiltonian
${\cal H}$,
%index{Hamiltonian@Hamiltonian}
or equivalently,
Hamilton's equation:
\begin{align*}
\frac{dq_{kn}}{dt}=
\frac{\partial {\cal H}}{\partial p_{kn}},
\quad
\frac{dp_{kn}}{dt}=-
\frac{\partial {\cal H}}{\partial q_{kn}},
\;\;
(k=1,2,3, \;\; n=1,2,\ldots,N).
\tag{\color{black}{9.1}}
%{\color{black}{9.1}}%POIUYTREWQ
\end{align*}

Fix $E>0$.
And define
the measure
$\nu_{{}_E} $
on
the energy surface
${{\Omega}}_{{}_E}$
($\equiv$
$\{ ({}q, p{}) \in {\mathbb R}^{6N}{}
\; | \;
{\cal H}({}q,p{}) = E \}$)
such that
\begin{align*}
%&
{{ \nu}_{{}_E} }({}B)
=
\int_B | \nabla {\cal H}({}q,p{}) |^{-1} d m_{6N-1}
%\tag{3}
%\\
%&
({}\forall B \in {\cal B}_{{{{\Omega}}_{{}_E}}},
\text{ the Borel field of }{{\Omega}}_{{}_E}
)
%\footnotemark
\end{align*}
where
$$
| \nabla {\cal H}({}q,p{}) |=
[\sum\limits_{n=1}^N
\sum\limits_{k=1,2,3}
\{
(\frac{\partial {\cal H}}{\partial p_{kn}})^2
+
(\frac{\partial {\cal H}}{\partial q_{kn}})^2
\}]^{1/2}
$$
and
$d m_{6N-1}$
is
the usual surface measure on
${{\Omega}}_{{}_E} $.
%index{flow@flow}

Liouville's theorem
%({}{{\it cf.}\cite{Kubo}})
says that
the measure
${ \nu}_{{}_E}$
is
invariant concerning
the flow
$\{ {{{}} \psi}^{{}_E}_t \}_{ - \infty  < t < \infty }$.
That is, it holds that

\begin{align*}
{\nu}_{{}_E}   (S{}) = 
{\nu}_{{}_E}   ({}\psi^{{}_E}_t (S{}){})
\qquad
({}0 {{\; \leqq \;}} \forall t < \infty,
\quad
\forall S
\in {\cal B}_{ {{\Omega}}_{{}_E} }
)
\tag{\color{black}{9.2}}
\end{align*}

\par
Putting
${C(\Omega)}=C_0(\Omega_{{}_E})$
$=C(\Omega_{{}_E})$
(from the compactness of $\Omega_{{}_E}$),
$T={\mathbb R}$,
$\omega_t =
(q(t),p(t))$,
$\phi_{t_1.t_2} = \psi_{t_2 -t_1}^E$,
$\Phi^*_{t_1. t_2 } \delta_{\omega_{t_1}}
=
\delta_{\phi_{t_1. t_2} (\omega_{t_1} )}
$
$(\forall \omega_{t_1} \in \Omega_{{}_E})$,
we define
the deterministic Markov relation
$\{\Phi_{t_1, t_2 }: C(\Omega_{{}_E}) \to C(\Omega_{{}_E})
\}_{(t_1.t_2) \in T_{\le}^2}$
in Axiom${}_{\text{\scriptsize c}}^{\text{\scriptsize pm}}$ 2.

\par
\vskip0.3cm
\par
\noindent
{\bf
%\large
About  \textcircled{\scriptsize 2}}

Now let us begin with
the well-known ergodic theorem.
({}{{\it cf.} ${{{}}}$\cite{Yosi}}).

For
example,
consider one particle
$P_1$.
Put
$
S_{P_1}=\{\omega \in \Omega_{{}_E}
\;|\;
$
a state $\omega$
such that
the particle $P_1$ always stays a corner of the box
$\}$.
Clearly, it holds that
$S_{P_1} \subsetneq \Omega_{{}_E}$.
Also,
if
$\psi^{{}_E}_t (
S_{P_1}
) \subseteq
S_{P_1}
$
$(0 {{\; \leqq \;}}\forall t < \infty )$,
then
the particle $P_1$
must always stay a corner.
This contradicts
\textcircled{\scriptsize 2}.
Therefore,
\textcircled{\scriptsize 2}
means the following:
\par
\begin{itemize}
\item[
\textcircled{\scriptsize 2}$'$
]
[Ergodic property]:
If
a compact set
$
S(\subseteq \Omega_{{}_E},
S \not= \emptyset )$
satisfies
$\psi^{{}_E}_t (S) \subseteq S$
$(0 {{\; \leqq \;}}\forall t < \infty )$,
then
it holds that
$S=\Omega_{{}_E}$.
%$\frac{ { \nu}_{{}_E} }{ { \nu}_{{}_E} ({}{{\Omega}}_{{}_E}{}) }$
\end{itemize}
%\END{itemize}
%\Index{normalized invariant measure@normalized invariant measure}
%\Index{invariant state@invariant state}
\par
\noindent
The ergodic theorem
%{}{({\it cf.} \cite{Kren})}
says that
the above
\textcircled{\scriptsize 2}$'$
is equivalent to the following equality:
\begin{align*}
&
\displaystyle{
{
%%%%
\mathop
{\int_{\Omega_{{}_E}} f( \omega ) {\overline \nu}_{{}_E}   ({}d \omega )}_{
{\text{\scriptsize ((state) space average)}}
%}
%}
}
%%%%
}
}
=
\displaystyle{
{
%%%%
\mathop
{
\lim_{ T \to \infty}
\frac{1}{T}
\int_\alpha^{\alpha+T} f( \psi^{{}_E}_{t} (\omega_0) )
dt
%%\int_\Omega f( \omega ) {\overline \nu}_{{}_E}   ({}d \omega )
}_{
{\text{\scriptsize (time average)}}
%}
%}
}
%%%%
}
}
\\
&
\qquad
(\forall \alpha \in {\mathbb R},
\forall f \in C( {{\Omega}}_{{}_E} ), \quad \forall 
\omega_0 \in {{\Omega}}_{{}_E} )
%%P%\TAG{9.2}
\tag{9.3}
\end{align*}
where the normalized measure
$ {\overline \nu}_{{}_E}$
is defined by
such that
${\overline \nu}_{{}_E}$
$=$
$\frac{ { \nu}_{{}_E} }{ { \nu}_{{}_E} ({}{{\Omega}}_{{}_E}{}) }$.

%we have the
%normalized measure space
%$({}{{\Omega}}_{{}_E} , {\cal B }_{{{\Omega}}_{{}_E} } ,
%{\overline \nu}_{{}_E}  {})$.

After all,
the ergodic property says that
if
$T$
is sufficiently large,
it holds that
\begin{align*}
\displaystyle{
{
%%%%
\mathop
{\int_{\Omega_{{}_E}} f( \omega ) {\overline \nu}_{{}_E}   ({}d \omega )}_{
{\text{\scriptsize }}
%}
%}
}
%%%%
}
}
{{\approx \;}}
\displaystyle{
{
%%%%
\mathop
{
%\lim_{ T \to \infty}
\frac{1}{T}
\int_\alpha^{\alpha +T} f( \psi^{{}_E}_{t} (\omega_0) )
dt
%%\int_\Omega f( \omega ) {\overline \nu}_{{}_E}   ({}d \omega )
}_{
{\text{\scriptsize }}
%}
%}
}
%%%%
}
}.
%\tag{5}
\end{align*}
\par
%mmmmmmmmmmm_{{}

Put
${\overline m}_{{}_T}(dt) = \frac{dt}{T}$.
The probability space
$([\alpha, \alpha+ T], {\cal B}_{[\alpha, \alpha + T]}, {\overline m}_{{}_T})$
(or equivalently,
$([0,T], {\cal B}_{[0,T]}, {\overline m}_{{}_T})$
)
is called
a (normalized)
{\bf first staying time space},
also,
the probability space
$({}{{\Omega}}_{{}_E} , {\cal B}_{ {{\Omega}}_{{}_E} }   , {\overline \nu}_{{}_E}  {})$
is called
a
(normalized){\bf second staying time space}.
Note that
these mathematical probability spaces
are not related to
{\lq\lq}probability".
%({\it cf.}
%Section 3.2).

%BFBF
\par
\noindent
%\it
%Remark 1.
%\rm
%[About \textcircled{\scriptsize 2}$'$].
%In \cite{Ishi2, Ishi3},
%we started form the mathematical statement \textcircled{\scriptsize 2}$'$.
%In this paper,
%this is improved by
%the phenomenological statement
%\textcircled{\scriptsize 2}.
%

\vskip0.3cm%%%%%%
%%%%%%%%3-4-3-4-3-4-3-4-
\par
\noindent
{\bf
%\large
About \textcircled{\scriptsize 3}
and
\textcircled{\scriptsize 4}}
\par
\noindent
\rm
\par
Put
${D}_N = \{ 1,2,\ldots, N ({}{{\approx}} 10^{24}{}) \} $.
For each
$k$
$({}\in {D}_N )$,
define the coordinate map
${{X}}_k{}: {{\Omega}}_{{}_E} ({}\subset {\mathbb{R}}^{6N}{}) \to {\mathbb{R}}^6 $
such that
\begin{align*}
{{X}}_k(\omega)
=
{{X}}_k
(q,p)
=
{{X}}_k (({} q_{1n}, q_{2n}, q_{3n} ,
p_{1n} , p_{2n} , p_{3n}{})_{n=1}^N{})
=
({} q_{1k}, q_{2k}, q_{3k} ,p_{1k} , p_{2k} , p_{3k}{})
%\tag{6} 
\end{align*}
for all
$
\omega=
(q,p)$
$=$
$({} q_{1n},$
$ q_{2n},$
$ q_{3n} ,$
$  p_{1n} ,$
$ p_{2n} , p_{3n}{})_{n=1}^N $
$\in$
$ {{\Omega}}_{{}_E} ({}\subset {\mathbb{R}}^{6N}{})$.
%P%\TAG{9.3}

Also,
for any subset
${D}$
$({}\subseteq {D}_N {{=}}$
$ \{1,2,$
$\ldots,N$
$ ({}{{\approx}} 10^{24}{}) \}{})$,
define the distribution map
${R}_{{D}}^{({}\cdot{})} $
$:
{{\Omega}}_{{}_E} $
$({}\subset {\mathbb{R}}^{6 N}{})  $
$ \to {\cal M}_{+1} ({}{\mathbb{R}}^6{}) $
such that
\begin{align*}
{R}_{D}^{({}q, p{}) }
=
\frac{1}{\sharp [{}{D}] }
\sum\limits_{ k \in {D} }
\delta_{
{{X}}_k ({}q,p{})
}
\quad
(
\forall
(q,p{})
\in  {{\Omega}}_{{}_E} ({}\subset {\mathbb{R}}^{6 N}{})   )
%P%\TAG{9.4}
%\tag{7}
\end{align*}
where
${\sharp [{}{D}] }$
is the number of the elements
of the set ${D}$.
\par
Let
$\omega_0 (
\in \Omega_{{}_E})
$
be a state.
For each
$n$
$(\in {D}_N )$,
we define the map
$Y_n^{\omega_0}: [0, T] \to {\mathbb R}^6$
such that
\begin{align*}
Y_n^{\omega_0} (t) =
{{X}}_n ( \psi^{{}_E}_t (\omega_0) )
\qquad
(\forall t \in [0,T]).
%P%\TAG{9.5}
%\tag{7} 
\end{align*}
And,
we regard
$\{Y_n^{\omega_0}\}_{n=1}^N$
as
random functions on
the probability space
$([0,T], {\cal B}_{[0,T]}, {\overline m}_{{}_T})$.
Then,
\textcircled{\scriptsize 3}
and
\textcircled{\scriptsize 4}
respectively
means
\begin{itemize}
\item[\textcircled{\scriptsize 3}$'$]
$\{Y_n^{\omega_0}\}_{n=1}^N$
is a {\it sequence with the
approximately
identical distribution
concerning time.}
In other words,
there exists
a normalized measure
$\rho_{{}_E}$
on ${\mathbb R}^6$
$(${}i.e.,
$\rho_{{}_E} \in {\cal M}^m_{+1} ({}{\mathbb R}^6{})${}$)$
such that:
\begin{align*}
&
{\overline m}_{{}_T}( \{ t \in [0,T] \;: \; Y_n^{\omega_0} ( t) \in \Xi \} )
{{\approx \;}}
\rho_{{}_E}(\Xi )
%\tag{8}
%\\
%&
\quad
(\forall \Xi \in {\cal B}_{{\mathbb R}^6}, n=1,2,\ldots, N)
%P%\TAG{9.6}
\end{align*}
\item[\textcircled{\scriptsize 4}$'$]
$\{Y_n^{\omega_0}\}_{n=1}^N$
is {\it
approximately independent},
in the sense that,
for
any
$
{D}_0 \subset  \{ 1,2,$
$\ldots,$
$N ({}{{\approx}} 10^{24}{})\}
$
such that
$1 {{\; \leqq \;}}
\sharp[{}{D}_0{}] \ll N{}
$
(
that is,
$\frac{\sharp[{}{D}_0{}] }{N}
{\approx} 0
$
),
it holds that
\begin{align*}
&
{\overline m}_{{}_T}(
\{
t
\in [0,T]
:
Y_{k}^{\omega_0} ( t) \in \Xi_{k} (\in {\cal B}_{{\mathbb R}^{6}} ),
{k} \in {D}_0 \})
\\
{{\approx}}
&
\bigtimes_{{k} \in {D}_0 }  {\overline m}_{{}_T}(
\{
t
\in [0,T]
:
Y_{k}^{\omega_0} ( t) \in \Xi_{k} (\in
{\cal B}_{{\mathbb R}^{6}}
 ) \}).
%P%\TAG{9.7}
%\tag{10}
\end{align*}
\end{itemize}

\par

\par
The following important remark was missed in \cite{Ishi2, Ishi3}.
This is the advantage of our method
in comparison with Ruelle's method ({\it cf.}{}{\cite{Ruel}}).

%\hfill{$///$}%%BFBFbfbf
\vskip0.5cm
Thus,
the law of large numbers says that,
putting
$D_0
(\subset D_N)$
such that
$1 \ll
\sharp[{}D_0{}] \ll N{}
$
(that is,
$\frac{1}{\sharp[{}D_0{}] } \doteqdot 0 \doteqdot \frac{\sharp[{}D_0{}] }{N}$
),
fot almost time
$t$
$(\in [0,T])$,
we see that
\begin{align*}
\frac{1}{\sharp [{}D_0] }
\sum\limits_{ k \in D_0 }
\delta_{
Y_k^{\omega_0} (t)
}
{\doteqdot}
\rho_{{}_E}
\qquad
(\forall \omega_0 \in \Omega_{_E}),
\quad
\text{
(
by
\textcircled{\scriptsize 3}
and
\textcircled{\scriptsize 4}
{)}{
}}
%%P%\tag{9.8}
\tag{\color{black}{9.4}}
\end{align*}

Also, we see, by {}{(9.3)}, that,
for
${D}_0 (\subseteq {D}_N)$
such that
$1 \le
\sharp[{}{D}_0{}] \ll N{}
$,
\begin{align*}
&
{\overline m}_{{}_T}(
\{
t
\in [0,T]
\;:\;
Y_{k}^{\omega_0} ( t) \in \Xi_{k} (\in
{\cal B}_{{\mathbb R}^{6}}
 ),
{k} \in {D}_0 \})
=
{\overline m}_{{}_T}(
\{
t
\in [0,T]
:
{{X}}_{k}(\psi^{{}_E}_t ( \omega_0) \in \Xi_{k} (\in 
{\cal B}_{{\mathbb R}^{6}}
),
{k} \in {D}_0 \})
\\
=
&
{\overline m}_{{}_T}(
\{
t
\in [0,T]
:
\psi^{{}_E}_t ( \omega_0) \in (({{X}}_k)_{k\in {D}_0})^{-1}
(
\!
\bigtimes_{k\in {D}_0}
\!
\Xi_{k}  )
\})
{{\approx}}
\;
{\overline \nu}_{{}_E}
\big(
(({{X}}_k)_{k\in {D}_0})^{-1}
(\bigtimes_{k\in {D}_0 }\Xi_{k}  )
\big)
\\
\equiv
&
\big(
{\overline \nu}_{{}_E}
\circ
(({{X}}_k)_{k\in {D}_0})^{-1}
\big)
(\bigtimes_{k\in {D}_0 }\Xi_{k}  ).
%%%\tag{9} 
\end{align*}
Particularly,
putting
${D}_0=\{k\}$,
we see:
\begin{align*}
{\overline m}_{{}_T}(
\{
t
\in [0,T]
\;:\;
Y_k^{\omega_0} ( t) \in \Xi \})
{{\approx \;}}
({\overline \nu }_{{}_E} \circ {{X}}_k^{-1}
)(\Xi)
\qquad
\qquad
(\forall \Xi \in
{\cal B}_{{\mathbb R}^{6}}
 ).
\tag{9.5}
%%P%\TAG{9.9}
\end{align*}
%%%
\par
\vskip0.1cm
\par

Hence, 
the setences
%we can describe
%the
\textcircled{\scriptsize 3}
and
\textcircled{\scriptsize 4}
in ordinary langugae
can be tlanslated to
measurement theoretical 
sentence
as follows.

\par
\noindent
\bf \vskip0.3cm
\vskip0.3cm
%BFBF
\par
\noindent
Hypothesis 9.1[
\textcircled{\scriptsize 3} and \textcircled{\scriptsize 4}
]$\;\;$%POPOPO
\rm
\rm
%[\textcircled{\scriptsize 3}
%and
%\textcircled{\scriptsize 4}
%].
\rm
Put
${D}_N$
$=$
$\{ 1,2,$
$\ldots,$
$N ( {{\approx}} 10^{24}) \}$.
Let
${\cal H} $,
$E$,
${\nu_{{}_E}}$,
${\overline \nu}_{{}_E}$,
${{X}}_k:{{\Omega}}_{{}_E} \to {\mathbb{R}}^6$
be as in the above.
Then,
summing up
\textcircled{\scriptsize 3}
and
\textcircled{\scriptsize 4},
we say:
\begin{itemize}
\item[(b)]
$\{ {{X}}_k:\Omega_{{}_E} \to {\mathbb{R}}^6 \}_{k=1}^N$
is approximately
independent random variables
with the identical distribution
in the sense that
there exists $\rho_{{}_E}$
$(\in {\cal M}_{+1} ({\mathbb{R}}^6))$
such that
\begin{align*}
\bigotimes_{ k \in {D}_0 }
{\rho_{{}_E}}
(=\text{{\lq\lq}product measure{\rq\rq}})
{{\approx}}
\;
{\overline \nu}_{{}_E}
\circ
({} ({}{{X}}_k{})_{ k \in {D}_0 }{})^{-1}.
%%%%
%P%\TAG{9.10}
%%\tag{11}
\end{align*}
%\item[]
for all
${D}_0 \subset
{D}_N
%\{ 1,2,\ldots,N ({}{{\approx}} 10^{24}{})\}
$
and
$1 {{\; \leqq \;}}
$
$
\sharp[{}{D}_0{}] $
$\ll N{}
$.
%can be approximately regarded as an {\it independent sequence with the
%identical distribution on}
%$(\Omega_{{}_E} , {\cal B}_{ {{\Omega}}_{{}_E} }, \overline{\nu}_E)$.
%% $ ({}{\mathbb{R}}^6{})${}$)$.
\end{itemize}
Also,
a state
$(q,p) (\in \Omega_{{}_E})$
is called
an {\it
equilibrium state}
if it satisfies
${R}_{{D}_N}^{ ({}q  , p  {})} {{\approx}} \rho_{{}_E}$.
\rm
%\hfill{$///$}
\par

\par
\vskip0.5cm
\par
%\noindent
%{\bf
%%\large
%3.1.5.
%Ergodic Hypothesis}
\par
\rm
Now,
we have the following theorem
({\it cf.}{}{\cite{IErgo}}):
%\par
{
\bf
%\vskip0.3cm
%\vskip0.3cm
%BFBF
\par
\noindent
\par
\noindent
\bf
Theorem 9.2
[Ergodic hypothesis{\rm{(ergodic hypothesis)}}]}
%\footnote{
%}.
\baselineskip=18pt
\rm
%index{@Ergodic hypothesis}
{\rm}
Assume
{}{Hypothesis 9.1}
$($
or
equivalently,
\textcircled{\scriptsize 3}
and
\textcircled{\scriptsize 4}
$)$.
Then,
for any
$\omega_0 = (q(0), p(0))
\in
\Omega_{{}_E}$,
it holds that
\begin{align*}
[{R}_{{D}_N}^{ ({}q({}t)  , p ({}t) {})}
](\Xi)
{{\approx \;}}
{\overline m}_{{}_T}(
\{
t
\in [0,T]
\;:\;
Y_k^{\omega_0} ( t) \in \Xi \})
\qquad
(
\forall \Xi \in {\cal B}_{{\mathbb R}^6},
k=1,2,\ldots,
N
({}{{\approx}} 10^{24}{}){})
\tag{9.6}
\end{align*}
for almost
all
$t$.
That is,
$0 {{\; \leqq \;}}$
${\overline m}_{{}_T}(\{t \in [0,T] : \text{{}{\rm (9.6)} does not hold\}{\rm )}}$
%$\frac{\int_{\{t \in [0,T] : 35) does not hold\}}dt}{T}$
$\ll 1$.

\par
\noindent
\par
\noindent
{\it $\;\;\;\;${Proof.}}$\;\;$
%{\it $\;\;$ Proof.}
\rm
Let
${D}_0 \subset {D}_N$
such that
$1 \ll
\sharp[{}{D}_0{}]
\equiv N_0
\ll N{}
$
(that is,
$\frac{1}{\sharp[{}{D}_0{}] } {\approx} 0 {\approx} \frac{\sharp[{}{D}_0{}] }{N}$
).
Then,
from
{}{Hypothesis A},
the law of large numbers
{}{({\it cf.} \cite{Kolm})}
says that
\begin{align*}
{R}_{{D}_0}^{ ({}q({}t)  , p ({}t) {})}
{{\approx \;}} {\overline \nu }_{{}_E} \circ {{X}}_k^{-1}
\;
(\;
 {{\approx \;}} {\rho_{{}_E}} \;{})
\qquad
%%%5
%%%%%P%\TAG{9.12}
\tag{9.7}
\end{align*}
for almost all time $t$.
Consider the decomposition
${D}_N$
$=$
$\{ {D}_{ ({}1{})}  , {D}_{ (2 {})} ,\ldots,$
$ {D}_{ (L {})}  \}$.
(i.e.,
${D}_N=\bigcup_{l=1}^L {D}_{(l)}$,
${D}_{(l)} \cap {D}_{(l')}=\emptyset
\;\;
(l \not= l')
$
),
where
$\sharp [{}{D}_{ ({}l{})} {}] {{\approx}} N_0 $
$({}l=1,2,\ldots, L{})$.
From
{}{(9.7)},
it holds that,
for each
$k$
$({}= 1,2,\ldots,N $
$({}{{\approx}} 10^{24}{}){})$,
\begin{align*}
&
{R}_{{D}_N}^{ ({}q({}t)  , p ({}t) {})}
=
\frac{1}{N} \sum\limits_{l=1}^L
[
\sharp [{}{D}_{(l)}{}]
\times
{R}_{{D}_{(l)}}^{ ({}q({}t)  , p ({}t) {})}
]
\\
{{\approx \;}}
&
\frac{1}{N}
 \sum\limits_{l=1}^L
[
\sharp [{}{D}_{(l)}{}]
\times
{\rho_{{}_E}}
]
{{\approx \;}}
{\overline \nu }_{{}_E} \circ {{X}}_k^{-1}
\;
({}\;
{{\approx \;}} {\rho_{{}_E}}
\;),
%%%%{4.20}
\tag{9.8}
\end{align*}
for almost all time $t$.
%Thus, by {}{(10)},
%we get {}{(12)}.
Hence, the proof is completed.
%\footnote{
%%%%%%%%%%nnnnnnnnnnnkkkkk{k}{k}kkkkkk{n}n{}{n}{n}{n}{n}

\par
\vskip0.5cm
\par
\rm
\par
%\noindent

\par
\noindent
We believe that
Theorem 9.2 is just what should be represented by the
{\it
{\lq\lq}ergodic hypothesis{\rq\rq}}
as follows.
\par
\noindent
\bf
{{Corollary }}9.3
[{\rm{ergodic hypothesis}}]
\rm
\begin{align*}
&
\text{{\lq\lq population average of
$N$ particles at each $t$\rq\rq}}
\\
=
&
\text{{\lq\lq time average of one particle\rq\rq}}.
\end{align*}
%%%%%
Thus, we can assert that
the ergodic hypothesis is related to
equilibrium statistical mechanics.
%({\it cf.}
%the (b) in the abstract).
Here, the ergodic property
\textcircled{\scriptsize 2}$'$
and
the above
ergodic hypothesis
should not be confused.
Also, it should be noted that
the ergodic hypothesis does not hold
if the box
( containing particles )
is too large.

\par
\noindent
%BBBBBBBBBBBBBBBBBB%SBSBSBS
{\small%%{\footnotesize
\vspace{0.1cm}
\begin{itemize}
\item[$\spadesuit$] \bf {{}}{Note }9.2{{}} \rm
Let us explain Corollary 9.3
as allegory of
\textcolor{black}{{Note }9.1}
as follows.
%,
%
%\item[] \bf  \rm %%%BBBBBBBBBBBBBBBBBBBB
%\\
We see that,
at almost every time
%
%Every kindergartner has the same palatability.
%Therefore, the sum total time of each three play is the same.
%For example,
%for every kindergartner,
%we see that
\begin{align*}
\cases
\text{
the number of the kindergartner who is doing the swing is
}
&
\quad
\text{50 persons}
\\
\text{
the number of the kindergartner who is doing the sliding way is
}
&
\quad
\text{
50 persons
}
\\
\text{
the number of the kindergartner who is doing the sand play is
}
&
\quad
\text{
20 persons
}
\\
\endcases
\end{align*}
Here, we want to add the remark
about
the time interval $[0,T]$].
For example,
as one of typical cases,
consider the motion of $10^{24}$ particles in a cubic box
(whose long side is 0.3m).
It is usual to consider that
{\lq\lq}averaging velocity{\rq\rq}=$5\times10^{2} {\rm m}/{\rm s}$,
{\lq\lq}mean free path{\rq\rq}=$10^{-7}{\rm m}$.
And therefore,
the collisions rarely happen
among $\sharp[{}{D}_0{}]$ particles
in the time interval
$[0, T]$,
and therefore,
the motion is "almost independent".
%Thus,
%as in the above
%\textcircled{\scriptsize 4},
%we say
%{\lq\lq}approximately independent{\rq\rq}
%on the $[0,5]$.
For example,
putting $\sharp[{}{D}_0{}]=10^{10}$,
we can calculate
the number of times
a certain particle collides
with ${D}_0$-particles
in [0,T]
as
$(10^{-7} \times \frac{10^{24}}{10^{10}})^{-1}
\times {(5 \times 10^2)} \times T$
$\approx   5 \times 10^{-5} \times T$.
Hence,
in order to
expect that
\textcircled{\scriptsize 3}$'$
and
\textcircled{\scriptsize 4}$'$
hold,
it suffices to
consider that
$T
\approx
5$ seconds.
Also,
note that
the term
"ergodic"
may be used in various meanings.
For example,
the formula \textcolor{black}{(9.3)}
is also said to be ergodic.
However,
here,
we use it in the above sense.

\end{itemize}
}
%%BBBBBBBBBBBBBBBBBB%SBSBSBSSP
\noindent
\rm
\rm
\par

\rm
\par
\par
\noindent
%It is  to note that
\par
\noindent
%BBBBBBBBBBBBBBBBBB%SBSBSBS
{\small%%{\footnotesize
\begin{itemize}
\item[$\spadesuit$] \bf {{}}{Note }9.3{{}} \rm
%(ii):
% and ?
\rm
The entropy
$H(q,p)$
of a state
$(q,p)
(\in \Omega_{{}_E})
$
is defined by
\begin{align*}
H(q,p)
=
%k_B
k
\log
[
\nu_{_E}
(
\{(q',p')
\in
\Omega_{{}_E}:{R}_{{D}_N}^{ ({}q  , p )}
{{\approx \;}}
{R}_{{D}_N}^{ ({}q'  , p' )}
)
\}
)
]
\end{align*}
where
$$
k=
%\frac{
[\text{Boltzmann constant}]
/
%}{
(
{[\text{Plank constant}]^{3N}N!}
)
%}
$$
%$kB=[\text{Boltzmann constant}]$.
%%
Since
almost every state
in $\Omega_{{}_E}$
is
equilibrium,
the entropy of
almost every
state
is equal
$k \log \nu_{{}_E}(\Omega_{{}_E})$.
Therefore,
it is natural to assume that
the law of increasing entropy
holds.
\end{itemize}
}
%%BBBBBBBBBBBBBBBBBB%SBSBSBSSP
%\noindent
%\rm
%\rm
%\par
%
%

\subsubsection{The probabilistic aspect of
equilibrium statistical mechanics
---
Where does probability come?}%{Sec. 9.1.2}
\par

\par
\noindent
%{\large \bf 3. Probabilistic aspects of
%equilibrium
%statistical mechanics }
\par
In this section we shall study the probabilistic aspects
of
equilibrium statistical mechanics.
For completeness,
note that
\begin{itemize}
\item[(a)]
the argument in the previous section
is not related to
{\lq\lq}probability{\rq\rq}
\end{itemize}
since {}{Axiom${}_{\text{\scriptsize c}}^{\text{\scriptsize p}}$ 1}
does not appear
in {}{Section 3.1}.
Also,
recall the (E$_4$),
that is,
\it
there is no probability
without measurement.
\rm

Theorem 9.2 says
that
the
equilibrium statistical mechanical system
at almost all time $t$
can be regarded as:
\begin{itemize}
\item[(b)]
a box including about $10^{24}$
particles such as the number of the particles whose
states belong to
$\Xi $ $({}\in {\cal B}_{{\mathbb R}^6 }{})$
is given by
$
\rho_{{}_E} ({}\Xi{}) \times 10^{24}
$.
\end{itemize}
Thus, it is natural to
assume as follows.
\begin{itemize}
\item[(c)]
if we, at random, choose a particle from $10^{24}$ particles
in the box
at time $t$,
then the
probability
that
the state $(q_1, q_2, q_3,$
$ p_1, p_2, p_3)$
$(\in {\mathbb R}^6)$
of the particle belongs to
$\Xi $ $({}\in {\cal B}_{{\mathbb R}^6 }{})$
is given by
$
\rho_{{}_E} ({}\Xi{})
$.
\end{itemize}
In what follows, we shall represent this {}{(J)} in terms of measurements.
Define the observable
${\mathsf O}_0=({\mathbb R}^6 , {\cal B}_{{\mathbb R}^6}, F_0)$
in $C( {\Omega}_{{}_E} )$ such that
\begin{align*}
[F_0( \Xi ) ]( q,p)
=
[ {R}_{{D}_N}^{({}q, p{}) }](\Xi)
\Big(\equiv
\frac{\sharp[\{ k \;|\;{{X}}_k ({}q,p{})\in \Xi \} ]}{\sharp [{}{D}_N] }
\Big)
\quad
\qquad
(
\forall \Xi
\in
{\cal B}_{{\mathbb R}^6},
\forall
(q,p{})
\in  {{\Omega}}_{{}_E} ({}\subset {\mathbb R}^{6 N}{})   ).
%\tag{15}
\end{align*}
Thus, we have the measurement
${\mathsf M}_{C( {\Omega}_{{}E} )}( {\mathsf O}_0:=
({\mathbb R}^6 , {\cal B}_{{\mathbb R}^6}, F_0),
S_{[ \delta_{\psi_t ( q_{{}_0} , p_{{}_0})}]} )$.
Then we say,
by Axiom${}_{\text{\scriptsize c}}^{\text{\scriptsize p}}$ 1,
that
\begin{itemize}
\item[(d)]
the probability that the measured value obtained by
the measurement
${\mathsf M}_{C( {\Omega}_{{}E} )}( {\mathsf O}_0:=
({\mathbb R}^6 , {\cal B}_{{\mathbb R}^6}, F_0),$
$
S_{[ \delta_{\psi_t ( q_{{}_0} , p_{{}_0} )}]} )$
belongs to
$\Xi ( \in {\cal B}_{{\mathbb R}^6})$
is given by $\rho_{{}_E} ( \Xi )$. That is because
{}{Theorem 9.2} says that
$
[ F_0( \Xi ) ]( \psi_t ( q_{{}_0} , p_{{}_0}) )
%= [F_0( \Xi )]( \psi_t ( \omega_0 ))
$
$
\approx
\rho_{{}_E} ( \Xi )
$
$
\text{(almost every time $t$)}
$.
\end{itemize}
%which is just the measurement theoretical representation of
%the $(A_1')$.
\par

%\newpage
\par
\noindent
Also,
let
$\Psi^{{}_E}_t: C(\Omega_{_E}) \to C(\Omega_{_E}) $
be a deterministic Markov operator
determined
by
the
continuous map
$\psi^{{}_E}_t: \Omega_{{}_E} \to \Omega_{{}_E} $
({\it cf.} {}{Section 3.1.2}).
Then,
it
clearly holds
$\Psi^{{}_E}_t{\mathsf{O}_0}={\mathsf{O}_0}$.
And, we must take
a
${\mathsf{M}}_{C( \Omega_{{}_E} )}( {\mathsf{O}_0},
S_{[{(q(t_k),p(t_k))}]} )$
for each
time
$t_1,t_2,\ldots,t_k, \ldots,
t_n$.
However,
Interpretation
(E$_2$)
says that
it suffices to
take
the simultaneous measurement
${\mathsf{M}}_{C( \Omega_{{}_E} )}(\bigtimes_{k=1}^n {\mathsf{O}_0}
%{{=}}
%({\mathbb{R}}^6 , {\cal B}_{{\mathbb{R}}^6}, F_0)
,$
$
S_{[\delta_{(q(0),p(0))}]} )$.

\par
\vskip0.3cm
\par
\noindent
%\it
%Remark 4.
%\rm
%[The principle of equal a priori probabilities
%].
%\rm
%The {}{(J)}
%(or equivalently,
%{}{({D})})
%says
%{\lq\lq}choose a particle from $N$ particles
%in box{\rq\rq}$\!$,
%and not
%{\lq\lq}choose a state
%from the state space $\Omega_{{}_E}${\rq\rq}.
%Thus,
%as mentioned in the abstract,
%the principle of equal (a priori) probability
%is not
%related to
%our method.
%If we try to describe
%Ruele's method {}{\cite{Ruel}}
%in terms of measurement theory,
%we must use mixed measurement theory
%({\it cf}.
%\cite{Ishi2,Ishi6}).
%However, this trial will end in failure.
%{X}\pi\pi
%% {\lq\lq}probability{\rq\rq}$\!$.

%\pi\pi\piYXXXXXXYYYYYYYY
\rm
\par
\noindent
%BBBBBBBBBBBBBBBBBB%SBSBSBS
{\small%%{\footnotesize
\vspace{0.1cm}
\begin{itemize}
\item[$\spadesuit$] \bf {{}}{Note }9.4{{}} \rm
We devote ourselves to the linguistic world-view.
We first have a language called measurement theory.
And,
we tried to describe
the phenomena
\textcircled{\scriptsize 1}$\text{--} $\textcircled{\scriptsize 4}
in
\textcolor{black}{{Sec. 9.1.1}}.
And we construct
equilibrium statistical mechanics.
We are faithful to
the linguistic spirit
"language is before world".
In this sense,
we think that
our
equilibrium statistical mechanics
is not physics.
%,
%equilibrium statistical mechanics and {{{{h}}}} and ,
%\textcolor{black}{\cite{Toda}}.
\end{itemize}
}

%%%%%%%%
\subsection{
"{Dynamical system theory}" vs. "{Measurement theory}"
in equilibrium statistical mechanics
}%9.2
\rm
\par
Equilibrium statistical mechanics
is
to derive
thermo-dynamics
from
[Newtonian mechanics]$+$
[$\alpha$].

%{{{{h}}}} and ,
%That is,
%Newtonian mechanics and {{{{h}}}} and 
%
%{}
%,
%$\alpha$
%.
\par
Our method
({{measurement theory}})
is
%\ssmall
\begin{itemize}
\item[$\;\;$(a)]
$
\underset{\text{\scriptsize ({{{measurement theory}}})}}{\fbox{equilibrium statistical mechanics}}
:=
\overset{({Sec.6.4})}{\underset{\text{\scriptsize ("{{Newton}} equation
\textcircled{\scriptsize 1}"
and
\textcircled{\scriptsize 2}
\text{--}
\textcircled{\scriptsize 4}
)}}{\fbox{Axiom${}_{\text{\scriptsize c}}^{\text{\scriptsize pm}}$ 2(causality )}}
}
+
\overset{({Sec.2.2})}{
\underset{\text{\scriptsize (urn problem)}}{\fbox{Axiom${}_{\text{\scriptsize c}}^{\text{\scriptsize p}}$ 1}}
}
$
\end{itemize}
\baselineskip=18pt
\par
\noindent
\baselineskip=18pt
%\normalsize
where
${\overline \nu}_{_E}$
is the normalized staying time
derived from Newtonian mechanics.
%,
%(POI),
%Newtonian mechanics,
%probability 
% and .
%probability ,
%
%Axiom${}_{\text{\scriptsize c}}^{\text{\scriptsize p}}$ 1({{measurement}}
%(\textcolor{black}{\REF{2secAxiom 1}}))
%.
%That is,
%equilibrium statistical mechanics(a)
%probability ,
%urn problem(\textcolor{black}{Example 2.10})probability
% and {}
%%\baselineskip=18pt
\par
On the other hand,
the conventional method
({dynamical system theory}({\rm cf.}
\textcolor{black}{\cite{Ruel}}))
is
%\ssmall
\begin{itemize}
\item[$\;\;$(b)]
$
\;\;\;\;
\underset{\text{\scriptsize ({dynamical system theory})}}{\fbox{equilibrium statistical mechanics}}
\;\;$
\\
\\
$
:=
\underset{("{{Newton}}
\textcircled{\scriptsize 1}"
and
\textcircled{\scriptsize 2}
}{\fbox{\;\;Newtonian mechanics\;\;}}
+
\underset{\text{
\scriptsize(
probabilistic interpretation of ${\overline \nu}_E$)}}{\fbox{
equal probability }}
$
\end{itemize}
\normalsize \baselineskip=18pt
\par
\noindent
\baselineskip=18pt
Thus, we have
$$
\text{
(a) vs. (b)
}
$$
Here, recall Note 1.4.
This "vs."
must be  settled in future.
%
%
%
%
%,
%probability ${\overline \nu}_{_E}$
%probabilistic interpretation$\big($={\bf probability },
%That is,
%$(\Omega_{_E},$
%$ {\cal B}_{\Omega_{_E}},$
%$ {\overline \nu}_{_E})$
%, probability 
%(,
%trial(\textcolor{black}{{Sec. 4.1.2}(A$_3$)}))
% and 
%$\big)$
%{}
%{ordinary language} and ,
% and probability  and 
%,
%(b), .
%,
%(b),
%%probability 
%%Newtonian mechanics,
%{{{measurement theory}}}.
%,
%,
%{{measurement}}
%${\mathsf M}_{C (\Omega_E)}
%({{\mathsf O}}^{\FIN}, S_{[*]}({\overline \nu}_{_E}))$
% and  and 
% and .
%%index{ and @probability }
%,
%(b),
%,
%equilibrium statistical mechanics and ,
%{}That is,
%\BEGIN{itemize}
%\item[]
%%Heisenberg({{Proposition }}3.1) and ,
%equilibrium statistical mechanics(b),
%{ordinary language},
%%equilibrium statistical mechanics(b)
%{{{{h}}}} and  and 
%%
%\END{itemize}
% and .
%
%
%{{measurement theory}},
%{{measurement theory}}, probability ,
%%probability 
%%(That is,
%
%probability  and ,
%%
%%) and ,
% and probability .
%%{}
%%\par
%%,
%%
%% and ,
%,
%equilibrium statistical mechanics(b),
%,
%.
%(a) vs. (b)
%, {},
%,
%
%(a),
%equilibrium statistical mechanics(b) and 
%(\textcolor{black}{{Note }1.4}).
%\textcolor{black}{%index{@(equilibrium statistical mechanics)}}
%%
\par
\noindent
%BBBBBBBBBBBBBBBBBB%SBSBSBS
{\small%%{\footnotesize
\vspace{0.1cm}
\begin{itemize}
\item[$\spadesuit$] \bf {{}}{Note }9.5{{}} \rm
We do not agree that
there are several "equilibrium statistical mechanics".
Also,
we should hurry up
to
describe various sciences in terms of measurement theory.
For "psychological tests",
see \cite{KPyco}.
\end{itemize}
}
%%BBBBBBBBBBBBBBBBBB%SBSBSBSSP
\noindent
\rm
\rm
\par

\vskip0.9cm%%%%%%
%%%%%%%%

\vskip0.9cm%%%%%%
%%%%%%%%

\subsection{Is quantum mechanics physics, or engineering? }%9.3
\par
For the foregoing paragraph,
I considered that equilibrium statistical mechanics was one of the science.
In this section, I further promote this and argue about
\begin{itemize}
\item[]
Is quantum mechanics physics, or several sciences (engineering)?
\end{itemize}

\par
Of course, if I say that quantum mechanics is not physics,
you will think that I do not have common sense.
%,
However, Einstein was skeptical to quantum mechanics throughout life(
{\rm cf.}
\textcolor{black}{{\cite{Sell}}}).
At least, I can be say that quantum mechanics was not "Einstein's physics."
%.
It is more convenient to think "quantum mechanics is one of the science instead of physics" also for measurement theory.
Although I explain the reason below,
it becomes the argument about von Neumann's work
"mathematical basis of quantum mechanics \cite{Neum};(1932)" after all.
%index{@${{\cdot}}$}
Then, I recommend that you read the following,
after you refer to the Internet etc. about von Neumann's
-the man who was the cleverest at the 20-th century- genius.
\par
I think over the classification of the measurement theory
of \textcolor{black}{{Chap.$\;$1}({}Y)} as follows. :
\begin{itemize}
\item[(a)]
$
%{}
\underset{
\text{\scriptsize (sientific language)}
}{\text{measurement theory (=MT)}}
\text{}
$
\\
\\
$
\quad
\cases
\underset{
%(POI)
}{\text{quantum MT}}
\text{}
&
\xrightarrow[\text{\scriptsize description}]{\text{\scriptsize quantum phenomena}}
\text{quantum mechanics (=quantum engineering)}
\\
\\
\underset{
%(POI)
}{\text{classical MT}}
&
\xrightarrow[\text{\scriptsize description}]{\text{\scriptsize ordinary phenomena}}
%\xrightarrow[]{}
\text{statistical mechanics, economics,}\; \cdots
%\cases
\endcases
%\TAG{0.4}
$
\end{itemize}
That is, it is a position which we consider is a linguistic science language
rather than consider quantum measurement theory to be physics.
Namely,
\begin{itemize}
\item[]
I consider that Born's quantum measurement theory[Axiom(Q);Chapter 3.1.1]
should be regarded as
"mystic words ",
and I assert a linguistic science view.
%index{@}
\end{itemize}
This is the position of considering that a linguistic science language called
measurement theory
(= [Quantum measurement {theory}] +[Classic measurement {theory}] ) occurs first,
and considering that the theory (one of the science) which described the law in the micro world with the language is quantum mechanics.
In this position, quantum mechanics is one of the science
-
although it may be a coined word of this book, {\bf quantum engineering}
-.
%index{@}
If you may think like this (i.e., if you may think that quantum measurement theory is not "physics"
but a "language" as shown in (a)),
Copenhagen interpretation (\textcolor{black}{(U$_1$)--(U$_7$) of {Chap.$\;$1})}
can also be allowed to be wonderful theory (Section 3.3).
Rather, the author thinks as follows (Note 3.6).
\begin{itemize}
\item[(b)]
Copenhagen interpretation is not the thing of "physics"
but a thing for "language (measurement theory)."
\end{itemize}

%index{@${{\cdot}}$}
Mathematical formulation of quantum mechanics was performed by von Neumann in
"The mathematical basis of quantum mechanics \cite{Neum};(1932)"
-
formulation of the quantum mechanics by Hilbert space
-.
Supposing that is right, you should just be going to consider the following (c) in detail.
Namely,
\begin{itemize}
\item[({}c)]
Although physics, such as Newtonian mechanics, electromagnetism, and the theory of relativity,
is formalized in the mathematics of differential geometry,
why is quantum mechanics formalized in the mathematics of Hilbert space?
$\;\;$
Is quantum mechanics really physics?
%\footnote{
%(
%{\rm cf.}
%\textcolor{black}{{\cite{Sell}}}),
%, ,
%.
%}
\end{itemize}
Supposing it is formalized in different mathematics,
it is natural that we consider that it is the world describing method of a different category.
If that is right, I would like to come to think
\begin{itemize}
\item[]
$
\cases
\text{[The realistic method]:}
\\
$\;\; \qquad$
\text{
Physics should be formulized in differential geometry.
}
\\
\\
\text{[The linguistic method]:}
\\
$\;\; \qquad$
\text{
Various sciences should be formulized in Hilbert space theory.
}
\endcases
$
\\
\\
\end{itemize}
in world description.

Thinking in this way, in this book,
we think that von Neumann proposed in "The mathematical basis of quantum mechanics \cite{Neum}" ,
not only formulation of quantum mechanics but big work with one another more, i.e.,
\begin{itemize}
\item[({}d)]
Formulation of the linguistic side of quantum mechanics
(namely, quantum measurement theory of {(a)})
\end{itemize}
I think that \cite{Neum} is written by me so that it can read also not as the book of physics
but as a book of the linguistic describing method.
Therefore, although it becomes the upper repetition,
\begin{itemize}
\item[(e)]
Substantially, measurement theory was advocated in \cite{Neum}.
That is, although Heisenberg, Schr\"odinger, and Born
advocated the physics of quantum mechanics,
von Neumann formalized the linguistic side of quantum mechanics.
\end{itemize}
Just before \textcolor{black}{{Chap.$\;$1}(B)}, I wrote "In order to declare a law,
the language for describing the law must be prepared before that."
However, actual history is somewhat complicated, should be corrected as follows and should be written.
\begin{itemize}
\item[]
Heisenberg and others described "the law of quantum mechanics" in the imperfect language.
And von Neumann made the imperfect language perfect.
\end{itemize}

If that is right,
von Neumann's opinion is "changing into an idiom [\textcolor{black}{{Chap.$\;$1}({}M$_2$)}]"
i.e., reverse "Using an idiom" of
\begin{itemize}
\item[
%$\underset{{\text{\scriptsize ({Chap.{\;}}1{}({}M$_2$)}}
%}{\text{
(f$_1$)
]
%}$]
%$\underset{\text{\scriptsize ({Chap.{\;}}1{}({}M$_2$)}}
%
%$
%%$
%\qquad
%$
[({}M$_2$) in Chap. 1]:
$
\overset{\text{\scriptsize (word is connected with world)}}{\underset{\text{\scriptsize (physics)}}{\text{
\fbox{quantum mechanics(H)}}}}
\xrightarrow[\text{\scriptsize proverbalizing}]{}
\overset{\text{\scriptsize (
word is independent of world
)}}{\underset{(language)}{\text{\fbox{
measurement theory}}}}
$
\end{itemize}
%%%%%%proverb
Namely,
\begin{itemize}
\item[({}f$_2$)]
$\quad
%\qquad
$
$
\overset{\text{\scriptsize (word is connected with world)}}{\underset{\text{\scriptsize (variuos science)}}{\text{
\fbox{quantum engineering}}}}
\xleftarrow[\text{\scriptsize using proverb}]{}
\overset{\text{\scriptsize (
word is independent of world
)}}{\underset{(language)}{\text{\fbox{
measurement theory}}}}
$
\end{itemize}
That is, although it becomes re-described of (a),
\begin{itemize}
\item[({}g)]
In a language called $\underset{\text{\scriptsize (General quantum mechanics)}}{\text{Measurement theory}}$,
it thinks as follows.
\item[(g$_1$)]:
The theories which describe quantum phenomena are several sciences of quantum mechanics
(namely, quantum engineering).
\item[(g$_2$)]:
The theories which describe an everyday phenomenon are ordinary science
(for example, equilibrium statistical mechanics).
\end{itemize}
\normalsize
\baselineskip=18pt
If it thinks in this way,
I can think that quantum mechanics (quantum mechanics
based on the Copenhagen interpretation of \cite{Neum} at least)
is one of the various sciences
-
quantum engineering
-
instead of physics.
(\textcolor{black}{Note 3.6}).
Although this (g$_1$) may not be von Neumann's \cite{Neum} intention,
I would like to consider it like this in this book.

The above is a reason for "branch of quantum mechanics"
%\ssmall
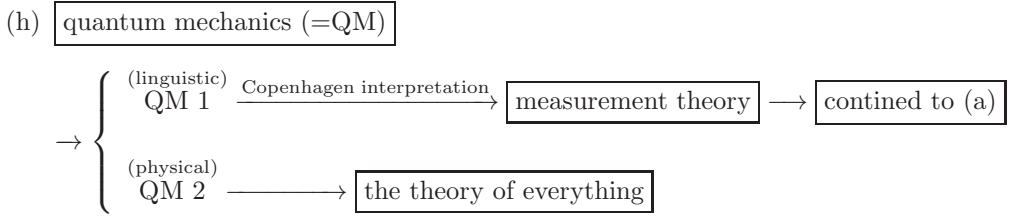
\begin{itemize}
\item[(h)]
$
\underset{}{\text{\fbox{quantum mechanics (=QM)}}}
$
\\
$
\xrightarrow[\; {{{}}}\;]{}
\cases
\overset{\text{\scriptsize (linguistic)}}{\text{QM 1}}
\xrightarrow[]{\text{\scriptsize Copenhagen interpretation}}
%\quad
%\dashbox{3}(40,17)[c]
\fbox{measurement theory}
\xrightarrow[]{\;\; \text{}\;\;}
%\quad
%\dashbox{3}(40,17)[c]
\fbox{contined to (a)}
%\dashbox{\text{}}
\\
\\
\overset{\text{\scriptsize (physical)}}{\text{QM 2}}
\xrightarrow[\qquad \qquad]{}
%\quad
%\dashbox{3}(70,17)[c]
\fbox{the theory of everything}
%\dashbox{3}(60,17)[c]{}
\endcases
$
%$\Big)$
\end{itemize}
\normalsize
\baselineskip=18pt
in \textcolor{black}{Fig 8.2}.
If that is right,
\begin{itemize}
\item[({}i)]
the quantum mechanics
(namely, quantum mechanics currently ordinarily taught at the university)
explained in the third chapter is not "quantum mechanics (2)" but quantum engineering.
That is, I would like to come to consider
$$
\text{
[quantum engineering] = [quantum mechanics based on the Copenhagen interpretation].
}
$$
\end{itemize}

Supposing that is right,
we would like to come to a conclusion
with "the language of measurement theory = science(Chapter 8.3(m))" as follows.
\begin{itemize}
\item[(j)]
\bf
Originally Copenhagen interpretation [\textcolor{black}{({{}}U$_1$)--({{}}U$_7$)}]
is a universal tenet for an understanding of
engineering, and science (not being a thing of quantum mechanics).
\end{itemize}
\rm
\par
\noindent
%BBBBBBBBBBBBBBBBBB%SBSBSBS
{\small%%{\footnotesize
\vspace{0.1cm}
\begin{itemize}
\item[$\spadesuit$] \bf {{}}Note 9.6{{}} \rm
Although "whether it is (f$_1$) or (f$_2$)?" may be an endless dispute,
we can assert one side of the following ($\sharp_1$) or ($\sharp_2$) anyway.
:
\begin{itemize}
\item[($\sharp_1$)]
If (f$_2$) is right,
quantum mechanics (= quantum engineering) and equilibrium statistical mechanics
are the greatest examples of application of measurement theory.
\item[($\sharp_2$)]
\textcolor{black}
If {(f$_1$)} is right,
measurement theory is the greatest example of application of quantum mechanics.
Although it is un-physical unique application,
considering the extensiveness of science(since most science is science),
the "greatest" description will be allowed.
%
% and 
%
\end{itemize}
%\footnote{
%{\textcolor{black}{FIG.}$\;$}1,
%,
%20
%(=),
%, ,
%${{\cdot}}$
% and .
%,
%, {\textcolor{black}{FIG.}$\;$}8.2.
%}.
Of course, in the position of this book, we will assert $(\sharp_1) $
---
"The measurement theory is more fundamental than quantum mechanics
(in the sense of (f$_2$))."
As mentioned before,
the worst understanding of
measurement theory
is
to consider
measurement theory
as
an ass in a lion's skin.
%,
%, .
%
%{}, 
\end{itemize}
}
%%BBBBBBBBBBBBBBBBBB%SBSBSBSSP
\noindent
\rm
\rm
\par

\par
\noindent
%It is  to note that
\par
\noindent
%BBBBBBBBBBBBBBBBBB%SBSBSBS
{\small%%{\footnotesize
\begin{itemize}
\item[$\spadesuit$] \bf {{}}Note 9.7{{}} \rm
von Neumann's "mathematical-basis of quantum mechanics\cite {Neum}"
is the book which can be read by various methods.
More precisely,
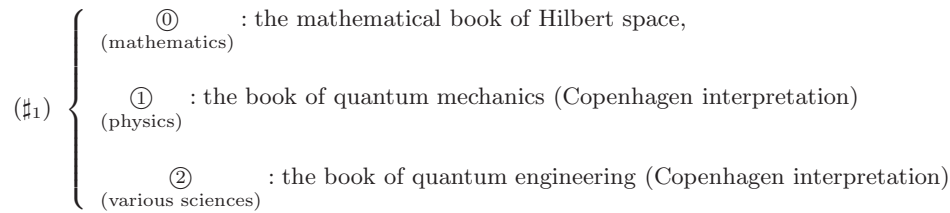
\begin{itemize}
\item[$(\sharp_1)$]
$
\cases
\underset{\text{\scriptsize (mathematics)}}{\textcircled{\scriptsize 0}}:
\text{the mathematical book of Hilbert space},
%\quad
\\
\\
\underset{\text{\scriptsize(physics)}}{\textcircled{\scriptsize 1}}:
\text{the book of quantum mechanics (Copenhagen interpretation)}
%\quad
\\
\\
\underset{\text{\scriptsize(various sciences)}}{\textcircled{\scriptsize 2}}:
\text{the book of quantum engineering (Copenhagen interpretation)}
%\underset{\text{\scriptsize(POI)}}{\textcircled{\scriptsize 2}:
%(
%)} and \\
%\hspace{0.6cm}  and 
\endcases
$
\end{itemize}
The reading of \textcircled{\scriptsize 1} was reading of the direction which von Neumann probably meant,
and though it was natural, it won a great success.
Development of a mathematics - Hilbert space theory and the operator algebra -
was greatly urged also to the reading of \textcircled {\scriptsize 0}.
It may be common sense
to evaluate \cite{Neum} from a viewpoint of
\textcircled{\scriptsize 0} and \textcircled{\scriptsize 1} (especially \textcircled {\scriptsize 1}),
and to conclude it as "great achievements (Bible of quantum mechanics)"
under the authority of mathematics and physics.
However,
\begin{itemize}
\item[$(\sharp_2)$]
In this book, as reading of \cite {Neum}, \textcircled {\scriptsize 2} is a right course,
and we can evaluate \cite {Neum} by this method still more greatly.
\end{itemize}
This is because (a) and (= \textcolor {black}{the draft amendment of {Chap.$\;$1} ({} Y)})
can be written as follows.
%{}:
\begin{itemize}
\item[($\sharp_3$)]
$
%{}
\text{enginering theory}=
$
\\
$
\quad
\underset{
\text{\scriptsize(generalized quantum theory)}
}{\text{measurement theory}}
\text{}
$
$
\cases
\underset{
\text{\scriptsize(linguistic world-view)}
}{\text{quantum measurement theory}}
\text{}
&\; \cdots \; \text{von Neumann}
\\
\\
\underset{
\text{\scriptsize (linguistic world-view)}
}{\text{classical measurement theory}}
\text{}
&\; \cdots \; \text{Fisher}
\endcases
%\TAG{0.4}
$
\end{itemize}
\par
\noindent
That is, even if it says to engineering and a mouthful,
it is vast and writes the genius of two poles
\begin{itemize}
\item[]
from Edison (master of invention) to von Neumann (advocate of Buddhist scripture \cite {Neum} of quantum engineering)
\end{itemize}
it is a broad field.
%,
%${{\cdot}}$
%%
%%20
%%
%
%%, 
% and .
%,
%,
%, ${{\cdot}}$
% and ,
%.
%,
However, if only a theoretical side (namely, upper $(\sharp_3) $) is considered,
in \textcircled{\scriptsize 6}Strong side$\text{ vs. }$Weak side of Note 2.9,
it can be written as
$$
\underset{\text{\scriptsize (the greatest physicist)}}{\text{Einstein}} 
\quad
\text{ vs. }
\quad
\underset{\text{\scriptsize (the greatest theoretical engineer)}}{\text{von Neumann(or,
Fisher)}}
%\footnotemark
$$
%\footnotetext{
%Although the word a "theoretical engineer" may be a coined word of this book,
%it will not be necessary to explain this to the reader who read so far and has advanced.
%%, \textcolor{black}{{\cite{Kalm}}}.
%}
\par
Although von Neumann's work is various,
I think that he was always siding with "the weak side (Section 2.4.1)."
If that is right,
I would like to consider \textcircled {\scriptsize 2} to be the reading of the right course
of \cite {Neum}.
%\footnote{

A fiction about Kant and von Neumann may be mixed too much
, in this section.
However, this is for compatibility with measurement theory (fiction of whole this book),
and wants a wise reader to enjoy a "fiction" in comparison with accepted theory.%}.
%(j),
%${{\cdot}}$
%, 20 and 
\end{itemize}
}
%%BBBBBBBBBBBBBBBBBB%SBSBSBSSP
\noindent
\rm
\rm
\par

\baselineskip=18pt
\font\twvtt = cmtt10 scaled \magstep2
\font\fottt = cmtt10 scaled \magstep4
\par

%BFBFAxiom${}_{\text{\scriptsize b}}^{\text{\scriptsize p}}$ 2
%%10 10 10
\baselineskip=18pt
%\vskip3.0cm
%\newpage
\section{Axiom${}_{\text{\scriptsize b}}^{\text{\scriptsize p}}$ 1
---
{{measurement (bounded type)}}
\label{Chap10}}%{Chap.{\;}}10{}
%\chapter[
%bounded type {{{measurement theory}}} (\textcolor{black}{Axiom${}_{\text{\scriptsize b}}^{\text{\scriptsize p}}$ 1}({{measurement}}))
%]{bounded type {{{measurement theory}}}
%\\
%{(}\textcolor{black}{Axiom${}_{\text{\scriptsize b}}^{\text{\scriptsize p}}$ 1}({{measurement}}){)}}
%%\vspace{-0.8cm}
\baselineskip=18pt
\noindent
\begin{itemize}
\item[{}]
{
\small
\baselineskip=15pt
\par%[Abstract].
\rm
$\;\;\;$
Classical {{{measurement theory}}} is classified as follows:
({{{}}}\textcolor{black}{{Chap.$\;$1}(Y)}):
\begin{align*}
classical {{{measurement theory}}}
\cases
\text{
continuous type {{{measurement theory}}}(\textcolor{black}{{{Chap.{\;}}2--8}})
}
\\
\text{
bounded type {{{measurement theory}}}(\textcolor{black}{{Chap.{\;}}I10.11})
}
\endcases
%TAG{10.1}
\end{align*}
Although continuous type {{{measurement theory}}}
us fundamental,
an exact {{{measurement}}}
is not generally defined.
% {} and simultaneous measurement.
Thus, we shall introduce
bounded type {{{measurement theory}}}
as follows.
%:
%(POI){{{measurement theory}}}, quantum mechanicsverbalizing,
\begin{align*}
%\dashbox{5}
\underset{\text{\scriptsize (language)}}{\text{{} $\fbox{bounded pure type {{{measurement theory}}}}$}}
:=
{
\overset{\text{\scriptsize [(bounded type )Axiom${}_{\text{\scriptsize b}}^{\text{\scriptsize p}}$ 1]}}
%\overset{\text{\scriptsize [(bounded type )Axiom${}_{\text{\scriptsize b}}^{\text{\scriptsize p}}$ 1\textcolor{black}{(Sec. \REF{10secAxiom 1})}]}}
{
\underset{\text{\scriptsize
[probabilistic interpretation]}}{\text{{} $\fbox{{{measurement}}}$}}}
}
+
{
\overset{\text{\scriptsize [(bounded type )Axiom${}_{\text{\scriptsize b}}^{\text{\scriptsize pm}}$ 2]}}
%\overset{\text{\scriptsize [(bounded type )Axiom${}_{\text{\scriptsize b}}^{\text{\scriptsize pm}}$ 2\textcolor{black}{(Sec. \REF{10secAxiom 2})}]}}
{
\underset{\text{\scriptsize [{{the Heisenberg picture}}]}}
{\text{{}$\fbox{ causality }$}}
}
}
%TAG*{$\displaystyle{\mathop{2)}_{(=10))}}$}
%%%%%%%%%%%%%2.2}CLAsSICAL
\end{align*}
%,
%, continuous type {{{measurement theory}}}
% and .
%
%
%bounded type {{{measurement theory}}}
%, 2\textcolor{black}{Axiom}
%,
In the chapter, we shall devote ourselves to the above axiom 1.
This is similar Axiom${}_{\text{\scriptsize b}}^{\text{\scriptsize p}}$ 1 (continuous pure type) in
\textcolor{black}{Chap. 2}.
Thus,
it is surely understandable.
%continuous type  and 
%, {}
}
\end{itemize}
\vskip1.0cm
\subsection{{{State}} and Observable
---
Primary quantity and Secondary quantity}%{Sec.10.1}
%\ssubsection{state space  and observable }
%index{@{Chap.{\;}}, {Chap.{\;}}}
\par
In continuous type {{{measurement theory}}}(\textcolor{black}{{Chap.{\;}}2--8}),
continuous function
(i.e., the element of $C(\Omega)$
)
plays an important  role.
On the other hand,
a bounded function is main in
bounded type {{{measurement theory}}}.
%(\textcolor{black}{{Chap.{\;}}IV})
%,
%{measurable function}
%.
%,
Since it is a small change,
readers can easily understand the bounded type measurement theory.
%
%
%
%,
%{measurable function}
%,
%(That is,
%$\lim$),
%{{{measurement theory}}}. {{{measurement theory}}}, continuous type {{{measurement theory}}}
% and .
%,
%continuous type {{{measurement theory}}}
%%(
%%\text{\textcolor{black}{{Chap.{\;}}II}}
%%)
% and ,
%bounded type {{{measurement theory}}}
%
% and .

\par
Consider a
locally compact space
$\Omega$
and
measure space
$(\Omega, {\cal B}_{\Omega}, \nu)$,
where
${\cal B}_{\Omega}$
is the Borel field
of $\Omega$,
that is,
the smallest $\sigma$-field that contains all open sets.
Further assume that
\begin{itemize}
\item[(a)]
$0 < \nu (U) {\; \leqq \;} \infty$
$( \forall$
open set $U$),
and
%index{@$\sigma$-}
$\nu$
is $\sigma$-finite.
%\footnote{
%Without loss of generality,
%we can assume that,
%$\Omega$
%is compact,
%and
%$\nu$ is finite).
%}.
%%}
\end{itemize}
%,
%$\Omega$
%(, $(\Omega, {\CAL B}_{\Omega}, \nu)$)
%state space  and .
\par
A Banach space
$L^r (\Omega, \nu)$
(where,
$r = 1, \infty $)
is the space composed of all complex valued
{measurable function}$f: \Omega\to {\mathbb R}$
with the norm
$\|f\|_{L^r (\Omega, \nu)} < \infty$
where
the norm
$\|f\|_{L^r (\Omega, \nu)} $
is defined by
\begin{align*}
\|f\|_{L^r (\Omega, \nu)}
=
\cases
{\displaystyle
\int_{\Omega} |f (\omega)|\, d\nu(\omega) }
\quad
&
\text{(when $r = 1$)}
\\
\\
\underset{\omega \in \Omega}{\text{\rm ess.sup} }|f (\omega)|
&
\text{(when $r = \infty$)}
\endcases
%\TAG{10.4}
\tag{\color{black}{10.1}}
%%%%%REDREDREDREDREDRE
\end{align*}
Here,
$$
\text{\rm ess.sup}_{\omega \in \Omega} |f (\omega)|
=
\sup \{ a \in {\mathbb R} \;|\;
\nu(\{ \omega \in \Omega \; :\; |f (\omega)| {\; \geqq \;}a
\;
%(\forall \omega \in \Omega)
\}) >0
\}
$$
%index{ess@$\text{\rm ess.sup}:$ }
% and .
%$L^r (\Omega, \nu)$
%,
%
%$L^r (\Omega)$
%$($
%, $L^r (\Omega, {\cal B}_{\Omega}, \nu)$
%$)$
% and  and .
%index{l^\infty@$L^\infty (\Omega, \nu)$,$L^\infty (\Omega, {\cal B}_{\Omega}, \nu)$}
%index{l^1@$L^1 (\Omega, \nu)$,$L^1 (\Omega, {\cal B}_{\Omega}, \nu)$}
\par
A function
$f$
$( \in L^\infty ( \Omega, \nu ))$
is said to be
essential continuous at
$\omega_0 (\in \Omega )$
if there exists a function
$g$
$( \in L^\infty ( \Omega, \nu ))$
that satisfies
the following
(b)
:
%index{@}
\begin{itemize}
\item[(b)]
$g : \Omega \to {\mathbb R}$
is continuous at
$\omega_0$
and
$\nu (\{ \omega \in \Omega \; | \; f( \omega ) \not= g( \omega ) \})=0$.
\end{itemize}
In this case,
the value $f(\omega_0)$
is defined
by
$g(\omega_0)$.

\par
\par
\vskip0.3cm
\par
\noindent
%\vskip-0.4cm%%%
%\begin{figure}[htbp]
\par
\noindent
%%%%%
\unitlength=0.28mm
%\unitlength=0.40mm
\begin{picture}(400,115)
\put(0,-15){
\put(27,18){0}
%\put(27,108){1}
\put(40,115){
%\LL{{}}
{not essentially continuous at $\omega_1$},
$\quad$
{essentially continuous at $\omega_2$}
%\omega_2$.
%$ 2^{\{ x_1, x_2, x_3 \} },$
%$ F)${FIG.$\;$}\RR
}
%\put(44,93){\tiny $[F(\{x_1\})](\omega)$}
%\put(170,88){\tiny $[F(\{x_2\})](\omega)$}
%\put(300,90){\tiny $[F(\{x_3\})](\omega)$}
%%%%%%%%%
\put(350,18){$(\Omega, \nu)$}
%\dottedline{3}(40,110)(340,110)
%\put(40,110){\line(1,0){300}}
\put(40,20){\line(0,1){100}}
%\linethickness{0.15mm}
%%\thicklines
\put(40,20){\line(1,0){300}}
%%%%%%%%%%\linethickness{0.15mm}
\put(150,80){\circle{4}}
\put(150,50){\circle*{4}}
\put(280,60){\circle{4}}
\put(280,80){\circle*{4}}
\thicklines
%\sspline(40,110)(60,108)(80,102)(100,80)
%(150,40)(200,30)(220,20)(240,20)
%\sspline(120,20)(130,20)(160,30)(250,50)
%(270,80)(280,100)(300,105)(340,110)
\spline(40,20)(60,22)(80,28)(100,50)
(150,80)
\spline(150,50)
(250,70)
(270,60)(279,60)
%(280,60)(300,75)(340,80)
\spline
(281,60)(300,75)(340,80)
\put(150,15){$\omega_1$}
\put(280,15){$\omega_2$}
}
\end{picture}
\begin{center}{Figure 10.1:
Essentially continuity
}
\end{center}
\par

\vskip0.3cm
%\vskip0.3cm

\rm
\par
\par
\noindent
%It is  to note that
\par
\noindent
%BBBBBBBBBBBBBBBBBB%SBSBSBS
{\small%%{\footnotesize
\begin{itemize}
\item[$\spadesuit$] \bf {{}}{Note }10.1{{}} \rm
It, of course,
holds
that
$C(\Omega) \subseteq L^\infty(\Omega ,\nu)$.
Also,
the space $L^\infty(\Omega ,\nu)$
is necessary
as a consequence
of quantum mechanics
( as mentioned in
\textcolor{black}{{Note }2.1}).
%, ( and ).
\end{itemize}
}
\par
Following \textcolor{black}{{Definition }2.1}
in
$C(\Omega)$,
we have the following definition:
\par
\noindent

%BFBF
\par
\noindent
{\bf {Definition }10.1
[{}{Observable}, state space, {{state}}, measured value space, measured value]}$\;\;$%POPOPO
%index{@observable }
%index{@}
%index{@state space }
%index{@{{state}}}
%index{@measured value }
%index{@measured value }
%%index{@probability }
\rm
The triple ${\mathsf{O}} {=}
(X, {\cal F}, F)$
is called an observable
in
$L^{\infty} (\Omega, \nu)$,
it it satisfies:
%t satisfies:
\begin{itemize}
\item[{\rm (i)}]
$X$is a set,
and
${\cal F}$
$( \subseteq {\cal P}(X) {{=}} \{ \Xi \; | \; \Xi \subseteq X \})$
is a $\sigma$-{{field}}.
\item[{\rm (ii)}]
The map $F : {\cal F}
\to
L^{\infty} (\Omega, \nu)$
satisfies::
\renewcommand{\footnoterule}{%
  \vspace{2mm}                      % 
  \noindent\rule{\textwidth}{0.4pt}   % , 
  \vspace{-5mm}
}
\begin{itemize}
\item[{\rm (a)}]
$\Xi \in {\cal F}$
$\Longrightarrow$
$F (\Xi) {\; \geqq \;}0$
$\;$
($\nu$-a.e.)
%\footnote{
%{\lq\lq
%a.e.
%\rq\rq}
%
%{\lq\lq}almost everywhere{\rq\rq},
%$F (\Xi) {\; \geqq \;}0$
%$\;$
%($\nu$-a.e.)
%$\Leftrightarrow$
%$\nu( \{ \omega \in \Omega \;|\; [F (\Xi)](\omega) < 0\})=0$.
%,
%%,
%,
%{\lq\lq}($\nu$-a.e.){\rq\rq} and 
%
%{\lq\lq}($\forall \omega \in \Omega $){\rq\rq}
% and 
%{\lq\lq}$({\roman a.e.}\;\; \omega \:)${\rq\rq}
% and  and .
%%index{almosteverywhere@a.e.}
%},
\item[{\rm (b)}]
$F (\emptyset) = 0$
and
$F (X) = 1$
($\nu$-a.e.),
\item[{\rm (c)}]
[Completely countability]:
For any countable decomposition
$\{\Xi_1, \Xi_2, \ldots, \Xi_n, \ldots\}$
$\bigm($
that is,
$\Xi = \bigcup\limits_{n=1}^\infty \Xi_n $, 
$\Xi_n \in {\cal F}, (n = 1, 2, \ldots)$,
$\Xi_m \cap \Xi_n =\emptyset \;\; (m \not= n )$
$\bigm)$
of
any $\Xi$
$(\in {\cal F} )
$,
it holds that
\begin{align*}
 \int_{\Omega} [F (\Xi)] (\omega) \rho (\omega) \, d\nu (\omega)
= \lim_{N \to \infty} \sum\limits_{n = 1}^{N}
\int_{\Omega} [F (\Xi_n)] (\omega) \rho (\omega)\, \nu (d \omega)
\;\;\;
%\\
%&
%\hspace{5cm}
( \forall \rho \in L^1 (\Omega, \nu))
\end{align*}
\end{itemize}
\end{itemize}
%\END{definition}
\par
\noindent
\rm
\par
\noindent
\rm
Here,
$\Omega$
(
or,
$(\Omega, {\cal B}_{\Omega}, \nu)$
)
is called a {\bf state space},
its element $\omega (\in \Omega )$
is called a state.
Also,
$X$
and
$x (\in X)$
is respectively called a {\bf measured value space}
and {\bf measured value }.
In addition,
if
${ F}({}\Xi{})$
$=$
$({ F}({}\Xi{}))^2$
holds for any
$ \Xi
({}\in {{\cal F}} {})$,
the
$(X, {{\cal F}} , { F} {})$
is called a projective observable.
%(
%,
%{\bf {observable }})
%index{@observable }
\par
\vskip0.2cm
\par
The following theorem is clear.
\par
\noindent
{\bf %2BFBF
{{Theorem }}10.2}$\;\;$%POPOPO
If ${\mathsf O}
$
${{=}}$
$(X, {\cal F}  , F{})$
is an observable in $C(\Omega)$,
the it is the observable in
$L^\infty( \Omega, {\cal B}_{\Omega},\nu)$.
%
%%\pa

\par
\vskip0.2cm
\par
The following examples
(
\textcolor{black}{Example 10.3$\text{--}$Example 10.5}
)
are similar to
the examples in
\textcolor{black}{{Chap.{\;}}2{}}.
\par
\noindent
{\bf %2BFBF
Example 10.3
[(i):{Exact observable} {}{\rm (cf. Example 2.5)}]}$\;\;$%POPOPO
%index{@{exact observable} }
Define the
{observable }
${\mathsf O}^{\FIN}
$
${{=}}$
$({}\Omega, {\cal B}_{\Omega}  , F^{\FIN}{})$
in
$L^\infty( \Omega, {\cal B}_{\Omega},\nu)$
such that
\par
\noindent
\begin{align*}
[F^{\FIN}(\Xi)](\omega)
=
\chi_{{}_\Xi} (\omega) =
\cases
1 \quad
& (
\omega \in \Xi ({}\in {\cal B}_{\Omega}{}) )\\
\\
0 &
(\omega \notin \Xi ({}\in {\cal B}_{\Omega}{})
).
\endcases
%%P%\tag{10.1}
\end{align*}
\par
\noindent
where
$\chi$
is the
{characteristic function}
%(\textcolor{black}{Appendix B.2(C)}).
This ${\mathsf O}^{\FIN}$
is called the exact observable.
%
%$
%${{=}}$
%$({}\Omega, {\cal B}_{\Omega}  , F^{\FIN}{})$
%in
%$L^\infty( \Omega, {\cal B}_{\Omega},\nu)$
%such that
%,  and {\bf {exact observable} } and :
As mentioned in
\textcolor{black}{Example 2.5},
the ${\mathsf O}^{\FIN}$
is no necessarily the observable in
$C(\Omega)$.

\rm

\par
\vskip0.2cm
\par
\normalsize
\baselineskip=18pt
\par
\noindent
{
\bf \vskip0.3cm
\vskip0.3cm
%BFBF
\par
\noindent
[{(ii):}Existence observable {\rm (cf. Example 2.2)}]}$\;\;$%POPOPO
%index{@existence observable }
As mentioned in \textcolor{black}{Example 2.2},
we can define
{\bf existence observable }
${\mathsf O}^{\roman{(exi)}} {{=}} (X , \{ \emptyset, X\},$
$
F^{\roman{(exi)}})$
in
$L^\infty (\Omega, \nu)$
such that
\begin{align*}
F^{\roman{(exi)}}({} \emptyset {}) = 0, \; \;
%F^{{\rm (exi)}}({}\{ 0 \}{}) {{=}} 0, \; \;
F^{\roman{(exi)}}({}
X{}) = 1, \; \;
%F^{{\rm (exi)}}({}\{ 0,1 \}{}) {{=}} 1 \; \;
\; \; (\text{$\nu$-a.e.}).\;
%\Big[
%\text{ resp.}
%\; \;
%F_0({}\{ 1 \}{}) \equv 1 \; \; \text{ in $L^\infty(\Omega_0{})$}
%\Big]
%P%\tag{10.2}
\end{align*}
%
%
%, {{measurement}}
%$
%{\mathsf M}_{L^\infty(\Omega)} ({\mathsf O}_{}
%{{=}}  ({}\{ 1\} , 2^{\{ 1\}}, $
%$F^{{\rm (exi)}}), $
%$ S_{[\omega]}{})
%$
%.
%\BEGIN{itemize}
%\ITEM[$(\sharp{})$]
%({}{{measurement}}
%${\mathsf M}_{L^\infty(\Omega)} ({\mathsf O}_{} ,  S_{[\omega]}{})$
%)measured value 
%$1$
%$({}\in \{ 0, 1 \}{})$
%probability 
%$1$
%{}
%
%{{}}measured value  {{}}$1$
%$({}\in \{ 0, 1 \})$.
%\END{itemize}
%\LL {{}}{{measurement}}
%${\mathsf M}_{L^\infty(\Omega)} ({\mathsf O}_{} ,$
%$  S_{[\omega]}{})$\RR
%{{}}
%%\LL to take no {{measurement}}\RR$\!\!\!,\;$
%%,
%{\LL}{\RR}$\!\!\!\!\; \;$
% and {}
\par
%\qed
\par
\noindent
\par
\noindent
{\bf %2BFBF
Example 10.4
[{}{{Normal observable}}{\rm (cf. Example 2.11)}{}]}$\;\;$%POPOPO
%index{@{{normal observable}}}
Put
$\Omega = {\mathbb R}(=\text{{{}}the real line})$
or,
$\Omega =
[a, b{}] $
(= the interval
$\; \subseteq {\mathbb R} )$.
Consider the
Lebesgue
measure
$m(d\omega) (= d \omega )$.
Assume that $\sigma>0$ is the standard deviation.
Define the {\bf {{normal observable}}}
${\mathsf O}_{G_\sigma} {{=}} ({\mathbb R}, {\cal B}_{\mathbb R}, G_{\sigma})$
in
$L^\infty (\Omega,m)$
such that
\begin{align*}
[G_{\sigma}(\Xi)] (\omega) = \frac{1}{\sqrt{2 \pi \sigma^2}}
\int_{\Xi} e^{- \frac{(x - \omega)^2}{2 \sigma^2}} dx
\quad (\forall \Xi \in {\cal B}_{\mathbb R}, \forall \omega \in \Omega)
%%%%%%3}
%P%\tag{10.3}
\end{align*}
This is also the observable in
$C(\Omega)$.
%(\textcolor{black}{
%%{\rm cf.}
%Example 2.11}).
\par
\vskip0.5cm
%\par
%\noindent
%\ssetlength{\unitlength}{0.7mm}
%%\BEGIN{picture}

%
%
%
%
%

\noindent
\par
\noindent
{\bf
%\vskip0.3cm
%\vskip0.3cm
%BFBF
\par
\noindent
Example 10.5
[Round observable{\rm (Example 2.6)}{}]}$\;\;$%POPOPO
%index{@observable }
\rm
%{(}{)}{{about-observable} } and ,
% and observable .
%.
Consider the state space $\Omega$
by the interval $ [0,100]$
with
the Lebesgue
measure
$d \omega$.
For each
$n  \in {\mathbb N}_{10}^{100}  {{=}} \{0,10,20,\ldots,100\}$,
define the function
$g_{n}:\Omega \to [0,1]$
such that
\begin{align*}
g_{n} (\omega)
=
\cases
0 & \quad (0 {{\; \leqq \;}}\omega {{\; \leqq \;}}n-5 ) \\
1
& \quad (n-5 {{\; < \;}}\omega {{\; \leqq \;}}n +5) \\
0 & \quad (n+5 {{\; < \;}}\omega {{\; \leqq \;}}100 )
\endcases
%P%\tag{10.4}
\end{align*}
%\omega\omegaxxxxxxxx
\par
\noindent
\par
\par
\noindent
\par
%\newpage
\par
\noindent
\par
\vskip0.9cm
\par
\noindent

\par
\noindent
% and ,
%$\{ g_n \}_{n \in {\mathbb N}_{10}^{100} }$
%$C([0,100])${the resosution the unity}{}
Here, define the
${\mathsf O}_{{\RND}}= (Y ({{=}} {\mathbb N}_{10}^{100})   , 2^Y, G_{{\RND}} )$
by
\begin{align*}
[G_{{\RND}}(\emptyset )](\omega ) = 0,
\quad
%&
[G_{{\RND}}(Y )](\omega ) = 1
\quad
%[F(\emptyset )](\omega ) = 0,
%\quad
[G_{{\RND}} (\Gamma )](\omega ) = \sum\limits_{n \in \Gamma } g_n (\omega )
\quad
(\forall \Gamma \in 2^Y=2^{{\mathbb N}_{10}^{100}  })
%P%\tag{10.5}
\end{align*}
Then,
the
${\mathsf O}_{{\RND}}= (Y( {{=}} {\mathbb N}_{10}^{100})   , 2^Y, G_{{\RND}} )$
is an projective observable in
$L^\infty ([0,100])$.
But, it is not an observable in
$C([0,100])$.
%(\textcolor{black}{Example 2.5}).

\rm
\baselineskip=18pt
\par
\noindent
\subsection{Axiom${}_{\text{\scriptsize b}}^{\text{\scriptsize p}}$ 1 ({{Bounded type measurement}})}%10.2
\label{10secAxiom 1}
%POIUYTREWQWERTYUIOPassoc
\subsubsection{Axiom${}_{\text{\scriptsize b}}^{\text{\scriptsize p}}$ 1 (bounded type)}%10.2.1
\par
%bounded type {{{measurement theory}}}\textcolor{black}{Axiom${}_{\text{\scriptsize b}}^{\text{\scriptsize p}}$ 1}({{measurement}}).
%continuous type {{{measurement theory}}} and  and ,
% and .
\par
\noindent
%\ssl
With any
\it
classical system
\rm
$S$,
a basic algebra
$ [C(\Omega), L^\infty (\Omega, \nu)]$
(or in short,
$L^\infty (\Omega, \nu )$
)
can be associated in which
measurement theory of that system can be formulated.
A
\it
state
\rm
of the system $S$
is represented by
a
state
$\omega (\in \Omega $,
i.e.,
a
\it
state space
${})$.
\rm
%a
%\it
%quantity
%\rm
%is
%represented by
%a self-adjoint element of ${\cal A}$,
%or generally,
%an $n$-tuple of the
%$(${}commutative $)$ self-adjoint elements of ${\cal A}$.
Also,
an
\it
observable
\rm
%$($
%which is a kind of generalization of a quantity
%$)$
\rm
is represented by
${\bold O}$
$\equiv$
$({}X , {\cal F} , F{})$
in
the
$L^\infty (\Omega, \nu )$.

%%index{idea@idea}
%%index{matter@matter}
%%index{form@form}
%index{@{{state}}}
%index{@observable }

The measurement
of
an observable
${\mathsf O}$
for
a system with a
{{}}{{state}}
$\omega$
is represented by
${\mathsf M}_{L^\infty(\Omega)} \big({}{\mathsf O} , S_{[\omega] } \big)$
(or,,
${\mathsf M}_{L^\infty(\Omega, \nu)} \big({}{\mathsf O} , S_{[\omega] } \big)$
).
%%index{measurement21@${\mathsf M}_{L^\infty(\Omega)}
%\big({}{\mathsf O} , S_{[{}\omega] } \big)$:
%(bounded type ){{measurement}}}
Also,
by
{{measurement}}
${\mathsf M}_{L^\infty(\Omega)} \big({}{\mathsf O} , S_{[{}\omega] } \big)$,
a
{\bf measured value }$x$
$({}\in X)$
is obtained.

\par

%\newpage
\rm

\vskip0.5cm
\par
By the same way of
Axiom${}_{\text{\scriptsize b}}^{\text{\scriptsize p}}$ 1 [{{(continuous type ) measurement}}]
in {Sec.2.2},
we can introduce Axiom${}_{\text{\scriptsize b}}^{\text{\scriptsize p}}$ 1
[{{(bounded type ) measurement}}]
as follow.
%%index{1a@Axiom${}_{\text{\scriptsize b}}^{\text{\scriptsize p}}$ 1 [{}{{measurement}}(bounded type )]}
\par
\noindent
\par
\noindent

\vskip0.5cm
\rm
%\newpage
%%index{measurement12@${\mathsf M}_{C (\Omega)} \big({}{\mathsf O},
%$
%$
%S_{[{}\ast] }(\nu)
%\big)$:{(}continuous type {}){{measurement}}}
%index{@{{measurement}}}
%
%\BEGIN{itembox}[c]
%\newpage
\par
\noindent
\begin{center}
{\bf
Axiom${}_{\text{\scriptsize b}}^{\text{\scriptsize p}}$ 1 (measurement :
bounded pure type)
}
\label{rule901}
\label{axiombp1}
\end{center}
%\END{document}
\par
\noindent
%\vskip0.1cm
\par
\noindent
\fbox{\parbox{155mm}{
\begin{itemize}
\item[]
Consider a measurement
${\mathsf M}_{L^\infty(\Omega)} \big({}{\mathsf O}{{=}} $
$(X, $
${\cal F} , F{})  , $
$
S_{[{}\omega] } \big)$
formulated in
a {basic algebra}
$[C(\Omega ),  L^\infty (\Omega, \nu)]$.
The probability that
a
{{}}measured value
$ x$
$({}\in X  {})$
obtained by
the
{{measurement}}
${\mathsf M}_{L^\infty(\Omega)}  \bigl({}{\mathsf O}  , S_{[{}\omega{}] } \bigl)$
belongs to
$ \Xi $
$({}\in  {\cal F}{})$
is give by
$[F(\Xi)](\omega)$,
if
$F(\Xi)$
is essentally continuous at
$\omega$.
%\END{itemize}

\end{itemize}
}
}
\par
\vskip0.5cm
\par
\noindent

%
%
%
%\BEGIN{itembox}[c]{
%\bf
%Axiom${}_{\text{\scriptsize b}}^{\text{\scriptsize p}}$ 1 [(bounded pure type ){ {measurement}}{}
%}
%\label{axiomcp1b}
%\label{rule901}
%\label{axiombp1}
%%index{1@Axiom${}_{\text{\scriptsize b}}^{\text{\scriptsize p}}$ 1[{}{{measurement}}{}(bounded type )]}
%%\BEGIN{itemize}
%%\item[(i)]
%%{{measurement}}.
%%\item[]
%%%(ii)]
%%{basic algebra}
%%$ L^\infty (\Omega, \nu)$
%%%$C^*$-
%%$L^\infty (\Omega, \nu)$
%%{{measurement}}
%Consider a measurement
%${\mathsf M}_{L^\infty(\Omega)} \big({}{\mathsf O}{{=}} $
%$(X, $
%${\cal F} , F{})  , $
%$
%S_{[{}\omega] } \big)$
%formulated in
%a {basic algebra}
%$[C(\Omega ),  L^\infty (\Omega, \nu)]$.
%The probability that
%a
%{{}}measured value
%$ x$
%$({}\in X  {})$
%obtained by
%the
%{{measurement}}
%${\mathsf M}_{L^\infty(\Omega)}  \bigl({}{\mathsf O}  , S_{[{}\omega{}] } \bigl)$
%belongs to
%$ \Xi $
%$({}\in  {\cal F}{})$
%is give by
%$[F(\Xi)](\omega)$,
%if
%$F(\Xi)$
%is essentally continuous at
%$\omega$.
%%\END{itemize}
%\END{itembox}
%

It is a matter of course that
Axiom${}_{\text{\scriptsize b}}^{\text{\scriptsize p}}$ 1 should be use
according to
the Copenhagen interpretation
[\textcolor{black}{[(U$_1$)--(U$_7$)}]
in Chap. 1.

%\noindent
%\normalsize
%\baselineskip=18pt
\vskip0.3cm
\vskip0.3cm
%BFBF
\par
\noindent
{\bf Remark 10.6}$\;\;$%POPOPO
When
$F(\Xi)$
is not essentially continuous
at
$\omega (\in \Omega )$,
the
sample probability space
$(X,{\cal F}, [F(\cdot)](\omega))$
is not
a mathematical probability space.
%(\textcolor{black}{Appendix B.5(B)}).
% and {Remark }(cf. \textcolor{black}{\cite{Keio}}).

\rm
\subsubsection{Simple examples
---
Urn problem and so on}%{Sec.10.2.2}
\par
\baselineskip=18pt

The following examples
(
\textcolor{black}{Example 10.7, Example 10.8}
)
are similar to
the examples in
\textcolor{black}{{Chap.{\;}}2{}}.

The bounded type {{{measurement theory}}}
is similar to
continuous type.
Thus, we introduce only typical examples.
%\textcolor{black}{Axiom${}_{\text{\scriptsize b}}^{\text{\scriptsize p}}$ 1}.

\par
\noindent
\par
\noindent
{
\bf %2BFBF
Example 10.7
[The exact measurement of temperature{{}}{\rm
(\textcolor{black}{{Sec.1.2}(V)})}]}$\;\;$%POPOPO
\rm
Consider the exact measurement
for the water with
the temperature
$\omega_0 \SD$.
Put
$\Omega = X=[0,100]$.
Let
Lebesgue measure space $(\Omega, {\cal B}_{\Omega}, m )$
be a state.
Consider the
{exact observable}
${\mathsf O}^{\FIN}$
$=$
$(X, {\cal B}_{X}, F^{\FIN})$
in
$L^\infty ( \Omega , m )$.
Then, wa say that
\begin{itemize}
\item[]
Assume that
a measured value $x_0 (\in X {{=}} \Omega )$
is obtained by]the exact
{{{measurement}}}
${\mathsf M}_{L^\infty(\Omega)}( {\mathsf O}^{\FIN}, S_{[\omega_0]})$.
Then,
we can surely say
that
$x_0=\omega_0$.
\end{itemize}
That is because:
\begin{itemize}
\item[]
Let
$D (\subseteq \Omega =X)$
be any open set such that
$\omega_0 \in D$.
Then,
the probability that
$x_0 \in D$
is given by
$[F^{\FIN}(D)](\omega_0)$
$=$
$\chi_{_D} (\omega_0)=1$.
Therefore, from the arbitrarity of
$D$,
we see that
$x_0=\omega_0$.
\end{itemize}
This completes the proof.
\qed
%
%\def\BC{\left[\begin{array}{ll}}
%\def\EC{\end{array}\right.}
%%\rm
%%%\vskip1.0cm
%%\par
%\par
%For example,
%assuming that
%$\omega_0 =35 \SD$,
%As similar as and 
%\textcolor{black}{{Chap.$\;$1}(V)}),
%we see that
%\BEGIN{itemize}
%\item[]
%When an {\bf observer}
%measures by the exact thermometer ${\mathsf O}^{\FIN}$
%for the
%water
%(i.e.,
%{\bf {measuring object}}
%)
%with
%the {\bf state}
%(
%the 35
%$\SD$
%),
%the
%{\bf probability}
%that
%{\bf measured value }[35]
%is obtained
%is
%$1$.
%\END{itemize}
%%

\par
\vskip0.2cm
\par
The following is
a bounded type {{measurement version}}
of
\textcolor{black}{Example 2.10}.
{}
%\par
%\noindent
%index{@urn problem}
{\bf
%\vskip0.3cm
%\vskip0.3cm
% BFBF
\par
\noindent
Example 10.8
[\textcolor{black}{Example 2.10}
{{}}[Bounded type {{measurement version}}
of the urn problem]}$\;\;$%POPOPO
\rm
There are two urns ${U}_1$
and
${U}_2$.
The urn ${U}_1$ [resp. ${U}_2$]
contains
8 white and 2 black balls
[resp.
4 white and 6 black balls]
%where
%$N$ is sufficiently large number.
({\it cf.} \textcolor{black}{Fig. 2.5}).
\par
\noindent
[I]:
Here,
consider the following phenomenon:
%{\lq\lq}measurement
%\textcolor{black}{(B$_1$)}":
\begin{itemize}
\item[(a)]
Pick out one ball at random
from the
urn $U_2$.
Then
the probability that
the ball is white
is given by
$0.4$.
\end{itemize}
This statement in ordinary language
will be translated to the statement (b) of measurement theory
in what follows.
%
%(The term
%"at random" will be often omitted in this book.)
%In measurement theory,
%the {\lq\lq}measurement
%\textcolor{black}{(B$_1$)}{\rq\rq}
%is formulated as follows:
%%%{U}_1{\omega}_2{\omega}_3%%%{\omega}_1{\omega}_2{\omega}_3
%%%{\omega_1{\omega_2{\omega_3%%%{\omega_1{\omega_2{\omega_3
%\BEGIN{align*}
%\omega_1 = [{}8:2], \quad
%\omega_2 = [{}4:6], \quad
%\omega_3 = [{}1:9]. \quad
%\END{align*}
%Thus,

Consider the state space
$\Omega =\{ \omega_1. \omega_2 \}$
with the discrete metric $d_D$
and with the measure $\nu$
such that
$$
\nu(\{ \omega_1 \})=1,
\qquad
\nu(\{ \omega_2 \})=1.
$$
(It is also possible to assume that $\nu(\{ \omega_1 \})=2$
and
$\nu(\{ \omega_2 \})=3$
).
Further,
assume that
\begin{align*}
U_1 \quad \cdots \quad
&
\text{{\lq\lq}the urn with the state $\omega_1
${\rq\rq}}
\\
U_2 \quad \cdots \quad
&
\text{{\lq\lq}the urn with the state $\omega_2
${\rq\rq}}
%\\
%U_3 \quad \cdots \quad
%&
%\text{{\lq\lq}the urn with the state $\omega_3${\rq\rq}}
%\tag{1.46} 
\end{align*}
And thus,
we consider the following identification:
%the state space
%$\Omega$
%by
%$\Omega = \{ {\omega}_1 , {\omega}_2 \}$.
%That is,
%we assume the identification;
\begin{align*}
U_1 \approx \omega_1, \quad
U_2 \approx \omega_2, \quad
%U_3 \approx \omega_3. \quad
%\tag{1.47} 
\end{align*}
\par
\noindent
%%TUBO
Put
{\lq\lq}$w${\rq\rq} = {\lq\lq white\rq\rq}$\!\!,\;$
{\lq\lq}$b${\rq\rq} = {\lq\lq black\rq\rq}$\!\!\;$,
and put
$X=\{w,b\}$.
%\BEGIN{itemize}
%\item[$(Q)$]
%\it
%Which is the chosen urn, ${\omega}_1$ or ${\omega}_2${}?
%\END{itemize}
%\rm
%
%%\rm
%
%\noindent
%[{\it Answer}].
%We regard $\Omega$
%$\big(\equiv$
%$\{ \omega_1 , \omega_2 \}
%\big)$ as the state space.
And define the observable
${\mathsf O}
\big(\equiv (X \equiv \{w,b\}, 2^{\{w,b\}}, F)
\big)$
in $L^\infty (\Omega, \nu)$
by
\begin{align*}
[F(\{w\})](\omega_1) = 0.8,& \qquad \qquad [F(\{b\})](\omega_1) = 0.2,
\\
{}[F(\{w\})](\omega_2) = 0.4,& \qquad \qquad [F(\{b\})](\omega_2) = 0.6.
%
%\tag{1.48}
\end{align*}
Thus,
we have the measurement
%
%
%we see that
%\BEGIN{align*}
%\text{\textcolor{black}{(B$_1$)}}
%=
${\mathsf M}_{L^\infty ({}\Omega{}, \nu )} ({}{\mathsf O} ,
S_{[{} {\omega_2}]})$

Here,
\textcolor{black}{Axiom${}_{\text{\scriptsize b}}^{\text{\scriptsize p}}$ 1(page \pageref{axiombp1})}
says that
\begin{itemize}
\item[(b)]
the probability
that
a measured value ${b}$
is obtained by
${\mathsf M}_{C({}\Omega{})} ({}{\mathsf O} ,
S_{[{} \delta_{\omega_2}]})$
is given
by
\begin{align*}
F(\{b\})(\omega_1) = 0.6
%\tag{1.50} 
\end{align*}
\end{itemize}
\par

\noindent
\par
\vskip0.5cm
\par
Now we can present the following theorem.
\def\FIN{{\roman{(exa)}}}
\def\EXI{{\roman{(exi)}}}

{\bf
%\vskip0.3cm
%\vskip0.3cm
%BFBFFIFINFINFINPPPPPPPPP
\par
\noindent
{{Theorem }}10.9
[Exact {{measurement}}]}$\;\;$%POPOPO
\rm
Consider the exact observable
${\mathsf O}^{\FIN}=(X,{\cal F}, F^{\FIN})$
in
$L^\infty(\Omega,\nu)$.
Assume that
a
measured value
$x$
$(\in X)$
is obtained by
the exact
{{measurement}}
${\mathsf M}_{L^\infty (\Omega, \nu)}({\mathsf O}^{\FIN}, S_{[\omega_0]} )$.
The,
the probability
that
$x=\omega_0$
is equal to
$1$.
\par
\noindent
{\it $\;\;\;\;${Proof.}}$\;\;$
The proof is the same as the proof in Example 10.7.
Let
$D (\subseteq \Omega =X)$
be any open set such that
$\omega_0 \in D$.
Then,
the probability that
$x_0 \in D$
is given by
$[F^{\FIN}(D)](\omega_0)$
$=$
$\chi_{_D} (\omega_0)=1$.
Therefore, from the arbitrarity of
$D$,
we see that
$x_0=\omega_0$.
This completes the proof.
\qed

\par

\rm
\subsection{System quantity
---
The origin of observables}%{Sec.10.3}
%%index{@system quantity(POI)}
\baselineskip=18pt\par
\par
In classical mechanics,
the term
"observable"
usually means
the continuous real valued function on
a state space
(that is,
{physical quantity}{)}
\rm
\rm
%,
%({} 2.7{})
An observable in measurement theory
is characterized as the natural generalization
of
the physical quantity.
This will be explained
in the following examples.
\par
\noindent
%\vskip0.3cm
%BFBF
\renewcommand{\footnoterule}{
  \vspace{2mm}                      % 
  \noindent\rule{\textwidth}{0.4pt}  
  \vspace{-3mm}
}
\par
\noindent
{\bf Example 10.10
[System quantity]}$\;\;$%POPOPO
%That is,
%{basic algebra}
Let ${L^\infty ( \Omega, \nu )}$
be a basic algebra.
A continuous real valued function
${\widetilde f}{}: \Omega \to {\mathbb R}$
(
or generally,
a measurable real valued function
${\widetilde f}{}: \Omega \to {\mathbb R}^n$
)
is called a system quantity
(or in short, quantity)
on $\Omega$.
Define the projective observable
${\mathsf O}=({}{\mathbb R}, {\cal B}_{\mathbb R} , F{})$
in
${L^\infty ( \Omega, \nu )}$
such that
\begin{align*}
[{}F({}\Xi{})] (\omega)
=
\left\{
\begin{array}{l}
1 \quad \text{when } \omega \in {\widetilde f}^{-1}({}\Xi{}) \
\\
\\
0 \quad \text{when } \omega \notin {\widetilde f}^{-1}({}\Xi{})
\end{array}
\right.
\qquad
(
\forall \Xi \in {\cal B}_{\mathbb R}
%,
%
)
%P%\tag{10.13}
\end{align*}
Here,
note that
\begin{align*}
{\widetilde f}
(\omega)
=
%&
\lim_{N \to \infty }
\sum\limits_{n=-N^2}^{N^2} \frac{n}{N}
\left[F \bigm( [\frac{n}{N},\frac{n+1}{N} ) \bigm)
\right](\omega)
=
\int_{\mathbb R} \lambda [F({}d \lambda{})
](\omega)
%\Big)
\qquad
%
%%%
%P%\tag{10.14}
\end{align*}
Thus, we have the following identification:
\begin{align*}
\underset{{\text{\scriptsize ({}system quantity on $\Omega$)}}}{{\widetilde f}}
{{\longleftrightarrow}}
\underset{{\text{\scriptsize (projective observable in $L^\infty(\Omega,\nu)$})}}{{\mathsf O}=({}{\mathbb R}, {\cal B}_{\mathbb R} , F{})
}
%{{\text{  in  }
%L^\infty(\Omega{},\nu) }
%\atop{}}.
%%%%P%\tag{10.15}
\tag{\color{black}{10.2}}
%%%%%REDREDREDREDREDRE
\end{align*}
%%%%{2.32}
\par
\noindent
This ${\mathsf O}$
is called the observable representation of
a system quantity${\widetilde f}$.
%index{@observable }
Therefore, we say that
%,
%34) and 36), genera
\begin{itemize}
\item[(a)]
An observable in measurement theory
is characterized as the natural generalization
of
the physical quantity.
\end{itemize}
Here,
recall the identification
\textcolor{black}{(3.2)},
which is the quantum version of
the identification
\textcolor{black}{(10.2)}.
%{Remark }.

\vskip0.2cm
\renewcommand{\footnoterule}{%
  \vspace{2mm}                      % 
  \noindent\rule{\textwidth}{0.4pt}   % , 
  \vspace{-5mm}
}
\noindent
{\bf %2BFBF
Example 10.11
[Position {observable }, momentum {observable },
energy {observable }{}]}$\;\;$%POPOPO
%index{@observable }
%index{@observable }
%index{@{observable }}
Consider Newtonian mechanics
in
the {basic algebra}
$L^\infty (\Omega, \nu)$.
For simplicity,
consider the two dimensional
space
$$
\Omega
=
{\mathbb R}_q\times{\mathbb R}_p 
{{=}}
\{ (q,p)=(
\text{position},
\text{momentum})
\; | \; q,p\in{\mathbb R}\}
$$
%\renewcommand{\footnoterule}{
%  \vspace{2mm}                      % 
%  \noindent\rule{\textwidth}{0.4pt}  
%  \vspace{-3mm}
%}
%%{{ and }}
%%$({\mathbb R}_q^s\times{\mathbb R}_p^s)$
%{}\footnote{
%,
%$1$-,
%$\Omega={\mathbb R}^6
%=\{ (q_x, q_y, q_z, p_x, p_y ,p_z )\}$.
%$N$-,
%$\Omega ={\mathbb R}^{6N}$.
%}.
%%index{@}
The following quantities are fundamental:
%{}
\begin{align*}
(\sharp_1):
&
{\widetilde q}:\Omega \to {\mathbb R},
\quad
&{\widetilde q}
(q,p)=&q \quad (\forall (q,p) \in \Omega )
\\
(\sharp_2):
&
{\widetilde p}:\Omega \to {\mathbb R},
\quad
&{\widetilde p}(q,p)=&p \quad (\forall (q,p) \in \Omega )
\\
(\sharp_3):
&
{\widetilde e}:\Omega \to {\mathbb R},
\quad
&
{\widetilde e}
(q,p)=&
{
\text{[potential energy ]}+
\text{[kinetic energy ]}
}
\\
&
&\quad =&
\underset{\text{\scriptsize (Hamiltonian)}}{U(q)+ \frac{p^2}{2m} }
\qquad (\forall (q,p) \in \Omega )
%P%\tag{10.16}
\end{align*}
where, $m$ is the mass of a particle
%, $g$
Under the identification (10.2),
$(\sharp_1)$
and
$(\sharp_2)$
is respectively called
a position observable
and a momentum observable.
%
%\textcolor{black}{(10.2)}
%,
%
%{observable }, {observable },
%{observable }
% and .
%\qed
\par
\noindent
%%%%%%%%%

\par
\vskip0.5cm
\rm
\par
%\textcolor{black}{{Note }10.2}
%,
%,
%\textcolor{black}{{Chap. 3}1} and 
%
% and ,  and 
%
%,

% and .
In what follows,
we shall introduce
Schr\"odinger
equation.
\par
\noindent
%BBBBBBBBBBBBBBBBBB%SBSBSBS
{\small%%{\footnotesize
\vspace{0.1cm}
\begin{itemize}
\item[$\spadesuit$] \bf {{}}{Note }10.2{{}} \rm
The Hamiltonian
${\cal H}(q,p)=\frac{p^2}{2m}+ U(q) $
produces classical and quantum kinetic equations.
\begin{itemize}
\item[$(\sharp_1)$]
Classical case:[Newtonian equation]
$$
\text{"the simple case of (9.1)"}=
\cases
\frac{dp}{dt}=-\frac{{\cal H}(q,p)}{\partial q}=- \frac{dU}{dq}
\\
\frac{dq}{dt}=\frac{{\cal H}(q,p)}{\partial p}=\frac{p}{m}
\endcases
$$
\item[$(\sharp_1)$]
quantum case:[Schr\"odinger equation]
%we obtain the Schr\"odinger equation:
%%index{Scgredinger@Schr\"odinger equation}
\begin{align*}
{ \hbar \sqrt{-1}}
\frac{\partial u_t(q)}{\partial t }
=
{\cal H}({}q,  \frac{\hbar}{\sqrt{-1}} \frac{ \partial}{
{}
\partial q })u_t(q)
=
- \frac{\hbar^2}{2 m} \frac{ \partial^2 u_t(q)}{ \partial q^2 }
+
U(q) u_t(q)
%\TAG{8}
\end{align*}
\end{itemize}
The solution
$\{ u_t \}_{t \in {\mathbb R}}$
of Schr\"odinger equation
represents
the
{{state}} change.
\end{itemize}
}

\par
\vskip0.6cm
\par
For each
$k =1,2,\ldots,n$,
consder an
{observable } ${\mathsf O}_k$
$=$
$(X_k , $
${\cal F}_k , $
$F_k{})$
in
$L^\infty(\Omega,\nu)$.
And consider
the simultaneous observable $\bigtimes_{k=1}^n {\mathsf O}_k$
$=$
$({}\bigtimes_{k\in K } X_k ,$
$ \bigstimes_{k=1}^n{\cal F}_k , $
$
\bigtimes_{k=1}^n  F_k
)$,
which is defined by the same way of
\textcolor{black}{{Definition }2.14}.

The following theorem is clear.
%%index{urn problem@urn problem}
{\bf
%\vskip0.3cm
%\vskip0.3cm
%BFBF
\par
\noindent
{{Theorem }}10.12
[Exact {{measurement}}
and system quantity]}$\;\;$%POPOPO
{}\rm{}
Let
${\mathsf O}^{\roman{(exa)}}_0=(X,{\cal F}, F^{\roman (exa)})$
(i.e.,
$(X,{\cal F},$
$F^{\roman (exa)})=(\Omega,{\cal B}_{\Omega}, \chi )$
)
be the exact observable in
$L^\infty(\Omega,\nu)$.
Let
${\mathsf O}_1=({\mathbb R},{\cal B}_{\mathbb R}, G)$
be the observable
that is induced by
a quantity
${\widetilde g}:\Omega \to {\mathbb R}$
as in \textcolor{black}{Example 9.6}.
Consider the simultaneous observable
${\mathsf O}^{\roman{(exa)}}_0 \bigtimes
{\mathsf O}_1
$.
Let
$(x,y)$
$(\in X \times {\mathbb R})$
be a measured value obtained by
the simultaneous measurement
${\overline{\mathsf M}}_{L^\infty (\Omega, \nu)}(
{\mathsf O}^{\roman{(exa)}}_0\bigtimes
{\mathsf O}_1
, S_{[\delta_\omega]} )$.
Then,
we can surely believe
that
$x=\omega$,
and
$y= {\widetilde g}(\omega)$.
%\END{The}
%BFBF
\par
\noindent
{\it $\;\;\;\;${Proof.}}$\;\;$
\rm
Let
$D_0
(\in {\cal B}_{\Omega} )$
be any open set
such that
$\omega (\in \Omega
{{=}}
X )$.
Also,
let
$D_1 (\in {\cal B}_{\mathbb R})$
be any open set
such that
${\widetilde g}(\omega)
\in$
$D_1$.
The probability
that
a measured value
$(x,y)$
obtained by
the measurement
${\overline{\mathsf M}}_{L^\infty (\Omega, \nu)}(
{\mathsf O}^{\roman{(exa)}}_0
\bigtimes
{\mathsf O}_1
, S_{[\delta_\omega]} )$
belongs to
$D_0 \times D_1$
is given by
$\chi_{_{D_0}}(\omega) \cdot \chi_{_{{\widetilde g}^{-1}(D_1)}} (\omega )=1$.
Since
$D_0$
and $D_1$
are arbitrary,
we can surely believe
that
$x=\omega$
and
$y={\widetilde g}(\omega)$.
\qed

%\vskip0.3cm

%\qed

\subsection{{{{Measurement theoretical }}}Kolmogorov extension theorem}%10.4
\par
We consider that
\begin{itemize}
\item[(a)]
The utility of
Kolmogorov extension theorem
in probability theory
is due to
\textcolor{black}{{{Theorem }}10.13}
({{{measurement theory}}} version of Kolmogorov extension theorem)
\end{itemize}
%That is, , ,
%%,
%{{{measurement theory}}}Kolmogorov extension theorem
%, {{measurement theory}}.
This will be asserted in this section.

\par
In this section we study
\LL Kolmogorov's extension theorem\RR
in the measurement theory.
%({}$W^*$-algebraic{}) Statistical MT.
It is generally said that
Kolmogorov's extension theorem
is most fundamental in
Kolmogorov's probability theory.
That is because
this theorem
assures
the existence of
a probability space
({}i.e.,
sample space{}).
On the other hand,
our theorem
(=
Theorem 10.1, i.e.,
Kolmogorov's extension theorem in measurement theory)
assures the
existence of
a measurement
({}or,
observable{}).
Recall the our spirit
$\Big($see Remark ({}in $\S2.3$(I){})$\Big)$:
\begin{itemize}
\item[$(\sharp)$]
there is no probability without measurements.
\end{itemize}
Thus,
in measurement theory,
the concept of
\LL measurement\RR is more fundamental
than
that of
\LL sample space\RR$\!\!\!\!.\; \;$
Therefore,
this theorem
({}i.e.,
Kolmogorov's extension theorem in measurement theory)
is very important in
measurement theory.
%We first introduce
%\LL the
%$W^*$-algebraic generalization of Kolmogorov's extension theorem\RR$\!\!\!\!.\; \;$
That is,
this theorem
({}= Theorem 10.1{}) is essential to measurement theory
just like
Kolmogorov's extension theorem
is so in his probability theory.
Using this theorem,
we can define
\LL particle's trajectory\RR by
\LL the sequence of measured values\RR$\!\!\!\!.\; \;$
And further we prove:
\begin{itemize}
\item[(i)]
the existence of \LL particle's trajectory\RR in Newtonian mechanics,
\item[(ii)]
the existence of Brownian motion.
\end{itemize}
Thus,
we can understand
the difference between
the concepts of
\LL particle's trajectory\RR
and
\LL state's evolution\RR
in both classical and quantum mechanics.
%The readers will see
%that,
%from the mathematical point of view,
%the $W^*-$algebraic formulation
%is more handy
%than
%the $C^*-$algebraic formulation.

\par
%\vskip0.5cm
\par

\baselineskip=18pt\par

\par
   Let ${\widehat \Lambda}$ be an index set.
For each $\lambda\in{\widehat \Lambda}$, consider a set $X_\lambda$.
For any subsets
$\Lambda_{1} \subseteq \Lambda_{2}({} \subseteq {\widehat \Lambda}{})$,
$\pi_{\Lambda_{1},\Lambda_{2}}$ is the natural projection such that:
\begin{align*}
\pi_{\Lambda_{1},\Lambda_{2}}:
\mathop{\mbox{\Large $\times$}}_{\lambda\in\Lambda_{2}}X_{\lambda}
\longrightarrow
\mathop{\mbox{\Large $\times$}}_{\lambda\in\Lambda_{1}}X_{\lambda}.
\end{align*}
Especially, put $\pi_{\Lambda}=\pi_{\Lambda,{\widehat \Lambda}}$.
For each $\lambda\in{\widehat \Lambda}$, consider an observable
$(X_{\lambda},{\cal F}_{\lambda},F_{\lambda})$ in
% $W^{*}$-algebra
$L^\infty(\Omega, \nu)$.
Note that the  quasi-product observable
$\overline{\bold O}$
$\equiv$
$(${}$\mbox{\Large $\times$}_{\lambda\in{\widehat \Lambda}}X_{\lambda},$
$\mbox{\Large $\times$}_{\lambda\in{\widehat \Lambda}} {\cal
F}_{\lambda},$
$F_{\widehat \Lambda}${}$)$
of
$\{$
$(X_{\lambda}, {\cal F}_{\lambda},$ $F_\lambda)$
$\; | \;$ ${\lambda}\in{\widehat \Lambda}$
$\}$
is characterized as the observable such that:
\begin{align*}
F_{\widehat \Lambda} ({}\pi_{\{ \lambda \}}^{-1} ({}\Xi_{\lambda}{}))
=
F_{\lambda}(\Xi_{\lambda})
\qquad
({}
\forall \Xi_\lambda \in {\cal F}_\lambda,
\forall {\lambda }\in{\widehat \Lambda}{}),
\tag{10.2}
\end{align*}
though
the existence and the uniqueness of a quasi-product
observable are not guaranteed in general.
The following theorem says something about
the existence and uniqueness of
the quasi-product observable.

\par
Let ${\widetilde \Lambda}$
be a set.
For each
$\lambda\in{\widetilde \Lambda}$,
consider a set $X_\lambda$.
For any subset
$\Lambda_{1} \subseteq \Lambda_{2}({} \subseteq {\widetilde \Lambda}{})$,
define the natural map
%index{@}
$
\pi_{\Lambda_{1},\Lambda_{2}}:
\bigtimes_{\lambda\in\Lambda_{2}}X_{\lambda}
\longrightarrow
\bigtimes_{\lambda\in\Lambda_{1}}X_{\lambda}
$
by
% {{}}natural operator such that:
\begin{align*}
\bigtimes_{\lambda\in\Lambda_{2}}X_{\lambda}
\ni
(x_\lambda )_{\lambda \in \Lambda_2 }
\mapsto
(x_\lambda )_{\lambda \in \Lambda_1}
\in
\bigtimes_{\lambda\in\Lambda_{1}}X_{\lambda}
%
%\longrightarrow
%\bigtimes_{\lambda\in\Lambda_{1}}X_{\lambda}.
%%%
%P%\tag{10.17}
\tag{\color{black}{10.3}}
\end{align*}
%, $\pi_{\widetilde \Lambda}=
%\pi_{\widetilde \Lambda,{\widetilde \Lambda}}$ and .
%$\lambda\in{\widetilde \Lambda}$,
%${L^\infty(\Omega, \nu)}$
%
%{observable }
%$(X_{\lambda},{\cal F}_{\lambda},F_{\lambda})$
%.
%{quasi-product }{observable }
%${\mathsf O}$
%${{=}}$
%$(${}$\mbox{\Large $\times$}_{\lambda\in{\widetilde \Lambda}}X_{\lambda},$
%$\mbox{\Large $\times$}_{\lambda\in{\widetilde \Lambda}} {\cal
%F}_{\lambda},$
%$F_{\widetilde \Lambda}${}$)$
%(of
%$\{$
%$(X_{\lambda}, {\cal F}_{\lambda},$ $F_\lambda)$
%$\; | \;$ ${\lambda}\in{\widetilde \Lambda}$
%$\}$)
%{observable }:
%\BEGIN{align*}
%F_{\widetilde \Lambda} ({}\pi_{\{ \lambda \}}^{-1} ({}\Xi_{\lambda}{}))
%=
%F_{\lambda}(\Xi_{\lambda})
%\qquad
%({}
%\forall \Xi_\lambda \in {\cal F}_\lambda,
%\forall {\lambda }\in{\widetilde \Lambda}{}),
%%%
%TAG{10.19}
%\END{align*}
%%though
%%{{}}  and {{}}uniqueness of a {quasi-product }{observable }
%are .
\par
\vskip1.0cm
\par
The following theorem
guarantees
the existence and uniqueness
of
{observable }.
It should be noted that
this is due to
\textcolor{black}{{Chap.$\;$1}the Copenhagen interpretation(U$_4$)}),
i.e.,
"only one measurement is permitted".
\par
\vskip0.2cm
\par
%index{@Kolmogorov extension theorem}
\noindent
%BFBF-
{\bf {{Theorem }}10.13
[{}{{{measurement theoretical version }}}of
Kolmogorov extension theorem
\rm
(
\textcolor{black}{{\rm cf.}
\cite{Keio}{}}
)
\bf
)
]}$\;\;$%POPOPO
\rm
For each $\lambda\in{\widehat \Lambda}$, consider
   a Borel measurable space
   $({}X_\lambda , {\cal F}_{\lambda}{})$,
   where
   $X_\lambda$
   is
   a separable complete metric space.
Define the set
${\cal P}_0 ({\widehat \Lambda})$
such as
$ {\cal P}_0 ({\widehat \Lambda}) \equiv
\{
\Lambda \subseteq {\widehat \Lambda} \;  | \;
\Lambda ~\mbox{\rm is finite  }
\}
$.
      Assume that the family of
     the observables
     $\bigl\{$
     ${\overline{\bold O}}_{\Lambda} \equiv$
     $(${}     $\mathop{\mbox{\Large $\times$}}_{\lambda\in\Lambda}X_{\lambda},
$
     $\mathop{\mbox{\Large $\times$}}_{\lambda\in\Lambda}{\cal
F}_{\lambda},$
     $F_{\Lambda}$
     $)$
     $~|~$
     $\Lambda\in{\cal P}_0 ({\widehat \Lambda})$
     $\bigr\}$
     in $L^\infty(\Omega, \nu)$
     satisfies the following {\lq\lq consistency
condition\rq\rq}$:$
       \begin{itemize}
       \item[$\bullet$]
       for any $\Lambda_1$, $\Lambda_2$
$\in$
${\cal P}_0 ({\widehat
\Lambda})$
       such that $\Lambda_1 \subseteq \Lambda_2$,
       \end{itemize}
\begin{align*}
       F_{\Lambda_2}
       \bigl({}
       \pi_{\Lambda_{1},\Lambda_{2}}^{-1}({\Xi}_{\Lambda_1}{})
       \bigr)
       =
       F_{\Lambda_1}
       \bigl({}{\Xi}_{\Lambda_1} \bigr)
       \quad
       ({}\forall {\Xi}_{\Lambda_1} \in
       \mathop{\mbox{\Large $\times$}}_{\lambda\in\Lambda_1}
       {\cal F}_{\lambda}{}).
\tag{10.4} \end{align*}
\par
\noindent
Then, there uniquely exists the observable
${\widetilde{\bold O}}_{\widehat{\Lambda}}$
$\equiv$
$\bigl(\mathop{\mbox{\Large $\times$}}_{\lambda\in{\widehat
\Lambda}}X_{\lambda}, $
$\mathop{\mbox{\Large $\times$}}_{\lambda\in{\widehat \Lambda}}
{\cal
F}_{\lambda},$
${\widetilde F}_{\widehat \Lambda}
\bigr)$
in $L^\infty(\Omega, \nu)$ such that:
\begin{align*}
       {\widetilde F}_{\widehat \Lambda}
       \bigl({}
       \pi_{\Lambda}^{-1}({\Xi}_{\Lambda}{})
       \bigr)
       =
       F_{\Lambda}  \bigl({}{\Xi}_{\Lambda} \bigr)
       \quad
       ({}\forall {\Xi}_{\Lambda} \in
       \mathop{\mbox{\Large $\times$}}_{\lambda\in\Lambda}
       {\cal F}_{\lambda},~
       \forall\Lambda\in{\cal P}_0 ({\widehat \Lambda}){}).
       %%\t
\tag{10.5}
\end{align*}
\par %\hfill$\blacksquare$
\par
\noindent
%------------------End of-----------------
\rm
\par
\noindent
%End of-
\rm
% of TH--
\par
\noindent
\renewcommand{\footnoterule}{
  \vspace{2mm}                      % 
  \noindent\rule{\textwidth}{0.4pt}  
  \vspace{-3mm}
}
{\it $\;\;\;\;${Proof.}}$\;\;$
For the proof, see {\textcolor{black}{\cite{Keio}}}.
%{\textcolor{black}{\cite{Keio}}},
%quantum {{measurement theory}}
%.
%probability Kolmogorov extension theorem(
%\textcolor{black}{Appendix B.6(H)})\footnote{
%\textcolor{black}{{Note }1.1}
%,
%\textcolor{black}{},
%\textcolor{black}{{{Theorem }}10.13}(POI),
%$$
%, Kolmogorov extension theorem and 
%?
%$$
% and . ,{\;}Kolmogorov extension theorem
%{{measurement}}.{}
%}.
\qed
%LLLLLLLLLLLLLLLLLLLLLLLLLLLLLLP

\baselineskip=18pt
\par
%,
%:
\par
\noindent
{\bf
% BFBF
Corollary 10.14
[Infinite simultaneous observable ]}$\;\;$%POPOPO
\rm
%Assume that
%${\cal A}$
%$=C(\Omega)$
%or
%${\cal A}$
%$=B_c(V)$.
For each
$k \in K \equiv  \{ 1,2,..., |K| \}$,
consider
an observable ${\mathsf O}_k$
$\equiv$
$({}X_k , {\cal F}_k , F_k{})$
in $L^\infty(\Omega, \nu)$.
If the commutativity condition:
%index{commutativity condition@commutativity condition}
\begin{align*}
F_{k_1} ({}\Xi_{k_1} {}) F_{k_2} ({}\Xi_{k_2} {})
=
F_{k_2} ({}\Xi_{k_2} {}) F_{k_1} ({}\Xi_{k_1} {})
\quad
({}\forall \Xi_{k_1} \in {\cal F}_{k_1} , \;
\forall \Xi_{k_2} \in {\cal F}_{k_2} ,
\;
k_1 \not= k_2)
\label{n33}
\end{align*}
holds,
then we can uniquely construct
a product observable
$\widetilde{\mathsf O}$
$\equiv$
$({}\bigtimes_{k \in K} X_k , \bigstimes_{k \in K} {\cal F}_k   , $
$ {\widetilde  F} \equiv \bigtimes_{k \in K} F_k {})$
such that:
\begin{align*}
{\widetilde F }({}\Xi_1 \bigtimes \Xi_2 \bigtimes \cdots \bigtimes \Xi_{|K|}{})
=
F_1 ({}\Xi_1{})
F_2 ({}\Xi_2{})
\cdots
F_{|K|}  ({}\Xi_{|K|} {}).
%\label{n34}
\end{align*}
The product observable is also called a {\it
simultaneous observable}.
\par
\noindent
\rm
(ii):
{}\rm{}
Even if $K$ is infinite,
the
product observable similarly exists.
%Note that
%the uniqueness
%({}of quasi-product observables{})
%is not guaranteed even under the above commutativity condition.
%Also, note that
%the product observable
%$\bigtimes_{k=1}^n{\mathsf O}_k$
%always exists
%for any
%${\mathsf O}_k$
%in
%a commutative $C^*$-algebra

Consider a basic structure
$[C(\Omega );{L^\infty ( \Omega, \nu )}]$.
Let ${\widetilde \Lambda}$
be a set.
For each
$\lambda\in{\widetilde \Lambda}$,
assume that
$X_\lambda$
is a {separable complete metric space},
${\cal F}_{\lambda}{}$
is its Borel field.
For each
$\lambda \in {\widetilde \Lambda} $,
consider
an
{observable } ${\mathsf O}_\lambda$
${{=}}$
$(X_\lambda , {\cal F}_\lambda , F_\lambda{})$
in
$L^\infty ( \Omega, \nu )$.
Then,
a simultaneous observable
$\widehat{\mathsf O}$
${{=}}$
$({}\bigtimes_{\lambda \in {\widetilde \Lambda}} X_\lambda , \bigstimes_{\lambda \in {\widetilde \Lambda}} {\cal F}_\lambda   , $
$ {\widehat  F} {{=}} \bigtimes_{\lambda \in {\widetilde \Lambda}} F_\lambda {})$
uniquely
exists.
That is,
for any {finite set}$\Lambda_0 (\subseteq {\widetilde \Lambda} )$,
it holds that
\begin{align*}
{\widehat F } \big({}
(\bigtimes_{\lambda \in \Lambda_0}
\Xi_\lambda)
\times
(\bigtimes_{\lambda \in {\widetilde \Lambda}\setminus \Lambda_0}
X_\lambda)
\big)
=
\bigtimes_{\lambda \in \Lambda_0}
F_\lambda ({}\Xi_\lambda{})
\qquad
(\forall
\Xi_\lambda
\in {\cal F}_\lambda,
\forall \lambda \in \Lambda_0
)
%P%\tag{10.20}
\end{align*}
\par
\par
%\qed
%$C({}\Omega{})$.
\par
\noindent
%\END{The}%BFBF
{\it $\;\;\;\;${Proof.}}
\rm
The proof is a direct consequence of
Theorem 10.13.
Thus, it is omitted.
\qed
%\hfill{$///$}%%BFBFbfbf

\rm
\par
\par
\noindent
%It is  to note that
\par
\noindent
%BBBBBBBBBBBBBBBBBB%SBSBSBS
{\small%%{\footnotesize
\begin{itemize}
\item[$\spadesuit$] \bf {{}}{Note }10.3{{}} \rm
if "continuous type measurement" is compared with
"bounded type measurement",
the former may be essential,
but
the latter is handy from the mathematical point of view.
\end{itemize}
}

\par
\noindent
%BBBBBBBBBBBBBBBBBB%SBSBSBS
{\small%%{\footnotesize
\begin{itemize}
\item[$\spadesuit$] \bf {{}}{Note }10.4{{}} \rm
In {basic algebra} $[C(\Omega );{L^\infty ( \Omega, \nu )}]$,
a
{mixed state}
$\rho$
$(\in L^1 (\Omega ))$
is defined by
\begin{align*}
\rho {\; \geqq \;}0,
\quad
\int_\Omega \rho(\omega ) \nu (d \omega )=1
\end{align*}
Then we have the following.
\end{itemize}
\par
\noindent
\begin{center}
{\bf
Axiom${}_{\text{\scriptsize b}}^{\text{\scriptsize m}}$ 1 
({{measurement (bounded $\cdot$ mixed type )}}) 
}
\label{rule902}
\end{center}
%\END{document}
\par
\noindent
%\vskip0.1cm
\par
%\noindent
\fbox{\parbox{155mm}{
\begin{itemize}
\item[]
Consider a mixed measurement
${\mathsf M}_{L^\infty (\Omega, \nu)} \big({}{\mathsf O}{{=}}
$
$ ({}X, {\cal F} , F{}),
$
$
S_{[{}\ast] }(\rho)
\big)$
in
{basic algebra}
$
L^\infty (\Omega, \nu)
$.
Then, the probability $P(\Xi)$ that
a measured value obtained by
${\mathsf M}_{L^\infty (\Omega, \nu)} \big({}{\mathsf O}{{=}}
$
$ ({}X, {\cal F} , F{}),
$
$
S_{[{}\ast] }(\rho)
\big)$
belongs to
$ \Xi(\in  {\cal F}{})$
is given by
\begin{align*}
P(\Xi)
=
\int_\Omega
[F({}\Xi {})](
\omega)
\rho(
\omega)
\nu({}d \omega{})
%\
\end{align*}
\end{itemize}
}
}
\par
\vskip0.5cm
\par
\noindent

%
%
%\BEGIN{itembox}[c]{
%\bf Axiom${}_{\text{\scriptsize b}}^{\text{\scriptsize m}}$ 1 ({{measurement (bounded $\cdot$ mixed type )}}) 
%%\textcolor{black}{(bounded type )Axiom${}_{\text{\scriptsize b}}^{\text{\scriptsize p}}$ 1}{(}{{measurement}}
%%{\rm
%%(mixed measurement)})
%%}
%}
%\label{rule902}
%\label{axiombm1}
%%index{1@Axiom${}_{\text{\scriptsize b}}^{\text{\scriptsize p}}$ 1[{}{{measurement}}{}(bounded type )]}
%%\label{rule902}
%%%index{ and @
%%%{statistics}probability 
%%%}
%%%%index{1@Axiom${}_{\text{\scriptsize b}}^{\text{\scr
%%iptsize p}}$ 1[{{measurement}}(bounded type )]}
%%\BEGIN{itemize}
%%%\item[(i)]
%%%{{measurement}}.
%%\item[]
%Consider a mixed measurement
%${\mathsf M}_{L^\infty (\Omega, \nu)} \big({}{\mathsf O}{{=}}
%$
%$ ({}X, {\cal F} , F{}),
%$
%$
%S_{[{}\ast] }(\rho)
%\big)$
%in
%{basic algebra}
%$
%L^\infty (\Omega, \nu)
%$.
%Then, the probability $P(\Xi)$ that
%a measured value obtained by
%${\mathsf M}_{L^\infty (\Omega, \nu)} \big({}{\mathsf O}{{=}}
%$
%$ ({}X, {\cal F} , F{}),
%$
%$
%S_{[{}\ast] }(\rho)
%\big)$
%belongs to
%$ \Xi(\in  {\cal F}{})$
%is given by
%\BEGIN{align*}
%P(\Xi)
%=
%\int_\Omega
%[F({}\Xi {})](
%\omega)
%\rho(
%\omega)
%\nu({}d \omega{})
%%\
%\END{align*}
%%
%%$F({}\Xi{})$
%%
%%\END{itemize}
%\END{itembox}
%%%index{measurement22@${\mathsf M}_{L^\infty(\Omega)}
%%\big({}{\mathsf O} , S_{[{}\ast] }(\rho) \big)$:
%%(bounded type ){{measurement}}}
%\par
%\vskip0.2cm
%\par
%
%\END{itemize}
%}
%\par
%\par
%\noindent
%
%

%
%
%%11 11 11 11 11 11 11
%
%
%\vskip3.0cm
%\newpage
%11 11 11 11 11 11 11
\section{Axiom${}_{\text{\scriptsize b}}^{\text{\scriptsize pm}}$ 2 - causality  (bounded type)
\label{Chap11}}%{Chap.{\;}}11{}
%%\vspace{-0.8cm}
%\chapter[
%{} (\textcolor{black}{Axiom}2(POI))
%]{{}
%\baselineskip=18pt\par
\noindent
\begin{itemize}
\item[{}]
{
%\footnotesize
\small
\par%[Abstract].
\rm
$\;\;\;\;$
Same as continued type measurement theory, bounded type measurement theory is formalized as follows:
Thus, we shall introduce
bounded type {{{measurement theory}}}
as follows.
%:
%(POI){{{measurement theory}}}, quantum mechanicsverbalizing,
\begin{align*}
%\dashbox{5}
\underset{\text{\scriptsize (language)}}{\text{{} $\fbox{bounded pure type {{{measurement theory}}}}$}}
:=
{
\overset{\text{\scriptsize [(bounded type )Axiom${}_{\text{\scriptsize b}}^{\text{\scriptsize p}}$ 1]}}
%\overset{\text{\scriptsize [(bounded type )Axiom${}_{\text{\scriptsize b}}^{\text{\scriptsize p}}$ 1\textcolor{black}{(Sec. \REF{10secAxiom 1})}]}}
{
\underset{\text{\scriptsize
[probabilistic interpretation]}}{\text{{} $\fbox{{{measurement}}}$}}}
}
+
{
\overset{\text{\scriptsize [(bounded type )Axiom${}_{\text{\scriptsize b}}^{\text{\scriptsize pm}}$ 2]}}
%\overset{\text{\scriptsize [(bounded type )Axiom${}_{\text{\scriptsize b}}^{\text{\scriptsize pm}}$ 2\textcolor{black}{(\REF{10secAxiom 2})}]}}
{
\underset{\text{\scriptsize [{{the Heisenberg picture}}]}}
{\text{{}$\fbox{ causality }$}}
}
}
%TAG*{$\displaystyle{\mathop{2)}_{(=10))}}$}
%%%%%%%%%%%%%2.2}CLAsSICAL
\end{align*}
In this chapter, I explain the bounded type Axiom${}_{\text{\scriptsize b}}^{\text{\scriptsize pm}}$ 2.
Since there is a close resemblance between it and "continuation and pure type measurement theory of \textcolor{black}{Chapter 6}
(\textcolor{black}{Axiom${}_{\text{\scriptsize c}}^{\text{\scriptsize pm}}$} 2)", you should be able to read easily.
However, you should be cautious of a tree semi-ordered set $T$ not being necessarily a finite set.
Bounded type measurement theory has a field including a vast domain, and when writing out, there is no end.
Therefore, in this chapter, it extracted only to Zeno's paradox and the argument of the circumference of it.
\vskip0.5cm
}
\end{itemize}
\normalsize 
\baselineskip=18pt
\subsection{Zeno's paradox
---
Flying arrow is not moving. }%11.1
\rm
\par
In this section, I explain the meaning of Zeno's paradox.
%index{@}
\rm
\subsubsection{Movement function method}%11.1.1
\par
In this book, movement function method is considered as follows.
%\par
%\noindent
Movement function method is the quantitative describing method of "movement and change",
and if it is called a {\bf time-position function method}, you may not have misunderstanding.
%%index{@()}
%%index{ and @}
Namely,
%\textcolor{black}{1.1({}D$'_0$)}(i) and ,
\begin{itemize}
\item[({}A$_1$)]
In order to express movement and change, $x$ of a position of a particle is denoted by function of time $t$.
That is, the time-position function $x=q(t)$ expresses change of the position of each time
(for example, a position, height, academic ability, GDP of a country, etc.) .
\end{itemize}
Of course, we studied the movement function method
in elementary school
as follows.
%I would like to ask for the time-position function $q(t)$, and, in
\begin{align*}
\cases
\frac{q(t_2)-q(t_1)}{t_2-t_1}=v,\quad
&
\text{this is called a speed if ii does not depend on ($t_1,t_2$
$(t_1 < t_2 )$}
%\\
%&
%\text{, {\bf []=[]/[]} and )}
%\qquad
\\
q(0)=a
&{}
\endcases
\end{align*}
The time-position function $q(t)=vt +a$ can be found from easy calculation.
We think
 \begin{itemize}
\item[({}A$_2$)]
Movement function method (= time-position function method) 
is a kind of world-view
such as
\begin{align*}
\text{
\bf
describe {movement and change} quantitatively with a time-position function}
\end{align*}
\end{itemize}
If I have a worldly way of speaking,
"See movement and change through prejudice called movement function method."
%
%BBBBBBBBBBBBBBBBBB%SBSBSBS
\par
\noindent
{\small%%{\footnotesize
\vspace{0.1cm}
\begin{itemize}
\item[$\spadesuit$] \bf {{}}11.1{{}} \rm
For completeness,
again recall
\par
\noindent
$\;\;$
$\underset{\text{\scriptsize (Chap. 1)}}{\text{\normalsize (X$_1$)}}$
$\overset{
}{\underset{\text{\scriptsize (before science)}}{
\text{
\fbox
{{\textcircled{\scriptsize 0}}
widely {ordinary language}}
}
}
}
$
$
\underset{\text{\scriptsize }}{\text{$\Longrightarrow$}}
$
$
\underset{\text{\scriptsize (Chap. 1(O))}}{\text{{world-description}}}
\cases
&
\!\!\!\!\!
\underset{\text{\scriptsize (world is before language)}}{
{\textcircled{\scriptsize 1}}
{\text{realistic method} \qquad}
}
%}
\\
\\
&
\!\!\!\!\!
\underset{{\text{\scriptsize (language is before world)}}}{
\text{\textcircled{\scriptsize 2}}
{\text{linguistic method}}
}
%{\text{\textcircled{\scriptsize 2}{linguistic method}(e.g., {{measurement theory}})}}
\endcases
$
\\
\\
Out problem is
\begin{itemize}
\item[]
Where do we discuss
the movement function method,
\textcircled{\scriptsize 0},
\textcircled{\scriptsize 1}
or
\textcircled{\scriptsize 2}
?
\end{itemize}
Usually, we may argue in \textcircled{\scriptsize 0} as elementary and junior high school students' problem.
%and to argue in \textcircled{\scriptsize 1}, if Laplace's demon(\textcolor{black}{Note 5.2} is carried out.
%%index{@}
However, if we persist in \textcolor{black}{Position 3.5 } of this book,
naturally
\begin{itemize}
\item[]
Movement function method should be described in \textcircled{\scriptsize 2}.
\end{itemize}
\end{itemize}
}
%%BBBBBBBBBBBBBBBBB
\par
\noindent

\renewcommand{\footnoterule}{%
  \vspace{2mm}                      % 
  \noindent\rule{\textwidth}{0.4pt}   % , 
  \vspace{-5mm}
}

\subsubsection{Zeno's paradox --- Why is it paradox?---}%11.1.2
\par
\rm
\textcolor{black}
As mentioned in
\textcolor{black}{problem 5.1},
% also described,
Zeno(BC490-BC430) considered the following problem about 2500 years ago.:
%index{@}
\par
\noindent
%%BFBFBF
{\bf
Problem 11.1
[Problem 5.1
(same as the arrow which is flying (movement function method))]}$\;\;$%POPOPO
\begin{itemize}
\item[({}B$_1$)][Problem]:
Is the flying arrow moving?
$\;\;$
Here, of course, "flying arrow" 
is one symbol of movement and change
such as
"running tortoise"
"the growth of rice",
"the flying migratory birds", "the growth of economy of the country"
and so on.
%,
%\footnote{
%,
%,
%,
%.
%,
%${{\cdot}}$, . }
\item[({}B$^1_2$)][Zeno's answer]
Assume that the arrow is flying.
This arrow has stopped at the time of when at that moment.
If it has stopped at the time of when at the moment, it will always have stopped,
and the arrow will have stopped, therefore it will not move.
\qed
\item[({}B$^2_2$)][The answer using movement function method]
%I consider a time-position function method(=movement function method\textcolor{black}{({}A$_2$)})
%-
%how to denote the position of the arrow in time $t$ by the value of a time-position function $q(t)$
%-.
In every time $t$, the position of the arrow becomes settled as a value of a time-position function $q(t)$.
However, we cannot conclude that
time-position function $x=q(t) $ is a constant function (i.e., function with a constant value).
After all, we say "The flying arrow which is moving."
\qed

\item[({}B$_3$)][Problem]
Of course, the problem which Zeno raised is the next.
\begin{itemize}
\item[]
Which shall you choose between Zeno's answer\textcolor{black}{({}B$^1_2$)}
and
a common sense answer\textcolor{black}{({}B$^2_2$)}?
$\;\;$
Or what is the basis as which you choose a common sense answer\textcolor{black}{({}B$^2_2$)}?
$\;\;$
Furthermore, a time-position function method (= movement function method) will not be a physical law,
and the empirical validation of it will also be impossible.
Although that is right, why do you believe and use a time-position function method?
$\;\;$
What is the basis?
\end{itemize}
If it puts in another way ,
\begin{itemize}
\item[($\sharp$)]
What is the best linguistic science language which includes a time-position function method (= movement function method)?
\end{itemize}
\end{itemize}
%c%index{@(POI)}

I will summarize the above and will write in the style of a fiction.

%\newpage
\par
\noindent
\begin{center}
{\bf
Zeno's paradox
}
%\label{rule902}
\end{center}
%\END{document}
\par
\noindent
%\vskip0.1cm
\par
\noindent
\fbox{\parbox{155mm}{
\begin{itemize}
\item[({}C$_1$)]
About 2500 years ago, Zeno showed us "the perfect logic of the arrow which flies\textcolor{black}{
(B$^1_2$)
}."
\item[({}C$_2$)]
We refuted it by "movement function method\textcolor{black}{({}A$_2$)}" like \textcolor{black}{(B$^2_2$)},
and answered with confidence "It is an easy problem, Mr. Zeno."
\item[({}C$_3$)]
At this time, Zeno brought forth a counterargument as follows immediately. :
\begin{itemize}
\item[]
Movement function method is mystic words which cannot carry out empirical validation.
Why do you believe and use such random and irresponsible movement function method?
$\;\;$
Is it science?
%index{@}
\end{itemize}
% and .
\item[({}C$_4$)]
We have to reply something to Zeno.
And, of course, we have continued thinking earnestly 2500 years.
However, we cannot yet answer.
%\END{itemize}
%%%%%%%%%%%
\end{itemize}
}
}
\par
\vskip0.5cm
\par
\noindent

%
%
%
%
%
%
%
%
%
%
%
%
%
%
%
%
%
%
%
%\BEGIN{itembox}[c]{\bf Zeno's paradox}
%\label{}
%\BEGIN{itemize}
%\item[({}C$_1$)]
%About 2500 years ago, Zeno showed us "the perfect logic of the arrow which flies\textcolor{black}{
%(B$^1_2$)
%}."
%
%
%\item[({}C$_2$)]
%We refuted it by "movement function method\textcolor{black}{({}A$_2$)}" like \textcolor{black}{(B$^2_2$)},
%and answered with confidence "It is an easy problem, Mr. Zeno."
%\item[({}C$_3$)]
%At this time, Zeno brought forth a counterargument as follows immediately. :
%\BEGIN{itemize}
%\item[]
%Movement function method is mystic words which cannot carry out empirical validation.
%Why do you believe and use such random and irresponsible movement function method?
%$\;\;$
%Is it science?
%%index{@}
%\END{itemize}
%% and .
%\item[({}C$_4$)]
%We have to reply something to Zeno.
%And, of course, we have continued thinking earnestly 2500 years.
%However, we cannot yet answer.
%\END{itemize}
%\END{itembox}

Of course, we become silent in \textcolor{black}{({}C$_4$)}
because we are caught by the realistic method.
Moreover, it is because we feel inferior in asserting dignifiedly the mathematics of the form where it was somehow buried into ordinary language
-
statistics and dynamical system theory
-.
Namely,
\begin{itemize}
\item[(C$_5$)]
It is because we cannot have the confidence to which we retort to Zeno
"We can answer easily in statistics and dynamical system theory(that is, as the easy case of an equation of state (1.1))."
\end{itemize}
Probably, it is good although the price
for which statistics and dynamical system theory have depended on two authority (Mathematics and Application) has turned.

However, in the position of this book which considers two classifications of world description
(\textcolor{black}{{Chap.$\;$1} (O) }):
\\
\\
$\qquad$
$
\underset{\text{\scriptsize (Chap. 1(O))}}{\text{{world-description}}}
\cases
&
\!\!\!\!\!
\underset{\text{\scriptsize (world is before language)}}{
{\textcircled{\scriptsize 1}}
{\text{realistic method} \qquad}
}
%}
\\
\\
&
\!\!\!\!\!
\underset{{\text{\scriptsize (language is before world)}}}{
\text{\textcircled{\scriptsize 2}}
{\text{linguistic method}}
}
%{\text{\textcircled{\scriptsize 2}{linguistic method}(e.g., {{measurement theory}})}}
\endcases
$
\\
it is as follows.
\begin{itemize}
\item[({}C$_6$)]
We are not in the extravagant situation
where we choose the logic of Zeno\textcolor{black}{({}B$^1_2$)}
and the common sense answer\textcolor{black}{(B$^1_2$)}.
That is, we know only a language called measurement theory as a language which describes science,
and we cannot but describe by it.
And then, what was described is "the world of the arrow which flies."
(\textcolor{black}{Chapter 8.1(m)}).
\end{itemize}
Although I will answer this in \textcolor{black}{Answer 11.11},
some preparations of (\textcolor{black}{Section 11.2--Section 11.4}) are needed before that.
%,
%.
%
\rm
\par
\noindent
%BBBBBBBBBBBBBBBBBB%SBSBSBS
{\small%%{\footnotesize
\vspace{0.1cm}
\begin{itemize}
\item[$\spadesuit$] \bf {{}}Note 11.2{{}} \rm
Although Zeno's paradox has some types elsewhere("Achilles and a tortoise", "dichotomy", "stadium", etc.),
"the arrow which flies" expresses the essence of the problem exactly and is the first masterpiece.
However, since "Achilles and the tortoise" may be more famous, I will also describe this.
The line of argument of Zeno about "the paradox of Achilles and a tortoise" is as follows. :,
\begin{itemize}
\item[]
I consider competition of Achilles and a tortoise.
Let the start point of a tortoise (a late runner) be the front from the starting point of Achilles (a quick runner).
Suppose that both started simultaneously.
If Achilles tries to pass a tortoise, Achilles has to go to the place in which a tortoise is present now.
However, then, the tortoise should have gone ahead more.
Achilles has to go to the place in which a tortoise is present now further.
Even Achilles continues this infinite, he can never catch up with a tortoise.
%\qed
\end{itemize}
Generally, you may suppose "Achilles and a tortoise are the problems of infinite geometrical progression."
That is, the time-position function of Achilles and a tortoise is set to $x=q_1(t)=vt$
and $y=q_2(t)=\gamma vt +a$, respectively.
($0 < \gamma v < v$,
$a >0$).
Here, you may think that you can solve by calculating the solution $s_0=\frac{a}{(1- \gamma ) v}$ to $q_1(s_0)=q_2(s_0)$
by the infinite geometrical progression
\begin{align*}
s_0=\frac{a}{v}(1+\gamma + \gamma^2 + \gamma^3 +... )= \frac{a}{(1- \gamma ) v}
\end{align*}.

However, if it is a problem finished now,
they will say that the philosophers who have continued considering Zeno's paradox 2500 years are foolish.
\renewcommand{\footnoterule}{%
  \vspace{2mm}                      % 
  \noindent\rule{\textwidth}{0.4pt}   % , 
  \vspace{-2mm}
Of course, most philosophers must be excellent and it is
$(\sharp)$
---
{
\bf
(to introduce metaphysics (except for mathematics) as a scientific (= material studies') base)
}\footnote{
This was a theme of this book.
}
---
of the problem \textcolor{black}{({}B$_3$)} that they have continued investigating.
%
%
%
% and .
%,
%${{\cdot}}$
%${{\cdot}}$.  and , ?
%8.1(m).
%%index{@}
It is because Zeno has asked
\begin{itemize}
\item[($\sharp_1$)]
Why do you believe and use movement function method which cannot carry out empirical validation?
\item[($\sharp_2$)]
Why don't you speak (C$_5$) with confidence?
\item[($\sharp_3$)]
Investigate $(\sharp)$ of the problem \textcolor{black}{({}B$_3$)}.
\end{itemize}
etc.,
so the answer of the infinite geometrical progression using movement function method does not become the answer.
}
\end{itemize}

\subsection{{{{Causal operator}}, predual {{causal operator}}, {{deterministic causal map}}}%{Sec.11.2}
}
\def\PAR{\roman{par}}
\rm
\normalsize
\baselineskip=18pt

\par
Like
\textcolor{black}{{Chap.{\;}}10{}},
assume
a state space $\Omega$
(i.e.,
locally compact space $\Omega$
)
and
a measure space
$(\Omega, {\cal B}_{\Omega}, \nu)$
satisfying
the condition (a)
in
\textcolor{black}{{}{{{}}}{Sec.10.1}}.

\par
\noindent
\vskip0.3cm
\vskip0.3cm
%BFBF
\par
\noindent
{\bf {Definition }11.2
[{{Causal operator}}, {{deterministic causal map}}]}$\;\;$%POPOPO
Let
$[C(\Omega_1),$
$L^\infty(\Omega_1, \nu_1)]$
and
$[C(\Omega_2),$
$L^\infty(\Omega_2, \nu_2)]$
be basic structures.
A continuous linear operator
$\Phi_{1,2}: L^\infty (\Omega_2, \nu_2 ) \to L^\infty (\Omega_1, \nu_1 )$
is called
a
{\bf {{causal operator}}},
if it satisfies the following
(i)
and
(ii):
%%final%index{@{{causal operator}}}
\begin{itemize}
\item[(i)]
for any
$f_2 (\in C(\Omega_2 ))$,
there exists
$\Phi_{1,2}f_2 \in C(\Omega_1)$,
and further,
the
operator
$\Phi_{1,2}: C (\Omega_2) \to C (\Omega_1)$
is the causal operator
in the sense of
\textcolor{black}{{Definition }6.2}.
\item[(ii)]
There exists a continuous linear operator
$[\Phi_{1,2}]_*: L^1 (\Omega_1, \nu_1 ) \to L^1 (\Omega_2, \nu_2 )$
such that
\begin{align*}
&
\int_{\Omega_1}
[\Phi_{1,2} f_2]
(\omega_1)
\cdot
\rho_1(\omega_1)
\; \nu_1(d\omega_1)
=
\int_{\Omega_2}
f_2
(\omega_2)
\cdot
\left([\Phi_{1,2}]_* \rho_1 \right)
(\omega_2)
\; \nu_2(d\omega_2)
%
%&{}_{{}_{{{L^1 (\Omega_1, \nu_1)}}}} \Big\langle \rho_1 ,
%\Phi_{1,2} f_2 \Big\rangle{}_{{}_{{{L^\infty (\Omega_1, \nu_1)}}}}
%=
%{}_{{}_{{{L^1 (\Omega_2, \nu_2)}}}}\Big\langle [\Phi_{1,2}]_* \rho_1 ,
%f_2 \Big\rangle{}_{{}_{{{L^\infty (\Omega_2, \nu_2)}}}}
\\
&
\qquad
\qquad
({}\forall \rho_1 \in {L^1 (\Omega_1, \nu_1)},
\;\;
\forall f_2 \in {L^\infty (\Omega_2, \nu_2)}{})
%P%\tag{11.1}
\end{align*}
This $[\Phi_{1,2}]_*$
is called the
{\bf
pre-dual {{causal operator}}
}.
%%final%index{faipre@$[\Phi_{1,2}]_*$:dual {{causal operator}}}
\end{itemize}
%%final%index{@dual {{causal operator}}}
In addition,
the
{{causal operator}}$\Phi_{1,2}: L^\infty (\Omega_2, \nu_2 ) \to L^\infty (\Omega_1, \nu_1 )$
is called a {\bf {{deterministic causal operator}}},
if there exists
a
{continuous map }$\phi_{1,2}:\Omega_1 \to \Omega_2 $
such that
%%final%index{@{{deterministic causal operator}}}
\begin{align*}
(\Phi_{1,2}f_2)(\omega_1 )
=
f_2 ( \phi_{1,2}(\omega_1 ))
\qquad
({\roman a.e.}\; \omega_1,
\quad
\forall f_2 \in L^\infty ( \Omega_2 ))
\end{align*}
%\END{itemize}
%\END{itemize}
Also,
the
{continuous map }$\phi_{1,2}:\Omega_1 \to \Omega_2 $
is said to be a {\bf {{deterministic causal map}}}.
%%final%index{@{{deterministic causal map}}}
\rm
\rm
\par
%\qed
%\par
%${L^\infty (\Omega_1, \nu_1)}$
%{{ and }}
%${L^\infty (\Omega_2, \nu_2)}$
%.
%
%

\par
\par
\noindent
\vskip0.3cm
\vskip0.3cm
%BFBF
\par
\noindent
{\bf
{{Theorem }}11.3
[{{{Causal operator}}} and observable ]}$\;\;$%POPOPO
For any observable
$({}X , {\cal F} , F_2{})$
in $L^\infty(\Omega_2, \nu_2)$,
the $({}X , {\cal F} , \Phi_{1,2} F_2{})$
is an observable in $L^\infty(\Omega_1, \nu_1)$,
which is denoted by
$\Phi_{12} {\mathsf O}_2$.
%\END{The}
%BFBF
\par
\noindent
\rm
{\it $\;\;\;\;${Proof.}}$\;\;$
it is easy to see that,
for any countable decomposition
$\{ \Xi _j \} _{j=1} ^{\infty} $
of $\Xi $,
({}$\Xi _j$, $\Xi $ $ \in {{\cal F}}  ${}),
it holds that
\begin{align*}
\allowbreak
\allowdisplaybreaks
&
\int_{\Omega_1}
\rho_1 ( \omega_1)
\cdot
[\Phi_{1,2} F_2(\Xi )]
( \omega_1)
\; \nu_1( d \omega_1)
=
%& \hspace{3cm}
\int_{\Omega_2}
([\Phi_{1,2}]_* \rho_1)( \omega_2)
\cdot
[F_2(\Xi )] ( \omega_2)
\; \nu_2( d \omega_2)
\\
%\hspace{1.5cm}
=
&
\int_{\Omega_2}
([\Phi_{1,2}]_* \rho_1)( \omega_2)
\cdot
[F_2(
\bigcup\limits_{i=1}^\infty \Xi_i
 )] ( \omega_2)
\; \nu_2( d \omega_2)
%\\
%& \hspace{1.5cm}
=
\lim_{N \to \infty }\int_{\Omega_2}
([\Phi_{1,2}]_* \rho_1)( \omega_2)
\cdot
\sum\limits_{i=1}^N
[F_2(
\Xi_i
 )] ( \omega_2)
\; \nu_2( d \omega_2)
\\
%\hspace{1.5cm}
=
&
\lim_{N \to \infty }
\int_{\Omega_1}
\rho_1 ( \omega_1)
\cdot
\sum\limits_{i=1}^N
[\Phi_{1,2} F_2(
\Xi_i
)
]
( \omega_1)
\; \nu_1( d \omega_1)
%P%\tag{11.2}
\end{align*}
Thus,
by \textcolor{black}{{Definition }10.1},
we get the proof.
\qed
\par
\noindent

%
%\Psi\Phi
%
%
%

%\noindent
\vskip0.3cm
\vskip0.3cm
%BFBF
\par
\noindent
%\Phi\Phi\; \nu
{\bf
{{Theorem }}11.4}$\;\;$%POPOPO
Any
{{deterministic causal operator}}
$\Phi_{1,2}: L^\infty(\Omega_2, \nu_2 )
\to L^\infty (\Omega_1, \nu_1 )$
%
% and .
%$\Phi_{1,2}$
%$:{L^\infty (\Omega_2, \nu_2)} \to {L^\infty (\Omega_1, \nu_1)}$
satisfies
\rm
%That is,
\begin{itemize}
\item[]
%\BEGIN{itemize}
%\ITEM
%(i)
$\Phi_{1,2} (f_2)
\cdot
\Phi_{1,2} ({}g_2{})$
$=$
$\Phi_{1,2} ({}f_2 \cdot
g_2{}) \qquad $
(
$\forall f_2, \forall g_2$
$\in {L^\infty (\Omega_2, \nu_2)}$)
%\ITEM[(ii)]
%$({}\Phi_{1,2} (f_2){})^*$
%$=$
%$\Phi_{1,2} ({}f_2^* {})$
%$\qquad (
%\forall
%f_2$
%$\in$
%${L^\infty (\Omega_2, \nu_2)})
%$.
\end{itemize}
\par
\noindent
\rm
{\it $\;\;\;\;${Proof.}}$\;\;$
The proof is the same as that of
\textcolor{black}{{{Theorem }}6.5}.
We can omit it.
% and , .
\qed
%\Phi\Psi
\par
\noindent
\vskip0.3cm
\vskip0.3cm
%BFBF
\par
\noindent
{\bf {{Theorem }}11.5
[{Continuous map } and {{deterministic causal map}}]}$\;\;$%POPOPO
%
%
%
%
%%%%\Phi\Psi\Psi\Phi
%
%\vskip0.3cm
%\vskip0.3cm
%%BFBF
%\par
%\noindent
%\BEGIN{The} \label{Theorem 11.4}
\sf
[Continuous map and deterministic causal map in classical systems].
\\
\rm
%(i):
%\ssl
%If
%$\Phi_{1,2}: L^\infty ({}\Omega_2 , \nu_2{}) \to
%L^\infty ({}\Omega_1 , \nu_1{})$
%be a deterministic causal operator,
%there exists a continous map $\phi_{1,2}$ from $\Omega_1$ into $\Omega_2$
%such that:
%\BEGIN{align*}
%[\Phi_{1,2}f_2](\omega_1)
%=
%f_2(\phi_{1,2}(\omega_1))
%\quad
%(
%\text{ a.e. } \omega_1
%\in \Omega_1,
%\forall f_2 \in
%L^{\infty} (\Omega_2,\nu_2 )
% ).
%\tag{11.5}
%\END{align*}
%\rm
%(ii):
{}\rm{}
%}$\;\;$%POPOPO
Let
$(\Omega_1, {\cal B}_{\Omega_1}, \nu_1)$
and
$(\Omega_2, {\cal B}_{\Omega_2}, \nu_2)$
be measure spaces.
Assume that
a continuous map
$\phi_{1,2}:\Omega_1 \to \Omega_2$
satisfies:
\begin{align*}
%&
%D_2 \in {\CAL B}(\Omega_2),
%\nu_2 ( D_2 ) > 0
%\Longrightarrow
%\nu_1( \phi_{1,2}^{-1} ( D_2 )) > 0
%\\
%&
D_2 \in {\cal B}_{\Omega_2},
\; \;
\nu_2 ( D_2 ) = 0
\quad
\Longrightarrow
\quad
\nu_1 ( \phi_{1,2}^{-1} ( D_2 )) = 0.
\end{align*}
Then,
the
continuous map
$\phi_{1,2}:\Omega_1 \to \Omega_2$
is deterministic,
that is,
the
operator ${\Phi}_{1,2}
:L^{\infty} (\Omega_2, \nu_2) \to L^{\infty} (\Omega_1,
\nu_1)$
defined by
\textcolor{black}{(11.5)}
is
a
deterministic
causal operator.
%\BEGIN{align*}
%&
%[\Phi_{1,2}f_2](\omega_1)
%=
%f_2(\phi_{1,2}(\omega_1))
%\quad
%(
%\text{ a.e. } \omega_1
%\in \Omega_1,
%\forall f_2 \in
%L^{\infty} (\Omega_2,\nu_2 )
% )
%%P%\TAG{11.3}
%\END{align*}
%\END{The}
%BFBF
\par
\noindent
\rm
{\it $\;\;\;\;${Proof.}}$\;\;$
%(i):
%It is obvious from the definition of
%the deterministic causal operator.
%\\
%(ii):
For
each
${\overline \rho}_1$
$\in L^1(\Omega_1 , \nu_1 )$,
define a measure
$\mu_2$
on
$(\Omega_2 , {\cal B}_{\Omega_2})$
such that
\begin{align*}
\mu_2 ( D_2 ) = \int_{\phi_{1,2}^{-1} ( D_2 ) } {\overline \rho}_1 (\omega_1)
\; \nu_1 ( d \omega_1 )
\qquad
(\forall D_2 \in {\cal B}_{\Omega_2}
)
\end{align*}
Then,
it suffices to
consider
the Radon-Nikodym derivative
$[\Phi_{1,2}]_* ({\overline \rho}_1)
=
{d \mu_2}/{d \nu_2 }$.
%({\it cf.} Appendix).
That is because
\begin{align*}
D_2 \in {\cal B}_{\Omega_2},
\; \;
\nu_2 ( D_2 ) = 0
\quad
\Longrightarrow
\quad
\nu_1 ( \phi_{1,2}^{-1} ( D_2 )) = 0
\quad
\Longrightarrow
\quad
\mu_2 ( D_2 ) = 0
\end{align*}
%( \phi_{1,2}^{-1} ( D_2 )) = 0
%\par
Thus,
by
the Radon-Nikodym
theorem,
we get
a continuous linear operator
$[{\Phi}_{1,2}]_\ast
:L^{1} (\Omega_1, \nu_1) \to L^{1} (\Omega_2,
\nu_2)$.
\qed
%\footnote{
%r(r{}$B%C%b{}(B2qdr
%$
%D_2 \in {\cal B}_{\Omega_2},
%\nu_2 ( D_2 ) = 0
%
%
%
%
%

%\rm
%\par
%\noindent
%\bf \vskip0.3cm
%BFBF

%%%%%%%%%%%%%%%%%%%%%%%%%%
%{\bf
%{\Large
%3.11.
\def\FIN{{\roman{(exa)}}}
\def\EXI{{\roman{(exi)}}}
\rm
\subsection{Bounded type Axiom${}_{\text{\scriptsize b}}^{\text{\scriptsize pm}}$ 2 (causality )}%{Sec.11.3}
\label{10secAxiom 2}
%}
%
%index{@semi-ordered tree}
%index{tree@$(T,{{\; \leqq \;}}), (T(t_0),{{\; \leqq \;}})$:semi-ordered tree}

\rm
Let $(T,\le)$ be a tree-like partial ordered set, i.e., a partial ordered set such that
\begin{align*}
\text{
{\lq\lq$t_1 {{\; \leqq \;}}t_3$
and
$t_2 {{\; \leqq \;}}t_3$\rq\rq} $\Longrightarrow$
{\lq\lq$t_1 {{\; \leqq \;}}t_2$
or
$t_2 {{\; \leqq \;}}t_1$\rq\rq}\!}
\end{align*}
Put $T^2_\le = \{ (t_1,t_2) \in T^2{}: t_1 \le t_2 \}$.
An element $t_0 \in T$ is called a {\it root} if
$t_0 \le t$ ($\forall t \in T$) holds.
If $T$ has the root
$t_0$,
we sometimes denote
$T$ by
$T(t_0)$.
$T' (\subseteq T )$
is called
{\it lower bounded}
if
there exists an element
$t_i (\in T)$
such that
$t_i {{\; \leqq \;}}t $
$(\forall t \in T')$.
Therefore,
if
$T$ has the root,
any $T' (\subseteq T )$
is lower bounded.
We always assume that
$T$ is complete,
that is,
for any
$T' (\subseteq T )$
which is lower bounded,
there exists
an
element
${\roman Inf}_T ( T' ) (\in T )$
that satisfies the following
(i)
and (ii):
%index{ZZZinfT@${\roman Inf}_T:$infimun}
\begin{itemize}
\rm
\item[(i)]
{}\rm{}
${\roman Inf}_T ( T' ) {{\; \leqq \;}}t \qquad ( \forall t \in T' )$
\rm
\item[(ii)]
{}\rm{}
If
$s {{\; \leqq \;}}t \; \; ( \forall t \in T' )$,
then
it holds that
$s {{\; \leqq \;}}{\roman Inf}_T ( T' )$
\end{itemize}
%%\END{Def}
%BFBF
\hfill{{$///$}}%%BFBFbfbfbf
\rm
%
%\par
%\vskip0.5cm
%\par
\par
In this book,
we are not concerned with
the topology
(or metric)
of the $T$.
\par

%
%
%\textcolor{black}{6.2}(continuous type {{measurement theory}}causality )
%bounded type {{measurement theory}}.
%$(T,{{\; \leqq \;}})$ {{}} semi-ordered tree, That is,  
%\BEGIN{align*}
%\text{
%{\lq\lq$t_1 {{\; \leqq \;}}t_3$
%
%$t_2 {{\; \leqq \;}}t_3$\rq\rq} $\Longrightarrow$
%{\lq\lq$t_1 {{\; \leqq \;}}t_2$
%
%$t_2 {{\; \leqq \;}}t_1$\rq\rq}\!}
%\END{align*}
% and .
%,
%{(}That is,
%bounded type {{{measurement theory}}}{)}
%,
%$T$
%{finite set} and  and .
%$T^2_{\leqq}= \{ (t_1,t_2) \in T^2{}: t_1 {{\; \leqq \;}}t_2 \}$
% and .
% $t_0 \in T$,
%$t_0 {{\; \leqq \;}}t$ ($\forall t \in T$)
%
% and ,
%{\bf } and .
%%index{ and @}
%%bounded type {{{measurement theory}}},
%semi-ordered tree$T$ and ,
%, ,
%%$t_0$
%%
%%,
%$T$
%
%$T{(t_0)}$
% and .
%\par
% and , \textcolor{black}{{Chap.{\;}}6{}}
%,
%,
%$T(0)=\{ t \in {\mathbb R}
%\;|\;
%t {\; \geqq \;}0 \}$
% and 
%$T(1)=\{1,2,\ldots
%\}$
%.
%\par
%$T' (\subseteq T )$
%,  and ,
%$t_i {{\; \leqq \;}}t $
%$(\forall t \in T')$
% and 
%$t_i \in T$
% and .
%,
%$T$,
%$T' (\subseteq T )$
%, {}
%\par
%$T$(semi-ordered tree) and .
%That is,
%$T' (\subseteq T )$
%,
%
%(i) and
%(ii)${\roman Inf}_T ( T' ) (\in T )$ and .
%%index{infT@${\roman Inf}_T:$}
%\BEGIN{itemize}
%\item[(i)]
%${\roman Inf}_T ( T' ) {{\; \leqq \;}}t \qquad ( \forall t \in T' )$
%\item[(ii)]
%
%$s {{\; \leqq \;}}t \; \; ( \forall t \in T' )$
%,
%$s {{\; \leqq \;}}{\roman Inf}_T ( T' )$.
%\END{itemize}
%%
%%\par
%%\vskip0.5cm
%%\par
%,
%, $T$${{\cdot}}$
%.
%\par

\par
\vskip0.2cm
\par
\noindent
%\vskip0.3cm
%\vskip0.3cm
%BFBF
\par
\noindent
{\bf
{Definition }11.6
[{{Sequential causal operator}}, sequential observable{}]}$\;\;$%POPOPO
$\;$
%index{@sequential observable}
\rm
%The pair
%$\overline{\bold S}({{\overline \rho}_{t_0}}) \equiv $
%$[{}{\overline S}({{\overline \rho}_{t_0}}),
A family
$\{ \Phi_{t_1,t_2}{}: $
$L^\infty(\Omega_{t_2}, \nu_{t_2}) \to L^\infty(\Omega_{t_1}, \nu_{t_1}) \}_{(t_1,t_2) \in T_{\le}^2}$
is called a
{\it causal relation}
(or,
{\it
causal relation}
),
%{\it general system with an initial state}
%${\overline S}({{\overline \rho}_{t_0}})$
if it satisfies the following conditions {\rm (i) and (ii)}.
\begin{itemize}
\item[{\rm (i)}]
With each $t \;(\in T)$, a basic algebra $L^\infty(\Omega_t, \nu_t)$ is associated.
%\item[{\rm (ii)}]
%Let $t_0 \;(\in T)$ be the root of $T$.
%And, assume that a system $S$ has the normal
%state ${{\overline \rho}_{t_0}} \;(\in {\frak S}^n((L^\infty(\Omega, \nu)PP_{t_0})_*))$
%at $t_0$, that is, the initial state is equal to
%${{\overline \rho}_{t_0}}$.
\item[{\rm (ii)}]
For every $(t_1,t_2) \in T_{\le}^2$,
a causal operator
$\Phi_{t_1,t_2}{}: L^\infty(\Omega_{t_2}, \nu_{t_2}) \to L^\infty(\Omega_{t_1}, \nu_{t_1})$
is defined such that
$\Phi_{t_1,t_2} \Phi_{t_2,t_3} = \Phi_{t_1,t_3}$
holds for all $(t_1,t_2)$, $(t_2,t_3) \in T_\le^2$.
\end{itemize}
\noindent
%The family
%$\{ \Phi_{t_1,t_2}{}: $
%$L^\infty(\Omega, \nu)PP_{t_2} \to L^\infty(\Omega, \nu)PP_{t_1} \}_{(t_1,t_2) \in T^2_\le}$
%is also called a
%{\lq\lq causal relation among systems\rq\rq}$\!.$
Let an observable ${}{{}\mathsf O}_t \equiv (X_t, {\cal F}_t, F_t)$ in $L^\infty(\Omega_t, \nu_t)$ be given for each $t \in T$.
The pair
$[{}\{ {{}\mathsf O}_t \}_{ t \in T} ,
\{  \Phi_{t_1,t_2}{}: $
$L^\infty(\Omega_{t_2}, \nu_{t_2}) \to L^\infty(\Omega_{t_1}, \nu_{t_1}) \}_{(t_1,t_2) \in T^2_\le }$
$]$
is called
a
\it
sequential observable
\rm
which is denoted by
$[{{}\mathsf O}_T{}]$,
i.e.,
$[{{}\mathsf O}_T{}]$
$\equiv$
$[{}\{ {{}\mathsf O}_t \}_{ t \in T} ,
\{  \Phi_{t_1,t_2}{}: $
$L^\infty(\Omega_{t_2}, \nu_{t_2}) \to L^\infty(\Omega_{t_1}, \nu_{t_1}) \}_{(t_1,t_2) \in T^2_\le }$
$]$.
\hfill{{$///$}}%BFBFbfbf
%\END{Def}

\rm

%
%$L^\infty(\Omega, \nu)PP_2S}_{[\rho_0^p]}{S}_{[\rho^p_{0}]}

Let
$(T{(t_0)}, {{\; \leqq \;}})$
be a tree with the root
$t_0$.
For each
$t \in T$,
define the separable complete metric space
$X_t$,
and
the
Borel field
$ {\cal B}_{X_t}$,
and further,
define
the observable
${\mathsf O}_t {{=}} (X_t, {\cal F}_t, F_t)$
in
${L^\infty(\Omega_t, \nu_t)}$.
That is,
we have
a sequential observable
$[{}{\mathbb O}_{T(t_0)}{}]$
$=$
$[{}\{ {\mathsf O}_t \}_{ t \in T} ,
\{  \Phi_{t_1,t_2}{}: $
$L^\infty(\Omega_{t_2}, \nu_{t_2})
 \to L^\infty(\Omega_{t_1}, \nu_{t_1})  \}_{(t_1,t_2) \in T^2_{\leqq}}$
$]$.

\par
Here, define,
$\overline{\cal P}_0(T)$
$( =
\overline{\cal P}_0(T(t_0))
\subseteq {\cal P}(T) )$
such that
\par
\noindent
\begin{align*}
&
\overline{\cal P}_0(T(t_0))
\\
{{=}}
&
\{ {T'} \subseteq T \;|\; {T'}
\text{ is finite}, t_0 \in T'
\text{ and
satisfies
}
{\roman Inf}_{T'} S = {\roman Inf}_T S
\;\;
(\forall S \subseteq T')
\}
\end{align*}
%\END{align*}
Let
${T'{(t_0)}} \in \overline{\cal P}_0(T{(t_0)} )$.
Since
$(T'(t_0), {{\; \leqq \;}})$
is finite,
we can
put
$({T'} {{=}} \{ t_0, t_1,\ldots , t_N \},$
$ \pi{}: {T'} \setminus \{t_0\} \to T'{})$,
where
$\pi$ is a parent map.
\par
\noindent
\par
Now, consider the sequential observable
$[{}\{ {\mathsf O}_t \}_{ t \in {T'}} ,
\{  \Phi_{\pi(t), t }{}: $
$L^\infty(\Omega_t, \nu_t) \to
L^\infty(\Omega_{\pi(t)}, \nu_{\pi(t)}) \}_{ t \in {T'} \setminus \{t_0 \} }$
$]$.
For each
$s$
$({}\in {T'}{})$,
putting
$T_s =\{ t \in T' \;|\; t {\; \geqq \;}s \}$,
define the observable
$\widehat{\mathsf O}_s {{=}} (\bigtimes_{t \in T_s } X_t, $
$\bigtimes_{t \in T_s } {\cal F}_t, {\widehat F}_s)$
in
$
L^\infty(\Omega_t, \nu_t)
$
such that
%such that:
\par
\noindent
\begin{align*}
\widehat{\mathsf O}_s
=
\begin{cases}
{\mathsf O}_s
\quad
&
\text{({}$ s \in {T'} \setminus \pi ({T'}) $\; \text{ and }{})}
\\
\\
{\mathsf O}_s
{\times}
({}{
\underset{{t \in \pi^{-1} ({}\{ s \}{})}}{\bigtimes}
} \Phi_{ \pi(t), t}
\widehat {\mathsf O}_t{})
\quad
&
\text{({}$ s \in \pi ({T'}) ${}\; \text{ and })}
\end{cases}
%\tag{\textcolor{black}{11.7}}
\end{align*}
And further,
iteratively,
we get
$\widehat{\mathsf O}_{t_0}{{=}} (\bigtimes_{t \in T' } X_t, $
$\bigtimes_{t \in T' } {\cal F}_t, {\widehat F}_{t_0})$,
which is also denoted by
$\widehat{\mathsf O}_{{T'}}{{=}} (\bigtimes_{t \in T' } X_t, $
$\bigtimes_{t \in T' } {\cal F}_t, {\widehat F}_{T'})$.

\par

%index{natural projection@natural projection}
For any subsets
$T_{1} \subseteq T_{2}({} \subseteq {{T}}{})$,
define the natural projection
$
\pi_{T_{1},T_{2}}:
\bigtimes_{{t} \in T_{2}} X_{{t}}
\longrightarrow
\bigtimes_{{t} \in T_{1}} X_{{t}}
$
by
%\textcolor{black}{(10103)} and ,
\begin{align*}
\mathop{\mbox{\Large $\times$}}_{t\in T_{2}}X_{t}
\ni
(x_t )_{t \in T_2 }
\mapsto
(x_t )_{t \in T_1}
\in
\mathop{\mbox{\Large $\times$}}_{t\in T_{1}}X_{t}
%
%\longrightarrow
%\mathop{\mbox{\Large $\times$}}_{\lambda\in\Lambda_{1}}X_{\lambda}.
%%%
%P%\tag{11.17}
\end{align*}

%, $\pi_{T}=\pi_{T,{{T}}}$ and .
%${t}\in{{T}}$, ${L^\infty(\Omega, \nu)}$
%{}
%$(X_{{t}},{\cal F}_{{t}},F_{{t}})$
%.

%
%].}
%\rm

\par
%$[{}{\mathbb O}_{T(t_0)}{}]$
%$=$
%$[{}\{ {\mathsf O}_t \}_{ t \in T} ,
%\{  \Phi_{t_1,t_2}{}: $
%${L^\infty (\Omega_{t_2}, \nu_{t_2})} \to {L^\infty (\Omega_{t_1},
%\nu_{t_1})} \}_{(t_1,t_2) \in T^2_{\leqq}}$
%$]$
%{} and .
%%.
Assume that the observables
     $\bigl\{$
$\widehat{\mathbb O}_{T'}{{=}} (\bigtimes_{t \in T' } X_t, $
$\bigtimes_{t \in T' } {\cal F}_t, {\widehat F}_{T'})$
%
%
%     ${{{}\mathsf O}}_{{T_0}} {{=}}$
%     $(${}     $\mathop{\mbox{\Large $\times$}}_{{t}\in{T_0}}X_{{t}},
%$
%     $\mathop{\mbox{\Large $\times$}}_{{t}\in{T_0}}{\cal
%F}_{{t}},$
%     $F_{{T_0}}$
%     $)$
    $~|~$
     ${T'}\in{\overline{\cal P}_0}({{T}})$
     $\bigr\}$
in
$L^\infty(\Omega_{t_0}, \nu_{t_0})$
satisfy the following consistency condition,
that is,
%index{consistency condition@consistency condition}
       \begin{itemize}
       \item[]
for any
$T_1,
T_2$
($\in$
${\overline{\cal P}_0}({
{T}})$)
such that
$T_1 \subseteq T_2$,
it holds that
       \end{itemize}
\begin{align*}
       {\widehat F}_{T_2}
       \bigl({}
       \pi_{T_{1},T_{2}}^{-1}({\Xi}_{T_1}{})
       \bigr)
       =
       {\widehat F}_{T_1}
       \bigl({}{\Xi}_{T_1} \bigr)
       \quad
       ({}\forall {\Xi}_{T_1} \in
       \bigtimes_{{t} \in T_1}
       {\cal F}_{{t}}{})
       %%
%P%\tag{11.6}
\end{align*}
\par
\noindent
Then,
by
\textcolor{black}{Theorem 10.13}[
Kolmogorov extension theorem in measurement theory],
there uniquely
exists the observable
${\widehat{\mathsf O}}_{{{T}}}$
${{=}}$
$\bigl(\mathop{\mbox{\Large $\times$}}_{{t} \in
{T}}X_{{t}}, $
$\mathop{\mbox{$\bigstimes$}}_{{t}\in{{T}}}
{\cal
F}_{{t}},$
${\widehat F}_{{T}}
\bigr)$
in
$L^\infty(\Omega_{t_0}, \nu_{t_0})$
such that:
\begin{align*}
       {\widehat F}_{{T}}
       \bigl({}
       \pi_{{T_0, T}}^{-1}({\Xi}_{{T_0}}{})
       \bigr)
       =
       {\widehat F}_{{T_0}}  \bigl({}{\Xi}_{{T_0}} \bigr)
       \quad
       ({}\forall {\Xi}_{{T_0}} \in
       \mathop{\mbox{$\bigstimes$}}_{{t}\in{T_0}}
       {\cal F}_{{t}},~
       \forall{T_0}\in{\overline{\cal P}_0}({{T}}){})
       %%\t
%P%\tag{11.7}
\end{align*}
%\qed
\par
\noindent
%------------------End of-----------------
\rm
%---------- of TH--------------
\par
\noindent
%%{{T}}{\widehat T_}{t}\lambda\lambda
%%{{T}}{\widehat {T_0}_}{t}\lambda\lambda
%%{{T}}{\widehat \Lambda_}{t}\lambda\lambda
This observable
${\widehat{\mathsf O}}_{{{T}}}$
${{=}}$
$(\mathop{\mbox{\Large $\times$}}_{{t} \in
{T}}X_{{t}}, $
$\mathop{\mbox{$\bigstimes$}}_{{t}\in{{T}}}
{\cal
F}_{{t}},$
${\widehat F}_{{T}}
)$
is called the realization of
the sequential observable
$[{}{\mathbb O}_{T(t_0)}{}]$
$=$
$[{}\{ {\mathsf O}_t \}_{ t \in T} ,
\{  \Phi_{t_1,t_2}{}: $
$L^\infty(\Omega_{t_2}, \nu_{t_2})
$
$\to L^\infty(\Omega_{t_1}, \nu_{t_1}) \}_{(t_1,t_2) \in T^2_{\leqq}}$
$]$.
%index{realizedcausalobservable@realized causal observable}
%K

\par
Summing up the essential part of the above argument,
we can propose the following axiom, which corresponds to
Axiom${}_{\text{\scriptsize c}}^{\text{\scriptsize pm}}$ 2 (Causality: page \pageref{axiomcpm2}).

%\vskip0.5cm
\par

\vskip0.1cm
\rm

\par
\noindent
%\vskip0.2cm
%
%%corresponds to
%
%{}
%Cf
%%index{2a@Axiom${}_{\text{\scriptsize b}}^{\text{\scriptsize pm}}$ 2[causality (bounded type )]}
%\large
\par
%\vskip0.2cm6.8
%\vskip0.1cm
%\par
%
%

%\vskip0.5cm

\par
\noindent
\begin{center}
{\bf
Axiom${}_{\text{\scriptsize b}}^{\text{\scriptsize pm}}$ 2
(causality : bounded type ) }
%\label{rule902}
\label{axiombpm2}
\end{center}
%\END{document}
\par
\noindent
%\vskip0.1cm
\par
\noindent
\fbox{\parbox{155mm}{
\begin{itemize}
\item[(i)] A chain of causalities
\\
A chain of causalities
is represented by a,
%({{the Heisenberg picture}})
%\\
{{sequential causal operator}}
$$
\{ \Phi_{t_1,t_2}{}:
%$
%$
{L^\infty (\Omega_{t_2}, \nu_{t_2})} \to {L^\infty (\Omega_{t_1}, \nu_{t_1})} \}_{(t_1,t_2) \in T^2_{\leqq}}
$$
\item[(ii)]
{{Realized causal}} observable
%]
%\hspace{-10mm}
\\
A sequential observable
$$
[{}{\mathsf O}_{T(t_0)}{}]
%$
%$
{{=}}
%$
%$
[{}\{ {\mathsf O}_t \}_{ t \in T} ,
\{  \Phi_{t_1,t_2}{}:
%$
%$
{L^\infty (\Omega_{t_2}, \nu_{t_2})} \to {L^\infty (\Omega_{t_1}, \nu_{t_1})} \}_{(t_1,t_2) \in T^2_{\leqq}}
%$
%${C (\Omega_{t_2})} \to {C (\Omega_{t_1})} \}_{(t_1,t_2) \in T^2_{\leqq}}
%$
%$
]
$$
is realized by its
{{realized causal}} observable
$
\widehat{\mathsf O}_{T{}}
{{=}} (\bigtimes_{t \in T} X_t, \bigstimes_{t \in T } {\cal F}_t,
{\widehat F}_T)
$
\end{itemize}
}
}
\par
\vskip0.5cm
\par
\noindent

%
%
%%KKKKKKKKKKKKKKKKKKKKKKKKKKKKKK{
%\BEGIN{itembox}[c]{
%{\bf
%Axiom${}_{\text{\scriptsize b}}^{\text{\scriptsize pm}}$ 2
%(causality (bounded type )) }
%%{\bf
%%(bounded type )\textcolor{black}{Axiom${}_{\text{\scriptsize b}}^{\text{\scriptsize pm}}$ 2}
%%[causality {\rm{(causality)}}]}
%}
%%%index{1@
%%Axiom${}_{\text{\scriptsize b}}^{\text{\scriptsize p}}$ 1[{{{measurement theory}}}]
%%}
%\label{rule1001}
%\label{axiombpm2}
%%index{2@Axiom${}_{\text{\scriptsize b}}^{\text{\scriptsize pm}}$ 2[{}causality {}(bounded type )]}
%%%index{ and @
%%{statistics}probability 
%%}
%%\BEGIN{itemize}
%%\item[(i)]
%(i):A chain of causalities
%\\
%A chain of causalities
%is represented by a,
%%({{the Heisenberg picture}})
%%\\
%{{sequential causal operator}}
%$$
%\{ \Phi_{t_1,t_2}{}:
%%$
%%$
%{L^\infty (\Omega_{t_2}, \nu_{t_2})} \to {L^\infty (\Omega_{t_1}, \nu_{t_1})} \}_{(t_1,t_2) \in T^2_{\leqq}}
%$$
%\\
%%\item[
%(ii):{{Realized causal}} observable
%%]
%%\hspace{-10mm}
%\\
%A sequential observable
%$$
%[{}{\mathsf O}_{T(t_0)}{}]
%%$
%%$
%{{=}}
%%$
%%$
%[{}\{ {\mathsf O}_t \}_{ t \in T} ,
%\{  \Phi_{t_1,t_2}{}:
%%$
%%$
%{L^\infty (\Omega_{t_2}, \nu_{t_2})} \to {L^\infty (\Omega_{t_1}, \nu_{t_1})} \}_{(t_1,t_2) \in T^2_{\leqq}}
%%$
%%${C (\Omega_{t_2})} \to {C (\Omega_{t_1})} \}_{(t_1,t_2) \in T^2_{\leqq}}
%%$
%%$
%]
%$$
%is realized by its
%{{realized causal}} observable
%$
%\widehat{\mathsf O}_{T{}}
%{{=}} (\bigtimes_{t \in T} X_t, \bigstimes_{t \in T } {\cal F}_t,
%{\widehat F}_T)
%$
%%\hfill$13)$
%%\END{itemize}
%\END{itembox}
%

\par

\par
Thus,
we have the{{{measurement theory}}}(bounded$\cdot$pure type)
as follows.
%:
\begin{align*}
%\dashbox{5}
\underset{\text{\scriptsize (scientific language)}}{\text{{}
$\fbox{
\text{{{measurement theory}}(bounded$\cdot$pure type)}
}$}}
:=
{
\overset{\text{\scriptsize [Axiom${}_{\text{\scriptsize b}}^{\text{\scriptsize p}}$ 1]}}
{
\underset{\text{\scriptsize
[probabilistic interpretation]}}{\text{{} $\fbox{(pure){{measurement}}}$}}}
}
+
{
\overset{\text{\scriptsize [Axiom${}_{\text{\scriptsize b}}^{\text{\scriptsize pm}}$ 2]}}
{
\underset{\text{\scriptsize [{{the Heisenberg picture}}]}}
{\text{{}$\fbox{ causality }$}}
}
}
\end{align*}

\par
Therefore, we say that
%
%\textcolor{black}{{(}bounded type {)}Axiom${}_{\text{\scriptsize 
%b}}^{\text{\scriptsize p}}$ 1}
%{Sec.$\;$}\textcolor{black}{(\REF{axiomcp1})}
% and ,
%%,
\begin{itemize}
\item[]
The probability that
a
measured value
$( x_t)_{t \in T}$
obtained by
a {{measurement}}
$
{\mathsf M}_{L^\infty(\Omega_{t_0}, \nu_{{}_{t_0}}
)}  ({}\widehat{\mathsf O}_{T{}}
{{=}}
$
$ (\bigtimes_{t \in T} X_t,$
$ \bigstimes_{t \in T } {\cal F}_t,
{\widehat F}_T), S_{[\omega_{t_0}]}{})
$
belongs to
${\widehat \Xi}$
$(\in $
$\bigstimes_{t \in T} {\cal F}_t )$
is given by
$
[{\widehat F}_T(
{\widehat \Xi}
)](
\omega_{t_0}
)$,
if
$
{\widehat F}_T
({\widehat \Xi}
)$
is essentially continuous at
$
\omega_{t_0}
(\in \Omega_{t_0})
$.
\end{itemize}

\par
\noindent
%BBBBBBBBBBBBBBBBBB%SBSBSBS
{\small%%{\footnotesize
\vspace{0.1cm}
\begin{itemize}
\item[$\spadesuit$] \bf {{}}{Note }11.3{{}} \rm
By an analogy of
Remark 6.19,
we also get
the measurement theory
(bounded$\cdot$mixed type)
as follows.
That is,
\begin{align*}
\!\!\!
\underset{\text{\scriptsize (scientific language)}}{\text{{}
$\fbox{
\text{
measurement theory (bounded$\cdot$mixed type)}
}$}}
%\\
:=
%&
%\!\!\!\!
%\!\!\!\!\!\!
{
\overset{\text{\scriptsize [Axiom${}_{\text{\scriptsize b}}^{\text{\scriptsize p}}$ 1]}}
{
\underset{\text{\scriptsize
[probabilistic interpretation]}}{\text{{} $\fbox{(mixed){{measurement}}}$}}}
}
\!\!
+
\!\!
{
\overset{\text{\scriptsize [Axiom${}_{\text{\scriptsize b}}^{\text{\scriptsize pm}}$ 2]}}
{
\underset{\text{\scriptsize [{{the Heisenberg picture}}]}}
{\text{{}$\fbox{ causality }$}}
}
}
%\TAG*{$\displaystyle{\mathop{1)}_{(=10))}}$}
%%%%%%%%%%%%%CLAsSICAL
\end{align*}
% and . Axiom${}_{\text{\scriptsize b}}^{\text{\scriptsize p}}$ 1,2{Note }10.4,{Sec.11.3}.
\end{itemize}
}

\par
\noindent
%{\bf {Definition }11.7}
%[].

\renewcommand{\footnoterule}{%
  \vspace{2mm}                      % 
  \noindent\rule{\textwidth}{0.4pt}   % , 
  \vspace{-5mm}
}

\subsection{{Is Brownian motion a motion?}}%11.4
%index{@Brownian motion}
\par

It is natural to consider that
%
%In this section,
%we do not consider that
%Newtonian mechanics is within the  world view,
%but investigate Newtonian mechanics
%in the view-point such that
\begin{itemize}
\item[(A)]
Brownian motion should be understood in measurement theory.
%\hfill{\roman [FN]}
\end{itemize}
Let us explain it as follows.

%we proposed
%
%
%
%
%
%In this section,
%from the measurement theoretical point of view,
%we review the Brownian motion.
%section,
%probability$B(t,\lambda)$
%------
%\textcolor{black}{6.24(})continuous
%------
%, ,
%
%measurement{}\footnote{
%=
%,
% and ,
%,
%,
% and .
%%index{@}
%}.

\par
Let
$(\Lambda, {\cal F}_\Lambda , P)$
be a probability space.
For each
$\lambda \in \Lambda $,
define
the real-valued continuous function
$B(\cdot ,\lambda):
T({{=}} [0, \infty)) \to {\mathbb R}$
such that,
for any
$t_0=0 < t_1 < t_2 < \cdots < t_n $,
\begin{align*}
&
P(\{ \lambda \in \Lambda \;|\;
B(t_k,\lambda) \in \Xi_k \in
{\cal B}_{\mathbb R}
\;\;
(k=1,2,\ldots, n )
\})
\\
=&
%\int_{\Xi_0 } {{\rho}_0}({}\omega_0{})
%\Big(
\int_{\Xi_1 }
\Big(\cdots
(\int_{\Xi_{t_{n-1}}}
(\int_{\Xi_{t_n}}
\bigtimes_{k=1}^{n}
G_{ \sqrt{t_{k} - t_{k-1}} } ({}\omega_{k} - \omega_{k-1}{})
d\omega_n)
d\omega_{n-1})
\cdots
\Big)
d\omega_1
\tag{\color{black}{11.1}}
%%%%%REDREDREDREDREDRE
%\Big)
%(
%\omega_0
%) 
%%%%\TAG{11.9}
\end{align*}
where,
$\omega_0
\in {\mathbb R}$,
$
d\omega_k$
is the Lebesgue measure
on ${\mathbb R}$,
$G_{\sqrt{t}}(q)
   =
\frac{1}{\sqrt{2\pi t}}\mbox{\rm exp}\left[{}- \frac{q^2}{2t} \right]$.
%
%\par
%,
%measurement{}$\; \Longrightarrow \;$,
%.
%
%Let
%$({\mathbb R} , {\cal B}_{\mathbb R},m)$
%be the Lesbegue measure space.
%%
%$[C(
%{\mathbb R}
%), L^{\infty}({\mathbb R} ,m)]$
%,
\par
\noindent
\par
Now consider the diffusion equation:
\begin{align*}
  \frac{\partial{\rho}_t (q)}{\partial t}
  =
  \frac{\partial^2{\rho}_t (q)}{\partial q^2},
\qquad
(\forall q \in {\mathbb R},
\forall t \in T {{=}} [0,\infty )
\;)
%
%P%\TAG{11.10}
\end{align*}
%$t (\in [0,\infty))$
%, statespace${\mathbb R}_t$
%$=$
%${\mathbb R} $
% and .
%\BEGIN{align*}
%  {\rho}_{0}
%\in
%         \{ {\rho}\in
%         L^1({}{\mathbb R} ,m{})
%
%%%
%%\TAG{11.29} \end{align*}
%\par
%\noindent
By the solution
$\rho_t$,
we get
predual operator
$\{
[{}\Phi_{t_1, t_2}{}]_*$
$:$
$ L^1({}{\mathbb R} ,dq{})$
$\to$
$ L^1({}{\mathbb R},dq{})\}$
as follows.
That is,
for each
$\rho_{t_1} \in L^1({}{\mathbb R} ,m{})$,
define
%,
\begin{align*}
\big({}[{}\Phi_{t_1, t_2}{}]_{*} ({}{\rho}_{t_1}{}) \big) (q)
=
{\rho}_{t_2}({}q)
=
\int_{-\infty}^{\infty} {\rho}_{t_1}(y) G_{\sqrt{t_2 - t_1} }(q -
y) m({}dy{})
\;\;
(\forall q \in {\mathbb R},
{}\forall ({}t_1 , t_2{}) \in T^2_{\leqq} {})
%P%\TAG{11.11}
\end{align*}
%$G_{t}(q)$ is the Normal function, that is,.
%The {\lq\lq state's evolution\rq\rq} is,
%of course,
%represented by
%the
%Schr\"odinger picture
%$\{ [{}\Phi_{ t_1 , t_2 }{}]_* \; | \; ({}t_1, t_2{}) \in 
%{\mathbb R}^2_{{}_{{\; \leqq \;}}} \}$.
\par
\par
For simplicity, we put
$(\Omega, {\cal B}, d \omega{})$
$=$
$({\mathbb R}_q, {\cal B}({\mathbb R}_q), dq)$.
And therefore,
put $({\cal N}, {\cal N}_*)$ $=$ $(L^{\infty}(\Omega, d \omega),
L^1(\Omega, d \omega))$.
Putting
$\Phi_{t_1 , t_2 }$
$=$
$({}[{}\Phi_{ t_1 , t_2 }{}]_*{})^*$,
we get
the causal relation
$\{  \Phi_{ t_1 , t_2 }  \; | \; ({}t_1, t_2{}) \in  {T}^2_{{}_{\le}} \}$.
For each
$t \in T$,
consider the exact observable
${\mathsf O}_t^\FIN = ( \Omega , {\cal B}_\Omega, F^\FIN )$
in
$L^\infty( \Omega, d \omega )$.
Thus, we get
the sequential causal exact observable
$[{\mathbb O}_T]$
$
=
[\{{\mathsf O}_t^\FIN \}_{t\in T} ;$
$
\{  \Phi_{ t_1 , t_2 }  \; | \; ({}t_1, t_2{}) \in  {T}^2_{{}_{\le}} \}
]$.
The Kolmogorov extension theorem
(Theorem 9.21)
says that
${\mathbb O}_T$
has the realized
causal observable
$\widehat{\mathsf O}_{t_0}$
$=$
$({\Omega}^T, {\cal B}({\Omega}^T),
\widehat{F}_{t_0}
)$
in
$L^\infty ( \Omega ,d \omega )$.

Assume that
\begin{itemize}
\item[]
$\quad$
a measured value
${\widehat \omega}$
$(=(\omega_t)_{t\in T  } \in\Omega^{T })$
is obtained
by
${\overline{\mathsf M}}_{L^\infty({}\Omega) } (\widehat{\mathsf O}_{
t_0 },$
$
S_{[\delta_{\omega_0}]} )$.
%({}{{\overline \rho}_0}{}){})$.
\end{itemize}
Let $D= \{t_1,t_2,\cdots,t_n\}$ be a finite subset of
$T$,
where $t_0=0 < t_1 < t_2 < \cdots < t_n $.
Put
${\widehat\Xi}=\mbox{\Large $\bigtimes$}_{t\in T
}^D\Xi_t$
$\bigl(\in{\cal B}^{ {\overline{\mathbb R}^+} }\bigl)$ where
$\Xi_t=\Omega$
$(\forall t\notin D)$.
Then, by Axiom${}_{\text{\scriptsize b}}^{\text{\scriptsize pm}}$ 2, we see
\begin{itemize}
\item[]
$\quad$
the probability that
${\widehat \omega} ({}=(\omega_t)_{t \in T}) $
belongs to the set
${\widehat\Xi} \equiv \mbox{\Large $\bigtimes$}_{t\in {T} }^D\Xi_t$
is given by
$
[\widehat{F}_{t_0}(
\mbox{\Large $\bigtimes$}_{t\in {T} }^D \Xi_t
)](\omega_0)
$
\end{itemize}
where
\begin{align*}
&
[\widehat{F}_{t_0}(
\mbox{\Large $\bigtimes$}_{t\in {T} }^D \Xi_t
)](\omega_0)
\\
=
&
\Big(F({}\Xi_{0})
\Phi_{ 0 , t_1 }
\Big(F({}\Xi_{t_1})
\cdots
\Phi_{ t_{n-2}  , t_{n-1}}
\Big({}
F({}\Xi_{t_{n-1}})
\bigl({}\Phi_{t_{n-1} , t_n} F({}\Xi_{t_n} {}) \bigl)
\Big)
\cdots
\Big)
(\omega_0)
\\
=&
%\int_{\Xi_0 } {{\overline \rho}_0}({}\omega_0{})
%\Big(
\int_{\Xi_1 }
\Big(\cdots
(\int_{\Xi_{t_{n-1}}}
(\int_{\Xi_{t_n}}
{\bigtimes}_{k=1}^{n}
G_{ t_{k} - t_{k-1} } ({}\omega_{k} - \omega_{k-1}{})
d\omega_n)
d\omega_{n-1})
\cdots
\Big)
d\omega_1
%\Big)
%d \omega_0  .
\tag{11.2}
\end{align*}
which is equal to
the
\textcolor{black}{(11.1)}.
\par
\noindent
Thus, we say that
\begin{itemize}
\item[]
%$\quad$
The Brownian motion $B(t,\lambda)$
is not a motion but
a measured value.
( Some may recall Parmenides' saying:
\it
There are no {\lq\lq}plurality",
but only {\lq\lq}one".
And therefore, there is no movement.
\rm
)
\end{itemize}
%%\Psi\Phi

\par
\noindent
%BBBBBBBBBBBBBBBBBB%SBSBSBS
{\small%%{\footnotesize
\vspace{0.1cm}
\begin{itemize}
\item[$\spadesuit$] \bf {{}}{Note }11.4{{}} \rm
The above argument gives
an answer to
the problem
(\textcolor{black}{{Chap.$\;$1}(F$_5$)(={Note }1.1$(\sharp_2)$})
),
i.e.,
\begin{itemize}
\item[]
Why is a mathematical theory
(i.e.,
Brownian motion,
stochastic process)
useful?
\end{itemize}
That is,
%(b),
\begin{itemize}
\item[]
Behind Brownian motion,
the world-view
(called classical {{measurement theory}}) is hidden
\end{itemize}
In this sense,
Nelson's probabilistic quantization
may be
the confusion of the order of things.
Also, recall
\textcolor{black}{{Note }2.17}
as follows.
Therefore, for example,

\begin{table}[h]%%b h(here) t p
\begin{center}
\begin{tabular}{l|ll}
       mathematics        &  & $\quad$  world-description method \\
\hline
%%%%
\text{differential geometry}
&
&$\quad$ 
\text{the theory of relativity}
\\
\text{differential equation}
&
&$\quad$ 
\text{{{Newton}} mechanics,
electromagnetism
}
\\
\text{Hilbert space}
&
&$\quad$
\text{quantum mechanics}
	\\
$\underset{\text{\scriptsize (Hilbert space)}}{\text{probability theory}}$
&
&$\quad$
{measurement theory}
	\\
\end{tabular}
%%\caption{Axiom}
%\BEGIN{center}
%	0.1: Axiom
%\END{center}
\end{center}
\end{table}
%\END{itemize}
\end{itemize}
}

\subsection{Exact measurement of deterministic {{sequential causal operator}}
and {{the Schr\"odinger{ picture}}}}%{Sec.11.5}
%%\ssubsubsection{{{the Schr\"odinger{ picture}}}}
\par
The Copenhagen interpretation
---
{Chap.$\;$1}(U$_4$)
%\textcolor{black}{\REF{2secAxiom 1}}
%\textcolor{black}{Axiom${}_{\text{\scriptsize b}}^{\text{\scriptsize p}}$ 1}(i
---
says that
"only one measurement is permitted",
which
implies
"State does not change".

\par
However,
as mentioned in Sec.6.4.4,
as a convenient method,
we sometimes use
%,  and ,
the state change
due to
{{the Schr\"odinger{ picture}}}.

\par
%\noindent
%\vskip0.3cm
\vskip0.3cm
%BFBF
\par
\noindent
{\bf {Definition }11.7}$\;\;$%POPOPO
{\bf [{{State}} change --- {{the Schr\"odinger{ picture}}}]}
Let
$
\{  \Phi_{t_1,t_2}{}: $
${L^\infty (\Omega_{t_2}, \nu_{t_2})}$
$ \to {L^\infty (\Omega_{t_1})},$
$ {\nu_{t_1})} \}_{(t_1,t_2) \in T^2_{\leqq}}$
be
a deterministic causal relation
with
the deterministic causal maps
$\phi_{t_1,t_2 }:\Omega_{t_1} \to \Omega_{t_2}$
$(\forall
{(t_1,t_2) \in T^2_{\leqq}}
)$.
Let
$\omega_{t_0} \in \Omega_{t_0}$
be an initial state.
Then,
the
$\{ \phi_{t_0,t} (\omega_{t_0} ) \}_{t \in T }$
(or,
$\{ \delta_{\phi_{t_0,t} (\omega_{t_0} )} \}_{t \in T }$
is called the Schr\"odinger picture representation.

\par
\vskip0.2cm
\par
The following is similar to
\textcolor{black}{{{Theorem }}6.18}
%bounded type {{measurement theory}}{}
%\par
%\noindent
%\vskip0.3cm
%\vskip0.3cm
%BFBF
\par
\noindent
{\bf {{Theorem }}11.8
[Deterministic {{sequential causal operator}}{{realized causal}} observable
]} %POPOPO
Let
$(T(t_0), {{\; \leqq \;}})$
be a tree with the root
$t_0$.
Let
$[{}{\mathbb O}_T{}]$
$=$
$[{}\{ {\mathsf O}_t \}_{ t \in T} ,
\{  \Phi_{t_1,t_2}{}: $
${\L^\infty (\Omega_{t_2}, \nu_{t_2})} \to {L^\infty (\Omega_{t_1}, \nu_{t_1})} \}_{(t_1,t_2) \in T^2_{\leqq}}$
$]$
be a deterministic sequential observable.
Then,
the realization
$\widehat{\mathsf O}_{{t_0}{}} $
$
\equiv ({\bigtimes}_{t \in T} X_t,{{\bigstimes}_{t \in T} {\cal F}_t},
{\widehat F}_{t_0})
$
is represented by
\begin{align*}
\widehat{\mathsf O}_{{t_0}{}} = \bigtimes_{t\in T} \Phi_{{t_0},t} {\mathsf O}_t
%P%\TAG{7.13}
%\tag{11.7}
\end{align*}
That is, it holds that
\begin{align*}
&
[\widehat{F}_{t_0} (
\bigtimes_{t\in T} \Xi_t \
%times \Xi_1 \times \cdots \times \Xi_N
)]
(\omega_{t_0} ) = \bigtimes_{t\in T} [\Phi_{{t_0},t} {F}_t (\Xi_t )](\omega_{t_0} )
= \bigtimes_{t\in T} [{F}_t (\Xi_t )](\phi_{{t_0},t} \omega_{t_0} )
%P%\TAG{7.14}
\\
&
\quad \qquad \quad \qquad \quad \qquad \quad \qquad
%\qqquad
(\forall \omega_{t_0} \in \Omega_{t_0}, \forall \Xi_t \in {\cal F}_t )
%\tag{11.8} 
\end{align*}
%{t_0}{t_0}00000000000
\par
\noindent
%\END{The}
%BFBF
\rm
\par
\noindent
\noindent
\rm
{\it $\;\;\;\;${Proof.}}$\;\;$
The proof
is similar to
that of
\textcolor{black}{Theorem 6.18}.
\qed
\par
\noindent

\par
\noindent
\vskip0.3cm
%\vskip0.3cm
%BFBF
\par
\noindent
{\bf {{Theorem }}11.9}
Let
$[{}{\mathbb O}_{T(t_0)}]$
${{=}}$
$[{}\{ {\mathsf O}^{\FIN}_t \}_{ t \in T} ,
\{  \Phi_{t_1,t_2}{}: $
${L^\infty (\Omega_{t_2}, \nu_{t_2})} \to {L^\infty (\Omega_{t_1}, \nu_{t_1})} \}_{(t_1,t_2) \in T^2_{\leqq}}$
%${C (\Omega_{t_2})} \to {C (\Omega_{t_1})} \}_{(t_1,t_2) \in T^2_{\leqq}}$
$]$
be a deterministic sequential exact observable,
which has
the deterministic causal maps
$\phi_{t_1,t_2 }:\Omega_{t_1} \to \Omega_{t_2}$
$(\forall
{(t_1,t_2) \in T^2_{\leqq}}
)$.
And
let
${\widehat{\mathsf O}}_{{{t_0}}}$
${{=}}$
$(\mathop{\mbox{\Large $\times$}}_{{t} \in
{T}}X_{{t}}, $
$\mathop{\mbox{\Large $\times$}}_{{t}\in{{T}}}
{\cal
F}_{{t}},$
${\widehat F}_{{T}}
)$
be its realized causal observable
in $L^\infty(\Omega_{t_0}, \nu_{t_0} )$.
Assume that
the measured value
$(x_t )_{t\in T }$
is obtained by
${\overline{\mathsf M}}_{L^\infty(\Omega_{t_0})}  ({}\widehat{\mathbb O}_{T{}}$
$
{{=}}
$
$(\bigtimes_{t \in T} X_t,
\mathop{\mbox{\Large $\times$}}_{{t}\in{{T}}}
{\cal
F}_{{t}}
,$
$ {\widehat F}_0), S_{[\omega_{t_0}]}{})$.
Then,
we surely believe that
\begin{align*}
x_t = \phi_{t_0,t } (\omega_{t_0} )
\qquad
(\forall t \in T )
%P%\tag{11.13}
\end{align*}
{}\rm{}
Thus,
we say that,
as far as a deterministic sequential observable,
\begin{itemize}
\rm
\item[]
{}\rm{}
exact measured value $(x_t)_{t\in T}$
=
the Schr\"odinger picture representation
$(\phi_{t_0,t } (\omega_{t_0} ))_{t\in T}$
\end{itemize}
%\END{The}
%BFBF
\par
\noindent
\rm
\noindent
\rm
{\it $\;\;\;\;${Proof.}}$\;\;$
Let
$D=\{t_1,t_2,\ldots,t_n\}
(\subseteq T)$
be any finite subset
of $T$.
Put
${\widehat\Xi}=\mbox{\Large $\bigtimes$}_{t\in {T}
}^D\Xi_t$
$=$
$(\bigtimes_{t\in D} \Xi_t ) \times (\bigtimes_{t\in T\setminus D} X_t ) $,
where
$\Xi_t$
$\subseteq$
$X_t ( = \Omega_t)$
is an open set
such that
$\phi_{t_0, t}
(\omega_{t_0})
\in \Xi_t$
$(\forall t\in D)$.
%
%$\Xi_t=\Omega_t$
%$(\forall t\notin D)$
Then,
we see that
\begin{itemize}
\item[]
the probability that
the measured value
$(x_t )_{t\in T }$
belongs to
${\widehat\Xi}=\mbox{\Large $\bigtimes$}_{t\in {T}
}^D\Xi_t$
is equal to
$1$.
%:
\end{itemize}
That is because
\textcolor{black}{Theorem 11.8} says that
{
%\footnotesize
\begin{align*}
%{\omega_{t_0}}
&
\bigl({}{\widehat F}_{T} ({\widehat\Xi}) \bigr)
({\omega_{t_0}})
%   =
%&
%\Big({}
%F^{\FIN}({}\Xi_{0})
%\Phi_{ t_0 , t_1 }
%\Big(F^{\FIN}({}\Xi_{t_1})
%\cdots
%\Phi_{ t_{n-2}  , t_{n-1}}
%\Big({}
%F^{\FIN}({}\Xi_{t_{n-1}})
%\bigl({}\Phi_{t_{n-1} , t_n} F^{\FIN}({}\Xi_{t_n} {}) \bigl)
%\Big)
%\cdots
%\Big)({\omega_{t_0}})
%   \\
=
\Big(\bigtimes_{k=1}^n \bigl({}\Phi_{t_0 , t_k} F^{\FIN}({}\Xi_{t_k} {}) \bigl)
\Big)({\omega_{t_0}})
%\text{ ({}because each $\Phi_{t_{k-1} , t_k }$ is {}) }
\\
=
&
\Big(\bigtimes_{k=1}^n F^{\FIN} ({}\phi_{t_0, t_k}^{-1}({}\Xi_{t_k} {}) \bigl)
\Big)({\omega_{t_0}})
=
\bigtimes_{k=1}^n \chi_{{}_{\Xi_{t_k}}}({\phi_{t_0, t_k}
(\omega_{t_0})})
=1
%   = &
%\int_{\Omega}
%\Big(\bigtimes_{k=0}^n
%\chi_{ \phi_{t_0 , t_k }^{-1}({}\Xi_{t_k})}
%(\omega)
%\Big)
%{\omega_{t_0}}(\omega)d\omega.
%P%\tag{11.14}
\end{align*}
}
\par
\noindent
\rm
Thus, from the arbitrarity of
$\Xi_t$,
we surely believe that
\begin{itemize}
\item[(c)]
$
(x_t )_{t\in T }
= \phi_{t_0 , t} ({}\omega_{t_0}{})
\qquad
({}\forall t \in
T{})
$
\end{itemize}
\qed

\par
\noindent
%BBBBBBBBBBBBBBBBBB%SBSBSBS
{\small%%{\footnotesize
\vspace{0.1cm}
\begin{itemize}
\item[$\spadesuit$] \bf {{}}{Note }11.5{{}} \rm
Note that
"(b)
$\Leftrightarrow$(c)"
in the above.
That is,
(b) is the definition of (c).
\end{itemize}
}

\par
\vskip0.2cm
\par
The following is a consequence of
\textcolor{black}{{{Theorem }}10.12}
and
\textcolor{black}{{{Theorem }}11.9}.
%
%\par
%\noindent
%\vskip0.3cm
\vskip0.3cm
%BFBF
\par
\noindent
{\bf Corollary 11.10}$\;\;$%POPOPO
[Quantity and exact observable].
{}\rm{}
For each
$t \in T(t_0)$,
consider the exact observable
${\mathsf O}^{\FIN}_t=(X, {\cal F}_t, F^{\FIN})
(=(\Omega_t , {\cal B}_t , \chi ))$
in
$L^\infty(\Omega_t, \nu_t)$
and
a quantity
$g_t:\Omega_t \to {\mathbb R}$
on
$\Omega_t$.
Let
${\mathsf O}'_t = ({\mathbb R}, {\cal B}_{\mathbb R}, G_t)$
be
the observable representation of
the quantity
$g_t$
in
$L^\infty (\Omega_t)$.
Assuming the simultaneous observable
${\mathsf O}^{\FIN}_t \times{\mathsf O}'_t$,
define the
sequential observable
$[{}{\mathbb O}_{T(t_0)}]$
${{=}}$
$[{}\{
{\mathsf O}^{\FIN}_t \times{\mathsf O}'_t
 \}_{ t \in T} ,
\{  \Phi_{t_1,t_2}{}: $
${L^\infty (\Omega_{t_2}, \nu_{t_2})} \to {L^\infty (\Omega_{t_1}, \nu_{t_1})} \}_{(t_1,t_2) \in T^2_{\leqq}}$
%${C (\Omega_{t_2})} \to {C (\Omega_{t_1})} \}_{(t_1,t_2) \in T^2_{\leqq}}$
$]$.
Let
$\phi_{t_1,t_2 }:\Omega_{t_1} \to \Omega_{t_2}$
$(\forall
{(t_1,t_2) \in T^2_{\leqq}}
)$
be the deterministic causal map.
Let
${\widehat{\mathsf O}}_{{{t_0}}}$
${{=}}$
$\bigl(\mathop{\mbox{\Large $\times$}}_{{t} \in
{T}}(X_{{t}}\times {\mathbb R}), $
$\mathop{\mbox{\Large $\times$}}_{{t}\in{{T}}}
({\cal
F}_{{t}}\times {\cal B}_{\mathbb R}),$
${\widehat F}_{{t_0}}
\bigr)$
be the realized causal observable.
Thus,
we have the measurement
${\overline{\mathsf M}}_{L^\infty(\Omega_{t_0})}  ({}\widehat{\mathsf O}_{t_0{}},$
$S_{[\omega_{t_0}]}{})$.
Let
$(x_t , y_t)_{t\in T }$
be the measured value
obtained by the
measurement
${\overline{\mathsf M}}_{L^\infty(\Omega_{t_0})}  ({}\widehat{\mathsf O}_{t_0{}},$
$S_{[\omega_{t_0}]}{})$.
Then, we can surely believe that
\begin{align*}
x_t = \phi_{{t_0},t } (\omega_{t_0} )
\;\;
\text{ and }
\;\;
y_t = g_t(\phi_{{t_0},t } (\omega_{t_0} ))
\qquad
(\forall t \in T )
%P%\tag{11.15}
\end{align*}
\rm
%\END{Cor}
%BFBF
\par
%\no

\vskip0.5cm
\noindent
\par
\noindent

%

%%BBBBBBBBBBBBBBBBBB%SBSBSBS{\mathsf O}
\par
\noindent
{\small%%{\footnotesize
\begin{itemize}
\item[$\spadesuit$] \bf {{}}{Note }11.6{{}} \rm
As mentioned in \textcolor{black}{{Note }1.7},
"linguistic monism"
is not yet established,
or it may not exist.
If it exists,
it may have the following form:
\footnotesize
\begin{itemize}
\item[$(\sharp)$]
$
\overset{\text{\scriptsize }}{\text{{}
$\fbox{
monistic {linguistic world-description method}
}$}}
:=
\!\!
{
\overset{\text{\scriptsize }}
{
\underset{\text{\scriptsize
[{Sec.1.2.2}]}}{\text{{} $\fbox{monistic description}$}}}
}
+
{
%\overset{\text{\scriptsize [11.10$\{g_t(\phi_{{t_0},t } (\omega_{t_0} ))\}_{t\in T}$]}}
\underset{\text{\scriptsize [Corollary 11.10]}}
{
%\underset{\text{\scriptsize [{{the Schr\"odinger{ picture}}}]}}
{\text{{}
$\fbox{
causality
$
:\{
\phi_{t_1,t_2}
\}_{t_1 { \leqq } t_2 }
$
}$
}}
}
}
$
%\TAG*{$\displaystyle{\mathop{1)}_{(=10))}}$}
%%%%%%%%%%%%%CLAsSICAL
\end{itemize}
\small
However,
it may be regarded as the abbreviation of
measurement theory.
That is,
the ($\sharp$)
is
absorbed into measurement theory.
%{linguistic method}(={{measurement theory}})
% and .
%(\textcolor{black}{{Chap.$\;$1}(F$'_3$)}).
\end{itemize}
}
%%BBBBBBBBBBBBBBBBBB%SBSBSBSS
\par
\noindent

%\normalsize

Before reading Answer 11.11
(
Zeno's paradox({flying arrow})
),
confirm
Standing point
3.5
in Chap. 3.
That is,
%\BEGIN{itemize}
%\item[$(\sharp_1)$]
%{{measurement theory}}
%,
%{{measurement theory}}
%
%\END{itemize}
%{} and 
\begin{itemize}
\item[(d)]
The theory
described in ordinary language
should be described in measurement theory.
That is because
almost ambiguous problems
are due
to
the lack of
"world-view".

the fact
that
\end{itemize}

\baselineskip=18pt

\par
\noindent
\vskip0.3cm
%BFBF
\par
\noindent
{\bf Answer11.11
[Answer to {{Problem }}11.1: Zeno's paradox({flying arrow}) ]}$\;\;$%POPOPO
%%index{ and @Newtonian mechanics}
Let us answer to {{Problem }}11.1({flying arrow})(cf. \textcolor{black}{\cite{IZeno}}).
As mentioned in \textcolor{black}{{Sec.11.1}({}C$_6$)},
There is only the method of describing by measurement theory.
Therefore,
in Corollary 11.10, putting
\begin{align*}
q(t)=g_t(\phi_{{t_0},t } (\omega_{t_0} ))
\end{align*}
we get
the time-position function
$q(t)$.
Thus,
it suffices to discuss
\textcolor{black}{{{Problem }}11.1(B$_2^2$)}
by the time-position function.
%.
\begin{itemize}
\item[(e)]
If Zeno asks
"Why do you use measurement theory?",
it suffices to
answer
"We have only measurement theory".
%\BEGIN{itemize}
%\item[$(\sharp )$]
%{{measurement theory}} and .
%,
%.
%.
%\END{itemize}
% and .
\end{itemize}
%,
%\textcolor{black}{{{Problem }}11.1}
%\textcolor{black}{({}B$_3$)
%$(\sharp)$} and ,
%{{measurement theory}},
%paradox
% and .
%,
%{{measurement theory}} and {linguistic world-description method}.
% and ,
%\BEGIN{itemize}
%\item[]
%{flying arrow} and 
%(\textcolor{black}{{Sec.8.1}(m)}),
%
%\END{itemize}
%%,
%%,
%% and .
%%%,
%, scientific language
%
%
% and (D)
%
%
%.
%Answer and .
\qed
\par
\noindent
%BBBBBBBBBBBBBBBBBB%SBSBSBS
{\small%%{\footnotesize
\vspace{0.1cm}
\begin{itemize}
\item[$\spadesuit$] \bf {{}}{Note }11.7{{}} \rm
%\item[] \bf  \rm %%%BBBBBBBBBBBBBBBBBBBB
%\\
%paradox,
%2500,
%,
%,
%,
%\BEGIN{itemize}
%\item[]
%\textcolor{black}{Answer11.11}, , ?$\;\;$?$\;\;$
%{}?
%\END{itemize}
% and .
%,
%\textcolor{black}{Answer11.11}(paradoxAnswer),
%{{measurement theory}} and 
%.
%%\footnote{
%,
%.
%{}paradox,
%(G$_1$),
% and .
%}.
% and  and 
%.
%,
%%.
%, Answer11.11
% and .
%%Answer11.11,
%%,
%,
Thus,
we add
"Zeno's paradoxes"
to
\textcolor{black}{{Sec.9.3}(a)},
as follows.
\begin{itemize}
\item[$\underset{\text{({Sec.9.3})}}{\text (a)}$]
$
%{{{measurement theory}}}
\underset{
%({linguistic scientific language})
}{\text{{{measurement theory}}}}
\text{}
\!\!
\cases
\underset{
%({linguistic scientific language})
}{\text{quantum}}
\text{}
&
\!\!
\xrightarrow[\text{\scriptsize
}]{\text{\scriptsize
quantum phenomena}}
\text{(quantum engineering)}
\\
\\
\underset{
%({linguistic scientific language})
}{\text{classical}}
&
\!\!
\xrightarrow[]{\text{\scriptsize
ordinary phenomena}}
\text{(statistical mechanics, economics, }
\\
&
 \hspace{2.5cm}
$\quad$
\text{\bf {flying arrow}}\cdots)
%\cases
\endcases
%\TAG{0.4}
$
\end{itemize}
%{flying arrow} and 
%
%2500{{unsolved problem}} and 
%.
%%,
%% and 
%%{}
% and 
%(e)$(\sharp )${{measurement theory}} and  and .
%%\BEGIN{itemize}
%\item[$(\sharp)$]
%, ?
%\BEGIN{itemize}
%\item[$(a)$]
%paradox, 
% and 
%\item[$(b)$]
%\textcolor{black}{Answer11.11}
%\item[$(c)$]
%\textcolor{black}{Answer11.11}Answer
%\END{itemize}
%{}
%\END{itemize}
%,
%%{{measurement theory}} and .
%\BEGIN{itemize}
%\item[$(\sharp_2)$]
%{{measurement theory}}(That is,
%linguistic world-view{FIG.$\;$}8.2(={FIG.$\;$}11.1)),
%2500?
%$\;\;$
%That is,
%{{measurement theory}},
%scientific language
%?
%% and ?
%\END{itemize}
% and  and .
%%,
%(b)(c)
%
%((c), (b), 
% and ),
%(a),
% and 
%.
%%,
%%$(\sharp_1)$--$(\sharp_3)$
%%,
%.
%{statistics}${{\cdot}}${dynamical system theory}
%,
%,
%{}
\end{itemize}
}
%{}3{}, .
% and 
%%BBBBBBBBBBBBBBBBB
%\BEGIN{itemize}
%\ITEM[(B)]
%$
%\c

\par
\noindent
%BBBBBBBBBBBBBBBBBB%SBSBSBS
{\small%%{\footnotesize
\begin{itemize}
\item[$\spadesuit$] \bf {{}}{Note }11.8{{}} \rm
%\item[] \bf  \rm %%%BBBBBBBBBBBBBBBBBBBB
%\\
In quantum {{measurement theory}},
the time-position function does not exist
(that is,
the trajectory of a particle is meaningless.
For the further argument,
see
\cite{Iint2}.
\end{itemize}
}

\normalsize

\normalsize
\baselineskip=18pt
\par
\noindent
\vskip0.3cm
%\vskip0.3cm
%BFBF
\par
\noindent
{\bf Example 11.12
[Newtonian mechanics in measurement theory]}$\;\;$%POPOPO
%%final%index{ and @Newtonian mechanics}
Let
$T={\mathbb R}$
be the time axis.

Newtonian equation
\textcolor{black}{(9.1)}
on
the
state space $\Omega$
determines
the
{continuous map }
$\phi_{t_1,t_2}:
\Omega_{t_1}(=\Omega)
\to
\Omega_{t_2}(=\Omega)
$
$(t_1 {{\; \leqq \;}}t_2)$.
{{The formula }}\textcolor{black}{(9.2)}
says that
there exists
the measure $\nu(=\nu_t)$
on
the
state space $\Omega(=\Omega_{t})$
such that
\begin{align*}
\nu(D_{t_1})=
\nu( \phi_{t_1,t_2}^{-1}( D_{t_1}))
\qquad
(\forall D_{t_1} \in {\cal B}_{\Omega_{t_1}})
\tag{11.3}
\end{align*}
Therefore,
by \textcolor{black}{{{Theorem }}11.5},
%{continuous map }
%$\phi_{t_1,t_2}:
%\Omega_{t_1}(=\Omega)
%\to
%\Omega_{t_2}(=\Omega)
%$
%$(t_1 {{\; \leqq \;}}t_2)$
%,
we get
a sequential
{{deterministic causal map}}
$\{
\phi_{t_1,t_2}:
\Omega_{t_1}(=\Omega)
\to
\Omega_{t_2}(=\Omega)
\}_{t_1 {{\; \leqq \;}}t_2}
$.
%$(POI)$
Thus, in Newtonian mechanics,
\textcolor{black}{{{Theorem }}11.9}
says that
\begin{itemize}
\item[]
the exact measured value sequence
=
{{state}} change due to {{the Schr\"odinger{ picture}}}
\end{itemize}
Thus we say,
from the measurement theoretical point of view,
that
{{Newtonian}} equation
does not represent the motion
but
the time series of exact measured values.
In this sense,
Newtonian mechanics
can be regarded as one of
various sciences
and not physics.

\par
\noindent
%BBBBBBBBBBBBBBBBBB%SBSBSBS
{\small%%{\footnotesize
\vspace{0.1cm}
\begin{itemize}
\item[$\spadesuit$] \bf {{}}{Note }11.9{{}} \rm
%(i):
%{{Newton}}
%?
%%\\
%%(ii):
%%, ?
%\item[] \bf  \rm %%%BBBBBBBBBBBBBBBBBBBB
%\\
In the above argument,
Newtonian equation has two aspects
as
%,
%{{{measurement theory}}}, {{Newton}},
%measured value 
%{}
%%, , 
%%(\textcolor{black}{{{{}}}{Sec.6.1.2}})
%% and .
%,
%Newtonian mechanics and {physics}
%, ,
%{{Newton}},
%.
%% and ,
%% and
%%.
That is,
\begin{align*}
%Newtonian mechanics
\text{motion}
\cases
\textcircled{\scriptsize 1}:
\text{
{{Newtonian equation}} in physics
}&{\cdots}
\text{{{state}} change}
\\
\\
\textcircled{\scriptsize 2}:
\text{
{{Newtonian equation}}
in measurement theory}&{\cdots}
\text{
exact measured value sequence
}
\endcases
\end{align*}

%
%{}\textcolor{black}{{Note }1.5},
%Newtonian mechanics{physics} and ,
%%,
%, {{measurement theory}} and .
%Newtonian mechanics{the theory of relativity}
% and ,
%\textcolor{black}{{Chap.{\;}}9{}}equilibrium statistical
%mechanics and ,
%Newtonian mechanics, 2(\textcircled{\scriptsize 1} and
%\textcircled{\scriptsize 2}).
%\\
%(ii):
%$\Omega$
%$=$
%%({}or state{})
%${\mathbb{R}}^{6N}$
%
%$({}q , p{})$
%$\bigl({{=}}
%(\text{{\LL}{\RR},{\LL}{\RR}})=
%$
%$({} q_{1n}, q_{2n}, q_{3n} ,$
%$  p_{1n} , p_{2n} ,$
%$ p_{3n}{})_{n=1}^N $
%$\bigl)$
%,
%-observable
%$Q: {\mathbb{R}}^{6N} \to {\mathbb{R}}^{3N}$
%
%\BEGIN{align*}
%Q(
%({} q_{1n}, q_{2n}, q_{3n} , p_{1n} , p_{2n} ,p_{3n}{})_{n=1}^N
%)
%=
%({} q_{1n}, q_{2n}, q_{3n} )_{n=1}^N
%)
%\BEGIN{align*}
% and ,
%%index{@(POI)}
% and ,
%\BEGIN{align*}
%
%=
%-system quantity(POI)measured value 
%\BEGIN{align*}
% and .
\end{itemize}
}
%{}3{}, .
% and 
%%BBBBBBBBBBBBBBBBB

\par
\noindent
%It is  to note that
%\END{align*}
\par
\noindent
\vskip0.3cm
%\vskip0.3cm
%BFBF
\par
\noindent
{\bf Example 11.13
[$1+1=2 \; $?
]}
%\footnote{
%,{\;}
%%\newblock {9--23}
%{Ishikawa, S., Kikuchi, K., Nakamura, M.}
%\newblock {\em 
%Elementary school mathematics in quantitative language, }
%{\rm Far east journal of mathematical education }
%{Vol.2(2)}
%{165--180}
%{(2009)}
%.{}
%%QWERT: IElem
%}.
$\;\;$%POPOPO
%\par
%\noindent
It is a famous anecdote that; when Thomas Edison,
the greatest-ever inventor, was a school child, he made the question:
``{\it Why does $1+1=2$ hold?}'' on his teacher, and
made the teacher embarrassed.
Although we do not know his real intention,
we consider, from the measurement theoretical point of view,
that this question is not so trivial.
Consider the following
\textcircled{\scriptsize 1}
--
\textcircled{\scriptsize 3}:
\begin{itemize}
\item[\textcircled{\scriptsize 1}:]
What is $1$ plus $1$?
Is $1+1=2$
true?
\item[\textcircled{\scriptsize 2}:]
Assume that a particle $A$ with the mass
$1 kg$
ans
a
particle $B$ with the mass
$1 kg$
are are combined.
Then,
how weight the combined particle
(i.e.,
$A$+$B$)
?
%{dynamical system theory}, .
\item[\textcircled{\scriptsize 3}:]
Assume that
the exact measured value of the mass of a particle
$A$
is $1 kg$
and
the exact measured value of the mass of a particle
$B$
is $1 kg$.
Then,
the exact measured value of the mass of
the combined particle
(i.e.,
$A$+$B$)
is equal to
$2 kg$.
Is it true?
\end{itemize}
{}

\par
\noindent
{\bf Answer}:
From the mathematical point of view,
the equality in \textcircled{\scriptsize 1}
is merely a mathematical rule.
In physics, the
\textcircled{\scriptsize 2}
is just the law of conservation of mass.
Thus
we focus on the
measurement theoretical aspect
\textcircled{\scriptsize 3}, in which the equality is not trivial.
The proof is as follows.

\par
\noindent
%[
%, . ,
%.
%,
%measured value .
%\END{itemize}
%}
%%SS
\par
\noindent
[
The proof of \textcircled{\scriptsize 3}]:
Consider the Lebesgue measure space
$( \Omega_1, {\Cal B}_{\Omega_1},$
$
 \nu_1)$
$=$
$(\mathbb{R}, {\cal B}_{\mathbb R}, m )$
and its product measure space
$( \Omega_0 , {\cal B}_{\Omega_0}, \nu_0)$
$=$
$(\mathbb{R}^2, {\cal B}_{{\mathbb{R}}^2}, m^2 )$.
Let
$\Omega_0$
and
$\Omega_1$
be state spaces.
Put
\begin{align*}
&
(\text{the mass of a particle $A$},
\text{
the mass of a particle $A$
})
\in \Omega_0
({{=}} \mathbb{R} \times \mathbb{R})
\end{align*}
(for simplicity,
assume that
negative mass is possible).
By the law of conservation of mass,
define
the {continuous map }$\phi_{0,1}: \Omega_0  \to \Omega_1$
by
$\phi_{0,1}(\alpha, \beta) = \alpha+\beta$
$\;$
$( \forall (\alpha, \beta) \in {\mathbb{R}} \times{\mathbb{R}} )$.
The following is clear:
\begin{align*}
D_1 \in
{\cal B}_{\Omega_1} (={\cal B}_{\mathbb R}),
\;\;
m(D_1)=0
\;\;
\Longrightarrow
m^2 ( \phi_{0,1}^{-1} ( D_1 ) )=0
\end{align*}
Thus, {{Theorem }}11.5
says that
the {continuous map }$\phi_{0,1}: \Omega_0  \to \Omega_1$
is
a {{deterministic causal map}}.
And thus,
the
{{deterministic causal operator}}
$\Phi_{0,1}: L^\infty( \Omega_1 , \nu_1) \to L^\infty( \Omega_0 ,\nu_0)$
is defined by
\begin{align*}
[\Phi_{0,1}(f_1)](\omega_0 )
=
f_1( \phi_{0,1}(\omega_0 ) )
\quad
(\forall f_1 \in L^\infty ( \Omega_1 , \nu_1),
\;\;
\text{a.e. $\omega_0$})
\end{align*}
Consider the
{exact observable}
${\mathsf{O}}^{\FIN}_0$
in
$ L^\infty( \Omega_0 , \nu_0)$
and
the {exact observable}
${\mathsf{O}}^{\FIN}_1$
in
$ L^\infty( \Omega_1 , \nu_1)$.
And consider the {{realized causal}} observable
$\widetilde{\mathsf{O}}_0$
$=$
${\mathsf{O}}^{\FIN}_0 \times \Phi_{0,1}{\mathsf{O}}^{\FIN}_1$.
Taking the
{{measurement}}
${{\mathsf{M}}}_{L^\infty(\Omega_0, \nu_0)}(
{\mathsf{O}}^{\FIN}_1 \times \Phi_{0,1}{\mathsf{O}}^{\FIN}_1 , S_{[
(\alpha, \beta )]})$,
we obtain
a
measured value $((\alpha_0,\beta_0),\gamma_0)$.
Then,
by
\textcolor{black}{{{Theorem }}11.9},
we see, with
probability $1$,
$$
(\alpha_0,\beta_0)
=
(\alpha,\beta)
\text{ and }
\gamma_0 = \alpha+\beta
$$
Therefore,
we see that
$
\alpha_0
+
\beta_0
=
\gamma_0
$.
\qed
\par
%\noindent
%%BBBBBBBBBBBBBBBBBB%SBSBSBS
%{\small%%{\footnotesize
%\BEGIN{itemize}
%\item[$\spadesuit$] \bf {{}}{Note }11.10{{}} \rm
%% and , \textcolor{black}{Example 11.13} and .
%\textcolor{black}{Example 11.13} and 
%(, ), 
%$(\sharp_1)$
%(=
%\textcolor{black}{3.1($\sharp_4$)}
%).
%%\textcolor{black}{({\rm cf.}
%%\cite{IElem})}.
%\BEGIN{itemize}
%\item[$(\sharp_1)$]
%$A$1.
%$B$1 and ,
%%.
%,
%$C$ and ,
%%.
%$C$,
%2 and .
%%That is,
%%1+1=2
%%(\textcolor{black}{{Sec.11.5}}){}
%\END{itemize}
%, quantum {{measurement theory}},
%
%({\rm cf.}
%\textcolor{black}{\cite{QYuka}})
%,
%$(\sharp_1)$.
% and ,  and ,
%,  and ,
% and  and  and .
%,
%
%(Answer11.11 and Example 11.13\textcircled{\scriptsize 3}),
%%\textcolor{black}{{Note }7.1$(\sharp)$} and .
%% and ,
%\BEGIN{itemize}
%\item[$(\sharp_2)$]
%, ?
% and
%, \textcircled{\scriptsize 3}$1+1=2$?
%
%\END{itemize}
% and {}
%\END{itemize}
%}

%===================================
%12 12 12 12 12 12 12 12
%\part{ and }
\vskip3.0cm
\newpage
\section{Realistic world-view and Linguistic world-view
\label{Chap12}
}%{Chap.{\;}}12{}
%\ssubsection{ and }%{Chap.{\;}}12{}
%%\vspace{-0.8cm}
%\baselineskip=18pt\par
\noindent
\begin{itemize}
\item[{}]
{
\small
\baselineskip=15pt
\small
\par%[Abstract].
\rm
$\;\;\;\;$
This chapter describes the conclusion of measurement theory in there comparison of
mathematics, physics, and various sciences(engineering).
%There may be some readers who read in order of
%{\small
%%\BEGIN{itemize}
%%\item[]
%[Preface]$\rightarrow$
%[Postscript]$\rightarrow$
%[Chapter.1]$\rightarrow$
%[Last Chapter]$\rightarrow$
%$\cdots$
%%\END{itemize}
%}
%{
%\small
%%Although not necessarily written supposing such a reader,
%this last chapter can be read as a result, read so.
%}
}
\end{itemize}

\par
\subsection{Mathematics, Physics, Various Sciences${{\cdot}}$Engineering}
\subsubsection{Language}
\par
If the ecology of various animals is observed, it will be clear
that
the base of language was due to
intimidation ${{\cdot}}$ solidarity ${{\cdot}}$ reproduction.
Language was one of the strongest arms for the survival and breeding.
Such a time have continued for millions of years.
Of course, the greatest incidents happened in "the history of language",
for example,
"a rhythm and a song", "logical structure",
 a "quantity concept", "grammar", "tense", a "character", etc.
However, it was too long years ago,
we cannot specify the contribution person's name.

\subsubsection{Mathematics - The language independent of the world.  }
\par
When
human beings began to have some confidence in "survival and breeding",
people who get interested in the "quantity concept" in ordinary language unusually have appeared.
Probably, it is good also considering the pioneer as Pythagoras
---
everything is a number.
This flow was inherited to Archimedes (BC287 -- BC 212 years), Euler (1707 -- 1783), and a gauss (1777 - 1855), and the "quantitative portion" in ordinary language was strengthened rapidly.
That is, the mathematical achievement was accumulated rapidly.
However, language makes survival and breeding the origin as mentioned above,
%,  and ,
\begin{itemize}
\item[(a)]
Ordinary language was not devised
in order to tell mathematics.
%{}
\end{itemize}
It will be Cantor (1845 year--1918 year) that noticed, if it considers from now on.
Therefore, Cantor made the special language - set theory (that is, "the mathematical origin is set") - for describing mathematics.
Of course, this was inherited to Hilbert (1862 -- 1943) or Godel (1906 -- 1978), and brought about the prosperity of modern mathematics.
However, it should be careful that 
the axiomatization of mathematics
clarifies that mathematics is independent of
our real world,
as symbolized by Hilbert's words "a point may be a chair 
 and a line may be a desk".
Though that is right, the custom
to consider that
statistics and dynamical system
is a part of ordinary language
continues to go out of use as a convenient and easy way.
\vskip2.0cm
\subsubsection{Physics - realistic science view (the world is before language) }
\par
If a time is traced back, although Demokritos (around BC370 - BC460 )
---
The origin of everything is an atom
---
may be famous, I will start with modern science, for example.
Of course, work of Galileo (1564 -- 1642) and Kepler (1571 year--1630 year) is admired.
However, language makes survival and breeding the origin,
\begin{itemize}
\item[(b)]
Ordinary language was not devised
in order to tell physical phenomena.
%{}
\end{itemize}
Newton (1642 -- 1727) found out it.
Therefore, Newton made the languages of special
---
Newtonian mechanics
--- 
for describing a classical mechanics phenomenon.
This was inherited to Maxwell (1831 -- 1879), Einstein (1879 -- 1955), etc., and brought about a great success of the realistic science view.
And imprinting, such as "realistic science view = science", has become common by the great success.
\vskip2.0cm
\subsubsection{Various sciences (engineering) - linguistic science view (language is before world.)
}
\par
When you understand various matters in the world, it is always a given occasion, and if thought individually, naturally you think that it is troublesome.
Therefore, the plan to consider is decided previously, and when various matters are faced, I would like to come to depend on the method of considering along with the plan.
That is, in order to have understood various matters in the world, when there was "form of common thinking", many philosophers must have believed and investigated.
The pioneer is Plato (BC 427 -- BC 347 )- Idealism -, and he advocated the prototype of dualism and idealism.
This was inherited to Descartes (1596 -- 1650) and Kant (1724 -- 1804), and the spirit of "language is before the world".
- Linguistic science view  - was further established very much by linguistic philosophy.
It is as having seen in \textcolor{black}{Chapter 8} that these formed the philosophical (= world description) main stream.

However, in respect of the technology of an understanding of various matters in the world, probably, 
we have to stress the importance of
"the method of the classical mechanics world view".
Although we can not say the name of the founders,
Fischer (1890 -- 1962),
%index{@Fisher}
Robert Wiener (1894 -- 1964),
%index{@}
Kolmogorov (1903 -- 1987 year 1),
%index{@}
 Kalman (1930 --),
%index{@}
etc. should be mentioned especially.
% and .
The two
(i.e.,
"the form of thinking(philosophy)" and "the classical mechanics world view (technology)")
have been independent.
From this unfortunate fact, various sciences were classified into the category of "the weak thing" in
\textcolor{black}{Section 2.4.1}.
The reason which has fallen into such a situation is reasonable for thinking that it originates in
% and ,
%
\begin{itemize}
\item[(c)]
Ordinary language was not devised
in order to tell usual scientific phenomena
\end{itemize}
Although I do not know whether von Neumann (1903 -- 1957) was conscious of this (c), as a result, von Neumann proposed the quantum measurement theory
--- at the meaning of Section 9.3, dualism idealism ---
 based on Copenhagen interpretation in "the mathematical basis of quantum mechanics \cite{Neum}"
And von Neumann's work was inherited and the languages of special make for describing the scientific phenomena of an everyday scale
- Measurement theory  - were proposed in this book.
\par
\vskip1.0cm
\par
%\END{document}
Two things unexpected as follows happened to Measurement theory.
\begin{itemize}
\item[(d$_1$)]
Suppose that the method of "considering along with the plan when the plan considered previously is decided and various matters are faced" as mentioned above was adopted.
Since various sciences are various,
it is a matter of course that
even if such a plan exists, it is an non-quantitative plan and the number of the plans increases considerably moreover.
However, the plan of Measurement theory was only two quantitative plans
(i.e., Axioms 1 and 2).
%%%\footnote{
%%%The reason of the question "why it ended with two" is that it did not extend a problem to philosophy of life, ethical philosophy, etc.
%%%However, I think that it (\textcolor{black}{Note 2.4}
%%described like) is the greatest discovery on histor
%%y of science supposing it sets one more to three in addition and can perform a still
%%more powerful language. }
\item[(d$_2$)]
The family line of Measurement theory was quantum mechanics.
And measurement theory has been saddled with "two kinds of absurd character"as stated repeatedly.
Namely,
% and .
%, =(8.1) and ,
\begin{itemize}
\item[$(\sharp)$]
"Absurd character" of Measurement Theory
\\
$\qquad$
$
\cases
\text{Idealism} &\; \cdots \; \text{Linguistic science view}
\\
\text{Dualism} &\; \cdots \; \text{Copenhagen interpretation (i.e.,
dualism)}
\endcases
$
\end{itemize}
%index{ and @}
That such measurement theory is materialized is a mystery which is hard to believe.
However, since this is a proposal of this book, the elucidation of this mystery must be left to readers as "homework to readers."
The author does not know the thing beyond having stated with \textcolor{black}{Section 9.3}.
\end{itemize}
% and ,  and .
%${{\cdot}}$,
%${{\cdot}}$${{\cdot}}$,
%,
%.
%, , 
%${{\cdot}}$${{\cdot}}$
% and 
% and .
%
%
%\END{document}
\vskip2.0cm
\subsubsection{Mathematics, Physics, Various sciences (engineering)}
\par
The above is summarized and the next is obtained.
(\textcolor{black}{Note 2.10, Note 9.6 }).
\begin{itemize}
\item[(e)]
$
\cases
\textcircled{\scriptsize 0}:
\text{Mathematical language}
% &
%\text{\hspace{-5.0cm}}
\cdots
\text{Set theory}
\\
\quad
\text{\footnotesize (The proposal and solution of mathematical outstanding problems )}
\\
\textcircled{\scriptsize 1}:
\text{languages of physics}
% &
%\text{\hspace{-5.0cm}}\\\; \cdots \; 
\cdots
\text{Newtonian mechanics, electromagnetism,...
% the theory of relativity, ...
}
\\
\quad
\text{\footnotesize (Realistic science view: Theory which explains what God made )}
\\
\textcircled{\scriptsize 2}:
\text{language of engineering}
% &
%\text{\hspace{-5.0cm}}\; \cdots \; 
\cdots
\text{Measurement theory}
\\
\quad
\text{\footnotesize (Linguistic science view: The language for making about the same robot as a scientist )}
\endcases
$
\end{itemize}
About \textcircled{\scriptsize 0} and \textcircled{\scriptsize 2}, following some will need to keep in mind that there is probably no complaint about physics\textcircled{\scriptsize 1}.
Even if it does not know set theory etc., mathematical outstanding problems may be solved, and even if it does not know Measurement theory (for example, set theory does not seem to have been indispensable in order to solve the four colors problem), "about the same robot as a scientist" may be made(\textcolor{black}{Note 2.10}).
Even if it says so,
\begin{itemize}
\item[]
If the foundation is established with languages of special make, "rapid development" should become possible and each language (set theory and measurement theory) of special make must be required also for \textcircled{\scriptsize 0} and \textcircled{\scriptsize 2}.
\end{itemize}

\par
\noindent

%}
%\END{itemize}
%
%13 13 13
\vskip2.0cm
\newpage
\section{Conclusions
\label{Chap13}
}% and 
\markboth{Conclusions}{}
%\vspace{-1.3cm}
\baselineskip=18pt
It is a matter of course, ordinary language is 
the greatest invention of mankind
However, science cannot be ripened only with ordinary language.
For that purpose, the world describing method is indispensable.
And,
it is always continuing evolving and developing toward the direction of "saving of thinking."
Supposing that is right, I claimed "evolution and development of 
world description" of 
the following figure (= \textcolor{black}{\textcolor{black}{Fig.}} 8.2).
\par
\vskip-0.2cm
\par
\noindent
%\begin{figure}[htbp]
\par
\noindent
{\small
\begin{picture}(500,9)
\put(0,0){
$\underset{\text{\scriptsize (what is world?)}}{\text{\fbox{\textcircled{\scriptsize 0}
ordinary language}}}
\xrightarrow[\text{\scriptsize Heracreitos}]{\text{\scriptsize Parmenides}}
\overset{{{{{}}}Sec. 6.1, {{{}}}Sec. 11.1.1}}{\underset{\text{
\scriptsize (Zeno's paradoxes)}}{
\text{\fbox{how to describe motion}}}}
\xrightarrow[]{}
\!\!\!\!
\overset{{{{}}}Sec. 6.1}{\underset{\text{
\scriptsize (Galilro, Bacon, DEcartes)}}{\text{\fbox{
{\bf causality}}}}}
\!\!\!\!
\xrightarrow[]{}
$}
%}}
%%%%%%%%%%%%%%%%%%
\end{picture}
}
\vskip0.8cm
%%---------------------------  and {FIG.$\;$}    ------------------
%%BBBBBBBBBBBBBBBBBB%SBSBSBS
%{\small%%{\footnotesize
%\vskip0.5cm
%\caption{${{\cdot}}${FIG.$\;$}}
%\END{figure}
%\par
\noindent
%\BEGIN{FIGUre}[htbp]
%%
\unitlength=0.25mm
%\unitlength=0.33mm
{\small
\begin{picture}(500,256)
\put(0,95){$
\xrightarrow[]{}
%\;\;
%\underset{({world-description})}{} %
\cases
\overset{{{{}}}}{\underset{}{\text{\fbox{Kant}}}}
\xrightarrow[{Sec.8.1}]{\qquad \quad \quad \qquad}
\overset{{{{}}}}{\underset{}{\text{\fbox{{linguistic philosohy}}}}}
\xrightarrow[{Sec.8.1}]{\qquad \quad \qquad \quad \;\;\;\;\;}
%\xrightarrow[  \quad \;\;\;\;\;\;\;]{}
\\
\\
\qquad
\qquad
\quad
%\qquad
\overset{\text{(classical mechanical world-view)}}{\text{
\fbox{{statistics}:({\bf trial})}
}}
\xleftarrow[\;\; \quad \text{\scriptsize abbreviation} \quad ]{}
\!\!\!\!
\overset{\text{\scriptsize (language)}}{\underset{\text{\scriptsize (idealism)}}{\text{\fbox{
{\bf measurement theory}}}}}
\\
\\
\\
\underset{}{\text{\fbox{Newton}}}
\xrightarrow[]{}
\!\!
\cases
\underset{\text{\scriptsize ({the causal world-view})}}{\text{\fbox{{state equation method}(Chap. 1(E$_1$))}}}
\\
\\
\underset{{Sec.3.1}}{\text{\fbox{quantum mechanics}}}
\xrightarrow[]{}
\cases
\underset{\text{\scriptsize (linguistic)}}{(1)}\xrightarrow[\;\;\; \; \;\;{Sec.3.2} \;\;\;\;\;\;]{\;\;
\;\;}
\\
\\
\\
%\\
\underset{\text{\scriptsize (realistic)}}{(2)}
\xrightarrow[\;\;\; \quad \quad]{}
\endcases
\\
\\
\underset{\text{\scriptsize (Einstein)}}{\text{\fbox{{the theory of relativity}}}}
\;\;
\xrightarrow[]{\qquad \quad \quad \quad \;\;\;\;\;\;\;\;\;\;\;}
\endcases
\endcases
$
}
%%%%%%%%%%
\put(420,69){\vector(0,1){72}}%%%%%%
\put(396,246){\vector(0,-1){25}}
\put(190,130){\dashline{3}(0,0)(0,35)}
\put(192,145){\footnotesize ({missing link})}
%%%%%%%%%%%%%%%%%%%%%%%%
%%%%%%%%%%%%%%%%%%%%%%%%%%%%%%
\put(45,-20)
{
{
\put(295,73)
{
$
{
\left.\begin{array}{ll}
{
\left.\begin{array}{ll}
\;\;
\\
\;\;
\\
\;\;
\\
\;\;
\\
\;\;
\\
\;\;
\end{array}\right.
}
\\
\\
\\
\\
{
\left.\begin{array}{ll}
\;\;
\\
\;\;
\\
\;\;
\quad
\end{array}\right\}
}
\xrightarrow[]{}
\!\!\!
\overset{\text{\scriptsize ({physics})}}{\underset{\text{\scriptsize (realism)}}{\text{\fbox{
{the theory of everything}}}}}
\end{array}\right.
}
$}
\put(325,-10)
%\put(315,-5)
{\dashline{3}(10,65)(0,65)(0,-35)(222,-35)(222,65)(200,65)
%{PHYSICS
\put(15,59){\footnotesize [{\bf {realistic world-view}}(unsolved)]}
}
%%%%%%%%%
%%%%%%%%%%%%%%%%%%%%%%%%%%
\put(325,150)
{\dashline{3}(35,88)(0,88)(0,10)(222,10)(222,88)(200,88)
%LINGUISTIC
\put(45,85){\footnotesize [{\bf linguistic world-view}]}
}
%%%%%%%%
}
}
\end{picture}
}
\vskip1.5cm
\begin{center}{Figure 13.1:
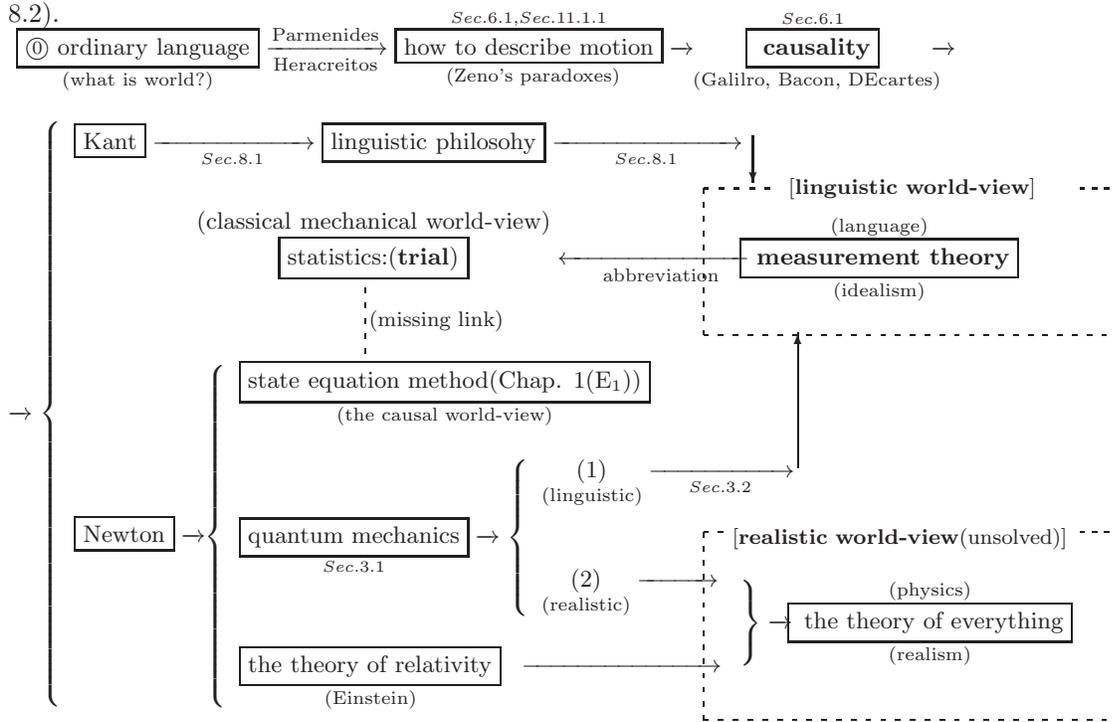
The  development of the {world-descriptions}(=\textcolor{black}{Fig.} 8.2)}
%\caption{{world-description}${{\cdot}}${FIG.$\;$}}
\end{center}
%}
%-----------------  and {\textcolor{black}{FIG.}$\;$}  ------------------------------

%
%\

%\vskip0.5cm
%\caption{${{\cdot}}${FIG.$\;$}}
%\END{figure}
%%}
%%-----------------  and {FIG.$\;$}  ------------------------------

\par
\noindent
If this is summarized (i.e., if the "beginning" and the "last" are written),
%\ssmall
we say that
\par
\noindent
$\;\;$
$\underset{\text{\scriptsize (Chap. 1)}}{\text{(X$_1$)}}$
$
\overset{
}{\underset{\text{(before science)}}{
\text{
\fbox
{
{\textcircled{\scriptsize 0}}
widely {ordinary language}}
}
}
}
$
\par
\noindent
$\qquad$
$\qquad$
$
\underset{\text{\scriptsize }}{\text{$\Longrightarrow$}}
$
$
\underset{\text{\scriptsize (Chap. 1(O))}}{\text{{world-description}}}
\cases
%\textcircled{\scriptsize 2}:
&
\!\!\!\!\!\!
%\underset{\scriptsize
%\text{}}
{\text{\textcircled{\scriptsize 2}{linguistic scientific language (measurement theory)}}}
\\
&{\text{\scriptsize (The language for making about the same robot as a scientist)}}
\\
\\
&
\!\!\!\!\!\!
%\textcircled{\scriptsize 1}:
%\underset{\scriptsize
%\text{}}
{\text{\textcircled{\scriptsize 1}realistic scientific language (the theory of everything)}}
\\
&{\text{\scriptsize
(Theory which explains what God made )
}}
\endcases
$

\par
\noindent

\vskip0.5cm

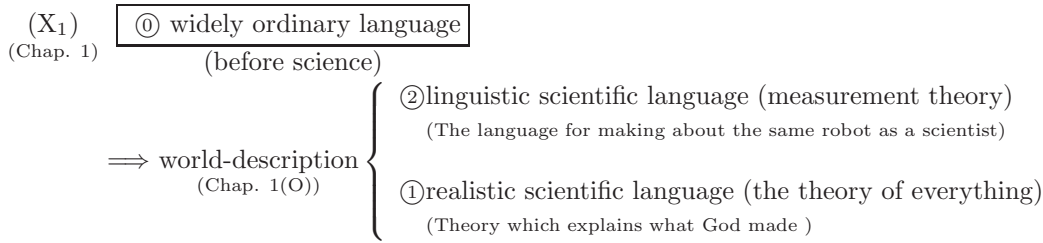
\normalsize
\baselineskip=18pt
\par
Although there was 3000 years of history in world description, a major event didnot necessarily break out frequently.
It has occurred only about at most 10 times, and,
as fa as
 about the linguistic describing method,
the major event (discovery of = mystic words) has occurred only 5 times.
%index{@}
Namely,
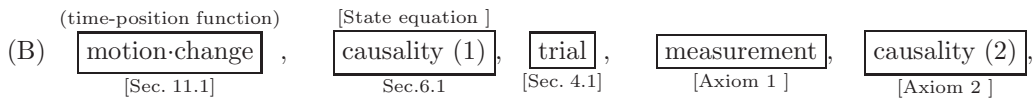
\begin{itemize}
\item[(B)]
$
\overset{\text{\scriptsize (time-position function)}}{
\underset{[{\text{\scriptsize Sec.$\;$11.1}}]}{\text{\fbox{{{motion$\cdot$change}}}}}},
%\underset{[11.11]}{\text{\fbo
\;\;
\quad
%\underset{}
\underset{\text{\scriptsize{Sec.6.1}}}
{
\overset{{
{\text{\scriptsize [State equation \textcolor{black}{
%({\text{Sec.$\;$}}\REF{6secAxiom2})
}]}
%{\text{\scriptsize [(1.1), Axiom2\textcolor{black}{({\text{Sec.$\;$}}\REF{6secAxiom2})}]}
}}}
{\text{\fbox{causality (1)}}}
},
\;\;
\underset{[{\text{\scriptsize Sec.$\;$4.1}}]}{\text{\fbox{{{trial}}}}},
\;\;
\quad
\underset{{
{\text{\scriptsize [Axiom 1 \textcolor{black}{
%({Sec.$\;$}\REF{2secAxiom1})
}]}
}}}
{\text{\fbox{measurement}}},
\quad
%\underset{}
\underset{{
{\text{\scriptsize [Axiom 2 \textcolor{black}{
%({\text{Sec.$\;$}}\REF{6secAxiom2})
}]}
%{\text{\scriptsize [(1.1), Axiom2\textcolor{black}{({\text{Sec.$\;$}}\REF{6secAxiom2})}]}
}}}
{\text{\fbox{causality (2)}}},
$
\end{itemize}
Even if
we have to add 
the gradually ripe process
of
dualistic idealism
(\textcolor{black}{{Sec.$\;$}8.1}), it can be said that it was quite peaceful history.
If \textcolor{black}{Fig.} 13.1 is followed, we would like to come to claim a scenario called
\begin{itemize}
\item[(C)]
$
\qquad
\qquad
\qquad
$
{
\bf
the happy end of a big tale}
\end{itemize}
%index{@}
in the sense that
both the linguistic world-view and the realistic world-view
are compatible.
% (materialism) "the face stood."
%. 
%({}A),
%,
%, \footnote[1]{
%, .
%}.

\vskip0.5cm
Of course, the above is only one scenario.
In order to make this scenario steadfast (or it overthrows), I would like to come to pursue the following problem:
% which it continued asking in this book whole volume.
\begin{itemize}
\item[(D)]
Why does the metaphysics of Measurement theory
hold?
%Or although it thinks that a leading linguistic science language 
%different from measurement {theory} is proposed rapidly, 
%it will be in a many children 100 house state, 
%and there cannot be no sole wins of Measurement theory, how is it?
\end{itemize}
%There is no hand that readers do not investigate (D) because 
%the direction which is called "limit of the limit = 
%engineering and science of measurement theory", 
%and is not repelled is carrying out in how as (m) of Section 8.1 described.
The answers were given to various miscellaneous mistily problems in this book.
For example,
\begin{itemize}
\item[({}E)]
$
\underset{(\textcolor{black}{{\text{\scriptsize Sec.$\;$2.3.3, Note 6.2}}})}{
\text{[What is sapce (or, time, causality, probability
?]}}
$,
$
\underset{\text{\scriptsize (Theorem 3.4)}}{\text{[Heisenberg's uncertaty relation]}}$,
\\
\\
$
\underset{\text{\scriptsize (Theorem 5.11)}}{\text{[sillogizm]}}$,
$\quad$
$
\underset{\text{\scriptsize (Theorem 6.21)}}{\text{[the priciple of equal weight]}}
$,
\\
\\
$
\underset{\text{\scriptsize ({Chap.{\;}}9{})}}{[\text{equilibrium statistical mechanics]}}
$,
$
\underset{\text{\scriptsize ({Chap.{\;}}11{})}}{\text{[1+1=2]}}
$,
$
\underset{\text{\scriptsize ({Chap.{\;}}11{})}}{\text{[Zeno's paradoxes]}}
$
\end{itemize}
were clarified.
\begin{itemize}
\item[(F)]
The answer
to
the question
"Why can the problems in (E) be solved?"
is clear.
That is because
these problems are the same problem,
i.e.,
the problem
"Propose the language for describing
these problems!".
\end{itemize}
Thus,
the problems in (E)
are easy exercises in measurement theory.
%, {}.
Though that was right, possibly readers had the following comment.
% and .
\begin{itemize}
\item[(G)]
The opinion of this book - establishment of a linguistic science view  - was understood once.
However, 
what the author did
is
that
a variety of miscellaneous mistily problems only come back to 
the one biggest mistily problem
such as
$$
\text{
Is it possible that metaphysics is introduced as a base of various sciences?
}
$$
\end{itemize}
However, it is the same as that of Newtonian mechanics, electromagnetism, or the theory of relativity,
that is, we believe that
\begin{itemize}
\item[(H)]
A scientific theory is returning various miscellaneous mysteries to one big mystery.
\end{itemize}
\par
Since mathematics and physics are the learning of God (namely, learning common to whole creation people) ,
the reason for the formation is substituted for a word of a "miracle", 
and the rest may be what to leave to the high alien.
However, measurement {theory} is man's learning (namely, learning depending on man's recognition and linguistic competence),
so man may reply to (D).
%Probably he may ask, and may want to already have come to investigate (D), 
%and the watch which an author learns may be shortly thought, 
%if it is a reader who finished reading this book(that is, it is if it 
%is the reader who understood that Measurement theory was not 
%"an ass in a lion's skin (i,e.,
%quantum mechanical skin)."
%).
\par
\vskip0.5cm
\par
Although it was above, possibly the author's interest ( namely, about (D)) was written too much.
%,
%\BEGIN{itemize}
%\item[(I)]
%(D) and Answer and ,
%,
%.
% and (2.4.2(e)) and
%
%\END{itemize}
% and {}
What was necessary was to have said only the following things, supposing that was right.
\begin{itemize}
\item[(I)]
Measurement theory was not made for "the happy end (C) of a big tale" (
or for the rehabilitation of the Descartes-Kant philosophy).
If an author's spirit is written honestly,
\begin{itemize}
\item[]
In the time of the engineering which will continue four hundreds years from now on,
in order to defeat the battle which risked survival of human beings(\textcolor{black}{Sec. 2.4.2(e) , (f)})
, measurement theory was made as
\bf the scripture of engineering and science.
\rm(Note 9.7)
\end{itemize}
I solved outstanding and ambiguous problems (E) since I wanted to have 
the courage to add this spirit to this final chapter.
\end{itemize}
If that is right, what should be performed now is the next.
\begin{itemize}
\item[(J)]
Under the belief (F), in a language called measurement theory, phenomena are described rapidly and engineering and science are developed rapidly.
\end{itemize}
Since measurement theory is a language,
\begin{align*}
\text{
Measurement theory
is valueless if not used.
}
\end{align*}

\par

\par
\noindent

%EEEEEEEEEEEEEEEEEEEEEEEEEEE

%\clearpage
%-
% The References -
%-
%\renewcommand{\bibname}{\LARGE }
%{0.3}
%\markboth{}{}
%%%%%%%%%%%%%%%
\bibliography{MYREF}
\bibliographystyle{jplain}

%\pagestyle{empty}
%\printindex
%\END
%%EEEEEEEEEEEEEEEEEEEEEEEEEEEEEEE
%KKKKKKKKKKKKKKKKKKKKKKKKKKKKK

\end{document}